%% file: 0000rev.tex
\documentstyle[epsf,psfig,12pt]{article}
\begin{document}
\include{00Title}       
\include{01Intro}
\include{02Hamil}
\include{04Conti}
\include{05Dilcq}

\include{06Nucle}

\include{07Exclu}

\include{08Vacuu}
\include{09Renor}

\include{10Chira}

\include{11Prosp}
\begin{appendix}
\include{12Appen}

\end{appendix}

\include{13Refer}
\end{document}

%% file: 00Title.tex
\title{Quantum Chromodynamics \\
          and Other Field Theories \\  
          on the Light Cone}

\author{ Stanley J. Brodsky, \\
Stanford Linear Accelerator Center \\ 
Stanford University, Stanford,California~~94309\\
\\ 
Hans-Christian Pauli \\ 
Max-Planck-Institut f\"ur Kernphysik \\ 
D-69029 Heidelberg \\ 
\\
Stephen S. Pinsky \\
Ohio State University \\
Columbus, Ohio 43210\\}

\date{28 April 1997}
\maketitle
\vfill 
\noindent  Preprint MPIH-V1-1997 

\newpage
\begin{abstract}
In recent years light-cone quantization of quantum field 
theory has emerged as a promising method for solving 
problems in the strong coupling regime.
The approach has a number of unique features 
that make it particularly appealing, most notably, 
the ground state of the free theory
is also a ground state of the full theory. 

We discuss the light-cone quantization of gauge theories 
from two perspectives: as a calculational tool for 
representing hadrons as QCD bound-states of 
relativistic quarks and gluons, and also as a
novel method for simulating  quantum field theory 
on a computer. 
The light-cone Fock state expansion of wavefunctions 
provides a precise definition of the parton model and 
a general calculus for hadronic matrix elements. 
We present several new applications of light-cone 
Fock methods, including calculations of exclusive weak 
decays of heavy hadrons, and intrinsic heavy-quark
contributions to structure functions. 
A general non-perturbative method for numerically 
solving quantum field theories, 
``discretized light-cone quantization", 
is outlined and applied to several gauge theories. 
This method is invariant under the large class of
light-cone Lorentz transformations, 
and it can be formulated such that ultraviolet regularization 
is independent of the momentum space discretization. 
Both the bound-state spectrum and the corresponding
relativistic light-cone wavefunctions can be obtained 
by matrix diagonalization and related techniques. 
We also discuss the construction of the light-cone Fock 
basis, the structure of the light-cone vacuum, 
and outline the renormalization techniques
required for solving gauge theories within the 
Hamiltonian formalism on the light cone. 
\end{abstract}

\newpage
\tableofcontents

%% file: 01Intro.tex
\section{Introduction}
\setcounter{equation}{0}
One of the outstanding central problems in particle physics is the 
determination of 
the structure of hadrons such as the proton and neutron in 
terms of their fundamental quark and gluon degrees of freedom.  
Over the past twenty years two fundamentally different pictures
of hadronic matter have developed. 
One, the constituent quark model (CQM) \cite{zwe64}, or the quark parton
model \cite{fey69,fey72}, is closely related to experimental observation. The
other,  quantum chromodynamics (QCD) is based on a
covariant non-abelian quantum field theory.
The front form of QCD \cite{gel64} appears to be  the only hope of reconciling 
these two. 
This elegant approach to quantum field theory is a Hamiltonian 
gauge-fixed formulation that avoids many of the most difficult
problems in the equal-time formulation of the theory.  
The idea of deriving a front form constituent quark model from
QCD actually dates from the early seventies, 
and there is a rich literature on the
subject \cite{buc70,dea73,efw73,bel74,des74,leu74a,leu74b,mel74,%
osb74,cah75,cat76,ida75a,ida75b,ida75c}.  
The main thrust of this review will be to discuss the
complexities that are unique to this formulation of 
QCD, and other quantum field theories, in varying degrees of detail. 
The goal is to present a self-consistent framework
rather than trying to cover the subject exhaustively. 
We will attempt to present sufficient background material 
to allow the reader to see some of the advantages and
complexities of light-front field theory.  
We will, however, not undertake to review all of
the successes or applications of this approach. 
Along the way we clarify some obscure or
little-known aspects, and offer some recent results.

The light-cone wavefunctions encode the hadronic properties
in terms of their quark and gluon degrees of freedom, 
and thus all hadronic properties can be derived from them.  
In the CQM, hadrons are relativistic bound states of a few
confined quark and gluon quanta. 
The momentum distributions of quarks making up the
nucleons in the CQM are well-determined  experimentally 
from deep inelastic lepton scattering measurements, 
but there has been relatively little progress in computing the
basic wavefunctions of hadrons from first principles. 
The bound state structure of hadrons plays a critical role 
in virtually every area of particle physics phenomenology.  
For example, in the case of the nucleon form factors and open
charm photo production the cross sections depend not only 
on the nature of the quark currents, but also on the coupling 
of the quarks to the initial and final hadronic states.  
Exclusive decay processes will be studied intensively at
$B$-meson factories. They depend not only on the
underlying weak transitions between the quark flavors, 
but also the wavefunctions which describe how $B$-mesons 
and  light hadrons are assembled in terms of their quark 
and gluon constituents.  
Unlike the leading twist structure functions measured in deep 
inelastic scattering, such exclusive channels are sensitive to 
the structure of the hadrons at the amplitude level and to the 
coherence between the contributions of the various 
quark currents and multi-parton amplitudes. 
In electro-weak theory, the central unknown required for
reliable calculations of weak decay amplitudes are the 
hadronic matrix elements. 
The coefficient functions in the operator product expansion 
needed to compute many types of experimental quantities are 
essentially unknown and can only be estimated at this point. 
The calculation of form factors and exclusive scattering processes, 
in general, depend in detail on the basic amplitude structure 
of the scattering hadrons in a general Lorentz frame. 
Even the calculation of the magnetic moment of a proton
requires wavefunctions  in a boosted frame. 
One thus needs a practical computational method for QCD 
which not only determines its spectrum, but which can
provide also the non-perturbative hadronic matrix elements 
needed for general calculations in hadron physics.

An intuitive approach  for solving relativistic bound-state 
problems would be to solve the gauge-fixed Hamiltonian
eigenvalue problem. The natural gauge for light-cone Hamiltonian 
theories is the light-cone gauge $A^+=0$.  In this physical
gauge the gluons have only two physical transverse degrees 
of freedom.  One imagines that there is an expansion in 
multi-particle occupation number Fock states. 
The solution of this problem is clearly a formidable task, 
and if successful, would allow one  to calculate the
structure of hadrons in terms of their fundamental degrees 
of freedom.
But even in the case of  the simpler abelian quantum theory
of electrodynamics very little is known about the nature of the 
bound state solutions in the strong-coupling domain. 
In the non-abelian quantum theory of chromodynamics 
a calculation of bound-state structure has to deal with many 
difficult aspects of the theory simultaneously:
confinement, vacuum structure, spontaneous  breaking of 
chiral symmetry (for massless quarks), and  describing a
relativistic many-body system with unbounded particle number. 
The analytic problem of describing QCD bound states is 
compounded not only by the physics of confinement, 
but also by the fact that the wave function of a composite of
relativistic constituents has to describe systems of an arbitrary 
number of quanta with arbitrary momenta and helicities.  
The conventional Fock state expansion based on equal-time 
quantization becomes quickly intractable because of the 
complexity of the vacuum in a relativistic quantum field theory. 
Furthermore, boosting such a wavefunctions from the hadron's 
rest frame to a moving frame is as complex a problem as 
solving the bound state problem itself. 
In modern textbooks on quantum field
theory \cite{itz85,mut87} 
one therefore hardly finds any trace of a Hamiltonian. 
This reflects the contemporary conviction that the concept of
a Hamiltonian is  old-fashioned  and littered with all  kinds 
of almost intractable difficulties.  
The presence of the square root operator in the equal-time 
Hamiltonian approach presents severe mathematical difficulties. 
Even if these problems could be solved, the eigensolution is 
only determined in its rest system as note above. 

Actually the action and the Hamiltonian principle in some sense are 
complementary, and both have their own virtues.  
In solvable models they can be translated into each other. 
In the absence of such, it depends on the kind of problem  
one is interested in: The action method is particularly suited 
for calculating cross sections, while the Hamiltonian method is more 
suited for calculating bound states. 
Considering composite systems,  systems of many constituent 
particles subject to their own  interactions, the Hamiltonian
approach seems to be  indispensable in describing the 
connections  between the constituent quark model, 
deep inelastic scattering, exclusive process,  {\it etc}. 
In  the CQM, 
one always describes mesons as made of a quark and an 
anti-quark, and baryons as made of three quarks 
(or three anti-quarks). These constituents are bound by 
some phenomenological potential which is tuned to account 
for  the hadron's properties such as masses, decay rates 
or magnetic moments. 
The CQM does not display any visible 
manifestation of spontaneous chiral symmetry breaking; 
actually, it totally prohibits such a symmetry since the constituent 
masses are large on a hadronic scale, typically of the order of 
one-half of a meson mass or one-third of a baryon mass.  
Standard values are 330 MeV for the up- and
down-quark, and 490 MeV for the strange-quark, 
very far from the 'current' masses of a few (tens) MeV. 
Even the ratio of the up- or down-quark masses to the
strange-quark mass is vastly different in the two pictures. 
If one attempted to incorporate a bound gluon into the model, 
one would have to assign to it a mass at least of the order of 
magnitude of the quark mass, in order to limit its impact 
on the classification scheme.
But a gluon mass violates the gauge-invariance of QCD. 

Fortunately ``light-cone quantization'', which can be
formulated independent  of the Lorentz frame, offers an 
elegant avenue of escape. 
The square root operator does not appear, 
and the vacuum structure is relatively simple. 
There is no spontaneous  creation of massive 
fermions in the light-cone quantized vacuum. 
There are, in fact, many reasons to quantize relativistic field
theories at fixed light-cone time. 
Dirac \cite{dir49}, in 1949, showed that in this so called 
``front form''  of Hamiltonian dynamics a maximum number 
of Poincare\'e generators become independent of the 
interaction , including certain Lorentz boosts. 
In fact, unlike the traditional equal-time Hamiltonian 
formalism, quantization on a plane tangential to the 
light-cone ( null plane)  can be formulated without 
reference to a specific Lorentz frame. 
One can construct an operator whose eigenvalues are the  invariant 
mass squared `` $M^2$.
The eigenvectors describe bound states of arbitrary 
four-momentum and invariant mass $M$ and allow  the
computation of scattering amplitudes and other dynamical 
quantities. The most remarkable feature of this approach, 
however, is the apparent simplicity of the light-cone vacuum. 
In many theories the vacuum state of the free Hamiltonian 
is also an eigenstate of the total light-cone Hamiltonian. 
The Fock expansion constructed on this vacuum state  
provides a complete relativistic many-particle basis for 
diagonalizing the full theory. 
The simplicity of the light-cone Fock representation as
compared to that in equal-time quantization is directly linked
to the fact that the physical vacuum state has a much 
simpler structure on the light cone   
because the Fock vacuum is an exact eigenstate of the 
full Hamiltonian.  
This follows from the fact that the total light-cone momentum $P^+ > 0$ 
 and it is conserved.  
This means that all constituents in a physical eigenstate 
are directly related to that state, and not to disconnected
vacuum fluctuations. 

 In the Tamm-Dancoff method (TDA) and
sometimes also in the method of discretized  light-cone
quantization (DLCQ), one approximates the field theory by 
truncating the Fock space. Based on the success of the 
constituent quark models, the assumption is that a few 
excitations describe the essential physics and that adding 
more Fock space excitations only refines the initial approximation.  
 Wilson \cite{wwh94} has stressed the 
point that the success of the Feynman parton model provides 
hope for the eventual success of the front-form methods. 

One of the most important tasks in hadron physics is to 
calculate the spectrum and the wavefunctions of  physical 
particles from  a covariant theory, as mentioned.
The method of `Discretized Light-Cone Quantization' has
precisely this goal. 
Since its first formulation \cite{pab85a,pab85b}  
many problems have been resolved but some remain open. 
To date DLCQ has proved to be one of the most powerful
tools available for solving bound state problems in quantum
field theory \cite{phw90,bmp93}.

Let us review briefly the difficulties. As with conventional 
non-relativistic many-body theory  one starts out with a 
Hamiltonian.  The kinetic energy
is a one-body operator and thus simple.  
The potential energy is at least
a two-body operator and thus complicated. 
One has solved the problem if one has found one or several 
eigenvalues and eigenfunctions  of the Hamiltonian equation. 
One always can expand the eigenstates in terms of products 
of single particle states. These single particle wavefunctions
are solutions of an arbitrary `single particle Hamiltonian'.  
In the Hamiltonian matrix for a two-body interaction most 
of the matrix-elements vanish, since a 2-body Hamiltonian 
changes the state of up to 2 particles. The structure of the 
Hamiltonian is that one of a finite penta\--diagonal bloc matrix. 
The dimension within a bloc, however, is infinite to start with. 
It is made finite by an artificial cut-off,  for example on the 
single particle quantum numbers. A finite matrix, however, 
can be diagonalized on a computer: the
problem becomes `approximately soluble'. 
Of course, at the end one must verify that the physical results  
are (more or less) insensitive to the cut-off(s) and other 
formal parameters.~-- 
Early calculations in one space dimension \cite{pau84}, where
this procedure was actually carried out in one space dimension,
showed rapid converge to the exact eigenvalues. 
The method was successful in 
generating  the exact  eigenvalues and eigenfunctions for
up to 30 particles. From these early calculations it was clear
that Discretized Plane Waves  are a manifestly useful tool 
for many-body problems. In this review we will display the 
extension of this method (DLCQ) to various quantum field
theories \cite
{epb87,elp89,els94,elk96,hbp90,hor90,%
kap93,kap94,kpp94,kap92,pab85a,pab85b,%
pkp95,tbp91,bap96,klp90,pam95,pab96,pau96,saw85,saw86}.

The first studies of model field theories had disregarded 
the so called `zero modes', the space-like constant
field components  defined in a finite spatial volume 
(discretization) and quantized at equal light-cone time. 
But subsequent studies have shown that they can 
support certain kinds of vacuum structure. The long range 
phenomena of spontaneous symmetry breaking  
\cite{hkw91a,hkw91b,hkw91c,bpv93,piv94,hpv95,rob93} 
as well as the topological structure \cite{kap94,kpp94} 
can in fact be reproduced when they are included carefully. 
The phenomena are realized in quite different ways.  For
example, spontaneous breaking of $Z_2$ symmetry 
($\phi \rightarrow -\phi$) in the $\phi^4$-theory in 1+1
dimension occurs via a {\em constrained} zero mode of 
the scalar field \cite{bpv93}.  There the zero mode satisfies 
a nonlinear constraint equation that relates
it to the dynamical modes in the problem.  
At the critical coupling a bifurcation of the solution occurs
\cite{hkw92a,hkw92b,rob93,bpv93}.  
In formulating the theory, one must choose one of them.
This choice is analogous to what in the conventional language 
we would call the choice of vacuum state. 
These solutions lead to new operators in the Hamiltonian
which break the $Z_2$ symmetry at and beyond the critical 
coupling. The various solutions contain $c$-number pieces 
which produce the possible vacuum expectation
values of $\phi$. The properties of the strong-coupling phase 
transition in this model are reproduced, including its 
second-order nature and a reasonable value for the
critical coupling\cite{bpv93,piv94}.  
One should emphasize that solving the constraint equations 
really amounts to determining  the Hamiltonian ($P^-$)
and possibly other Poincare\'e generators, while the
wave function of the vacuum remains simple.
In general,  $P^-$ becomes very complicated when the 
constraint zero modes are included, and this  
in some sense is the price to pay to have a formulation 
with a simple vacuum, combined with possibly finite 
vacuum expectation values. Alternatively, it should be
possible to think of  discretization as a cutoff which 
removes states with $0<p^+<\pi/L$,  and the zero mode
contributions to the Hamiltonian as effective interactions that
restore the discarded physics. In the light-front  power  
counting \`a la Wilson it is clear that there will be a huge
number of allowed operators. 

Quite separately, Kalloniatis {\it et al. } \cite{kap94} has shown that
also a {\em dynamical} zero mode arises in a pure $SU(2)$
Yang-Mills theory in 1+1  dimensions.  
A complete fixing of the gauge leaves the theory with one 
degree of freedom, the zero mode of the vector potential
$A^+$. The theory has a discrete spectrum of zero-$P^+$ 
states corresponding to modes of the flux loop around the 
finite space.  Only one state has a zero eigenvalue of
the energy $P^-$, and is the true ground state of the theory. 
The non-zero eigenvalues are proportional to the length 
of  the spatial box, consistent with the flux loop picture.  
This is a direct result of the topology of the space. 
Since the theory considered there was a purely topological 
field theory, the exact solution was identical to that in the 
conventional equal-time approach on the analogous 
spatial topology \cite{het93}.

Much of the work so far performed has been for theories 
in $1+1$ dimensions. 
For these theories there is much success to report.  
Numerical solutions have been obtained for a 
variety of gauge theories including $U(1)$ and $SU(N)$ 
for $N=1,2,3$ and 4 
\cite{hor90,hbp90,hor91,hor92,klp90}; 
Yukawa \cite{ghp93}; and to some extent $\phi^4$ 
\cite{hav87,hav88}. 
A considerable amount of analysis of $\phi^4$  
\cite{hav87,hav88,hkw91a,hkw91b,hkw91c,hkw92a,hkw92b,hei96b} 
has been performed and a fairly complete discussion of 
the Schwinger model has been presented
\cite{epb87,elp89,els94,mcc91,hkw92b,hei96b,ltl91}. 
The long-standing problem
in reaching high numerical accuracy towards the 
massless limit has been resolved recently \cite{van96}. 

The extension of this program to physical theories in 3+1 
dimensions is a formidable computational task because of 
the much larger number of degrees of freedom.
The amount of work is therefore understandably smaller; 
however, progress is being made.  
Analyses of the spectrum and light-cone wavefunctions 
of positronium in QED$_{3+1}$ have been made by Tang
{\it et al.} \cite{tbp91} and Krautg\"artner  {\it et al.} \cite{kpw92}. 
Numerical studies on positronium have provided the Bohr, 
the fine, and the hyperfine structure with very good
accuracy \cite{trp96}.  
Currently, Hiller, Brodsky, and Okamoto \cite{hbo95} are 
pursuing a non perturbative calculation of the lepton 
anomalous moment in QED using the DLCQ method. 
Burkardt \cite{bur94} and more recently van de Sande and
Dalley  \cite{bur94,vab96,vad96,dav96}   have recently
solved scalar theories with  transverse dimensions by 
combining a Monte Carlo lattice method with DLCQ,
taking up an old suggestion of Bardeen and Pearson
\cite{bap76,bpr80}.   Also of interest is recent work of
Hollenberg and Witte \cite{how94}, who have shown 
how Lanczos tri-diagonalization can be combined with a
plaquette expansion to obtain an analytic extrapolation 
of a physical system to infinite volume. 
The major  problem one faces here is a reasonable 
definition of an effective interaction including the 
many-body amplitudes \cite{pau93,pau96}. 
There has been considerable work focusing on  the 
truncations required to reduce the space of states 
to a manageable level 
\cite{phw90,per94a,per94b,wwh94}.  
The natural language for this discussion is that of the 
renormalization group, with the goal being to
understand the kinds of effective interactions that occur 
when states are removed, either by cutoffs of some kind 
or by an explicit Tamm-Dancoff truncation.  
Solutions of the resulting effective Hamiltonian can then 
be obtained  by various means, for example using 
DLCQ or basis function techniques.  
Some calculations of the spectrum of heavy quarkonia 
in this approach have recently been 
reported \cite{brp95}.
Formal work on renormalization in $3+1$ dimensions  
\cite{mps91} has yielded some positive results but many 
questions remain.  
More recently, DLCQ has been applied to new variants 
of QCD$_{1+1}$ with quarks in the adjoint representation, 
thus obtaining color-singlet eigenstates analogous to 
gluonium states \cite{dkb94,pab96,vab96}. 

The physical nature of the light-cone Fock representation 
has important consequences for the description of 
hadronic states.  As to be discussed in greater detail in 
sections~\ref{sec:continuum} and \ref{sec:impact}, 
one can compute electro-magnetic and weak form factors 
rather directly from an  overlap of light-cone 
wavefunctions $\psi_n(x_i, k_{\perp_i},\lambda_i)$
\cite{dly70,leb80,shb90}.  
Form factors are generally constructed from hadronic 
matrix elements of the current 
$\langle p \vert j^\mu(0) \vert p + q \rangle$.  
In the interaction picture one can identify the fully 
interacting Heisenberg current $J^\mu$ with the free 
current $j^\mu$ at the spacetime point $x^\mu = 0$. 
Calculating matrix elements of the current $j^+=j^0+j^3$
in a frame with $q^+=0$, only diagonal matrix elements 
in particle number $n^\prime = n$ are needed.  
In contrast, in the equal-time theory one must also
consider off-diagonal matrix elements and fluctuations 
due to particle creation and annihilation in the vacuum.  
In the non-relativistic limit one can make contact with
the usual formulas for form factors in Schr\"odinger 
many-body theory.

In the case of inclusive reactions, the hadron and nuclear 
structure functions are the probability distributions 
constructed from integrals and sums over the absolute
squares $\vert \psi_n \vert^2 $. In the far off-shell
domain of large parton virtuality,  one can use perturbative 
QCD to derive the asymptotic fall-off of the Fock amplitudes, 
which then in turn leads to the QCD evolution equations for
distribution amplitudes and structure functions. 
More generally, one can prove factorization theorems for 
exclusive and inclusive reactions which separate the hard
and soft momentum transfer regimes, 
thus obtaining rigorous predictions for the leading power 
behavior contributions to large momentum 
transfer cross sections.  
One can also compute the far off-shell amplitudes within 
the light-cone wavefunctions where heavy quark pairs 
appear in the Fock states.  Such states persist over a time
$\tau \simeq P^+/{\cal M}^2$ until they are materialized 
in the hadron collisions. As we shall discuss in 
section~\ref{sec:exclusive}, this leads to a number of 
novel effects in the hadroproduction of heavy quark 
hadronic states \cite{bhm92}.

A number of properties of the light-cone wavefunctions 
of the hadrons are known from both phenomenology 
and the basic properties of QCD.  
For example, the endpoint behavior of light-cone
wave and structure functions can be determined 
from perturbative arguments and Regge arguments.  
Applications are presented in Ref.\cite{bbs94}.  
There are also correspondence principles.  
For example, for heavy quarks in the non-relativistic limit, 
the light-cone formalism reduces to conventional
many-body Schr\"odinger theory.  
On the other hand, we can also build effective 
three-quark models which encode the static properties 
of relativistic baryons. 
The properties of such wavefunctions are discussed 
in section~\ref{sec:impact}.

We will review  the properties of vector and axial vector
non-singlet charges and compare the space-time with 
their light-cone realization. 
We will show that the space-time and light-cone axial 
currents are distinct; this remark is at the root of the 
difference between the chiral properties of QCD in the
two frames. We show in the free quark model in a 
light-front frame is chirally symmetric in the $SU(3)$ limit 
whether the common mass is zero or not. In QCD chiral
symmetry is broken both explicitly and dynamically. 
This reflected in the light-cone by the fact that the 
axial-charges are not conserve even in the chiral limit. 
Vector and axial-vector charges annihilate the Fock 
space vacuum and so are { \it bona fide} operators. 
They form an $SU(3) \otimes SU(3)$ algebra and 
conserve the number of quarks and anti-quarks 
separately when acting on a hadron state. 
Hence they classify hadrons, on the basis of their 
valence structure, into multiplets which are not mass
degenerate. This classification however turns out to be 
phenomenologically deficient.
The remedy of this situation is unitary transformation 
between the charges and the physical generators 
of the classifying $SU(3) \otimes SU(3) $ algebra.

Although we are still far from solving QCD explicitly, 
it now is the right time to give a presentation
of the light-cone activities to a larger community.
The front form can contribute to the physical insight and 
interpretation of experimental results. We therefore will combine
a certain amount of pedagogical presentation of canonical
field theory with the rather abstract and theoretical questions
of most recent advances. This attempt can neither be
exhaustive nor complete, but we have in mind that
we ultimately have deal with the true physical questions 
of experiment.

We will use two different metrics in this review.
The literature is about evenly split in their use. We have, for the most 
part, used the metric that was used in the original work being reviewed.
We label them the LB convention and the KS convention and discuss 
them in more detail in chapter II and the appendix.

%% file: 02Hamil.tex
\section{Hamiltonian Dynamics}
\label{sec:Hamiltonian}
\setcounter{equation}{0}

What is a Hamiltonian? Dirac \cite{dir58} defines the 
Hamiltonian $ H $ as that
{\it operator}  whose action on the state vector 
$\vert\ t \ \rangle $ of a physical system has the same 
effect as taking the partial derivative with respect to time 
$ t $, {\it i.e.} 
\begin {equation} 
  H \ \vert\ t \ \rangle = i {\partial \over \partial t} 
      \ \vert\ t \ \rangle 
\ . \end   {equation} 
Its expectation value is a {\it constant of the motion},  
referred to shortly as the `energy' of the system.  
We will not consider pathological constructs where a 
Hamiltonian depends explicitly on time. 
The concept of an energy has developed over many centuries 
and applies irrespective of whether one deals 
with the motion of a non-relativistic particle in classical
mechanics or  with a non-relativistic wave function in the
Schr\"odinger equation,  and it  generalizes almost 
unchanged to a relativistic and covariant field theory.   
The Hamiltonian operator $ P _0 $ is a
constant of the motion which acts as the displacement 
operator in time $x ^0\equiv t$
\begin {equation} 
   P _0 \ \vert\ x ^0 \ \rangle = i {\partial \over \partial x ^0} 
      \ \vert\ x ^0\ \rangle 
\ . \end   {equation} 
This definition applies also in the front form,
where  the `Hamiltonian' operator $ P _+ $ is a the constant of 
the motion whose action on the state vector, 
\begin {equation} 
   P _+ \ \vert\ x ^+ \ \rangle = i {\partial \over \partial x ^+} 
      \ \vert\ x ^+ \ \rangle 
\,,\end   {equation} 
has the same effect as the partial with respect
to `light-cone time' $ \displaystyle x ^+ \equiv  (t+ z) $. 
In this chapter we elaborate on these concepts and operational 
definitions to some detail for a relativistic theory, 
focusing on covariant gauge field theories.
For the most part the LB convention is used however many of the 
results are convention independent.
 
\subsection{Abelian Gauge Theory: Quantum Electrodynamics}
\label{sec:qed}

The prototype of a field theory is Faraday's and Maxwell's 
electrodynamics \cite{max54}, which is gauge invariant as first
pointed by  Hermann Weyl \cite{wey29}.

The non-trivial set of Maxwells equations has the four components
\begin {equation}          \partial _\mu  F  ^{\mu \nu} = g  J ^\nu 
\ . \label{eq:2.4}\end   {equation} 
The six components of the electric and magnetic fields are 
collected into the antisymmetric {\it electro-magnetic field} tensor
$ F ^{\mu \nu} \equiv \partial ^\mu A ^\nu   - \partial ^\nu A ^\mu $
and expressed in terms of the {\it vector potentials} $ A  ^\mu $
describing vector bosons with a strictly vanishing mass.
Each component is a real valued operator function of the three 
space coordinates $ x ^k = (x, y, z) $ and of the time $ x ^0 = t $. 
The space-time coordinates are arranged into the 
vector $x ^\mu$ labeled by the {\it Lorentz indices} 
($ \kappa, \lambda, \mu , \nu = 0, 1, 2, 3$ ).
The Lorentz indices are lowered by the {\it metric tensor}
$ g _{\mu \nu} $ and raised by $ g ^{\mu \nu} $ with
$ g _{\kappa \mu} g ^{\mu \lambda} =  \delta _\kappa ^\lambda$.
These and other conventions are collected in 
Appendix~\ref{app:General}.
The coupling constant $ g $ is related to the dimensionless 
{\it fine structure constant} by 
\begin{equation} 
   \alpha = { g ^2 \over 4 \pi \hbar c}
\,. \end{equation} 
The antisymmetry of $ F ^{\mu\nu} $ implies a vanishing 
four-divergence of the current $ J ^\nu(x) $, {\it i.e.} 
\begin {equation}
              \partial _\mu J ^\mu = 0
\,. \label{eq:2.5} \end   {equation} 
In the equation of motion, the time derivatives of the vector potentials are expressed as
functionals of the fields and their space-like derivatives,
which in the present case  are of second order in the time, 
like $\partial _0  \partial _0 A ^\mu = f [ A^\nu, J ^\mu ] $. 
The Dirac equations 
\begin {equation} 
  \left( i \gamma ^\mu \partial _\mu -  m\right) \Psi  
   = g \gamma ^\mu  A _\mu \Psi 
\,,\label{eq:2.6} \end   {equation} 
for given values of the vector potentials $A _\mu $, 
define the time-derivatives of the 
four complex valued spinor components $ \Psi _\alpha (x) $ 
and their adjoints $ \overline \Psi _\alpha (x) = 
 \Psi _\beta ^\dagger(x) \,(\gamma ^0) _{\beta \alpha}  $, 
and thus of the current 
$ J ^\nu \equiv \overline \Psi \gamma ^\nu \Psi = 
 \overline \Psi _\alpha \gamma ^\nu _{\alpha \beta} \Psi _\beta$.
The mass of the fermion is denoted by $m$, 
the four {\it Dirac matrices} by
$ \gamma ^\mu = (\gamma ^\mu) _{\alpha \beta} $.
The {\it Dirac indices} $\alpha$ or $\beta$ enumerate the 
components from 1 to 4, doubly occuring indices are implicitly 
summed over without reference to their lowering or raising. 

The combined set of the Maxwell {\it and} Dirac equations is closed. 
The combined set of the 12 coupled differential equations in 3+1 
space-time dimensions is called {\it Quantum Electrodynamics} (QED).

The trajectories of physical particles extremalize the {\it action}.
Similarly, the equations of motion in a field theory like
Eqs.(\ref{eq:2.4}) and (\ref{eq:2.6}) extremalize 
the action {\it density}, usually referred to as the
{\it Lagrangian} ${\cal L} $.
The Lagrangian of Quantum Electrodynamics (QED) 
 \begin {equation} 
     {\cal L}  = 
   - {1 \over 4}  F  ^{\mu \nu}  F _{\mu \nu} + {1 \over2} \left[ 
      \overline \Psi \left( i \gamma ^\mu D _\mu  -  m\right) \Psi  
   + \ {\rm h.c.} \ \right]
\ , \label {eq:2.7} \end   {equation}
with the  {\it covariant derivative} 
$D _\mu = \partial _\mu - ig  A _\mu $, 
is a local and hermitean operator, classically a real function
of space-time $x ^\mu$. This almost empirical fact can be cast 
into the familiar and canonical {\it calculus of variation} 
as displayed in many text books \cite{bjd65,itz85}, 
whose essentials shall be recalled briefly.

The Lagrangian for QED is a functional of the twelve components 
$ \Psi _\alpha (x) $, $\overline \Psi _\alpha (x) $, $ A _\mu (x) $
and their space-time derivatives. Denoting them collectively 
by $ \phi _r (x)$ and $ \partial _\mu \phi _r (x)$ one has thus 
${\cal L} = {\cal L} \left[\phi _r, \partial _\mu \phi _r \right]$.
Crucial is that ${\cal L}$ depends on space-time 
only through the fields.
Independent variation of the action with respect to 
$ \phi _r$ and $\partial _\mu \phi _r $, 
\begin {equation} 
  \delta _{\phi} \int d x ^0 d x ^1 d x ^2 d x ^3\ {\cal L}  (x) = 0 
\ , \label{var-lagrangian} 
\end {equation} 
results in the 12 {\em equations of motion}, 
the {\it Euler equations}  
\begin {equation} 
   \partial _\kappa \pi _r ^\kappa
  - \delta {\cal L}  / \delta \phi _r = 0
   \ , \qquad\quad {\rm with } \quad
   \pi _r ^\kappa [\phi ] \equiv {\delta  {\cal L} \over \delta 
   \left(\partial _\kappa \phi _r \right)} 
  \ , \label{eq:2.9}\end{equation}
for $r = 1,2, \dots 12 $. The {\it generalized momentum fields} 
$ \pi _r ^\kappa [\phi ] $ are introduced here for convenience 
and later use, with the argument $[\phi ]$ usually suppressed 
except when useful to emphasize the field in question. 
The Euler equations symbolize the most compact form of 
equations of motion. 
Indeed, the variation with respect to the vector potentials 
\begin {equation} 
  {\delta {\cal L} \over 
   \delta \left(\partial _\kappa A _\lambda \right)} 
   \equiv \pi ^{\lambda\kappa} [A] 
   = - F ^{\kappa \lambda} 
  \qquad {\rm and} \qquad
  {\delta {\cal L} \over 
   \delta A _\lambda } \equiv g  J ^\lambda = 
  g \overline \Psi \gamma ^\lambda \Psi
\  \label {eq:2.10} \end{equation}
yields straightforwardly the Maxwell equations (\ref{eq:2.4}), 
and varying with respect to the spinors 
\begin{equation} 
 \pi _\alpha ^\kappa   [\psi] \equiv 
  {\delta {\cal L} \over 
   \delta \left(\partial _\kappa \Psi _\alpha \right)} = 
   {i \over 2} \overline \Psi _\beta \gamma ^\kappa _{\beta \alpha } 
  \qquad {\rm and} \qquad
  {\delta {\cal L} \over 
   \delta \Psi _\alpha } = 
  - {i \over 2} \partial _\mu \overline \Psi _\beta 
                \gamma   ^\mu _{\beta \alpha } +
  g \overline \Psi _\beta \gamma ^\mu _{\beta \alpha } A _\mu - 
  m \overline \Psi _\alpha 
\  \label {eq:2.11} \end{equation} 
and its adjoints give the Dirac equations (\ref{eq:2.6}).

The canonical formalism is particularly suited for discussing
the { symmetries} of a field theory. 
According to a theorem of Noether \cite{itz85,noe18} every continuous 
symmetry of the Lagrangian is associated with a four-current 
whose four-divergence vanishes. This in turn implies 
a {\it conserved charge} as a {\it constant of motion}.
Integrating the current $ J ^\mu$ in Eq.(\ref{eq:2.5}) over a 
three-dimensional surface  of a { hypersphere}, 
embedded in four dimensional space-time, 
generates a conserved charge.   
The surface element $ d \omega _\lambda $ and the 
(finite) volume $\Omega$ are defined most conveniently 
in terms of the totally antisymmetric tensor 
$ \epsilon _{\lambda \mu \nu \rho}$
($ \epsilon _{0123} = 1$)
\begin {equation} 
  d \omega _\lambda = {1 \over 3 !} 
  \epsilon _{\lambda \mu \nu \rho} d x ^\mu d x ^\nu d x ^\rho
  \qquad {\rm and }\quad
  \Omega = \int d \omega_0 = \int d x ^1 d x ^2 d x ^3
\ , \end   {equation} 
respectively. 
Integrating Eq.(\ref{eq:2.5}) over the hyper-surface 
specified by $x ^0 = const$ reads then
 \begin {equation} 
   {\partial \over \partial x ^0} 
   \int _\Omega d x ^1 d x ^2 d x ^3  J ^0 (x) + 
   \int _\Omega d x ^1 d x ^2 d x ^3  \Biggl[ 
   \ {\partial \over \partial x ^1}  J ^1 (x) + 
     {\partial \over \partial x ^2}  J ^2 (x) + 
     {\partial \over \partial x ^3}  J ^3 (x) \Biggr] = 0
 \ . \end   {equation} 
The terms in the square bracket reduce to surface terms 
which vanish if the boundary conditions are carefully defined. 
Under that {\it proviso} the charge 
\begin   {equation} 
  Q = \int d \omega _0  \ J ^0 (x) = 
      \int _\Omega d x ^1 d x ^2 d x ^3  \ J ^0 (x^0,x^1,x^2,x^3) 
\,,\end   {equation} 
is independent of time $x^0$ and a constant of the motion. 

Since ${\cal L}$ is frame-independent, there must be ten 
conserved four-currents. 
Here they are 
\begin {equation} 
     \partial _\lambda   T ^{\lambda \nu} = 0\,,
    \qquad {\rm and} \quad 
    \partial _\lambda   J ^{\lambda , \mu \nu} = 0 
\,,\nonumber \end   {equation} 
where the {\it energy-\-momentum}  $T ^{\lambda \nu}$ and 
the {\it boost\--angular-\-momentum} stress tensor 
$J ^{\lambda , \mu \nu}$ are respectively,
\begin {equation} 
   T ^{\lambda \nu} = 
   \pi _r ^\lambda \partial ^\nu \ \phi _r 
    - g ^{\lambda \mu}  {\cal L} 
 \ ,  \qquad {\rm and} \quad
   J ^{\lambda , \mu \nu} =
   x ^\mu T ^{\lambda \nu} - x ^\nu T ^{\lambda \mu} + 
   \pi _r ^\lambda \ \Sigma _{rs} ^{\mu \nu} \ \phi _s 
 \ . \label {eq:2.16} \end   {equation} 
As a consequence the Lorentz group has ten `conserved 
charges', the ten constants of the motion
 \begin {eqnarray}
    P  ^\nu & = & \int _\Omega d \omega_0 
    \bigl(\pi _r ^0 \partial ^\nu \phi _r - g ^{0 \mu} {\cal L} \bigr)
    \ ,  \nonumber \\  {\rm and} \quad \qquad 
    M  ^{\mu \nu} & = & \int _\Omega d \omega_0 
   \bigl( x ^\mu T ^{0 \nu} - x ^\nu T ^{0 \mu} 
        + \pi _r ^0  \Sigma _{rs} ^{\mu \nu} \phi _s (x) \bigr)
\,, \label {eq:2.17} \end    {eqnarray}
the 4 components of {\it energy-momentum}  and the 
6 {\it boost-angular} momenta, respectively.
The first two terms in  $ M ^{\mu\nu}$ corresponds to the 
orbital and the last term to the spin part of angular momentum.
The spin part $\Sigma$ is either 
\begin {equation} 
   \Sigma _{\alpha \beta} ^{\mu \nu} = 
   {1 \over 4} \left[\gamma ^\mu , \gamma ^\nu \right] _{\alpha \beta} 
   \qquad \quad {\rm or} \quad\qquad
   \Sigma _{\rho \sigma} ^{\mu \nu} = 
   g ^\mu _\rho g ^\nu _\sigma- g ^\mu _\sigma g ^\nu _\rho
\ , \end   {equation}
depending on whether $ \phi _r$ refers to spinor or to vector
fields, respectively. In the latter case, we substitute 
$\pi _r ^\lambda \rightarrow \pi ^{\rho\lambda} = \delta {\cal L} 
/ \delta \left(\partial _\lambda  A_\rho \right)$ and
$\phi_s \rightarrow A^\sigma$.
Inserting Eqs.(\ref{eq:2.10}) and (\ref{eq:2.11}) one gets
for gauge theory the familiar expressions \cite{bjd65}
\begin {equation} 
   J ^{\lambda , \mu \nu} =
   x ^\mu T ^{\lambda \nu} - 
   x ^\nu T ^{\lambda \mu} + {i\over8} \overline \Psi 
   \left(  \gamma ^\lambda 
   [ \gamma ^\mu, \gamma ^\nu ] +
   [ \gamma ^\nu, \gamma ^\mu ] 
   \gamma ^\lambda \right)  \Psi  +
   A ^\mu F ^{\lambda\nu} -
   A ^\nu F ^{\lambda\mu}    
\ .  \end   {equation} 
The symmetries will be discussed further in 
section~\ref{sec:symmetries}.

In deriving the energy-momentum stress tensor 
one might overlook that $ \pi _r ^\lambda [\phi] $ 
does {\it not necessarily} commute with $ \partial ^\mu \phi _r $.
As a rule, one therefore should symmetrize in the boson 
and anti-symmetrize in the fermion fields, {\it i.e. } 
\begin {eqnarray}
  \pi _r ^\lambda [\phi ] \partial ^\mu \phi _r & \longrightarrow & 
  {1 \over2} \left( \pi _r ^\lambda [\phi ] \partial ^\mu \phi _r 
           + \partial ^\mu \phi _r \pi _r ^\lambda [\phi ] \right) 
\ , \nonumber \\
  \pi _r ^\lambda [\psi ] \partial ^\mu \psi _r & \longrightarrow & 
  {1 \over2} \left( \pi _r ^\lambda [\psi ] \partial ^\mu \psi _r 
    - \partial ^\mu \psi _r \pi _r ^\lambda [\psi ] \right)
 \ , \end   {eqnarray}
respectively, but this will be done only implicitly.

The Lagrangian $ {\cal L} $ is invariant under local gauge 
transformations, in general described by a unitary and space-time 
dependent matrix operator $ U ^{-1} (x) = U ^\dagger (x)$.
In QED, the dimension of this matrix is 1 with the most general form 
$ \displaystyle U (x) = {\rm e} ^{-ig \Lambda (x)} $.
Its elements form the abelian group $U(1)$, 
hence abelian gauge theory.
If one substitutes the spinor and vector fields in 
$ F ^ {\mu\nu} $ and 
$ \overline \Psi _\alpha D _\mu \Psi _\beta $
according to
\begin{eqnarray}
  \widetilde \Psi _\alpha & = & U \ \Psi _\alpha
\ , \nonumber \\ {\rm and} \qquad \quad
  \widetilde   A _ \mu   & = & U A _\mu U ^\dagger + 
   {i\over g} \bigl( \partial _\mu U \bigr) U ^\dagger
\  , \label {eq:2.20} \end{eqnarray}
one verifies their invariance under this transformation,
as well as that of the whole Lagrangian.
The Noether current associated with this symmetry 
is the $ J ^\mu $ of Eq.(\ref{eq:2.10}). 

A straightforward application of the variational principle, 
Eqs.(\ref{eq:2.10}) and (\ref{eq:2.11}), 
does not yield immediately manifestly gauge 
invariant expressions. Rather one gets 
\begin {equation} 
   T ^{\mu\nu} =  F^{\mu\kappa} A _\kappa
  + {1\over2} \bigl[ \overline \Psi  i\gamma^\mu \partial^\nu \Psi  
  + \ {\rm h.c.} \bigr] 
 - g^{\mu\nu} {\cal L} \ . \end   {equation} 
However using the Maxwell equations one derives the identity 
\begin {equation} 
  F^{\mu\kappa} \ \partial^\nu  A_\kappa =
  F^{\mu\kappa} F^\nu_{\phantom{\nu} \kappa } 
  + g  J^\mu A^\nu 
  + \partial_\kappa\bigl(F^{\mu\kappa} A^\nu \bigr) 
\ .  \end   {equation}
Inserting that into the former gives 
\begin {equation}
    T ^{\mu\nu} =  F ^{\mu\kappa} 
    F _\kappa^{\phantom{\kappa}\nu} 
   + {1\over2} \bigl[ i \overline \Psi \gamma ^\mu D ^\nu \Psi  
   + \ {\rm h.c.} \bigr] 
  - g^{\mu\nu} {\cal L}  
  + \partial_\kappa
    \bigl( F ^{\kappa\mu} A ^\nu \bigr) 
\ . \end {equation}
All explicit gauge dependence resides in the last term in the 
form of a four-divergence.
One can thus write 
\begin {equation}
    T ^{\mu\nu} =  F ^{\mu\kappa} 
    F _\kappa^{\phantom{\kappa}\nu} 
   + {1\over2} \bigl[ i \overline \Psi \gamma ^\mu D ^\nu \Psi  
   + \ {\rm h.c.} \bigr] 
  - g^{\mu\nu} {\cal L}  
\ , \label{eq:2.23}\end {equation}
which together with energy-momentum 
\begin {equation} 
    P^\nu = \int_\Omega \! d\omega _0 \ \biggl( F^ {0\kappa} 
    F _\kappa^{\phantom{\kappa}\nu} 
    - g^{0\nu} {\cal L}  + {1\over2} 
   \Bigl[ i \overline \Psi  \gamma ^0  D ^\nu \Psi  
  + \ {\rm h.c.} \Bigr] \biggr) 
\label{eq:2.24} \end {equation}
is {\it manifestly gauge invariant}.

\subsection{Non-Abelian Gauge Theory: Quantum Chromodynamics}
\label{sec:qcd} 

For the gauge group SU(3), one replaces each local gauge field 
$ A^\mu (x)$ by the $3\times 3$ {\it matrix} ${\bf A} ^\mu (x) $, 
\begin {equation}
   A ^\mu \longrightarrow
  ({\bf A}^\mu)_{cc^\prime} = {1\over2} \pmatrix{
   {1\over\sqrt3} A^\mu_8+ A^\mu_3  
  & A^\mu_1-i A^\mu_2    
  & A^\mu_4-i A^\mu_5 \cr
    A^\mu_1+i A^\mu_2    
  &{1\over\sqrt3} A^\mu_8- A^\mu_3  
  & A^\mu_6-i A^\mu_7 \cr
    A^\mu_4+i A^\mu_5    
  & A^\mu_6+i A^\mu_7      
  &-{2\over\sqrt3} A^\mu_8 \cr} 
\,.\label {eq:2.25} \end {equation}
This way one moves from Quantum Electrodynamics to 
{\it Quantum Chromodynamics} with the eight real valued 
{\it color vector potentials} $ A ^\mu _a$ enumerated by the
{\it gluon index} $a = 1, \dots, 8$. 
These matrices are all {\it hermitean} and 
{\it traceless} since the trace can always be absorbed into 
an Abelian U(1) gauge theory. 
They belong thus to the class of
{\it special unitary $3\times 3$ matrices} SU(3).
In order to make sense of expressions like 
$ \overline \Psi {\bf A} ^\mu \Psi $ the {\it quark fields} $\Psi (x)$ 
must carry a {\it color index} $c = 1,2,3$ which are usually 
suppressed as are the Dirac indices
in the color triplet spinor $\Psi _{c, \alpha } (x)$.

More generally for SU(N), the vector potentials $ {\bf A}^\mu $
are hermitian and traceless $N \times N$ matrices.
All such matrices can be parametrized  
$ {\bf A} ^\mu \equiv T ^a _{cc ^\prime} A ^\mu _a$.
The color index $c$ (or $c^\prime$) runs now from 1 to $n_c$, 
and correspondingly the gluon index $a$ (or $r,s,t$) 
from 1 to $n_c^2-1$. Both are implicitly summed, 
with no distinction of lowering or raising them.
The color matrices $ T ^a_{cc ^\prime}$ obey 
\begin {equation}
  \Bigl[ T ^r, T ^s \Bigr] _{cc ^\prime} 
  = i f ^{rsa}  T _{cc ^\prime} ^a 
  \qquad\quad {\rm and } \qquad
  {\rm Tr}\ \bigl(T ^r  T ^s \bigr) = {1\over2}\delta _r ^s 
\ . \label{eq:2.26}\end   {equation}
The {\it structure constants} $\displaystyle f ^{rst}$ are tabulated 
in the literature \cite{itz85,mut87,nac86} for SU(3). 
For SU(2) they are the totally antisymmetric tensor
$\displaystyle \epsilon _{rst}$, since 
$T ^a = {1\over 2} \sigma ^a $ with $\sigma ^a$ being the Pauli 
matrices. For SU(3), the $ T ^a $ are related to the Gell-Mann 
matrices $\lambda ^a$ by $ T ^a = {1\over 2} \lambda ^a$.
The gauge invariant Lagrangian density for QCD or SU(N) is 
\begin {eqnarray}
    {\cal L} & = & - {1\over2} 
    {\rm Tr} \bigl({\bf F}^{\mu\nu} {\bf F}_{\mu \nu} \bigr)
 + {1\over2} \bigl[ \overline \Psi \bigl(i\gamma^\mu {\bf D} _\mu 
   - {\bf m} \bigr)  \Psi   + \ {\rm h.c.} \bigr] 
\ , \nonumber \\
             & = & - {1\over4}  F^{\mu\nu}_a  F_{\mu \nu}^a
  + {1\over2} \bigl[ \overline \Psi \bigl(i\gamma^\mu  
   {\bf D} _\mu -  {\bf m} \bigr) \Psi   + \ {\rm h.c.} \bigr] 
\ ,  \end {eqnarray}
in analogy to Eq.(\ref{eq:2.7}).
The unfamiliar factor of 2 is because of the trace convention in 
Eq.(\ref{eq:2.26}). 
The mass matrix ${\bf m} = m \delta_{cc^\prime}$ is diagonal
in color space. The matrix notation is particularly
suited for establishing gauge invariance according to 
Eq.(\ref{eq:2.20}) with the unitary operators $ {\bf U}$ 
now being $N \times N $ matrices, hence {\it non-Abelian} gauge theory.
The latter fact generates an extra term in
the {\it color-electro-magnetic} fields
\begin {eqnarray}
   {\bf F}^{\mu\nu} \equiv \partial^\mu  {\bf A}^\nu  
   - \partial^\nu  {\bf A}^\mu 
   + i g \bigl[ {\bf A}^\mu, {\bf A}^\nu\bigr] 
\ , \nonumber \\
{\rm or} \qquad\quad
   F^{\mu\nu}_a \equiv \partial^\mu  A^\nu_a
   - \partial^\nu  A^\mu_a  
   - g  f^{ars}  A^\mu_r  A^\nu_s  
\ ,\label{eq:2.31}\end {eqnarray}
but such that $F ^{\mu\nu} $ remains antisymmetric in the 
Lorentz indices. The covariant derivative {\it matrix} finally is
$ {\bf D} ^\mu _{cc^\prime} =
  \delta_{cc^\prime} \partial^\mu 
  + i g {\bf A}^\mu _{cc^\prime} $.
The variational derivatives are now
\begin {equation} {\delta  {\cal L} \over \delta 
   (\partial _\kappa  A _\lambda ^r)} = -  F ^{\kappa\lambda} _r 
   \ \  {\rm and} \quad
   {\delta  {\cal L} \over \delta  A _\lambda ^r} =  -gJ ^\lambda _r 
    , \ \  {\rm with} \ %
   J ^\lambda _r =  \overline \Psi \gamma ^\lambda  T^a \Psi 
   + f^{ars} F^{\lambda \kappa}_r A_\kappa^s 
\ , \label{eq:2.29} \end {equation}
in analogy to Eq.(\ref{eq:2.10}), 
and yield the {\it color-Maxwell equations} 
\begin {equation}
   \partial_\mu  {\bf F}^{\mu \nu} = g {\bf J}^\nu \ ,
   \qquad\ {\rm with}\quad   
   {\bf J}^\nu = \overline \Psi \gamma^\nu  T^a \Psi   T^a 
  + {1\over i} \bigl[ {\bf F}^{\nu\kappa}, {\bf A}_\kappa\bigr]
\ \label{eq:2.30} . \end   {equation}
The {\it color-Maxwell current} is {\it conserved}, 
\begin {equation}
         \partial _\mu {\bf J} ^\mu = 0 
. \end   {equation}
Note that the { color-fermion current} 
$j^\mu_a=\overline \Psi \gamma^\nu  T^a \Psi$
is {\it not trivially conserved}.
The variational derivatives with respect to 
the spinor fields like Eq.(\ref{eq:2.11}) 
give correspondingly the { color-Dirac equations}
\begin {equation}
   \bigl(i \gamma^\mu{\bf D} _\mu -{\bf m}\bigr)\Psi=0
.\label{eq:2.35}\end   {equation}
Everything proceeds in analogy with QED. 
The color-Maxwell equations allow  for the identity
\begin {equation}
  F^{\mu\kappa}_a \partial^\nu  A_\kappa^a =
  F^{\mu\kappa}_a  F^\nu_{\phantom{\nu} \kappa,a} 
 + g  J^\mu_a  A^\nu_a
 + g  f^{ars} F^{\mu\kappa}_a  A^\nu_r  A_\kappa^s
  + \partial_\kappa\bigl(F^{\mu\kappa}_a  A^\nu_a \bigr)
\ . \end{equation}
The energy-momentum stress tensor becomes
\begin {equation}
   T ^{\mu\nu} = 2{\rm Tr} \bigl({\bf F}^{\mu\kappa} 
   {\bf F}_\kappa^{\phantom{\kappa}\nu}\bigr)
  + {1\over2} \bigl[ i \overline \Psi \gamma^\mu  {\bf D} ^\nu \Psi  
  + \ {\rm h.c.} \bigr]  - g^{\mu\nu} {\cal L}  
  - 2\partial_\kappa
  {\rm Tr} \bigl({\bf F}^{\mu\kappa} {\bf A}^\nu \bigr) 
\ .\end {equation} 
Leaving out the four divergence,
$T ^{\mu\nu}$ is { manifestly gauge-invariant} 
\begin {equation}
    T ^{\mu\nu} = 2{\rm Tr} \bigl({\bf F}^{\mu\kappa} 
   {\bf F}_\kappa^{\phantom{\kappa}\nu}\bigr)
  + {1\over2} \bigl[ i \overline \Psi \gamma^\mu  {\bf D} ^\nu \Psi  
  + \ {\rm h.c.} \bigr]  - g^{\mu\nu} {\cal L}  
\ , \end {equation} 
as are the generalized momenta \cite{jam80}
\begin {equation} 
    P^\nu = \int_\Omega \! d\omega_0 
    \ \biggl( 2 {\rm Tr} ({\bf F}^{0\kappa} 
    {\bf F}_\kappa^{\phantom{\kappa}\nu})
  - g^{0\nu} {\cal L}  + {1\over2} 
  \Bigl[ i \overline \Psi \gamma^0  {\bf D} ^\nu \Psi  
  + \ {\rm h.c.} \Bigr] \biggr) 
\ . \label{eq:2.39}\end {equation}
Note that all this holds for SU(N), in fact it holds 
for $d+1$ dimensions.

\subsection{Parametrization of Space-Time}
\label{sec:space_time} 

Let us  review some aspects of canonical field theory. 
The Lagrangian determines both, the equations of motion
and the constants of motion. The equations of motion
are differential equations. 
Solving differential equations one must give initial data.
On a hypersphere in four-space, characterized 
by a fixed initial `time' $x ^0=0$, one assumes to know all 
necessary field components $ \phi _r ( x ^0 _0, \underline x ) $. 
The goal is then to generate the fields for { all} space-time 
by means of the differential equations of motion.

Equivalently, one can propagate the initial configurations 
forward or backward in time with the Hamiltonian.
In a classical field theory, particularly one in which every field 
$\phi _r$ has a { conjugate momentum} 
$ \pi _r [\phi ] \equiv \pi _r ^0 [\phi ] $, see Eq.(\ref{eq:2.9}), 
one gets from the constant of motion $P _0$ to the Hamiltonian 
$P _0$ by substituting the velocity fields 
$\partial _0 \phi _r$ with the canonically conjugate momenta 
$\pi _r$, thus $P_0 = P_0 [\phi,\pi] $.
Equations of motion are then given in terms of the classical 
{ Poisson brackets} \cite {gol50} 
\begin {equation}
   \partial _0 \phi _r = \Bigl\{ P_0 , \phi _r \Bigr\} _{\rm cl}
   \qquad {\rm and} \quad
   \partial _0 \pi  _r = \Bigl\{ P_0 , \pi  _r \Bigr\} _{\rm cl}
\ . \end   {equation}
They are discussed in greater details in
Appendix~\ref{sec:Bergmann}.
Following Dirac \cite{dir58,dir64,dir81}, the
transition to an operator formalism 
like quantum mechanics is consistently achieved by 
replacing the classical Poisson brackets of two { functions}
$ A $ and $ B $ by the `quantum Poisson brackets', 
the commutators of two { operators} $ A $ and $ B $ 
\begin {equation}
   \Bigl\{ A , B \Bigr\} _{\rm cl} \longrightarrow 
   {1 \over i \hbar} \Bigl[ A , B \Bigr] _{x ^0 = y ^0}
\ , \end   {equation}
and correspondingly by the anti-commutator for two fermionic fields.
Particularly one substitutes the basic Poisson bracket 
\begin {equation}
   \Bigl\{ \phi _r (x) , \pi _s (y) \Bigr\} _{\rm cl}
   = \delta _{rs} \ \delta ^{(3)} ( \underline x - \underline y )
\end   {equation}
by the basic commutator 
\begin {equation}
   {1 \over i \hbar} 
   \Bigl[ \phi _r (x) , \pi _s (y) \Bigr] _{x ^0 = y ^0}
   = \delta _{rs} \ \delta ^{(3)} ( \underline x - \underline y )
\ . \label {eq:2.38} \end   {equation}
The time derivatives of the operator fields are then given by
the Heisenberg equations, see Eq.(\ref{eq:2.109}).

In gauge theory like QED and QCD one cannot proceed so 
straightforwardly as in the the above canonical procedure,
for two reasons: (1) Not all of the fields have
a conjugate momentum, that is not all of them are independent; 
(2) Gauge theory has redundant degrees of freedom.
There are plenty of conventions how one can `fix the gauge'. 
Suffice it to say for the moment that `canonical quantization' 
applies only for the independent fields. 
In Appendix~\ref{sec:Bergmann} we will review the 
Dirac-Bergman procedure for handling dependent degrees
of freedom, or for `quantizing under constraint.'

Thus far time $t$ and space $\underline x$ was treated 
as if they were completely separate issues. 
But in a covariant theory, time and space are only different 
aspects of  four dimensional space-time. 
One can however generalize the concepts
of space and of time in an operational sense.
One can define `space' as that hypersphere in four-\-space 
on which one chooses
the initial field configurations in accord with microcausality.
The remaining, the fourth coordinate can be thought of being 
kind of normal to the hypersphere and understood as `time'.
Below we shall speak of space-like and time-like coordinates,
correspondingly.

These concepts can be grasped more formally by conveniently 
introducing { generalized coordinates} $ \widetilde x  ^\nu $. 
Starting from a baseline parametrization of space time
like the above $ x ^\mu$ \cite {bjd65} with a given metric tensor 
$ g ^{\mu\nu}$ whose elements are all zero except 
$ g ^{00} = 1$, $ g ^{11} = -1$, $ g ^{22} = -1$, and $ g ^{33} = -1$,
one parametrizes space-time by a certain functional 
relation
\begin {equation}      \widetilde x  ^\nu = \widetilde x  ^\nu (x ^\mu)
\ . \end   {equation}
The freedom in choosing $ \widetilde x  ^\nu (x ^\mu)$ is 
restricted only by the condition that the inverse 
$ x ^\mu (\widetilde x  ^\nu)$ exists as well. 
The transformation conserves the arc length, thus
$ (ds) ^2 = g _{\mu \nu} dx ^\mu dx ^\nu 
  = \widetilde g _{\kappa \lambda} 
  d \widetilde x ^\kappa d \widetilde x ^\lambda $.
The metric tensors for the two parametrizations are 
then related by 
 \begin {equation} 
 \widetilde g _{\kappa \lambda} 
  = \left( {\partial x ^\mu \over
 \partial \widetilde x ^\kappa} \right) 
 g _{\mu \nu} \left( {\partial x ^\nu \over
 \partial \widetilde x ^\lambda} \right)  
 \ . \label{eq:2.48}\end   {equation} 
The two four-volume elements are related by the Jacobian
$ {\cal J} (\widetilde x) 
= \vert \vert \partial x / \partial \widetilde x \vert \vert $, 
particularly 
$ d ^4 x =  {\cal J} (\widetilde x ) \ d  ^4 \widetilde x $.
We shall keep track of the Jacobian only implicitly.
The three-volume element $ d \omega _0 $ is treated 
correspondingly. 

All the above considerations must be independent of this 
reparametrization. The fundamental expressions like 
the Lagrangian can be 
expressed in terms of either $x $ or $ \widetilde x$. 
There is however one subtle point.
By matter of convenience one defines the hypersphere as that
locus in four-\-space on which one sets the `initial conditions'
at the same `initial time', or on which one `quantizes' the system 
correspondingly in a quantum theory. The hypersphere is thus 
defined as that locus in four-\-space with the same value
of the `time-like' coordinate $ \widetilde x ^0$, {\it i.e.}  
$ \widetilde x ^0 (x ^0, \underline x ) = const $.
Correspondingly, the remaining coordinates 
are called `space-like' and denoted by the spatial three-vector
$ \underline {\widetilde x} 
  = (\widetilde x  ^1 , \widetilde x  ^2 , \widetilde x  ^3 ) $.
Because of the (in general) more complicated metric, 
cuts through the four-\-space characterized by 
$\widetilde x ^0 = const $ are quite different 
from those with $\widetilde x _0 = const $. 
In generalized coordinates the covariant and contravariant 
indices can have rather different interpretation, and one 
must be careful with the lowering and rising of the Lorentz 
indices. For example, only 
$ \partial _0 = 
{\partial / \partial \widetilde x ^0} $
is a `time-derivative' and only $ P _0$ a `Hamiltonian',
as opposed to $\partial ^0$ and $ P ^0$ which in general
are completely different objects. 
The actual choice of $ \widetilde x (x) $ is a matter of 
preference and convenience.

\subsection{Forms of Hamiltonian Dynamics}
\label{sec:forms-of-hamiltonian} 

\begin{figure}
\epsfxsize=160mm\epsfbox{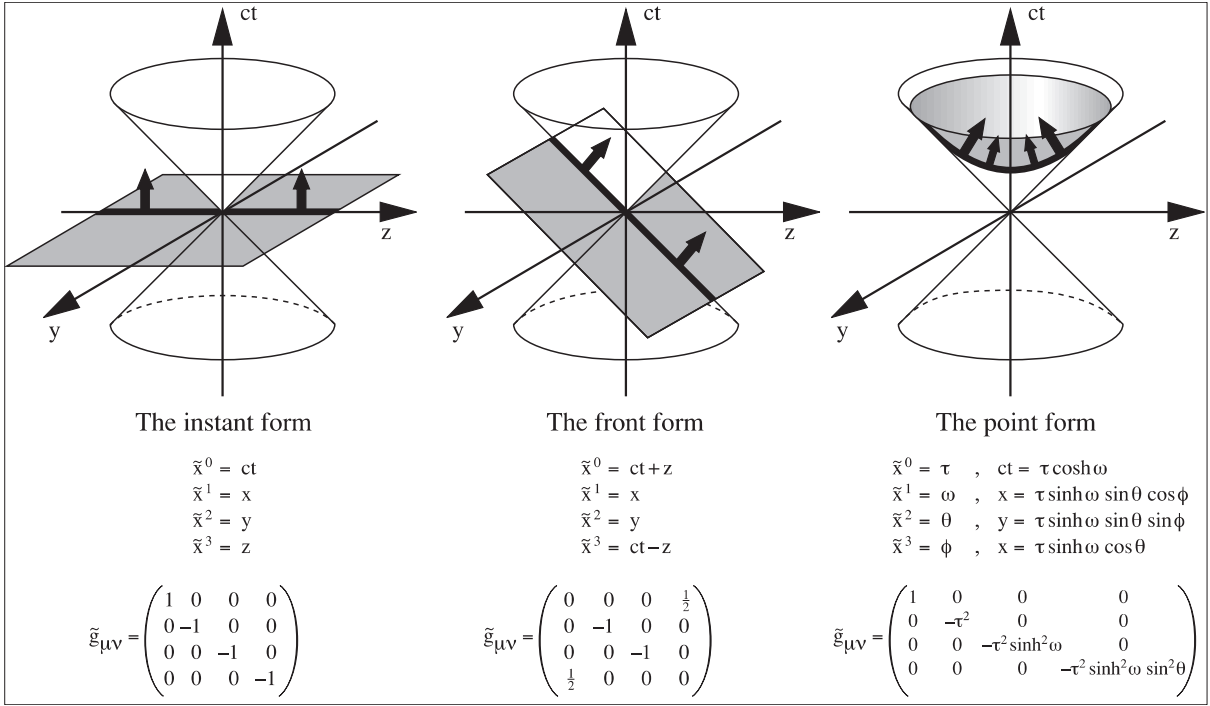}
\caption{\label{fig:bir2}
    Dirac's three forms of Hamiltonian dynamics.
}  \end{figure}

Obviously, one has many possibilities to parametrize
space-time by introducing some  generalized coordinates
$ \widetilde x (x) $. 
But one should exclude all those which are accessible
by a Lorentz transformation. Those are included anyway in a 
covariant formalism. This limits considerably the freedom 
and excludes for example almost all rotation angles.
Following Dirac \cite{dir49} there are no more than
three basically different parametrizations. They are 
illustrated in Figure~\ref{fig:bir2}, and 
cannot be mapped on each other by a Lorentz transform. 
They differ by the hypersphere on which the fields are 
initialized, and correspondingly one has different ``times''.
Each of these space-time parametrizations has thus
its own Hamiltonian, and correspondingly Dirac \cite{dir49}
speaks of  the three {\em forms of Hamiltonian dynamics}:
The {\em instant form} is the familiar one, 
with its hypersphere given by $t=0$. 
In the {\em front form} the hypersphere is a tangent plane 
to the light cone.
In the {\em point form} the time-like coordinate is 
identified with the eigentime of a physical system
and the hypersphere has a shape of a hyperboloid.

Which of the three forms should be prefered?
The question is difficult to answer, in fact it is ill-posed.
In principle, all three forms should yield the same physical
results, since physics should not depend on how one
parametrizes the space (and the time). If it depends 
on it, one has made a mistake.
But usually one adjusts parametrization to  the nature of the 
physical problem to simplify the amount of practical work. 
Since one knows so little on the
typical solutions of a field theory, it might well be worth
the effort to admit also other than the conventional 
``instant'' form.

The bulk of research on field theory implicitly uses the 
instant form, which we do not even attempt to summarize. 
Although it is the  conventional choice for quantizing field 
theory, it has many practical disadvantages. 
For example, given the wavefunctions of an $n$-electron 
atom at an initial time $t=0$,  $\psi_n(\vec x_i,t=0)$, 
one can use the Hamiltonian $H$ to 
evolve $\psi_n(\vec x_i,t)$ to later times $t$.
However, an experiment which specifies the initial 
wave function would require the simultaneous measurement 
of the positions of all of the bounded electrons.
In contrast, determining the initial wave function at fixed 
light-cone time $\tau=0$ only requires an experiment which 
scatters one plane-wave laser beam, since the signal reaching 
each of the $n$ electrons, along the light front, at the 
same light-cone time $\tau = t_i+z_i/c$.

A reasonable choice of $ \widetilde x (x)$ is restricted
by microcausality: a light signal emitted from any point 
on the hypersphere must not cross the hypersphere.
This holds for the instant or for the point form, 
but the front form seems to be in trouble. The light cone
corresponds to light emitted from the origin and touches 
the front form hypersphere at $(x,y) = (0,0)$. 
A signal carrying actually information moves with the
group  velocity always smaller than the phase velocity $ c$. 
Thus, if no information is carried by the signal, 
points on the light cone are unable to communicate. 
Only when solving problems in one-space and one-time dimension,
the front form initializes fields only on the characteristic.  
Whether this generates problems for pathological cases 
like massless bosons (or fermions) is still under debate.

Comparatively little work is done in the point form 
\cite {fhj73,grs74,som73}. Stech and
collaborators \cite {grs74} have investigated the free 
particle, by analyzing the Klein-Gordon and the 
Dirac equation.  As it turns out, the orthonormal functions 
spanning the Hilbert space for these cases are rather 
difficult to work with. 
Their addition theorems are certainly more complicated
than the simple plane waves states applicable in the instant 
or the front form. 

The front form has a number of advantages which we will
review in this article. Dirac's legacy had been forgotten
and re-invented  several times, thus the approach carries 
names as different as 
{\em Infinite-Momentum Frame}, 
{\em Null-Plane Quantization}, 
{\em Light-Cone Quantization}, or
most unnecessarily 
{\em Light-Front Quantization}.
In the essence these are the same. 

The infinite-momentum frame first appeared in the work
of Fubini and Furlan \cite{fuf65} in connection with current 
algebra as the limit of a reference frame moving with 
almost the speed of light.
Weinberg \cite {wei66} asked whether this limit might be 
more generally useful. He considered the infinite
momentum limit of the old-fashioned perturbation diagrams 
for scalar meson theories and showed that the vacuum 
structure of these theories simplified in this
limit. Later Susskind \cite {suf67,sus68} showed that the 
infinities which occur among the generators
of the Poincare\'e group when they are boosted to a 
fast-moving reference frame can be scaled or subtracted
out consistently. The result is essentially a change of the
variables. Susskind used the new variables to draw attention
to the (two-dimensional) Galilean subgroup of the Poincare\'e 
group. He pointed out that the simplified vacuum structure 
and the non-relativistic kinematics of theories at infinite
momentum might offer potential-theoretic intuition in
relativistic quantum mechanics.
Bardakci and Halpern \cite{bah68} further analyzed the
structure of the theories at infinite momentum.
They viewed the infinite-momentum limit as a change of
variables from the laboratory time $t$ and space 
coordinate $z$ to a new ``time'' $\tau=(t+z)/\sqrt{2}$ and 
a new ``space'' $\zeta=(t-z)/\sqrt{2}$.
Chang and Ma \cite{chm69} considered the Feynman 
diagrams for a $\phi^3$-theory and quantum electrodynamics
from this point of view and where able to demonstrate the 
advantage of their approach in several illustrative calculations.
Kogut and Soper \cite{kos70} have examined the formal
foundations of quantum electrodynamics in the 
infinite-momentum frame, and interpret the 
infinite-momentum limit as the change of variables thus 
avoiding limiting procedures. The time-ordered perturbation
series of the S-matrix is due to them, see also
\cite{bks71,sop71,kos70,kos73}.
Drell, Levy, and Yan \cite {dly69,dly70,dly70b,dry70} 
have recognized that the formalism could  serve as kind 
of natural tool for formulating the quark-parton model. 

Independent of and almost simultaneous with the 
infinite-momentum frame is the work on 
{\em Null Plane Quantization} by Leutwyler \cite{leu68,leu69}, 
Klauder, Leutwyler, and Streit \cite{kll69}, and by
Rohrlich \cite{roh71}. 
In particular they have investigated the stability of the so
called `little group' among the Poincare\'e generators
\cite {leu74a,leu74b,leu74c,les78}. 
Leutwyler recognized 
the utility of defining quark wavefunctions to give an 
unambiguous meaning to concepts used in the parton model. 

The later  developments using the infinite-momentum frame
have displayed that the naming is somewhat unfortunate
since  the total momentum is finite and since the front form 
needs no particular Lorentz frame. 
Rather it is {\em frame-independent} and covariant.
{\em Light-Cone Quantization} seemed to be more appropriate.
Casher \cite{cas76} gave the first construction of the
light-cone Hamiltonian for non-Abelian gauge theory 
and gave an overview of important considerations 
in light-cone quantization.
Chang, Root, and Yan \cite{cry73a,cry73b,chy73,cha76}
demonstrated the equivalence of light-cone
quantization with standard covariant Feynman analysis. 
Brodsky, Roskies and Suaya \cite{brs73} calculated 
one-loop radiative corrections and demonstrated
renormalizability. Light-cone Fock methods were used by 
Lepage and Brodsky in the analysis of exclusive processes 
in QCD \cite {leb79a,leb79b,leb80,lbh83,
brl89,nam84}.
In all of this work was no citation of Dirac's work.
It did reappear first in the work of Pauli and Brodsky
\cite{pab85a,pab85b}, who explicitly diagonalize a 
light-cone Hamiltonian 
by the method of {\em Discretized Light-Cone Quantization},
see also Section~\ref{sec:DLCQ}. 
Light-{\em Front} Quantization appeared first in 
the work of Harindranath and Vary \cite{hav87,hav88}
adopting the above concepts without change. 
Franke and 
collaborators \cite {apf93,fnp81a,fnp81b,fnp82,prf89}, 
Karmanov \cite{kar80,kar81},  and Pervushin \cite{per90} 
have also done important work on light-cone quantization. 
Comprehensive reviews can be found in:
\cite{lbh83,brl89,brp91,jix93,brs95,gla95,bur95a}

\subsection{Parametrizations of the front form}
\label{sec:frontform} 

If one were free to parametrize the front form, one would
choose it most naturally as a {\em real rotation} of the 
coordinate system, with an angle $\varphi = \pi/4$.
The `time-like' coordinate would then
be $x ^+ = \widetilde x ^0$ and the `space-like' coordinate 
$x ^- = \widetilde x ^3$, or collectively
 \begin {equation} 
 \pmatrix {x ^+ \cr x ^- \cr} = {1 \over \sqrt 2} 
 \pmatrix {1 & -1 \cr  1 & \phantom{-} 1 \cr} 
 \pmatrix{ x ^0 \cr x ^3 \cr}  
 \qquad\quad {\rm and} \qquad g _{\alpha\beta} =
 \pmatrix { 0 & 1 \cr  1 & 0 \cr} 
 \ . \end{equation}
The metric tensor $g^{\mu\nu}$ obviously transforms
according to Eq.(\ref{eq:2.48}), and
the Jacobian for this transformation is unity.

But this has not what has been done, starting way back 
with Bardakci and Halpern \cite{bah68} and continuing
with Kogut and Soper \cite{kos70}. Their definition
corresponds  to a rotation of the coordinate system by
$\varphi = -\pi/4$ {\em and an reflection} of $x^-$. The
Kogut-Soper convention (KS)  \cite{kos70} is thus: 
 \begin {equation} 
 \pmatrix {x ^+ \cr x ^- \cr} = {1 \over \sqrt 2} 
 \pmatrix { 1 & \phantom{-}1 \cr  1 &  -1 \cr} 
 \pmatrix{ x ^0 \cr x ^3 \cr}  
 \qquad\quad {\rm and} \qquad g _{\alpha\beta} =
 \pmatrix { 0 & 1 \cr  1 & 0 \cr} 
 \ , \end{equation}
see also Appendix~\ref{app:Kogut-Soper}.
It is often convenient to distinguish longitudinal Lorentz 
indices $\alpha$ or $\beta$ ($+,-$)  
from the transversal ones $i$ or $j$ ($1,2$), 
and to introduce transversal vectors by 
$ \vec x _{\!\perp} = ( x^1, x ^2)$.
The KS-convention is particularly suited for theoretical 
work, since the raising and lowering of the Lorentz indices
is simple. 
With the totally antisymmetric symbol
\begin{equation}  
   \epsilon_{+12}^{\phantom{+12}+}=1
   \ , \quad{\rm thus}\quad
   \epsilon_{+12-}=1
\ ,\end{equation}
the volume integral becomes
\begin{equation}  
   \int\!d\omega_+ = \int\!dx^-d^2x_{\!\perp} 
   = \int\!dx_+d^2x_{\!\perp}
\ .\end{equation}
One should emphasize that
$ \partial _+ = \partial ^- $ is a time-like  derivative 
$\partial /\partial  x ^+ = \partial /\partial  x _-$
as opposed to $ \partial _- = \partial ^+ $, which is a
space-like derivative 
$\partial /\partial  x ^- = \partial /\partial  x _+$. 
Correspondingly, $ P _+ = P ^- $ is the Hamiltonian
which propagates in the light-cone time $x^+$, 
while $ P _- = P ^+ $ is the longitudinal space-like
momentum. 

In much of the practical work, however,
one is bothered with the $\sqrt{2}$'s scattered all over the
place. At the expense of having various factors of 2, 
this is avoided in the Lepage-Brodsky (LB)
convention \cite{leb80}: 
\begin {equation} 
        \pmatrix {x ^+ \cr x ^- \cr} = 
        \pmatrix { 1 & \phantom{-}1 \cr  1 &  -1 \cr} 
        \pmatrix{ x ^0 \cr x ^3 \cr}  
        \ , \qquad{\rm thus} \quad 
        g ^{\alpha\beta} = \pmatrix { 0 & 2 \cr  2 & 0 \cr} 
        \quad {\rm and} \quad 
        g _{\alpha\beta} =
        \pmatrix { 0 & {1\over2} \cr  {1\over2} & 0 \cr} 
\ , \end{equation}
see also Appendix~\ref{app:Lepage-Brodsky}.
Here, $ \partial _+ = {1\over2}\partial ^- $ is a time-like  
and $ \partial _- = {1\over2}\partial ^+ $ 
a space-like derivative.  The Hamiltonian is
$ P _+ = {1\over2} P ^- $, and $ P _- = {1\over2}P ^+ $ 
is the longitudinal momentum. 
With the totally antisymmetric symbol
\begin{equation}  
   \epsilon_{+12}^{\phantom{+12}+}=1
   \ , \quad{\rm thus}\quad
   \epsilon_{+12-}= {1\over2}
\ ,\end{equation}
the volume integral becomes
\begin{equation}  
   \int\!d\omega_+ = {1\over2}\int\!dx^-d^2x_{\!\perp} 
   = \int\!dx_+d^2x_{\!\perp}
\ .\end{equation}
We will use both the LB-convention and the KS-convention in 
this review, and indicate in each section 
which convention we are using.

The transition from the instant form to the front form is
quite simple: In all the equations found in sections
\ref {sec:qed} and \ref {sec:qcd} one has to substitute
the ``0'' by the ``+'' and the ``3'' by the ``-''.
Take as an example the QED four-momentum 
in Eq.(\ref {eq:2.24}) to get
\begin {eqnarray} 
    P_\nu & = & \int_\Omega \! d\omega _0 
    \ \biggl( F^{0\kappa}      F _{\kappa\nu} 
   +{1\over4} g^0_\nu F ^{\kappa\lambda} F _{\kappa\lambda} 
   +{1\over2} \Bigl[i\overline\Psi\gamma^0 D_\nu\Psi
   +\ {\rm h.c.} \Bigr] \biggr) 
\ ,\nonumber \\
   P_\nu & = & \int_\Omega \! d\omega _+ 
   \ \biggl( F^ {+\kappa} F _{\kappa\nu}
   +{1\over4}g^+_\nu F^ {\kappa\lambda} F_{\kappa\lambda}
   +{1\over2}\Bigl[i\overline\Psi \gamma^+ D _\nu \Psi 
   +\ {\rm h.c.} \Bigr] \biggr) 
\ ,\label{eq:2.53} \end {eqnarray}
also in KS-convention.
The instant and the front form look thus almost identical. 
However after having worked out the Lorentz algebra, 
the expressions for the instant and front form Hamiltonians 
are drastically different:  
\begin {eqnarray} 
   P _0  & = & {1\over2} 
   \int_\Omega \! d\omega _0 
   \ ( \vec E ^2 + \vec B ^2 )
   \quad + {1\over2} \int_\Omega \! d\omega _0 \Bigl[ 
   i \overline \Psi  \gamma ^+  D _0\Psi +\ {\rm h.c.}\Bigr]
\ ,\nonumber \\
   P _+  & = &  {1\over2} \int_\Omega \! d\omega _+ 
   \ ( E _\parallel ^2 + B_\parallel ^2)
   \quad + {1\over2} 
   \int_\Omega \! d\omega _+ \Bigl[ 
   i \overline\Psi\gamma^+ D _+\Psi +\ {\rm h.c.}\Bigr]
\ ,\label{eq:2.52}\end {eqnarray}
for the the instant and the front form energy, respectively.
In the former one has to deal with all three components of the
electric and the magnetic field, in the
latter only with two of them, namely with the longitudinal
components 
$ E _\parallel = {1\over2} F ^{+-}= E_z $
and 
$ B _\parallel = F ^{1 2} = B_z$.
Correspondingly, energy-momentum for 
non-abelian gauge theory is
\begin {eqnarray} 
    P_\nu & = & \int_\Omega \! d\omega _0 
    \ \biggl(F^{0\kappa}_a F ^a_{\kappa\nu} +
    {1\over4} g^0_\nu
    F^{\kappa\lambda}_a F_ {\kappa\lambda}^a +
    {1\over2}\Bigl[i\overline\Psi\gamma^0 
    T^a D^a_\nu\Psi  
  + \ {\rm h.c.} \Bigr] \biggr) 
, \nonumber \\
    P_\nu & = & \int_\Omega \! d\omega _+ 
    \ \biggl(F^{+\kappa}_a F ^a_{\kappa\nu} 
    + {1\over4} g^+_\nu
    F^{\kappa\lambda}_a F_ {\kappa\lambda}^a
    + {1\over2}\Bigl[i\overline\Psi\gamma^+
    T^a D^a_\nu\Psi  
  + \ {\rm h.c.} \Bigr] \biggr) 
\ .\label{eq:2.60} \end {eqnarray}
These expressions are exact but not yet very useful,
and we shall come back to them in later sections.
But they are good enough to discuss their symmetries
in general.

\subsection{The Poincare\'e symmetries in the front form}
\label{sec:symmetries} 

The algebra of the four-energy-momentum $ P^\mu= p ^\mu $ 
and four-angu\-lar-momen\-tum  
$ M^{\mu \nu} = x ^\mu p^\nu - x ^\nu p ^\mu $ 
for free particles \cite {bar48,sch61,tun85,wig39}
with the basic commutator 
$ {1\over i\hbar}[x ^\mu, p _\nu ] = \delta ^\mu _\nu $ is
 \begin {eqnarray}
     {1 \over i \hbar}
     \left[ P ^\rho, M^{\mu \nu} \right]  
     &=& 
     g ^{\rho \mu}  P  ^\nu - g ^{\rho \nu}  P  ^\mu \ , \quad
             {1 \over i \hbar}
     \left[ P ^\rho,  P ^\mu \right]  = 0 \  ,  
 \nonumber \\  
     \quad {\rm and} \ \quad {1 \over i \hbar}
     \left[ M ^{\rho \sigma} ,  M ^{\mu \nu} \right]  
     &=& 
     g  ^{\rho \nu} M ^{\sigma \mu}
     + g  ^{\sigma \mu} M ^{\rho \nu} 
     - g  ^{\rho \mu}  M ^{\sigma \nu}
     - g  ^{\sigma \nu}  M ^{\rho \mu}   
\  .\label{eq:group} \end  {eqnarray}
It is postulated that the generalized momentum 
operators satisfy the same commutator relations.
They form thus a group and  act as propagators
in the sense of the Heisenberg equations
\begin{eqnarray} 
   {1 \over i \hbar} \left[ P ^\nu ,\phi_r(x)\right]  
   &=& i\partial ^\nu \phi _r (x)  
\nonumber \\ {\rm and} \quad 
   {1\over i\hbar}\left[ M ^{\mu\nu},\phi_r(x)\right] 
   &=& \left( x ^\mu \partial ^\nu - 
   x ^\nu \partial ^\mu \right) \phi _r (x) +
   \Sigma ^{\mu\nu} _{rs} \phi _s 
\ .\label{eq:2.109} \end{eqnarray} 
Their validity for the front form was verified by
Chang, Root and Yang \cite{chy73,cry73a,cry73b}, 
and partially even before that by Kogut and Soper 
\cite{kos70}. Leutwyler and others have made
important contributions
\cite{leu68,leu69,leu74a,leu74b,leu74c,les78}. 
The ten constants of motion $ P ^\mu $ and $ M ^{\mu \nu} $ 
are  observables, thus hermitean operators with real 
eigenvalues. It is advantageous to construct representations 
in which the constants of motion are diagonal. 
The corresponding Heisenberg equations, for example, 
become then almost trivial.
But one cannot diagonalize all ten constants of motion 
simultaneously because they do not commute. 
One has to make a choice.

The commutation relations Eq.(\ref{eq:group}) define a group. 
The group is isomorphous to the { Poincare\'e group},  
to the ten $4 \times 4$ matrices which generate
an arbitrary inhomogeneous Lorentz transformation. 
The question of how many and which operators can be 
diagonalized simultaneously turns out to be identical to 
the problem of classifying all irreducible unitary 
transformations of the Poincare\'e group.
According to Dirac \cite{dir49} one cannot find more than
seven mutually commuting operators.

It is convenient to discuss the structure of the Poincare\'e 
group \cite {sch61,tun85} in terms of the 
Pauli-Lubansky vector
$  V ^\kappa \equiv \epsilon  ^{\kappa \lambda \mu \nu} 
                 P _\lambda  M _{\mu \nu} $,
with $ \epsilon  ^{\kappa \lambda \mu \nu}$ being  the 
totally antisymmetric symbol in 4 dimensions.
$ V  $ is orthogonal to the generalized momenta, 
$  P _\mu  V ^\mu = 0$, and obeys  the algebra
\begin{eqnarray} 
      {1\over i \hbar} \left[ V ^\kappa ,  P ^\mu \right] 
      &=& 0 
\ , \nonumber \\
      {1 \over i \hbar} \left[ V ^\kappa ,  M ^{\mu \nu} \right] 
      &=&
      g ^{\kappa \nu}  V ^\mu -  g ^{\kappa \mu}  V ^\nu 
\ , \nonumber \\
      {1 \over i \hbar} \left[ V ^\kappa ,  V ^\lambda \right] 
      &=&
      \epsilon^{\kappa\lambda\mu\nu} V _\mu  P _\nu 
\ . \end{eqnarray} 
The two group invariants are 
the operator for the invariant mass-squared 
$  M^2  =  P ^\mu  P _\mu $
and the operator for intrinsic spin-squared
$  V^2  =  V ^\mu  V _\mu $.
They are Lorentz scalars and commute with all 
generators $  P ^\mu $ and $  M ^{\mu \nu} $, 
as well as with all $ V ^\mu$.
A convenient choice of the six mutually commuting 
operators is therefore for the front form:

\begin{tabular}{lll}
(1)   & the invariant mass squared 
      & $M^2 =  P ^\mu  P _\mu $,   \\
(2-4) & the three space-like momenta 
      & $P ^+$ and $\vec P _{\!\bot}$, \\
(5)   & the total spin squared     
      & $ S^2 = V ^\mu   V _\mu $,  \\
(6)   & and one component of $ V $, say $ V ^+$,
      called\qquad  & $S_z$. 
\end{tabular}

\noindent
There are other equivalent choices.
In constructing a representation which diagonalizes 
simultaneously the six mutually commuting operators 
one can proceed consecutively, in principle, by 
diagonalizing one after the other. 
At the end, one will have realized the old dream 
of Wigner \cite {wig39} and of Dirac \cite{dir49} to 
classify physical systems with the quantum numbers of
the irreducible representations of the Poincare\'e group.

Inspecting the definition of  boost-angular momentum 
$M_{\mu\nu}$ in Eq.(\ref{eq:2.17}) one identifies which 
components are dependent on the interaction and which not.
Dirac \cite{dir49,dir64} calls them complicated and simple, or 
dynamic and kinematic, or  Hamiltonians and Momenta, respectively. 
In the instant form, the three components of the 
boost vector $K_i = M_{i0}$ are dynamic, and
the three component of angular momentum
$J_i=\epsilon_{ijk} M_{jk}$ are kinematic. 
The cyclic symbol $\epsilon_{ijk} $ is 1, if the  
space-like indices $ijk$ are in cyclic 
order, and zero otherwise.

The front form is special in having four kinematic four kinematic
components of $M _{\mu\nu}$
($ M _{+-}, M _{12}, M _{1-}, M _{2-}$)  and 
two dynamic components ($ M _{+1}$ and $ M_{+2}$), 
as noted already by Dirac \cite{dir49}. One checks this 
directly from the defining equation (\ref{eq:2.17}).
Kogut and Soper \cite{kos70} discuss and interpret 
them in terms of the above boosts and angular momenta.
They introduce the transversal vector $\vec B_{\perp}$ 
with components 
\begin{equation}
      B_{\!\perp 1} = M_{+1} = {1\over\sqrt 2} (K_1+J_2)
      \qquad{\rm and}\quad
      B_{\!\perp 2} =M_{+2} = {1\over\sqrt 2} (K_2-J_1)
\ . \end{equation}
In the front form they are kinematic and boost the system
in $x$ and $y$-direction, respectively. The kinematic 
operators 
\begin{equation}
      M_{12} = J_3
      \qquad{\rm and}\quad
      M_{+-} = K_3
\end{equation}
rotate the system in the $x$-$y$ plane and boost it
in the longitudinal direction, respectively. 
In the front form one deals thus with seven mutually 
commuting operators \cite{dir49}
\begin{equation}
      M_{+-}\ ,\ \vec B_{\perp}\ , \quad{\rm and\ all\ } P^\mu
\ ,\end{equation}
instead of the six in the instant form. 
The remaining two Poincare\'e
generators are combined into a transversal 
angular-momentum vector $\vec S_{\!\perp} $ with
\begin{equation}
      S_{\!\perp 1} = M_{1-} = {1\over \sqrt 2} (K_1-J_2) 
\quad {\rm and} \quad
      S_{\!\perp 2} = M_{2-} = {1\over \sqrt 2}(K_2+J_1) 
\ .\end{equation}
They are both dynamical, but commute  with each other 
and $M^2$. They are thus members of a dynamical subgroup 
\cite{kos70}, whose relevance has yet to be exploited.

Thus one can diagonalize the light-cone energy $P^-$ 
within a Fock basis where the constituents have fixed 
total $P^+$ and $\vec P_\perp$.
For convenience we shall define a `light-cone Hamiltonian'
as the operator
\begin{equation}
   H_{LC} = P^\mu P_\mu = P^- P^+ - \vec P^2_{\!\perp}
\ ,\label{eq:LC-hamiltonianI} \end{equation}
so that its eigenvalues correspond to the 
invariant mass spectrum $M_i$ of the theory.
The boost invariance of the eigensolutions of $H_{LC}$ 
reflects the fact that the boost operators $K_3$ 
and $\vec B_{\!\perp}$ are kinematical. 
In fact one can boost the system to an `intrinsic frame'
in which the transversal momentum vanishes
\begin{equation}
   \vec P_{\!\perp} = \vec 0
\ ,\qquad{\rm thus}\quad    H_{LC} = P^- P^+ 
\ .\label{eq:LC-hamiltonianII} \end{equation}
In this frame, the longitudinal component of the
Pauli-Lubansky vector reduces to the longitudinal
angular momentum $J_{3}= J_{z}$, which allows for
considerable reduction of the numerical work
\cite{trp96}. The transformation to an arbitrary
frame with finite values of $\vec P_{\!\perp}$ 
is then trivially performed.

The above symmetries imply the very important aspect
of the front form that both the Hamiltonian 
and all amplitudes obtained in light-cone perturbation 
theory (graph by graph!) are manifestly invariant under 
a large class of Lorentz transformations:

\begin{tabular}{llll}
(1)& boosts along the 3-direction:
   & $p^+ \rightarrow C_{\parallel}^{\phantom{-1}} p^+$
   & $\vec p_{\!\perp}\rightarrow \vec p_{\!\perp}$
\\
 & & $p^- \rightarrow C_{\parallel}^{-1} p^-$
\\
(2)& transverse boosts:
   & $p^+ \rightarrow p^+\ $
   & $\vec p_{\!\perp} \rightarrow \vec p_{\!\perp}
     + p^+ \vec C_{\!\perp}\ $ 
\\
 & & $p^- \rightarrow p^- 
     + 2\vec p_{\!\perp}\cdot \vec C_{\!\perp}
     + p^+ \vec C_{\!\perp}^2$ 
\\
(3)& rotations in the $x$-$y$ plane:
   & $p^+ \rightarrow p^+\ ,$
   & $\vec p_{\!\perp}^{\;2} \rightarrow 
         \vec p_{\!\perp}^{\;2}\ .$
\end{tabular}

\noindent
All of these hold for every single particle momentum 
$p^\mu$, and for any set of dimensionless c-numbers
$ C_{\parallel}$ and $ \vec C_{\!\perp}$.
It is these invariances which also lead to the frame 
independence of the Fock state wave functions.
  
If a theory is rotational invariant, then each eigenstate 
of the Hamiltonian which describes a state of nonzero mass 
can be classified in its rest frame by its spin eigenvalues
\begin{eqnarray}
      \vec J ^2 \left| P_0 = M ,\vec P = \vec 0 \right\rangle 
      &=&  s (s+1) \left|P^0= M, \vec P = \vec 0 \right\rangle
\ , \nonumber \\ {\rm and}\quad
      J_z \left|P^0= M, \vec P = \vec 0 \right\rangle 
      &=& s_z \left| P^0= M, \vec P = \vec 0 \right\rangle
\ .\end{eqnarray}
This procedure is more complicated in the front form since 
the angular momentum operator does not commute with 
the invariant mass-squared operator $M^2$.
Nevertheless, Hornbostel \cite{hor90,hor91,hor92}  
constructs light-cone operators 
\begin{eqnarray}
      \vec {\cal J} ^2
      &=& {\cal J}^2_3+\vec {\cal J} ^2_{\!\perp}
\ , \nonumber \\ {\rm with} \quad
      {\cal J}_3
      &=& J_3 + \epsilon_{i j}B_{\perp i}P_{\perp j}/P^+ 
\ , \nonumber \\ {\rm and} \quad
      {\cal J}_{\!\perp k} 
      &=& {1 \over M}\epsilon_{k \ell}
      (S_{\perp \ell} P^+ - B_{\perp \ell} P^- -
      K_3 P_{\perp \ell} + {\cal J}_3 
      \epsilon_{\ell m} P_{\perp m}) 
\ ,\end{eqnarray}
which, in principle, could be applied to an eigenstate 
$\left| P^+, \vec P_{\!\perp}\right\rangle$ 
to obtain the rest frame spin
quantum numbers. 
This is straightforward for ${\cal J}_3$ since it
is kinematical; in fact, ${\cal J}_3 = J_3$ in a frame with
$\vec P_{\!\perp} = \vec 0_{\!\perp}$. 
However, $\vec {\cal J}_{\!\perp}$ is dynamical and depends 
on the interactions. Thus it is generally difficult to explicitly 
compute the total spin of a state using light-cone quantization. 
Some of the aspects have been discussed by
Coester \cite{coe92} and 
collaborators \cite{cop82,cpc88}.
A practical and simple way has been applied by  
Trittmann \cite{trp96}. Diagonalizing the light-cone
Hamiltonian in the intrinsic frame for $J_z\neq0$, 
he can ask for $J_{max}$, the maximum eigenvalue of $J_z$
within a numerically degenerate multiplet of 
mass-squared eigenvalues.
The total `spin $J$' is then determined by $J=2J_{max}+1$, 
as to be discussed in Section~\ref{sec:DLCQ}. 
But more work on this question is certainly necessary, 
as well as on the discrete symmetries like parity and 
time-reversal and their quantum numbers for a particular 
state, see also Hornbostel \cite{hor90,hor91,hor92}. 
One needs the appropriate language for dealing with spin
in highly relativistic systems.

\subsection{The equations of motion 
    and the energy-momentum tensor}
\label{sec:equations-of-motion}

Energy-momentum for gauge theory had been given 
in Eq.(\ref{eq:2.60}).
They contain time-derivatives of the fields which can
be eliminated using the equations of motion.

\textbf{The color-Maxwell equations} are given 
in Eq.(\ref{eq:2.30}). They are four (sets of) equations for
determining the four (sets of) functions $A^\mu_a$. 
One of the equations of motion is removed by fixing the gauge
and we choose the light-cone gauge \cite{bas91}
\begin{equation} A ^+ _a = 0 
\ .\label{eq:LCgauge}\end{equation}
Two of the equations of motion express the time derivatives of the
two transversal components $\vec A ^a _{\!\perp} $ 
in terms of the other fields. Since the front form
momenta in Eq.(\ref{eq:2.60}) do not depend on them,
we discard them here. The fourth is the analogue of
the Coulomb equation or of the Gauss' law in the instant form, 
particularly $ \partial _\mu  F  _a^{\mu +} = g  J   _a^+$.
In the light-cone gauge the color-Maxwell charge density 
$J ^+_a$ is independent of the vector potentials, and the 
Coulomb equation reduces to  
\begin{equation}
   - \partial^+\partial_-  A  _a^-  
   - \partial^+ \partial_i  A  _{\!\perp a}^i   = g J^+_a
\ . \end{equation}
This equation involves only (light-cone) space-derivatives.
Therefore, it can be satisfied only, if one of the components
is a functional of the others. There are subtleties involved
in actually doing this, in particular one has to cope with 
the `zero mode problem', see for example \cite{pkp95}.
Disregarding this here, one inverts the equation by
\begin{equation}  
   A_+^a = \widetilde A_+^a + {g\over (i\partial^+)^2}\,J^+_a
\ .\label{eq:2.57}\end{equation}
For the free case ($g=0$), $A ^-$ reduces 
to $\widetilde A ^-$.
Following Lepage and Brodsky \cite{leb80}, one can 
collect all components which survive the limit $g\rightarrow0$
into the `free solution' $\widetilde A_a^\mu $, defined by
\begin {equation} 
   \widetilde A_+^a=   
   -{1\over\partial^+}\ \partial_i A^i_{\!\perp a} 
   \ ,\qquad{\rm thus}\quad
   \widetilde A_a^\mu = \left( 0,\vec A_{\!\perp a},
   \widetilde A^+_a \right) 
\ . \end {equation}
Its four-divergence vanishes by construction 
and the Lorentz condition 
$\partial _\mu \widetilde A ^\mu _a = 0$
is satisfied {\it as an operator}. 
As a consequence, $\widetilde A ^\mu _a $ is 
purely transverse.
The inverse space derivatives 
$\left(i\partial^+\right)^{-1}$ and 
$\left(i\partial^+\right)^{-2}$ are actually
Green's functions. 
Since they depend only on $x^-$, they are comparatively 
simple, much simpler than in the instant form where
$(\vec \nabla ^2)^{-1}$ depends on all three space-like 
coordinates. 

\textbf{The color-Dirac equations} are defined 
in Eq.(\ref{eq:2.35}) and are used here to express the 
time derivatives $\partial_+ \Psi $ as function of the 
other fields. After multiplication with
$\beta=\gamma^0$ they read explicitly
\begin{equation}   \bigl(
       i \gamma^0\gamma^+ T^a D^a _+ + 
       i \gamma^0\gamma^-  T^a D^a _-   +
       i \alpha^i_{\!\perp}   T^a D^a _{\!\perp i} \bigr) \Psi
       = m \beta\Psi
\ , \end{equation}
with the usual $\alpha^k=\gamma^0\gamma^k$, 
$k=1,2,3$. In order to isolate the  time derivative 
one introduces the 
projectors  $\Lambda _\pm = \Lambda ^\pm $
and projected spinors $ \Psi_\pm  = \Psi ^\pm $ by
\begin{equation}
    \Lambda _\pm 
   = {1\over2}(1\pm\alpha^3)     
   \qquad{\rm and}\qquad
   \Psi_\pm = \Lambda_\pm\Psi
\ . \label{eq:2.59}\end{equation}
Note that the raising or lowering of the projector labels
$\pm$ is irrelevant. 
The $ \gamma^0\gamma^\pm$ are obviously related to 
the $ \Lambda ^\pm$, but  differently in the KS- and
LB-convention 
\begin{equation}
       \gamma^0\gamma^\pm
       = 2 \,\Lambda ^\pm _{\rm LB} 
       = \sqrt{2}\,\Lambda ^\pm _{\rm KS} 
\ . \label{eq:2.71} \end{equation}
Multiplying the color-Dirac equation
once with $\Lambda^+$ and once with $\Lambda^-$,
one obtains a coupled set of spinor equations 
\begin{eqnarray} 
   2i\partial _+\Psi _+ &=&  \left( m \beta - 
   i\alpha_{\!\perp}^i T^a D^a _{\!\perp i} \right) \Psi _-  
   +  2g   A  _+^a T^a \Psi _+
\ , \nonumber \\ {\rm and} \quad 
   2i\partial _- \Psi _-  &=&  \left( m \beta - 
   i\alpha_{\!\perp}^i T^a D^a _{\!\perp i}\right)  \Psi_+
    + 2gA_-^a T^a \Psi _-
\ . \label{eq:2.61}\end {eqnarray} 
Only the first of them involves a time derivative.
The second is a constraint, similar to the above in the 
Coulomb equation. With the same proviso in mind, 
one defines 
\begin{equation} 
       \Psi _{-} =  {1\over 2i\partial_-}    \left( m\beta - 
       i\alpha_{\!\perp}^i T^a D^a _{\!\perp i} \right) \Psi_+
\ .\label{PsiMinus}\end{equation}
Substituting this  in  the former,  the time derivative is
\begin{equation} 
   2i\partial _+ \Psi _+  
   =  2g   A  _+^a T^a \Psi ^+ +   \left(m\beta -
   i\alpha_{\!\perp}^j T^a D^a _{\!\perp j}\right) 
   {1\over 2i\partial _-} \left( m\beta -
   i\alpha_{\!\perp}^i T^a D^a _{\!\perp i} \right) \Psi _+
\,.\label{eq:2.63}\end{equation} 
Finally, in analogy to the color-Maxwell case 
once can conveniently introduce the
free spinors $\widetilde\Psi 
= \widetilde\Psi _+ + \widetilde\Psi _- $ by
\begin {equation} 
   \widetilde\Psi = \Psi _+
   + \left(  m\beta - i\alpha ^i \partial  _{\!\perp i}\right) 
   {1\over 2i\partial _-} \Psi _+
\ .\label{eq:2.64}\end {equation}
Contrary to the full spinor, see for example Eq.(\ref{PsiMinus}), 
$\widetilde\Psi $ is independent of the interaction.
To get the corresponding relations for the KS-convention,
one substitutes the ``2'' by ``$\sqrt{2}$'' in accord with 
Eq.(\ref{eq:2.71}). 

The front form Hamiltonian according to Eq.(\ref{eq:2.60})
is 
\begin{equation}
   P _+ = \int_\Omega\!d\omega _+\ \biggl( F ^{+\kappa} 
   F _{\kappa+}  + 
   {1\over4} F^{\kappa\lambda}_a F_{\kappa\lambda}^a
   + {1\over2} \Bigl[ 
   i\overline\Psi\gamma^+ T ^a D ^a_+\Psi + 
   \ {\rm h.c.} \Bigr] \biggr) 
\ .\end{equation}
Expressing it as a functional of  the fields
will finally lead to  Eq.(\ref{eq:2.87}) below, but 
despite the straightforward calculation we display 
explicitly the intermediate steps. 
Consider first the energy density of the 
color-electro-magnetic fields
${1\over4} F^{\kappa\lambda} F_{\kappa\lambda}
+ F ^{+\kappa} F _{\kappa+}$.
Conveniently defining the abbreviations
\begin{equation} 
   B ^{\mu\nu}_a =f^{a b c} A ^\mu  _b A ^\nu _c
   \quad{\rm and}\quad
   \chi ^\mu_a   =f^{a b c} \partial ^\mu 
   A ^\nu_b A _\nu^c
\ ,\end{equation} 
the field tensors in Eq.(\ref{eq:2.31}) are rewritten as
$ F ^{\mu\nu}_a = \partial^\mu  A ^\nu_a
   - \partial^\nu  A ^\mu_a  - g  B ^{\mu\nu}_a $
and typical tensor contractions become
\begin{equation} 
   {1\over2} F ^{\mu\nu}_a F _{\mu\nu}^a 
   = \partial^\mu  A ^\nu_a \partial_\mu  A _\nu^a
   -  \partial^\mu  A ^\nu_a \partial_\nu  A _\mu^a
   + 2\chi^\mu_a A _\mu^a
   + {g^2\over2} B ^{\mu\nu}_a B _{\mu\nu}^a 
\ .\end{equation} 
Using 
$F ^{\alpha\kappa} F _{\alpha\kappa} 
= 2 F ^{+\kappa} F _{+\kappa}$,
the color-electro-magnetic energy density  
\begin{eqnarray} 
   {1\over4} F^{\kappa\lambda} F_{\kappa\lambda}
   + F ^{+\kappa} F _{\kappa+} 
   &=& {1\over4} F^{\kappa\lambda} F_{\kappa\lambda}
   - {1\over2} F ^{\kappa\alpha} F _{\kappa\alpha}  
   = {1\over4} F^{\kappa i} F_{\kappa i}
   - {1\over4} F ^{\kappa\alpha} F _{\kappa\alpha} 
\nonumber \\ 
   &=& {1\over4} F^{i j} F_{i j}
   -{1\over4} F^{\alpha\beta} F_{\alpha\beta}
\end{eqnarray} 
separates completely into a longitudinal ($\alpha,\beta$) 
and a transversal contribution ($i ,j $) \cite{pkp95}, see
also Eq.(\ref{eq:2.52}). 
Substituting $ A _+$ by Eq.(\ref{eq:2.57}),   
the color-electric and color-magnetic parts become
\begin{eqnarray} 
   {1\over4} F ^{\alpha\beta} F _{\beta\alpha} 
   &=& {1\over2} \partial^+ A_+\partial^+ A_+
   = {g^2\over2} J ^+ {1\over(i\partial^+)^2} J ^+ 
   + {1\over2}(\partial_i A ^i_{\!\perp})^2
   + g J ^+ \widetilde A _+
\ ,\nonumber \\
   {1\over4} F ^{i j} F _{i j} 
   &=& {g^2\over4} B ^{i j} B _{i j}  
   - {1\over2}(\partial_i A ^i_{\!\perp})^2 + \chi^i  A  _i
   + {1\over2} A ^j (\partial^i\partial_i) A _j
\ ,\label{eq:2.gluons}\end{eqnarray} 
respectively. The role of the different terms will be discussed
below. The color-quark energy density is
evaluated in the LS-convention. With 
$i\overline\Psi\gamma^+ D ^a_+T ^a \Psi  =
   i\Psi^\dagger\gamma^0 \gamma^+ D ^a_+T ^a \Psi $
and the projectors of Eq.(\ref{eq:2.59}) one gets first
$  i\overline\Psi\gamma^+ D ^a_+ T ^a \Psi  =
   i\sqrt{2} \Psi_+^\dagger D ^a_+ T ^a \Psi _+   $.
Direct substitution of  the time derivatives in
Eq.(\ref{eq:2.63})  gives then
\begin{equation}
   i\overline\Psi\gamma^+ D ^a_+ T ^a \Psi  
   =  \Psi_+^\dagger\left(m\beta -
   i\alpha_{\!\perp}^j D^a _{\!\perp j} T ^a \right) 
   {1\over \sqrt{2} i\partial _-} \left( m\beta -
   i\alpha_{\!\perp}^i D^b _{\!\perp i} T ^b \right) \Psi _+ 
\ .\end{equation}
Isolating the interaction in the covariant derivatives 
$ i T ^a D ^a _\mu = i \partial_\mu - g T ^a A ^a _\mu $
produces 
\begin{eqnarray}
   i\overline\Psi\gamma^+ D ^a_+ T ^a \Psi  &=&
   g\widetilde \Psi_+^\dagger\alpha_{\!\perp}^j  A^a _{\!\perp j} 
   T ^a \widetilde \Psi_- 
   + g\widetilde \Psi_-^\dagger\alpha_{\!\perp}^j  A^a _{\!\perp j} 
   T ^a \widetilde \Psi_+ 
\nonumber \\   
   &+& {g^2\over \sqrt{2}}\Psi_+^\dagger
   \alpha_{\!\perp}^j A^a _{\!\perp j} T ^a 
   {1\over i \partial _-} 
   \alpha_{\!\perp}^i A^b _{\!\perp i} T ^b \Psi _+
\nonumber \\   
   &+& {1\over \sqrt{2}} \Psi_+^\dagger\left(m\beta -
   i\alpha_{\!\perp}^j \partial_{\!\perp j} \right) 
   {1\over  i\partial _-} \left( m\beta -
   i\alpha_{\!\perp}^i \partial_{\!\perp i} \right) \Psi _+
\ .\end{eqnarray}
Introducing $\widetilde j ^\mu_a$ as the color-fermion 
part of  the total current $\widetilde J ^\mu_a$, that is  
\begin {equation} 
   \widetilde j ^\nu_a (x) 
   =\overline{\widetilde\Psi}\gamma^\nu T^a \widetilde\Psi  
   \,\quad{\rm with}\quad
   \widetilde J ^\nu_a (x) 
   = \widetilde j ^\nu_a (x) + \widetilde \chi ^\nu_a (x) 
\ ,\end {equation}
one notes that $ J ^+_a = \widetilde J ^+_a  $ when
comparing with  the defining equation (\ref{eq:2.64}). 
For the transversal parts holds obviously
\begin{equation}
   \widetilde j ^{\,i}_{\!\perp a}  = 
   \widetilde \Psi^\dagger\alpha_{\!\perp}^i  
   T ^a \widetilde \Psi  = 
   \widetilde \Psi_+^\dagger\alpha_{\!\perp}^i 
   T ^a \widetilde \Psi_- +
   \widetilde \Psi_-^\dagger\alpha_{\!\perp}^i  
   T ^a \widetilde \Psi_+  
\ .\end{equation}
With $\gamma^+ \gamma^+ = 0 $ one finds
\begin{eqnarray}
   \overline{\widetilde \Psi}\,\gamma^\mu \widetilde A _\mu 
   \gamma ^+\gamma^\nu\widetilde A _\nu\,\widetilde \Psi
   &=& \overline{\widetilde \Psi}
   \gamma^i_{\!\perp}\widetilde A _{\!\perp i}\,\gamma ^+
   \gamma^i_{\!\perp}\widetilde A _{\!\perp i}\,\widetilde \Psi
   =\widetilde\Psi ^\dagger
   \alpha^i_{\!\perp}\widetilde A _{\!\perp i}
   \gamma^+\gamma ^0 
   \alpha^j_{\!\perp}\widetilde A _{\!\perp j} 
   \,\widetilde \Psi 
\nonumber \\
   &=& \sqrt{2} \widetilde \Psi ^\dagger _+
   \,\alpha^i_{\!\perp}  \widetilde A _{\!\perp i} 
   \,\alpha^i_{\!\perp}  \widetilde A _{\!\perp i} 
   \,\widetilde \Psi_+
\ ,\end{eqnarray}
see also \cite{lbh83}.
The covariant time-derivative of the dynamic spinors 
$\Psi_\alpha$ is therefore
\begin{equation}
   i\overline\Psi\gamma^+ D ^a_+ T ^a \Psi  
   = g \widetilde j ^i _{\!\perp}\widetilde A  _{\!\perp i}
   + {g^2\over2} \overline{\widetilde \Psi}
   \gamma^\mu \widetilde{\bf A }_\mu
   {\gamma^+\over i\partial ^+} 
   \gamma^\nu\widetilde{\bf A }_\nu\,\widetilde \Psi
   + {1\over2}\overline{\widetilde \Psi} \gamma^+
      {m^2 -\nabla_{\!\perp} ^2\over i\partial ^+}\widetilde\Psi
\,\label{eq:2.quarks}\end{equation}
in terms of the fields  
$\widetilde A _\mu$ and $\widetilde\Psi _\alpha$.
One finds the same expression in LB-convention.  
Since it is a hermitean operator one can
add Eqs.(\ref{eq:2.gluons}) and (\ref{eq:2.quarks})
to finally get the front form Hamiltonian as a sum of
five terms 
\begin{eqnarray} 
   P _+
   &=& {1\over2}\int\!dx_+d^2x_{\!\perp} \biggl(
   \overline{\widetilde\Psi} \gamma^+
   {m^2 +(i\nabla_{\!\!\perp}) ^2 \over i\partial^+}
   \widetilde\Psi   +
   \widetilde A ^\mu_a (i\nabla_{\!\!\perp}) ^2 
   \widetilde A _\mu^a     \biggr)
\nonumber  \\ 
   &+& g  \int\!dx_+d^2x_{\!\perp}
      \ \widetilde J ^\mu_a \widetilde A _\mu^a        
\nonumber  \\ 
   &+& { g  ^2 \over 4} \int\!dx_+d^2x_{\!\perp}
      \ \widetilde B ^{\mu\nu}_a \widetilde B _{\mu\nu}^a 
\nonumber  \\ 
   &+& { g  ^2 \over 2} \int\!dx_+d^2x_{\!\perp}
      \ \widetilde J ^+_a 
     {1\over \left(i \partial ^+ \right)^2} \widetilde J ^+_a 
\nonumber \\
   &+&{ g  ^2 \over2} \int\!dx_+d^2x_{\!\perp}
    \ \overline{\widetilde\Psi}  \gamma ^\mu  T  ^a 
    \widetilde A ^a_\mu \ {\gamma ^+\over i\partial ^+}
    \left( \gamma ^\nu  T  ^b \widetilde A ^b_\nu 
    \widetilde\Psi  \right)
\ .\label{eq:2.87} \end{eqnarray}
Only the first term survives the limit $g\rightarrow 0$, 
hence  $P ^- \rightarrow\widetilde P ^-$,  referred to 
as the free part of the Hamiltonian.
For completeness, the space-like components 
of energy-momentum as given in Eq.(\ref{eq:2.60}) 
become
\begin{eqnarray}
   P_k &=&  \int\!dx_+d^2x_{\!\perp} \ \biggl( F^ {+\kappa} 
   F _{\kappa k} + 
   i \overline \Psi  \gamma ^+T^a D^a _k \Psi 
\nonumber \\       
   &=& \int\!dx_+d^2x_{\!\perp} \Bigl(
   \overline{\widetilde\Psi}\ \gamma^+ i\partial _k  
   \widetilde \Psi  
   + \widetilde A ^\mu_a \ \partial^+\partial _k  
   \widetilde A_\mu^a \Bigr)  
   \ ,\qquad{\rm for}\ k=1,2,-
\ .\label{eq:total-momenta}\end{eqnarray}
Inserting the free solutions as given below in
Eq.(\ref{eq:2.72}),  one gets for $\widetilde P  ^\mu 
= (P ^+, \vec P _{\!\bot} , \widetilde P ^-)$
\begin{equation}  
   \widetilde P  ^\mu = \sum _{\lambda,c,f} 
   \int\!dp^+d^2 \!p_{\!\bot}  \  p ^\mu \left( 
   \widetilde a^\dagger(q) \widetilde a(q) + 
   \widetilde b^\dagger(q) \widetilde b(q) + 
   \widetilde d^\dagger(q) \widetilde d(q) 
   \right)
\ ,\end{equation}
in line with expectation: In momentum representation 
the momenta $ \widetilde P ^\mu $ are diagonal
operators. 
Terms depending on the coupling constant are interactions
and in general are non-diagonal operators in Fock space.

Equations (\ref{eq:2.87}) and (\ref{eq:total-momenta})
are quite generally applicable: 
\begin{itemize}
\item  
They hold both in the Kogut-Soper and Lepage-Brodsky
convention. 
\item  
They hold for arbitrary non-abelian gauge theory $SU(N)$. 
\item   
They hold therefore also for QCD ($N=3$)
and are manifestly invariant under color rotations.
\item  
They hold for abelian gauge theory (QED), 
formally  by replacing the color-matrices 
$T^a_{c,c^\prime}$ with the unit matrix and by setting to 
zero the structure constants $f^{abc}$, 
thus $B^{\mu\nu}=0$ and $\chi^{\mu}=0$.
\item  
They hold for 1 time dimension and arbitrary
$d+1$ space dimensions, with $i=1,\dots,d$.
All what has to be adjusted is the volume integral
$\int\!dx_+d^2x_{\!\perp} $. 
\item  
They thus hold also for  the popular
toy models in 1+1 dimensions.
\item  
Last but not least, they hold for the `dimensionally
reduced models' of gauge theory, formally by setting to 
zero the transversal derivatives of the free fields, that is
$\vec\partial_{\!\perp} \widetilde \Psi _\alpha= 0$ and 
$\vec\partial_{\!\perp} \widetilde A _\mu= 0$.
\end{itemize}
Most remarkable, however, is that the relativistic
Hamiltonian in Eq. (\ref{eq:2.87}) is {\em additive} 
\cite{kos70} in the `kinetic' and the `potential' energy, 
very much like a non-relativistic Hamiltonian
\begin{equation}
   H = T + U
\ .\label{eq:2.91}\end{equation}
In this respect the front form is distinctly different  from  
the conventional instant form. With $H \equiv P_+$ 
the {\em kinetic energy}  
\begin{equation}
   T = \widetilde P _+ = {1\over2}\int\!dx_+d^2x_{\!\perp} 
   \biggl(\overline{\widetilde\Psi} \gamma^+
   {m^2 +(i\nabla_{\!\!\perp}) ^2 \over i\partial^+}
   \widetilde\Psi   +
   \widetilde A ^\mu_a (i\nabla_{\!\!\perp}) ^2 
   \widetilde A _\mu^a     \biggr)
\end{equation}
is the only term surviving the limit $ g \rightarrow 0$
in Eq.(\ref{eq:2.87}). 
The {\em potential energy} $U$ is correspondingly the 
sum of the four terms 
\begin{equation}
     U = V + W_1+W_2+W_3
\ .\label{eq:2.93}\end{equation}
Each of them has a different origin and interpretation.
The {\em vertex interaction}
\begin{equation}
    V = g  \int\!dx_+d^2x_{\!\perp} 
      \ \widetilde J ^\mu_a \widetilde A _\mu^a        
\label{eq:2.94}\end{equation}
is the light-cone analogue of the $J_\mu A^\mu$-structures 
known from covariant theories particularly electrodynamics.
It generates three-point-vertices describing 
bremsstrahlung and pair creation. However, since 
$\widetilde J ^\mu$ contains also the pure gluon part 
$\widetilde\chi^\mu$, it includes the three-point-gluon 
vertices as well. The {\em four-point-gluon interactions}
\begin{equation}
     W_1 = { g  ^2 \over 4} \int\!dx_+d^2x_{\!\perp} 
      \ \widetilde B ^{\mu\nu}_a \widetilde B _{\mu\nu}^a 
\label{eq:2.95}\end{equation}
describe the four-point gluon-vertices. They 
are typical for non-abelian gauge theory and come only
from the color-magnetic fields in Eq.(\ref{eq:2.gluons}). 
The {\em instantaneous-gluon interaction}
\begin{equation}
     W_2 = { g  ^2 \over 2} \int\!dx_+d^2x_{\!\perp} 
      \ \widetilde J ^+_a 
     {1\over \left(i \partial ^+ \right)^2} \widetilde J ^+_a 
\label{eq:2.96}\end{equation}
is the light-cone analogue of the Coulomb energy,
having  the same structure (density-propagator-density)
and the same origin, namely  Gauss' equation (\ref{eq:2.57}).
$W_3$ describes quark-quark, gluon-gluon, 
and quark-gluon instantaneous-gluon interactions. 
The last term, finally, is the 
{\em instantaneous-fermion interaction}
\begin{equation}
    W_3 = { g  ^2 \over2} \int\!dx_+d^2x_{\!\perp} 
    \ \overline{\widetilde\Psi}  \gamma ^\mu  T  ^a 
    \widetilde A ^a_\mu \ {\gamma ^+\over i\partial ^+}
    \left( \gamma ^\nu  T  ^b \widetilde A ^b_\nu 
    \widetilde\Psi  \right)
\ .\label{eq:2.97}\end{equation}
It originates from the light-cone specific decomposition 
of Dirac's equation (\ref{eq:2.61}) and 
has no counterpart in conventional theories.
The present formalism is however more symmetric:
The instantaneous gluons and the instantaneous fermions
are partners. This has some interesting  consequences, 
as we shall see below. Actually, the instantaneous
interactions were seen first by Kogut and Soper \cite{kos70} 
in the time-dependent analysis of the scattering
amplitude as remnants of choosing the light-cone gauge.

One should carefully distinguish the above front-form
Hamiltonian $H$ from the light-cone Hamiltonian $H_{LC}$, 
defined in Eqs.(\ref{eq:LC-hamiltonianI})
and (\ref{eq:LC-hamiltonianII}) as the operator of invariant
mass-squared. The former is the time-like component of 
a four-vector and therefore frame dependent. 
The latter is a Lorentz scalar
and therefore independent of the frame.
The former is covariant, the latter invariant under Lorentz
transformations, particularly under boosts.
The two are related to each other by multiplying $H$ with
a number, the eigenvalue of $2P^+$:
\begin{equation}
   H_{LC} = 2P^+ H  
\ .\label{eq:LC-hamiltonianIII} \end{equation}
The above discussion and interpretation of $H$ applies 
therefore also to $H_{LC}$. 
Note that matrix elements of the `Hamiltonian' have the 
dimension $<\!energy\!>^2$. 

\subsection{The interactions as operators acting in
Fock-space}  \label{sec:Fock-operators}

In Section~\ref{sec:equations-of-motion} the 
energy-momentum four-vector $P_\mu$
was expressed in terms of the free fields.
One inserts them into the expressions for the interactions 
and integrates over configuration space. 
The free fields are
\begin  {eqnarray} 
   \widetilde\Psi  _{\alpha cf} (x) &=&
   \sum _\lambda\!\int\!\!  
   {dp^+ d^2p_{\!\bot}\over \sqrt{2p^+(2\pi)^3}}
   \left(\widetilde b (q) u_\alpha (p ,\lambda ) e^{-ipx} +
   \widetilde d^\dagger (q)v_\alpha(p,\lambda) 
   e^{+ipx}\right)
,\nonumber \\
   \widetilde A _\mu^a (x) &=& 
   \sum _\lambda\!\int\!\!  
   {dp^+ d^2p_{\!\bot}\over \sqrt{2p^+(2\pi)^3}}
   \left(\widetilde a(q)\epsilon_\mu(p,\lambda) e^{-ipx} +
   \widetilde a^\dagger(q)\epsilon_\mu^\star(p,\lambda)
   e^{+ipx} \right) 
\,,\label{eq:2.72}\end {eqnarray} 
where the the properties of the $u_\alpha $, $v_\alpha $ 
and $\epsilon_\mu$ are given in the appendices and where
\begin {equation} 
   \left [\widetilde a(q),\widetilde a^\dagger(q^\prime)\right]  = 
   \left\{\widetilde b(q),\widetilde b^\dagger(q^\prime)\right\} = 
   \left\{\widetilde d(q),\widetilde d^\dagger(q^\prime)\right\} = 
   \delta (p^+-p^{+\,\prime}) 
   \delta ^{(2)}(\vec p_{\!\bot}-\vec p_{\!\bot} ^{\,\prime}) 
   \delta _\lambda ^{\lambda ^\prime} 
   \delta _c^{c^\prime} 
   \delta _f^{f^\prime} 
\ .\end {equation}
Doing that in detail is quite laborious. 
We therefore restrict ourselves here to a few instructive examples, 
the vertex interaction $V$, 
the instantaneous-gluon interaction $W_2$
and the instantaneous-fermion interaction $W_3$.

According to Eq.(\ref{eq:2.94}) the fermionic contribution 
to the vertex interaction is 
\begin{eqnarray}
   V _f &=& g  \int\!dx_+d^2x_{\!\perp} 
   \ \widetilde j ^\mu_a \widetilde A _\mu^a        
   = g  \int\!dx_+d^2x_{\!\perp}\left.
   \ \overline{\widetilde\Psi}(x)  \gamma ^\mu  T  ^a 
   \widetilde\Psi(x)  \widetilde A ^a_\mu (x) 
   \right\vert_{x^+=0}
\nonumber\\
   &=& {g\over\sqrt{(2\pi)^3}}
   \sum _{\lambda_1,\lambda_2,\lambda_3}
   \sum _{c_1,c_2,a_3}
   \int\! {dp^+_1 d^2p_{\!\bot 1}\over \sqrt{2p^+ _1}}
   \int\! {dp^+_2 d^2p_{\!\bot 2}\over \sqrt{2p^+ _2}}
   \int\! {dp^+_3 d^2p_{\!\bot 3}\over \sqrt{2p^+ _3}}
\nonumber\\
   &\times& \int\!{dx_+d^2x_{\!\perp} \over(2\pi)^3} 
   \left[\left(\widetilde b ^\dagger (q_1) \overline u_\alpha 
   (p_1,\lambda_1) e^{+ip_1x}  +    
   \widetilde d(q_1)\overline v_\alpha(p_1,\lambda_1)  
   e^{-ip_1x}\right)T^{a_3}_{c_1,c_2} \right.
\nonumber\\
   &\times& \phantom{T^{a_3}_{c_1,c_2}}
   \phantom{d^\dagger q}\gamma ^\mu _{\alpha\beta}
   \left.\left( 
   \widetilde d^\dagger (q_2)v_\beta(p_2,\lambda_2) 
   e^{+ip_2x} +
   \widetilde b (q_2) u_\beta (p_2 ,\lambda _2) 
   e^{-ip_2x} \right)\right]
\nonumber\\
   &\times& \phantom{\gamma ^\mu _{\alpha\beta}
  T^{a_3}_{c_1,c_2} d^\dagger q}
   \left.\left(\widetilde a^\dagger(q_3)\epsilon_\mu^\star
   (p_3,\lambda_3) e^{+ip_3x} +
   \widetilde a(q_3)\epsilon_\mu(p_3,\lambda_3) 
   e^{-ip_3x} \right) \right.
.\end{eqnarray}
The integration over configuration space produces
essentially Dirac delta-functions in the single particle
momenta, which reflect momentum conservation:
\begin{eqnarray}
   \int\!{dx_+\over2\pi} \ e^{i x_+\big(\sum_j p_j^+\big)} 
   &=& \delta\Big( \sum_j p _j^+\Big)
\ , \qquad{\rm and} \nonumber\\
   \int\!{d^2x_{\!\perp} \over(2\pi)^3}    
   \ e^{-i\vec x_{\!\perp} \big(\sum_j \vec p_{\!\perp j}\big)}  
   &=& \delta^{(2)}\Big(  \sum_j\vec p_{\!\perp j}\Big)
\end{eqnarray}
Note that the sum of these single particle momenta is
essentially the sum of the particle momenta minus the sum
of the hole momenta. Consequently, if a particular term
has {\em only creation or only destruction operators} as in
\begin{eqnarray}
   b^\dagger (q_1) d^\dagger (q_2) a^\dagger (q_3)
   \ \delta\Big( p _1^+ + p _2^+ + p _3^+\Big) \simeq 0
,\nonumber \end{eqnarray}
its contribution vanishes since the light-cone longitudinal 
momenta $p ^+$ are all positive and can not add to zero.
The case that they are exactly equal to zero is excluded 
by the regularization procedures discussed below in 
Section~\ref{sec:DLCQ}. As a consequence,  all
energy diagrams which generate the vacuum fluctuations 
in the usual formulation of quantum field theory are absent 
from the outset in the front form.

The purely fermionic part of the  {\em instantaneous-gluon interaction} given by
Eq.(\ref{eq:2.96}) becomes correspondingly 
\begin{eqnarray}
   W_{2,f} &=& { g  ^2 \over 2} \int\!dx_+d^2x_{\!\perp} 
   \ \widetilde j ^+_a 
   {1\over \left(i \partial ^+ \right)^2} \widetilde j ^+_a 
\nonumber\\
   &=& { g  ^2 \over 2} \int\!dx_+d^2x_{\!\perp} \left.
   \overline{\widetilde\Psi}(x)  \gamma ^+  T  ^a 
   \widetilde\Psi(x)  
   {1\over \left(i \partial ^+ \right)^2} 
   \ \overline{\widetilde\Psi}(x)  \gamma ^+  T  ^a 
   \widetilde\Psi(x)  
   \right\vert_{x^+=0}
.\nonumber \end{eqnarray}
\begin{eqnarray}
   W_{2,f} &=& { g  ^2 \over 2(2\pi)^3}
   \sum _{\lambda_j} \sum _{c_1,c_2,c_3,c_4}
   \int\! {dp^+_1 d^2p_{\!\bot 1}\over \sqrt{2p^+ _1}}
   \int\! {dp^+_2 d^2p_{\!\bot 2}\over \sqrt{2p^+ _2}}
   \int\! {dp^+_3 d^2p_{\!\bot 3}\over \sqrt{2p^+ _3}}
   \int\! {dp^+_4 d^2p_{\!\bot 4}\over \sqrt{2p^+ _4}}
\nonumber\\
   &\times& \int\!{dx_+d^2x_{\!\perp} \over(2\pi)^3} 
   \left[\left(\widetilde b ^\dagger (q_1) \overline u_\alpha 
   (p_1,\lambda_1) e^{+ip_1x}  +    
   \widetilde d(q_1)\overline v_\alpha(p_1,\lambda_1)  
   e^{-ip_1x}\right)T^{a}_{c_1,c_2} \right.
\nonumber\\
   &\times& \phantom{T^{a}_{c_1,c_2}d}\phantom{d^\dagger}
   \gamma ^+ _{\alpha\beta}
   \left. \left( 
   \widetilde d^\dagger (q_2)v_\alpha(p_2,\lambda_2) 
   e^{+ip_2x} +
   \widetilde b (q_2) u_\alpha (p_2 ,\lambda _2) 
   e^{-ip_2x} \right)\right]
\nonumber\\
   &\times& \phantom{d^\dagger dd}
  {1\over \left(i \partial ^+ \right)^2} 
   \left[\left(\widetilde b ^\dagger (q_3) \overline u_\alpha 
   (p_3,\lambda_3) e^{+ip_3x}  +    
   \widetilde d(q_3)\overline v_\alpha(p_3,\lambda_3)  
   e^{-ip_3x}\right)T^{a}_{c_3,c_4} \right.
\nonumber \\
   &\times& \phantom{T^{a}_{c_1,c_2}d}\phantom{d^\dagger} 
   \gamma ^+ _{\alpha\beta} \left. \left( 
   \widetilde d^\dagger (q_4)v_\beta(p_4,\lambda_4) 
   e^{+ip_4x} +
   \widetilde b (q_4) u_\beta (p_4 ,\lambda _4) 
   e^{-ip_4x} \right)\right]
.\end{eqnarray}
By the same reason as discussed above, there will
be no contributions from terms with only creation or
only destruction operators.
The instantaneous-fermion interaction, finally, 
becomes according to Eq.(\ref{eq:2.97})
\begin{eqnarray}
    W_3 &=& { g  ^2 \over2} \int\!dx_+d^2x_{\!\perp} 
    \ \overline{\widetilde\Psi}  \gamma ^\mu  T  ^a 
    \widetilde A ^a_\mu \ {\gamma ^+\over i\partial ^+}
    \left( \gamma ^\nu  T  ^b \widetilde A ^b_\nu 
    \widetilde\Psi  \right)
\nonumber \\
&=& { g  ^2 \over 2(2\pi)^3}
   \sum _{\lambda_j} \sum _{c_1,a_2,a_3,c_4}
   \int\! {dp^+_1 d^2p_{\!\bot 1}\over \sqrt{2p^+ _1}}
   \int\! {dp^+_2 d^2p_{\!\bot 2}\over \sqrt{2p^+ _2}}
   \int\! {dp^+_3 d^2p_{\!\bot 3}\over \sqrt{2p^+ _3}}
   \int\! {dp^+_4 d^2p_{\!\bot 4}\over \sqrt{2p^+ _4}}
\nonumber\\
   &\times& \int\!{dx_+d^2x_{\!\perp} \over(2\pi)^3} 
   \left[\left(\widetilde b ^\dagger (q_1) \overline u
   (p_1,\lambda_1) e^{+ip_1x}  +    
   \widetilde d(q_1)\overline v (p_1,\lambda_1)  
   e^{-ip_1x}\right)T^{a_2}_{c_1,c} \right.
\nonumber\\
   &\times& \phantom{T^{a_2}_{c_1,c}dd}\phantom{d^\dagger}
   \gamma ^\mu 
   \left.\left(   \widetilde a^\dagger(q_2)\epsilon_\mu^\star
   (p_2,\lambda_2) e^{+ip_2x} +
   \widetilde a(q_2)\epsilon_\mu(p_2,\lambda_2) 
   e^{-ip_2x} \right) \right.
\nonumber\\
   &\times& \phantom{ddd}{1\over i \partial ^+ } 
   \phantom{ddd}
   \left.\left(   \widetilde a^\dagger(q_3)\epsilon_\nu^\star
   (p_3,\lambda_3) e^{+ip_3x} +
   \widetilde a(q_3)\epsilon_\nu(p_3,\lambda_3) 
   e^{-ip_3x} \right) T^{a_3}_{c,c_4}\right.
\nonumber \\
   &\times& \phantom{T^{a_3}_{c,c_4}dd}\phantom{d^\dagger} 
   \gamma ^\nu \left. \left( 
   \widetilde d^\dagger (q_4)v (p_4,\lambda_4) 
   e^{+ip_4x} +
   \widetilde b (q_4) u (p_4 ,\lambda _4) 
   e^{-ip_4x} \right)\right]
.\end{eqnarray}
Each of the instantaneous interactions types has primarily
$2^4-2=14$ individual contributions, which will not be
enumerated in all detail.  
In Section~\ref{sec:DLCQ} complete tables of all
interactions will be tabulated in their final normal ordered
form, that is with all creation operators are to the left of the
destruction operators. All instantaneous interactions
like those shown above are four-point interactions and the 
creation and destruction operators appear in a natural 
order. According to Wick's theorem this `time-ordered' 
product equals to the normal ordered product plus the sum
of all possible pairwise contractions. The fully contracted
interactions are simple c-numbers which can be omitted
due to vacuum renormalization. The one-pair contracted
operators, however, can not be thrown away
and typically have a structure like
\begin{eqnarray}
   I(q)\,\widetilde b^\dagger (q) \widetilde b(q)
.\end{eqnarray}
Due to the properties of the spinors and polarization
functions $u_\alpha,v\alpha$ and $\epsilon_\mu$ they
become diagonal operators in momentum space. 
The coefficients $I(q)$ are kind of mass terms and 
have been labeled  as `self-induced inertias' \cite{pab85a}.
Even if they formally diverge, they are part of the operator
structure of field theory, and therefore should not be
discarded  but need careful regularization.  In
Section~\ref{sec:DLCQ} they will be tabulated as well.

%% file: 04Conti.tex
\section{Bound States on the Light Cone}
\label{sec:continuum}
\setcounter{equation}{0}

In principle, the problem of computing for quantum chromodynamics
the spectrum and the corresponding wavefunctions
can be reduced to diagonalizing the  light-cone Hamiltonian.
Any hadronic state must be an eigenstate  of the light-cone 
Hamiltonian, thus a bound state of mass $M$, which satisfies 
$ \left(M^2 - H_{\rm LC}\right)\vert M \rangle = 0 $. 
Projecting the Hamiltonian eigenvalue equation 
onto the various Fock states $\langle q\bar q\vert$,
$\langle q\bar qg\vert \ldots{}$  results in an infinite 
number of coupled integral  eigenvalue equations. 
Solving these equations is  equivalent to solving the field theory.  
The light-cone Fock basis is a very physical tool for discussing
these  theories because the vacuum state is simple and 
the wavefunctions can be written in terms of relative
coordinates  which are frame independent. In terms of the 
Fock-space wave function one can give exact expressions 
for the form factors and structure functions of physical states. 
As an example we evaluate these expressions 
with a perturbative wave function for the electron and calculate the 
anomalous magnetic  moment of the electron. 

In order to lay down the groundwork for upcoming non-perturbative 
studies, it is indispensable to gain control  over the perturbative 
treatment. We devote therefore a section to the
perturbative treatment of quantum electrodynamics and
gauge theory on the light cone. Light-cone perturbation
theory is really Hamiltonian perturbation theory, and we give
the complete set of rules which are the analogue s of 
Feynmans rules. We  shall demonstrate in
a selected example, that one gets  the same covariant and
gauge-invariant scattering amplitude as in Feynman theory. 
We also shall discuss
one-loop renormalization of QED in the Hamiltonian formalism.  

Quantization is done in the light-cone gauge,  and  the
light-cone time-ordered perturbation theory is developed
in the null-plane Hamiltonian formalism. 
For  gauge-invariant quantities, this is very loosely equivalent 
to the use of Feynman diagrams together with an
integration over $p^-$ by residues \cite{tho79a,tho79b}.  
The one-loop renormalization
of QED quantized on the null-plane looks very  different 
from the standard treatment. In addition to not being  
manifestly covariant, $x^+$-ordered perturbation theory 
is fraught with  singularities, even at tree level. 
The origin of  these unusual, 
``spurious'', infrared divergences is no mystery. 
Consider for example a free particle whose transverse 
momentum $\vec p_{\!\perp}= (p^1 ,p^2 )$  is fixed, and
whose third component $p^3$ is cut at some momentum 
$\Lambda$. Using the mass-shell relation,  
$p^-= (m^2 + \vec p^{\,2}_{\!\perp})/2 p^+$, one sees
that $p^+$ has a lower bound proportional to $\Lambda ^{-1}$. 
Hence the light-cone spurious  infrared  divergences are 
simply a manifestation of space-time ultraviolet  divergences. 
A great deal of work is continuing on how to treat  these 
divergences in a self-consistent manner \cite{wwh94}. 
{\it Bona fide} infrared  divergences are of course also
present, and can be taken care of as usual by giving the 
photon a small mass, consistent with light-cone 
quantization \cite{sop71}.

As a matter of practical experience, and quite opposed
to the instant form of the Hamiltonian approach, one gets
reasonable results even if  the infinite number of integral is 
equations  truncated.
The Schwinger model is particularly illustrative 
because in the instant form  this bound state has a very  
complicated structure in terms of Fock states while 
in the front form the bound state consists of a single 
electron-positron pair.  One might hope that a similar
simplification occurs in QCD.  
The Yukawa model is treated here 
in Tamm-Dancoff truncation  in 3+1  dimensions
\cite{ghp93,pin91b,pin92}. This model is particularly important 
because it features a number of the renormalization 
problems  inherent to the front form, and because it  
motivate  the approach of Wilson to be discussed later.

\subsection{The hadronic eigenvalue problem}

The first step is to find a language in which one can represent 
hadrons in terms of relativistic confined quarks and
gluons.  The Bethe-Salpeter formalism \cite{beh55,lsg91} 
has been the central method  for analyzing hydrogenic atoms 
in QED, providing a completely 
covariant procedure for obtaining bound state solutions. 
However, calculations using this method are extremely 
complex and appear to be intractable much beyond the 
ladder approximation. It also appears impractical to extend 
this method to systems with more than a few constituent
particles. A review can be found in \cite{lsg91}.
 
An intuitive approach  for solving relativistic bound-state 
problems would be to solve the Hamiltonian eigenvalue 
problem
\begin{equation}
   H\left|{\Psi}\right\rangle = \sqrt{M^2 + \vec P ^{\,2}}
   \left|{\Psi}\right\rangle
\end{equation}
for the particle's mass, $M$, and wave function,
$\left|\Psi\right\rangle$. Here, one imagines that
$\left|\Psi\right\rangle$ is an expansion in multi-particle
occupation number Fock states, and that the operators $H$ 
and $\vec P $ are second-quantized Heisenberg 
operators. Unfortunately, this method, as described by 
Tamm and Dancoff \cite{dan50,tam45}, is
complicated by its non-covariant structure and the 
necessity to first understand its complicated vacuum 
eigensolution over all space and time. The presence of the
square root operator also presents severe mathematical 
difficulties. Even if these problems could be solved, 
the eigensolution is only determined in its rest system
($\vec P = 0$); determining the boosted wave function
is as complicated as diagonalizing $H$ itself.
 
In principle, the front form approach works in the
same way.  One aims at solving  the Hamiltonian 
eigenvalue problem
\begin{equation}
   H\left|{\Psi}\right\rangle = 
  { M^2 + \vec P _{\!\perp}^{\,2}\over 2 P ^+}
   \left|{\Psi}\right\rangle
\ ,\label{eq:4.2}\end{equation}
which for several reasons is easier: 
Contrary to $ P _z$  the operator $ P ^+$ is positive,
having only positive eigenvalues. 
The square-root operator is absent, and the boost 
operators are kinematic, see Sect.~\ref{sec:symmetries}.
As discussed there, in both the instant and the front form, 
the eigenfunctions can be labeled with six numbers, the six
eigenvalues of 
the invariant mass M, of the three space-like momenta 
$ P ^+, \vec P _{\!\perp}$, and  of the generalized total
spin-squared  $ S ^2$ and its longitudinal projection $ S_z$,  
that is
\begin{equation}
   \left|\Psi\right\rangle = 
   \left|\Psi; M , P ^+,\vec P _{\!\perp}, S ^2, S _z ; h \right
   \rangle 
\ . \label{eq:4.3}\end{equation}
In addition, the eigenfunction is labeled by quantum numbers 
like charge, parity, or baryon number which specify a particular 
hadron $ h $.  The ket $\left|\Psi\right\rangle $ can be
calculated in terms of 
of a complete set of functions
$\left|\mu\right\rangle$ or $\left|\mu_n\right\rangle$, 

\begin{equation}
   \int\!d[\mu]
   \ \left|\mu\right\rangle\left\langle\mu\right|
   = \sum\limits_{n}\int\!d[\mu_n]
   \ \left|\mu_{n}\right\rangle\left\langle\mu_{n}\right|
   = \mathbf {1}
\ .\end{equation}
The transformation  between the complete set
of eigenstates $ \left|\Psi\right\rangle $ and the complete set
of basis states  $ \left| \mu_n \right\rangle $ are then 
$\left\langle\mu_n|\Psi\right\rangle $. 
The projections of $ \left|\Psi\right\rangle $ on 
$ \left| \mu_n \right\rangle $
are usually called the  {\em wavefunctions}  
\begin{equation}
    \Psi_{n/h\,(M,P ^+,\vec P _{\!\perp}, S ^2, S _z)}(\mu)
    \equiv \left\langle\mu_n|\Psi\right\rangle 
\ .\label{eq:4.5} \end{equation}
Since the values of $(M,P ^+,\vec P _{\!\perp}, S ^2, S _z)$  
are obvious in the context of a concrete case, we
convene to drop reference to them and write simply
\begin{equation}
   \left|\Psi\right\rangle 
   = \sum _n \int\! d[\mu_n] \ \left|\mu_n\right\rangle
    \Psi_{n/h} (\mu)
   \equiv  \sum _n \int\! d[\mu_n]
   \ \left|\mu_n\right\rangle\left\langle\mu_n|\Psi
   ; M , P ^+,\vec P _{\!\perp}, S ^2, S _z; h\right\rangle 
\ .\end{equation}
One constructs the complete basis of Fock states 
$ \left| \mu_n \right\rangle $ in the usual way by
applying products of free field creation operators to the 
vacuum state $\left| 0 \right\rangle$:
\begin{equation}
\begin{array}{lrllrl}
    n=0: \qquad
     & {}&\left| 0 \right\rangle\ ,   \nonumber \\
     n=1: \qquad\hfill
 & {}&\left|{q\bar q: k^+_i, \vec k _{\!\perp i},\lambda_i}
    \right\rangle  &=& 
    b^\dagger( q _1)\,
    d^\dagger( q _2)\,
&\left| 0 \right\rangle\ , \nonumber \\
   n=2: \qquad
   &{}&\left|{q\bar q g: k^+_i, \vec k _{\!\perp i},\lambda_i}
   \right\rangle &=& 
    b^\dagger( q _1)\,
    d^\dagger( q _2)\,
    a^\dagger( q _3)\, 
&\left| 0 \right\rangle\ , \nonumber \\
    n=3: \qquad
    & {}&\left|{g g: k^+_i, \vec k _{\!\perp i},\lambda_i}
    \right\rangle &=&  
    a^\dagger( q _1)\,
    a^\dagger( q _2)\,
&\left| 0 \right\rangle\ , \nonumber \\    
    \vdots & {} & \vdots & \vdots & \vdots & \left|0\right\rangle\ .
    \end{array}
\label{eq:4.7}\end{equation}
The operators
$b^\dagger(q)$, $d^\dagger(q)$ and $a^\dagger(q)$ 
create bare leptons (electrons or quarks), bare anti-leptons
(positrons or antiquarks) and bare vector bosons (photons or
gluons).
In the above notation on explicitly keeps track of only  the 
three continuous momenta  
$k^+_i$ and $\vec k _{\!\perp i}$ 
and of the discrete helicities $\lambda_i$.
The various Fock-space classes are conveniently labeled 
with a running index $n$. Each Fock state 
$\left|\mu_n \right\rangle 
= \big|{n: k^+_i, \vec k _{\!\perp i},\lambda_i}\big\rangle$ 
is an eigenstate of $P^+$ and $\vec P _{\!\perp}$. 
The eigenvalues are
\begin{equation}
   \vec P _{\!\perp} = \sum\limits_{i\in n} \vec k _{\!\perp i}
   \quad{\rm and}\quad
   P^+ = \sum\limits_{i\in n} k^+_i
   \ , \quad{\rm with}\quad k_i^+ > 0
\ .\label{eq:4.8}\end{equation}
The vacuum is has eigenvalue $0$, {\it i.e.}
$\vec P _{\!\perp} \left| 0 \right\rangle=\vec 0$ and 
$P^+ \left| 0 \right\rangle=0$. 

The restriction $k^+ > 0$ for massive quanta is a key difference
between light-cone quantization and ordinary equal-time
quantization. In equal-time quantization, the state of a parton is
specified by its ordinary three-momentum $\vec k =
(k_x,k_y,k_z)$. Since each component of $\vec k$  can
be either positive or negative, there exist zero total momentum 
Fock states of arbitrary particle number, and these will mix with 
the zero-particle state to build up the ground state, the
physical vacuum. However, in light-cone quantization each of the 
particles forming a zero-momentum state must have vanishingly 
small $k^+$.  
The free or Fock space vacuum $ \left|0\right\rangle$ is then
an exact  eigenstate of the full front form Hamiltonian $H$, in
stark contrast to the quantization at equal usual-time.
However, as we shall see later, the vacuum in QCD is
undoubtedly more complicated due to the possibility of 
color-singlet states with $P^+ = 0$ built on  
zero-mode massless gluon quanta \cite{gri78}, 
but as discussed in Section~\ref{sec:Vacuum},  
the physical vacuum is {\em still far simpler} than usually.

Since $k_i^+ > 0$ and $ P ^+ > 0$,  one can define 
boost-invariant longitudinal momentum fractions
\begin{equation}
   x_i = {k_i^+ \over P^+ }
   \ , \qquad {\rm with}\quad
   0 < x_i  < 1
\ ,\end{equation}
and adjust the notation. All  particles 
in a Fock state  $\left|\mu_n \right\rangle 
= \big|{n: x_i, \vec k _{\!\perp i},\lambda_i}\big\rangle$ 
have then four-momentum
\begin{equation}
   k^\mu_i \equiv (k^+,\vec k _{\!\perp},k^-) _i = 
   \left( x_i P^+, \vec k _{\!\perp i} ,
   \frac{m_i^2 + k^2 _{\!\perp i}}{x_i P^+} \right)
   \ ,\qquad{\rm for}\quad i=1,\dots,N_n
\ ,\end{equation}
and are ``on shell,'' $(k^\mu k_\mu)_i=m_i^2$.
Also the Fock state is ``on shell'' since one can
interpret
\begin{equation}
      \left(\sum_{i=1}^{N_n} k^-_i\right) P^+  
      - \vec P _{\!\perp}^{\,2}
      =  \sum_{i=1}^{N_n} 
      \left({ (\vec k _{\!\perp i} + x_i\vec P _{\!\perp})^{\,2}
      +m_i^2\over x}\right) - \vec P _{\!\perp}^{\,2}
      = \sum_{i=1}^{N_n} 
      \left({ \vec k_{\!\perp} ^{\,2}+m^2\over x}\right)_i 
\end{equation}
as its free invariant mass squared 
$\widetilde M ^2 
   = \widetilde P ^\mu \widetilde P _\mu$. 
There is some confusion over the terms `on-shell' and `off-shell' 
in the literature \cite{per94a}.
The single particle states are on-shell, as mentioned, but
the Fock states $\mu_n$ are off the energy shell
since $\widetilde M$ in general is different from the
bound state mass $M$ which appears in Eq.(\ref{eq:4.2}). 
In the intrinsic frame ($\vec P _{\!\perp} = \vec 0$),
the values of $x_i$ and 
$\vec k_{\!\perp i} $ are constrained by 
\begin{equation}
   \sum\limits_{i=1} ^{N_n}  x_i = 1
   \quad{\rm and}\quad
   \sum\limits_{i=1} ^{N_n}  \vec k _{\!\perp i} = \vec 0
\ ,\label{eq:4.12}\end{equation}
because of Eq.(\ref{eq:4.8}).
The phase-space differential
$d[\mu_n]$ depends on how one normalizes 
the single particle states. 
In the convention where commutators are normalized to 
a Dirac $\delta$-function, the phase space integration is 
\begin{eqnarray}
   \int\!d[\mu_n] \dots &=& \sum _{\lambda_i \in n} 
   \int \left[dx_i d^2 k_{\!\perp i} \right] \dots
\ , \qquad{\rm with} \nonumber \\ 
   \left[dx_i d^2 k_{\!\perp i} \right] 
   &=& \delta \Big(1-\sum_{j=1}^{N_n} x_j\Big)
   \delta^{(2)} \Big(\sum _{j=1}^{N_n} \vec k_{\perp j}\Big) 
   \ dx_1 \dots  dx_{N_n} 
   \ d^2 k _{\!\perp 1} \dots d^2 k _{\!\perp N_n}
\ .\label{eq:4.13}\end{eqnarray}
The additional Dirac $\delta$-functions account for the 
constraints (\ref{eq:4.12}).
The eigenvalue equation (\ref{eq:4.2}) therefore stands
for an infinite set of  coupled integral equations
\begin{equation}
\sum_{n^\prime} 
\int [d\mu^\prime_{n^\prime}]
\ \langle n: x_i, \vec k _{\!\perp i}, \lambda_i \vert H
\vert n^\prime: x_i^\prime, \vec k _{\!\perp i}^{\,\prime},
\lambda_i^\prime \rangle \,\Psi_{n^\prime/h}
(x_i^\prime, \vec k _{\!\perp i}^{\,\prime},\lambda_i^\prime)
=    { M^2 + \vec P _{\!\perp}^{\,2}\over 2 P ^+}
\Psi _{n/h}(x_i, \vec k _{\!\perp}, \lambda_i )
\ ,\label{eq:4.17}\end{equation}
for $n=1,\dots,\infty$.
The major difficulty  is not primarily the large number of coupled 
integral equations,  but rather that the above equations are 
ill-defined for very large values of the transversal momenta 
(`ultraviolet singularities') and for values of the 
longitudinal momenta close to the endpoints
$x\sim0$ or $x\sim1$ (`endpoint singularities'). 
One often has  to introduce cut-offs $\Lambda$ for to regulate 
the theory in some convenient way, and subsequently 
to renormalize it at a particular mass or momentum 
scale $Q$.  The corresponding wave function 
will be indicated by corresponding upper-scripts, 
\begin{eqnarray}
\Psi _{n/h}^{(\Lambda)}(x_i, \vec k _{\!\perp}, \lambda_i )
\qquad {\rm or}\qquad
\Psi _{n/h}^{(Q)}(x_i, \vec k _{\!\perp}, \lambda_i )
\ .\end{eqnarray}
Consider a pion in QCD with momentum 
$ P = (P^+,\vec P _{\!\perp})$ as an example. 
It is described by
\begin{equation}
   \left|{\pi: P}\right\rangle 
  = \sum_{n=1} ^{\infty} \int\!d[\mu_n] 
  \left|{n: x_i P^+, \vec k _{\!\perp i}
   + x_i \vec P _{\!\perp} , \lambda_i}\right\rangle\,
   \Psi_{n/\pi} (x_i,\vec k _{\!\perp i},\lambda_i)
\ ,\label{eq:4.19}\end{equation}
where the sum is over all Fock space sectors of 
Eq.(\ref{eq:4.7}).
The ability to specify wavefunctions simultaneously
in any frame is a special feature of light-cone quantization. 
The light-cone wavefunctions $\Psi_{n/\pi}$ 
do not depend on the total momentum, 
since $x_i$ is the longitudinal momentum 
fraction carried by the $i^{\rm th}$ parton  and 
$\vec k _{\!\perp i} $ is its momentum ``transverse'' to the 
direction of the meson; both of these are 
frame-independent quantities.  They are the probability
amplitudes to find a Fock state of  bare particles in the
physical pion.

\begin{figure} [t]
\epsfxsize=160mm\epsfbox{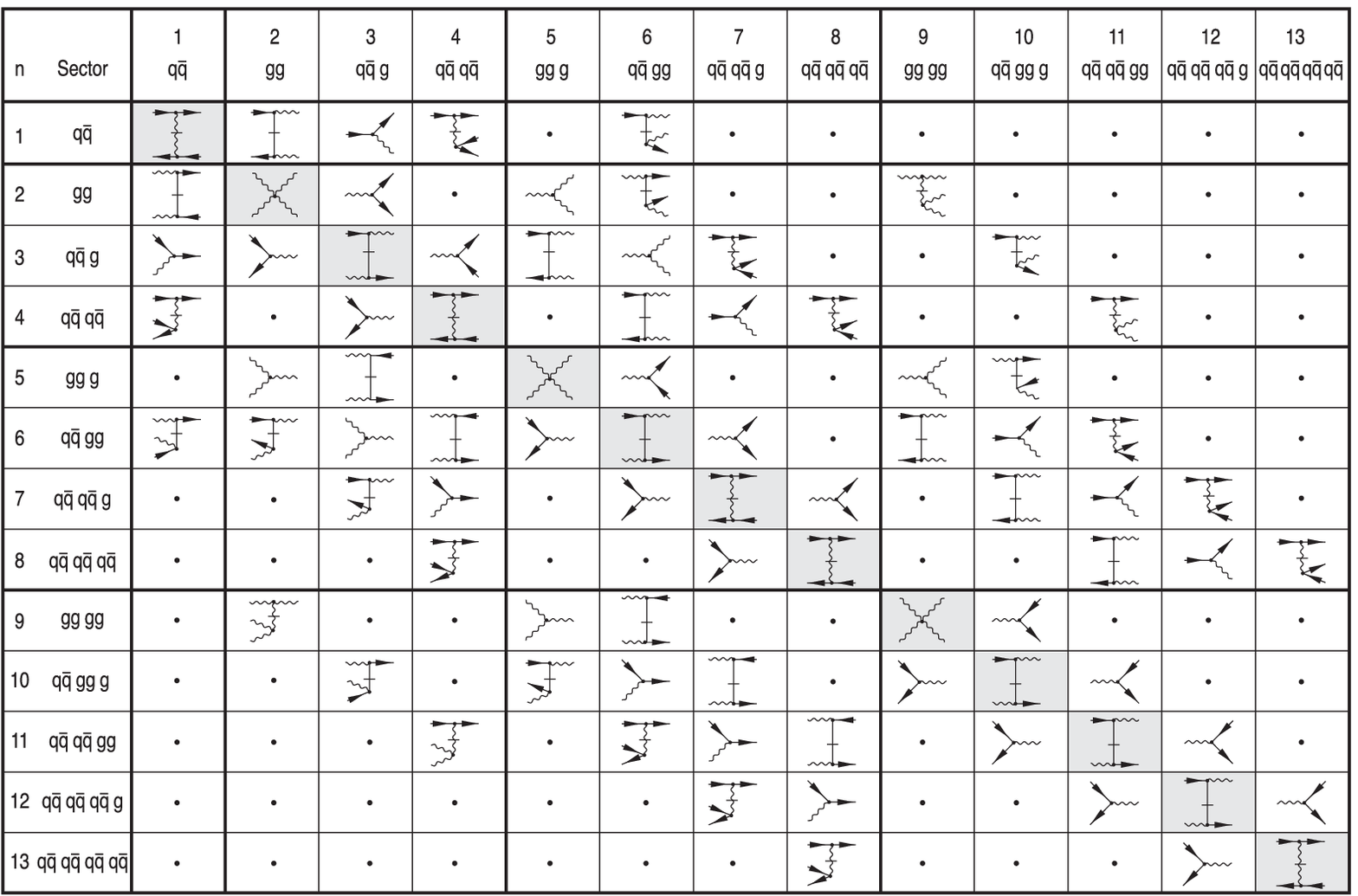}
\caption{\label{fig:holy-1} 
     The Hamiltonian matrix for a SU(N)-meson. 
     The matrix elements are represented by energy 
     diagrams. 
     Within each block they are all of the same type: 
     either vertex, fork or seagull diagrams.
     Zero matrices are denoted by a dot ($\cdot$).
     The single gluon is absent since it cannot be 
     color neutral.
}\end{figure}
More generally consider a meson in
SU(N). The kernel of the integral equation
(\ref{eq:4.17}) is illustrated in
Figure~\ref{fig:holy-1} 
in terms of the bloc matrix
$\langle n: x_i, \vec k _{\!\perp i}, \lambda_i \vert H
\vert n^\prime: x_i^\prime, \vec k _{\!\perp i}^{\,\prime},
\lambda_i^\prime \rangle $.
The structure of this matrix depends of course on the
way one has arranged the Fock space, see Eq.(\ref{eq:4.7}).
Note that most of the bloc matrix elements vanish 
due to the nature of the light-cone interaction as defined
in Eqs.(\ref{eq:2.93}). The vertex interaction
in Eq.(\ref{eq:2.94}) changes the particle number by one,
while the instantaneous interactions in Eqs.(\ref{eq:2.95}) 
to (\ref{eq:2.97}) change the particle number only up to two.

\subsection{The use of light-cone wavefunctions}
\label{sec:wavefunctions}

The infinite set of integral equations (\ref{eq:4.17}) 
is  difficult if not impossible to solve.  But 
given the light-cone wavefunctions $\Psi_{n/h} (x_i,
\vec k _{\!\perp i} , \lambda_i)$, one can compute  any  
hadronic quantity by convolution with the appropriate quark 
and gluon matrix elements. 
In many cases of practical interest it suffices to know
less information  than the complete wave function. 
As an example  consider
\begin{equation}
   G_{a/h}(x,Q) = \sum_{n} \int d[\mu_n]
   \ \left|\Psi_{n/h}^{(Q)}(x_i,\vec k_{\!\perp i} ,
   \lambda_i)\right|^2 
   \sum_{i} \delta(x-x_i) \ .\end{equation}
$G_{a/h}$ is a  function of one variable, characteristic for a 
particular hadron, and depends parametrically on the
typical scale $Q$. 
It gives the probability to find  in that hadron a particle with
longitudinal momentum fraction $x$, 
irrespective of the particle type, and irrespective of its spin, color, 
flavor or transversal momentum $\vec k_{\!\perp}$.
Because of wave function normalization the integrated 
probability is normalized to one.

One can ask also for conditional probabilities, 
for example for the probability to find a quark 
of a particular flavor $f$ and its momentum fraction $x$, 
but again irrespective of the other quantum numbers. 
Thus 
\begin{equation}
   G_{f/h}(x;Q) = \sum_{n} \int d[\mu_n]
   \ \left|\Psi_{n/h}^{(Q)}(x_i,\vec k_{\!\perp i} ,\lambda_i)
   \right|^2 \sum_{i}  \delta(x-x_i) \,\delta_{i,f}
\ .\end{equation}
The conditional probability is {\em not
normalized},  even if one sums over all flavors.
Such probability functions can be measured.
For exclusive cross sections one often needs only the 
probability amplitudes of the valence part 
\begin{equation}
   \Phi_{f/h}(x;Q) = \sum_{n} \int d[\mu_n]
   \ \Psi_{n/h}^{(Q)}(x_i,\vec k_{\!\perp i} ,\lambda_i)
   \sum_{i}  \delta(x-x_i) \,\delta_{i,f}
   \delta_{n,{\rm valence}}
   \ \Theta\left(\vec k_{\!\perp i}^{\,2} \leq Q^2\right)
\ .\end{equation}
Here, the transverse momenta are integrated up to momentum 
transfer $Q^2$.

The leading-twist  structure functions  
measured in deep  inelastic lepton scattering are 
immediately related to the above light-cone probability
distributions by 
\begin{equation}
   2M\,F_1(x,Q) = 
   {F_2(x,Q)\over x}\approx 
   \sum_f e_f^2\,G_{f/p}(x,Q)
\ .\end{equation}
This follows from the observation that deep inelastic lepton 
scattering in the Bjorken-scaling limit occurs if $x_{bj}$  
matches the light-cone fraction of the struck quark
with charge $e_f$. 
However, the light cone wavefunctions contain much more 
information for the final state of deep inelastic scattering, 
such as the multi-parton distributions, spin and flavor
correlations, and the spectator jet composition.

\begin{figure}
\begin{minipage} {80mm}
\epsfxsize=80mm\epsfbox{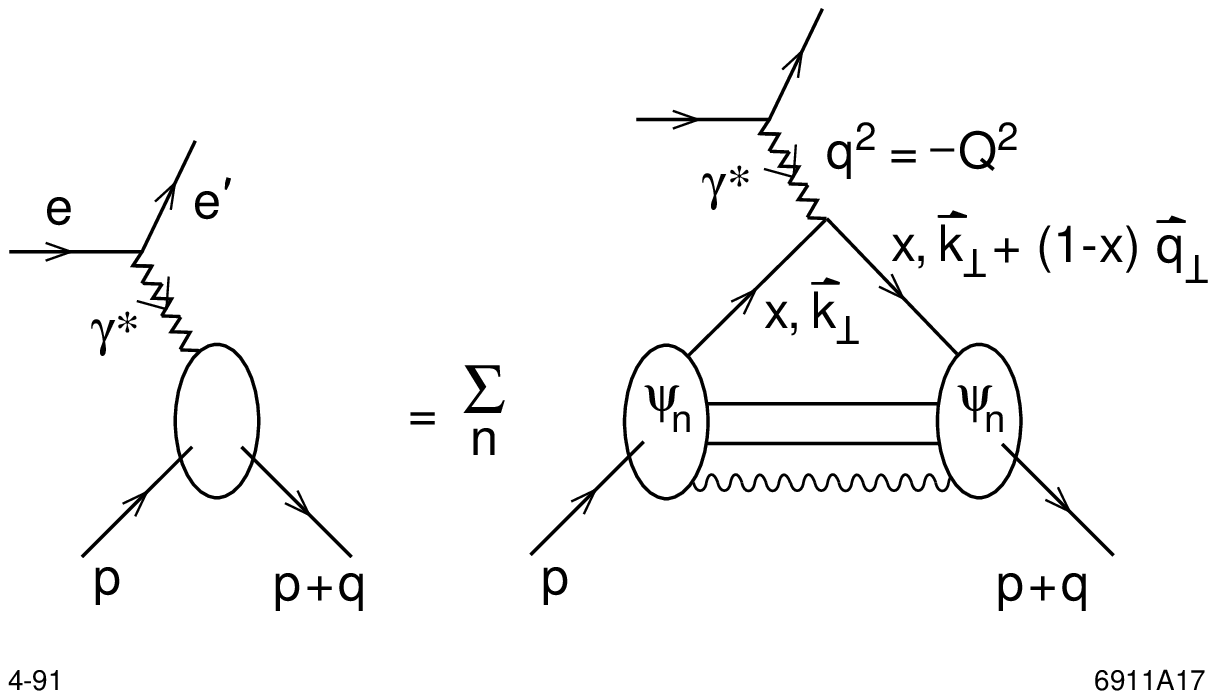} 
\caption{\label{fig:4.1} \sl
Calculation of the form factor of a bound state from the 
convolution of light-cone Fock amplitudes. The result is 
exact if one sums over all $\Psi_n$. }
\end{minipage}
\ \hfill
\begin{minipage} {70mm}
\epsfxsize=70mm\epsfbox{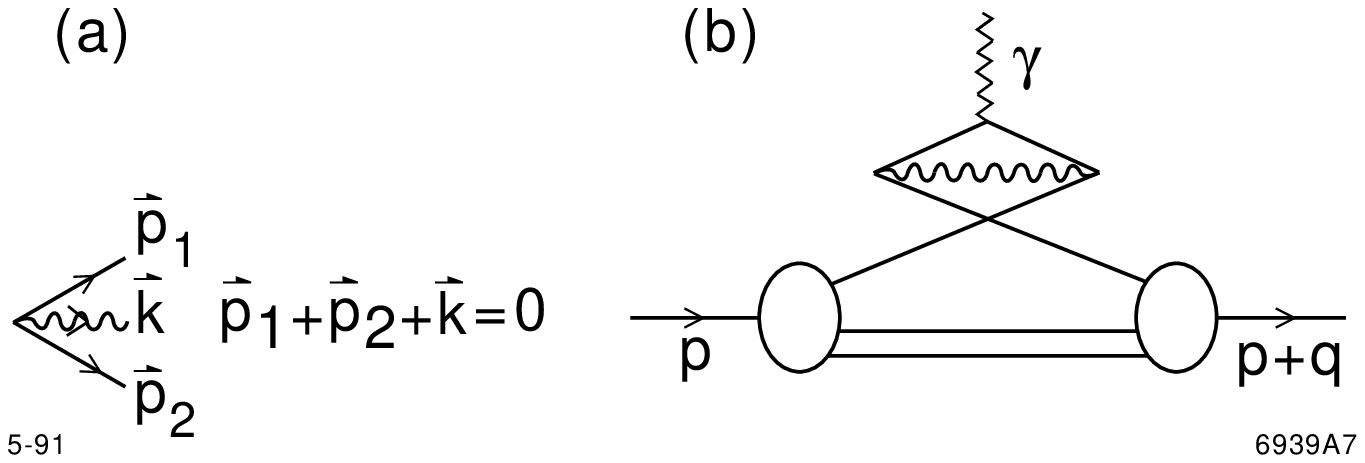} 
\caption{\label{fig:4.3} 
\sl (a) Illustration of a vacuum creation graph in time-ordered
perturbation theory. A corresponding contribution to the form 
factor of a bound state is shown in figure (b). } 
\end{minipage}
\end{figure}
 
One of the most remarkable simplicities
of the light-cone formalism is that one can write down 
exact  expressions for the electro-magnetic form factors. 
In the interaction picture one can equate the full Heisenberg 
current to the free (quark) current $J^\mu(0)$ described by
the free  Hamiltonian at $x^+=0$. 
As was first shown by Drell and Yan \cite{dry70},  it is
advantageous to  choose a special coordinate frame to compute
form factors, structure functions, and other current
matrix elements at space-like photon momentum. 
One then has to examine only the $ J ^+$ component to get
form factors like
\begin{equation}
   F_{S\rightarrow S^\prime}(q^2) 
   = \left\langle P^\prime,S^\prime\mid
   J^+ \mid P,S\right\rangle
   \ , \qquad {\rm with}\quad
   q_\mu = P_\mu^\prime - P_\mu 
\ .\label{eq:4.20}\end{equation}
This holds for any (composite) hadron of mass $M$, and any initial 
or final spins $S$ \cite{dry70,brd80}. 
In the Drell frame, as illustrated in Fig.~\ref{fig:4.1}, 
the photon's momentum is transverse 
to the momentum of the incident hadron
and the incident hadron can be directed along the $z$ 
direction, thus
\begin{equation}
   P^\mu = \left(P^+,\vec 0_\perp, {M^2\over P^+}\right)
   \ ,\qquad{\rm and}\quad
   q^\mu = \left(0, \vec q_\perp,{2q\cdot P \over P^+}\right)
\ .\end{equation}
With such a choice the four-momentum transfer is
$ -q_\mu q^\mu \equiv Q^2 = \vec q _{\!\perp} ^{\,2}$,
and the quark current can neither create pairs nor  
annihilate the vacuum.   
This is distinctly different from the conventional treatment,
where there are contributions from terms in which the current
is annihilated by the vacuum, as
illustrated in Fig.~\ref{fig:4.3}.
Front form kinematics allow to trivially boost the hadron's
four-momentum from $P$ to $P^\prime$, and therefore  
the space-like form factor for a hadron 
is just a sum of overlap integrals analogous to 
the corresponding non-relativistic formula \cite{dry70}: 
\begin{equation}    F _{S\rightarrow S^\prime}(Q^2) =
   \sum_{n} \sum_f e_f \int d[\mu_n]
   \,\Psi_{n,S^\prime}^\star (x_i,\vec \ell_\perp{}_i,\lambda_i)
   \,\Psi_{n,S} (x_i,\vec k_\perp{}_i,\lambda_i) 
\ .\label{eq:4.23}\end{equation}
Here $e_f$ is the charge of the struck quark, and
\begin{equation}
  \vec \ell _{\!\perp i} \equiv \cases{
  \vec k _{\!\perp i} - x_i \vec q _{\!\perp}  
+ \vec q _{\!\perp},   & \hbox{for the struck quark,} \cr
  \vec k _{\!\perp i} - x_i \vec q _{\!\perp},      
& \hbox{for all other partons.}\cr}
\label{eq:4.24}\end{equation}
This is particularly simple for a spin-zero hadron like a pion.
Notice that the transverse momenta appearing as arguments 
of the first wave function correspond not to the actual momenta 
carried by the partons but to the actual momenta minus 
$x_i\vec q _{\!\perp} $, to
account for the motion of the final hadron. 
Notice also that $\vec\ell _{\!\perp i} $ and 
$\vec k _{\!\perp i} $ become equal as 
$\vec q_\perp\rightarrow 0$, and that $F_\pi\rightarrow1$ 
in this limit due to wave function normalization.
In most of the cases it suffices to treat the problem 
in perturbation theory.

\subsection{Perturbation theory in the front form}
 
\begin{figure}
\fbox{\begin{minipage}{28mm}
\epsfxsize=25mm\epsfbox{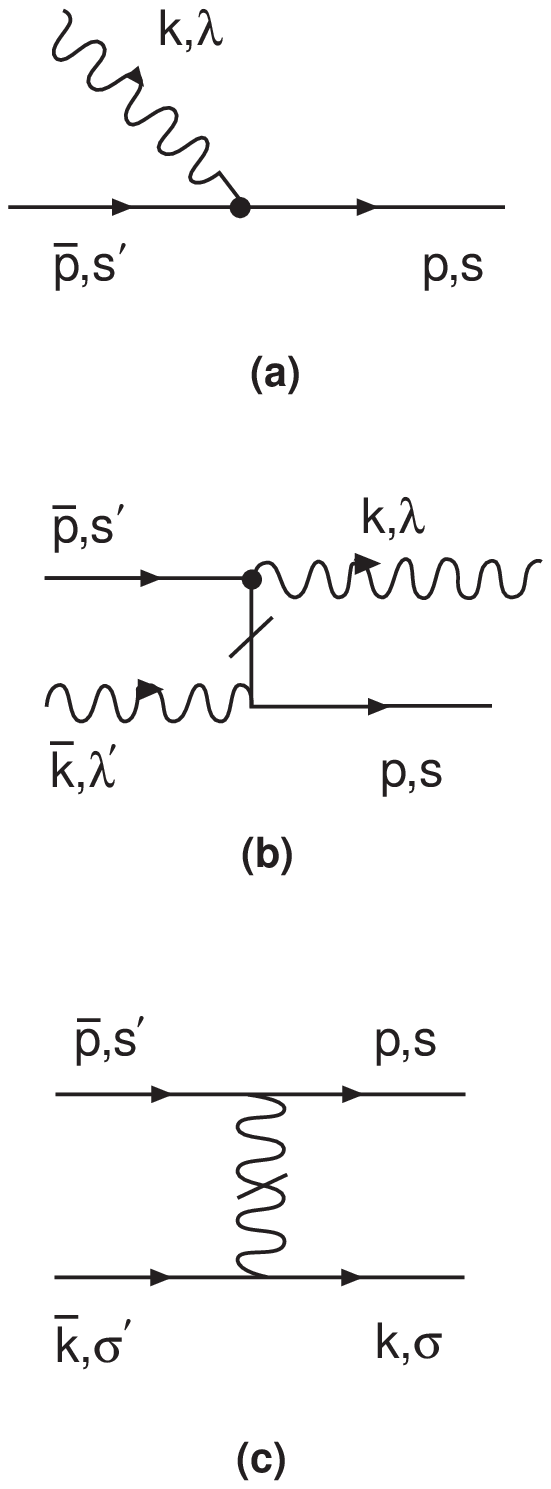}
\end{minipage} }
\hfill 
\fbox{\begin{minipage}{110mm}
\[  V ={e\over (2\pi)^{3/2}} 
    {\bar u(\bar p,s')\over \sqrt{2\bar p^+}}
    {/\!\!\!\epsilon^*(k,\lambda)\over\sqrt{2k^+}}
    {u(p,s)\over \sqrt{2p^+}}  \]
\[  W_f ={e^2 \over (2\pi)^3}
    {\bar u(\bar p,s')\over\sqrt{2{\bar p}^+}}
    {{/\!\!\!\epsilon}^*(\bar k,\lambda')
    \over\sqrt{2\bar k^+}}
    {\gamma^+  \over 2({\bar k}^+-p^+ )}
    {{/\!\!\!\epsilon}(k,\lambda) \over\sqrt{2k^+}}
    {u(p,s)\over\sqrt{2p^+}}    \]
\[  W_b=-{e^2 \over {(2\pi)^3 }}
    {{\bar u(\bar p,s') \gamma^+ u(p,s)} 
     \over \sqrt{2\bar p^+}
    \sqrt{2p^+}} {1  \over 2(\bar p^+-p^+ )^2}
    {{\bar u(\bar k,\sigma')
    \gamma^+ u(k,\sigma)} 
    \over \sqrt{2\bar k^+} 
    \sqrt{2k^+}}       \]    \qquad
\end{minipage}
}\caption{\label{fig:4.4} 
A few selected matrix elements of the QED front form 
Hamiltonian $ H = P_+$ in KS-convention. }
\end{figure}

The light-cone Green's functions $\widetilde G_{fi}(x^+)$ are
the probability amplitudes  that a state starting in Fock 
state $ \left| i\right\rangle$ ends up in Fock state 
$\left| f \right\rangle$ a (light-cone)  time $x^+$ later
\begin{eqnarray}
   \Big\langle f \mid \widetilde G(x^+) \mid i \Big\rangle \
   = \langle f | e^{-i P_+ x^+} | i \rangle 
   = \langle f | e^{-i H x^+} | i \rangle 
   = i \int {d\epsilon \over 2\pi} \ e^{-i\epsilon x^+}
   \ \left\langle f \mid G(\epsilon) \mid i \right\rangle 
.\end{eqnarray}
The Fourier transform 
$\left\langle f \mid G(\epsilon) \mid i \right\rangle $  
is usually called the resolvent of the Hamiltonian $H$
\cite{mes62}, {\it i.e.}
\begin{equation}
\left\langle f \mid G(\epsilon) \mid i \right\rangle 
   = \Bigl\langle f \left| {1\over \epsilon - H + i0_+}
   \right| i \Bigr\rangle 
   = \Bigl\langle f \left| {1\over \epsilon - H_0  - U + i0_+}
   \right| i \Bigr\rangle 
.\end{equation}
Separating the Hamiltonian $H=H_0 + U$ according to 
Eq.(\ref{eq:2.91}) into a free part  $ H_0 $ and an interaction 
$ U $, one can expand the resolvent into the usual series 
\begin{eqnarray}
   \left\langle f \mid G(\epsilon) \mid i \right\rangle 
   =&&\Bigl\langle f \left|{1\over \epsilon - H_0 +i0_+}
   + {1\over \epsilon - H_0 + i0_+} \, U \,
   {1\over \epsilon - H_0 + i0_+}\, + \right. \nonumber \\
   &&\left.\phantom{\langle f\mid}
   {1\over \epsilon - H_0 + i0_+} \, U \,
   {1\over \epsilon - H_0 + i0_+} \, U \,
   {1\over \epsilon - H_0 + i0_+} + \ \ldots
   \ \right| i \Bigr\rangle 
.\end{eqnarray}
The rules for $x^+$-ordered perturbation theory follow 
immediately when the resolvent of the free Hamiltonian
$(\epsilon - H_0)^{-1}$ is replaced by its spectral 
decomposition. 
\begin{equation}
   {1\over \epsilon - H_0 + i0_+} = \sum_{n}
   \int d[\mu_n]
   \left| n: k^+_i,\vec k_{\!\perp i},\lambda_i\right\rangle \,
   {1\over \epsilon - \sum\limits_i 
   \Big({k_{\!\perp}^{\,2} +m^2\over2k^+}\Big)_i +    i0_+} 
   \left\langle n:k^+_i,\vec k_{\!\perp i},\lambda_i\right| 
.\end{equation}
The sum becomes a sum over all states $n$ intermediate 
between two interactions $U$.
 
\begin{figure}
\fbox{\begin{minipage}{28mm}
\epsfxsize=25mm\epsfbox{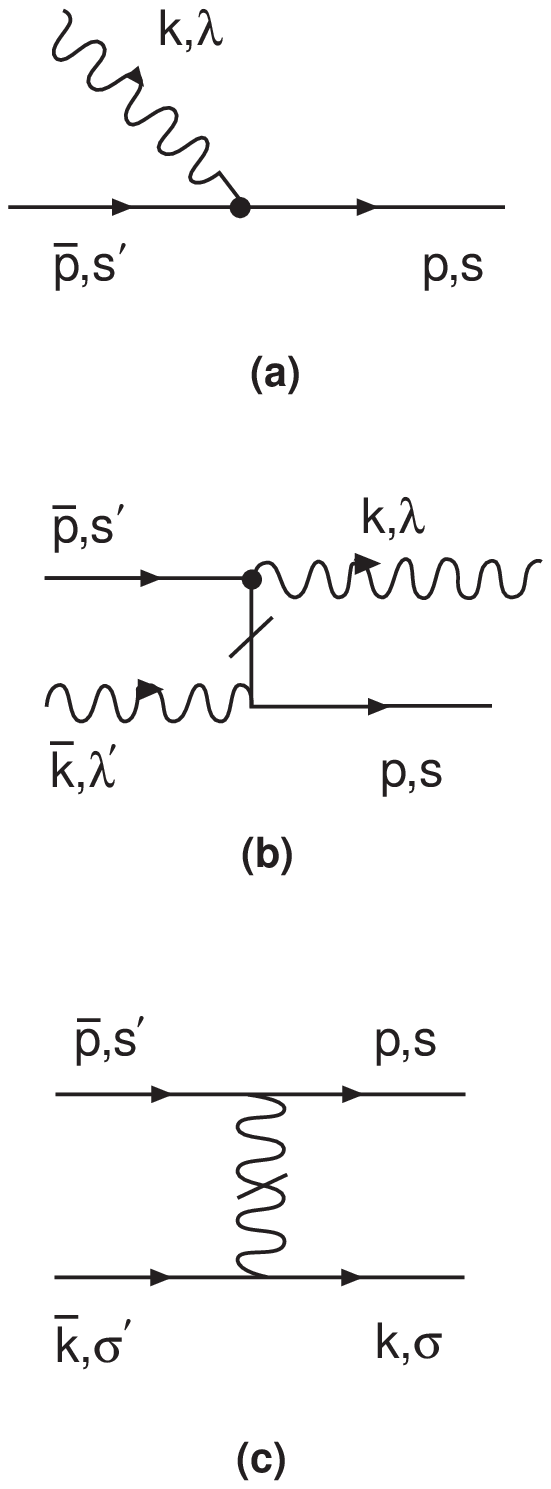}
\end{minipage}}
\hfill 
\fbox{\begin{minipage}{120mm}
\[  V ={g\over (2\pi)^{3/2}} 
    \,{\bar u(\bar p,s')\over \sqrt{2\bar p^+}}
    \,{/\!\!\!\epsilon^*(k,\lambda)\over\sqrt{2k^+}}
    \, {u(p,s)\over \sqrt{2p^+}}  
    \, T^{a_{\bar k}} _{c_{\bar p}c_{p}}  \]
\[  W_f ={g^2 \over (2\pi)^3}
    {\bar u(\bar p,s')\over\sqrt{2{\bar p}^+}}
    {{/\!\!\!\epsilon}^*(\bar k,\lambda') \over\sqrt{2\bar k^+}}
    {T_{c_{\bar p},c}^{a_{\bar k}}\,\gamma^+\,
      T_{c,c_{p}} ^{a_{k}} \over 2({\bar k}^+-p^+ )}
    {{/\!\!\!\epsilon}(k,\lambda) \over\sqrt{2k^+}}
    {u(p,s)\over\sqrt{2p^+}}  \]
\[  W_b=-{g^2 \over (2\pi)^3 }
    {\bar u(\bar p,s') \gamma^+ u(p,s) 
      \over \sqrt{2\bar p^+}\phantom{\gamma^+ }\sqrt{2p^+}}
    {T_{c_{\bar p},c}^{a_{\bar k}}
      T_{c,c_{p}} ^{a_{k}} \over (\bar p^+-p^+ )^2}
    {\bar u(\bar k,\sigma') \gamma^+ u(k,\sigma) 
    \over \sqrt{2\bar k^+} \phantom{\gamma^+}\sqrt{2k^+}}     
\]    \qquad
\end{minipage}
}\caption{\label{fig:4.4a} 
A few selected matrix elements of the QCD front form 
Hamiltonian $ H = P_+$ in LB-convention. }
\end{figure}

To calculate then $\langle f|(\epsilon)|i\rangle$ 
perturbatively, all $x^+$-ordered diagrams must be 
considered, the contribution from each graph computed 
according to the rules of old-fashioned Hamiltonian 
perturbation theory \cite{kos70,leb80}:
\begin{enumerate}
\item
Draw all topologically distinct $x^+$-ordered diagrams.
\item
Assign to each line a momentum $k^\mu$, a helicity
$\lambda$, as well as color and flavor, 
corresponding to a single particle on-shell, with
$k^\mu k_\mu = m^2$. With fermions (electrons or quark)
associate a spinor $u_\alpha(k,\lambda)$,
with antifermions $v_\alpha(k,\lambda)$,
and with vector bosons (photons or gluons)  a
polarization vector $\epsilon_\mu(k,\lambda)$.
These are given explicitly in App.~\ref{app:Lepage-Brodsky}
and \ref{app:Kogut-Soper}.
\item
For each vertex include factor  $V$ as given in 
Fig.~\ref{fig:4.4}  for QED and Fig.~\ref{fig:4.4a}  for QCD,
with further tables given in section~\ref{sec:DLCQ}.
To convert incoming into outgoing lines or vice
versa replace
\[u \leftrightarrow v \ , \qquad {\overline u} \leftrightarrow
-{\overline v} \ , \qquad \epsilon \leftrightarrow \epsilon^\ast\]
in any of these vertices. (See also items 8,9, and 10)
\item
For each intermediate state there is a factor
\[ {1\over \epsilon - \sum\limits_i 
    \Big({k_{\!\perp}^{\,2} +m^2\over2k^+}\Big)_i +    i0_+} 
\ , \] 
where $\epsilon = \widetilde P_{{\rm in}+} $ is the incident
light-cone energy.
\item 
To account for three-momentum conservation include 
for each intermediate state the delta-functions  
$\delta \Big( P ^+ - \sum_i k ^+_i \Big)$ 
and $\delta ^{(2)}\Big( \vec P _{\!\perp} 
- \sum_i \vec k _{\!\perp i} \Big)$. 
\item
Integrate over each internal $k$ with the weight
\[ \int d^2k_{\!\perp} \, dk^+ {\theta(k^+) \over (2\pi)^{3/2}} \]
and sum over internal helicities (and colors for gauge theories).
\item
Include a factor $-1$ for each closed fermion loop, 
for each fermion
line that both begins and ends in the initial state, 
and for each diagram in which fermion
lines are interchanged in either of the initial or final states.
\item
Imagine that every internal line is a sum
of a `dynamic' and an `instantaneous' line, and draw
all diagrams with $1,2,3,\dots$ instantaneous lines. 
\item
Two consecutive instantaneous interactions give a 
vanishing contribution.
\item
For the instantaneous fermion lines use the 
factor $W_f$ in Figs.~\ref{fig:4.4}  or \ref{fig:4.4a},  or the
corresponding  tables in Section~\ref{sec:DLCQ}.
For the instantaneous boson lines use the factor $W_b$.
\end{enumerate}
The light-cone Fock state representation can thus be used
advantageously in perturbation theory. 
The sum over intermediate Fock states is equivalent to 
summing all $x^+-$ordered diagrams and integrating over 
the transverse momentum and light-cone fractions
$x$. Because of the restriction to positive $x,$ diagrams
corresponding to vacuum fluctuations or those containing
backward-moving lines are eliminated. 

\subsection{Example 1: The $q\bar q$-scattering amplitude}
\label{sec:scattering}

The simplest application of the above rules is  
the calculation of the electron-muon scattering amplitude 
to lowest non-trivial order. But the quark-antiquark 
scattering is only marginally more difficult.
We thus imagine an initial $(q ,\bar q)$-pair with different 
flavors $f \neq \bar f$ to be scattered off each other by
exchanging a gluon.

Let us treat this problem as a pedagogical example
to demonstrate the rules.
Rule 1: There are two time-ordered diagrams associated with
this process. In the first one the gluon is emitted by
the quark and absorbed by the antiquark, and in the second
it is emitted by the antiquark and absorbed by the quark.
For the first diagram, we assign the momenta required 
in rule 2 by giving explicitly the initial and final Fock states
\begin  {eqnarray}
       \vert q ,\bar q \rangle 
       &=&{1\over \sqrt{n_c}} \sum _{c=1} ^{n_c}
       b^\dagger_{cf} (k_q,\lambda_q) 
       d^\dagger_{c\bar f}  (k_{\bar q},\lambda_{\bar q})
       \vert 0 \rangle 
\  ,  \\ 
       \vert q^\prime, \bar q^\prime \rangle 
       &=&{1\over \sqrt{n_c}} \sum _{c=1} ^{n_c}
       b^\dagger_{cf} (k_q^\prime,\lambda_q^\prime) 
       d^\dagger_{c\bar f}(k_{\bar q}^\prime,\lambda_{\bar q}^\prime)
       \vert 0 \rangle
\ ,\end {eqnarray} 
respectively. Note that both states are invariant under  $SU(n_c)$. The usual color singlets of
QCD are abstained by setting $n_c=3$.
The intermediate state
\begin  {eqnarray}
      \vert q^\prime, \bar q,g  \rangle 
&=& \sqrt{{2\over n_c^2-1}} 
       \sum _{c=1} ^{n_c} 
       \sum _{c^\prime=1} ^{n_c} 
       \sum _{a=1} ^{n_c^2-1} 
       T^a_{c,c^\prime}
       b^\dagger_{c\bar f} (k_q^\prime,\lambda_q^\prime) 
       d^\dagger_{c'\bar f}  (k_{\bar q},\lambda_{\bar q})
       a^\dagger_a  (k_{g},\lambda_{g})
       \vert 0 \rangle 
\  , \end {eqnarray} 
has `a gluon in flight'. Under that 
impact, the quark has changed  its momentum (and spin), while the antiquark as a spectator is still 
in its initial state.
At the second vertex, the gluon in flight is absorbed by the 
antiquark, the latter acquiring  its final values
($k'_{\bar q},\lambda'_{\bar q} $). Since the gluons
longitudinal momentum is positive, the diagram allows
only for  $k_q^{\prime+} < k_q^+ $.
Rule 3 requires at each vertex the factors
\begin {eqnarray} 
       \langle q ,\bar q \vert \,V\,
       \vert q^\prime, \bar q , g  \rangle 
       &=& {g\over (2\pi)^{{3\over2}}}
       \sqrt{{n_c^2-1\over 2n_c}}
       \ {\left[ \overline u (k_q,\lambda_q) 
       \,\gamma^\mu\epsilon_\mu (k_g,\lambda_g)\,
        u(k_q^\prime,\lambda_q^\prime)\right] 
        \over \sqrt{2k^+_{q}} \sqrt{2 k^+_{ g}} \sqrt{2k'^+_{q}}  }
\ ,\\
       \langle q^\prime, \bar q ,g  \vert \,V\,
       \vert q ^\prime ,\bar q ^\prime \rangle 
       &=& {g\over (2\pi)^{{3\over2}}}
       \sqrt{{n_c^2-1\over 2n_c}}
        \ {\left[ \overline u (k_{\bar q},\lambda_{\bar q}) 
        \,\gamma^\nu\epsilon^\star _\nu(k_g,\lambda_g)\,
        u(k_{\bar q}^\prime,\lambda_{\bar q}^\prime)\right] 
        \over \sqrt{2k^+_{\bar q}} \sqrt{2k^+_{g}}
        \sqrt{2k'^+_{\bar q}}} 
\  ,    \end {eqnarray} 
respectively. If one works with  color neutral Fock states,
all color structure reduce to
an overall factor $C$, with $C^2=(n_c^2-1)/2n_c$.  
This factor is the only difference between QCD and 
QED for this example.
For QCD  $C^2=4/3$ and for QED  $C^2=1$. 
Rule 4 requires the energy denominator $1/\Delta E$.
With the initial energy
\[
    \epsilon= \widetilde P _+ = {1\over2}\widetilde P ^-
   = (k_q+k_{\bar q})_+ = {1\over2}(k_q+k_{\bar q})^-  
\ ,\]
the energy denominator  
\begin{equation} 
   2\Delta E 
   = (k_q+k_{\bar q})^- - (k_g+k'_q+k_{\bar q})^-
   = -{Q^2 \over k^+_g} 
\label{eq:4s.7}\end{equation} 
can be expressed in terms of the Feynman four-momentum
transfers 
\begin{eqnarray}      
   Q^2 = k_g^+(k_g+k_q^\prime-k_q)^-
   \ ,\quad{\rm and}\quad
   \overline Q^2 = k_g^+(k_g+k_{\bar q}- k_{\bar q} ^\prime)^-
.\label{eq:4s.8}\end{eqnarray}
Rule 5 requires two Dirac-delta  functions, one at each 
vertex, to account for conservation of three-momentum. 
One of them is removed by the requirement of  rule 6, 
namely to integrate over all intermediate internal momenta and
the other remains in the final equation
(\ref{eq:4s.15}).  The momentum of the exchanged gluon is thus 
fixed by the external legs of the graph. 
Rule 6 requires that one sums over the gluon 
helicities.  The  polarization sum gives
\begin {eqnarray} 
        d_{\mu\nu}(k_g) \equiv \sum_{\lambda_g}
        \epsilon_\mu (k_g,\lambda_g)\,
        \epsilon^\star_\nu (k_g,\lambda_g)
        = -g_{\mu\nu} + 
        {k_{g,\mu}\eta_\nu + k_{g,\nu}\eta_\mu 
        \over k_g^\kappa \eta_\kappa } 
\ ,\label{eq:4s.9}\end {eqnarray} 
see Appendix~\ref{app:Lepage-Brodsky}.
The null vector $\eta^\mu$  has the components \cite{leb80}
\begin {equation} 
        \eta^\mu 
        = (\eta^+,\vec\eta_{\!\perp},\eta^-)
        = (0,\vec 0_{\!\perp},2)
\ , \end {equation}
and thus the properties $\eta^2\equiv\eta^\mu\eta_\mu=0$ 
and $k\eta=k^+$. 
In light cone gauge, we find for the $\eta$-dependent terms 
\begin {eqnarray} 
       \left( \sum_{\lambda_g} 
       \langle q ,\bar q \vert \,V\,\vert q', \bar q , g  \rangle 
       \langle q', \bar q ,g  \vert \,V\,\vert q',\bar q'\rangle\right)_\eta 
       &=& {(gC)^2 \over (2\pi)^3}
        \,{1\over 2k^+_g (k_g\eta) }\times
\nonumber\\
        \times\bigg\{
        {\left[\overline u (q)\gamma_\mu k_g^\mu
        u(q^\prime)\right] \over  \sqrt{4k^+_{q}k'^+_{q}}}
        {\left[\overline u (\bar q) \gamma_\nu\eta^\nu 
        u(\bar q ^\prime)\right]
        \over  \sqrt{4k^+_{\bar q} k'^+_{\bar q}} }
        &+&{\left[ \overline u (q)\gamma_\mu\eta^\mu 
        u(q ^\prime)\right] \over \sqrt{4k^+_{q}k'^+_{q}}}
        {\left[ \overline u (\bar q) 
        \gamma_\nu k_g^\nu u(\bar q ^\prime)\right]
        \over  \sqrt{4k^+_{\bar q} k'^+_{\bar q}}} \bigg\} 
\ .\label{eq:4s.11}\end {eqnarray} 
Next, we introduce four-vectors like
$ l^\mu_q = \left(k_g + k_q - k_q^\prime\right) ^\mu$.
Since its three-components vanish by momentum
conservation, $ l^\mu_q $ must be proportional 
to the null vector $\eta ^\mu$. With Eq.(\ref{eq:4s.8}) 
one gets
\begin{equation}      
       l^\mu_q = \left(k_g + k_q - k_q^\prime\right) ^\mu
       = {Q^2\over  2k_g^+}\ \eta ^\mu
       \quad{\rm and}\quad
       l_{\bar q} ^\mu = \left(k_g + k'_{\bar q} 
       - k_{\bar q}\right) ^\mu
       = {\overline Q^2\over  2k_g^+}\ \eta ^\mu
.\end{equation}
The well-known property of the Dirac spinors,  
$ (k_q - k'_q)^\mu\left[
   \overline u (k_q,\lambda_q) \,\gamma_\mu
   \,u(k'_q,\lambda'_q)\right] = 0 $,
yields then 
\[  \left[\overline u (q)\gamma_\mu k_g^\mu u(q^\prime)\right] 
  =\left[\overline u (q)\gamma_\mu \eta^\mu u(q^\prime)\right] 
  {Q^2\over 2k_g^+}
  =\left[\overline u (q)\gamma^+u(q^\prime)\right] 
  {Q^2\over 2k_g^+}      ,\]
and  Eq.(\ref{eq:4s.11}) becomes 
\begin {eqnarray} 
       \left(\sum_{\lambda_g} 
       \langle q ,\bar q \vert \,V\,
       \vert q^\prime, \bar q , g  \rangle 
       \langle q^\prime, \bar q ,g  \vert \,V\,
       \vert q ^\prime ,\bar q ^\prime \rangle \right)_\eta 
       =  {(gC)^2 \over (2\pi)^3} 
        \,{Q^2\over 2(k^+_{ g}  )^3 }
        \ {\left[ \overline u (q) \gamma^+ u(q ^\prime)\right] 
        \over \sqrt{4k^+_q k'^+_q} }
        {\left[ \overline u (\bar q) 
        \gamma^+ u(\bar q ^\prime)\right] 
        \over  \sqrt{4k^+_{\bar q} k'^+_{\bar q}} }. 
\end {eqnarray} 
Including the $g_{\mu\nu}$ contribution, 
the diagram of second order in $V$ gives thus 
\begin{eqnarray} 
   V{1\over \widetilde P_+ -H_0}V 
       =  {g^2 C^2\over (2\pi)^3} 
       \ {\left[ \overline u (k_q,\lambda_q) 
       \gamma^\mu u(k'_q,\lambda'_q)\right]
        \over \sqrt{4k^+_q k'^+_q} }
       {1\over  Q^2}
       {\left[ \overline u (k_{\bar q},\lambda_{\bar q}) 
       \gamma_\mu u(k'_{\bar q},\lambda'_{\bar q})
       \right]\over \sqrt{4k^+_{\bar q} k'^+_{\bar q}}}
\nonumber\\
       -{g^2 C^2\over (2\pi)^3} 
       \ {\left[ \overline u (k_q,\lambda_q) 
       \gamma^+ u(k'_q,\lambda'_q)\right]
        \over \sqrt{4k^+_q k'^+_q} }
       {1\over  \big( k^+_g \big) ^2}
       {\left[ \overline u (k_{\bar q},\lambda_{\bar q}) 
       \gamma^+ u(k'_{\bar q},\lambda'_{\bar q})
       \right]\over \sqrt{4k^+_{\bar q} k'^+_{\bar q}}}
,\label{eq:4s.14}\end{eqnarray} 
up to the delta-functions, and a step function
$\Theta(k'^+_q\leq k^+_q)$,
which truncates the final momenta $k'^+$.~-- 
Evaluating the second time ordered diagram, 
one gets
the same result up to the step function
$\Theta(k'^+_q\geq k^+_q)$. Using
\[   \Theta(k'^+_q\leq k^+_q)+\Theta(k'^+_q\geq k^+_q)=1
   \ ,\]
the final sum of all time-ordered
diagrams to order $g^2$ is Eq.(\ref{eq:4s.14}).
One proceeds with rule 8, by including consecutively
the instantaneous lines. In the present case, there is
only one. From Figure~\ref{fig:4.4}
we find
\begin {eqnarray} 
       \langle q ,\bar q \vert \,W_b\,
       \vert q', \bar q' \rangle 
       =  {g^2 C^2\over (2\pi)^3} 
       \ {\left[ \overline u (k_q,\lambda_q) 
       \gamma^+ u(k'_q,\lambda'_q)\right]
        \over \sqrt{4k^+_q k'^+_q} }
       {1\over  \big( k^+_{q} - k'^+_{q}\big) ^2}
       {\left[ \overline u (k_{\bar q},\lambda_{\bar q}) 
       \gamma^+ u(k'_{\bar q},\lambda'_{\bar q})
       \right]\over \sqrt{4k^+_{\bar q} k'^+_{\bar q}}}
\ .\label{eq:4s.14a}\end {eqnarray} 
Finally, adding up all contributions up to order $g^2$, the
$q\bar q$-scattering  amplitude becomes  
\begin{eqnarray} 
   W+V{1\over \widetilde P_+ -H_0}V 
        &=& {(gC)^2\over (2\pi)^3}
        \ {(-1)\over (k_q-k'_{\bar q}) ^2}
          \ \left[ \overline u (k_{q},\lambda_{q}) 
        \,\gamma^\mu\,
        u(k_{q}^\prime,\lambda_{q}^\prime)\right] 
        \left[ \overline u (k_{\bar q},\lambda_{\bar q}) 
        \,\gamma_\mu\,
        u(k_{\bar q}^\prime,\lambda_{\bar q}^\prime)\right] 
\nonumber\\  
        &\times&       
        {1\over \sqrt{k^+_q k^+_{\bar q}  k'^+_q k'^+_{\bar q}}}    
         \delta(P ^+ - P '^+)
        \delta^{(2)}(\vec P_{\!\perp} -\vec P'_{\!\perp})
\,.\label{eq:4s.15}\end{eqnarray}
The instantaneous diagram $W$ is thus cancelled exactly
against a corresponding term in the diagram of second order
in the vertex interaction $V$. 
Their sum gives the correct second order result.

\subsection{Example 2: Perturbative mass renormalization in
QED (KS)}  \label{sec:mass_ren}

 As an example for  light-cone perturbation theory
we follow here the work of 
Mustaki,  Pinsky,  Shigemitsu and Wilson \cite{mps91,pin91a}
to calculate the second order mass renormalization of the 
electron and the renormalization constants $Z_2$ and $Z_3$
in the KS convention.

\begin{figure}
\begin{minipage} {62mm}\fbox{
\epsfysize=65mm\epsfbox{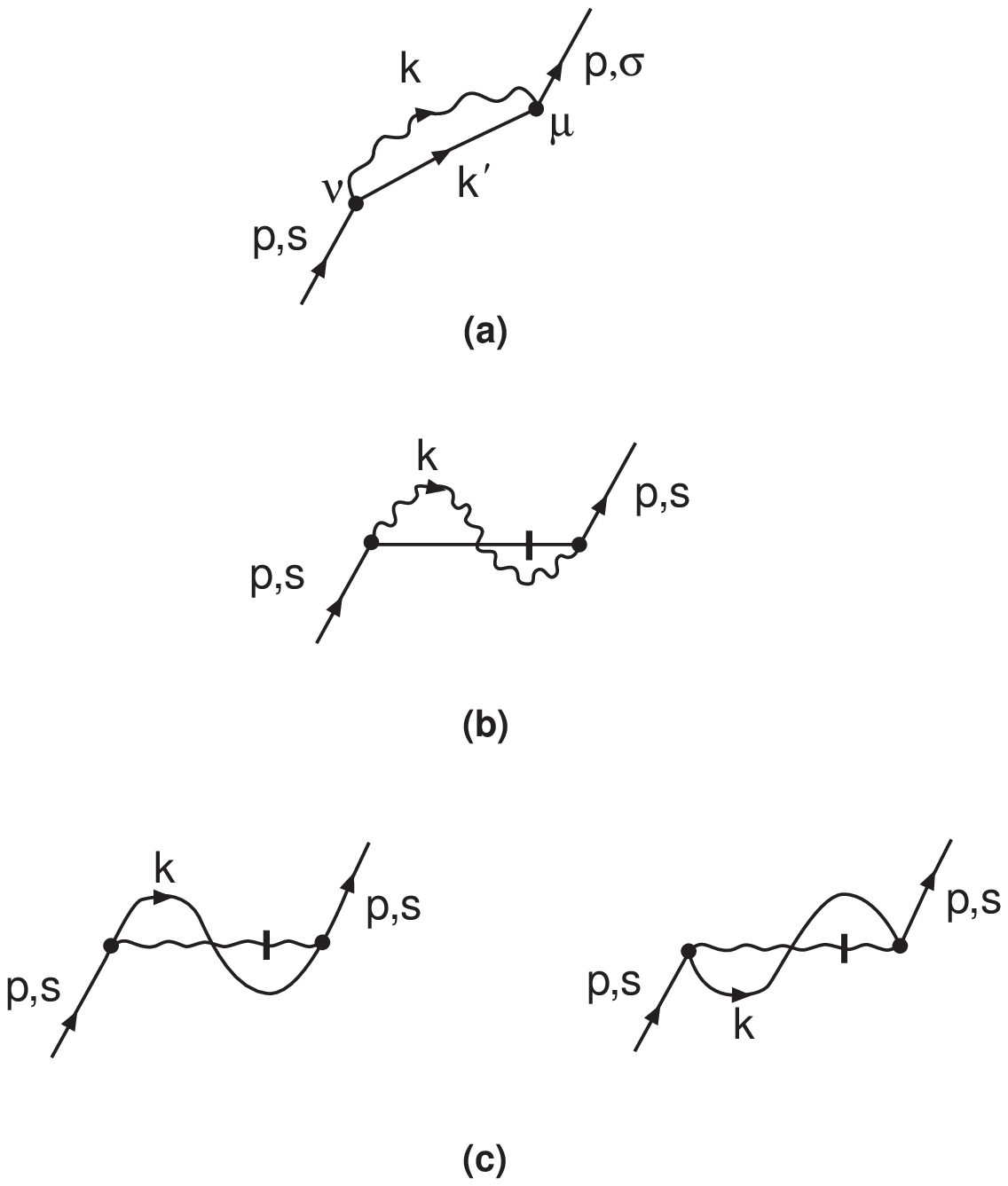}
}\caption{\label{fig:4.5} \sl 
One loop self energy correction for the electron. 
Time flows upward in these diagrams.} 
\end{minipage}
\ \hfill
\begin{minipage} {85mm}\fbox{
\epsfxsize=80mm\epsfbox{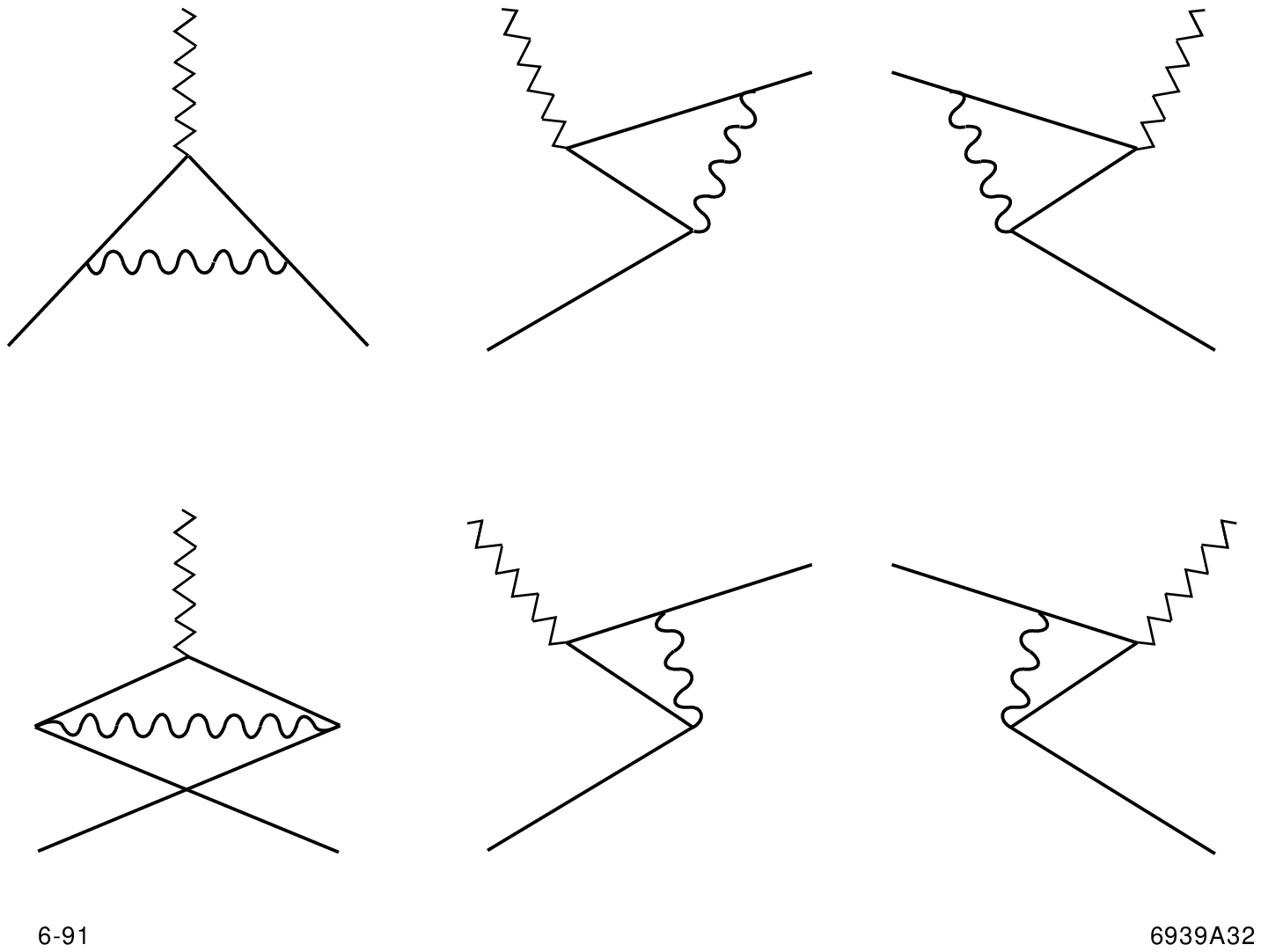} 
}\caption{\label{fig:4.6} \sl 
Time-ordered contributions to the electron's anomalous 
magnetic moment. In light-cone quantization  with 
$q^+=0,$ only the upper left graph needs to be 
computed to  obtain the Schwinger result.}
\end{minipage}
\end{figure}

Since all particles are on-shell in light-cone time-ordered 
perturbation theory,  the electron wave-function renormalization 
$Z_2$ must be obtained separately from the mass
renormalization  $\delta m$.  At order $e^2$, one finds three
contributions. First, the perturbation expansion
\begin {equation}
   T=W+V{1\over p_+ -H_0}V
\end {equation}
yields a second-order contribution in $V$,  as shown in
diagram (a) of Fig.\ref{fig:4.5}. The initial (or final) electron
four momentum is denoted by
\begin {equation}
   p^\mu =(p^+,\vec p_{\!\perp}, {p_{\!\perp}^{\,2} +m^2 
   \over 2p^+}) 
\ . \end {equation}
Second and third, one has first-order contributions from
$W_f$ and  $W_g$, corresponding to diagrams (b) and (c) of
the figure. In the literature \cite{pab85a,tbp91,brp91} these
two-point vertices have been called  ``seagulls'' or 
``self-induced inertias''.

One has to calculate the transition matrix amplitude 
$T_{pp}\delta_{s,\sigma}= \left\langle p,s\mid T 
\mid p,\sigma\right\rangle$
 between a free electron states with momentum and spin $(p,s)$ 
and one with  momentum and spin $(p,\sigma)$.  
The normalization of states as in Eq.(\ref{eq:4.3}) was thus far
\begin{equation}
   \langle p', s'\vert p,s\rangle  =\delta(p^+ -p'^+)\delta^2
   ( p_{\perp}- p_{\perp}')\delta_{s,s'}
\ ,\end{equation}
but for an invariant normalization it is better to use $\langle
{\tilde p,s} |\equiv\sqrt{2p^+}\,\langle {p,s}|$. Then one finds,
\begin {equation}
   2m\,\delta m\,\delta_{s\sigma}\equiv T_{\tilde p\tilde p}=
   2p^+ T_{pp}\ \Longrightarrow
   \ \delta m\,\delta_{s\sigma}={p^+ \over m}T_{pp}
\ .\end {equation}
The other momenta appearing in diagram (a) of 
Fig.\ref{fig:4.5} are 
\begin{eqnarray}
   k &=& (k^+,\vec k_{\!\perp},{k_{\!\perp}^2 \over 2k^+})
\ , \\ {\rm and }\quad
   k'&=& (p^+ -k^+,{( p_{\!\perp} - k_{\!\perp})^2 +m^2 
   \over 2(p^+ -k^+)}, \vec p_{\!\perp}- \vec k_{\!\perp}) 
\ . \end{eqnarray}
Using the above rules to calculate
$T_{pp}$, one obtains for the contribution from diagram (a) 
\begin{eqnarray}
   \delta m_a \,\delta_{s\sigma}= 
   e^2{1 \over m}\sum_{\lambda,{s'}}
   \int{d^2k_{\!\perp}\over(4\pi)^3} \int_{0}^{p^+} dk^+ 
   {[\bar u(p,\sigma)/\!\!\!\epsilon^*(k,\lambda) u(k',s') ]
   [\bar u(k',s')/\!\!\!\epsilon(k,\lambda) u(p,s)] \over 
   k^+ (p^+ -k^+) (p^- -k^- -k'^-)}.
\end{eqnarray}
It can be shown that
\begin{equation}
   [\bar u(p,\sigma)\gamma^\mu(/\!\!\!{k}'+m)
   \gamma^\nu u(p,s)]\,d_{\mu\nu}(k)=4
   \delta_{s\sigma}
   \bigg[\Big({2p^+ \over k^+}
   +{k^+ \over p^+ -k^+}\Big)(p\cdot k)-m^2\bigg]
\ ,\end{equation}
which leads to the expression given below for $\delta m_a$.
For diagram (b), one gets, using the rule for the instantaneous 
fermion,
\begin{eqnarray}
   \delta m_b & = &e^2{ p^+ \over 2m}
   \sum_\lambda \int {d^2 k_{\!\perp} \over(2\pi)^3} \int_{0}^{+\infty} dk^+
   {\bar u(p,s) /\!\!\!\epsilon^*(k,\lambda)
   \gamma^+ /\!\!\!\epsilon(k,\lambda) u(p,\sigma) \over 2p^+ 2k^+
   2(p^+-k^+)} \nonumber \\ 
   &= & e^2 {p^+ \over 2m} \int {d^2 k_{\!\perp}\over(2\pi)^3}
   \int_{0}^{+\infty}{dk^+\over k^+ (p^+ -k^+)} .
\end{eqnarray}
For diagram 1(c) one finds,
\begin{eqnarray}
   \delta m_c &=&{e^2 p^+ \over 2m}\sum_s\int {d^2
   k_{\!\perp} \over(2\pi)^3 } \int_{0}^{+\infty}{dk^+ \over 2 k^+}
   {\bar u(p,s) \gamma^+ \over \sqrt2 p^+}
   \bigg[{ u(k,s) \bar u(k,s)  \over 2(p^+ -k^+)^2}-
   {v(k,s) \bar v(k,s) \over   2(p^+ +k^+)^2}\bigg]
   {\gamma^+ u(p,\sigma) \over \sqrt2 p^+}
\nonumber \\
   &=&{e^2 p^+ \over 2m}\int {d^2k_{\!\perp}
   \over(2\pi)^3} \bigg[\int_{0}^{+\infty} {dk^+\over (p^+-k^+)^2}\,
   -\int_{0}^{+\infty} {dk^+\over (p^+ +k^+)^2}\bigg]
\ .\end{eqnarray}
These integrals have potential singularities at $k^+ =0$ 
and $k^+ = p^+$, as well as an ultra-violet divergence in 
$ k_{\!\perp}$. To regularize them, we introduce in a first step 
small cut-offs $\alpha$ and $\beta$:
\begin{equation}
   \alpha<k^+<p^+ -\beta
\ ,\end{equation}
and get rid of the pole at $k^+ =p^+$ in $\delta m_b$ and 
$\delta m_c$ by a principal value prescription. One
obtains then
\begin{eqnarray}
   \delta m_a &=&{e^2 \over 2m}\int {d^2 k_{\!\perp}\over(2\pi)^3}
   \bigg[\int_{0}^{p^+}{dk^+ \over k^+}{m^2 \over p\cdot k}
   -2\Big({p^+\over\alpha}-1\Big)
   -\ln\Big({p^+ \over\beta}\Big)\bigg]
\ ,\nonumber \\
   \delta m_b &=&{e^2\over 2m}\int {d^2 k_{\!\perp}\over(2\pi)^3}
   \ln\Big({p^+ \over\alpha}\Big)
\ ,\\ \nonumber
   \delta m_c &=&{e^2 \over m}
   \int {d^2 k_{\!\perp}\over(2\pi)^3}
   \bigg({p^+ \over\alpha}-1\bigg)
\ ,\end{eqnarray}
where
\begin{equation}
   p\cdot k={m^2 (k^+)^2 +(p^+)^2  k_{\!\perp}^2 \over 2p^+ k^+}
\ .\end{equation}
Adding these three contributions yields
\begin{equation}
   \delta m={e^2 \over 2m}\int {d^2 k_{\!\perp}\over(2\pi)^3}
   \bigg[\int_{0}^{p^+}{dk^+ \over k^+}{m^2 \over p\cdot k}
   +\ln\Big({\beta \over\alpha}\Big)\bigg]
\ .\label{eq:2.regu}\end{equation}
Note the cancelation of the most singular infrared divergence.

To complete the calculation, we present two possible regularization
procedures:\\
\noindent\textbf{1. Transverse dimensional regularization.}
The dimension of transverse space, $d$, is continued 
from its physical value of $2$ to $2+\epsilon$ and all
integrals  are replaced by
\begin{equation}
   \int d^2 k_{\!\perp}\to(\mu^2)^\epsilon\int d^d k_{\!\perp}
,\end{equation}
using $\epsilon=1-d/2$ as a small quantity. One thus gets
\begin{eqnarray}
   (\mu^2)^\epsilon\int d^dk_{\!\perp}
   \left(\vec k^{\,2}_{\!\perp}\right)^\alpha
   &=& 0,\qquad{\rm for}\quad\alpha\ge0
, \nonumber\\
   (\mu^2)^\epsilon\int d^dk_{\!\perp}
   {1\over\vec k^{\,2}_{\!\perp}+M^2}
   &=&  \left({\mu^2\over M^2}\right)^\epsilon{\pi\over\epsilon}
, \nonumber\\
   (\mu^2)^\epsilon\int d^d k_{\!\perp}
   {1\over (\vec k^{\,2}_{\!\perp}+M^2)^2}
   &=& \left({\mu^2\over M^2}\right)^\epsilon
   {\pi\over M^2}
, \nonumber\\
   (\mu^2)^\epsilon\int d^d k_{\!\perp}
   {\vec k^{\,2}_{\!\perp}\over \vec k^{\,2}_{\!\perp}+M^2}
   &=& -\left({\mu^2\over M^2}\right)^\epsilon
   {\pi M^2 \over\epsilon}
.\end{eqnarray}
In this method,  $\alpha$ and $\beta$ in Eq.(\ref{eq:2.regu})
are treated as  constants. Dimensional regularization gives
zero for the logarithmic term, and for the remainder
\begin {equation}
   \delta m={e^2 m\over (2\pi)^3}\int^{1}_{0}dx
   \int {d^2  k_{\perp}\over  k_{\perp}^{\,2} + m^2 x^2}
\ ,\end {equation}
with $x\equiv(k^+ /p^+)$, the above integral yields 
\begin {equation}
   \delta m={e^2 m\over 8\pi^2 \epsilon} 
\end {equation}
as the final result.

\textbf{2. Cut-offs.}
In this method \cite{leb80,tbp91,brp91}, one restricts the
momenta of any  intermediate Fock state by means of 
the invariant condition
\begin {equation}
\widetilde P^2 = \sum_i \left(
{m^2+k_{\!\perp}^2\over x}\right)_i \leq\Lambda^2
\ ,\label{cut}\end {equation}
where $\widetilde P$ is the free total four-momentum 
of the intermediate state, and  where $\Lambda$ is a 
large cut-off.  Furthermore, one assumes that all 
transverse momenta are smaller than a certain cut-off  
$\Lambda_{\perp}$, with
\begin {equation}
   \Lambda_{\perp}\ll \Lambda
\ .\end {equation}
In the case of diagram (a) of Fig.~\ref{fig:4.4}, 
Eq.~(\ref{cut}) reads 
\begin{equation}
   { k_{\perp}^2 \over k^+}
   +{( p_{\perp}- k_{\perp})^2 +m^2\over p^+
   -k^+} < \Lambda'
\ , \qquad{\rm with\quad}
   \Lambda'\equiv{\Lambda^2 + p_{\perp}^2 \over p^+}
\ .\end{equation}
Hence
\begin{equation}
   \alpha={ k_{\perp}^2 \over \Lambda'}
\ ,\quad
   \beta={( p_{\perp}-k_{\perp})^2 +m^2 \over\Lambda'}
   \quad\Longrightarrow\quad
   {\beta \over \alpha}
   ={(p_{\perp}- k_{\perp})^2 +m^2 \over k_{\perp}^2}
\ .\end{equation}
In \cite{mps91} it is shown that
\begin{equation}
   \int d^2 k_{\perp}\,\ln\Big({\beta\over\alpha}\Big)
   =\int d^2k_{\perp}\,\int\limits_{0}^{p^+}
   {dk^+ \over p^+}{m^2 \over p\cdot k}
\ . \end{equation}
Now
\begin{equation}
   \delta m={e^2 \over 2m}
   \int {d^2 k_{\perp}\over(2\pi)^3}
   \int_{0}^{p^+}dk^+ {m^2 \over p\cdot k}
   \Big({1\over p^+}+{1\over k^+}\Big)
\ .\end{equation}
Upon integration, and dropping the finite part, one finds 
\begin {equation}
   \delta m={3e^2 m\over 16\pi^2}
   \ln\Big({\Lambda_{\perp}^2 \over m^2}\Big)
\ ,\end {equation}
which is of the same form as the standard result \cite{bjd65}. 
Since $\delta m$ is not by itself a measurable quantity, 
there is no contradiction in finding different results. 
Note that the seagulls are necessary for obtaining
the conventional result.

Finally, the wave-function renormalization $Z_2$, at order
$e^2$, is given by 
\begin {equation}
   1-Z_2 = \sum_{m}\!'\,  
   { \Big| \left\langle p\mid V\mid m\right\rangle \Big|^2
   \over (p_+ - \widetilde P_{+,m })^2}
\ , \end {equation}
where $\widetilde P_{+,m }$ is the free total energy of the
intermediate state $m$. Note that this expression is the 
same as one of the contributions to $\delta m$, 
except that here the denominator is squared. 
One  has thus
\begin {eqnarray}
   (1-Z_2)\delta_{s\sigma} &=& {e^2 \over p^+}
   \int {d^2 k_{\perp}\over(4\pi)^3} 
   \int_{0}^{p^+}{dk^+\over k^+ (p^+ -k^+)}
   {\bar u(p,\sigma)\gamma^\mu 
   (\gamma^\alpha k_{\alpha}'+m)\gamma^\nu u(p,s) 
   d_{\mu\nu}(k)\over (p^- -k^- -k'^-)^2} 
\nonumber \\
   &=& {e^2 \delta_{s\sigma}\over (2\pi)^3}\int_{0}^{1}dx
   \int {d^2 k_{\perp}\over
   k_{\perp}^2 +m^2 x^2} \bigg[{2(1-x) k_{\perp}^2 
   \over x( k_{\perp}^2 +m^2 x^2)}+x\bigg]
\ ,\end {eqnarray}
which is the same result as obtained by Kogut and Soper
\cite{kos70}. Naturally this  integral is both infrared and 
ultraviolet divergent.  Using the above rules, one gets
\begin {equation}
   Z_2 (p^+)=1+{e^2\over8\pi^2\epsilon}\bigg[{3\over 2}
   -2\ln\Big({p^+ \over\alpha}\Big)\bigg]
  +{e^2\over(2\pi)^2}\ln\Big({p^+ \over\alpha}\Big)
   \bigg[1-2\ln\Big({\mu \over m}\Big)
   -\ln\Big({p^+ \over\alpha}\Big)\bigg]
\ ,\end {equation}
where $\mu^2$ is the scale introduced by dimensional  
regularization. Note that $Z_2$ has an unusual  dependence 
on the longitudinal momentum, not found in the conventional
instant form. But this may vary with the choice of
regularization.  A similar $p^+$  dependence  was
found for scalar QED by Thorn \cite{tho79a,tho79b}. 

In \cite{mps91} the full renormalization of front form QED
was carried out to the one-loop level.  
Electron and photon mass corrections were evaluated,
as well as the wave function  renormalization constants 
$Z_2$ and $Z_3$, and the vertex correction $Z_1$. 
One feature that distinguishes the  front form from the 
instant form results is that the ultraviolet-divergent parts 
of $Z_1$ and $Z_2$ exhibit momentum dependence.  
For physical quantities such as the renormalized 
charge $e_R$, this  momentum dependence cancels
due to the Ward identity 
$Z_1(p^+,p'^+)=\sqrt{Z_2(p^+)Z_2(p'^+)}$.
On the other hand,  momentum-dependent  renormalization 
constants imply nonlocal counter terms. Given that the tree 
level Hamiltonian is nonlocal in $x^{-}$, it is actually not 
surprising to find counter terms exhibiting non-locality.  
As mentioned in \cite{wwh94}, the  power  counting works 
differently here in the front than in the instant form. 
This is  already indicated by the presence of four-point 
interactions in the  Hamiltonian.
The momentum dependence in $Z_1$ and $Z_2$ is another  
manifestation of unusual power counting laws. It will be  
interesting to apply them systematically in the case of QED. 
Power counting  alone does not provide information about 
cancelation of divergences  between diagrams. It is therefore
important to gain more insight into the mechanism of 
cancelation in cases where one does expect this to occur 
as in the calculation of the electron mass  shift. 

\subsection{Example 3: The anomalous magnetic moment}

The anomalous magnetic moment of the electron
had been calculated in the front form by
Brodsky, Roskies and Suaya \cite{brs73}, using 
the method of alternating denominators.  
Its calculation is a transparent example of calculating
electro-magnetic form factors for both elementary and 
composite systems \cite{bks71,brd80} as presented in 
Section~\ref{sec:wavefunctions}, and for applying
light-cone perturbation theory.
Langnau and Burkardt \cite{bul91a,bul91b,lab93a,lab93b} 
have calculated the anomalous magnetic moment at very 
strong coupling, by combining this method with 
discretized light-cone quantization, see below.
We choose light-cone coordinates corresponding to the 
Drell frame, Eq.(\ref{eq:4.17}), and denote as in the
preceding section the electron's  four-momentum and 
spin with $(p,s)$. 
In line with Eq.(\ref{eq:4.20}), the Dirac and Pauli form
factors can be identified  from the spin-conserving and 
spin-flip current matrix elements:

\begin{eqnarray}
   {\cal M}^+_{\uparrow\uparrow} 
   &=&\Big\langle p+q,\uparrow\Big|\frac{J^+(0)}{p^+}
   \Big|p,\uparrow\Big\rangle 
   = 2F_1(q^2) 
\ , \label{eq7}\\[5pt]
   {\cal M}^+_{\uparrow\downarrow} 
   &=&\Big\langle p+q,\uparrow\Big|\frac{J^+(0)}{p^+}
   \Big|p,\downarrow\Big\rangle 
   = - 2(q_1-iq_2) \ \frac{F_2(q^2)}{2M} 
\ ,\label{eq8} \end{eqnarray}
where $\uparrow$ corresponds to positive spin projection 
$s_z=+ \frac{1}{2}$ along the $z$-axis.
The mass of the composite system $M$ is of course the 
physical mass $m$ of the lepton. 
The interaction of the current $J^+(0)$ conserves the 
helicity of the struck constituent fermion 
$(\bar u_{\lambda^\prime}\gamma^+u_\lambda)/k_+
=2\delta_{\lambda\lambda^\prime}$.   
Thus, one has from Eqs.~(\ref{eq:4.23}), (\ref{eq7}) and  
(\ref{eq8}) 
\begin{eqnarray}
   F_1(q^2) = \frac{1}{2} {\cal M}^+_{\uparrow\uparrow} 
   &=& \sum_{j} e_j\int [d\mu_n]\,
   \psi^{*(n)}_{p+q,\uparrow}\left(x,
   k_{\!\perp},\lambda\right)\,
   \psi^{(n)}_{p,\uparrow}\left(x, k_{\!\perp},\lambda\right) 
,\label{eq9} \\ 
   -\left(\frac{q_1-iq_2}{2M}\right) F_2(q^2)  
   ={1\over2} {\cal M}^+_{\uparrow\downarrow}
   &=& \sum_{j} e_j \int [d\mu_n]\,
   \psi^{*(n)}_{p+q,\uparrow} (x,k_{\!\perp},\lambda)\,
   \psi^{(n)}_{p,\downarrow} (x, k_{\!\perp},\lambda) 
.\label{eq:4.i10} \end{eqnarray}
In this notation, the summation over all contributing Fock 
states ($n$) and helicities  ($\lambda$) is assumed, and 
the reference to single particle states  $i$ in the Fock states 
is suppressed. Momentum conservation is used to eliminate 
the explicit reference to the momentum of the struck lepton  
in Eq.(\ref{eq:4.24}). Finally, the leptons wave function 
directed along the  final direction $p+q$ in the current 
matrix element  is denoted as
\[  \psi^{(n)}_{p+q,s_z} (x, \vec k_{\!\perp}, \lambda) 
     = \Psi_{n/e\,(p+q, s ^2, s _z)}(x_i,\vec k_{\!\perp i},
     \lambda_i) \ .\]
One recalls that $F_1(q^2)$ evaluated in the limit  
$q^2\rightarrow 0$  with $F_1\rightarrow 1$ is equivalent 
to wave function  normalization 
\begin{equation}
   \int[d\mu]\,\psi^*_{p\uparrow}\,\psi_{p\uparrow}=1
   \qquad{\rm and}\quad
   \int[d\mu]\,\psi^*_{p\downarrow}\,\psi_{p\downarrow}=1  
\ .\label{eq:4.i15}\end{equation}
The anomalous moment $a=F_2(0)/F_1(0)$ can be 
determined from the coefficient linear in $q_1-iq_2$ from
$\psi^*_{p+q}$ in Eq. (\ref{eq:4.i10}). 
Since according to Eq.(\ref{eq:4.24})
\begin{equation}
   \frac{\partial}{\partial q_{\!\perp}}\, \psi^*_{p+q}
   \equiv - \sum_{i\ne j}x_i \frac{\partial}{\partial
   k_{\!\perp i}}\, \psi^*_{p+q}
\ ,\end{equation}
one can, after integration by parts, write explicitly
\begin{equation}
\frac{a}{M} = - \sum_je_j\int [d\mu_n]\, \sum_{i\ne j} 
\psi^*_{p \uparrow}\ x_i \Big( 
\frac{\partial}{\partial k_1}+i\frac{\partial}{\partial k_2}
\Big)_i \psi_{p\downarrow}
\ .\label{eq:4.i14}\end{equation}
The anomalous moment can thus be expressed in terms 
of a local matrix element at zero momentum transfer, 
(see also with Section~\ref{sec:impact} below).
It should be emphasized that Eq.(\ref{eq:4.i14}) is exact,
valid for the anomalous element of actually {\em any}
spin-$\frac{1}{2}$-system.

As an example for the above perturbative formalism,
one can evaluate the electron's anomalous moment to
order $\alpha$ \cite{brs73}. 
In principle, one would have to account for all $x^+$-ordered
diagrams as displayed in Figure~\ref{fig:4.6}. 
But most of them do not contribute, because either the 
vacuum fluctuation graphs vanish in the front form or
they vanish because of using the Drell frame.
Only the diagram in the upper left corner of Figure~\ref{fig:4.6}
contributes the two electron-photon Fock states with spins 
$|{1\over2}\lambda_e,\lambda_\gamma\rangle
   =| -\frac{1}{2},1\rangle$ and $|\frac{1}{2},-1\rangle$:
\begin{eqnarray}
   \psi_{p\downarrow} 
   &=& \frac{e/\sqrt x} {M^2-\frac{k^2_{\!\perp}+\lambda^2}
   {x}-\frac{k^2_{\!\perp}+\widehat m^2}{1-x}}\times\cases{ 
   \sqrt 2\  \frac{(k_1-ik_2)}{x}, & for 
   $|-\frac{1}{2}\rangle \rightarrow|-\frac{1}{2},1\rangle,$\cr 
   \sqrt 2\  \frac{M(1-x)-\widehat m}{1-x}, & for
   $|-\frac{1}{2}\rangle\rightarrow|\frac{1}{2},-1\rangle,$\cr}
\label{eq:4.i16} \\
   \psi^*_{p\uparrow} 
   &=& \frac{e/\sqrt x} {M^2-\frac{k^2_{\!\perp}+\lambda^2}
   {x}-\frac{k^2_{\!\perp}z+\widehat m^2}{1-x}}\times\cases{
   -\sqrt 2\  \frac{M(1-x)-\widehat m}{1-x}, & for
   $|-\frac{1}{2},1\rangle \rightarrow|\frac{1}{2}\rangle,$\cr 
   -\sqrt 2\  \frac{(k_1-ik_2)}{x}, & for
   $|\frac{1}{2},-1\rangle\rightarrow|\frac{1}{2}\rangle.$\cr}
\label{eq:4.i17}\end{eqnarray}
The quantities to the left of the curly bracket in 
Eqs.(\ref{eq:4.i16}) and (\ref{eq:4.i17}) are the matrix 
elements of
\[ {\bar u(p+k,\lambda)\over p^+-k^+)^{1/2}}
\,\gamma^\mu\epsilon^*_\mu    
(k,\lambda'') \,{u(p,\lambda')\over (p^+)^{1/2}}      
\qquad \hbox{and}\quad     
{\bar u (p,\lambda)\over(p^+)^{1/2}}
\,\gamma^\mu\epsilon_\mu (k,\lambda'')
\,{u(p-k,\lambda')\over (p^+-k^+)^{1/2}} 
\ ,\]
respectively, where $k^\mu\epsilon_\mu (k,\lambda)=0$
and in light-cone gauge $\epsilon^+(k,\lambda)=0$.
In LB-convention holds
$\vec \epsilon _{\!\perp}(\vec k_{\!\perp},\lambda)
  \longrightarrow \vec \epsilon _{\!\perp}(\vec k_{\!\perp},\pm)
  =\pm (1/\sqrt2)(\widehat x\pm i\widehat y)$, see also 
Appendix~\ref{app:Lepage-Brodsky} \cite{bks71}.
For the sake of generality, we let the intermediate lepton 
and boson have mass $\widehat m$ and $\widetilde m$,
respectively. 
Substituting (\ref{eq:4.i16}) and (\ref{eq:4.i17}) into 
Eq.(\ref{eq:4.i14}),
one finds that only the $|-\frac{1}{2},1\rangle$
intermediate state actually contributes to $a$, 
since terms which involve differentiation of the 
denominator of $\psi_{p\downarrow}$ cancel.  
One thus gets \cite{brd80}
\begin{eqnarray}
   a &=& 4M\, e^2\int \frac{d^2k_{\!\perp}}{16\pi^3}\int^1_0dx
   \ \frac{\left[\widehat
   m-(1-x)M\right]/x(1-x)}{\left[M^2-(k^2_{\!\perp}+\widehat
   m^2)/(1-x)-(k^2_{\!\perp}+\widetilde m^2)/x\right]^2} 
\ ,\nonumber \\
   &=& \frac{\alpha}{\pi}\int^1_0dx
   \ \frac{M\left[\widehat m-M(1-x)\right]x(1-x)}
   {\widehat m^2x+\widetilde m^2(1-x)-M^2x(1-x)} 
\ ,\end{eqnarray}
which, in the case of QED $(\widehat m=M, \widetilde m=0)$ 
gives the Schwinger result $a=\alpha/2\pi$ \cite{brs73}.
As compared to Schwinger the above is an almost trivial
calculation.

The general result (\ref{eq:4.i14}) can also be written 
in matrix form:
\begin{equation}
\frac{a}{2M} = - \sum_j e_j \int \left[dx\,d^2k_{\!\perp}\right]\,
\psi^\star \vec S_{\!\perp}\cdot \vec L_{\!\perp}\psi 
\ ,\label{eq20}\end{equation}
where $\vec S_{\!\perp}$ is the spin operator for the total 
system and $\vec L_{\!\perp}$ is the generator of ``Galilean'' 
transverse boosts \cite{bks71} on the light cone, {\it i.e.}  
$\vec S_{\!\perp}\cdot  \vec L_{\!\perp} 
  = (S_+L_-+S_-L_+)/2$ where $S_\pm = (S_1\pm iS_2)$ 
is the spin-ladder operator and
\begin{equation}
L_\pm = \sum_{i\ne j} x_i \left(\frac{\partial}{\partial k_{
i}} \mp i\, \frac{\partial}{\partial k_{2i}}\right)
\label{eq21}
\end{equation}
(summed over spectators) in the analog of the angular 
momentum operator $ \vec r \times  \vec p$.  
Eq.(\ref{eq:4.i14}) can also be written simply as an expectation 
value in impact space.

The results given in Eqs. (\ref{eq9}), (\ref{eq:4.i10}), and
(\ref{eq:4.i14}) may also be convenient for calculating the anomalous
moments and form factors of hadrons in quantum chromodynamics
directly from the quark and gluon wavefunctions 
$\psi(x,\vec k_{\!\perp},\lambda)$.  These wave functions 
can also be used to calculate the structure functions and 
distribution amplitudes which control large momentum 
transfer inclusive and exclusive processes.  The charge 
radius of a composite system can also be written in the 
form of a local, forward matrix element:
\begin{equation}
    {\partial F_1(q^2)\over \partial q^2}\Bigg|_{q^2=0} 
    = - \sum _j e_j \int \left[dx\,d^2k_{\!\perp}\right]
    \,\psi^\star_{p,\uparrow} \Big( \sum_{i\ne j} 
    x_i\,{\partial\over \partial k_{\!\perp i}} \Big)^2\,
    \psi_{p,\uparrow}   
\ . \label{eq22}\end{equation}

We thus find that, in general, any Fock state $\left|n\right\rangle$
which couples to both $\psi^*_\uparrow$ and $\psi_\downarrow$
will give a contribution to the anomalous moment.  
Notice that because of rotational symmetry in the 
$\widehat x$- and $\widehat y$-direction, the
contribution to $a=F_2(0)$ in Eq. (\ref{eq:4.i14}) 
always involves the form $(a,b=1,\ldots,n)$
\begin{equation}
    M\sum_{i\ne j} \psi^*_\uparrow
    \ x_i\, {\partial\over \partial     k_{\!\perp i}}\ \psi_\downarrow 
    \sim \mu M 
    \ \rho(\vec k^{\,a}_{\!\perp}\cdot \vec k^{\,b}_{\!\perp})
\ ,\label{eq23}\end{equation}
compared to the integral (\ref{eq:4.i15}) for wave-function
normalization which has terms of order
\begin{equation}
   \psi^*_\uparrow \psi_\uparrow 
   \sim  \vec k^{\,a}_{\!\perp}\cdot \vec k^{\,b}_{\!\perp}
   \ \rho(\vec k^{\,a}_{\!\perp}\cdot \vec k^{\,b}_{\!\perp})
   \quad{\rm and}\quad   \mu^2 
   \,\rho(\vec k^{\,a}_{\!\perp}\cdot \vec k^{\,b}_{\!\perp}) 
\ .\label{eq24} \end{equation}
Here $\rho$ is a rotationally invariant function of the
transverse momenta and $\mu$ is a constant with 
dimensions of mass.  Thus, by order of magnitude
\begin{equation}
   a = {\cal O}\, \Big({\mu M\over \mu^2+
   \langle \vec k^{\,2}_{\!\perp}\rangle}\Big)
\label{eq25}\end{equation}
summed and weighted over the Fock states.  In the case of a
renormalizable theory, the only parameters $\mu$ with the
dimension of mass are fermion masses.  
In super-renormalizable theories, $\mu$ can be proportional 
to a coupling constant $g$ with dimension of mass.

In the case where all the mass-scale parameters of the 
composite state are of the same order of magnitude, 
we obtain $a={\cal O}(MR)$ as in Eq. (\ref{eq:4.13}), 
where $R= \langle \vec k^{\,2}_{\!\perp}\rangle^{-1/2}$ 
is the characteristic size of the Fock state.  
On the other hand, in theories where
$\mu^2 \ll \langle \vec k^{\,2}_{\!\perp}\rangle$, 
we obtain the quadratic relation $a ={\cal O}(\mu\, MR^2)$.
Thus composite models for leptons can avoid conflict with the
high-precision QED measurements in several ways.
\begin{itemize}
\item
There can be strong cancelations between the contribution of
different Fock states.
\item
The parameter $\mu$ can be minimized.  For example, in a
renormalizable theory this can be accomplished by having 
the bound state of light fermions and heavy bosons.  
Since $\mu \geq M$, we then have $a \geq {\cal O}(M^2R^2)$.
\item
If the parameter $\mu$ is of the same order as the other mass 
scales in the composite state, then we have a linear condition 
$a = {\cal O}(MR)$.
\end{itemize}

\subsection{1+1 Dimensional: Schwinger Model (LB)} 
 
Quantum electrodynamics in one-space and one-time
dimension (QED$_{1+1}$) with massless charged fermions  
is known as the Schwinger model. It is one of the very few
models of field theory which can be solved analytically 
\cite{los71,sch62a,sch62b,col73,cjs75,col76}.  The charged particles are 
confined because the Coulomb interaction in one space 
dimension is linear in the relative distance, and there is
only one physical particle, a massive neutral scalar particle 
with no self-interactions.  
The Fock-space content of the physical states depends
crucially on the coordinate system and on the gauge.
It is only in the front form that a simple
constituent picture emerges \cite{ber77,mcc91,maz93}. 
It is the best example of the type of simplification that 
people hope will occur for  QCD in physical space-time.
Recent studies of similar model with massive fermion 
and for non-abelian theory where the fermion is in the 
fundamental and adjoint representation show however 
that many  properties are unique to the Schwinger model 
\cite{grk96,nak83}. 

The Schwinger model in Hamiltonian front form field 
theory was studied first by Bergknoff\cite{ber77}.
The description here follows him closely, as well as Perry's
recent lectures \cite{per94a}.  There is an
extensive literature on this subject: 
DLCQ \cite{epb87,yuh91}, 
lattice gauge theory \cite{crh80}, 
light-front integral equations\cite{mah89}, 
and light-front Tamm-Dancoff approaches\cite{mop93}
have used the model for testing the various methods.

Bergknoff showed that the physical boson in the Schwinger 
model in light-cone gauge is a pure electron-positron state.  
This is an amazing result in a strong-coupling theory of
massless bare particles, and it illustrates how a constituent 
picture may arise in QCD.  
The kinetic energy vanishes in the massless limit, 
and the potential energy is minimized by a wave function 
that is flat in momentum space.
One might expect that since a linear potential produces 
a state that is as localized as possible in position space.  

Consider first the massive Schwinger model.
The finite fermion mass $m$ is a parameter to be set to 
zero, later. The Lagrangian for the theory takes the same
form as  the QED Lagrangian, Eq.(\ref{eq:2.7}).
Again one works in the light-cone gauge $A^+=0$, and
uses the same projection operators $\Lambda_\pm$  
as in Section~\ref{sec:Hamiltonian}.
The analogue of Eq.(\ref{eq:2.61}) becomes now simply
\begin{equation}
   i \partial^- \psi_+ = -i m \psi_- + e A^- \psi_+ 
   \ ,\quad{\rm and}\quad
   i \partial^+ \psi_- = i m \psi_- 
\ .\end{equation}
The equation for $\psi_+$ involves the light-front time 
derivative $\partial^-$, so $\psi_+$ is a dynamical degree
of freedom that must be quantized.  On the other hand, 
the equation for $\psi_-$ involves only spatial derivatives, 
so $\psi_-$ is a constrained degree of freedom that should 
be eliminated in favor of $\psi_+$.  
Formally,
\begin{equation}
   \psi_-={m \over \partial^+} \psi_+ 
\ .\end{equation}
It is necessary to specify boundary condition in order to invert 
the operator $\partial^+$.  If we had not chosen a finite mass
for the  fermions then both $\psi_+$ and $\psi_-$ would be 
independent degrees of freedom and we would have
to specify initial conditions for both. Furthermore, in the front
form, it has  only been possible to calculate the condensate 
$\langle 0 |\psi \overline{\psi} |0 \rangle$ for the
Schwinger model by identifying it as the coefficient of the 
linear term in the mass expansion of matrix element of the 
currents \cite{ber77}. Due to the gauge, one component is
fixed to $A^+=0$, but the other component  $A^-$ of the
gauge field  is also a constrained degree of freedom. It 
can be formally eliminated by the light-cone analogue of Gauss's law:
\begin{equation}
   A^- = -{4 e \over \bigl(\partial^+\bigr)^2} 
   \psi_+^\dagger \psi_+ 
\ .\end{equation}
One is left with a single dynamical degree of freedom,
$\psi_+$, which is canonically quantized at $x^+=0$, 
\begin{equation}
   \bigl\{\psi_+(x^-),\psi_+^\dagger(y^-)\bigr\} 
   = \Lambda_+\delta(x^--y^-) 
\ .\end{equation}
similar to what was done in QED.
The field operator at $x^+=0$, expanded in terms of 
the free particle creation and annihilation operators,
takes the very simple form
\begin{eqnarray}
   \psi_+(x^-) &=& \int_{k^+ > 0} {dk^+ \over 4\pi} \left[ 
   b_k e^{-i k\cdot x} + d_k^\dagger e^{i k \cdot x} \right] 
\ ,\nonumber \\ {\rm with}\qquad 
   &&\bigl\{d_k,d_p^\dagger\} 
   = \bigl\{b_k,b_p^\dagger\} 
   = 4 \pi \delta(k^+-p^+) 
\ .\label{eq:4.free}\end{eqnarray}
The canonical Hamiltonian $H= P_+={1\over2}P^-$
is divided into the three parts
\begin{equation}
   H = H_0 + H_{0}^\prime + V^\prime 
\ .\end{equation}
These Fock-space operators are obtained 
by inserting the free fields in Eq.(\ref{eq:4.free})
into the canonical expressions in Eq.(\ref{eq:2.87}).
The free part of the Hamiltonian becomes
\begin{equation}
   H_0=\int_{k>0} {dk \over 8\pi} \Biggl({m^2 \over k}\Biggr)
   \bigl(b_k^\dagger b_k+d_k^\dagger d_k\bigr) \;.
\end{equation}
$H_{0}^\prime$ is the one-body operator which is obtained
by normal  ordering the interaction, {\it i.e.}
\begin{equation}
  H'_0={e^2 \over 4\pi} \int_{k>0} {dk \over 4\pi} 
  \int_{p>0}    dp \left( {1 \over (k-p)^2} - {1 \over(k+p)^2} \right)
  \left( b_k^\dagger b_k+d_k^\dagger d_k\right) 
\ . \end{equation}
The divergent momentum integral  is regulated by the 
momentum cut-off, $|k-p|>\epsilon$. One finds 
\begin{equation} 
   H'_0={e^2 \over 2\pi} \int{dk \over 4\pi} \left({1 \over
   \epsilon} - {1 \over k} + {\cal O}(\epsilon) \right)
   \left( b_k^\dagger b_k + d_k^\dagger d_k \right)
\ .\end{equation}
The normal-ordered interactions is
\begin{eqnarray}
  V'&=&4\pi e^2 \int {dk_1 \over 4\pi}\,\dots
   \,{dk_4 \over 4\pi} \ \delta(k_1+k_2-k_3-k_4) 
   \left\{ {2 \over (k_1-k_3)^2} 
   b_1^\dagger d_2^\dagger d_4 b_3 \right.
\nonumber \\
   &+&\left.{2 \over (k_1+k_2)^2} 
   b_1^\dagger d_2^\dagger d_3 b_4 
   -{1 \over (k_1-k_3)^2}
    \bigl(b_1^\dagger b_2^\dagger b_3 b_4 
   +d_1^\dagger d_2^\dagger d_3 d_4\bigr) 
   + \ \dots\, \right\} 
.\end{eqnarray}
The interactions that involve the creation or annihilation 
of electron-positron pairs are not displayed.
The first term in $V'$ is the electron-positron interaction.
The longitudinal momentum cut-off  requires
$|k_1-k_3| > \epsilon$ and leads to the potential 
\begin{eqnarray}
   v(x^-) = 4 q_1 q_2 \int\limits_{-\infty}^\infty 
   {dk \over 4\pi}\ \theta(|k|-\epsilon) 
    \,e^{-{i \over 2} k x^-} 
   = q_1 q_2 \left[ {2 \over \pi \epsilon} 
   - |x^-| + {\cal O}(\epsilon) \right] 
\ .\end{eqnarray}
This potential contains a linear Coulomb potential that we
expect in two dimensions, but it also contains a divergent 
constant, being negative for unlike charges and positive for
like charges.

In charge neutral states the infinite constant in $V'$ is 
{\it exactly} canceled by the divergent `mass' term in 
$H'_0$. This Hamiltonian assigns an infinite energy 
to states with net charge, and a finite energy as,  
$\epsilon \rightarrow 0$, to charge zero states.
This does not imply that charged particles are confined, 
but that the linear potential prevents charged particles 
from moving to arbitrarily large separation except as 
charge-neutral states.  

One should emphasize that even though the interaction
between charges is long-ranged, there are no van der 
Waals forces in 1+1dimensions.  It is a simple geometrical 
calculation to show that all long range forces between two 
neutral states cancel exactly.  This does not happen in 
higher dimensions, and if we use long-range two-body 
operators to implement confinement we must also find
many-body operators that cancel the strong long-range 
van der Waals interactions.

Given the complete Hamiltonian in normal-ordered form 
we can study bound states.  A powerful tool for getting 
started is the variational wave function.  In this case, one 
can begin with a state that contains a single 
electron-positron pair
\begin{equation}
   |\Psi(P)\rangle = \int_0^P {dp \over 4\pi}\ \phi(p) 
   \,b_p^\dagger d_{P-p}^\dagger |0\rangle 
\ .\end{equation}
The normalization of this state is
$ \langle \Psi(P')|\Psi(P)\rangle = 4\pi P \delta (P'-P) $.
The expectation value of the one-body operators in the 
Hamiltonian is
\begin{equation}
   \langle\Psi|H_0+H'_0|\Psi\rangle 
   = {1 \over 2P} \int {dk\over 4\pi} \left[
   {m^2-e^2/\pi \over k}+ {m^2-e^2/\pi \over P-k}
   + {2 e^2 \over \pi \epsilon} \right] |\phi(k)|^2 
\ ,\end{equation}
and the expectation value of the normal-ordered 
interaction is
\begin{equation}
   \langle\Psi|V'|\Psi\rangle 
   = -{ e^2 \over P} \int' {dk_1 \over 4\pi} 
   {dk_2 \over 4\pi} \left[g{1 \over (k_1-k_2)^2}
   +{1 \over P^2}\right]\phi^*(k_1) \phi(k_2) 
\ .\end{equation}
The prime on the last integral indicates that the range of 
integration in which $|k_1-k_2|<\epsilon$ must be removed.  
By expanding the integrand about $k_1=k_2$, one can 
easily confirm that the $1/\epsilon$ divergences cancel.

The easiest case to study is the massless Schwinger model.  
With $m=0$, the energy is minimized when
\begin{equation}
   \phi(k)=\sqrt{4\pi} 
\ .\end{equation}
The invariant-mass squared, $M^2 = 2PH$, 
becomes then finally   
\begin{equation}
   M^2={e^2 \over \pi} 
\ .\end{equation}
This type of simple analysis can be used to show that 
{\em this electron-positron state is actually the exact 
ground state of the theory} with momentum $P$, and 
that bound states do not interact with one another 
\cite{per94a}. 

It is intriguing that for massless fermions, 
the massive bound states is a simple bound state of an 
electron and a positron when the theory is formulated in
the front form using the light-cone gauge.  
This is not true in other gauges and coordinate systems. 
This happens because the charges screen one another 
perfectly, and this may be the way a constituent picture 
emerge in QCD. On the other hand there are many 
differences between two and four dimensions. In two
dimensions for example the coupling has the dimension 
of mass making it natural for the the bound state mass 
to be proportional to coupling in the massless limit. 
On the other hand, in four dimensions the coupling is
dimensionless and the bound states in a four 
dimensional massless theory must acquire a mass through 
dimensional transmutations. A simple  model of how this 
might happen is discussed in the renormalization of the 
Yukawa model and in some simple models in the section 
on renormalization.

\subsection{3+1 Dimensional: Yukawa model}

Our ultimate aim is to study the bound state problem in QCD. 
However light-front QCD is plagued with divergences 
arising from both small longitudinal momentum and large 
transverse momentum. To gain experience with the novel 
renormalization programs that this requires, it is useful
to study a simpler model . The two-fermion bound-state
problem in the 3+1 light-front Yukawa model has many of
the non-perturbative problems of QCD while still being
tractable in the Tamm-Dancoff approximation. 
This section follows closely the work in 
\cite{ghp93,pin91b,pin92}. The problems that were 
encountered in this calculation are typical of any
$(3+1)$ dimensional non-perturbative calculation and 
laid the basis for Wilson's current Light Front program
\cite{wil65,wil70,wwh94,peh91,per93a,pew93} which will be briefly discussed in the
section on renormalization.

The Light-Front Tamm-Dancoff method  (LFTD) is 
Tamm-Dancoff truncation of the Fock space in light-front 
quantum field theory and was  proposed 
\cite{phw90,tbp91} to overcome some of the problems in 
the equal-time Tamm-Dancoff method \cite{bmp93}.  
In this approach one introduces a
longitudinal momentum cut-off $\epsilon$ to remove all 
the troublesome vacuum diagrams. The bare vacuum 
state is then an eigenstate of the Hamiltonian. One can 
also introduce  a transverse momentum cut-off
$\Lambda$ to regulate ultraviolet divergences. 
Of course the particle truncation and momentum cut-offs 
spoil Lorentz symmetries. In a properly renormalized 
theory one has to remove the  cut-off dependence from 
the observables and recover the lost Lorentz symmetries. 
One has avoided the original vacuum problem but now 
the construction of a properly renormalized Hamiltonian 
is a nontrivial problem.  In particular the light-front
Tamm-Dancoff approximation breaks rotational invariance 
with  respect to the two transverse directions. 
This is visible in the spectrum which does not exhibit the 
degeneracy associated with the total angular momentum
multiplets. It is seen that renormalization has sufficient 
flexibility to restore the degeneracy.

Retaining only two-fermion and two-fermion, one-boson 
states one obtains a two-fermion bound state problem 
in the lowest order Tamm-Dancoff truncation. 
This is accomplished by eliminating the three-body sector
algebraically which leave an integral equation for the 
two-body state. This bound state equation has both 
divergent self-energy and divergent one-boson exchange 
contributions.  In the renormalization of the  one-boson
exchange divergences the  self-energy corrections are
ignored. Related work can be found in 
\cite{gla84,kpw92,wor92}. 

Different counter terms are introduced for to renormalize the
divergences associated with one-boson exchange.  The
basis for these counter terms is easily understood, and 
uses a momentum space slicing called the High-Low 
analysis. It was introduced by Wilson \cite{wil76} and is
discussed in detail for a simple one dimensional model 
in the section on renormalization.

To remove the self-energy divergences one first introduces a
sector-dependent mass counter term which removes the 
quadratic divergence. The remaining logarithmic divergence 
is removed by a redefinition of the coupling constant. 
Here one faces the well-known problem of triviality:  For a
fixed renormalized coupling the bare coupling becomes 
imaginary beyond a certain ultraviolet cut-off. This was 
probably seen first in the Lee model \cite{lee54} and then 
in meson-nucleon scattering using the equal-time
Tamm-Dancoff method \cite{dad55}.

The canonical light-front Hamiltonian for the 3+1 dimensional 
Yukawa model is given by
\begin{equation} 
   P^- = {1 \over 2} \int dx^- d^2x_{\!\perp} [ 2 i \psi_-^\dagger  
   \partial^+ \psi_- + m_{B}^2 \phi^2 + \partial_{\!\perp}   
   \phi \cdot \partial_{\!\perp}   \phi ]
 .\end{equation}
The equations of motion are used to express 
$\psi_-$ in terms of $\psi_+$, {\it i.e.}
\begin{equation} 
   \psi_- = {1 \over i \partial^+} 
   [ i \alpha_{\!\perp}\cdot\partial_{\!\perp} + \beta(m_{F} 
   + g \phi)] \psi_+
\ .\end{equation}
For simplicity the two fermions are taken to be of different 
flavors, one denoted by $b_\sigma$ and the other by
$B_\sigma$. We divide the Hamiltonian $P^- $ into  
$P^-_{free}$ and $P^-_{int}$,  where
\begin{eqnarray}
   P^-_{free} &=& \int [ d^3k]\ {m_B^2 + k^2 \over k^+}
   \ a^\dagger (k)a(k) 
\nonumber \\ 
   &+& \sum_{\sigma} \int [ d^3k] 
   \ {m_F^2 + k^2 \over k^+} 
   \Big [ b_{\sigma}^\dagger  (k) b_{\sigma}(k)  
   + B_{\sigma}^\dagger  (k) B_{\sigma}(k) \Big ] 
\ , \\ {\rm and}\quad
   P^-_{int}\ &=& g \sum_{\sigma_1, \sigma_2}  
   \int  [d^3 k_1] \int [d^3 k_2] \int [d^3 k_3 ] 
   \ 2 (2 \pi )^3  \delta^3(k_1 - k_2 - k_3) 
\nonumber \\
   &&\times
   \Big [(b^\dagger_{\sigma_1}(k_1) b_{\sigma_2}(k_2) 
   + B^\dagger_{\sigma_1}(k_1) B_{\sigma_2}(k_2))  
   a(k_3)    {\bar u_{\sigma_1}}(k_1) u_{\sigma_2}(k_2) 
\nonumber \\
   &&+\ (b^\dagger_{\sigma_2}(k_2) b_{\sigma_1}(k_1) 
   + B^\dagger_{\sigma_2}(k_2) B_{\sigma_1}(k_1))
   a^\dagger(k_3){\bar u_{\sigma_2}}(k_2) 
   u_{\sigma_1}(k_1) \Big ]  
\ .\end{eqnarray}
Note that the instantaneous interaction was dropped
from $P^-_{int}$ for simplicity. The fermion number 2 
state that is an eigenstate of $P^-$ with momentum $P$ 
and helicity $\sigma$ is denote as $ \mid \Psi(P, \sigma) 
\rangle $ . The wave function is normalized in the 
truncated Fock space, with
\[ \langle \Psi(P', \sigma') \mid \Psi(P, \sigma) \rangle  
   = 2 (2 \pi)^3  P^+ \delta^3(P - P') \delta_{\sigma \sigma'} 
\ .\]
In the lowest order Tamm-Dancoff truncation one has
\begin{eqnarray} 
   \vert  \Psi (P, \sigma) \rangle 
   =  \sum_{\sigma_1 \sigma_2} \int [d^3k_1] \int [ d^3k_2] 
   \ \Phi_2(P,\sigma \mid k_1 \sigma_1,k_2 \sigma_2)
   \ b^\dagger_{\sigma_1} (k_1) B^\dagger_ {\sigma_2}(k_2) 
   \vert 0\rangle 
\nonumber \\  \nonumber 
   + \sum_{\sigma_1 \sigma_2} \int [d^3k_1] 
   \int [ d^3k_2] \int [d^3k_3]\ \Phi_3(P,\sigma \mid k_1
   \sigma_1,    k_2 \sigma_2,k_3) b^\dagger _ {\sigma_1}  (k_1) 
   \ B^\dagger _ {\sigma_2} (k_2)  a^\dagger (k_3)
   \vert 0\rangle 
,\end{eqnarray}
where $ \Phi_2 $ is the two-particle and $ \Phi_3 $ the 
three-particle amplitude, and where $ \vert0 \rangle$ is the
vacuum state. For notational convenience one introduces the
amplitudes $ \Psi_2$ and $\Psi_3$  by
\begin{equation} 
   \Phi_2(P, \sigma \mid k_1 \sigma_1,k_2 \sigma_2)  
   = 2 (2 \pi)^3  P^+ \delta^3(P - k_1 - k_2) \sqrt{x_1 x_2}
   \Psi_{2 }^{\sigma_1 \sigma_2}(\kappa  _1x_1,\kappa  _2 x_2)  ,
\end{equation} 
and
\begin{equation} 
   \Phi_3(P, \sigma \mid k_1 \sigma_1,k_2 \sigma_2,k_3) 
   = 2 (2 \pi)^3 P^+ \; \delta^3(P - \Sigma k_i )
\sqrt{x_1x_2x_3} \Psi_{3 }^{\sigma_1 \sigma_2}(
\kappa  _1 x_1, \kappa  _2 x_2, \kappa  _3 x_3)
.\end{equation}
As usual, the intrinsic variables  are $x_i$ and 
$\kappa_{i} = \vec \kappa_{\!\perp i}$
\[   k_i ^\mu = \left( x_i P^+,\vec \kappa_{\!\perp i}, 
  {\vec \kappa_{\!\perp i}^{\,2} +m^2 \over x_i P^+}
   \right)  \ ,\]
with $\sum_i x_i = 1$ and 
$\sum_i \vec \kappa_{\!\perp i}= 0$. By projecting the
eigenvalue equation
\begin{equation}  
   (P^+ P^- - P_{\!\perp}^2) \vert\Psi\rangle   
   = M^2 \vert\Psi\rangle 
\end{equation}
onto a set of free Fock states, one obtains two
coupled integral equations:
\begin{eqnarray} 
  \bigg[ M^2 &-& {m_{F}^2 + (\kappa_1 )^2 \over x_1} -
             {m_{F}^2 + (\kappa_1 )^2 \over x_2} \bigg] 
  \Psi_2^{\sigma_1 \sigma_2}(\kappa_1  ,x_1 ) 
\nonumber \\         
 &=&\ {g \over 2(2 \pi)^3} \sum_{s_1} 
 \int { dy_1 d^2q_1 \over\sqrt{(x_1-y_1) x_1\;y_1}} 
 \Psi_3^{s_1,\sigma_2}(q_1  ,y_1;\kappa_2 ,x_2)
 {\bar u_{\sigma_1}}(\kappa_1,x_1 ) u_{s_1}(q_1,y_1 ) 
\nonumber \\           
  &+&\ {g\over 2(2 \pi)^3}\sum_{s_2}
  \int {dy_2d^2q_2 \over\sqrt{(x_2-y_2) x_2 \; y_2}}
  \Psi_3^{\sigma_1 s_2}(\kappa_1  ,x_1 ; q_2  ,y_2) 
  {\bar u_{\sigma_2}}(\kappa_2,x_2 ) u_{s_2}(q_2 ,y_2)
\ ,\end{eqnarray}
and
\begin{eqnarray} 
   \Big[ M^2 &&- {m_{F}^2 + (\kappa_1  )^2 \over x_1} -
   {m_{F}^2 + (\kappa_2  )^2 \over x_2} - 
   {m_{B}^2 + (\kappa_1  +\kappa_2  )^2 \over x_3 }  
   \Big] \Psi_3^{\sigma_1 \sigma_2}
   (\kappa_1  ,x_1 ; \kappa_2  ,x_2) 
\nonumber \\         
   &&= \ g \sum_{s_1} 
   {\Psi_2^{s_1 \sigma_2} (-\kappa_2 ,x_1+x_3) 
   \over \sqrt{x_3\; x_1(x_1+x_3)}}
   \ {\bar u_{\sigma_1}}(\kappa_1,x_1)
   u_{s_1}(-\kappa_2,x_1+x_3) 
\nonumber \\           
   &&+\ g\sum_{s_2} { \Psi_2^{\sigma_1 s_2}(\kappa_1,x_1)
   \over \sqrt{x_3\; x_2(x_3+x_2) }}
   \ {\bar u_{\sigma_2}}(\kappa_2,x_2)
   u_{s_2}(-\kappa_1,x_2+x_3)   
\ .\end{eqnarray}
After eliminating $\Psi_3$ one ends up with an integral 
equation for $\Psi_2$ and the eigenvalue $M^2$:
\begin{eqnarray}   
   &&M^2\ \Psi_2^{\sigma_1\sigma_2}(\kappa  ,x)  
   = \left({m_F^2 + \kappa^2\over x(1-x)}+ [S.E.]
   \ \right)\,\Psi_2^{\sigma_1\sigma_2} (\kappa  ,x)  
\nonumber  \\      
   &&\phantom{M^2}
   + {\alpha \over 4\pi^2}\, \sum_{s_1,s_2}\int\! dy \,d^2q 
   \ K(\kappa,x;q,y;\omega)_{\sigma_1\sigma_2;s_1s_2}
   \,\Psi_2^{s_1s_2}(q,y)   \quad + {\rm\ counterterms}   
\ ,\label{eq:4.int}\end{eqnarray}
where $\alpha ={g^2 /4\pi}$ is the fine structure constant.
The absorption of the boson on the same fermion gives rise
to the self-energy term $[S.E.]$, the one by the other fermion
generates an effective interaction, or the boson-exchange
kernel $K$,
\begin{equation}
   K(\kappa  ,x;q  ,y;\omega)_{\sigma_1\sigma_2;s_1s_2}  
   = { \left[\bar u( \kappa,  x;\sigma_1)u( q,  y;s_1)\right]
   \left[\bar u(-\kappa,1-x;\sigma_2)u(-q,1-y;s_2) \right]
   \over (a +2(\kappa \cdot q)) \sqrt{x(1-x)y(1-y)}} 
\ ,\end{equation}
with
\begin{eqnarray}
    a  &=& |x-y|\left\{\omega - {1\over 2} \left[{{m_F^2 + k^2}
    \over  {x(1-x)}} + {{m_F^2 + q^2} \over {y(1-y)}} \right] 
    \right\}  - m_B^2 + 2 m_F^2  
\nonumber \\ 
   &-& {{m_F^2 + k^2} \over 2} \left[{y \over x} 
   + {1-y\over 1-x} \right] - {{m_F^2 + q^2} \over 2 }
   \left[{x \over y} + {1-x\over 1-y} \right] 
\ ,\end{eqnarray}
with $k = |\kappa|$ and $\omega\equiv M^2$. 
Possible counter terms will be discussed below.

Since $\sigma=\uparrow,\downarrow$ one faces thus
$4\times4=16$ coupled integral equations in the three
variables $x$ and $\vec \kappa_{\!\perp}$.
But the problem is simplified considerably by exploiting 
the rotational symmetry around the $z$-axis.
Let us demonstrate that shortly.
By Fourier transforming over the angle $\phi$, one 
introduces first states $\Phi$ with good total spin-projection 
$S_z=\sigma_1+\sigma_2\equiv m$,
\begin{equation} 
   \Psi_2^{\sigma_1\sigma_2}(\kappa  ,x) 
  = \sum_m {{e^{im\phi}} \over \sqrt{2\pi}}
  \,\Phi_{\sigma_1\sigma_2}(k,x;m)             
\, \end{equation}
and uses that second to redefine the kernel:
\begin{equation}
   V(k,x,m;q,y,m';M^2)_{\sigma_1\sigma_2;s_1s_2}  
   =  \int {{d\phi\;d\phi'} \over {2\,\pi}}e^{-im\phi}\,e^{im'\phi'}
   \,K(k,\phi,x;q,\phi',y;M^2)_{\sigma_1\sigma_2;s_1s_2}   
. \end{equation}
The $\phi$-integrals can be done analytically.  
Now recall that neither $S_z$ nor $L_z$ is conserved;  
only $J_z = S_z + L_z$ is a good quantum number.  
In the two-particles sector of
spin-${1\over2}$ particles the spin projections are limited
to $|S_z|\leq1$, and thus, for $J_z$ given, one has to
consider only the four amplitudes 
$\Phi_{\uparrow\uparrow}(k,x;J_z-1)$, 
$\Phi_{\uparrow\downarrow}(k,x;J_z)$, 
$\Phi_{\downarrow\uparrow}(k,x;J_z)$, and
$\Phi_{\downarrow\downarrow}(k,x;J_z+1)$.
Rotational symmetry allows thus to reduce the number of 
coupled equations from 16 to 4, and the number of 
integration variables from 3 to 2. 
Finally, one always can add and subtract the states, introducing
\begin{eqnarray} 
   \Phi_t^\pm(k,x) &=& {1 \over \sqrt2}
   \left(\Phi_{\uparrow\uparrow}(k,x;J_z-1)
   \pm \Phi_{\downarrow\downarrow}(k,x;J_z+1)\right)     
\nonumber \\                  
   \Phi_s^\pm(k,x) &=& {1 \over \sqrt2}
   \left(\Phi_{\uparrow\downarrow}(k,x;J_z)
   \pm \Phi_{\downarrow\uparrow}(k,x;J_z)\right)   
\ . \end{eqnarray}
The integral equations couple the sets $(t^-, s^+)$ and
$(t^+,s^-)$. For $J_z = 0$, the `singlet' and the `triplet' states
un-couple completely, and one has to solve only two pairs
of two coupled integral equations.
In a way, these reductions are quite  natural and
straightforward, and have been applied independently 
also by \cite{kpw92} and most recently by \cite{trp96}.

Next let us discuss the structure of the {\em integrand}
in Eq.(\ref{eq:4.int}) and analyze eventual divergences.
Restrict first to $J_z=0$, and consider 
$\left[\bar u( x,\kappa;\sigma_1)u(y,q;s_1)\right]$
for large $q$, taken from the tables in
Section~\ref{sec:DLCQ}. They are such, that the kernel 
$K$ becomes independent of $|q |$ in the limit $|q |
\rightarrow \infty$. Thus, unless $\Phi$ vanishes faster than 
$|q |^{-2}$, the $q$-integral potentially diverges.
In fact, introducing an ultraviolet cut-off $\Lambda$ to 
regularize the $|q |$-dependence, 
the integrals involving the singlet wavefunctions 
$\Phi_s^\pm$ diverge logarithmically with $\Lambda$.
In the $J_z=1$ sector one must solve a system of four coupled 
integral  equations.  One finds that the kernel 
$V_{\uparrow\uparrow,\downarrow\downarrow}$ 
approaches the same limit $-f(x,y)$  as  $q$ becomes 
large relative to $k$. 
All other kernels fall off  faster with $q$. 
For higher values of $J_z$, the integrand converges
since the wavefunctions fall-off faster than $|q  |^{-2}$.
Counter terms are therefore needed only for $J_z =0$ 
and $J_z = \pm 1$. These boson-exchange
counter terms have no analogue in equal-time 
perturbation theory, and will be discussed below.

These integral equations are solved numerically, using
Gauss-Legendre quadratures to evaluate  the $q$ and $y$ 
integrals. Note that the eigenvalue $M^2$
appears  on both the left and right hand side of the 
integral equation.  One handles this  with choosing 
some `starting point' value $\omega$ on the r.h.s. 
By solving the resulting matrix eigenvalue problem 
one obtains the eigenvalue $M^2(\omega)$. Taking that
as the new starting point value, one iterates the procedure
until $M^2(\omega)=\omega$ is numerically fulfilled  
sufficiently well.

For the parameter values $1\leq\alpha\leq2$  and
$2\leq m_F/m_B \leq 4 $ one finds only two stable bound
states, one for each  $|J_z|\leq1$.  In the corresponding 
wavefunctions, one observes a dominance of the spin-zero 
configuration $S_z = 0$.   The admixture  from higher values 
of $L_z$ increases gradually  with increasing $\alpha$, but 
the  predominance of $L_z =0$  persists also  when 
counter terms are included in the calculation. 
With  the above parameter choice no bound states  have been 
found numerically for $J_z>1$. They start to appear only
when $\alpha$ is significantly increased.

The above bound state equations are regularized,
how are they renormalized?
In the section on renormalization, below,  we shall show in 
simple one-dimensional models that it is possible to add
counter terms to the integral equation of this type that 
completely remove all the cut-off dependence from both 
the wavefunctions and the bound state spectrum.  
In these one-dimensional models the finite part of the 
counter term contains an arbitrary dimensionful scale $\mu$ 
and an associated arbitrary constant.  In two-dimensional
models the arbitrary constant becomes an arbitrary function. 
The analysis presented here is based on the methods used in
the one-dimensional models.  It is convenient to subdivide
the study of these counter terms into two categories. 
One is called the asymptotic counter terms, and the other is called the
perturbative counter terms.

Studies of the simple models  and the general power
counting arguments show that integral equations should be
supplemented by a counter term of the form,
\begin{equation}
   G\left(\Lambda\right) \int q\, dq\, dy
   \, F \left( x,y\right) \phi \left(q,y\right)
.\end{equation}
For the Yukawa model one has not been able to solve 
for $G(\Lambda) \, F(x,y)$ exactly such that it removes 
all that cut-off dependence.  One can, however
estimate $G(\Lambda)\, F(x,y)$ perturbatively.  
The lowest order (order $\alpha^2$) perturbative 
counter terms correspond to the box graphs in the integral 
equation thus they are called the ``box counter terms''  (B.C.T). 
Applying it to the Yukawa model,  one finds that the integral 
equation should be modified according to
\begin{equation}
   V(k,x;q,y;\omega) \longrightarrow V(k,x;q,y;\omega) -
   V^{B.C.T.}(x,y)            .\end{equation}
$V^{B.C.T.}(x,y)$ contains an undetermined parameter`$C$'.
Redoing the bound state mass calculations with this counter term 
one finds that the cut-off independence of the solutions is greatly
reduced.  Thus one has an  (almost) finite calculation
involving arbitrary parameters, $C$ for each sector.  
Adjusting the $C$'s allows us to move eigenvalues around only 
in a limited way. It is possible however to make the $J_{z}=1$ state 
degenerate with either of the two $J_{z}=0$ states. The
splitting among the two $J_{z}=0$ states remains small. 

One can also  eliminate divergences non perturbatively by
subtracting the large transverse momentum limit of the kernel. 
This type of counter term we call the asymptotic counter term. 
In the Yukawa model one is only able to employ
such counter terms in the $J_z =0$ sector. One then has
\begin{equation}
V(k,x;q,y;M^2)^{s+,s+}\longrightarrow
V(k,x;q,y;M^2)^{s+,s+}+f(x,y)            ,\end{equation}
\begin{equation}
V(k,x;q,y;M^2)^{s-,s-}\longrightarrow
V(k,x;q,y;M^2)^{s-,s-}-f(x,y)            .\end{equation}
One can find an extra interaction allowed by power counting in
the LC-Hamiltonian that would give rise to more terms. One
finds that with the asymptotic counter term  the cut-off
dependence has been eliminated for the (t-,s+) states and 
improved for the (t+,s-) states.  We also find that this
counter term  modifies the large $k$ behavior of the 
amplitudes $\Phi(k,x)$ making  them fall off faster than
before.
  
The asymptotic counter term, as it stands, does not include 
any arbitrary constants that can be tuned to renormalize the 
theory to some experimental  input.  This differs from the case 
with the box counter term where such a constant appeared.  
One may, however, add an adjustable piece which in general
involves an arbitrary function of longitudinal momenta.  This is
motivated by the simple models discussed in the section on 
renormalization. One replaces:
\begin{equation}
f(x,y) \longrightarrow  f(x,y) -{{G_{\mu}}\over{1+{G_{\mu}\over
6}\,ln({\Lambda\over\mu})}} 
\end{equation}
$G_{\mu}$ and the scale $\mu$ are not independent.
A change in $\mu$ can be compensated by adjusting $G_{\mu}$ such that 
${1 \over G_\mu} - {1 \over 6}\,ln(\mu) = constant $.  This `$constant$' is
arbitrary and plays the role of the constant `C' in the box counter term.  One
finds that by adjusting the constant a much wider range of possible
eigenvalues can be covered, compared to the situation with the box
counter term.   

Consider now the effects of the self-energy term,
$[S.E.]$.  Note that in the bound state problem 
the self-energy is a function of the bound state energy
$M^2$.  The most severe ultraviolet divergence in $[S.E.]_{(M^2)}$ is a 
quadratical divergence.  One eliminates this divergence by subtracting 
at the threshold $M^2 = M^2_0 \equiv (m_F^2 + k^2)/(x(1-x))$
\begin{equation}
  [S.E.]_{(M^2)}  \longrightarrow 
  [S.E.]_{(M^2)} - [S.E.]_{(M^2_0)}  \\         
   \equiv \;g^2\,\bigl(M^2 - M^2_0 \bigr)\,\sigma_{(M^2)}  
\end{equation}
$\sigma_{(M^2)}$ is still logarithmically divergent.  The remaining
logarithmically divergent piece corresponds to wave  function
renormalization of the two fermion lines. One finds:
\begin{equation} 
\sigma _{log div. part}  = \Bigl({{\partial [S.E.]} \over {\partial
M^2}}\Bigr)_ {log div.} \equiv - W \left(\Lambda \right)  
\end{equation}
One can absorb this divergence into a new definition of 
the coupling constant.  After the subtraction 
(but ignoring all `boson-exchange' counter terms) the integral
equation becomes
\begin{equation}
 \bigl( M^2 -M^2_0 \bigr)\,\Psi_2^{\sigma_1\sigma_2}
(\vec{k},x)  = {\alpha \over 4\pi^2}\;\bigl[ B.\;E. \bigr] +
{\alpha\over 4\pi^2}\, \bigl( M^2 - M^2_0
\bigr)\,\sigma_{(M^2)}\,\Psi_2^{\sigma_1\sigma_2} (\vec{k},x) 
,\end{equation}
where $[B.E.]$ stands for the term with the kernel $K$.
Rearranging terms one finds ( all spin indices suppressed)
\begin{equation}
\bigl(M^2 - M^2_0 \bigr)\,\bigl[1 + {\alpha\over 4\pi^2}\,W \left(\Lambda
\right) \bigr] \,\Psi =  {\alpha\over 4\pi^2}\,\bigl[ B.\;E.\;\bigr] +
{\alpha\over 4\pi^2}\,  \bigl(M^2- M^2_0 \bigr) \, \bigl(\sigma + W
\left(\Lambda\right) \bigr)\,\Psi          
\ .\end{equation}
The r.h.s. is now finite. One must still deal with the 
divergent piece $W$ on the l.h.s. of the equation. Define
\begin{equation}
\alpha_R = { \alpha \over {1 + {\textstyle\strut \alpha \over
\textstyle\strut 4\pi^2}\,  W \left(\Lambda\right)}}              
\end{equation}
Then one can trade a $\Lambda$ dependent bare coupling 
$\alpha$ in favor of a finite renormalized coupling $\alpha_R\,$. 
One has
\begin{equation}
\bigl(M^2 - M^2_0 \bigr) \,\Psi = {{\alpha_R/4\pi^2} \over {1 -
\alpha_R/4\pi^2 (\sigma_{(M^2)} + W \left(\Lambda\right))}}
\,\bigl[ B.\;E.\;\bigr]   
.\end{equation}
One sees that the form of the equation is identical to 
what was solved earlier (where all counter terms were ignored) 
with $\alpha$ replaced by $\alpha_R\Bigl/[1 - \alpha_R/4\pi^2(\sigma
+~W)]$.   One should note that $\sigma$ is a function of $x$ and $k$, and
therefore effectively changes the kernel.  In lowest order Tamm-Dancoff the
divergent parts of $[S.E.]$ can hence be absorbed into a renormalized mass and
coupling.  It is however not clear whether this method will work in higher
orders. 

Inverting the equation  for $\alpha_R$ one has
\begin{equation}
\alpha(\Lambda) = {{\alpha_R}  \over { 1 - {\textstyle\strut
\alpha_R\over \textstyle\strut 4\pi^2} \,W(\Lambda)}}             
.\end{equation}
One sees that for every value of $\alpha_R$ other than $\alpha_R =
0$  there  will be a cut-off $\Lambda$ at which the denominator vanishes and
$\alpha$ becomes infinite . This is just a manifestation of 'triviality' in
this model.  The only way the theory can be sensible for arbitrarily  large
cut-off $\Lambda \rightarrow\infty$, is when $\alpha_R \rightarrow 0$.  In
practice this means that for fixed cut-off there will be an upper  bound on
$\alpha_R$. 

%% file: 05Dilcq.tex
\section{Discretized Light-Cone Quantization}
\label {sec:DLCQ}
\setcounter{equation}{0}

Constructing even the lowest state, the `vacuum', of a 
Quantum Field Theory has been so notoriously difficult that the 
conventional Hamiltonian approach was given up altogether  
long ago in the Fifties, in favor of action oriented approaches.
It was overlooked that Dirac's `front form' of Hamiltonian 
dynamics \cite {dir49} might have less severe problems.
Of course, the action and the Hamiltonian forms of dynamics
are equivalent to each other, but the action approach is 
certainly more suitable for deriving cross sections, 
while the Hamiltonian approach is more convenient when 
considering the structure of bound states in atoms,
nuclei, and hadrons. 
In fact, in the front form with periodic boundary 
conditions one can combine the aspects of a simple vacuum 
\cite{wei66} and a careful treatment of the infrared degrees 
of freedom. This method is called
`Discretized Light-Cone Quantization' (DLCQ) \cite {pab85a}
and has three important aspects: 
\begin {itemize}
\item[(1)] 
    The theory is formulated in a Hamiltonian approach;
\item[(2)] 
    Calculations are done in momentum representation; 
\item[(3)] 
    Quantization is done at equal light-cone 
    rather than at equal usual time.
\end {itemize}
As a method, `Discretized Light-Cone Quantization' has the 
ambitious goal to calculate the spectra and wavefunctions of 
physical hadrons from a covariant gauge field theory.
The conversion of this non-perturbative method into a reliable 
tool for hadronic physics is beset with many difficulties 
\cite {gla95}. Their resolution will continue to take time.
Since its first formulation \cite {{pab85a,pab85b}}
many problems have been resolved but many remain, as we shall see.
Many of these challenges are actually not peculiar to the front form 
but appear also in conventional Hamiltonian dynamics.
For example, the renormalization program for a quantum field 
theory has been formulated thus far only in order-by-order 
perturbation theory. Little work has been done on formulating
a non-perturbative Hamiltonian renormalization \cite{phw90,wwh94}.

At the beginning, one should emphasize 
a rather important aspect of periodic boundary conditions:
All charges are strictly conserved.
Every local Lagrangian field theory has vanishing 
four-divergences of some `currents' of the form
$ \partial _\mu J ^\mu = 0 $. 
Written out explicitly this reads
\begin {equation}
    \partial _+ J ^+ + 
    \partial _- J ^- = 0 
\,. \label {eq:continuity}
 \end {equation}
The restriction to 1+1 dimension suffices for the argument.
The case of 3+1 dimensions is a simple generalization.
The `charge' is defined by
\begin {equation}
    Q (x ^+) \equiv
    \int \limits _{-L} ^{+L} dx ^-\,J ^+ ( x ^+ , x ^- )
\,.\end {equation}
Conservation is proven by 
integrating Eq.(\ref{eq:continuity}),  
\begin {equation}
{d \over dx ^+} Q (x ^+) = 0
\,, \end {equation}
provided that the terms from the boundaries {\it vanish}, {\it i.e.}
\begin {equation}
    J ^+ ( x ^+ , L ) - J ^+ ( x ^+ , -L ) = 0 
\,. \end {equation}
This is precisely the condition for periodic boundary conditions.
If one does not use periodic boundary conditions, then one 
has to ensure that all fields tend to 
vanish `sufficiently fast' at the boundaries.
To guarantee the latter is much more difficult than taking
the limit $ L \rightarrow \infty$ at the end of a calculation.
Examples of conserved four-currents are the components of the
energy-momentum stress tensor with $\partial _\mu\Theta ^{\mu\nu} = 0$, 
the conserved `charges' being the four components of the energy-momentum
four-vector $ P ^\nu $. 

Discretized Light-Cone Quantization applied to abelian and non-abelian
quantum field theories faces a number of problems only part
of which have been resolved by recent work. 
Here is a rather incomplete list:
\begin {itemize}
\item[(1)]
Is the front form of Hamiltonian dynamics equivalent to the instant form? 
Does one get the same results in both approaches?~--
Except for a class of problems involving massless left-handed fields,
it has been established that all explicit calculations with the front form yield
the same results as in the instant form, provided the latter are
available and reliable.
\item[(2)]
One of the major problems is to find a suitable and appropriate gauge. 
One has to fix the gauge before one can formulate the Hamiltonian.
One faces the problem of quantizing a quantum field theory
`under constraints'.~-- Today one knows much better how to cope 
with these problems, and the Dirac-Bergman method is discussed
in detail in Appendix~\ref{sec:Bergmann}.
\item[(3)]
Can a Hamiltonian matrix be properly renormalized with a cut-off
such that the physical results are independent of the cut-off?
Hamiltonian renormalization theory is only now starting to be 
understood.
\item[(4)]
In hadron phenomenology the aspects of isospin and chirality
play a central role. In DLCQ applied to QCD they have not 
been tackled yet. 
\end {itemize}
In this section we shall give a number of concrete examples 
where the method has been successful. 

\subsection {Why Discretized Momenta?}
\label {sec:why-dis-mom}

Not even for a conventional quantum many-body problem has 
one realized the goal of rigorously diagonalizing a Hamiltonian.
How can one dare to address to a field theory, 
where not even the particle number is conserved?

Let us briefly review the difficulties for a
conventional non-relativistic many-body theory.
One starts out with a many-body
Hamiltonian $ H  =  T + U $.
The kinetic energy $ T$
is usually a one-body operator and thus simple.
The potential energy $U $ is at least a two-body operator
and thus complicated.
One has solved the problem if one has found
one or several eigenvalues and eigenfunctions
of the Hamiltonian equation, $ H \Psi  =  E \Psi  $.
One always can expand the eigenstates
in terms of products of single particle states
$\langle \vec x \vert m \rangle $, which usually belong to a
complete set of ortho-normal
functions of position $\vec x$, labeled by a quantum number $m$.
When antisymmetrized, one refers to them as `Slater-determinants'.
All Slater-determinants with a fixed particle number
form a complete set.

\begin{figure}
\centering
\epsfysize=85mm\epsfbox{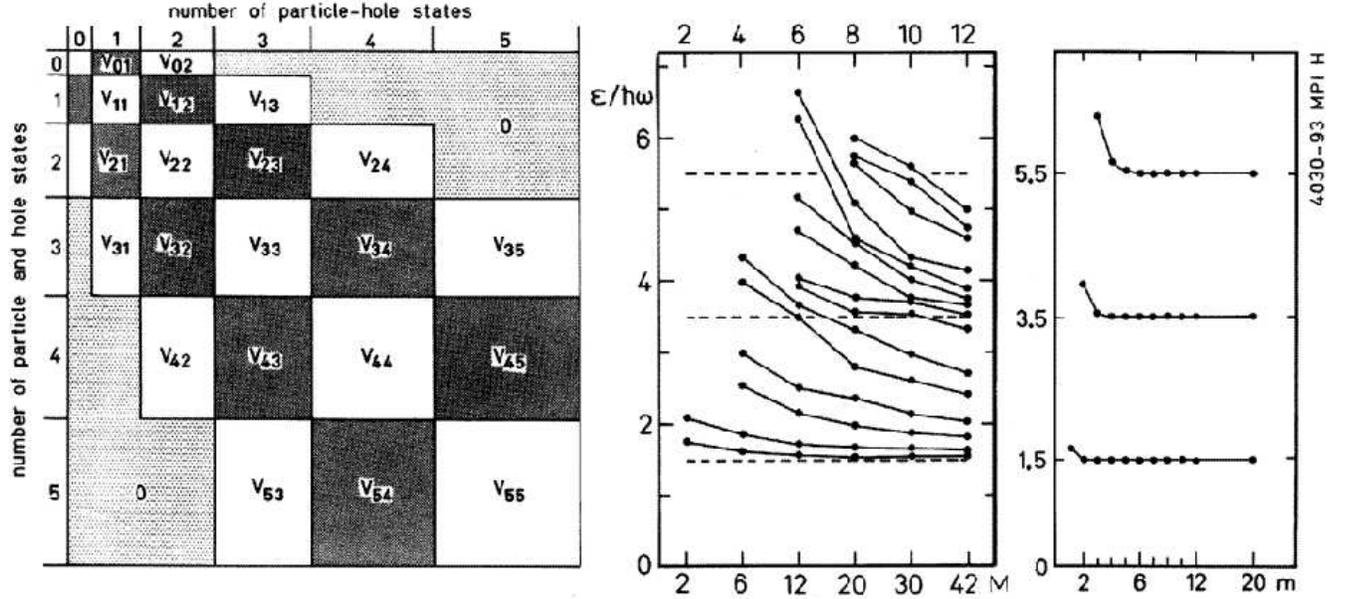}
\caption{\label{fig-kyf-1}
    Non-relativistic many-body theory. 
}\end{figure}

One can proceed as follows. In the first step one chooses
a complete set of single particle wave functions.
These single particle wave functions are solutions of an arbitrary
`single particle Hamiltonian' and its selection is a science of its own.
In the second step, one defines one (and only one) reference state,
which in field theory finds its analogue as the `Fock-space vacuum'.
All Slater determinants can be classified relative to this
reference state as 1-particle-1-hole (1-ph) states, 
2-particle-2-hole (2-ph) states, and so on.
The Hilbert space is truncated at some level.
In a third step, one calculates the Hamiltonian matrix
within this Hilbert space.

In Figure~\ref{fig-kyf-1},
the Hamiltonian matrix for a two-body interaction
is displayed schematically. Most of the matrix-elements vanish,
since a 2-body Hamiltonian changes the state by up to 2 particles.
Therefore the structure of the Hamiltonian is a 
{\it finite penta\--diagonal bloc matrix}. 
The dimension within a bloc, however, is infinite.
It is made finite by an {\it artificial cut-off} on the
kinetic energy, {\it i.e.}
on the single particle quantum numbers $m$.
A finite matrix, however, can be diagonalized on a computer:
the problem becomes `approximately soluble'.
Of course, at the end one must verify that the physical results
are reasonably insensitive to the cut-off(s) and other
formal parameters.

This procedure was actually carried out in one space dimension
\cite {pau84} with two different sets of single-particle functions,
\begin {equation}
   \langle x \vert m \rangle = N_m H_m \Big( {x\over L} \Big)
   \exp {\left\{-{1\over 2} \Big( {x\over L} \Big) ^2 \right\}}
   \quad\quad{\rm and} \quad
   \langle x \vert m \rangle
 = N_m \exp {\left\{ i m{x\over L}\pi\right\}}
\ . \label{eold} \end {equation}
The two sets are the eigenfunctions to the harmonic oscillator
($ L \equiv \hbar / m \omega  $) with its Hermite polynomials 
$H_m$, and the eigenfunctions of the momentum of a 
free particle with periodic boundary conditions.
Both are suitably normalized ($N_m $), and both depend
parametrically on a characteristic length parameter $L$.
The calculations are particularly easy for particle number $2$,
and for a harmonic two-body interaction.
The results are displayed  in Figure~\ref{fig-kyf-1},
and surprisingly different.
For the plane waves, the results converge rapidly
to the exact eigenvalues
$E= {3\over2}, {7\over2}, {11\over2}, \dots$,
as shown in the right part of the figure.
Opposed to this, the results with the oscillator states
converge extremely slowly. Obviously, the larger part
of the Slater determinants is wasted on building
up the plane wave states of center of mass motion from the
Slater determinants of oscillator wave functions.
It is obvious, that the plane waves are superior, since they
account for the symmetry of the problem, namely Galilean covariance.
For completeness one should mention that the approach
with discretized plane waves was successful in getting the exact
eigenvalues and eigenfunctions for up to 30 particles
in one dimension \cite {pau84}
for harmonic and other interactions.

From these calculations, one should conclude:
\begin {itemize}
\item[(1)] 
Discretized plane waves are a useful tool for many-body problems.
\item[(2)] 
Discretized plane waves and their Slater determinants are denumerable,
and thus allow the construction of a Hamiltonian matrix.
\item[(3)] 
Periodic boundary conditions generate good wavefunctions
even for a `confining' potential like the harmonic oscillator.
\end {itemize}
A numerical `solution' of the many-body problem is thus possible
at least in one space dimension.
Periodic boundary conditions should  also be applicable 
to gauge field theory.

\subsection {Quantum Chromodynamics in 1+1 dimensions (KS)}
\label {sec:opo-gat}
DLCQ \cite {pab85a} in 1-space and 1-time dimension had been 
applied first to Yukawa theory \cite {pab85a,pab85b} followed by 
an application to QED \cite {epb87} and to QCD \cite{hbp90}. 
However, before we go into the technical details, let us first see how 
much we can say about the theory without doing any calculations.
With only one space dimension there are no rotations --- hence no 
angular momentum. The Dirac equation is only a two-component equation.
Chirality can still formally be defined.
Secondly, the gauge field does not contain any dynamical degree
of freedom (up to a zero mode which will be discussed in a later section)
since there are no transverse dimensions. 
This can be understood as follows.
In four dimensions, the $A^\mu$ field has four components. 
One is eliminated by fixing the gauge. 
A second component corresponds to the static Coulomb field 
and only the remaining two transverse components
are dynamical (their `equations of motion' contain a time derivative).
In contrast, in $1+1$ dimensions, one starts with only two components 
for the $ A ^\mu$-field. 
Thus, after fixing the gauge and eliminating the Coulomb part, 
there are no dynamical degrees of freedom left. 
Furthermore, in an axial gauge  the nonlinear
term in the only non vanishing component of $F ^{\mu \nu}$ drops out,
and there are no gluon-gluon interactions. 
Nevertheless, the theory confines quarks. 
One way to see that is to analyze the solution to the Poisson equation 
in 1 space dimension which gives rise to a linearly rising potential. 
This however is not peculiar to QCD$_{1+1}$. Most if not all 
field theories confine in 1+1 dimensions.       

\begin{figure}
\unitlength1cm   \centering 
\begin{minipage}[t]{80mm}
\epsfysize=95mm\epsfbox{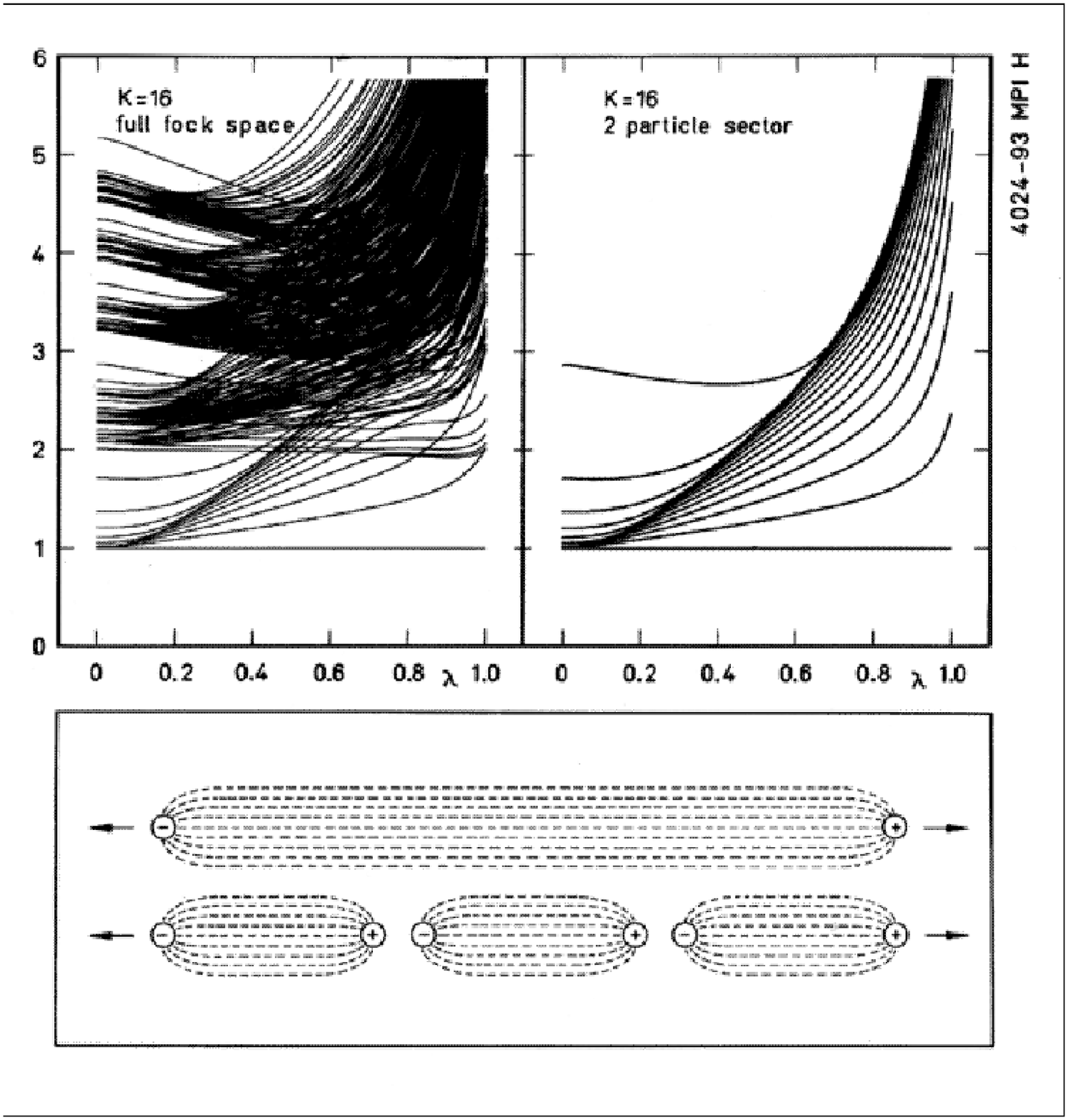}
\end{minipage}
\hfill
\begin{minipage}[t]{80mm}
\epsfysize=95mm\epsfbox{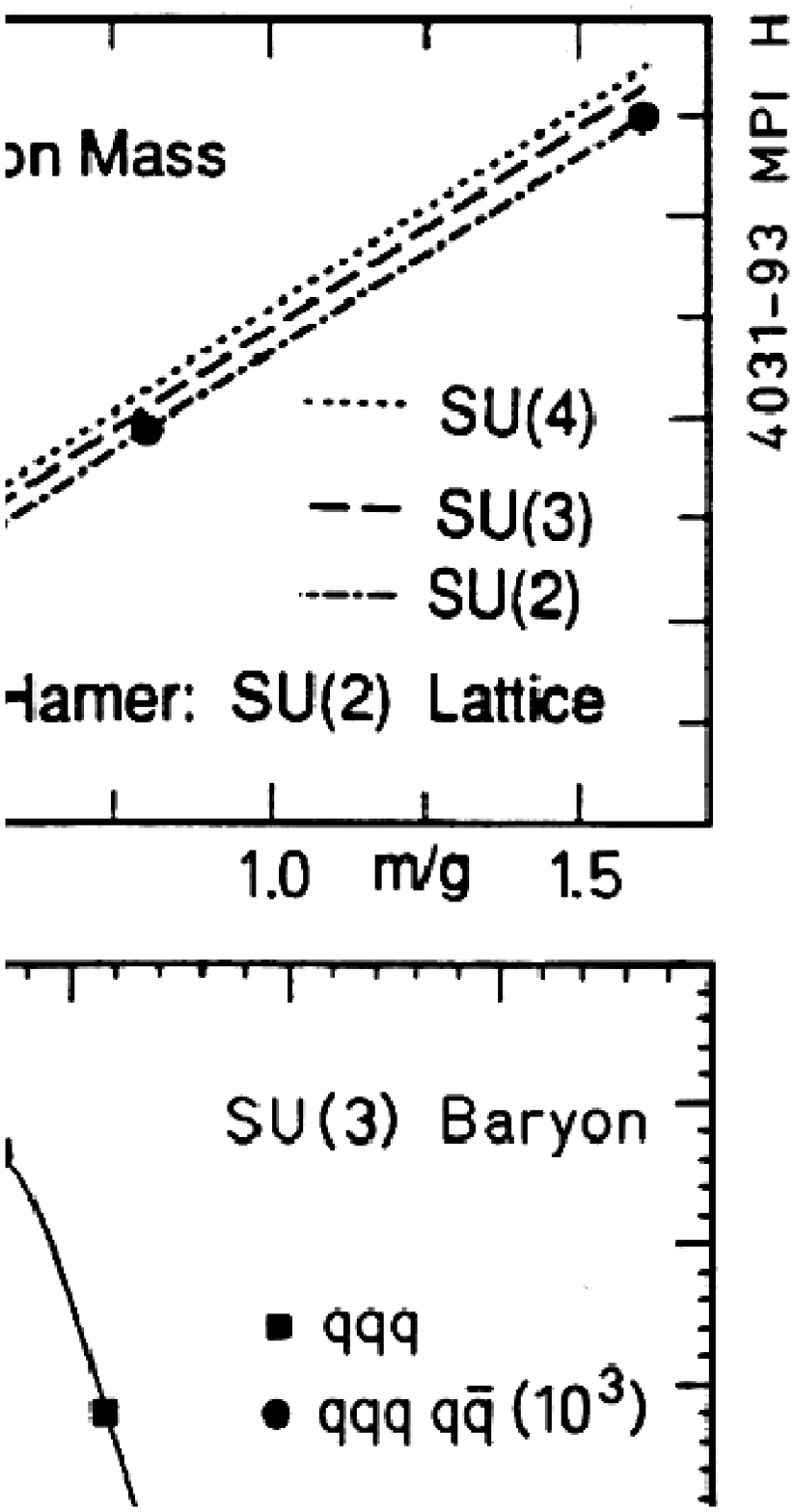}
\end{minipage}
\caption{\label{fig-kyf-4}
    Spectra and wavefunctions in 1+1 dimension, taken from [136,224]
    Lattice results are 
    from [192,193,194].
}\end{figure}

In $1+1$ dimensions quantum electro\-dynamics \cite {epb87} 
and quantum chromo\-dynamics  \cite {hbp90} 
show many similarities, both from the technical and from the 
phenomenological point of view. 
A plot like that on the left side in Figure~\ref{fig-kyf-4} was first 
given by Eller 
for periodic boundary conditions on the fermion fields \cite {{epb87}}, 
and repeated recently for anti-periodic ones \cite {els94}. 
For a fixed value of the resolution, it shows the full mass
spectrum  of QED in the charge zero sector for all values of the 
coupling constant and the fermion mass, parametrized by  
$\lambda = (1+ \pi (m/g)^2) ^{-{1\over2}}$. 
It includes the free case $\lambda=0$ ($g=0$) and the 
Schwinger model $\lambda=1$ ($m=0$). 
The eigenvalues $M_i$ are plotted in units where the mass 
of the lowest `positronium' state has the numerical value 1. 
All states with $M>2$ are unbound.
In the right part of Figure~\ref{fig-kyf-4} some of the results of
Hornbostel \cite {hbp90} one the spectrum and the 
wavefunctions for QCD are displayed. 
Fock states in non-abelian gauge theory SU(N) 
can be made color singlets for any order of the gauge group 
and thus one can calculate mass spectra for mesons and
baryons for almost arbitrary values of N. In the upper light
part of the figure the lowest mass eigenvalue of a meson
is given for $N=2,3,4$. 
Lattice gauge calculations to compare with are available only for 
$N=2$  and for the lowest two eigenstates; the agreement is very 
good. In the left lower part of the figure the
structure function of a baryon is plotted versus (Bj\o rken-)$x$ for
$m/g=1.6$. With DLCQ it is possible to calculate also higher 
Fock space components. As an example, the figure includes the
probability to find a quark in a $qqq\,q\bar q\,$-state.

Meanwhile, many calculations have been done for 1+1 dimension, 
among them those by 
Eller {\it et al.} \cite{epb87,elp89}, 
Hornbostel {\it et al.} \cite{hbp88,hbp90,hor90,hor91,hor92}, 
Antonuccio {\it et al.} \cite{and95,and96a,and96b,anp97}, 
Burkardt {\it et al.} \cite{bur89,bul91a,bul91b,bur93,bur94}, 
Dalley {\it et al.} \cite{dak93,dav96,vad96}, 
Elser {\it et al.} \cite{els94,elk96,hek95}, 
Fields {\it et al.} \cite{fgv95,npf97,vfp94,vfp96}, 
Fujita {\it et al.} \cite{fuo93,fio94,fus95a,fus95b,fht96a,fht96b,otf95,tof91}, 
Harada {\it et al.} \cite{hot94,hot95,hao96,hot96,hot97}, 
Harindranath {\it et al.} \cite{ghp93,hav87,hav88,hap91,peh91,zah93a,zah93b,zah93c},
Hiller {\it et al.} \cite{brh83,hil91a,hil91b,abh90,hbo95,hpv95,mah89,swh93,wih93}, 
Hollenberg {\it et al.} \cite{hhw91,how94}, 
Itakura {\it et al.} \cite{ita96a,ita96b,itm97}, Pesando {\it et al.} \cite{pes95a,pes95b}, 
Kalloniatis {\it et al.} \cite{elk96,hek95,kap93,kap94,kar94,kpp94,kal95,kar96,pkp95,pik96}, 
Klebanov {\it et al.} \cite{bdk93,dak93,dkb94}, 
McCartor {\it et al.} \cite{mcc88,mcc91,mcr92,mcr94,mcc94,mcr95,mcr97}, 
Nardelli {\it et al.} \cite{bkk93,bgn96,ban97}, 
van de Sande {\it et al.} \cite{bpv93,dav96,hpv95,piv94,vap92,vap95,vab96,van96,vad96},
Sugihara {\it et al.} \cite{smy94,suy96,suy97}, 
Tachibana {\it et al.} \cite{tac95}, 
Thies {\it et al.} \cite{ltl91,tho93},  
Tsujimaru {\it et al.} \cite{hot93,kty95,ntt95,tsy97}, 
and others \cite{aat97,jir95,kou95,mis96,rud94,sri94}, 
Aspects of reaction theory can be studied now.
Hiller \cite {hil91a}
for example has calculated the total annihilation cross section
$R_{e\bar e}$ in 1+1 dimension, with success. 

We will use the work of Hornbostel \cite {hbp90} as an example 
to demonstrate how DLCQ works.

Consider the light-cone gauge, $A^+ =0$, with the gauge group SU(N).
In a representation in which $\gamma^5$ is diagonal
one introduces the chiral components of the fermion spinors:
\begin {equation}
    \psi _\alpha = \pmatrix {\psi _L \cr \psi _R \cr}
\,, \end {equation}
The usual group generators for SU(N) are the $T^a= {\lambda^a \over 2}$.
In a box with length $2L$ one finds
\begin {equation}
   \psi_L(x) = - {im \over 4} \int_{-L}^{+L} dy^-
   \epsilon \left( x^--y^-\right) \psi_R(x^+,y^-)
\end {equation}
and
\begin {equation}
   A^{-a}(x) = - {g\over 2} \int_{-L}^{+L} dy^- \left|x^--y^-\right|
   \psi_R^\dagger T^a \psi_R (x^+,y^-) \ .
\end {equation}
The light-cone momentum and light-cone energy operators are
\begin {equation}
   P^+= \int_{-L}^{+L} dx^- \psi_R^\dagger \partial_-\psi_R
\,,  \label{lcmom} \end {equation}
and
\begin {eqnarray}
   P^-
   = &-& {im^2 \over 4} \int_{-L}^{+L}dx^- \int_{-L}^{+L}dy^-
   \psi_R^\dagger (x^-) \epsilon ( x ^- - y ^-) \,\psi_R(y^-)
\nonumber \\
   &-& {g^2\over 2} \int_{-L}^{+L}dx^- \int_{-L}^{+L}dy^-
    \psi_R^\dagger T^a \psi_R(x^-) \left| x^--y^- \right|
    \psi_R^\dagger T^a \psi_R(y^-)
\,, \label{lcen} \end {eqnarray}
respectively. Here, $\psi$ is subject to the canonical 
anti-commutation relations. 
For example, for anti-periodic boundary conditions one can expand
\begin {equation}
   \psi_R(x^-)_c = {1\over \sqrt{2L}}
   \sum_{n={1\over 2}, {3\over 2},...}^{\infty}
    \left( b_{n,c} e^{-i{n\pi \over L}x^-}
     + d_{n,c}^\dagger  e^{i{n\pi \over L}x^-} \right)
\ ,\end {equation}
where
\begin {equation}
   \left\{ b_{n,c_1}^\dagger, b_{m,c_2} \right\} =
   \left\{ d_{n,c_1}^\dagger, d_{m,c_2} \right\} =
   \delta_{c_1,c_2} \delta_{n,m}
\ ,\end {equation}
with all other anticommutators vanishing.
Inserting this expansion into the expressions for $P^+$,
Eq.(\ref{lcmom}), one thus finds
\begin {equation}
   P^+ = \left( {2\pi \over L } \right)
   \sum_{n={1\over 2}, {3\over 2},...}^{\infty}
   n \left( b_{n,c}^\dagger b_{n,c} + d_{n,c}^\dagger d_{n,c} \right)
\ .\end {equation}
Similarly one finds for $P^-$ of Eq.(\ref{lcen})
\begin {equation}
   P^- = \left( {L \over 2\pi } \right)\left( H_0 + V\right)
\ ,\end {equation}
where
\begin {equation}
   H_0 =
   \sum_{n={1\over 2}, {3\over 2},...}^{\infty} {m^2\over n}
   \left( b_{n,c}^\dagger b_{n,c} + d_{n,c}^\dagger d_{n,c} \right)
  \end {equation}
is the free kinetic term, and the interaction term $V$ is given by
\begin {equation}
   V = {g^2\over \pi} \sum_{k=-\infty}^\infty
   j^a(k) {1\over k^2} j^a(-k)
\ , \label{vdis} \end {equation}
where
\begin {equation}
   j^a(k) =
   T^a_{c_1,c_2} \sum_{n=-\infty}^{\infty}
   \left( \Theta(n) b_{n,c1}^\dagger + \Theta(-n) d_{n,c1} \right)
   \left( \Theta(n-k) b_{n-k,c2}
 + \Theta(k-n) d_{k-n,c2}^\dagger \right) .
\end {equation}
Since we will restrict ourselves to the color singlet sector,
there is no problem from $k=0$ in Eq.(\ref{vdis}), since $j^a(0)=0$
acting on color singlet states.
Normal ordering the interaction  (\ref {vdis}) gives an
diagonal operator piece
\begin {equation}
   V = :V: \ +\ {g^2 C_F\over \pi} 
   \sum_{n={1\over 2}, {3\over 2},...}^{\infty}
   { I_n \over n} 
   \left( b_{n,c}^\dagger b_{n,c} + d_{n,c}^\dagger d_{n,c} \right)
\ ,\end {equation}
with the `self-induced inertia' 
\begin {equation}
   I_n = - {1\over 2n} + \sum_{m=1}^{n+{1\over2}} {1\over m^2}
\,.\end {equation}
The color  factor is $C_F = {N^2-1 \over 2N}$.
The explicit form of  the normal ordered piece $:V:$ can be found
in \cite{hbp90} or in the explicit tables below in this section.
It is very important to keep the self-induced inertias from the
normal ordering, because they are needed to cancel the infrared
singularity in the interaction term in the continuum limit.
Already classically, the self energy of one single quark is
infrared divergent because its color electric field extends
to infinity. The same infrared singularity (with opposite sign)
appears in the interaction term. They cancel for color
singlet states, because there the color electric field is nonzero
only inside the hadron. Since the hadron has a finite size, the
resulting total color electric field energy must be infrared finite.

The next step is to actually solve the equations of motions in the
discretized space. Typically one proceeds as follows: Since $P^+$ 
and $P^-$ commute they can be diagonalized simultaneously.
Actually, in the momentum representation, $P^+$ is already 
diagonal, with eigenvalues proportional to $2\pi/L$.
Therefore the {\em harmonic resolution} $K$ \cite {epb87},
\begin{equation}
      K = {L\over 2\pi} P^+
\,,\end{equation}
determines the size of the Fock space and thus the dimension of the 
Hamiltonian matrix, which simplifies the calculations considerably. 
For a given $K=1,2,3,\dots$, there are only a finite number of Fock 
states due to the positivity condition on the light-cone momenta. 
One selects now one value for $K$ and constructs all color singlet 
states. In the next step one can either diagonalize $H$ in the full 
space of states with momentum $K$ (DLCQ approximation) or 
in a subspace of that space (for example with a Tamm-Dancoff 
approximation).
The eigenvalue $E_i(K)$ correspond to invariant masses
\begin {equation}
   M_i^2(K) \equiv 2 P ^+ _i P ^- _i = K E _i(K)
  \end {equation}
where we indicated the parametric dependence of the eigenvalues 
on $K$. 

Notice that the length $L$ drops out in the invariant mass, 
and that one gets a spectrum for {\em any value of } $K$.
Most recent developments in string theory, the so called 
`M(atrix)-theory' \cite{sus97},  emphasizes this aspect, but for
the present one should consider the solutions to be physical
only in the continuum limit  $K\rightarrow\infty$. 

Of course there are limitations
on the size of matrices that one can diagonalize (although the
Lanczos algorithm allows quite impressive sizes \cite{hil91a}).
Therefore what one typically does is to repeat the calculations
for increasing values of $K$ and to extrapolate observables to
$K\rightarrow \infty$. The first QCD$_2$ calculations in that direction
were performed in Refs.\cite {hbp90} and \cite {bur89}. In these
pioneer works it was shown that the numerics actually converged
rather quickly (except for very small quark masses, where ground state
mesons and ground state baryons become massless) since
the lowest Fock component dominates these hadrons
(typically less than one percent of the momentum is carried by
the sea component).
One does not know of any simple explanation of this surprising result,
except the rather intuitive argument that these ground state hadrons
are very small and pointlike objects cannot radiate.
Due to these fortunate circumstances a variety of phenomena could
be investigated. For example Hornbostel studied hadron masses
and structure functions for various $N$ which showed very simple 
scaling behavior with $N$. 
A correspondence with the analytic work 
of Einhorn \cite{ein76} for meson form factors in QCD$_{1+1}$
was also established.
Ref.\cite {bur89} focused more on nuclear
phenomena. There it was shown that two nucleons in
QCD$_{1+1}$ with two colors and two flavors form a loosely
bound state --- the ``deuteron''. Since the calculation was
based entirely on quark degrees of freedom it was possible
to study binding effects on the nuclear structure function
(``EMC-effect'').
Other applications of include a study of ``Pauli-blocking''
in QCD$_{1+1}$. Since quarks are fermions, one would expect
that sea quarks which have the same flavor as the
majority of the valence quarks (the up quarks in a proton)
are suppressed compared to those which have the minority
flavor (the down quarks in a proton) --- at least if isospin
breaking effects are small. However, an explicit calculation
shows that the opposite is true in QCD$_2$! This so called
``anti Pauli-blocking'' has been investigated
in Ref. \cite {bul91a,bul91b}, where one can also find an intuitive
explanation.

\subsection {The Hamiltonian operator in 3+1 dimensions (BL)}
\label {sec:ham-matrix}

Periodic boundary conditions on ${\cal L} $ can be realized by
periodic boundary conditions on the vector potentials $A _\mu$
and anti-periodic boundary conditions on the spinor fields,
since ${\cal L} $ is bilinear in the $\Psi _\alpha$.
In momentum representation one expands these fields into plane
wave states $ e^{-ip _\mu x ^\mu} $,
and satisfies the boundary conditions by {\it discretized momenta}
\begin {eqnarray}
    p _- 
 &=& \cases{ {\pi\over \scriptstyle L} n ,
             &with $n =
              {\scriptstyle1\over\scriptstyle2},
              {\scriptstyle3\over\scriptstyle2},\dots,
              \infty\ $ for fermions, \cr
              {\pi\over L} n  ,
             &with $n = 1,2,\dots,\infty\ $ for bosons, \cr}
\nonumber \\
   \quad{\rm and}\ \vec p _{\!\bot}
 &=& {\pi\over L_{\!\bot}} \vec n _{\!\bot},
  \ \quad{\rm with} \ n_x,n_y = 0,\pm1,\pm2,\dots,\pm\infty
  \ \quad{\rm for}\ {\rm both} ,
\label{enapbc} \end {eqnarray}
at the expense of introducing two artificial length parameters,
$ L$ and $ L_{\!\bot}$. They also define
the normalization volume $ \Omega \equiv 2L(2L_{\!\bot})^2$.

\begin{table} [t]
\begin{center}
\begin{tabular}{|ll|} 
\hline
\begin{tabular}{l}  
\epsfysize=32ex\epsfbox{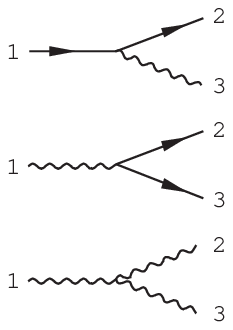}
\\ 
\end{tabular}
&
\begin{tabular}{l}                
  $\displaystyle  V_{1\phantom{,3}} =  
  {\Delta_V\over \sqrt{k^+_1k^+_2 k^+_3}}
  \ (\bar u _1 /\!\!\!\epsilon_3T^{a_3}u_2)$
\\ 
  $\displaystyle V_{2\phantom{,3}} =  
  {\Delta_V \over \sqrt{k^+_1k^+_2 k^+_3}}
  \ (\bar v_2 /\!\!\!\epsilon^\star_1T^{a_1}u_3)$
\\ 
  $\displaystyle V_{3,1} =  
  {iC_{a_2a_3}^{a_1}\,\Delta_V \over \sqrt{k^+_1k^+_2 k^+_3}}
  \ (\epsilon^\star_1k_3)\ (\epsilon_2\epsilon_3) $
\\ 
  $\displaystyle V_{3,2} =  
  {iC_{a_2a_3}^{a_1}\,\Delta_V \over \sqrt{k^+_1k^+_2 k^+_3}}
  \ (\epsilon_3k_1)\ (\epsilon^\star_1\epsilon_2) $
\\ 
  $\displaystyle V_{3,3} =  
  {iC_{a_2a_3}^{a_1}\,\Delta_V \over \sqrt{k^+_1k^+_2 k^+_3}}
  \ (\epsilon_3k_2)\ (\epsilon^\star_1\epsilon_2) $
\\ 
\end{tabular}
\\ \hline
\end{tabular}
\caption [verspi] {\label {tab:verspi} 
   The vertex interaction in terms of Dirac spinors.
   The matrix elements $V_{n}$ are displayed on the right, 
   the corresponding (energy) graphs on the left.
   All matrix elements are proportional to 
   $\Delta_V = \widehat g \delta (k^+_1| k^+_2 +_3) 
   \delta ^{(2)} (\vec k _{\!\perp,1} | \vec k _{\!\perp,2} +
   \vec k _{\!\perp,3} ) $,
   with $\widehat  g= g P^+/\sqrt{\Omega} $. 
   In the continuum limit,  see Sec.~\protect{\ref{sec:Retrieving}}, 
   one uses $\widehat  g = g P^+/\sqrt{2(2\pi)^3} $.
}\end{center}
\end{table}

More explicitly, the free fields are expanded as
\begin {eqnarray}
  \widetilde \Psi _\alpha (x)   = { 1\over \sqrt{\Omega } }
  \sum_q { 1\over \sqrt{ p^+} }
  \left(   b_q u_\alpha (p,\lambda) e^{ -ipx}
         + d^\dagger _q v_\alpha (p,\lambda) e^{ipx} \right) ,
\nonumber \\
  \quad{\rm and}\quad
  \widetilde A  _\mu (x) = { 1\over \sqrt{\Omega } }
  \sum_q { 1\over \sqrt{ p^+} }
  \left(   a _q \epsilon_\mu       (p,\lambda) e^{ -ipx}
         + a^\dagger _q \epsilon_\mu^\star (p,\lambda) e^{ipx} \right) ,
\label{enexdi} \end {eqnarray}
particularly for the two transversal vector potentials
$ \widetilde A  ^i \equiv   \widetilde A  ^i _{\!\bot}$,
($i=1,2$). The light-cone gauge and the light-cone Gauss equation,
{\it i.e.}  %
$ A  ^+ = 0$ and
$ A  ^- = {2 g  \over (i\partial ^+)^2} \, J  ^+
     \ -\  {2 \over (i\partial ^+) } \, i\partial _j A _{\!\perp}  ^j $,
respectively, complete the specification of the
vector potentials $A ^\mu$. 
The subtlety of the missing {\it zero-mode} $n = 0$ in the
expansion of the $\widetilde A  _{\!\bot}$ will be discussed below.
Each denumerable single particle state `$q$' is specified by 
at least six quantum numbers, {\it i.e.}
\begin {equation}
q = \{ q | n,\, n_x,\, n_y, \,\lambda, \,c , \,f \} 
\ . \label {eq:state-q} \end {equation}
The quantum numbers denote the three discrete momenta
$n, n _x, n _y$, the two helicities $\lambda = (\uparrow, \downarrow) $,
the color index $c=1,2,\dots, N _C$, and the flavor index 
$f = 1,2, \dots, N_F $. For a gluon state, the color index is 
replaced by the glue index $a=1,2, \dots, N _C ^2-1$ and the
flavor index absent. Correspondingly, for QED the color- and flavor 
index is absent. 
The creation and destruction operators like $a^\dagger _q$ and 
$a _q$ create and destroy single particle states $q$, and obey
(anti-) commutation relations like
\begin {equation}
  \big[  a _q, a^\dagger _{q^\prime} \big]  =
  \big\{ b _q, b^\dagger _{q^\prime} \big\} =
  \big\{ d _q, d^\dagger _{q^\prime} \big\} =
  \delta_{q,q^\prime}
\ .\end {equation}
The Kronecker symbol is unity only if all six quantum numbers
coincide. The spinors $u_\alpha $ and $v_\alpha$, and the 
transversal polarization vectors $\vec \epsilon _{\!\bot} $
are the usual ones, and can be found in \cite {{brp91}} 
and in the appendix.

\begin{table}
\begin{center}
\begin{tabular}{|ll|} 
\hline 
\begin{tabular}{l}  
\epsfysize=54ex\epsfbox{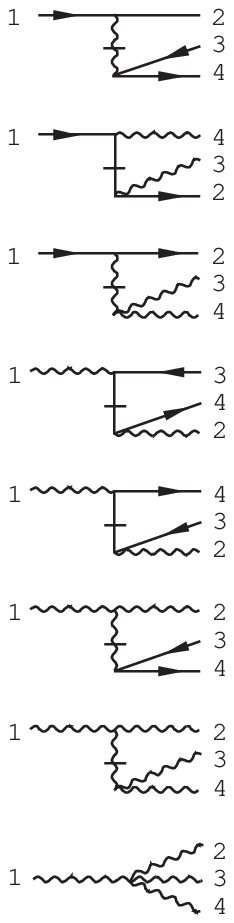}
\end{tabular}
&
\begin{tabular}{l}  
  $\displaystyle F_{3,1} = + 
  {2\Delta\over \sqrt{k^+_1k^+_2 k^+_3 k^+_4}}
  {(\bar u _1T^a\gamma^+u_2)\ (\bar v_3\gamma^+T^au_4)
   \over(k^+_1-k^+_2)^2}$
\\                                           
  $\displaystyle F_{5,1} = +
  {\Delta\over \sqrt{k^+_1k^+_2 k^+_3 k^+_4}}
  {(\bar u _1T^{a_4}/\!\!\!\epsilon_4\gamma^+
   /\!\!\!\epsilon_3T^{a_2}u_2)
   \over( k^+_1-k^+_4)}$
\\                                           
  $\displaystyle F_{5,2} = -
  {2k^+_3\Delta\over \sqrt{k^+_1k^+_2 k^+_3 k^+_4}}
  {(\bar u _1T^a\gamma^+u_2)
  \ (\epsilon_3iC^a\epsilon_4)\over( k^+_1-k^+_2)^2}$
\\                                           
  $\displaystyle F_{7,1} = +
  {\Delta\over \sqrt{k^+_1k^+_2 k^+_3 k^+_4}}
  {(\bar v _3T^{a_1}/\!\!\!\epsilon^\star_1\gamma^+
  /\!\!\!\epsilon_2T^{a_2}u_4)
  \over( k^+_1-k^+_3)}$
\\                                           
  $\displaystyle F_{7,2} = -
  {\Delta\over \sqrt{k^+_1k^+_2 k^+_3 k^+_4}}
  {(\bar v _3T^{a_2}/\!\!\!\epsilon_2\gamma^+
  /\!\!\!\epsilon^\star_1T^{a_1}u_4)
  \over( k^+_1-k^+_4)}$
\\                                           
  $\displaystyle F_{7,3} = +
  {2(k^+_1+k^+_2)\Delta\over \sqrt{k^+_1k^+_2 k^+_3 k^+_4}}
  {(\bar v _3T^a\,\gamma^+\,u_4)   
   \ (\epsilon^\star_1iC^a\epsilon_2)  \over( k^+_1-k^+_2)^2}$
\\                                           
  $\displaystyle F_{9,1} = +
  {2k^+_3(k^+_1+k^+_2)\Delta\over 
  \sqrt{k^+_1k^+_2 k^+_3 k^+_4}\hfill}
  \ {(\epsilon^\star_1C^a\epsilon_2)
  \   (\epsilon_3C^a\epsilon_4)  \over( k^+_1-k^+_2)^2}$
\\                                           
  $\displaystyle F_{9,2} = +
  {2\Delta\over \sqrt{k^+_1k^+_2 k^+_3 k^+_4}}
  \ (\epsilon^\star_1\epsilon_3)\ (\epsilon_2\epsilon_4) 
  \ C^a_{a_1a_2} C^a_{a_3a_4}$
\end{tabular}
\\  \hline
\end{tabular}
\caption [forspi] {\label {tab:forspi} 
   The fork interaction in terms of Dirac spinors.
   The matrix elements $F_{n,j}$ are displayed on the right, 
   the corresponding (energy) graphs on the left.
   All matrix elements are proportional to 
   $\Delta = \widetilde g^2  \delta (k^+_1 | k^+_2 + k^+_3+k^+_4) 
   \delta ^{(2)} (\vec k _{\!\perp,1} | \vec k _{\!\perp,2} +
                  \vec k _{\!\perp,3} + \vec k _{\!\perp,4} ) $, 
   with $\widetilde g^2 = g^2 P^+/(2\Omega) $. 
   In the continuum limit,  see Sec.~\protect{\ref{sec:Retrieving}}, 
   one uses $\widetilde g^2 = g^2 P^+/(4(2\pi)^3) $. 
}\end{center}
\end{table}

\begin{table}
\begin{center}
\begin{tabular}{|ll|} 
\hline 
\begin{tabular}{l}  
\epsfysize=86ex\epsfbox{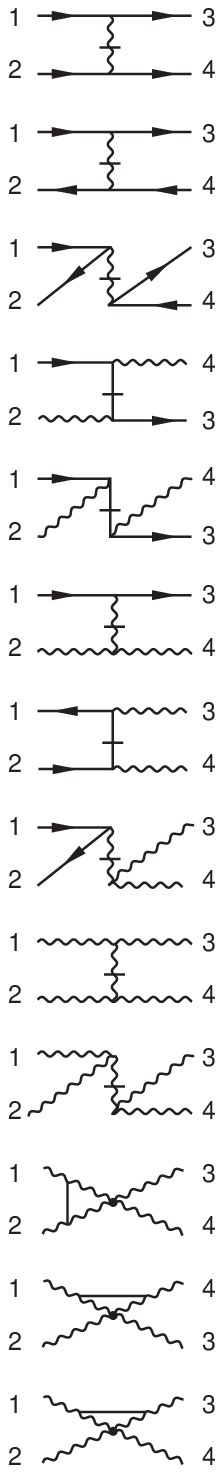}
\\ 
\end{tabular}
&
\begin{tabular}{l}  
  $\displaystyle S_{1,1} = - 
  {\Delta\over \sqrt{k^+_1k^+_2 k^+_3 k^+_4}}
  {(\bar u _1T^a\gamma^+u_3)\ (\bar u _2\gamma^+T^au_4)
   \over(k^+_1-k^+_3)^2}$
\\ 
  $\displaystyle S_{3,1} = +
  {2\Delta\over \sqrt{k^+_1k^+_2 k^+_3 k^+_4}}
  {(\bar u _1T^a\gamma^+u_3)\ (\bar v _2\gamma^+T^av_4)
  \over(k^+_1-k^+_3)^2}$
\\ 
  $\displaystyle S_{3,2} = -
  {2\Delta\over \sqrt{k^+_1k^+_2 k^+_3 k^+_4}}
  {(\bar v_2T^a\gamma^+u_1)\ (\bar v _4\gamma^+T^au_3)
  \over(k^+_1+k^+_2)^2}$
\\ 
  $\displaystyle S_{5,1} = +
  {\Delta\over \sqrt{k^+_1k^+_2 k^+_3 k^+_4}}
  {(\bar u _1T^{a_4}/\!\!\!\epsilon_4\gamma^+
   /\!\!\!\epsilon^\star _2T^{a_2}u_3)
   \over( k^+_1-k^+_4)}$
\\ 
  $\displaystyle S_{5,2} = +
  {\Delta\over \sqrt{k^+_1k^+_2 k^+_3 k^+_4}}
  {(\bar u _1T^{a_2}/\!\!\!\epsilon^\star _2\gamma^+
  /\!\!\!\epsilon_4T^{a_4}u_3)
  \over( k^+_1+k^+_2)}$
\\ 
  $\displaystyle S_{5,3} = +
  {2(k^+_2+k^+_4)\Delta\over \sqrt{k^+_1k^+_2 k^+_3 k^+_4}}
  {(\bar u _1T^a\gamma^+u_3)
  \ (\epsilon_2^\star iC^a\epsilon_4)\over( k^+_1-k^+_3)^2}$
\\ 
  $\displaystyle S_{7,1} = +
  {\Delta\over \sqrt{k^+_1k^+_2 k^+_3 k^+_4}}
  {(\bar u _1T^{a_3}/\!\!\!\epsilon_3\gamma^+/\!\!\!\epsilon_4
   T^{a_4}v_2) \over( k^+_1-k^+_3)}$
\\ 
  $\displaystyle S_{7,2} = -
  {(k^+_3-k^+_4)\Delta\over \sqrt{k^+_1k^+_2 k^+_3 k^+_4}}
  {(\bar u _1T^a\,\gamma^+\,v_2)   
   \ (\epsilon_3iC^a\epsilon_4)  \over( k^+_1+k^+_2)^2}$
\\ 
  $\displaystyle S_{9,1} = -
  {(k^+_1+k^+_3)(k^+_2+k^+_4)\Delta\over 
  \sqrt{k^+_1k^+_2 k^+_3 k^+_4}\hfill}
  \ {(\epsilon^\star_1C^a\epsilon_3)
   \ (\epsilon^\star_2C^a\epsilon_4)  \over( k^+_1-k^+_3)^2}$
\\ 
  $\displaystyle S_{9,2} = +
  {2k^+_3k^+_4\Delta\over \sqrt{k^+_1k^+_2 k^+_3 k^+_4}}
   \ {(\epsilon^\star_1C^a\epsilon^\star_2)
    \ (\epsilon_3C^a\epsilon_4)  \over( k^+_1+k^+_2)^2}$
\\ 
  $\displaystyle S_{9,3} = +
  {\Delta\over \sqrt{k^+_1k^+_2 k^+_3 k^+_4}}
  \ (\epsilon^\star_1\epsilon^\star_2)\ (\epsilon_3\epsilon_4) 
  \ C^a_{a_1a_3} C^a_{a_2a_4}$
\\ 
  $\displaystyle S_{9,4} = +
  {\Delta\over \sqrt{k^+_1k^+_2 k^+_3 k^+_4}}
  \ (\epsilon^\star_1\epsilon_3)\ (\epsilon^\star_2\epsilon_4) 
  \ C^a_{a_1a_2} C^a_{a_3a_4}$
\\ 
  $\displaystyle S_{9,5} = +
  {\Delta\over \sqrt{k^+_1k^+_2 k^+_3 k^+_4}}
  \ (\epsilon^\star_1\epsilon_3)\ (\epsilon^\star_2\epsilon_4) 
  \ C^a_{a_1a_4} C^a_{a_3a_2}$
\\ 
\end{tabular}
\\  \hline
\end{tabular}
\caption [seaspi] {\label {tab:seaspi} 
   The seagull interaction in terms of Dirac spinors.
   The matrix elements $S_{n,j}$ are displayed on the right, 
   the corresponding (energy) graphs on the left.
   All matrix elements are proportional to 
   $\Delta = \widetilde g^2  \delta (k^+_1 + k^+_2 | k^+_3+k^+_4) 
   \delta ^{(2)} (\vec k _{\!\perp,1} + \vec k _{\!\perp,2} |
                  \vec k _{\!\perp,3} + \vec k _{\!\perp,4} ) $, 
   with $\widetilde g^2 = g^2 P^+/(2\Omega) $. 
   In the continuum limit one,  see Sec.~\protect{\ref{sec:Retrieving}}, 
   uses $\widetilde g^2 = g^2 P^+/(4(2\pi)^3) $. 
}\end{center}
\end{table}

Finally, after inserting all fields in terms of the expansions
in Eq.(\ref{enexdi}), one performs the space-like integrations
and ends up with the light-cone energy-momenta
$ P  ^\nu = P  ^\nu \big( a _q, a^\dagger _q,
        b _q, b^\dagger _q, d _q, d^\dagger _q \big) $
as operators acting in Fock space. 
The space-like components of $P^\nu$ are simple  and  diagonal, 
and its time-time like component complicated and off-diagonal. 
Its Lorentz-invariant contraction 
\begin {equation}
    H_{LC} \equiv P^\nu P_\nu = P^+ P^- - \vec P_{\!\bot} ^2
\end {equation}
is then also off-diagonal. 
For simplicity it  is  referred  to as the {\em light-cone 
Hamiltonian} $H_{LC} $, and often abbreviated as
$H=H_{LC} $. 
It carries the dimension of an invariant mass squared.
In a frame in which $P_{\!\bot} = 0$, it reduces to
$H=P^+ P^-$. 
It is useful to give its general structure in terms of
Fock-space operators. 

\begin{table} [t]
\caption [uudir] {\label {tab:uudir}
   {\em Matrix elements of Dirac spinors} 
   $\bar u(p){\cal M}u(q)$. \par}
\begin{tabular}{|c|c|c|} 
\hline 
    $\displaystyle \rule[2ex]{0ex}{2ex} {\cal M} \rule[-3ex]{0ex}{2ex} $ 
  & $\displaystyle {1\over \sqrt{p^+q^+}} (\bar u(p){\cal M}u(q)) 
    \ \delta_{\lambda_p,\lambda_q} $
  & $\displaystyle {1\over \sqrt{p^+q^+}} (\bar u(p){\cal M}u(q)) 
    \ \delta_{\lambda_p,-\lambda_q} $
\\ \hline \hline
   $\displaystyle \rule[1ex]{0ex}{2ex}\gamma^+\rule[-0.0ex]{0ex}{2ex} $    
  &$2$   &$0$
\\ 
   $\displaystyle \rule[2ex]{0ex}{2ex} \gamma^- \rule[-2.5ex]{0ex}{2ex} $ 
  &$\displaystyle {2\over p^+ q^+}
     \left(\vec p_{\!\perp} \!\cdot\!\vec q_{\!\perp} + m^2 
     +i\lambda_q \vec p_{\!\perp}\!\wedge\!\vec q_{\!\perp}\right)$ 
  &$\displaystyle {2m\over p^+ q^+} \left( 
     p_{\!\perp}(\lambda_q) - q_{\!\perp}(\lambda_q) 
     \right) $
\\ 
   $\displaystyle \rule[2ex]{0ex}{2ex} 
    \vec\gamma_{\!\perp}\!\cdot\!\vec a_{\!\perp} 
    \rule[-2.5ex]{0ex}{2ex} $ 
  &$\displaystyle \vec a_{\!\perp}\!\cdot\!\left(
    {\vec p_{\!\perp}\over p^+}+{\vec q_{\!\perp}\over q^+}
    \right) - i\lambda_q 
    \vec a_{\!\perp}\!\wedge\!\left(
    {\vec p_{\!\perp}\over p^+}-{\vec q_{\!\perp}\over q^+}
    \right) $ 
  &$\displaystyle - a_{\!\perp}(\lambda_q)\left(
     {m\over p^+} - {m\over q^+} \right) $
\\ \hline
   $\displaystyle \rule[2ex]{0ex}{2ex} 1\rule[-2.5ex]{0ex}{2ex} $ 
  &$\displaystyle {m\over p^+}+{m\over q^+}$
  &$\displaystyle 
       {p_{\!\perp}(\lambda_q)\over q^+} 
     - {q_{\!\perp}(\lambda_q)\over p^+} $
\\ \hline 
   $\displaystyle \rule[1.5ex]{0ex}{2ex} 
    \gamma^-\,\gamma^+\,\gamma^- \rule[-2.5ex]{0ex}{2ex} $ 
  &$\displaystyle {8\over p^+ q^+}
     \left(\vec p_{\!\perp} \!\cdot\!\vec q_{\!\perp} + m^2 
     +i\lambda_q \vec p_{\!\perp}\!\wedge\!\vec q_{\!\perp}\right)$
  &$\displaystyle {8m\over p^+ q^+} \left( 
     p_{\!\perp}(\lambda_q) - q_{\!\perp}(\lambda_q) 
     \right) $
\\ 
   $\displaystyle \rule[1.0ex]{0ex}{2ex} \gamma^-\,\gamma^+\,
     \vec\gamma_{\!\perp}\!\cdot\!\vec a_{\!\perp}
     \rule[-2.0ex]{0ex}{2ex}$
  &$\displaystyle {4\over p^+} 
     \left(\vec a_{\!\perp}\!\cdot\!\vec p_{\!\perp}- i\lambda_q 
     \vec a_{\!\perp}\!\wedge\!\vec p_{\!\perp}\right) $
  &$\displaystyle - {4m\over p^+}a_{\!\perp}(\lambda_q)$
\\ 
   $\displaystyle \rule[2ex]{0ex}{2ex} 
    \vec a_{\!\perp}\!\cdot\!\vec\gamma_{\!\perp}
     \,\gamma^+\,\gamma^-\rule[-2.5ex]{0ex}{2ex} $
  &$\displaystyle {4\over q^+} 
     \left(\vec a_{\!\perp}\!\cdot\!\vec q_{\!\perp} + i\lambda_q 
     \vec a_{\!\perp}\!\wedge\!\vec q_{\!\perp}\right) $
  &$\displaystyle \phantom{-}{4m\over q^+}
    a_{\!\perp}(\lambda_q)$
\\ 
   $\displaystyle \vec a_{\!\perp}\!\cdot\!\vec\gamma_{\!\perp}
   \,\gamma^+\,\vec\gamma_{\!\perp}\!\cdot\!\vec b_{\!\perp}$
  &$\displaystyle 2\Big(\vec a_{\!\perp}\!\cdot\!\vec b_{\!\perp} 
   + i\lambda_q\vec a_{\!\perp}\!\wedge\!\vec b_{\!\perp}\Big)$
  &$\displaystyle \rule[1.0ex]{0ex}{2ex} 0 \rule[-1.5ex]{0ex}{2ex} $
\\  \hline \hline
   \multicolumn{3}{|l|} { {\rm Notation:\ }\hfill 
   $\displaystyle \rule[1.0ex]{0ex}{2ex} \lambda = \pm 1$, \quad
   $\displaystyle a_{\!\perp}(\lambda) 
     = -\lambda a_ x - i a_y $, \quad
   $\displaystyle \vec a_{\!\perp}\!\cdot\!\vec b_{\!\perp} 
     = a_x b_ x + a_y b_y $, \quad
   $\displaystyle \vec a_{\!\perp}\!\wedge\!\vec b_{\!\perp} 
     = a_x b_y - a_y b_ x $.}
\\ \multicolumn{3}{|l|} { {\rm Symmetries:\ } 
   \hfill $\displaystyle \bar v(p) \,v(q) 
   = - \bar u(q) \,u(p)$, \quad
   $\displaystyle \bar v(p) \,\gamma^\mu\,v(q) 
   =   \bar u(q) \,\gamma^\mu\,u(p)$,} 
\\ \multicolumn{3}{|l|} {\hfill \rule[-1.5ex]{0ex}{2ex} 
   $\displaystyle \bar v(p) 
   \,\gamma^\mu\gamma^\nu\gamma^\rho\,v(q) 
   =   \bar u(q) 
   \,\gamma^\rho\gamma^\nu\gamma^\mu\,u(p)$.}
\\ \hline 
\end{tabular}
\end{table}

\subsubsection{A typical term of the Hamiltonian operator} 

As an example we consider a typical term in the Hamiltonian,
{\it i.e.}
\[ P^-_{g}= { g ^2 \over 4} \int \!\! d p^+ d^2 p_\perp
\ \widetilde B _{\mu\nu}^a \widetilde B ^{\mu\nu}_a 
\,.\]   
Inserting the free field solutions $\widetilde A^\mu_a$, 
one deals with  $2^4=16$ terms, see also Eq.(\ref{eq:2.95}) 
in Section~\ref{sec:Fock-operators}. 
They can be classified according to their operator structure, 
and belong to one of the six classes
\begin{eqnarray}
a_{q_1} a_{q_2} a_{q_3} a_{q_4} 
\ , \qquad
a^\dagger_{q_4}a^\dagger_{q_3}a^\dagger_{q_2}a^\dagger_{q_1}
\ , \nonumber\\
a^\dagger_{q_1}a_{q_2} a_{q_3} a_{q_4} 
\ , \qquad
a^\dagger_{q_4} a^\dagger_{q_3} a^\dagger_{q_2} a_{q_1}
\ , \nonumber\\
a_{q_1}^\dagger a^\dagger_{q_2} a_{q_3} a_{q_4} 
\ , \qquad
a^\dagger_{q_4} a^\dagger_{q_3} a_{q_2} a_{q_1} 
\ . \nonumber\end{eqnarray} 
In the first step, we pick out only those terms with
one creation and three destruction operators. 
Integration over the space-like coordinates produces 
a product of three Kronecker delta functions
$\delta (k^+_1|k^+_2 + k^+_3+k^+_4) 
   \delta ^{(2)} (\vec k _{\!\perp 1} | \vec k _{\!\perp 2} +
                  \vec k _{\!\perp 3} + \vec k _{\!\perp 4} ) $, 
as opposed to the Dirac delta functions in 
Section~\ref{sec:Fock-operators}.
\begin{table} [t]
\caption [uvdir] {\label {tab:uvdir}
   {\em Matrix elements of Dirac spinors}
   $\bar v(p){\cal M}u(q)$. \par}
\begin{tabular}{|c|c|c|} 
\hline 
    $\displaystyle \rule[2ex]{0ex}{2ex} {\cal M} \rule[-3ex]{0ex}{2ex} $ 
  & $\displaystyle {1\over \sqrt{p^+q^+}} (\bar v(p){\cal M}u(q)) 
    \ \delta_{\lambda_p,\lambda_q} $
  & $\displaystyle {1\over \sqrt{p^+q^+}} (\bar v(p){\cal M}u(q)) 
    \ \delta_{\lambda_p,-\lambda_q} $
\\ 
   \hline \hline
   $\displaystyle \rule[1ex]{0ex}{2ex}\gamma^+\rule[-0.0ex]{0ex}{2ex} $    
   &$0$ &$2$   
\\ 
   $\displaystyle \rule[2ex]{0ex}{2ex} \gamma^- \rule[-2.5ex]{0ex}{2ex} $ 
  &$\displaystyle {2m\over p^+ q^+} 
    \left( p_{\!\perp} (\lambda_q) +
           q_{\!\perp} (\lambda_q) \right) $
  &$\displaystyle  {2\over p^+ q^+}
     \left(\vec p_{\!\perp} \!\cdot\!\vec q_{\!\perp} - m^2 
     +i\lambda_q \vec p_{\!\perp}\!\wedge\!\vec q_{\!\perp}\right)$ 
\\ 
   $\displaystyle \rule[2ex]{0ex}{2ex} 
    \vec\gamma_{\!\perp}\!\cdot\!\vec a_{\!\perp} 
    \rule[-2.5ex]{0ex}{2ex} $ 
  &$\displaystyle a_{\!\perp}(\lambda_q)\left(
     {m\over p^+} + {m\over q^+} \right) $
  &$\displaystyle \vec a_{\!\perp}\!\cdot\!\left(
    {\vec p_{\!\perp}\over p^+}+{\vec q_{\!\perp}\over q^+}
    \right) - i\lambda_q 
    \vec a_{\!\perp}\!\wedge\!\left(
    {\vec p_{\!\perp}\over p^+}-{\vec q_{\!\perp}\over q^+}
    \right) $ 
\\ \hline
   $\displaystyle \rule[2ex]{0ex}{2ex} 1\rule[-2.5ex]{0ex}{2ex} $ 
  &$\displaystyle 
     {p_{\!\perp}(\lambda_q)\over p^+} + 
     {q_{\!\perp}(\lambda_q)\over q^+}  $
  &$\displaystyle -{m\over p^+}+{m\over q^+}$
\\ \hline 
   $\displaystyle \rule[1.5ex]{0ex}{2ex} 
    \gamma^-\,\gamma^+\,\gamma^- \rule[-2.5ex]{0ex}{2ex} $ 
  &$\displaystyle {8m\over p^+ q^+} 
    \left( p_{\!\perp}(\lambda_q) 
         + q_{\!\perp}(\lambda_q) \right) $
  &$\displaystyle {8\over p^+ q^+}
     \left(\vec p_{\!\perp} \!\cdot\!\vec q_{\!\perp} - m^2 
     +i\lambda_q \vec p_{\!\perp}\!\wedge\!\vec q_{\!\perp}\right)$
\\ 
   $\displaystyle \rule[1.0ex]{0ex}{2ex} \gamma^-\,\gamma^+\,
     \vec\gamma_{\!\perp}\!\cdot\!\vec a_{\!\perp}
     \rule[-2.0ex]{0ex}{2ex}$
  &$\displaystyle {4m\over p^+} a_{\!\perp}(\lambda_q)$
  &$\displaystyle {4\over p^+} 
     \left(\vec a_{\!\perp}\!\cdot\!\vec p_{\!\perp}- i\lambda_q  
     \vec a_{\!\perp}\!\wedge\!\vec p_{\!\perp}\right) $
\\ 
   $\displaystyle \rule[2ex]{0ex}{2ex} 
    \vec a_{\!\perp}\!\cdot\!\vec\gamma_{\!\perp}
     \,\gamma^+\,\gamma^-\rule[-2.5ex]{0ex}{2ex} $
  &$\displaystyle {4m\over q^+}
    a_{\!\perp}(\lambda_q)$
  &$\displaystyle {4\over q^+} 
     \left(\vec a_{\!\perp}\!\cdot\!\vec q_{\!\perp} + i\lambda_q 
     \vec a_{\!\perp}\!\wedge\!\vec q_{\!\perp}\right) $
\\ 
   $\displaystyle \vec a_{\!\perp}\!\cdot\!\vec\gamma_{\!\perp}
   \,\gamma^+\,\vec\gamma_{\!\perp}\!\cdot\!\vec b_{\!\perp}$
  &$\displaystyle \rule[1.0ex]{0ex}{2ex} 0 \rule[-1.5ex]{0ex}{2ex} $
  &$\displaystyle 2\Big(\vec a_{\!\perp}\!\cdot\!\vec b_{\!\perp} 
   + i\lambda_q\vec a_{\!\perp}\!\wedge\!\vec b_{\!\perp}\Big)$
\\  \hline \hline
   \multicolumn{3}{|l|} { {\rm Notation:\ }\hfill 
   $\displaystyle \rule[-2ex]{0ex}{5ex} \lambda = \pm 1$, \quad
   $\displaystyle a_{\!\perp}(\lambda) 
     = -\lambda a_ x - i a_y $, \quad
   $\displaystyle \vec a_{\!\perp}\!\cdot\!\vec b_{\!\perp} 
     = a_x b_ x + a_y b_y $, \quad
   $\displaystyle \vec a_{\!\perp}\!\wedge\!\vec b_{\!\perp} 
     = a_x b_y - a_y b_ x $.}
\\ \hline 
\end{tabular}
\end{table}
The Kronecker delta functions are conveniently defined by 
\begin{equation}
    \delta (k^+|p^+)
= {1\over 2L}\int\limits_{-L}^{+L}\!dx^- e^{+i(k_--p_-)x^-}
= {1\over 2L}\int\limits_{-L}^{+L}\!dx^- e^{+i(n-m){\pi x^-\over L}}
= \delta _{n,m}
\,,\label{eq:kronecker}\end{equation}
and similarly for the transversal delta functions.
One gets then 
\begin{eqnarray}
    P^+P ^-_{g} &=& {g^2 P^+ \over 8L(2L_{\!\perp})^2}  
    \sum_{q_1, q_2} \sum_{q_3, q_4} 
    {1\over \sqrt{k^+_1k^+_2 k^+_3 k^+_4} } 
    \ C^a_{a_1a_2} C^a_{a_3a_4} \times
\nonumber\\ 
    &\Big(& a_{q_1}^\dagger a_{q_2} a_{q_3} a_{q_4} 
    \ (\epsilon^\star_1\epsilon_3)\ (\epsilon_2\epsilon_4) 
    \ \delta (k^+_1 | k^+_2 + k^+_3+k^+_4) 
    \delta ^{(2)} (\vec k _{\!\perp 1}  | \vec k _{\!\perp 2} +
                             \vec k _{\!\perp 3} + \vec k _{\!\perp 4} ) 
\nonumber\\ 
    &{+}& a_{q_1} a_{q_2}^\dagger  a_{q_3} a_{q_4} 
    \ (\epsilon_1\epsilon_3)\ (\epsilon_2^\star\epsilon_4) 
    \ \delta (k^+_2 | k^+_3+k^+_4 + k^+_1) 
    \delta ^{(2)} (\vec k _{\!\perp 2}  | \vec k _{\!\perp 3} +
                             \vec k _{\!\perp 4} + \vec k _{\!\perp 1} ) 
\nonumber\\ 
    &{+}& a_{q_1} a_{q_2} a_{q_3}^\dagger  a_{q_4} 
    \ (\epsilon_1\epsilon_3^\star)\ (\epsilon_2\epsilon_4) 
    \ \delta (k^+_3 | k^+_4 + k^+_1 + k^+_2) 
    \delta ^{(2)} (\vec k _{\!\perp 3}  | \vec k _{\!\perp 4} +
                             \vec k _{\!\perp 1} +\vec k _{\!\perp 2} ) 
\nonumber\\ 
    &{+}& a_{q_1} a_{q_2} a_{q_3} a_{q_4}^\dagger  
    \ (\epsilon_1\epsilon_3)\ (\epsilon_2\epsilon_4^\star) 
    \ \delta (k^+_4 | k^+_1 + k^+_2+k^+_3) 
    \delta ^{(2)} (\vec k _{\!\perp 4}  | \vec k _{\!\perp 1} +
                             \vec k _{\!\perp 2} + \vec k _{\!\perp 3} ) 
\ \Big).\nonumber \end{eqnarray}
Introduce  for convenience the function of $4\times5$ variables
\begin{equation}
    F_{9,2} (q_1;q_2,q_3,q_4) = 
    {2\Delta\over \sqrt{k^+_1k^+_2 k^+_3 k^+_4}}\ \Big(
    \epsilon_\mu^\star(k_1,\lambda_1)
    \epsilon^\mu(k_3,\lambda_3)\Big)\ \Big(
    \epsilon_\nu(k_2,\lambda_2)
    \epsilon^\nu(k_4,\lambda_4)\Big) 
    \ C^a_{a_1a_2} C^a_{a_3a_4}
\ ,\label{F92_disc}\end{equation}
with the overall factor $\Delta$ containing the Kronecker deltas
\begin{equation}
    \Delta(q_1;q_2,q_3,q_4) = 
    \widetilde g^2  \ \delta (k^+_1| k^+_2 + k^+_3+k^+_4) \,
    \delta ^{(2)} (\vec k _{\!\perp 1} | \vec k _{\!\perp 2} +
                  \vec k _{\!\perp 3} + \vec k _{\!\perp 4} ) 
\end{equation}
and, as an abbreviation, the `tilded coupling constant' $\widetilde g$,
\begin{equation}
    \widetilde g^2 = {g^2 P^+\over 2\Omega }
\ .\label{disc_coup}\end{equation}
The normalization volume is as usual
$\Omega = 2L(2L_{\!\perp})^2$. 

\begin{table} 
\begin{center}
\begin{tabular}{|ll|} 
\hline 
\begin{tabular}{@{}l@{}}  
\epsfysize=55ex\epsfbox{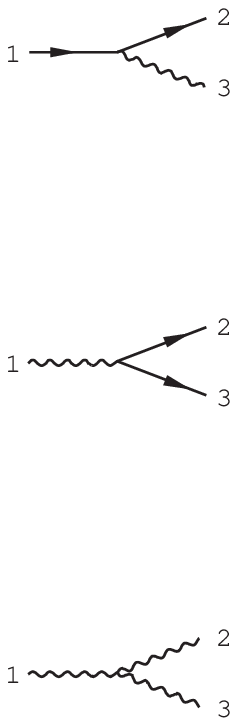}
\end{tabular}
&
\begin{tabular}{@{}r@{}l@{}l@{\,}l@{}}                
      $\displaystyle V_1 
      = \hfill+ \Delta_V \sqrt{{1\over x_3}} m_{\rm F}
      \bigg[ {1\over x_1}-{1\over x_2}\bigg] $
  & $ \delta^{\phantom{-}\lambda _2}_{-\lambda _1} 
      \delta^{\lambda _3}_{\lambda _1}$
  & $ \delta^{f_2}_{f_1} $
  & $ T^{a_3}_{c_1c_2} $ 
\\ 
      $\displaystyle  + \Delta_V \sqrt{{2\over x_3}}
      \ \vec\epsilon_{\!\perp,3}\cdot
      \bigg[\bigg({\vec k\!_\perp\over x}\bigg)_{\!3} -
            \bigg({\vec k\!_\perp\over x}\bigg)_{\!2} \bigg]$
  & $\ \delta^{\phantom{\ }\lambda _2}_{+\lambda _1}
       \delta^{\lambda_3}_{\lambda _1}\ $
  & $ \delta^{f_2}_{f_1} $
  & $ T^{a_3}_{c_1c_2} $ 
\\ 
      $\displaystyle + \Delta_V \sqrt{{2\over x_3}}
      \ \vec\epsilon_{\!\perp,3}\cdot
      \bigg[\bigg({\vec k\!_\perp\over x}\bigg)_{\! 3} -
      \bigg({\vec k\!_\perp\over x}\bigg)_{\! 1}\bigg] $
  & $ \delta^{\phantom{\ }\lambda _2}_{+\lambda _1}
    \delta^{\phantom{\ }\lambda _3}_{-\lambda _1} $
  & $ \delta^{f_2}_{f_1} $
  & $ T^{a_3}_{c_1c_2} $ 
\\ \hline
      $\displaystyle V_2 
      = \hfill + \Delta_V \sqrt{{1\over x_1}} \ m_{\rm F}
      \bigg[{1\over x_2}+{1\over x_3}\bigg] $
  & $ \delta^{\phantom{\ }\lambda _3}_{+\lambda _2}
    \delta^{\phantom{\ }\lambda _3}_{+\lambda _1} $
  & $ \delta^{f_3}_{f_2} $
  & $ T^{a_1}_{c_2c_3} $ 
\\ 
      $\displaystyle + \Delta_V \sqrt{{2\over x_1}}\
      \vec\epsilon_{\!\perp,1}\cdot
      \bigg[\bigg({\vec k\!_\perp\over x}\bigg)_{\!1} -
      \bigg({\vec k\!_\perp\over x}\bigg)_{\!3} \bigg] $
  & $ \delta^{\phantom{\ }\lambda _3}_{-\lambda _2}
      \delta^{\phantom{\ }\lambda _3}_{-\lambda _1} $
  & $ \delta^{f_3}_{f_2} $
  & $ T^{a_1}_{c_2c_3} $ 
\\ 
      $\displaystyle + \Delta_V \sqrt{{2\over x_1}}
      \ \vec\epsilon_{\!\perp,1}\cdot
      \bigg[\bigg({\vec k\!_\perp\over x}\bigg)_{\!1} -
      \bigg({\vec k\!_\perp\over x}\bigg)_{\!2}\bigg] $
  & $ \delta^{\phantom{\ }\lambda _3}_{-\lambda _2}
      \delta^{\phantom{\ }\lambda _3}_{+\lambda _1} $
  & $ \delta^{f_3}_{f_2} $
  & $ T^{a_1}_{c_2c_3} $ 
\\ \hline
      $\displaystyle  V_3 
      = \hfill - \Delta_V \sqrt{ {x_3\over 2 x_1 x_2} }
        \ {\vec\epsilon_{\!\perp,1}}^{\,\star}\cdot \bigg[
          \bigg({\vec k\!_\perp\over x}\bigg)_{\!1}
         -\bigg({\vec k\!_\perp\over x}\bigg)_{\!3}\bigg] $
  & $ \delta^{\phantom{\ }\lambda _3}_{-\lambda _2} $
  & $ $
  & $ i C^{a_1}_{a_2 a_3} $ 
\\ 
      $\displaystyle - \Delta_V \sqrt{{x_1\over 2 x_2 x_3} }
      \ \vec\epsilon_{\!\perp,3}\cdot \bigg[
          \bigg({\vec k\!_\perp\over x}\bigg)_{\!3}
         -\bigg({\vec k\!_\perp\over x}\bigg)_{\!2}\bigg] $
  & $ \delta^{\lambda _2}_{\lambda _1} $
  & $ $
  & $ i C^{a_1}_{a_2a_3} $ 
\\ 
       $\displaystyle  - \Delta_V \sqrt{ {x_2\over 2 x_3 x_1} }
       \ \vec\epsilon_{\!\perp,3}\cdot \bigg[
         \bigg({\vec k\!_\perp\over x}\bigg)_{\!3}
        -\bigg({\vec k\!_\perp\over x}\bigg)_{\!1}\bigg]$
  & $ \delta^{\lambda _2}_{\lambda _1} $
  & $ $
  & $ i C^{a_1}_{a_2a_3} $ 
\end{tabular}
\\ \hline                   
\end{tabular}
\caption [vermat] {\label {tab:vermat} 
   The explicit matrix elements of the vertex interaction.
    The vertex interaction in terms of Dirac spinors.
   The matrix elements $V_{n}$ are displayed on the right, 
   the corresponding (energy) graphs on the left.
   All matrix elements are proportional to 
   $\Delta_V = \widehat g \delta (k^+_1| k^+_2 +_3) 
   \delta ^{(2)} (\vec k _{\!\perp,1} | \vec k _{\!\perp,2} + \vec k _{\!\perp,3} ) $,
   with $\widehat  g = g P^+/\sqrt{\Omega} $. 
   In the continuum limit,  see Sec.~\protect{\ref{sec:Retrieving}}, 
   one uses $\widehat  g= g P^+/\sqrt{2(2\pi)^3} $.
}\end{center}
\end{table}

Next, consider terms with two creation and two destruction 
operators. There are six of them. Relabeling the indices  leaves one
with three different terms. After normal ordering the operator parts on
arrives at
\begin{equation}
     P^+P ^-_{g} = 
    \sum_{q_1, q_2} \sum_{q_3, q_4} \Big(
    S_{9,3} (q_1, q_2; q_3, q_4)  +
    S_{9,4} (q_1, q_2; q_3, q_4)  +
    S_{9,5} (q_1, q_2; q_3, q_4)  \Big)
    \ a_{q_1} ^\dagger a_{q_2} ^\dagger a_{q_3} a_{q_4} 
\ . \end{equation}
The matrix elements $S_{9,3}$, $S_{9,4}$ and $S_{9,5}$ 
can be found in Table~\ref{tab:seaspi}. By the process of
normal ordering one obtains additional diagonal operators,
the self-induced inertias, which are absorbed into the contraction
terms as tabulated in Table~\ref{tab:conmat}, below.

Relabel the summation indices in the above equation 
and  get identically:
\begin{eqnarray}
     P^+P ^-_{g} = {1\over 4}
    \sum_{q_1, q_2} \sum_{q_3, q_4} 
    F_{9,2} (q_1; q_2, q_3, q_4) \ &\Big(&
    a_{q_1} ^\dagger a_{q_2} a_{q_3} a_{q_4} +
    a_{q_2} a_{q_1} ^\dagger a_{q_4} a_{q_3} 
\nonumber\\ 
   &+&
    a_{q_3} a_{q_4} a_{q_1} ^\dagger a_{q_2} +
    a_{q_4} a_{q_3} a_{q_2} a_{q_1} ^\dagger \Big)
\,. \nonumber\end{eqnarray}
After normal ordering, the contribution to
the Hamiltonian becomes
\begin{equation}
     P^+P ^-_{g} = 
    \sum_{q_1, q_2} \sum_{q_3, q_4} 
    F_{9,2} (q_1; q_2, q_3, q_4) 
    \ a_{q_1} ^\dagger a_{q_2} a_{q_3} a_{q_4} 
\,.\label{eq:Pg_disc} \end{equation}
This `matrix element' $F_{9,2} $ can also be found 
in Table~\ref{tab:forspi}.

Finally, focus on terms with only creation or only destruction 
operators. Integration over the space-like coordinates leads to
a product of three Kronecker delta's
\begin{equation}
   \delta (k^+_1 +k^+_2 + k^+_3+k^+_4 | 0) 
   \delta ^{(2)} (\vec k _{\!\perp 1} + \vec k _{\!\perp 2} +
                  \vec k _{\!\perp 3} + \vec k _{\!\perp 4} | \vec 0 ) 
\ , \end{equation}
as a consequence of momentum conservation. 
With $ k^+ = n \pi/(2L)$ and $n$ positive one has thus 
\begin{equation}
   \delta (n_1+n_2 + n_3+n_4 | 0) \equiv 0
. \end{equation}
The sum of posive numbers  can never add up to zero. 
This is the deeper reason why all parts of the light-cone
Hamiltonian with {\em only creation operators} or 
{\em only destruction operators} are {\em strictly zero} in
DLCQ, for any value of the harmonic resolution. 
Therefore the vacuum state cannot couple to any Fock state
by the Hamiltonian, rendering the Fock-space vacuum
identical with the physical vacuum. 
``The vacuum is trivial''.
This holds in general, as long as one disregards the impact of
zero modes, particularly gauge zero modes, see for example 
\cite{kap93,kap94,kar94,kpp94,kal95,pkp95}.

\subsubsection{Retrieving the continuum limit}
\label{sec:Retrieving}

The strictly periodic operator functions in Eq.(\ref{enexdi})
become identical with those of Eq.(\ref{eq:2.72}) in the
continuum limit. In fact, using Eq.(\ref{enapbc}) they
can be translated into each other on a one to one level.
The key relation is the connection between sum and integrals
\begin{eqnarray}
      \int\! dk^+ f(k^+,\vec k _{\!\perp}) \Longleftrightarrow
      {\pi\over 2L}\sum_{n} f(k^+,\vec k _{\!\perp})
\quad{\rm and}\quad
      \int\! dk _{\!\perp i} f(k^+,\vec k _{\!\perp}) \Longleftrightarrow
      {\pi\over L_{\!\perp}}\sum_{n_{\!\perp i}} f(k^+,\vec k _{\!\perp})
\,,\end{eqnarray}
which can be combined to yield
\begin{eqnarray}
      \int\! dk^+ d^2\vec k _{\!\perp}\ f(k^+,\vec k _{\!\perp}) \Longleftrightarrow
      {2(2\pi)^3\over\Omega}\sum_{n,n_{\!\perp}} f(k^+,\vec k _{\!\perp})
\,.\label{eq:int_sum}\end{eqnarray}
Similarly, Dirac delta and Kronecker delta functions are
related by 
\begin{eqnarray}
      \delta(k^+)\ \delta^{(2)}(\vec k _{\!\perp}) \longleftrightarrow
      {\Omega\over 2(2\pi)^3}\ \delta(k^+|0)\ \delta^{(2)}
      (\vec k _{\!\perp}|\vec0) 
\,.\label{eq:dir_kro}\end{eqnarray}
Because of  that, in order to satisfy the respective commutation 
relations, the boson operators $\widetilde a $
and $a$ must be related by
\begin{eqnarray}
      \widetilde a (k) \longleftrightarrow
      \sqrt{{\Omega\over 2(2\pi)^3}}\ a (k) 
\,.\label{eq:bos_fer}\end{eqnarray}
and correspondingly for the fermion operators.
Substituting the three relations Eqs.(\ref{eq:int_sum}),
(\ref{eq:dir_kro}), and (\ref{eq:bos_fer}) into
Eq.(\ref{eq:Pg_disc}), for example, one gets straightforwardly
\begin{eqnarray}
     P^+P ^-_{g} &=& 
    \int\! dk^+_1 d^2\vec k _{\!\perp 1}
    \int\! dk^+_2 d^2\vec k _{\!\perp 2}
    \int\! dk^+_3 d^2\vec k _{\!\perp 3}
    \int\! dk^+_4 d^2\vec k _{\!\perp 4}
\nonumber\\ 
    &\times&
    \sum_{a_1, a_2,a_3, a_4} 
    \sum_{\lambda_1, \lambda_2,\lambda_3, \lambda_4} 
    F_{9,2} (q_1; q_2, q_3, q_4) 
    \ \widetilde a_{q_1} ^\dagger \widetilde a_{q_2} 
    \widetilde a_{q_3} \widetilde a_{q_4} 
\,,\label{eq:Pg_cont} \end{eqnarray}
with the matrix element $F_{9,2}$ formally defined as in 
Eq.(\ref{F92_disc}), except that here holds
\begin{equation}
    \Delta(q_1;q_2,q_3,q_4) = 
    {g^2 P^+\over 4(2\pi)^3 }\ \delta (k^+_1+ k^+_2 + k^+_3+k^+_4) \,
    \delta ^{(2)} (\vec k _{\!\perp 1} + \vec k _{\!\perp 2} +
                  \vec k _{\!\perp 3} + \vec k _{\!\perp 4} ) 
\,.\end{equation}
Of course, one has formally to replace sums by integrals, 
Kronecker delta by Dirac delta functions, and single particle
operators by their tilded versions.
But as a net effect, all what one has to do is to replace the 
tilded coupling constant by
\begin{equation}
    \widetilde g^2 = {g^2 P^+\over 4(2\pi)^3 }
\,,\label{cont_coup}\end{equation}
in order to get from the discretized expressions in the
tables like \ref{tab:verspi}, \ref{tab:forspi} or \ref{tab:seaspi} 
to those in the continuum limit.

\subsubsection{The explicit Hamiltonian for QCD}

Unlike in the instant form, the front form Hamiltonian for the
interacting theory  is additive in the free part $T$ of the 
non-interacting theory and the interaction $U$,
\begin{equation}
H = P^+P^- = T + U
\ .\end{equation}
The {\em kinetic energy} $T$ is the only part of $H$ which does 
not depend on the coupling constant 
\begin{equation}
T = \sum_{q} \ {m_q^2 + \vec k_{\!\bot} ^2 \over x} 
\Big( a_q^\dagger a_q + b_q^\dagger b_q 
+ d_q^\dagger d_q \Big)
\ .\end{equation}
It is a diagonal operator.
The interaction $U$ breaks up into about 12 types of matrix
elements, which are grouped  here into four parts
\begin{equation}
U= V + F + S + C
\ . \end{equation}
We shall discuss them one after another

The {\em vertex interaction} $V$, 
\begin{eqnarray}
   V  &=& \sum_{q_1,q_2,q_3} \bigg[
   b^\dagger_1 b_2 a_3\ V_{1}(1;2,3) -
   d^\dagger_1 d_2 a_3\ V_{1}^{\,\star}(1;2,3)  
   +    {\rm h.c.}   \bigg]
\nonumber \\
   &+& \sum_{q_1,q_2,q_3} \bigg[
   a^\dagger_1d_2b_3 \ V_{2}(1;2,3) +
   a^\dagger_1a_2a_3 \ V_{3}(1;2,3)  
   +    {\rm h.c.}   \bigg]
\ , \end{eqnarray}
operates only between Fock states whose particle number
differs by 1.
The operator aspects of $V$ are carried by the creation and 
destruction operators. The {\em matrix elements} 
$V_{i}(1;2,3)$ are $c$-numbers and 
functions of the various single-particle momenta 
$k^+,\vec k_{\!\perp}$, helicities, color and flavors, 
being tabulated in 
Tables~\ref{tab:verspi}  and \ref{tab:vermat}.
One should emphasize that the graphs in these tables
are {\em energy graphs but no Feynman diagrams}. Like in
all of this review they symbolize {\em matrix elements of
the Hamiltonian} but not some scattering amplitudes. 
They conserve, for example, three-momentum of the particles, 
but opposed to Feynman diagrams they do not conserve their
four-momentum.

\begin{table} 
\begin{center}
\begin{tabular}{|ll|} 
\hline 
\begin{tabular}{@{}l@{}}  
\epsfysize=45ex\epsfbox{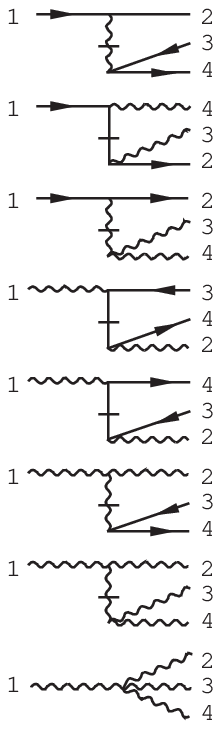}
\end{tabular}
&
\begin{tabular}{@{}l@{\,}l@{\,}l@{\,}l@{}}                
      $\displaystyle F_{3,1} = 
      {2\Delta\over (x_1-x_2)^2}$
  & $\delta^{\lambda_2}_{\lambda_1}\delta^{\phantom{\ }
      \lambda_4}_{-\lambda_3} $
  & $\displaystyle \delta^{f_2}_{f_1} \delta^{f_4}_{f_3} $
  & $\displaystyle T^a_{c_1c_2} T^a_{c_3c_4}$ 
\\ \hline
      $\displaystyle F_{5,1} = 
      {\Delta\over(x_1-x_4)} {1\over\sqrt{x_3x_4}} $
  & $ \delta^{\lambda_2}_{\lambda_1}
      \delta^{\phantom{\ }\lambda_4}_{-\lambda_3}
      \delta^{\phantom{\ }\lambda_4}_{+\lambda_1} $
  & $ \delta^{f_2}_{f_1}  $
  & $T^{a_3}_{c_1c} T^{a_4}_{cc_2}$ 
\\
      $\displaystyle F_{5,2} = 
     {\Delta\over (x_1-x_2)^2}\sqrt{{x_3\over x_4}}$
  & $ \delta^{\lambda_2}_{\lambda_1}
      \delta^{\phantom{\ }\lambda_4}_{-\lambda_3}$
  & $ \delta^{f_2}_{f_1}  $
  & $i T^a_{c_1c_2} C^a_{a_3 a_4} $ 
\\ \hline
      $\displaystyle F_{7,1}  = 
      {\Delta\over (x_1-x_3)} {1\over\sqrt{x_1x_2}} $
  & $\displaystyle \ \delta^{\lambda_2}_{\lambda_1}
      \delta^{\phantom{\ }\lambda_4}_{-\lambda_3}
      \delta^{\phantom{\ }\lambda_4}_{-\lambda_1}\ $
  & $\delta^{f_4}_{f_3}$
  & $T^{a_1}_{c_3c}T^{a_2}_{cc_4}$
\\
      $\displaystyle F_{7,2} = 
      {-\Delta\over (x_1-x_4)}{1\over\sqrt{x_1x_2}} $
  & $ \delta^{\lambda_2}_{\lambda_1}
      \delta^{\phantom{\ }\lambda_4}_{-\lambda_3}
      \delta^{\phantom{\ }\lambda_4}_{+\lambda_1} $
  & $ \delta^{f_4}_{f_3}$
  & $T^{a_2}_{c_3c}T^{a_1}_{cc_4}$
\\
      $\displaystyle F_{7,3} = 
      {\Delta\over (x_1-x_2)^2}{(x_1+x_2)\over\sqrt{x_1x_2}} $
  & $ \delta^{\lambda_2}_{\lambda_1}
      \delta^{\phantom{\ }\lambda_4}_{-\lambda_3} $
  & $ \delta^{f_4}_{f_3}$
  & $ iC^a_{a_1 a_2} T^a_{c_3c_4}$
\\ \hline
      $\displaystyle F_{9,1}  = 
      {\Delta\over 2(x_1-x_2)^2} \hfill
      {(x_1+x_2)x_3\over\sqrt{x_1x_2x_3x_4}} $
  & $ \delta^{            \lambda_2}_{ \lambda_1}
      \delta^{\phantom{\ }\lambda_4}_{-\lambda_3} $
  & $ $
  & $ C^a_{a_1 a_2} C^a_{a_3 a_4} $ 
\\
      $\displaystyle F_{9,2}  = 
               {\Delta\over2\sqrt{x_1x_2x_3x_4}} $
  & $ \delta^{\lambda_3}_{ \lambda_1}
      \delta^{\lambda_4}_{ \lambda_2}$
  & $ $
  & $ C^a_{a_1 a_2} C^a_{a_3 a_4} $ 
\end{tabular}
\\ \hline
\end{tabular}
\caption [format] {\label {tab:format} 
   {\em The matrix elements of the fork interaction}.~--- 
   The matrix elements $F_{n,j}$ are displayed on the right, 
   the corresponding (energy) graphs on the left.
   All matrix elements are proportional to 
   $\Delta = \widetilde g^2  \delta (k^+_1 | k^+_2 + k^+_3+k^+_4) 
   \delta ^{(2)} (\vec k _{\!\perp,1} | \vec k _{\!\perp,2} +
                  \vec k _{\!\perp,3} + \vec k _{\!\perp,4} ) $, 
   with $\widetilde g^2 = g^2 P^+/(2\Omega) $. 
   In the continuum limit,  see Sec.~\protect{\ref{sec:Retrieving}}, 
   one uses $\widetilde g^2 = g^2 P^+/(4(2\pi)^3) $. 
}\end{center}
\end{table}
%

\begin{table} 
\begin{center}
\begin{tabular}{|ll|} 
\hline 
\begin{tabular}{@{}l@{}}  
\epsfysize=72ex\epsfbox{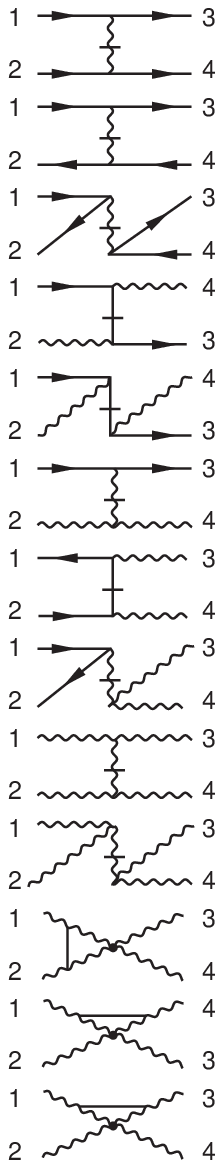}
\end{tabular}
&
\begin{tabular}{@{}l@{\,}l@{\,}l@{\,}l@{}}                
      $\displaystyle 
      S_{1,1} = { -\Delta\over (x_1 - x_3)^2}$
  & $ \delta^{\lambda_3}_{\lambda_1} 
      \delta^{\lambda_4}_{\lambda_2} $
  & $ \delta^{f_3}_{f_1}\delta^{f_4}_{f_2} $
  & $ T^a_{c_1c_3} T^a_{c_2c_4}$ 
\\ 
      $\displaystyle 
      S_{3,1}  = {2\Delta\over(x_1-x_3)^2}$
  & $ \delta^{\lambda_3}_{\lambda_1} 
      \delta^{\lambda_4}_{\lambda_2} $
  & $ \delta^{f_3}_{f_1}\delta^{f_4}_{f_2}  $
  & $ T^a_{c_1c_3} T^a_{c_4c_2}$ 
\\ 
      $\displaystyle 
      S_{3,2}  =  {-2\Delta\over (x_1+x_2)^2} $
  & $\delta^{\phantom{\ } \lambda_2}_{-\lambda_1}
      \delta^{\phantom{\ } \lambda_4}_{-\lambda_3}$
  & $\delta^{f_2}_{f_1}\delta^{f_4}_{f_3}  $
  & $T^a_{c_1c_2} T^a_{c_4c_3}$ 
\\  
     $\displaystyle 
     S_{5,1}  =  {\Delta\over x_1-x_4 }{1 \over \sqrt{x_2x_4} }$
  & $\ \delta^{\lambda_3}_{\lambda_1}
     \delta^{\lambda_4}_{\lambda_2}
     \delta^{\lambda_2}_{\lambda_1}\ $
  & $\delta^{f_3}_{f_1}$
  & $T^{a_4}_{c_1c}T^{a_2}_{cc_3}$ 
\\ 
      $\displaystyle S_{5,2} = 
      {\Delta\over x_1 + x_2 }{ 1 \over \sqrt{x_2x_4} }$
  & $\delta^{\lambda_3}_{\lambda_1} 
     \delta^{\lambda_4}_{\lambda_2}
     \delta^{\phantom{\ } \lambda_2}_{-\lambda_1} $
  & $\delta^{f_3}_{f_1}$
  & $T^{a_2}_{c_1c}T^{a_4}_{cc_3}$ 
\\ 
      $\displaystyle 
      S_{5,3}  = {\Delta\over (x_1-x_3)^2 }
       { (x_2+x_4) \over \sqrt{x_2x_4} } $
  & $\delta^{\lambda_3}_{\lambda_1}
     \delta^{\lambda_4}_{\lambda_2} $
  & $\delta^{f_3}_{f_1}$
  & $iT^a_{c_1c_3} C^a_{a_2a_4} $ 
\\ 
      $\displaystyle 
      S_{7,1}  =  {\Delta\over x_1-x_3 }{1\over\sqrt{x_3x_4}}$
  & $\delta^{\phantom{\ }\lambda_2}_{-\lambda_1}
     \delta^{\phantom{\ }\lambda_4}_{-\lambda_3}
     \delta^{\lambda_3}_{\lambda_1}$
  & $\delta^{f_2}_{f_1}$
  & $T^{a_3}_{c_1c}T^{a_4}_{cc_2}$ 
\\ 
       $\displaystyle 
       S_{7,2}  =  {\Delta\over (x_1+x_2)^2 }
        {(x_3-x_4)\over \sqrt{x_3x_4} } $
  & $\delta^{\phantom{\ }\lambda_2}_{-\lambda_1}
     \delta^{\phantom{\ }\lambda_4}_{-\lambda_3} $
  & $\delta^{f_2}_{f_1}$
  & $iT^a_{c_1c_2} C^a_{a_3a_4} $ 
\\ 
      $\displaystyle 
      S_{9,1}  = { -\Delta(x_1 + x_3)\over 4(x_1 - x_3)^2 }
      {(x_2 + x_4) \over \sqrt{ x_1 x_2 x_3 x_4} } $
  & $\delta^{\lambda_3}_{\lambda_1} 
     \delta^{\lambda_4}_{\lambda_2} $
  & $ $
  & $C^a_{a_1 a_3} C^a_{a_2 a_4} $ 
\\ 
      $\displaystyle S_{9,2}  = {\Delta\over 2(x_1+x_2)^2 }
      \sqrt{ {x_1 x_3 \over x_2 x_4} } $
  & $\delta^{\phantom{\ }\lambda_2}_{-\lambda_1}
     \delta^{\phantom{\ }\lambda_4}_{-\lambda_3} $
  & $ $
  & $C^a_{a_1a_2} C^a_{a_3a_4} $ 
\\ 
      $\displaystyle 
      S_{9,3}  =  {\Delta\over4\sqrt{ x_1 x_2 x_3 x_4}}$ 
  & $\delta^{\lambda_3}_{\lambda_1} 
     \delta^{\lambda_4}_{\lambda_2} $
  & $ $
  & $C^a_{a_1a_2} C^a_{a_3a_4} $ 
\\ 
      $\displaystyle 
      S_{9,4}  = {\Delta\over4\sqrt{ x_1 x_2 x_3 x_4}}$
  & $\delta^{\lambda_3}_{\lambda_1}
     \delta^{\lambda_4}_{\lambda_2} $
  & $ $
  & $ C^a_{a_1a_4} C^a_{a_3a_2} $ 
\\ 
      $\displaystyle 
      S_{9,5} = {\Delta\over4\sqrt{ x_1 x_2 x_3 x_4}}$
  & $\delta^{\phantom{\ }\lambda_2}_{-\lambda_1}
     \delta^{\phantom{\ }\lambda_4}_{-\lambda_3} $
  & $ $
  & $C^a_{a_1 a_3} C^a_{a_2 a_4} $ 
\end{tabular}
\\  \hline
\end{tabular}
\caption [seamat] {\label {tab:seamat} 
   The matrix elements of the seagull interaction. 
   The matrix elements $S_{n,j}$ are displayed on the right, 
   the corresponding (energy) graphs on the left.
   All matrix elements are proportional to 
   $\Delta = \widetilde g^2  \delta (k^+_1 + k^+_2 | k^+_3+k^+_4) 
   \delta ^{(2)} (\vec k _{\!\perp,1} + \vec k _{\!\perp,2} |
                  \vec k _{\!\perp,3} + \vec k _{\!\perp,4} ) $, 
   with $\widetilde g^2 = g^2 P^+/(2\Omega) $. 
   In the continuum limit,  see Sec.~\protect{\ref{sec:Retrieving}}, 
   one uses $\widetilde g^2 = g^2 P^+/(4(2\pi)^3) $. 
}\end{center}
\end{table}

The {\em fork interaction} $F$,
\begin{eqnarray}
   F &=& \sum_{q_1,q_2,q_3,q_4}  \bigg[ (
   b_1 ^\dagger b_2 d_3 b_4 +
   d_1 ^\dagger d_2 b_3 d_4 ) \ F_3(1;2,3,4)  +  
   a_1 ^\dagger a_2 d_3 b_4    \,F_7(1;2,3,4) \bigg] 
\nonumber \\
   &+&    \sum_{q_1,q_2,q_3,q_4}  \bigg[ ( 
   b_1 ^\dagger b_2 a_3 a_4 +
   d_1 ^\dagger d_2 a_3 a_4)\ F_5(1;2,3,4) + 
   a_1 ^\dagger a_2 a_3 a_4 \,F_9(1;2,3,4)  \bigg]
\nonumber \\
   &+&    {\rm h.c.}   
\end{eqnarray}
changes the particle number  by 2. In other words, the 
operator is has non-vanishing (Fock-space) matrix elements only 
if  the particle number of the Fock states differs exactly by 2,
For all other cases, the matrix elements vanish strictly. 
The matrix elements $F_i(1;2,3,4)$ are explicitly tabulated in
Tables~\ref{tab:forspi}  and \ref{tab:format}.

The {\em seagull interaction} $S$ is
\begin{eqnarray}
      S  &=& \sum_{q_1,q_2,q_3,q_4} \bigg[ (
      b_1^\dagger b_2^\dagger b_3 b_4 +
      d_1^\dagger d_2^\dagger d_3 d_4 )\ S_1(1,2;3,4) +
      b_1^\dagger d_2^\dagger b_3 d_4   \ S_3(1,2;3,4) 
     \bigg] 
\nonumber \\
      &+&  \sum_{q_1,q_2,q_3,q_4} \phantom{\bigg[}
      (b_1^\dagger a_2^\dagger b_3 a_4 +
      d_1^\dagger a_2^\dagger d_3 a_4 )\ S_5(1,2;3,4) 
      \phantom{\bigg]}
\nonumber \\
      &+& \sum_{q_1,q_2,q_3,q_4} \ \bigg[ (
      b_1^\dagger d_2^\dagger a_3 a_4 +
      a_4^\dagger a_3^\dagger d_2 b_1 )\ S_7(1,2;3,4)  +
      a_1^\dagger a_2^\dagger a_3 a_4 \ S_9(1,2;3,4)  \bigg] 
\  .\end{eqnarray}
By definition, it has the same number of creation and destruction
operators, and consequently can act only in between Fock states
which have the same particle  number.  
The matrix elements $S_{i}(1,2;3,4) $ are tabulated and graphically
represented in Tables.~\ref{tab:seaspi}  and \ref{tab:seamat}.

\begin{table} 
\begin{center}
\begin{tabular}{|ll|} 
\hline 
\begin{tabular}{@{}l@{}}  
\epsfysize=33ex\epsfbox{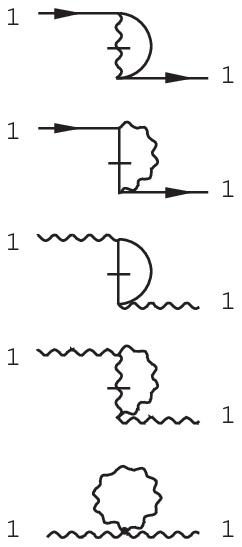}
\end{tabular}
&
\begin{tabular}{@{}l@{}}
    $\displaystyle 
    I_{1,1}(q_1) = \overline g^2 {(N_c^2-1)\over2N_c}
    \sum_{x,\vec k\!_\perp} \Big[
    {x_1\over(x_1-x)^2} - {x_1\over(x_1+x)^2}  \Big] $ 
\\
    $\displaystyle 
    I_{1,2}(q_1) = \overline g^2 {(N_c^2-1)\over4N_c}
    \sum_{x,\vec k\!_\perp} 
    \Big[{x_1\over x_1-x}+{x_1\over x_1+x}\Big]{1\over x}$ 
\\ \hline
    $\displaystyle 
    I_{2,1}(q_1) = \overline g^2 {N_f\over2}
    \sum_{x,\vec k\!_\perp} \Big[
    {1\over(x_1-x)} - {1\over(x_1+x)} \Big] $ 
\\ 
    $\displaystyle 
    I_{2,2}(q_1) = \overline g^2 {N_c\over4}
    \sum_{x,\vec k\!_\perp} \Big[
    {(x_1+x)^2\over(x_1-x)^2} + 
    {(x_1-x)^2\over(x_1+x)^2}  \Big]{1\over x}$ 
\\ 
    $\displaystyle 
    I_{2,3}(q_1) = \overline g^2 {N_c\over2}
    \sum_{x,\vec k\!_\perp} {1\over x} $
\end{tabular}
\\  \hline
\end{tabular}
\par
\caption [contraction] {\label {tab:conmat} 
   The matrix elements of the contractions.
   The  self-induced inertias $I_{n,j}$ are displayed on the right, 
   the corresponding (energy) graphs on the left.
   The number of colors and flavors is denoted by 
   $N_c$ and $N_f$, respectively.
   In the discrete case, one uses
   $\overline g^2 = 2g^2/(\Omega P^+) $, in the continuum limit
   In the continuum limit,  see Sec.~\protect{\ref{sec:Retrieving}}, 
   one uses $\overline g^2 = g^2 /(2\pi)^3 $. 
}\end{center}
\end{table}

The {\em contractions} or {\em self-induced inertias} $C$
\begin{eqnarray}
    C = \sum_{q} \ {I_{q}\over x} \Big( 
    a^\dagger_{q} a_{q} +    b^\dagger_{q} b_{q} 
+ d^\dagger_{q} d_{q} \Big)
    \Longleftrightarrow \sum_{\lambda,c,f} \int\! dk^+ d^2\vec k _{\!\perp}
    \ {I_{q}\over x} \Big( 
    \widetilde a^\dagger_{q} \widetilde a_{q} 
+ \widetilde b^\dagger_{q} b_{q} 
+ \widetilde d^\dagger_{q} d_{q} \Big)
\ , \end{eqnarray}
arise due to bringing $P^-$, or more specifically $S$, into the
above normal ordered form \cite {pab85a,pab85b}. Since they are
(diagonal) {\em operators} they are part of the operator
structure and should not be omitted from the outset. 
However, their structure allows to interpret them as
mass  terms which often can be absorbed into the mass counter 
terms which usually are introduced in the process of regulating
the theory, see below.
They are tabulated below in Tables~\ref{tab:conmat}. 

\subsubsection{Further evaluation of the Hamiltonian matrix elements}

The light-cone Hamiltonian matrix elements in
Figures~\ref{tab:verspi}, \ref{tab:forspi}, and \ref{tab:seaspi} 
are expressed in terms of the Dirac spinors $u_\alpha(k,\lambda)$ 
and $v_\alpha(k,\lambda)$, and polarization vectors
$\epsilon_\mu(k,\lambda)$, which can be found in 
Appendix~\ref{app:General}
and \ref{app:Lepage-Brodsky}.
This representation is particularly useful for perturbative
calculations as we have seen in 
Section~\ref{sec:continuum}.
Very often however, the practitioner needs these matrix
elements  as explicit functions of
the single particle momenta $k^+$ and $\vec k_{\!\perp}$
Their calculation is straightforward but cumbersome, 
even if one uses as a short-cut the tables of Lepage and
Brodsky \cite{leb80} on spinor contractions like 
$\overline u_\alpha \Gamma_{\alpha\beta } u_\beta$. 
We include them here in updated form, particularly 
for $\overline u \Gamma u$ in Table~\ref{tab:uudir}
and for $\overline u \Gamma v$ in Table~\ref{tab:uvdir}.
Using the general symmetry relations between spinor matrix
elements in Appendix~\ref{app:General}.
These tables are all one needs in practice.

For convenience we include also tables with the explicit
expressions. Inserting the spinor matrix elements of
Tables~\ref{tab:uudir}and \ref{tab:uvdir}
into the matrix elements of 
Figures~\ref{tab:verspi}, \ref{tab:forspi}, and
\ref{tab:seaspi}, one obtains those in 
Figures~\ref{tab:vermat}, \ref{tab:format}, and
\ref{tab:seamat}, correspondingly.
Figure~\ref{tab:vermat} compiles the expressions 
for the vertex interaction,
Figure~\ref{tab:format} those for the fork interaction, and
Figure~\ref{tab:seamat} those for the seagull interaction.
The contraction terms, finally, are collected in
Figure~\ref{tab:conmat}.

One should emphasize like in
Section~\ref{sec:equations-of-motion}
that all of these tables and figures hold for QED as well as for
non-abelian gauge theory SU(N) including QCD.
With a grain of salt, they even hold for arbitrary 
n-space and 1-time dimension.
Using the translation keys in Section~\ref{sec:Retrieving}, 
the matrix elements in all of these figures can be translated
easily into the continuum formulation.

\subsection {The Fock space and the Hamiltonian matrix}
\label {sec:construct-matrix}

The Hilbert space for the single particle creation and destruction
operators is the {\em Fock space}, {\it i.e.}  the complete set
all possible {\em Fock states}
\begin {equation}
     \vert \Phi _i \rangle = N _i
     \ b ^\dagger _{q_1} b ^\dagger _{q_2} \dots b ^\dagger _{q_N}
     \ d ^\dagger _{q_1} d ^\dagger _{q_2} \dots 
       d ^\dagger _{q_{\overline N}}
     \ a ^\dagger _{q_1} a ^\dagger _{q_2} \dots
       a ^\dagger _{q_{\widetilde N}} \vert 0 \rangle
\,.\label{eFSE} \end {equation}
The normalization constant $ N _i$ is uninteresting in this context.
They are the analogue to the Slater-determinants
of section~\ref{sec:why-dis-mom}.
As consequence of discretization, the Fock states are orthonormal, 
$\langle \Phi _i \vert \Phi _j \rangle = \delta _{i,j} $, and denumerable.
Only one Fock state, the reference state or 
{\em Fock-space vacuum} $ \vert 0 \rangle $,
is annihilated by all destruction operators.

\begin{figure}
\centerline{
\epsfysize=100mm\epsfbox{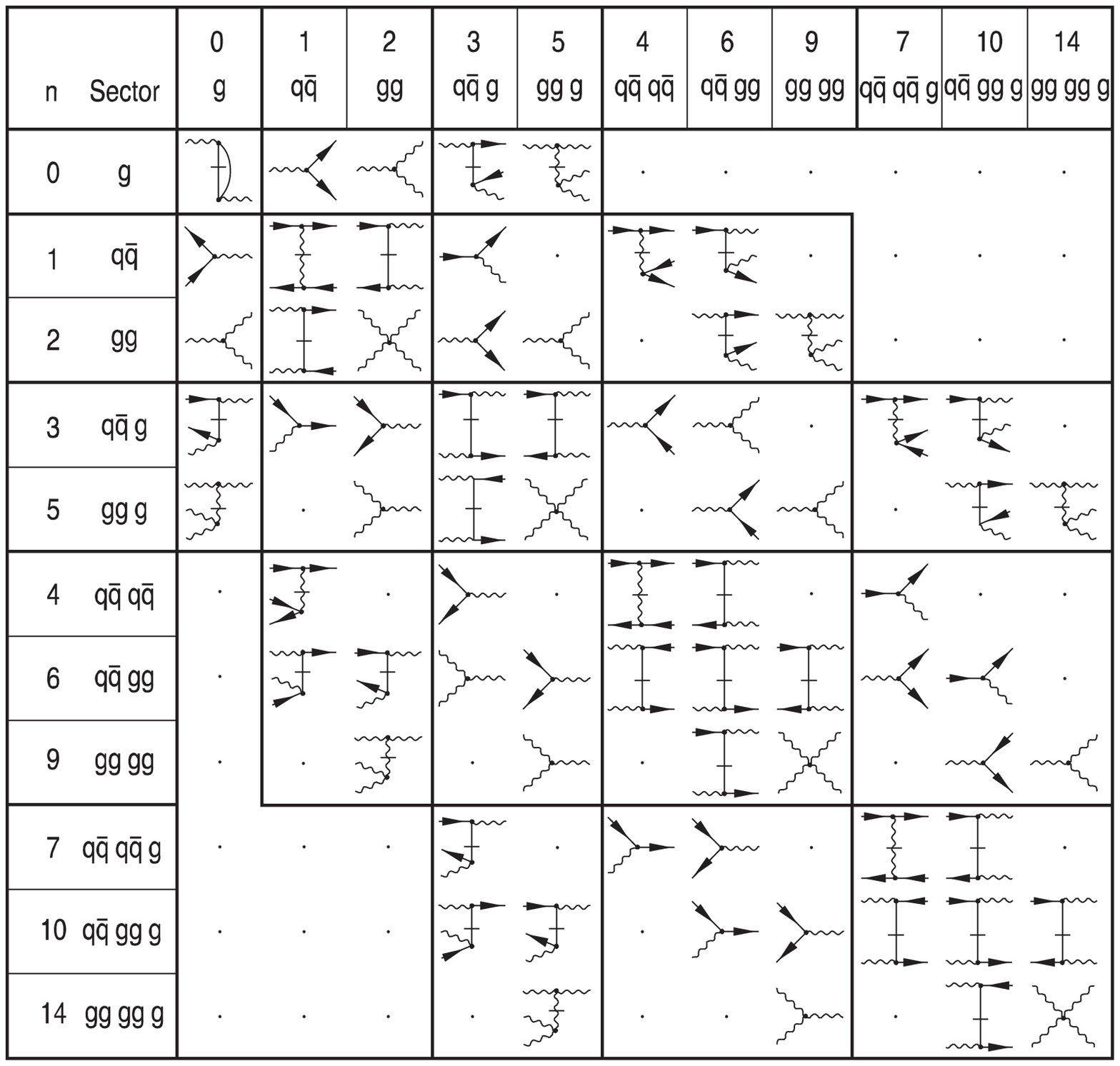}
} \caption{\label{fig-kyf-3}
    The Hamiltonian matrix for a meson. 
    Allowing for a maximum parton number 5, 
    the Fock space can be divided into 11 sectors. 
    Within each sector there are many Fock states 
    $ \vert \Phi _i \rangle $. 
     The matrix elements are represented by diagrams, which 
     are characteristic for each bloc.  
     Note that the figure mixes aspects of QCD 
     where the single gluon is absent and of QED which has no 
     three-photon vertices.
}  \end{figure}

It is natural to decompose the Fock space into sectors, labeled with
the number of quarks, antiquarks and gluons,
 $ N $,  $ \overline N $ and $ \widetilde N $, respectively. 
Mesons (or positronium) have total charge
$Q=0$, and thus $ N = \overline N $. These sectors can be 
arranged arbitrarily, and de-numerated differently.
A particular example was given in Figure~\ref{fig:holy-1}.
In Figure~\ref{fig-kyf-3}, the Fock-space sectors are arranged 
according to total particle number $ N + \overline N + \widetilde N $.

Since all components of the energy momentum commute with each other,
and since the space-like momenta are diagonal in momentum
representation, all Fock states must have the same {\it value} of
$ P ^+ = \sum _{\nu } p^+ _{\nu } $ and
$ \vec P  _{\!\bot}
  = \sum _{\nu } (\vec p _{\!\bot} )_{\nu } $,
with the sums running over all partons $\nu \in n_c$ in a particular
Fock-space class.
For any fixed $P  ^+$ and thus for any fixed resolution $K$, 
the number of Fock-space classes is finite.
As a consequence, the DLCQ-Hamiltonian matrix has a
{\it finite number of blocs}, as illustrated in Figure~\ref{fig-kyf-3}.
For the example, the maximum parton number is 5, 
corresponding to 11 sectors.
However, within each sector,  the number of Fock states is 
still unlimited and must be regularized (see below).

Since the Fock states are denumerable, one can associate
a non-diagonal {\it matrix} with the light-cone energy {\it operator}
$ P  ^- \big( a _q, a^\dagger _q,
        b _q, b^\dagger _q, d _q, d^\dagger _q \big) $  \cite {{brp91}}.
The matrix elements are strictly zero when the parton number
of rows and columns differs by more than two units.
From the outset, the light-cone energy matrix has a
tri-diagonal bloc structure, very similar to a
non-relativistic Hamiltonian with pair interaction,
see Figure~\ref{fig-kyf-1}.
When the parton number differs by two, the matrix elements
correspond to `fork interactions'.
The `vertex interaction' connects states which differ by one parton.
Finally `seagull interaction' conserves
parton number.
In Figure~\ref{fig-kyf-3}, these interactions are represented
by graphs. 
Two words of caution are in order:
(1) As always when dealing with light-cone
quantization \cite {brl89,brs73,leb80,lbh83}, 
these graphs are energy {\it not Feynman} diagrams.
The partons are `on-shell'.
The interaction conserves three- but not four-momentum.
(2) Figure~\ref{fig-kyf-3} refers to both QED and QCD.
In QCD, the single
gluon state is absent, since a gluon cannot be in a
color-singlet state. In QED, there are no three photon-vertices.

\textbf{Fock-space regularization.}
In an arbitrary frame,
each particle is on its mass-shell $ p^2 = m^2$.
Its four-momentum is $ p^\mu = (p^+, \vec p _{\!\bot}, p^- )$ 
with $p^- = (m^2 + \vec p ^{\,2}_{\!\bot}) / p^+ $.
For the free theory ($g=0$), the total four-momentum is
$ P ^\mu  _{free} = \sum _{\nu } p^\mu _{\nu } $
where the index $\nu$ runs over all particles in 
a particular Fock state $ | \Phi _i > $ of class $n_c$.
The index $i$ will be suppressed in the subsequent 
considerations. The components of the free four-momentum are thus
\begin {equation}
   P ^+ = \sum _{\nu \in n _c } \left( p ^+ \right) _\nu
     \,,\quad 
   \vec P _\bot = \sum _{\nu \in n _c } 
     \left( \vec p _{\!\bot} \right) _\nu
     \,,\quad 
   P ^- _{free} = \sum _{\nu \in n_c } \bigg(
     {m ^2 + \vec p ^{2} _{\!\bot} \over p ^+ } \bigg) _{\nu }
\,.  \label {eq:free-momenta} \end {equation}
For the space-like components  $ P ^k = P ^k _{free} $.
We now introduce the {\it intrinsic} momenta 
$x$ and $\vec k _{\!\bot}$ by
\begin {equation} 
      x _{\nu } = {p^+ _{\nu } \over P ^+}
      \,,\quad{\rm and}\quad
      ( \vec p _{\!\bot} )_{\nu }  = ( \vec k _{\!\bot} )_{\nu }
    + x_{\nu } \vec P  _{\!\bot} 
\,. \end {equation} 
The first two of the Eqs.(\ref{eq:free-momenta}) become the constraints
\begin {equation}
     \sum _{\nu } x_{\nu } = 1 
     \,,\qquad{\rm and}\quad
     \sum _{\nu } (\vec k _{\!\bot} )_{\nu } = 0 
\,,  \end {equation}
while the free invariant mass of each Fock state becomes
\begin {equation}
     M ^2 _{n_c} 
     = \sum _{\nu \in n_c } \bigg(
     {m ^2 + \vec k ^{\,2} _{\!\bot} \over x } \bigg) _{\nu }
\,. \label {eq:free-mass} \end {equation}
Note that the free invariant mass squared has a {\it minimum}
with respect to $ \vec k _{\!\bot} $ and $x$, {\it i.e.}
\begin {equation}
    \overline M  ^2_{n_c} 
  \equiv 
    {\rm min} \left( 
    \sum _{\nu \in n_c } \bigg(
    {m ^2 + \vec k ^{2} _{\!\bot} \over x } \bigg) _{\nu }
    \right)
    \simeq \sum _{\nu \in n_c } (m) _{\nu }
\,. \end {equation}

The free invariant mass-squared $ M ^2 $
plays the same role in DLCQ as the kinetic energy $ T $
in non-relativistic quantum mechanics, 
see section~\ref{sec:why-dis-mom}. 
The analogy can be used to regulate the Fock space:
A Fock state is admitted only when its kinetic energy 
is below a certain cut-off, {\it i.e.}
\begin {equation}
     \sum _{\nu \in n_c } \bigg(
     {m ^2 + \vec k ^{2} _{\!\bot} \over x } \bigg) _{\nu }
     \leq \Lambda ^2 _{n_c} 
\,. \label {eq:regularization} \end {equation}
Since only Lorentz scalars appear, this regularization 
is Lorentz- but not necessarily gauge invariant.
The constants $ \Lambda  _{n_c} $ with the dimension of a 
 $<\!\!mass\!\!>$  are at our disposal. 
DLCQ has an option for having as many 
`regularization parameters' as might be convenient. 
Three different conventions come to mind:
\begin {itemize} 
\item[(1)]
The {\it universal cut-off} $\Lambda _{n_c} \equiv \Lambda $
corresponds to Brodsky-Lepage
regularization \cite {{brl89},{brs73},{leb80},{lbh83}}.
It has the advantage to cut out 
Fock-space classes with many massive particles.
\item[(2)]
The {\it dynamic cut-off} 
$ \Lambda ^2 _{n_c} \equiv \overline M  ^2_{n_c} + \Lambda ^2 $
removes the `frozen mass' and truncates the momenta
in each state without acting like a sector regulator; 
\item[(3)]
The {\it sector-dependent dynamic cut-off} 
$ \Lambda ^2 _{n_c} \equiv \overline M  ^2_{n_c} 
+ (N + \overline N + \widetilde N ) \Lambda ^2 $
accounts for the number of particles in a Fock-state.
The maximum transversal momentum of a parton is then
approximately independent of the class.
\end {itemize} 
Other cut-offs have also been proposed \cite{phw90,wwh94}.

\subsection {Effective interactions in 3+1 dimensions}
\label {sec:tamm-dancoff}

Instead of an infinite set of coupled integral equations like in
Eq.(\ref{eq:4.17}), the eigenvalue equation 
$H \vert \Psi\rangle = E\vert\Psi\rangle$ 
leads in DLCQ to a strictly finite set of coupled matrix equations
\begin {equation} 
      \sum _{j=1} ^{N} 
      \ \langle i \vert H \vert j \rangle 
      \ \langle i \vert \Psi\rangle 
      = E\ \langle n \vert \Psi\rangle 
\, \qquad {\rm for\ all\ } i = 1,2,\dots,N 
\,.\label{eq:4.56}\end {equation} 
The rows and columns of the block matrices 
$\langle i \vert H \vert j \rangle $ are denumerated by the 
sector numbers $i,j=1,2,\dots N$, 
in accord with the  Fock-space sectors
in  Figures~\ref{fig:holy-1} or \ref{fig-kyf-3}.
Each sector contains many individual Fock states 
with different values of  $x$, $\vec p  _{\!\bot}$ and $\lambda$,
but due to Fock-space regularization ($\Lambda$), their
number is finite.

In principle one could proceed like in Sec.~\ref{sec:opo-gat}
for 1+1 dimension: One selects a particular value of the harmonic
resolution $K$  and the cut-off $\Lambda$, and diagonalizes
the finite dimensional Hamiltonian matrix by numerical methods.
But here then is the problem, the bottle neck of any field theoretic
Hamiltonian approach in physical space-time: 
The dimension of the Hamiltonian
matrix  increases exponentially fast with $\Lambda$. 
Suppose, the regularization procedure allows for 10 discrete
momentum states in each direction. A single particle has then 
about $10^3$ degrees of freedom.
A Fock-space sector with $n$ particles
has then roughly  $10^{n-1}$ different Fock states.
Sector 13 in Fig.~\ref{fig:holy-1} with its  8 particles has thus about  
$10^{21}$, and the $q\bar q$-sector about $10^3$ Fock states. 
One needs to develop effective interactions, which act in
smaller matrix spaces, and still are related to the full interaction.
Deriving an effective interaction can be understood as 
reducing the dimension in a matrix diagonalization problem 
from $10^{21}$ to say $10^{3}$!

Effective interactions are a well known tool in  many-body 
physics \cite{mof53}. In field theory
the method is known as the Tamm-Dancoff-approach.  
It was applied first  by Tamm \cite{tam45} and by Dancoff \cite{dan50} 
to Yukawa theory for describing the nucleon-nucleon interaction. 
For the front form, a considerable amount of work has been done thus far,
for instance by
Tang {\it et al.} \cite{tbp91},
Burkardt {\it et al.} \cite{bur95a,bur95b,bue96,bur96,buk97}, 
Fuda {\it et al.} \cite{fud87,fud90,fud91,fud92,fud94,fud95,fud96}, 
G{\l}azek {\it et al.} \cite{ghp93,wwh94}, 
Gubankova {\it et al.} \cite{guw97},
Hamer {\it et al.} \cite{thy95}, 
Heinzl {\it et al.} \cite{hew94,hei95,hei96a,hei96b,hei97}, 
Hiller {\it et al.} \cite{wih93}, 
Hollenberg {\it et al.} \cite{hhw91}, 
Jones {\it et al.} \cite{jop96a,jop96b},
Kalu\v za {\it et al.} \cite{kap92}, 
Kalloniatis {\it et al.} \cite{kar94,pnk96}, 
Krautg\"artner  {\it et al.} \cite{kpw92}, 
Prokhatilov {\it et al.} \cite{apf93}, 
Trittmann {\it et al.} \cite{trp96,trp97a,trp97b}, 
Wort  \cite{wor92}, 
Zhang  {\it et al.} \cite{zah93a,zah93b,zah93c}, 
and others 
\cite{brp96,dks95,gri92b,hye94,jic94,jhm95,kei94,kru93}, 
but the subject continues to be a challenge for QCD.
In particular one faces the problem of non-perturbative
renormalization, but progress is being made in recent work 
\cite{amm94,atw93,bur97,zha94}, 
particularly see the work by
Bakker {\it et al.} \cite{lib95a,lib95b,scb97}, 
Bassetto {\it et al.} \cite{acb94,bas93,bar93,bkk93,bas96}, 
Brisudova {\it et al.} \cite{brp95,brp96,bpw97}, as will
be discussed in Sec.~\ref{sec:renormalization}. 
 
Let us review it in short the general procedure \cite{mof53}
on which the Tamm-Dancoff approach \cite{dan50,tam45} is based.
The rows and columns of any Hamiltonian matrix can always be split
into two parts. One speaks of  the $ P $-space  
and of the rest, the $Q$-space $Q\equiv 1-P$.
The devision is arbitrary, but for to be specific let us identify
first the $P$-space with the $q\bar q$-space:
\begin {equation} 
      P = \vert 1 \rangle\langle 1 \vert 
      \quad{\rm and}\quad
      Q = \sum _{j=2} ^N  \vert j \rangle\langle j \vert 
\,.\label{eq:4.57}\end {equation}
Eq.(\ref{eq:4.56}) can then be rewritten conveniently 
as a $2\times2$ block matrix
\begin {equation} 
  \pmatrix{ \langle P \vert H \vert P \rangle 
          & \langle P \vert H \vert Q \rangle \cr 
            \langle Q \vert H \vert P \rangle 
          & \langle Q \vert H \vert Q \rangle \cr} 
  \ \pmatrix{ \langle P \vert\Psi \rangle \cr 
              \langle Q \vert\Psi \rangle \cr  } 
  = E  
  \ \pmatrix{ \langle P \vert\Psi \rangle \cr 
              \langle Q \vert\Psi \rangle \cr  }
 \,,\label{eq:4.58}\end {equation}
or  explicitly
\begin {eqnarray} 
   \langle P \vert H \vert P \rangle\ \langle P \vert\Psi\rangle 
 + \langle P \vert H \vert Q \rangle\ \langle Q \vert\Psi\rangle 
 &=& E \:\langle P \vert \Psi \rangle 
 \ ,  \label{eq:4.59}\\   {\rm and} \qquad
   \langle Q \vert H \vert P \rangle\ \langle P \vert\Psi\rangle 
 + \langle Q \vert H \vert Q \rangle\ \langle Q \vert\Psi\rangle 
 &=& E \:\langle Q \vert \Psi \rangle 
\,.\label{eq:4.60}\end {eqnarray}
Rewriting the second equation as
\begin {equation} 
      \langle Q \vert E  -  H \vert Q \rangle 
    \ \langle Q \vert \Psi \rangle 
  =   \langle Q \vert H \vert P \rangle 
    \ \langle P \vert\Psi\rangle 
,\label{eq:4.61}\end {equation}
one observes that the quadratic matrix 
$ \langle Q\vert E -  H \vert Q \rangle $ could be inverted 
to express the Q-space wavefunction $\langle Q \vert\Psi\rangle $
in terms of the $ P $-space wavefunction $\langle P \vert\Psi\rangle$. 
But the eigenvalue $ E $ is unknown at this point. 
To avoid that one solves first an other problem: One introduces
{\em the starting point energy} $\omega$ as a redundant
parameter  at disposal, and {\em defines the $ Q $-space resolvent} 
as the inverse of the block matrix 
$\langle Q \vert\omega- H \vert Q \rangle$,  
\begin {equation} 
     G _ Q (\omega)   =  
     {1\over\langle Q \vert\omega- H \vert Q \rangle} 
\,.\label{eq:4.62}\end{equation} 
In line with Eq.(\ref{eq:4.61}) one defines thus
\begin {equation} 
         \langle Q \vert \Psi \rangle 
  \equiv
         \langle Q \vert \Psi (\omega)\rangle 
  =    G _ Q (\omega) \langle Q \vert H \vert P \rangle 
         \,\langle P \vert\Psi\rangle 
\,,\label{eq:4.63}\end {equation} 
and inserts it into Eq.(\ref{eq:4.60}). 
This yields an eigenvalue equation 
\begin{equation} 
      H _{\rm eff} (\omega ) 
      \vert P\rangle\,\langle P \vert\Psi_k(\omega)\rangle =
      E _k (\omega )\,\vert\Psi _k (\omega ) \rangle 
\,,\label{eq:4.64}\end{equation} 
and defines unambiguously the effective interaction
\begin{equation} 
      \langle P \vert H _{\rm eff} (\omega) \vert P \rangle =
      \langle P \vert H \vert P \rangle +
      \langle P \vert H \vert Q \rangle 
      \:G _ Q (\omega) \:
      \langle Q \vert H \vert P \rangle 
\,.\label{eq:4.65}\end{equation} 
Both of them act only in the usually much smaller model space, the
$P$-space. The effective interaction is thus well defined: It is the 
original block matrix $\langle P\vert H\vert P\rangle$ 
plus a part where the system is scattered virtually into the 
$ Q $-space, propagating there by impact of the true interaction, 
and finally is scattered back into the $ P $-space: 
$\langle P \vert H \vert Q \rangle\:G _ Q (\omega) 
\:\langle Q \vert H \vert P \rangle$. Every numerical value of 
$\omega$ defines a different Hamiltonian and a different spectrum. 
Varying $\omega$ one generates a set of  
{\em energy functions} $ E _k(\omega) $. Whenever one finds 
a solution to the {\em fixpoint equation} \cite{pau81,zvb93}
\begin {equation}
      E _k (\omega ) = \omega 
\,,\label{eq:4.66}\end {equation}
one has found one of the true eigenvalues and
eigenfunctions of $H$,  by construction. 

It looks therefore as if one has mapped a difficult problem, the 
diagonalization of a large  matrix ($10^{21}$) onto a simpler problem,  
the diagonalization of a much smaller matrix  in the model space
($10^{3}$). But this true only in a restricted sense. 
One has to invert a matrix. The numerical inversion of a
matrix takes  about the same effort as its diagonalization. 
In addition,  one has to vary $\omega$ and solve the fixpoint 
equation (\ref{eq:4.66}).  The numerical work is thus rather 
larger than smaller as compared to a direct diagonalization. 
But the procedure is exact in principle. Particularly one can find 
all eigenvalues of the full Hamiltonian $H$,  
irrespective of how small one chooses the $ P $-space. 
Explicit examples for that can be found 
in \cite{pau81,pau96,zvb93}.

They key problem is how to get 
$\left(\langle Q\vert\omega-H\vert Q\rangle\right)^{-1}$,
the inversion of the Hamiltonian matrix in the Q-sector, 
as required by Eq.(\ref{eq:4.62}).
Once this is achieved, for example by an approximation, see below, 
the sparseness of the Hamiltonian matrix can be made use of
rather effectively: Only comparatively few block matrices 
$\langle P\vert H\vert Q\rangle$ differ from being strict zero 
matrices, see Figs.~\ref{fig:holy-1} or \ref{fig-kyf-3}.

In fact the sparseness of the Hamiltonian matrix can be made use
of even more effectively by introducing more than two projectors, as 
done in the method of  iterated resolvents \cite{pau93,pau96,pau96a}.
One easily recognizes  that Eq.(\ref{eq:4.58}) to Eq.(\ref{eq:4.65}) can 
be interpreted as the reduction  of the block matrix dimension  
from 2 to 1. But there is no need to identify the $P$-space with 
the lowest sector. One also can choose the $Q$-space identical 
with last sector and the $P$-space with the rest, $P=1-Q$: 
\begin{equation}
     P = \sum _{j=1} ^n  \vert j \rangle\langle j \vert\,,
     \quad{\rm with}\ 1\leq n\leq N\,, \quad{\rm and}\quad
     Q\equiv 1-P
\,.\end{equation}
The same steps as above reduce then the block matrix dimension 
from $N$ to $N-1$. The effective interaction acts in the now 
smaller space of $N-1$ sectors. This procedure can be repeated until one arrives 
at a block matrix dimension 1 where the procedure stops: 
The effective interaction in the Fock-space sector with only 
one quark and one antiquark is defined again unambiguously. 
More explicitly, suppose that in the course of this reduction one 
has arrived at  block matrix dimension $n$. Denote the corresponding 
effective interaction  $H_n (\omega)$. The eigenvalue problem 
corresponding to Eq.(\ref{eq:4.56}) reads then
\begin {equation} 
   \sum _{j=1} ^{n} \langle i \vert H _n (\omega)\vert j \rangle 
                      \langle j \vert\Psi  (\omega)\rangle 
   =  E (\omega)\ \langle i \vert\Psi (\omega)\rangle 
\ , \ \quad{\rm for}\ i=1,2,\dots,n
.\label{eq:4.70}\end {equation}
Observe that $i$ and $j$ refer here to sector numbers.  
Since one has started from the full Hamiltonian in the last 
sector, one has to convene that $H_{N}=H$. 
Now,  in analogy to Eqs.(\ref{eq:4.62}) and (\ref{eq:4.63}), 
define the resolvent of the effective sector Hamiltonian
$H_n(\omega)$ by
\begin {eqnarray} 
            G _ n (\omega)   
&=&   {1\over \langle n \vert\omega- H_n (\omega)\vert n \rangle} 
\,, \qquad {\rm and} \label{eq:4.71}\\  
            \langle n \vert \Psi (\omega)\rangle   
&=&   G _ n (\omega) 
   \sum _{j=1} ^{n-1} \langle n \vert H _n (\omega)\vert j \rangle 
   \ \langle j \vert \Psi (\omega) \rangle 
\,,\label{eq:4.72}\end {eqnarray}
respectively. The effective interaction 
in the  ($n -1$)-space becomes then \cite{pau93} 
\begin {equation}  
       H _{n -1} (\omega) =  H _n (\omega)
  +  H _n(\omega) G _ n  (\omega) H _n (\omega)
\label{eq:4.73}\end {equation}
for every block matrix  element 
$\langle i \vert H _{n-1}(\omega)\vert j \rangle$.  
To get the corresponding eigenvalue equation one
substitutes $n$ by $n-1$ everywhere in 
Eq.(\ref{eq:4.70}). Everything proceeds like in the above, 
including the fixed point equation  
$ E  (\omega ) = \omega $.
But one has achieved much more: Eq.(\ref{eq:4.73}) is a 
{\em recursion relation} which holds for all $1<n<N$!
Notice that the method of iterated resolvents requires only the 
inversion of the effective sector Hamiltonians 
$\langle n\vert H_n\vert n\rangle$. On a computer, this is an 
easier problem than the inversion of the full Q-space matrix
as in Eq.(\ref{eq:4.62}). Moreover, one can now make use of all 
zero block matrices in the Hamiltonian, as worked out in \cite{pau96}.

\begin{figure}
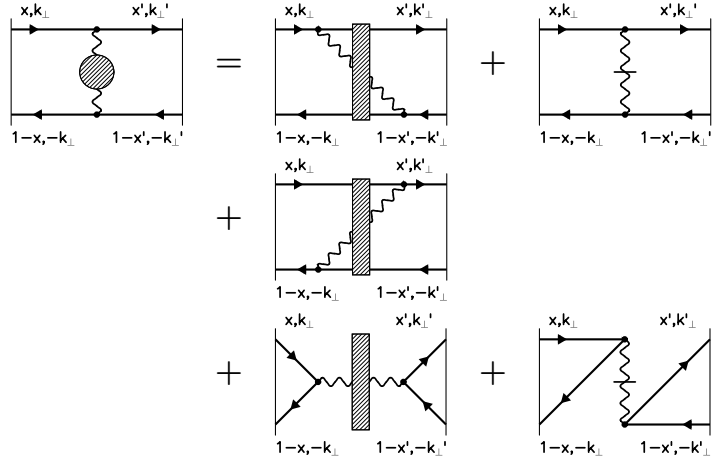

\centerline{\begin{tabular}{ccccccc}
\psfig{figure=all_effective.epsi,width=2.3cm,angle=-90}
&\hspace{0.0cm}\raisebox{1.0cm}[-1.4cm]{\bf =}&  
\psfig{figure=dyn_effective.epsi,width=2.3cm,angle=-90} 
&\raisebox{1.0cm}[-1.4cm]{\bf +} 
&\psfig{figure=seagull.epsi,width=2.3cm,angle=-90} \\
&\raisebox{1.0cm}[-1.4cm]{\bf +}&
\psfig{figure=dyn2_effective.epsi,width=2.3cm,angle=-90}
& &\\
&\raisebox{1.0cm}[-1.4cm]{\bf +}&
\psfig{figure=anni_effective.epsi,width=2.3cm,angle=-90}
&\raisebox{1.0cm}[-1.4cm]{\bf +}
& \psfig{figure=anni_seag.epsi,width=2.3cm,angle=-90}
\end{tabular}
}\caption{\label{fig:uwe_1}
     The graphs of the effective one-photon exchange interaction. 
     The effective interaction is a sum of the dynamic one-photon 
     exchange with both time orderings, 
     the instantaneous one-photon exchange, 
     the dynamic and the instantaneous annihilation interactions,
     all represented by energy graphs. 
     The hashed rectangles represent the effective photon or 
     the effective propagator $G_0$. 
     Taken from [427]
}\end{figure}

The Tamm-Dancoff approach (TDA) as used in the literature, 
however,  does not follow literally the 
outline given in Eqs.(\ref{eq:4.57}) to (\ref{eq:4.65}), 
rather one substitutes the `energy denominator'  in 
Eq.(\ref{eq:4.62}) according to
\begin{eqnarray}
       {1\over\langle Q\vert\omega-T-U\vert Q\rangle} &=&
       {1\over\langle Q\vert T^*-T-\delta U(\omega)\vert Q\rangle} 
       \Longrightarrow
       {1\over\langle Q\vert T^*-T\vert Q\rangle} 
\,,\label{eq:4.74}\\
       {\rm with}\quad \delta U(\omega) &=& \omega - T^* - U
\,.\nonumber\end{eqnarray}
Here, $T^*$ is not an operator but a $c$-number, denoting the 
mean kinetic energy in the P-space \cite{dan50,tam45}.
In fact, the two resolvents 
\begin {equation}
     G_Q(\omega)={1\over\langle Q\vert
     T^*-T-\delta U(\omega) \vert Q\rangle} 
     \quad{\rm and}\quad 
     G _0 =  {1\over \langle Q \vert T^*- T \vert Q \rangle} 
\,\end {equation} 
are identically related by
\begin{equation}  
     G_Q(\omega)=G _0+G_0\,\delta U(\omega) \,G_Q (\omega)   
\,\end {equation} 
or by the infinite series of perturbation theory
\begin{equation}  
    G _Q (\omega) =
    G _0 + G_0\,\delta U(\omega)\,G _0 
    +    G_0\,\delta U(\omega)\,G_0\,\delta U(\omega)\,G _0 
    + \dots 
\,\label{eq:4.69}\end {equation} 
The idea is that the operator $\delta U(\omega) $ in some sense 
is small, or at least that its mean value in the $Q$-space is 
close to zero, $\langle\delta U(\omega) \rangle\approx0$.
In such a case it is justified to restrict to the very first term in the
expansion, $    G _Q (\omega) =G _0$, as usually done in TDA.
Notice that the diagonal kinetic energy $T^*-T$ can be inverted 
trivially to get the resolvent $G_0$.

\subsection {Quantum Electrodynamics in 3+1 dimensions}
\label {sec:eff-int-pos}

Tang {\it et al.} \cite{tbp91} have first applied
DLCQ on the problem of quantum electrodynamics at strong 
coupling, followed later by Kalu\v za {\it et al.} \cite{kap92}. 
Both were addressing to get the positronium eigenvalue spectrum
as a test of the method. 
In either case the Fock space was truncated to include only the 
$q\bar q$ and $q\bar q\,g$ states. The so truncated DLCQ-matrix 
was  diagonalized numerically, with rather slow convergence of the 
results. Omitting the one-photon state 
$g$, they have excluded the impact of annihilation. Therefore, 
rather than `positronium', one should call such models 
`muonium with equal masses'. 
Langnau and Burkardt have calculated the anomalous
magnetic moment of the electron for very strong coupling
\cite{bul91a,bul91b,lab93a,lab93b}.
Krautg\"artner {\it et al.} \cite {kpw92,kap93b} proceed 
in a more general way, by using the effective interaction of the 
Tamm-Dancoff approach. A detailed analysis of the Coulomb 
singularity and its impact on numerical calculations in momentum 
representation has lead them to develop a Coulomb counter term 
technology, which did improve the rate  of numerical convergence 
significantly. It was possible now to reproduce quantitatively 
the Bohr aspects of the spectrum, as well as the fine and hyperfine 
structure.

One should emphasize that the aim of calculating the positronium
spectrum by a Hamiltonian eigenvalue equation is by no means a 
trivial problem. In the instant form, for example, the hyperfine interaction 
is so singular, that thus far the Hamiltonian eigenvalue equation 
has not been solved. The hyperfine corrections have only been 
calculated in the lowest non-trivial orders of perturbation theory,
see \cite{byg85}. One also notes, that the usual problems
with the recoil or the reduced mass are simply absent in a
momentum representation.

\begin{figure}
\unitlength1cm   
\begin{minipage}[t]{79mm}
\epsfxsize=78mm\epsfbox{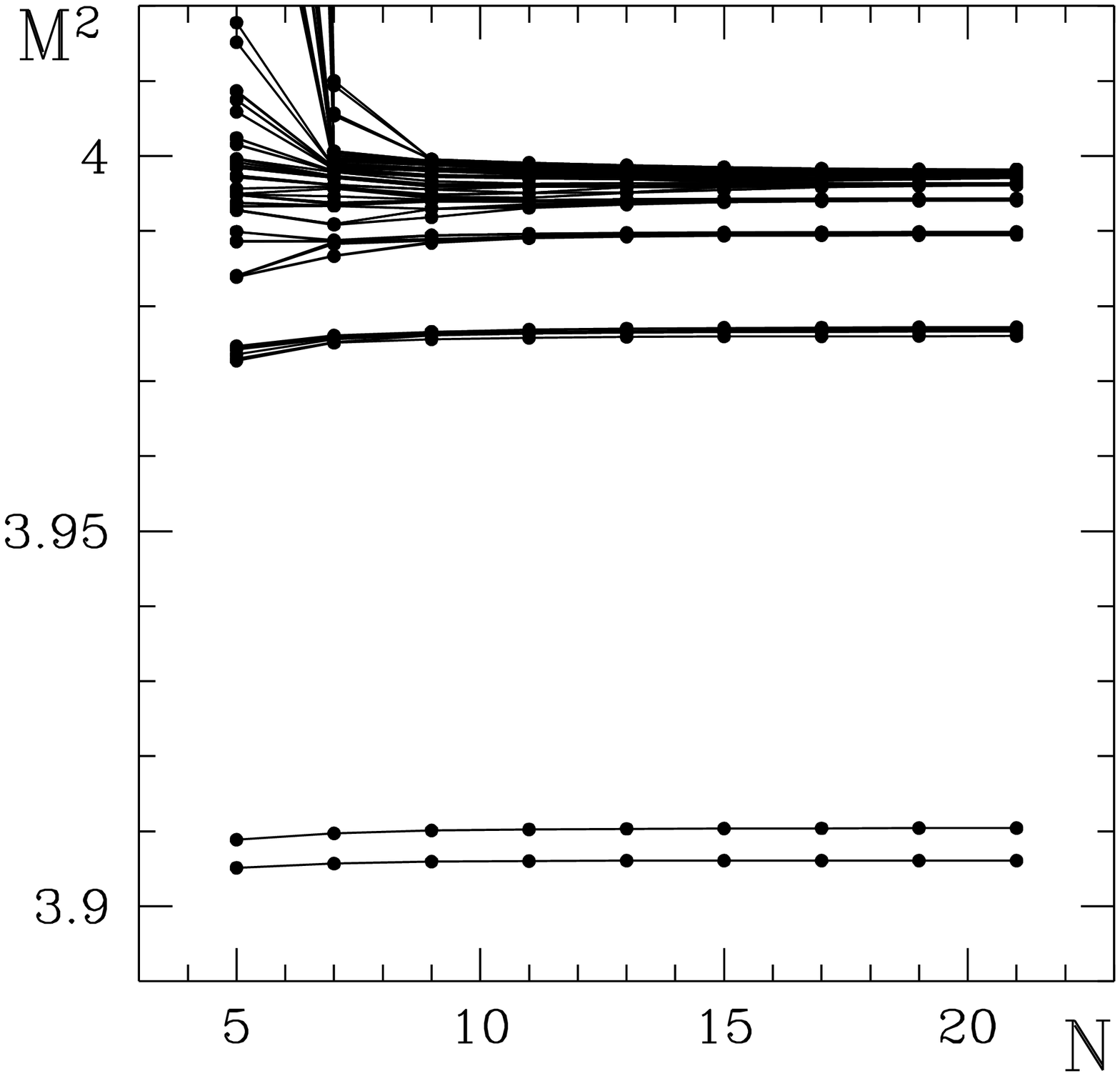}
\caption{\label{fig:uwe_2}
      Stability of positronium spectrum for $J_z=0$,  
      without the annihilation interaction. 
      Eigenvalues $M_i^2$ for $\alpha=0.3$ and $\Lambda=1$
      are plotted versus $N$, the number of integration Gaussian 
      points. Masses are in units of the electron mass.
      Taken from [426].
}\end{minipage}
\hfill
\begin{minipage}[t]{79mm}
\epsfxsize=78mm\epsfbox{Sin_fig11.epsi}
\caption{\label{fig:uwe_4} 
      The decrease of the $J_z=0$ singlet ground state 
      wavefunction with anti-parallel helicities as a function of the 
      momentum variable $\mu$ for $\alpha=0.3$ and 
      $\Lambda=1.0$. The six different curves correspond 
      to six values of $\theta$. Taken from [425].
}\end{minipage}
\end{figure} 

Although the Tamm-Dancoff approach was applied originally
in the instant form \cite{dan50,tam45}, one can translate it
easily into the front form. 
The approximation of Eqs.(\ref{eq:4.74}) and (\ref{eq:4.65})
give $G_Q(\omega)\approx G _0 $, and thus the virtual scattering
into the $Q$-space produces an additional $P$-space interaction,
the one-photon exchange interaction $VG _0 V$. Its  two time 
orderings are given diagrammatically in Figure~\ref{fig:uwe_1}.
The original $P$-space interaction is the kinetic energy, of course, 
plus the seagull interaction. Of the latter, we keep here only
the instantaneous-photon exchange and denote it as $W$, which is
represented by the first graph in Fig.~\ref{fig:uwe_1}.
Without the annihilation terms, the effective Hamiltonian is thus 
\begin{equation}
       H _{\rm eff} =  T + W + VG _0 V = T + U_{\rm eff}
\,.\end{equation}
The only difference is that the unperturbed energy is to be replaced 
by the mean kinetic energy $T^*$ as introduced in Eq.(\ref{eq:4.74}), 
which in the front form is given by
\begin{equation}
     T^* = {1\over 2}  \left( 
    {m_{q}^2 + \vec k_{\!\perp}^2 \over x } +
    {m_{\bar  q}^2 + \vec k_{\!\perp}^2 \over 1-x } +
    {m_{q}^2 + \vec k_{\!\perp}^{\prime\,2} \over x^\prime } +
    {m_{\bar  q}^2 + \vec k_{\!\perp}^{\prime\,2} \over 1-x^\prime } \right)
\,.\end{equation}
In correspondence to Eq.(\ref{eq:4s.7}), the energy denominator in the 
intermediate state of the $Q$-space
\begin {equation}
      T^*-T = -{Q^2\over \vert x-x^\prime\vert}
\end {equation}  
can now be expressed in terms $Q$, of the average four-momentum
transfer along the electron and the positron line, {\it i.e.}
\begin {equation}
        Q ^2=-{1\over 2} \left(
        (k_q-k _q^\prime)^2+(k_{\bar q} - k _{\bar q}^\prime)^2\right)
\,.\label{eq:4.80}\end {equation}  
As illustrated in Figure~\ref{fig:uwe_1}, the effective interaction 
$U_{\rm eff}$ scatters an electron with on-shell four-momentum 
$k_q$ and helicity $\lambda_q$ into a state with $k ^\prime_q$ 
and $\lambda_q^\prime$, and correspondingly the positron from
$k _{\bar q}$ and $\lambda_{\bar q}$ to
$k ^\prime_{\bar q}$ and $\lambda^\prime_{\bar q}$.
The evaluation of the so defined effective interaction
has been done explicitly in Sec.~\ref{sec:scattering}.

In the sequel we follow the more recent work of Trittmann {\it et al.} 
\cite{trp96,trp97a,trp97b}, where the Coulomb counter term 
technology was improved further to the extent that a calculation of all 
spin-parity multiplets of positronium was meaningful. In particular,
it was possible to investigate the important question to which extent
the members of the multiplets are numerically degenerate with $J_z$.
One recalls that the operator for the projection of total angular
momentum $J_z$ is kinematic in the front form, whereas total 
angular momentum $J^2$ is not, see Sec.~\ref{sec:symmetries}.

\begin{figure}
\unitlength1cm   \centering 
\begin{minipage}[t]{7cm}
\epsfxsize=70mm\epsfbox{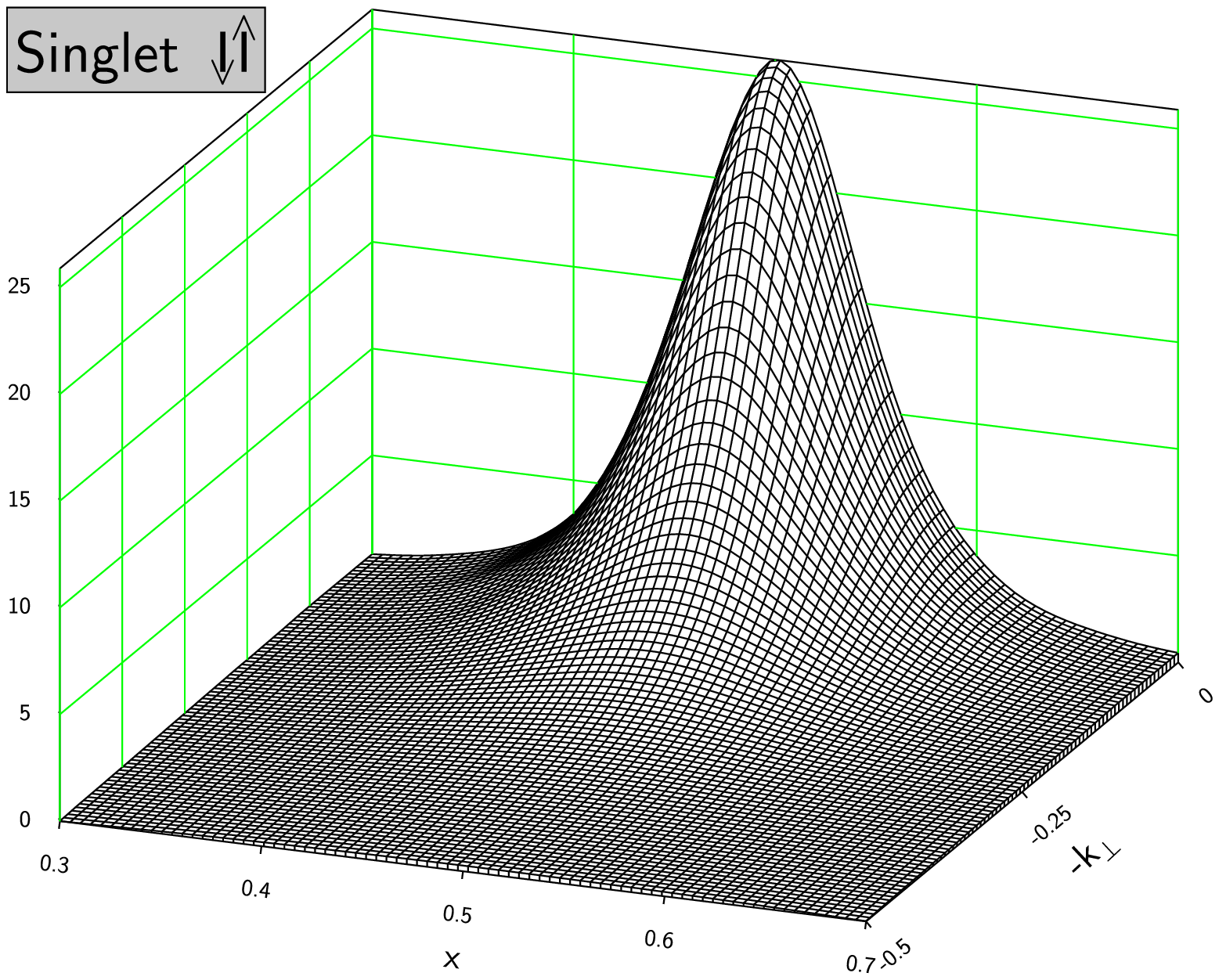}
\end{minipage}
\hfill
\begin{minipage}[t]{7cm}
\epsfxsize=70mm\epsfbox{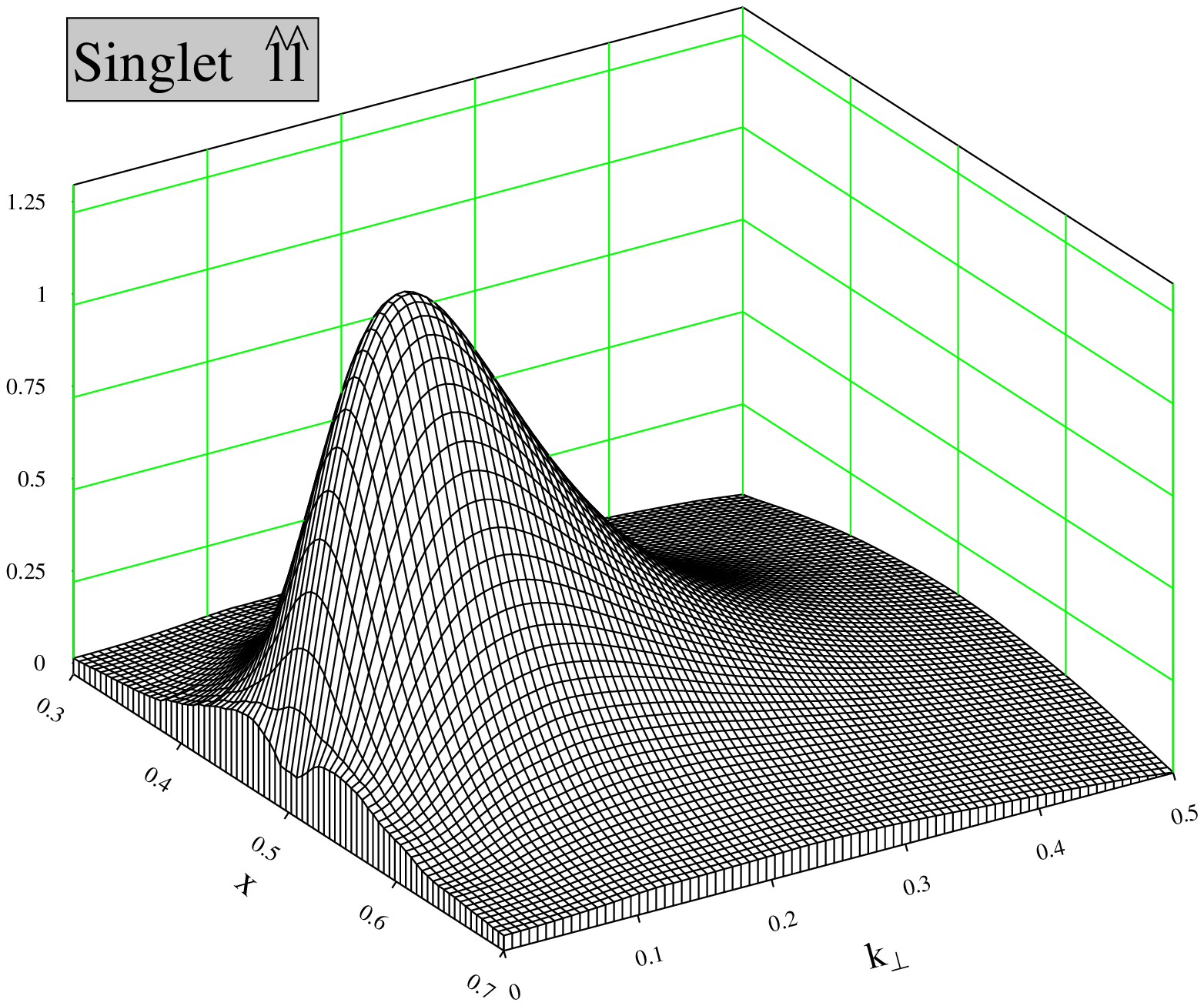}
\end{minipage}
\caption{\label{fig:mia_9}
    Singlet wavefunctions of positronium [426]. 
}\end{figure}

Up to this point it was convenient to work with DLCQ, and coupled matrix
equations. All spatial momenta $k^+$ and $\vec k _{\!\perp}$ 
are still discrete. But now that all the approximations have been done, 
one goes conveniently over to the {\em continuum limit} by converting
sums to integrals according to Eq.(\ref{eq:int_sum}).
In the continuum limit, the DLCQ-matrix equation is converted 
into an integral equation in momentum space, 
\begin{eqnarray} 
    M^2\langle x,\vec k_{\!\perp}; \lambda_{q},
    \lambda_{\bar q}  \vert \psi \rangle =
    \left[ {m_{q} + \vec k_{\!\perp}^2 \over x } +
    {m_{\bar  q} + \vec k_{\!\perp}^2 \over 1-x } \right]
    \langle x,\vec k_{\!\perp}; \lambda_{q},
    \lambda_{\bar q}  \vert \psi\rangle &&
\nonumber\\
    +\sum _{ \lambda_q^\prime,\lambda_{\bar q}^\prime}
    \!\int_D\!dx^\prime d^2 \vec k_{\!\perp}^\prime\,
    \langle x,\vec k_{\!\perp}; \lambda_{q}, \lambda_{\bar q}
    \vert U_{\rm eff}\vert x^\prime,\vec k_{\!\perp}^\prime; 
    \lambda_{q}^\prime, \lambda_{\bar q}^\prime\rangle\,
    \langle x^\prime,\vec k_{\!\perp}^\prime; 
    \lambda_{q}^\prime,\lambda_{\bar q}^\prime  
    \vert \psi\rangle &&
\,.\label{eq:4.81}\end{eqnarray}
The domain $D$ restricts integration in line with 
Fock-space regularization
\begin{equation}
        {m_{q}^2 +\vec k _{\!\bot}^{\,2}\over x } +
        {m_{\bar  q}^2 +\vec k _{\!\bot}^{\,2}\over 1-x }
        \leq (m_{q}+m_{\bar  q})^2 + \Lambda ^2 
\,.\label{eq:4.82}\end{equation}
The bras and kets refer to $q\bar q$ Fock states,
$\vert  x,\vec k_{\!\perp}; \lambda_{q},\lambda_{\bar  q}\rangle  =
b^\dagger(k_{q},\lambda_{q}) 
d^\dagger(k_{\bar  q},\lambda_{\bar  q}) \vert 0 \rangle$.
Goal of the calculation are the momentum-space
wavefunctions  $\langle x,\vec k_{\!\perp};  
\lambda_{q},\lambda_{\bar  q}\vert \psi\rangle $ and
the eigenvalues $M^2$.
The former are the probability amplitudes for finding the quark 
with helicity projection $\lambda_q$, longitudinal momentum 
fraction $x\equiv k_{q}^+/P^+$ and transversal momentum 
$ \vec k _{\!\bot}$, and simultaneously the antiquark with 
$\lambda_{\bar  q}$, $1-x$ and $-\vec k_{\!\bot}$.  
According to Eq.(\ref{eq:4s.15}), the effective interaction 
$U_{\rm eff} $ becomes
\begin {eqnarray} 
        U_{\rm eff} = - {1\over 4\pi^2}
        \,{\alpha\over Q  ^2} 
        \,{\left[ \overline u (k_q,\lambda_q) \,\gamma^\mu\,
        u(k_q^\prime,\lambda_q^\prime)\right] 
        \left[ \overline u (k_{\bar q},\lambda_{\bar q}) 
        \,\gamma_\mu\,
        u(k_{\bar q}^\prime,\lambda_{\bar q}^\prime)\right] 
        \over \sqrt{x(1 - x) x^\prime (1- x ^\prime)} } 
\,,\label{eq:4.83}\end {eqnarray}
with $\alpha \equiv g^2/4\pi$.
Notice that both the dynamic and the instantaneous 
one-photon exchange interaction in Eqs.(\ref{eq:4s.14}) 
and (\ref{eq:4s.14a}), respectively, contain a non-integrable 
singularity $\sim (x-x^\prime)^{-2}$, which cancel each other
in the final expressions, Eqs.(\ref{eq:4s.15}) or (\ref{eq:4.83}).
Only the square integrable `Coulomb singularity'
$1/Q^2$ remains, see also \cite{leb80}.

\begin{figure}
\centerline{
\epsfxsize=150mm\epsfbox{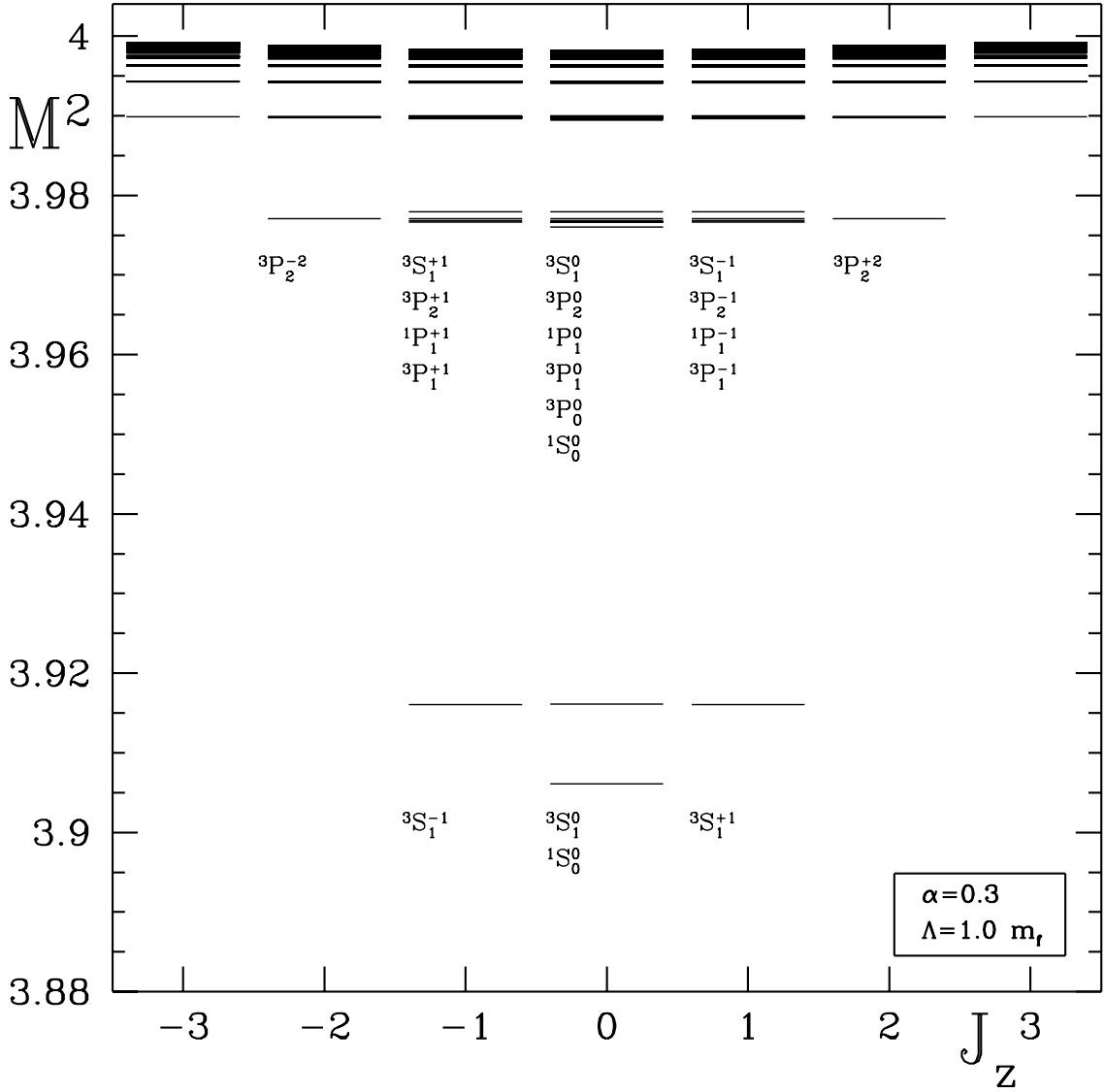}
}\caption{\label{fig:uwe_3} 
      Positronium spectrum for $-3\leq J_z\leq 3$, 
      $\alpha=0.3$ and $\Lambda=1$ including the 
      annihilation interaction. For an easier identification of
      the spin-parity multiplets, the corresponding non-relativistic 
      notation $^{3S+1} L _J ^{J_z}$ is inserted.
      Masses are given in units of the electron mass. 
      Taken from [427].
} \end{figure} 

In the numerical work \cite{kpw92,trp96} it is favorable to replace
the two transversal momenta $k_{\!\perp x}$ and $k_{\!\perp x}$ 
by the absolute value of $k_{\!\perp }$ and the angle $\phi$.
The integral equation is approximated by 
Gaussian quadratures, and the results are studied as a function 
of the number of integration points $N$,
as displayed in the left part of Figure~\ref{fig:uwe_2}.
One sees there, that the results stabilize themselves quickly.
All eigenvalues displayed have
the same eigenvalue of total angular momentum projection,
{\it i.e.}  $J_z=0$.
Since one calculates the values of an invariant mass squared,
a comparative large value of the fine structure constant
$\alpha = 0.3$ has been chosen. One recognizes
the ionization threshold at $M^2 \sim 4 m^2$,
the Bohr spectrum, and even more important, the fine structure.
The two lowest eigenvalues correspond to the singlet
and triplet state of positronium, respectively. The agreement
is quantitative, particularly for the physical value
of the fine structure constant $\alpha={1\over 137}$.
In order to verify this agreement, one needs a relative
numerical accuracy of roughly $10^{-11}$.
The numerical stability and precision is remarkable, indeed.
The stability with respect to the cut-off
$\Lambda $ has been studied also.

An inspection of the numerical wavefunctions
$\psi  ( x  , {\vec k}_{\!\bot} \,)$, 
as displayed for example in Figure~\ref{fig:mia_9}, 
reveals that they are strongly peaked around $ {\vec k}_{\!\bot} \sim 0$
and $ x \sim {1\over 2} $. Outside the region 
\begin {equation} 
     {\vec k}_{\!\bot} ^{\ 2} \ll  m^2 
     \,, \qquad{\rm and}\quad 
     ( x -{1\over2} )^2 \ll 1
\,, \label {eq:4.84}\end {equation}
they are smaller than the peak value by many orders of magnitude. 
Also, the singlet wave function with anti-parallel helicities is dominant
with more than a factor 20 over the component with parallel helicities.
The latter would be zero in a non-relativistic calculation.
Relativistic effects are responsible also that the 
singlet-($\uparrow\downarrow$) wavefunction is not rotationally 
symmetric. To see that is is plotted in Fig.~\ref{fig:uwe_4}
versus the of shell momentum variable $\mu$, defined by
\cite{kar81,saw85,saw86}
\begin{eqnarray}
      x &=& \frac{1}{2}\left(1 + \frac{\mu\cos\theta}{\sqrt{m^2+\mu^2}}\right)
\,,\label {eq:4.85}\\
      \vec{k}_{\perp} &=& (\mu\sin\theta\cos\varphi,\mu\sin\theta\sin\varphi)
\,,\label{eq:4.86}\end{eqnarray}
for different values of $\theta$. The numerically significant deviation,
however, occurs only for the very relativistic momenta $\mu\ge 10m$.

Trittmann {\it et al.} \cite{trp96,trp97a,trp97b} 
have also included the annihilation interaction
as illustrated in Fig.\ref{fig:uwe_1} and calculated numerically 
the spectrum for various values of $J_z$. 
The results are compiled in Figure~\ref{fig:uwe_3}. 
As one can see there, certain eigenvalues at $J_z=0$ are degenerate
to a numerically very high degree of freedom with certain mass
eigenvalues at other $J_z$. 
Consider the second lowest eigenvalue for  $J_z=0$. 
It is degenerate with the lowest eigenvalue for $J_z=\pm 1$,
and can thus be classified as a member of the triplet with $J=1$.
Correspondingly, the lowest eigenvalue for $J_z=0$ having
no companion can be classified as the singlet state with $J=0$.
Quite in general one can interpret degenerate multiplets
as members of a state with total angular momentum
$J=2J_{z,max}+1$. An inspection of the wavefunctions allows
to conclude whether helicity parallel or anti parallel 
is the leading component. In a pragmatical sense, one thus 
can conclude on the `total spin' $S$, and on `total orbital
angular momentum' $L$, although in the front form 
neither $J$, nor $S$ or $L$ 
make sense as operator eigenvalues.
In fact, they are not, see Sec.~\ref{sec:symmetries}.
But in this way one can make contact with the conventional 
classification scheme $^{3S+1} L _J ^{J_z}$, as inserted in the figure.
It is remarkable, than one finds {\em all states} which one 
expects \cite{trp97b}.

\subsection {The Coulomb interaction in the front form}
\label {sec:coul-pot}

The $j ^\mu j _\mu$-term in Eq.(\ref{eq:4.83}) represents 
retardation and mediates the
fine and hyperfine interactions. One can switch them off by
substituting the momenta by the equilibrium values, 
\begin{equation}
        \overline k_{\!\perp}=0, \quad{\rm and}\quad 
        \overline x = {m_q\over m_q+m_{\bar q}}
\,,\label{eq:4.87}\end{equation}
which gives by means of Table~\ref{tab:uudir}: 
\begin{equation}
        \left[ \overline u (k_q,\lambda_q) \,\gamma^\mu\,
        u(k_q^\prime,\lambda_q^\prime)\right] 
        \left[ \overline u (k_{\bar q},\lambda_{\bar q}) 
        \,\gamma_\mu\,
        u(k_{\bar q}^\prime,\lambda_{\bar q}^\prime)\right] 
        \Longrightarrow (m_q+m_{\bar q})^2
        \,\delta_{\lambda_q,\lambda_q^\prime}
        \,\delta_{\lambda_{\bar q},\lambda_{\bar q}^\prime}
\,.\label{eq:4.88}\end{equation}
The effective interaction in Eq.(\ref{eq:4.83}) simplifies 
correspondingly and becomes the Coulomb interaction in 
front form:
\begin {eqnarray} 
        U_{\rm eff} = - {1\over 4\pi^2}
        \,{\alpha\over Q  ^2} 
        \,{(m_q+m_{\bar q})^2\over\sqrt{x(1 - x)x^\prime(1-x^\prime)} } 
\,.\label{eq:4.89}\end {eqnarray}
To see that one performs a variable transformation from $x$ to 
$k_z(x)$. The inverse transformation \cite{pam95} 
\begin {equation} 
     x = x (k _z) = {k_z+E_1 \over E_1+E_2}
, \quad{\rm with} \quad
     E_i=\sqrt{m_i^2+\vec k_{\!\bot}^{\, 2}+\vec k_z^{\, 2}}
, \quad i=1,2
\,,\end {equation} 
maps the domain of integration $-\infty \leq k_z \leq \infty$ 
into the domain $0\leq x \leq 1$, and produces 
the equilibrium value for $k_z=0$, Eq.(\ref{eq:4.87}). 
One can combine $k_z$ and $\vec k_{\!\perp}$ into a 
three-vector $\vec k = (\vec k_{\!\perp},k_z)$.
By means of the identity
\begin {equation}
      x(1-x)={(E_1+k_z)(E_2-k_z)\over (E_1+E _2)^2}
\,,\end {equation}
the Jacobian of the transformation becomes straightforwardly
\begin {equation}
      { dx ^\prime \over \sqrt{x(1-x)\, x'(1-x')} } 
      = dk _z\:\left({1\over E_1}+{1\over E_2}\right)
      \:\sqrt { {( E _ 1 + k _z ' )( E _ 2 - k _z ' )
      \over ( E _1 + k _z   )( E _ 2 - k _z   ) } }
\,. \end {equation}
For equal masses $m_1=m_2=m$ (positronium), the kinetic
energy is
\begin {equation} 
     { m^2 + {\vec k} _{\!\bot} ^{\ 2} 
     \over x (1- x )} = 4 m^2 + 4 \vec k ^{\,2}  
\,, \end {equation} 
and the domain of integration Eq.(\ref{eq:4.82}) reduces to
$4\vec k ^{\,2} \leq \Lambda^2$. 
The momentum scale $\mu$ \cite{kar81,saw85,saw86}, 
as introduced in Eq.(\ref{eq:4.86}), identifies its self as
$\mu = 2\vert\vec k\vert$. As shown by \cite{pam95}, the
four-momentum transfer Eq.(\ref{eq:4.80}) can be exactly 
rewritten as
\begin {equation} 
     Q^2 = ({\vec k} - {\vec k}^{\,\prime})^2 
\,. \end {equation} 
Finally, after substituting the invariant mass squared eigenvalue
$M^2$ by an energy eigenvalue $E$, 
\begin {equation}
     M ^2 = 4 m^2 + 4 mE 
\,, \end {equation} 
and introducing a new wavefunction $\phi$,  
\begin {equation}
       \phi(\vec k) = 
      \langle x(k_z),\vec k_{\!\perp}; \lambda_{q},
      \lambda_{\bar q}  \vert \psi\rangle
      \:{1\over m}\sqrt{m^2 + \vec k_{\!\perp}^{\,2}}
\,,\label{eq:4.95}\end {equation} 
one rewrites Eq.(\ref{eq:4.81}) with Eq.(\ref{eq:4.89}) identically as
\begin {equation}
       \left(E -  {\vec k^{\,2}\over 2 m_r} \right)\,\phi(\vec k) 
       = - {\alpha\over 2\pi^2}\,{m\over\sqrt{m^2+ \vec k^{\,2} } }
      \int_ D  d^3 \vec k^{\,\prime} \;
      {1\over (\vec k - \vec k^{\,\prime} ) ^2}  \,\phi(\vec k^{\,\prime}) 
\,.\label{eq:4.96}\end {equation}
Since $ m _r  = m/2 $ is the reduced mass, this is the 
non-relativistic Schr\"o\-dinger equation in momentum 
representation for $k^2\ll m^2$, see also \cite{pam95}.

Notice that only retardation was suppressed to get this result. 
The impact of the relativistic treatment in the front form settles 
in the factor $(1+ k^2/m^2)^{-{1\over2}}$.
It induces a weak non-locality in the front form Coulomb potential.
Notice also that the solution of Eq.(\ref{eq:4.96}) is rotationally
symmetric for the lowest state. Therefore, the original front
form wavefunction $\langle x(k_z),\vec k_{\!\perp}\vert\psi\rangle$
in Eq.(\ref{eq:4.95}) cannot be rotationally symmetric. 
The deviations from rotational symmetry, however, are small and 
can occur only for $k^2\gg m^2$, as can be  
observed in Fig.~\ref{fig:uwe_4}.

%% file: 06Nucle.tex
\section{The Impact on Hadronic  Physics}
\label{sec:impact}
\setcounter{equation}{0}

In this chapter we discuss a number of novel applications of Quantum
Chromodynamics to nuclear structure and dynamics, such as the
reduced amplitude formalism for exclusive nuclear amplitudes. We
particularly emphasize the importance of light-cone Hamiltonian and
Fock State methods as a tool for describing the wavefunctions of
composite relativistic many-body systems and their interactions. We
also show that the use of covariant kinematics leads to nontrivial
corrections to the standard formulae for the axial, magnetic, and
quadrupole moments of nucleons and nuclei.

In principle, quantum chromodynamics can provide a  fundamental
description of hadron and nuclei structure and  dynamics in terms of
elementary quark and gluon degrees of freedom. In practice, the
direct application of QCD to hadron and nuclear phenomena is
extremely complex because of the interplay of non perturbative
effects such as color confinement and multi-quark coherence. Despite
these challenging theoretical difficulties, there  has been
substantial progress in identifying specific QCD effects in nuclear
physics. A crucial tool in these analyses is the use of relativistic
light-cone quantum mechanics and Fock state methods in order to
provide a tractable and consistent treatment of relativistic
many-body effects. In some applications, such as exclusive processes
at large momentum transfer, one can make first-principle predictions
using factorization theorems which separate  hard perturbative
dynamics from the non perturbative physics associated with hadron or
nuclear binding. In other applications, such as the passage of
hadrons through nuclear matter and the calculation of the axial,
magnetic, and  quadrupole moments of light nuclei, the QCD
description provides new insights which go well beyond the usual
assumptions of traditional nuclear physics.

\subsection{Light-Cone Methods in QCD}

In recent years  quantization of quantum
chromodynamics at fixed light-cone time $\tau=t-z/c$
has emerged as a promising method for solving
relativistic bound-state problems in the strong coupling regime
including nuclear systems.Light-cone quantization has a number of unique
features that make it appealing, most notably, the ground state of the
free theory is also a ground state of the full theory, and the Fock
expansion constructed on this vacuum state provides a complete relativistic
many-particle basis for diagonalizing the full theory. The
light-cone wavefunctions $\psi_n(x_i, k_{\perp i}, \lambda_i)$,
which describe the hadrons and nuclei in terms of their fundamental
quark and gluon degrees of freedom, are frame-independent. The
essential variables are the  boost-invariant light-cone momentum
fractions $x_i=p_i^+/P^+$, where $P^\mu $ and $p^\mu _i$ are the
hadron and quark or gluon momenta, respectively, with $P^\pm=P^0 \pm
P^z$. The internal transverse momentum variables $\vec k_{\perp i}$
are given by $\vec k_{\perp i}=\vec p_{\perp i}-x_i \vec P_\perp$
with the constraints $\sum \vec k_{\perp i}=0$ and $\sum x_i=1$.
{\it i.e.} , the light-cone momentum fractions $x_i$ and $
\vec k_{\perp i}$ are relative coordinates, and they describe the hadronic
system independent of its total four momentum $p^\mu.$
The entire spectrum of hadrons and nuclei and their scattering
states is given by the set of  eigenstates of the light-cone
Hamiltonian $H_{LC}$ of QCD. The Heisenberg problem takes the form:
\begin{equation}
H_{LC} \vert \Psi \rangle = M^2 \vert \Psi \rangle.
\end{equation}
For example, each hadron has the eigenfunction  $\vert \Psi_H
\rangle$ of $H_{LC}^{QCD}$ with eigenvalue $M^2=M^2_H.$ If we could
solve the light-cone Heisenberg problem for the proton in QCD, we
could then expand its eigenstate on the complete set of quark
and gluon eigensolutions  $\vert n \rangle = \vert uud \rangle ,
\vert uudg \rangle \cdots$ of the  free Hamiltonian $H_{LC}^{0}$
with the same global quantum numbers:
\begin{equation}
\vert \Psi_p\rangle
= \sum_n \vert n \rangle \psi_n(x_i, k_{\perp_i}, \lambda_i).
\end{equation}
The $\psi_n$ $(n = 3, 4, \ldots)$ are first-quantized amplitudes
analogous to the Schr\"odinger wave function, but it is Lorentz-frame
independent.  Particle number is generally not conserved in a
relativistic quantum field theory. Thus each eigenstate is
represented as a sum over Fock states of arbitrary particle number
and in QCD each hadron is expanded as second-quantized sums over
fluctuations of color-singlet quark and gluon states of different
momenta and number. The coefficients of these fluctuations are the
light-cone wavefunctions $\psi_n(x_i, k_{\perp i}, \lambda_i).$ The
invariant mass ${\cal M}$ of the partons in a given Fock state can
be written in the elegant form  ${\cal M}^2 = \sum_{i=1}^3 {\vec
k_{\perp i}^2+m^2\over x_i}$. The dominant configurations in the
wave function are generally those with minimum values of ${\cal
M}^2.$ Note that except for the case $m_i=0$ and $\vec k_{\perp
i}=\vec 0$, the limit $x_i \to 0$ is an ultraviolet limit;  {\it i.e.} \ it
corresponds to particles moving with infinite momentum in the
negative $z$ direction: $k^z_i \to - k^0_i \to - \infty.$

In the case of QCD in one space and one time dimensions,  the
application of discretized light-cone quantization \cite{brp91},
see Section~\ref{sec:DLCQ}, provides complete solutions 
of the theory,
including the entire spectrum of mesons, baryons, and nuclei, 
and their wavefunctions \cite{hbp90}. 
In the DLCQ method, one simply
diagonalizes the light-cone Hamiltonian for QCD on a discretized
Fock state basis. The DLCQ solutions can be obtained for arbitrary
parameters including the number of flavors and colors and quark
masses.  More recently, DLCQ has been applied to new variants of
QCD(1+1) with quarks in the adjoint representation, thus obtaining
color-singlet eigenstates analogous to gluonium states \cite{dak93}.

The DLCQ method becomes much more numerically intense when applied
to physical theories in $3 + 1$ dimensions; however, progress is
being made.  An analysis of the spectrum and light-cone
wavefunctions of  positronium in QED(3+1) is given in Ref.\cite{kpw92}.
Currently, Hiller, Okamoto, and Brodsky \cite{hbo95}\ are pursuing
a non perturbative calculation of the lepton anomalous moment in QED
using this method. Burkardt has recently solved scalar theories with
transverse dimensions by combining a Monte Carlo lattice method with
DLCQ \cite{bur94}.

Given the light-cone wavefunctions, $\psi_{n/H} (x_i, \vec
k_\perp{}_i, \lambda_i),$ one can compute virtually any  hadronic
quantity by convolution with the appropriate quark and gluon matrix
elements. For example,  the leading-twist structure functions
measured in deep inelastic lepton scattering are immediately related
to the light-cone probability distributions:
\begin{equation}
2M\,F_1(x,Q) = {F_2(x,Q)\over x}\approx \sum_a e_a^2\,G_{a/p}(x,Q)
\label {CAK} 
\end{equation}
where
\begin{equation}
  G_{a/p}(x,Q) = \sum_{n,\lambda_i} \int
   \overline{\prod_i}\ {dx_i d^2\vec k_\perp{}_i \over 16\pi^3}\,
  |\psi_{n}^{(Q)}(x_i,\vec k_\perp{}_i,\lambda_i)|^2 
   \sum_{b=a} \delta(x_b-x)
\label {CAL} 
\end{equation}
is the number density of partons of type $a$ with longitudinal
momentum fraction $x$ in the proton. This follows from the
observation that deep inelastic lepton scattering in the
Bjorken-scaling limit occurs if $x_{bj}$  matches the light-cone
fraction of the struck quark. (The $\sum_b$ is over all partons of
type $a$ in state $n$.) However, the light cone wavefunctions
contain much more information for the final state of deep inelastic
scattering, such as the multi-parton distributions, spin and flavor
correlations, and the spectator jet composition.
 
\begin{figure}\centerline{
\epsfysize=60mm\epsfbox{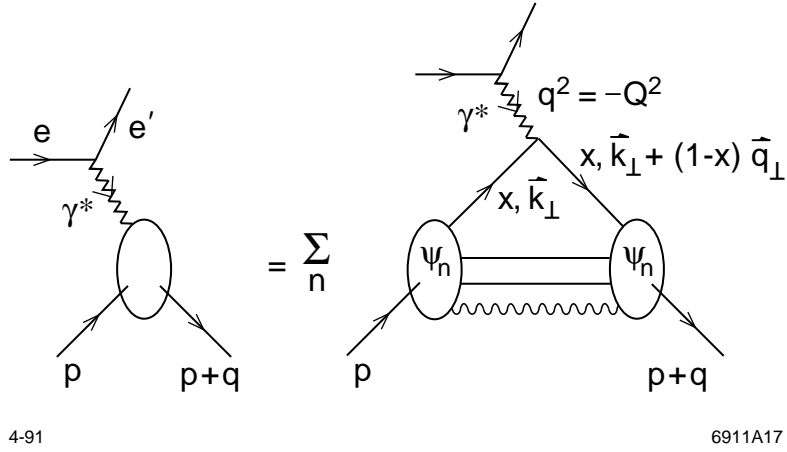} 
}\caption{\label{fig:LLF}
Calculation of the form factor of a bound state from the convolution
of light-cone Fock amplitudes. The result is exact if one sums over
all $\psi_n$.
}\end{figure}

As was first shown by Drell and Yan \cite{dry70},  it is
advantageous to  choose a coordinate frame where $q^+=0$ 
to compute form factors $F_i(q^2)$, structure functions, 
and other current matrix elements at space-like photon momentum. 
With such a choice the quark current cannot create 
or annihilate pairs, and 
$\left\langle{p' \vert j^{+}\vert p}\right\rangle$ 
can be computed as a simple overlap of Fock
space wavefunctions; all off-diagonal terms involving pair
production or annihilation by the current or vacuum vanish. In the
interaction picture one can equate the full Heisenberg current to
the quark current  described by the free Hamiltonian at $\tau=0.$
Accordingly, the form factor is easily expressed in terms of the
pion's light cone wavefunctions  by examining the $\mu = +$
component of this equation in a frame where the photon's momentum is
transverse to the incident pion momentum, with $\vec
q_\perp^{\,2}=Q^2=-q^2.$ The space-like form factor is then just a
sum of overlap integrals analogous to the corresponding
non-relativistic formula: \cite{dry70}  (See Fig. \ref{fig:LLF}.~)
\begin{equation}
   F (q^2) =
   \sum_{n,\lambda_i} \sum_a e_a \int
   \overline{\prod_i} \ {dx_i\,d^2\vec k_\perp{}_i \over 16\pi^3}
   \,\psi_{n}^{(\Lambda)*}(x_i,\vec \ell_\perp{}_i,\lambda_i)
   \,\psi_{n}^{(\Lambda)}(x_i,\vec k_\perp{}_i,\lambda_i) .
\label {CAI} 
\end{equation}
Here $e_a$ is the charge of the struck quark, $\Lambda^2\gg\vec
q_\perp^{\,2}$, and
\begin{equation}
  \vec \ell_\perp{}_i \equiv \cases{
  \vec k_\perp{}_i - x_i \vec q_\perp 
+ \vec q_\perp  & \hbox{for the struck quark} \cr
  \vec k_\perp{}_i - x_i \vec q_\perp        
& \hbox{for all other partons.}\cr}
\label {CAJ} 
\end{equation}
Notice that the transverse momenta appearing as arguments of the
first wavefunctions correspond not to the actual momenta carried by
the partons but to the actual momenta minus $x_i\vec q_\perp$, to
account for the motion of the final hadron. Notice also that $\vec
\ell_\perp$ and $\vec k_\perp$ become equal as $\vec
q_\perp\rightarrow 0$, and that $F_\pi\rightarrow 1$ in this limit
due to wavefunctions normalization. All of the various form factors
of hadrons with spin can be obtained by computing the matrix element
of the plus current between states of different initial and final
hadron helicities \cite{bjd65}.
 
\begin{figure}\centerline{
\epsfbox{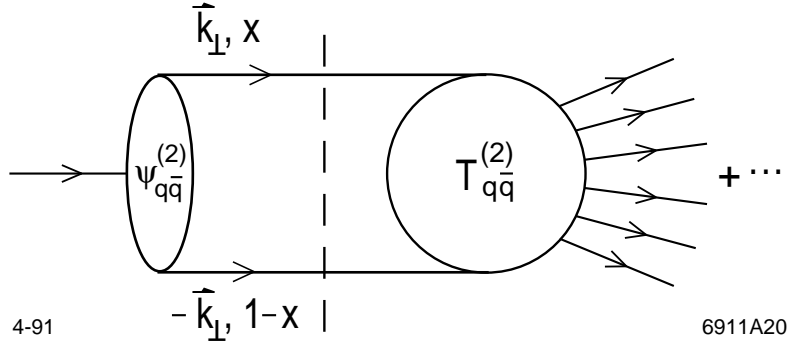} 
}\caption{\label{fig:LLI} 
Calculation of hadronic amplitudes in the light-cone Fock formalism.
}\end{figure}

As we have emphasized above, in principle, the light-cone
wavefunctions determine all properties of a hadron.  The general
rule for calculating an amplitude involving the wavefunctions
$\psi_n^{(\Lambda)}$, describing Fock state $n$ in a hadron with
$\underline P = (P+,\overrightarrow P_\perp)$, has the form
\cite{brl89} (see Fig. \ref{fig:LLI}):
\begin{equation}
  \sum_{\lambda_i} \int \overline{\prod_i}\ 
   {dx_i d^2\vec k_\perp{}_i\over\sqrt{x_i} 16\pi^3}\,
   \psi_{n}^{(\Lambda)}(x_i,\vec k_\perp{}_i,\lambda_i)\ 
    T_n^{(\Lambda)}
  (x_i P^+,x_i\overrightarrow P_\perp + \vec k_\perp{}_i,\lambda_i)
\label {CAF} 
\end{equation}
where $T^{(\Lambda)}_n$ is the irreducible scattering amplitude in
LCPTh with the hadron replaced by Fock state $n$. If only the
valence wavefunctions is to be used, $T_n^{(\Lambda)}$ is irreducible
with respect to the valence Fock state only; {\it e.g.}\
$T_n^{(\Lambda)}$ for a pion has no $q\bar q$ intermediate states.
Otherwise contributions from all Fock states must be summed, and
$T_n^{(\Lambda)}$ is completely irreducible.
 
The leptonic decay of the $\pi^\pm$ is one of the simplest processes
to compute since it involves only the $q\bar q$ Fock state. The sole
contribution to $\pi^-$ decay is from
\begin{eqnarray}
& &\left\langle{0\,\left|\, \overline\psi_u \gamma^+ (1-\gamma_5)
   \psi_d\,\right|\,\pi^-}\right\rangle =
            -\sqrt{2} P^+ f_\pi  \nonumber \\
& & =
   \int {dx\,d^2\vec k_\perp \over 16\pi^3}\ 
   \psi^{(\Lambda)}_{d\bar u}(x,\vec k_\perp)
   {\sqrt{n_c} \over \sqrt{2}} \Bigg\{
   { \bar{v}_\downarrow \over \sqrt{1-x}} \,\gamma^+ (1-\gamma_5)\,
   {u_\uparrow \over \sqrt{x}}\ 
    +\ (\uparrow\leftrightarrow\downarrow) \Bigg\}
\label {CAG} 
\end{eqnarray}
where $n_c = 3$ is the number of colors, $f_\pi \approx
\hbox{93~MeV}$, and where only the $L_z=S_z=0$ component of the
general $q\bar q$ wave function contributes. Thus we have
\begin{equation}
  \int {dx\,d^2\vec k_\perp \over 16\pi^3}\, 
  \psi^{(\Lambda)}_{d\bar u}(x,\vec k_\perp)
  = {f_\pi \over 2\sqrt{3}}.
\label {pinorm}
\end{equation}
This result must be independent of the ultraviolet cutoff $\Lambda$
of the theory provided $\Lambda$ is large compared with typical
hadronic scales. This equation is an important constraint upon the
normalization of the $d\bar u$ wave function. It also shows that
there is a finite probability for finding a $\pi^-$ in a pure $d\bar
u$ Fock state.
 
The fact that a hadron can have a non-zero projection on a Fock
state of fixed particle number seems to conflict with the notion
that bound states in QCD have an infinitely recurring parton
substructure, both from the infrared region (from soft gluons) and
the ultraviolet regime (from QCD evolution to high momentum). In
fact, there is no conflict. Because of coherent color-screening in
the color-singlet hadrons, the infrared gluons with wavelength
longer than the hadron size decouple from the hadron wave function.
 
The question of parton substructure is related to the resolution
scale or ultraviolet cut-off of the theory. Any renormalizable
theory must be defined by imposing an ultraviolet cutoff $\Lambda$
on the momenta occurring in theory. The  scale $\Lambda$ is usually
chosen to be much larger than the physical scales $\mu$ of interest;
however it is usually more useful to choose a smaller value for
$\Lambda$, but at the expense of introducing new higher-twist terms
in an effective Lagrangian: \cite{let87}~ 
\begin{equation} 
{\cal L}^{(\Lambda)} =
{\cal L}^{(\Lambda)}_0(\alpha_s(\Lambda), m(\Lambda)) +
\sum^N_{n=1}\left(1\over\Lambda\right)^n\delta
{\cal L}^{(\Lambda)}_n(\alpha_s(\Lambda),m(\Lambda)) +
{\cal O}\left(1\over\Lambda\right)^{N+1}
\end{equation}
where
\begin{equation} 
{\cal L}^{(\Lambda)}_0 = 
        -{1\over4}\, F^{(\Lambda)}_{a\mu\nu}
         F_n^{(\Lambda)a\mu\nu}+\overline\psi^{(\Lambda)}
         \left[i\,\not D^{(\Lambda)}-m(\Lambda)\right] 
         \psi^{(\Lambda)} \ .
\end{equation}
The neglected physics of parton momenta and substructure beyond the
cutoff scale has the effect of renormalizing the values of the input
coupling  constant $g(\Lambda^2)$ and the input mass parameter
$m(\Lambda^2)$ of the quark partons in the Lagrangian.
 
One clearly should choose $\Lambda$ large enough to avoid large
contributions from the higher-twist terms in the effective
Lagrangian, but small enough so that the Fock space domain is
minimized. Thus if $\Lambda$ is chosen of order 5 to 10 times the
typical QCD momentum scale, then it is reasonable to hope that the
mass, magnetic moment and other low momentum properties of the
hadron could  be well-described on a Fock basis of limited size.
Furthermore, by iterating the equations of motion, one can construct
a relativistic Schr\"odinger equation with an effective potential
acting on the valence lowest-particle number state wave function
\cite{leb79a,leb79b}.  Such a picture would explain the apparent success
of constituent quark models for explaining the hadronic spectrum and
low energy properties of hadron.
 
It should be emphasized that infinitely-growing  parton content of
hadrons due to the evolution of the deep inelastic structure
functions at increasing momentum transfer, is associated with the
renormalization group substructure of the quarks themselves, rather
than the ``intrinsic" structure of the bound state wave function
\cite{brs90,brs91}.  The fact that the light-cone kinetic energy
$\left\langle {\vec k_\perp^{\,2}+m^2\over x}\right\rangle$ of the
constituents in the bound state is bounded by $\Lambda^2$ excludes
singular behavior of the Fock wavefunctions at $x \to 0.$ There are
several examples where the light-cone Fock structure of the bound
state solutions is known. In the case of the super-renormalizable
gauge theory, $QED(1+1),$ the probability of having non-valence
states in the light-cone expansion of the lowest lying meson and
baryon eigenstates to be less than $10^{-3}$, even at very strong
coupling \cite{hbp90}.  In the case of QED(3+1), the lowest state of
positronium can be well described on a light-cone basis with two to
four particles, $\left| {e^+ e^-}\right\rangle,~ \left|{e^+ e^-
\gamma}\right\rangle,~ \left|{e^+ e^- \gamma \gamma}\right\rangle,$
and $\left|{e^+ e^- e^+ e^-}\right\rangle;$  in particular, the
description of the Lamb-shift in positronium requires the coupling
of the system to light-cone Fock states with two photons ``in
flight" in light-cone gauge. The ultraviolet cut-off scale $\Lambda$
only needs to be taken large compared to the electron mass. On the
other hand, a charged particle such as the electron does not have a
finite Fock decomposition, unless one imposes an artificial infrared
cut-off.
 
We thus expect that a limited light-cone Fock basis should be
sufficient to represent bound color-singlet states of heavy quarks
in QCD(3+1) because of the coherent color cancelations and the
suppressed amplitude for transversely-polarized gluon emission by
heavy quarks. However, the description of light hadrons is
undoubtedly much more complex due to the likely influence of chiral
symmetry breaking and zero-mode gluons in the light-cone vacuum. We
return to this problem later.
 
Even without solving the QCD light-cone equations of motion, we can
anticipate some general features of the behavior of the light-cone
wavefunctions. Each Fock component describes a system of free
particles with kinematic invariant mass squared:
\begin{equation}
{\cal M}^2 = \sum^n_i {\vec k_\perp^{\,2}{}_i + m^2_i\over x_i} ,
\end{equation}
On general dynamical grounds, we can expect that states with very
high ${\cal M}^2$ are suppressed in physical hadrons, with the
highest mass configurations computable from perturbative
considerations. We also note that $\ell n~ x_i = \ell n~
{(k^0+k^z)_i\over (P^0+P^z)} = y_i - y_P$ is the rapidity difference
between the constituent with light-cone fraction $x_i$ and the
rapidity of the hadron itself. Since correlations between particles
rarely extend over two units of rapidity in hadron physics, this
argues that constituents which are correlated with the hadron's
quantum numbers are primarily found with $x > 0.2 .$
 
The limit  $x \to  0$ is normally an ultraviolet limit in a
light-cone wave function. Recall, that in any  Lorentz frame, the
light-cone fraction is $x = k^+/p^+ =(k^0+k^z)/(P^0+P^z).$ Thus in a
frame where the bound state is moving infinitely fast in the
positive $z$ direction (``the infinite momentum frame"), the
light-cone fraction becomes the momentum fraction $x \to k^z/p^z.$
However, in the rest frame $\overrightarrow P = \overrightarrow 0,$
$x = (k^0 + k^z)/M.$  Thus  $x \to 0$ generally implies very large
constituent momentum $k^z \to - k^0 \to -\infty$ in the rest frame;
it is excluded by the ultraviolet regulation of the theory ---unless
the particle has strictly zero mass and transverse momentum.

If a particle has non-relativistic momentum in the bound state, then
we can identify $k^z \sim x M - m.$ This correspondence is useful
when one matches physics at the relativistic/non-relativistic
interface. In fact, any non-relativistic solution to the
Schr\"o\-din\-ger equation can be immediately written in light-cone form
by identifying the two forms of coordinates. For example, the
Schr\"odin\-ger solution for particles bound in a harmonic
oscillator potential can be taken as a model for the light-cone wave function
for quarks in a confining linear potential: \cite{leb80}~
\begin{equation}
\psi(x_i, \vec k_\perp{}_i)=A \exp (-b {\cal M}^2)
= \exp -\left( b \sum^n_i {k^2_{\perp i}+ m^2_i\over x_i}\right)\ .
\end{equation} 
This form exhibits the strong fall-off at large relative transverse
momentum and at the $x \to 0$ and $x \to 1$ endpoints expected for
soft non-perturbative solutions in QCD. The perturbative corrections
due to hard gluon exchange give amplitudes suppressed only by power
laws and thus will eventually  dominate wave function behavior over
the soft contributions in these regions. This {\it ansatz} is the
central assumption required to derive dimensional counting
perturbative QCD predictions for exclusive processes at large
momentum transfer and the $x \to 1$ behavior of deep inelastic
structure functions. A review is given in Ref. \cite{brl89}. A model
for the polarized and unpolarized gluon distributions in the proton
which takes into account both perturbative QCD constraints at large
$x$ and coherent cancelations at low $x$ and small transverse
momentum is given in Ref. \cite{brs90,brs91}.

The light-cone approach to QCD has immediate application to nuclear
systems: The formalism provides a covariant many-body description of
nuclear systems formally similar to non-relativistic many-body theory.

One can derive rigorous predictions for the leading power-law
fall-off of nuclear amplitudes, including the nucleon-nucleon
potential, the deuteron form factor, and the distributions of
nucleons within nuclei at large momentum fraction.  For example, the
leading electromagnetic form factor of the deuteron falls  as
$F_d(Q^2) = f(\alpha_s(Q^2))/(Q^2)^5$, where, asymptotically,
$f(\alpha_s(Q^2)) \propto \alpha_s(Q^2)^{5+\gamma}.$  The leading
anomalous dimension $\gamma$ is computed in Ref. \cite{bjl83}.

In general the six-quark Fock state of the deuteron is a mixture of
five different color-singlet states. The dominant color
configuration of the six quarks corresponds to the usual
proton-neutron bound state. However, as $Q^2$ increases, the
deuteron form factor becomes sensitive to  deuteron wave function
configurations where all six quarks overlap within an impact
separation $b^{\perp i} < {\cal O} (1/Q).$ In the asymptotic domain,
all five Fock color-singlet components acquire equal weight;  {\it i.e.} ,
the deuteron wave function becomes 80\%\ ``hidden color" at short
distances. The derivation of  the evolution equation for the
deuteron distribution amplitude is given in
Refs. \cite{bjl83,jib86}.

QCD predicts that Fock components of a hadron with a small color
dipole moment can pass through nuclear matter without
interactions \cite{bbg81,brm88}, 
see also \cite{mil97}.  Thus in the case
of large momentum transfer reactions where only small-size valence
Fock state configurations enter the hard scattering amplitude, both
the initial and final state interactions of the hadron states become
negligible. There is now evidence for QCD ``color transparency" in
exclusive virtual photon $\rho$ production for both nuclear coherent
and incoherent reactions in the E665 experiment at
Fermilab  \cite{fan93}.
as well as the original measurement at BNL in quasi-elastic $p
p$ scattering in nuclei  \cite{hep90}. The recent NE18
measurement of quasi-elastic electron-proton scattering  at SLAC
finds results which do not clearly distinguish between conventional
Glauber theory predictions and PQCD color
transparency  \cite{mae94}.

In contrast to color transparency, Fock states with large-scale
color configurations strongly interact with high particle number
production  \cite{bbf93}.

The traditional nuclear physics assumption that the nuclear form
factor factorizes in the form
$F_A(Q^2)= \sum_N F_N(Q^2) F^{\rm body}_{N/A}(Q^2)$,
where $F_N(Q^2)$  is the on-shell nucleon form factor is in general
incorrect.  The struck nucleon is necessarily off-shell, since it
must transmit momentum to align the  spectator nucleons along the
direction of the recoiling nucleus.

Nuclear form factors and scattering amplitudes can be factored in
the form given by the reduced amplitude formalism  \cite{brc76},
which follows from the cluster decomposition of the
nucleus in the limit of zero nuclear binding. The reduced form
factor formalism takes into account the fact that each nucleon in an
exclusive nuclear transition typically absorbs momentum $Q_N \simeq
Q/N.$  Tests of this formalism are discussed in a later section.

The use of covariant kinematics leads to a number of striking
conclusions for the electromagnetic and weak moments of nucleons and
nuclei. For example, magnetic moments cannot be written as the naive
sum $\mu = \sum \mu_i$ of the magnetic moments of the constituents,
except in the non-relativistic limit where the radius of the bound
state is much larger than its Compton scale: $R_A M_A \gg 1$. The
deuteron quadrupole moment is in general nonzero even if the
nucleon-nucleon bound state has no D-wave component  \cite{brh83}.
Such effects are due to the fact that even ``static"
moments have to be computed as transitions between states of
different momentum $p^\mu$ and $p^\mu + q^\mu$ with $q^\mu \to 0$.
Thus one must construct current matrix elements between boosted
states. The Wigner boost generates nontrivial corrections to the
current interactions of bound systems  \cite{brp60}.

One can also use light-cone methods to show that the proton's
magnetic moment $\mu_p$ and its axial-vector coupling $g_A$  have a
relationship independent of the assumed form of the light-cone
wave function  \cite{brs94}.  At the physical value of
the proton radius computed from the slope of the Dirac form factor,
$R_1=0.76$ fm, one obtains the experimental values for both $\mu_p$
and $g_A$;  the helicity carried by the valence $u$ and $d$ quarks
are each reduced by a factor  $\simeq 0.75$ relative to their
non-relativistic values. At infinitely small radius  $R_p M_p \to
0$, $\mu_p$ becomes equal to the Dirac moment, as demanded by the
Drell-Hearn-Gerasimov sum rule \cite{ger65,drh66}.
Another surprising fact is that as $R_1 \to 0,$ the
constituent quark helicities become completely disoriented and $g_A
\to 0$. We discuss these features in more detail in the following
section.

In the case of the deuteron, both the quadrupole and magnetic
moments become equal to that of an elementary vector boson in the
the Standard Model in the limit $M_d R_d \to 0.$ The three form
factors of the deuteron have the same ratio as that of the $W$ boson
in the Standard Model  \cite{brh83}.

The basic amplitude controlling the nuclear force, the
nucleon-nucleon scattering amplitude can be systematically analyzed
in QCD in terms of basic quark and gluon scattering subprocesses.
The high momentum transfer behavior of the amplitude from
dimensional counting is ${\cal M}_{pp \rightarrow pp} \simeq f_{pp
\rightarrow pp}(t/s)/t^4$ at fixed center of mass angle. A review is
given in Ref.\cite{brl89}.  The fundamental subprocesses,
including pinch contributions  \cite{lan74}, can be classified
as arising from both quark interchange and gluon exchange
contributions. In the case of meson-nucleon scattering, the quark
exchange graphs  \cite{bbg73}  can explain virtually
all of the observed features of large momentum transfer fixed CM
angle scattering distributions and ratios  \cite{car93}. The
connection between Regge behavior and fixed angle scattering in
perturbative QCD for quark exchange reactions is discussed in
Ref. \cite{brt93}.  Sotiropoulos and Sterman  \cite{sos94} have shown how
one can consistently interpolate from fixed angle scaling behavior to  the
$1/t^8$ scaling behavior of the elastic cross section in the $ s \gg -t$,
large $-t$ regime.

One of the most striking anomalies in elastic proton-proton
scattering  is the large  spin correlation $A_{NN}$ observed at
large angles  \cite{kri92}.  At $\sqrt s \simeq 5 $ GeV, the rate
for scattering with incident proton spins parallel and normal to the
scattering plane is four times larger than scattering with
anti-parallel polarization. This phenomena in elastic $p p$
scattering can be explained as the effect due to the onset of charm
production in the intermediate state at this energy  \cite{brt88}.
The intermediate state $\vert u u d u u d c \bar c
\rangle$ has odd intrinsic parity and  couples to the $J=S=1$
initial state, thus strongly enhancing scattering when the incident
projectile and target protons have their spins parallel and normal
to the scattering plane.

The simplest form of the nuclear force is the interaction between
two heavy quarkonium states, such as the $\Upsilon (b \bar b)$ and
the $J/\psi(c \bar c)$. Since there are no valence quarks in common,
the dominant color-singlet interaction arises simply  from the
exchange of two or more gluons, the analog of the van der Waals
molecular force in QED. In principle, one could measure the
interactions of such systems by producing pairs of quarkonia in high
energy hadron collisions. The same fundamental QCD van der Waals
potential also dominates the interactions of heavy quarkonia with
ordinary hadrons and nuclei. As shown in Ref.  \cite{lms92},
the small size of the $Q \bar Q$ bound state relative
to the much larger hadron sizes allows a systematic expansion of the
gluonic potential using the operator product potential.  The matrix
elements of multigluon exchange in the quarkonium state can be computed
from non-relativistic heavy quark theory. The coupling of the
scalar part of the interaction to  large-size hadrons is rigorously
normalized to the mass of the state via the trace anomaly. This
attractive potential dominates the interactions at low relative
velocity. In this way one establishes that the nuclear force between
heavy quarkonia and ordinary nuclei is attractive and sufficiently
strong to produce nuclear-bound quarkonium  \cite{brt90}.

\subsection{Moments of Nucleons and Nuclei in the Light-Cone Formalism}

Let us consider an effective three-quark light-cone Fock description
of the nucleon in which additional degrees of freedom (including
zero modes) are parameterized in an effective
potential.  After  truncation, one could in principle
obtain the mass $M$ and light-cone wave function of the three-quark
bound-states by solving the Hamiltonian eigenvalue problem. It is
reasonable to assume  that adding more quark and gluonic excitations
will only refine this initial approximation  \cite{phw90}.
In such a theory the constituent quarks will also
acquire effective masses and form factors. However, even without
explicit solutions, one knows that the helicity and flavor structure
of the baryon eigenfunctions will reflect the assumed global SU(6)
symmetry and Lorentz invariance of the theory. Since we do not have
an explicit representation for the effective potential in the
light-cone Hamiltonian $H^{\rm effective}_{\rm LC}$ for
three-quarks, we shall  proceed by making an ansatz for the momentum
space structure of the wave function $\Psi.$ As we will show below,
for a given size of the proton, the predictions and interrelations
between observables at $Q^2=0,$ such as the proton magnetic moment
$\mu_p$ and its axial coupling $g_A,$ turn out to be essentially
independent of the shape of the wave function  \cite{brs94}.

The light-cone model given in
Ref. \cite{sch93a,sch93b,sch93a}  provides a framework
for representing the general structure of the effective three-quark
wavefunctions for baryons. The wave function $\Psi$ is constructed as
the product of a momentum wave function, which is spherically
symmetric and invariant under permutations, and a spin-isospin wave
function, which is uniquely determined by SU(6)-symmetry
requirements.  A Wigner--Melosh \cite{wig39,mel74}
rotation is applied to the spinors, so that the wave function of the
proton is an eigenfunction of $J$ and $J_z$ in its rest
frame \cite{cop82,bmp93}.
To represent the range of uncertainty
in the possible form of the momentum wave function, we shall choose
two simple functions of the invariant mass ${\cal M}$ of the quarks:
\begin{eqnarray}
\psi_{\rm H.O.}({\cal M}^2 )=N_{\rm
H.O.}\exp(-{\cal M}^2/2\beta^2),
\psi_{\rm Power}({\cal M}^2) =N_{\rm Power}
(1+{\cal M}^2/\beta^2)^{-p}  
\end{eqnarray}
where $\beta$ sets the characteristic internal momentum scale.
Perturbative QCD predicts a nominal power-law fall off at large
$k_\perp$ corresponding to
$p=3.5$  \cite{leb80,sch93a,sch93b,sch93c,sch94}.  The
Melosh rotation insures that the nucleon has $j={ 1 \over 2}$ in its rest
system.  It has the matrix representation  \cite{mel74} 
\begin{equation}
R_M(x_i,k_{\perp i},m)={m+x_i {\cal M}-i\vec \sigma\cdot(\vec
n\times \vec  k_i)\over\sqrt{(m+x_i {\cal M})^2+\vec k_{\perp i}^2}
}\end{equation}
with $\vec  n=(0,0,1)$, and it becomes the unit matrix if the
quarks are collinear $ R_M(x_i,0,m)=1.$ Thus the internal transverse
momentum dependence of the light-cone wavefunctions also affects its
helicity structure  \cite{brp60}.

The Dirac and Pauli form factors $F_1(Q^2)$ and $F_2(Q^2)$ of the
nucleons are given by the spin-conserving and the spin-flip vector
current $J^+_V$ matrix elements ($Q^2=-q^2$)  \cite{brt88}
\begin{eqnarray}
F_1(Q^2) = \bigl< p+q,\uparrow | J^+_V |
p,\uparrow \bigr> , 
(Q_1-i Q_2) F_2(Q^2) = -2M bigl<
p+q,\uparrow | J^+_V | p, \downarrow
\bigr> \ . 
\end{eqnarray}
We then can calculate the anomalous magnetic moment $a=\lim_{Q^2\to
0} F_2(Q^2)$.  [The total proton magnetic moment is  $\mu_p = {e
\over 2M}(1+a_p).$] The same parameters as
in Ref. \cite{sch93b} are
chosen; namely $m=0.263$ GeV (0.26 GeV) for the up- and down-quark
masses, and $\beta=0.607$ GeV (0.55 GeV) for $\psi_{\rm Power}$
($\psi_{\rm H.O.}$) and $p=3.5$. The quark currents are taken as
elementary currents with Dirac moments ${e_q \over 2 m_q}.$ All of
the baryon moments are well-fit if one takes the strange quark mass
as 0.38 GeV. With the above values, the proton magnetic moment is
2.81 nuclear magnetons, the neutron magnetic moment is $-1.66$
nuclear magnetons. (The neutron value can be improved by relaxing
the assumption of isospin symmetry.) The radius of the proton is
0.76 fm;  {\it i.e.} , $M_p R_1=3.63$.

\begin{figure}\centerline{
\epsfysize=80mm\epsfbox{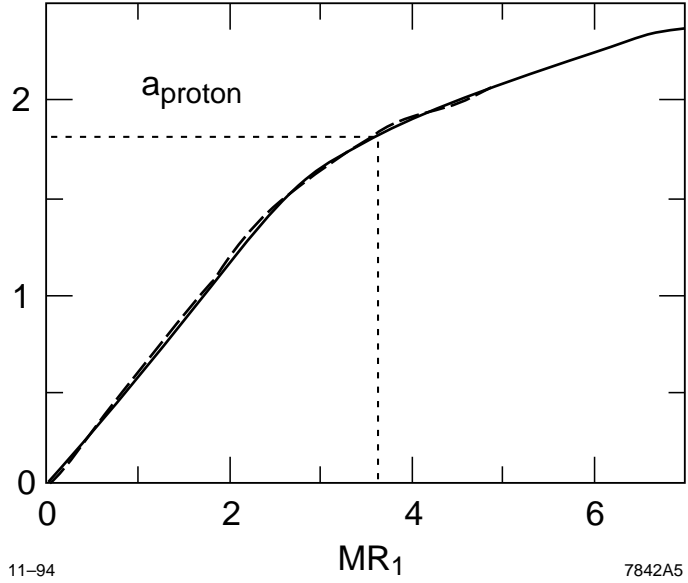}
}\caption{\label{fig:FA} 
The anomalous magnetic moment $a=F_2(0)$ of the proton as a function
of $M_p R_1$: broken line, pole type wave function; continuous line,
Gaussian wave function. The experimental value is given by the dotted
lines. The prediction of the model is independent of the
wave function for $Q^2=0$.
}\end{figure}

In Fig. \ref{fig:FA} we show the
functional relationship between the anomalous moment $a_p$ and its Dirac
radius predicted by the three-quark light-cone model. The value of $R^2_1
= -6 dF_1(Q^2)/dQ^2\vert_{Q^2=0}$ is varied by changing $\beta$ in the
light-cone wave function while keeping the quark mass $m$ fixed. The
prediction for the power-law wave function $\psi_{\rm Power}$ is given by
the broken line; the continuous line represents
$\psi_{\rm H.O.}$.
Figure~\ref{fig:FA} 
shows that when one plots the dimensionless
observable $a_p$  against the dimensionless observable $M R_1$ the
prediction is essentially independent of the assumed power-law or
Gaussian form of the three-quark light-cone wave function.  Different
values of $p > 2 $ also do not affect the functional dependence of
$a_p(M_p R_1)$ shown in Fig.~ \ref{fig:FA} 
In this sense the predictions of the
three-quark  light-cone model  relating the $Q^2 \to 0$ observables
are essentially model-independent. The only parameter controlling
the relation between the dimensionless observables in the light-cone
three-quark model is $m/M_p$ which is set to 0.28. For the physical
proton radius $M_p R_1=3.63$ one obtains the empirical value for
$a_p=1.79$ (indicated by the dotted lines in Fig. \ref{fig:FA}).

The prediction for the anomalous moment $a$ can be written
analytically  as $a=\langle \gamma_V \rangle a^{\rm NR}$, where
$a^{\rm NR}=2M_p/3m$ is the non-relativistic ($R\to\infty$) value
and $\gamma_V$ is given as   \cite{chc91}
\begin{equation}
 \gamma_V(x_i,k_{\perp
i},m)= {3m\over {\cal M}}\left[ {(1-x_3){\cal M}(m+x_3 {\cal
M})- \vec k_{\perp 3}^2/2\over (m+x_3 {\cal M})^2+\vec k_{\perp 3}^2}
\right] . 
\end{equation} 
The expectation  value $< \gamma_V >$ is
evaluated as
\begin{equation}
<{\gamma_V}> = {\int [d^3k]
\gamma_V |\psi|^2\over \int [d^3k] |\psi|^2} , 
\end{equation}
where $[d^3k]=d\vec k_1d\vec k_2d\vec
k_3\delta(\vec k_1+ \vec k_2+ \vec k_3)$. The third component of
$\vec k$ is defined as $k_{3i}= {1\over2}(x_i{\cal
M}-{m^2+\vec k_{\perp i}^2\over x_i {\cal M}})$. This measure
differs from the usual one used in Ref. \cite{leb80}  by the
Jacobian $\prod {dk_{3i}\over dx_i}$ which can be absorbed into the
wave function.

Let us take a closer look at the two limits $R \to \infty$ and $R\to
0$. In the non-relativistic limit we let $\beta \to 0$ and keep the
quark mass $m$ and the proton mass $M_p$ fixed. In this limit the
proton radius $R_1 \to \infty$ and $a_p \to 2M_p/3m = 2.38$ since
$< \gamma_V > \to 1$. (This differs slightly from
the usual non-relativistic formula $1+a=\sum_q {e_q\over e}
{M_p\over m_q}$ due to the non-vanishing binding energy which
results in $M_p \neq 3m_q$.). Thus the physical value of the
anomalous magnetic moment at the empirical proton radius $M_p
R_1=3.63$ is reduced by 25\% from its non-relativistic value due to
relativistic recoil and nonzero $k_\perp$ (The
non-relativistic value of the neutron magnetic moment is reduced by
31\%.).

To obtain the ultra-relativistic limit, we let $\beta \to \infty$
while keeping $m$ fixed.  In this limit the proton becomes pointlike
$(M_p R_1 \to 0)$ and the internal transverse momenta $k_\perp \to
\infty$. The anomalous magnetic momentum of the proton goes linearly
to zero as $a=0.43 M_p R_1$ since $< \gamma_V \rangle \to 0$.
Indeed, the Drell-Hearn-Gerasimov sum rule  \cite{ger65,drh66}
demands that the proton magnetic moment becomes equal to the Dirac moment
at small radius.  For a spin-$1\over 2$ system \begin{equation}
 a^2={M^2\over 2\pi^2\alpha}\int_{s_{th}}^\infty {ds\over
s}\left[ \sigma_P(s)-\sigma_A(s)\right], 
\end{equation}
where $\sigma_{P(A)}$ is
the total photo-absorption cross section with parallel (anti-parallel)
photon and target spins. If we take the point-like limit, such that
the threshold for inelastic excitation becomes infinite while the
mass of the system is kept finite, the integral over the
photo-absorption cross section vanishes and $a=0$  \cite{brt88}.
In contrast, the anomalous magnetic moment of the proton
does not vanish in the non-relativistic quark model as $R\to 0$. The
non-relativistic quark model does not take into account the fact that
the magnetic moment of a baryon is derived from lepton scattering at
nonzero momentum transfer;  {\it i.e.} , the calculation of a magnetic moment
requires knowledge of the boosted wave function. The Melosh
transformation is also essential for deriving the DHG sum rule and
low energy theorems of composite systems  \cite{brp60}.

\begin{figure} [t]
\epsfbox{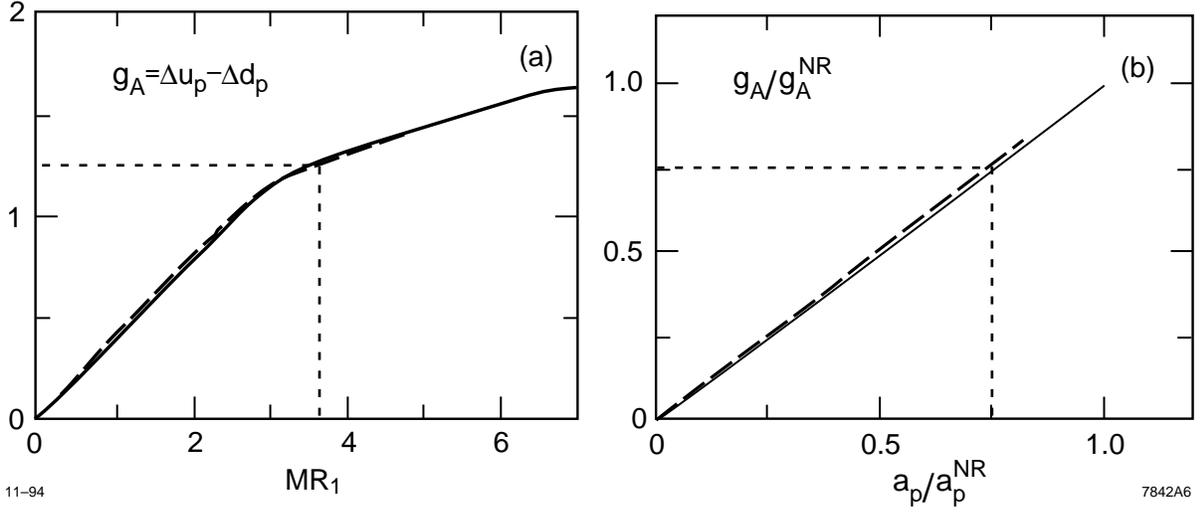}
\caption{\label{fig:FB}  
(a) The axial vector coupling $g_A$ of the neutron to proton
decay as a function of $M_p R_1$. 
The experimental value is given by the dotted lines. 
(b) The ratio $g_A/g_A(R_1 \to \infty)$ versus
$a_p/a_p(R_1 \to \infty)$ as a function of the proton radius $R_1$.
}\end{figure}

A similar analysis can be performed for the axial-vector coupling
measured in neutron decay. The coupling $g_A$ is given by the
spin-conserving axial current $J_A^+$ matrix element $ g_A(0) =
\langle p,\uparrow | J^+_A | p,\uparrow \rangle$. 
The value for $g_A$ can be written as $g_A=<\gamma_A \rangle g_A^{\rm
NR}$ with $g_A^{\rm NR}$ being the non-relativistic value of $g_A$ and
with $\gamma_A$ as {  \cite{chc91}, \cite{ma91}}
\begin{equation}
   \gamma_A(x_i,k_{\perp
i},m)={(m+x_3 {\cal M})^2-\vec k_{\perp 3}^2\over (m+x_3 {\cal
M})^2+\vec k_{\perp 3}^2} . 
\end{equation}
In Fig.~\ref{fig:FB} (a) since $< \gamma_A \rangle =0.75$.  
The measured value is $g_A= 1.2573\pm 0.0028$  \cite{pdg92}.
This is a 25\% reduction compared to the non-relativistic
SU(6) value $g_A=5/3,$ which is only valid for a proton with large
radius $R_1 \gg 1/M_p.$ As shown in Ref.\cite{ma91}, the Melosh
rotation generated by the internal transverse momentum spoils the
usual identification of the $\gamma^+ \gamma_5$ quark current matrix
element with the total rest-frame spin projection $s_z$, thus
resulting in a reduction of $g_A$.

Thus, given the empirical values for  the proton's anomalous moment
$a_p$ and radius $M_p R_1,$ its axial-vector coupling is
automatically fixed at the value $g_A=1.25.$   This prediction is an
essentially model-independent  prediction of the three-quark
structure of the proton in QCD. The Melosh rotation of the
light-cone wave function is crucial for reducing the value of the
axial coupling from its non-relativistic value 5/3 to its empirical
value. In Fig.~\ref{fig:FB}
(b)  we plot $g_A/g_A(R_1 \to \infty)$ versus
$a_p/a_p(R_1 \to \infty)$ by varying the proton radius $R_1.$ The
near equality of these ratios reflects the relativistic spinor
structure of the nucleon bound state, which is essentially
independent of the detailed shape of the momentum-space dependence
of the light-cone wave function. We emphasize that at small proton
radius the light-cone model predicts not only a vanishing anomalous
moment but also $ \lim_{R_1 \to 0} g_A(M_p R_1)=0. $ One can
understand this physically: in the zero radius limit the internal
transverse momenta become infinite and the quark helicities become
completely disoriented.  This is in contradiction with chiral models
which suggest that for a zero radius composite baryon one should
obtain the chiral symmetry result $g_A=1$.

The helicity measures $\Delta u$ and $\Delta d$ of the nucleon each
experience the same reduction as $g_A$ due to the Melosh effect.
Indeed, the quantity $\Delta q$ is defined by the axial current
matrix element
\begin{equation}
    \Delta q= \langle p,\uparrow | \bar
q\gamma^+\gamma_5 q | p,\uparrow \rangle ,
\end{equation}
and the value for $\Delta q$ can be written analytically as $\Delta
q= \langle \gamma_A \rangle \Delta
q^{\rm NR}$ with $\Delta q^{\rm NR}$ being the non-relativistic or naive
value of $\Delta q$ and with $\gamma_A$.

The light-cone model also predicts that the quark helicity sum
$\Delta\Sigma=\Delta u+\Delta d$ vanishes as a function of the
proton radius $R_1$. Since the helicity sum $\Delta\Sigma$ depends
on the proton size, and thus it cannot be identified as the vector
sum of the rest-frame constituent spins. As emphasized in
Refs. \cite{ma91,brp68}, the rest-frame spin sum is not a Lorentz invariant
for a composite system. Empirically, one measures $\Delta q$ from the
first moment of the leading twist polarized structure function
$g_1(x,Q).$  In the light-cone and parton model descriptions,
$\Delta q=\int_0^1 dx [q^\uparrow (x) - q^\downarrow (x)]$, where
$q^\uparrow (x)$ and $q^\downarrow (x)$ can be interpreted as the
probability for finding a quark or antiquark with longitudinal
momentum fraction $x$  and polarization parallel or anti-parallel to
the proton helicity in the proton's infinite momentum
frame  \cite{leb80}.
[In the infinite momentum there is no
distinction between the quark helicity and its spin-projection
$s_z.$] Thus $\Delta q$ refers to the difference of  helicities at
fixed light-cone time or at infinite momentum; it cannot be
identified with $q(s_z=+{1\over2})-q(s_z=-{1\over2}),$ the spin
carried by each quark flavor in the proton rest frame in the equal
time formalism.

Thus the usual SU(6) values $\Delta u^{\rm
NR}=4/3$ and $\Delta d^{\rm NR}=-1/3$ are only valid predictions for
the proton at large $M R_1.$ At the physical radius the quark
helicities are reduced by the same ratio 0.75 as $g_A/g_A^{\rm NR}$
due to the Melosh rotation. Qualitative arguments for such a
reduction have been given in Refs.  \cite{kar92} and \cite{fri90}.
For
$M_p R_1 = 3.63,$ the three-quark model predicts $\Delta u=1,$
$\Delta d=-1/4,$ and $\Delta\Sigma=\Delta u+\Delta d  = 0.75$.
Although the gluon contribution $\Delta G=0$ in our model, the
general sum rule  
\begin{equation}
 {1\over2}\Delta \Sigma +\Delta G+L_z= {1\over2}
\end{equation}
is still satisfied, since the Melosh transformation effectively
contributes to $L_z$.

Suppose one adds polarized gluons to the three-quark light-cone
model. Then the flavor-singlet quark-loop radiative corrections to
the gluon propagator will give an  anomalous contribution $\delta
(\Delta q)=-{\alpha_s\over2\pi}\Delta G$ to each light quark
helicity. The predicted value of $g_A = \Delta u -
\Delta d$  is of course unchanged. For illustration we shall choose
${\alpha_s\over2\pi}\Delta G=0.15$. The gluon-enhanced quark model
then gives the values in Table~\ref{tab:stan}, which agree well with the present
experimental values. Note that the gluon anomaly contribution to
$\Delta s$ has probably been overestimated here due to the large
strange quark mass. One could also envision other sources for this
shift of $\Delta q$ such as intrinsic flavor  \cite{fri90}. A
specific model for the gluon helicity distribution in the nucleon
bound state is given in Ref.\cite{bbs94}.

In summary, we have shown that relativistic effects are crucial
for understanding the spin structure of the nucleons. By plotting
dimensionless observables against dimensionless observables we
obtain model-independent relations independent of the momentum-space
form of the three-quark light-cone wavefunctions. For example, the
value of $g_A \simeq 1.25$ is correctly predicted from the empirical
value of the proton's anomalous moment. For the physical proton
radius $M_p R_1= 3.63$ the inclusion of the Wigner (Melosh) rotation
due to  the finite  relative transverse momenta of the three quarks
results in a  $\simeq 25\% $ reduction of the non-relativistic
predictions for the anomalous magnetic moment, the axial vector
coupling, and the quark helicity content of the proton. At zero
radius, the quark helicities become completely disoriented because
of the large internal momenta, resulting in the vanishing of $g_A$
and the total quark helicity $\Delta \Sigma.$

\begin{table}
\begin{tabular}{|crrrr|} \hline
&&&&\\
Quantity\qquad &\qquad NR &\qquad  $3q$ &\qquad $3q+g$
&\qquad Experiment\qquad \\[8pt] \hline
&&&&\\
$\Delta u$ & ${4\over3}$ & 1 & 0.85 & $0.83\pm 0.03$\\
&&&&\\
$\Delta d$ &$-{1\over3}$ & $-{1\over4}$ & --0.40 & $-0.43\pm 0.03$\\
&&&&\\
$\Delta s$ & 0 & 0 & --0.15   & $-0.10\pm 0.03$\\
&&&&\\
$\Delta \Sigma$ &1 & ${3\over4}$ & 0.30 & $0.31\pm 0.07$\\
&&&&\\
 \hline
\end{tabular}
\caption{\label{tab:stan}
Comparison of the quark content of the proton in the
non-relativistic quark model (NR), in the three-quark model (3q), in
a gluon-enhanced three-quark  model (3q+g), and with
experiment.}
\end{table}

\subsection{Applications to Nuclear Systems}
We can analyze a nuclear system in the same way as we did the nucleon in
the preceding section. The triton, for instance,  is modeled as a bound
state of a proton and two neutrons. The same formulae as in the preceding
chapter are valid (for spin-$1\over 2$ nuclei); we only have to use the
appropriate parameters for the constituents.

\begin{figure} [t]
\epsfysize=60mm\epsfbox{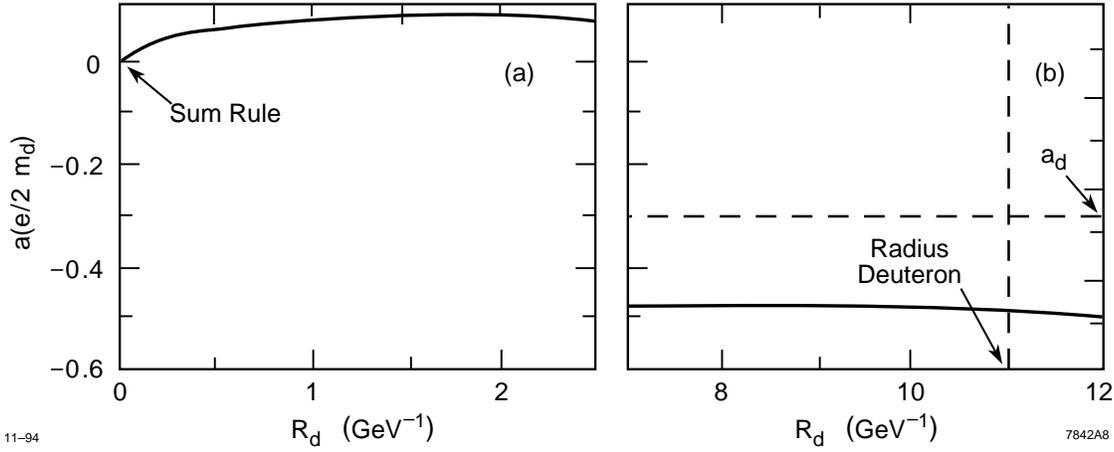}
\caption{\label{fig:FD} 
The anomalous moment  $a_d$ of the deuteron 
as a function of the deuteron radius $R_d$. 
In the limit of zero radius, 
the anomalous moment vanishes.
}\end{figure}

\begin{figure} [t]
\epsfbox{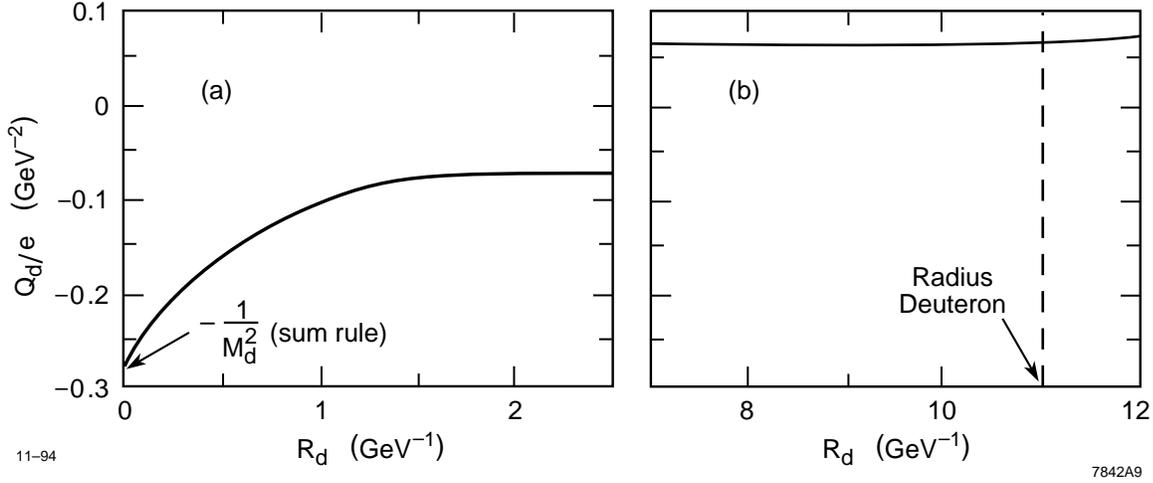}
\caption{\label{fig:FE}
The quadrupole moment $Q_d$ of the deuteron 
as a function of the deuteron radius $R_d$. 
In the limit of zero radius, the quadrupole
moment approaches its canonical value 
$Q_d=-e/M_d^2$. 
}\end{figure}

The light-cone analysis  yields nontrivial corrections to the
moments of nuclei. For example, consider the anomalous magnetic
moment $a_d$ and anomalous quadrupole moment $Q^a_d=Q_d+e/M_d^2$ of
the deuteron. As shown in
\cite{tun68}, these moments satisfy the sum rule
\begin{equation}
a_d^2 + {2 t \over M_d^2} (a_d + {M_d \over 2} Q_d^a )^2 = {1\over 4 \pi}
\int^\infty_{\nu_{th}^2} {d \nu^2 \over (\nu - t/4)^3}
({\rm Im} f_P(\nu,t) - {\rm Im}f_A(\nu,t)). 
\end{equation}
Here  $f_{P (A)}(\nu,t)$ is the non-forward Compton amplitude for
incident parallel (anti-parallel) photon-deuteron helicities. Thus,
in the pointlike limit where the threshold for particle excitation
$\nu_{th} \to \infty, $ the deuteron acquires the same
electro-magnetic moments $Q_d^a \to 0, a_d \to 0$ as that of the $W$
in the Standard Model \cite{brh83}.
The  approach to zero anomalous magnetic and
quadrupole moments for $R_d \to 0$  is shown in Figs. \ref{fig:FD} and
\ref{fig:FE}. Thus, even if the deuteron has no D-wave component, a
nonzero quadrupole moment arises from the relativistic recoil
correction. This correction, which is mandated by relativity, could
cure a long-standing discrepancy between experiment and the
traditional  nuclear physics predictions for the deuteron
quadrupole. Conventional nuclear theory predicts a quadrupole moment
of 7.233 GeV$^{-2}$ which is smaller than the experimental value
$(7.369 \pm 0.039)$ GeV$^{-2}$. The light-cone calculation for a
pure S-wave gives a positive contribution of 0.08 GeV$^{-2}$ which
accounts for most of the previous discrepancy.

\begin{figure}\centerline{
\epsfysize=80mm\epsfbox{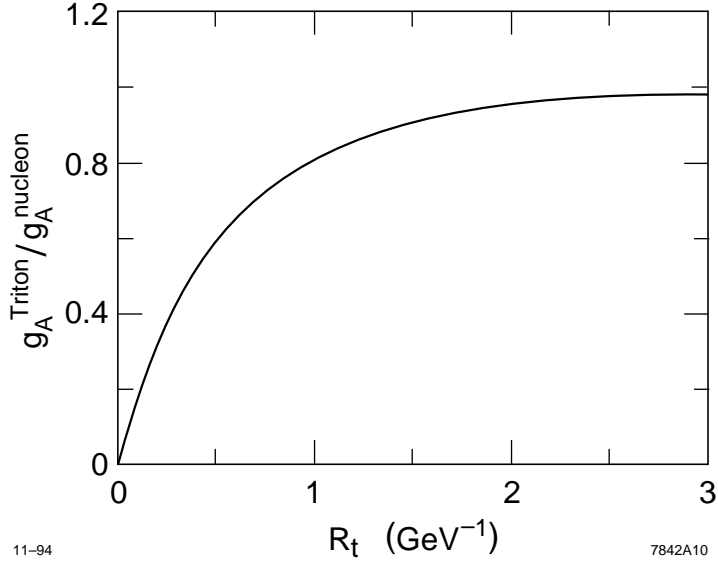}
}\caption{\label{fig:FF} 
The reduced Gamow-Teller matrix 
element for tritium decay as a
function of the tritium radius.
}\end{figure}

In the case of the tritium nucleus, the value of the Gamow-Teller
matrix element can be calculated in the same way as we calculated
the axial vector coupling $g_A$ of the nucleon in the previous
section. The correction to the non-relativistic limit for the S-wave
contribution is $g_A=\langle \gamma_A \rangle g_A^{\rm NR}$. For the
physical quantities of the triton we get $\langle \gamma_A \rangle
= 0.99$. This means that even at the physical  radius, we find a
nontrivial nonzero correction of order $-0.01$ to
$g^{\rm triton}_A/g_A^{\rm
nucleon}$ due to the relativistic recoil correction implicit in the
light-cone formalism. The Gamow-Teller matrix element is measured to
be $0.961 \pm 0.003$. The wave function of the tritium ($^3$H) is a
superposition of a dominant S-state and small D- and S'-state
components $\phi=\phi_S+\phi_{S'}+\phi_D$. The Gamow-Teller matrix
element in the non-relativistic theory is then given by
$g^{\rm triton}_A/g_A^{\rm
nucleon}=(|\phi_S|^2-{1\over 3}|\phi_{S'}|^2+{1\over 3}|\phi_D|^2)
(1+0.0589)=0.974$, where the last term is a correction due to meson
exchange currents. Figure~\ref{fig:FF}
shows that the Gamow-Teller matrix
element of tritium must approach zero in the limit of small nuclear
radius, just as in the case of the nucleon as a bound state of three
quarks.  This phenomenon is confirmed in the light-cone analysis.

\subsection{Exclusive Nuclear Processes}

One of the most elegant areas of application of QCD to nuclear
physics is the domain of large momentum transfer exclusive nuclear
processes \cite{cpc88}. Rigorous results  for the asymptotic properties of the
deuteron form factor at large momentum transfer are given in
Ref. \cite{bjl83}.  In the asymptotic limit $Q^2\rightarrow
\infty$ the deuteron distribution amplitude, which controls large
momentum transfer deuteron reactions, becomes fully symmetric among
the five possible color-singlet combinations of the six quarks. One
can also study the evolution of the ``hidden color'' components
(orthogonal to the $n p$ and $\Delta \Delta$ degrees of freedom)
from intermediate to large momentum transfer scales; the results
also give constraints on the nature of the nuclear force at short
distances in QCD. The existence of hidden color degrees of freedom
further illustrates the complexity of nuclear systems in QCD. It is
conceivable that six-quark $d^\ast$ resonances corresponding to
these new degrees of freedom may be found by careful searches of the
$\gamma^\ast d\rightarrow \gamma d$ and $\gamma^\ast d\rightarrow \pi d$
channels.

The basic scaling law for the helicity-conserving deuteron form
factor, $F_d(Q^2) \sim 1/Q^{10}$, comes from simple quark
counting rules, as well as perturbative QCD. One cannot expect this
asymptotic prediction to become accurate until very large $Q^2$
since the momentum transfer has to be shared by at least six
constituents. However, one can identify the QCD physics due to the
compositeness of the nucleus, with respect to its nucleon degrees of
freedom by using the reduced amplitude formalism \cite{bmp93}.
For example, consider the deuteron form factor in
QCD. By definition this quantity is the probability amplitude for
the deuteron to scatter from $p$ to $p+q$ but remain intact.

\begin{figure}
\epsfbox{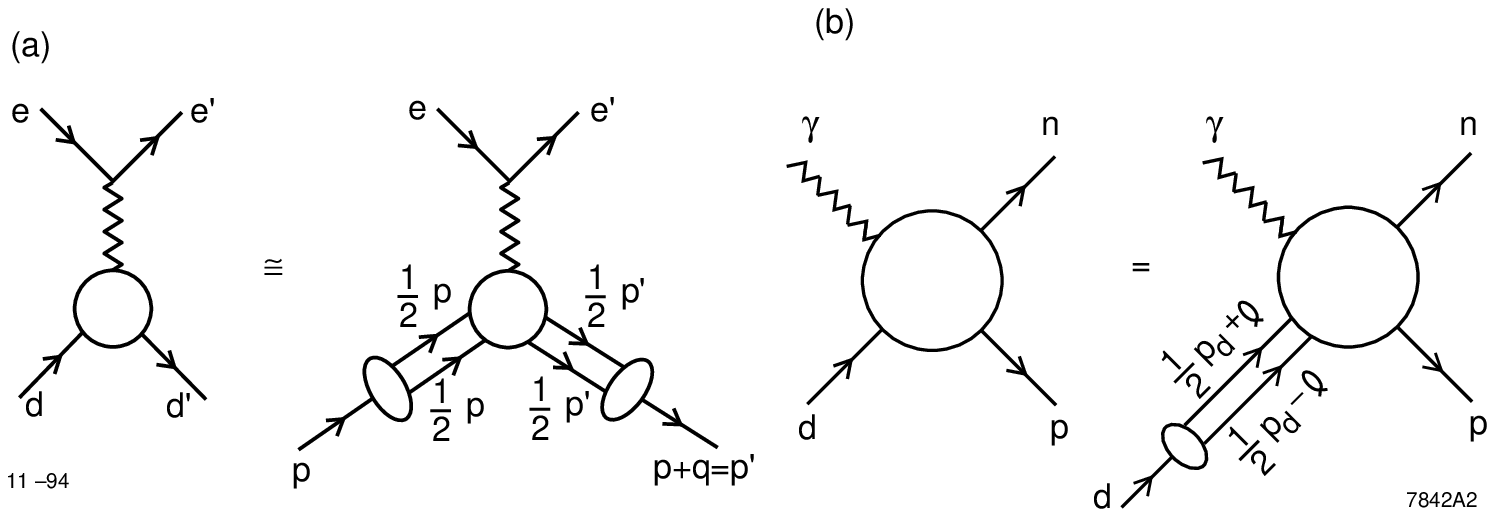}
\caption{\label{fig:RED}
(a) Application of the reduced amplitude formalism to 
the deuteron form factor at large momentum transfer. 
(b) Construction of the reduced nuclear amplitude for 
two-body inelastic deuteron reactions.
}\end{figure}

\begin{figure}\centerline{
\epsfysize=100mm\epsfbox{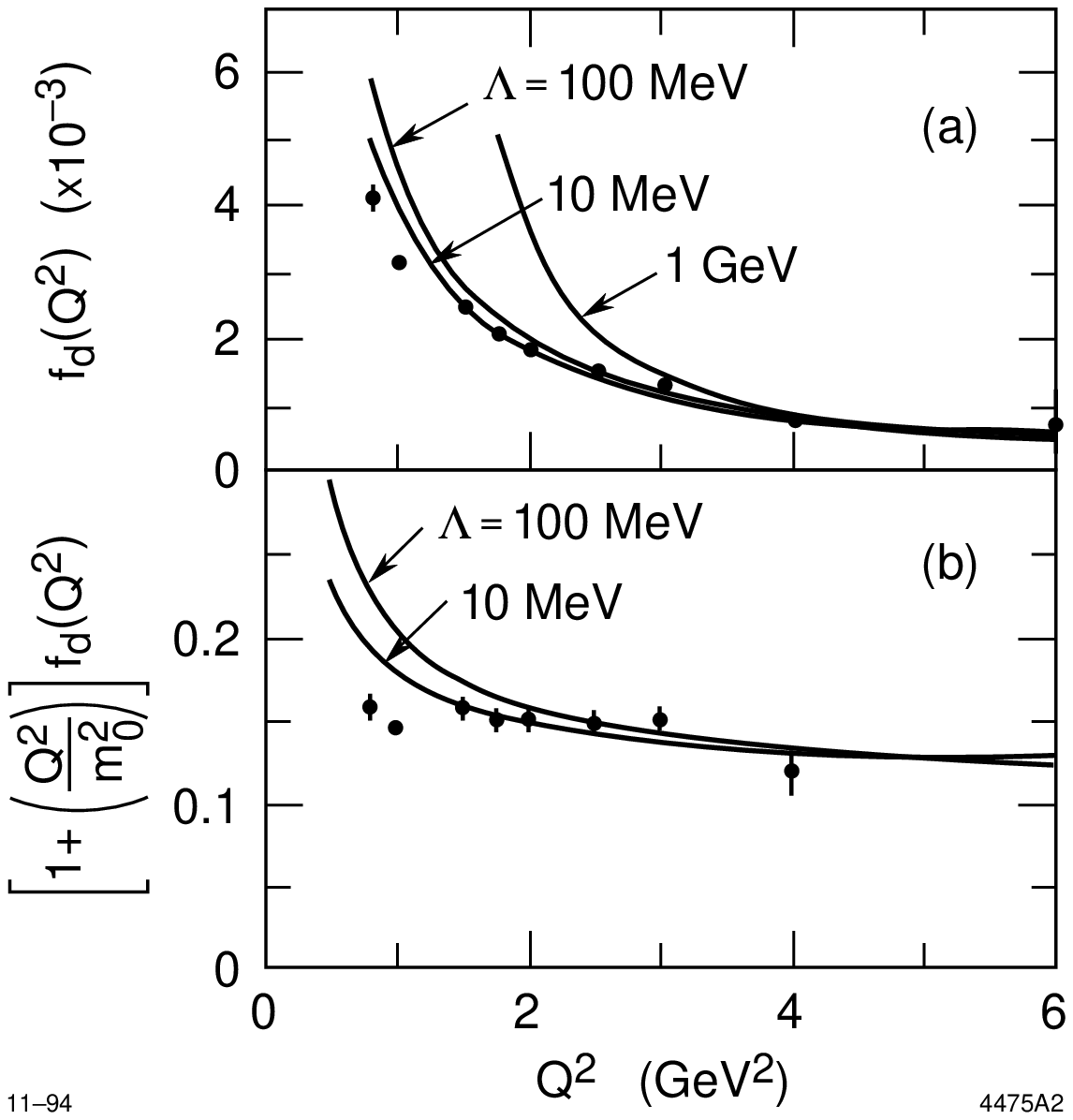}
}\caption{\label{fig:FYP}
Scaling of the deuteron reduced form factor.
}\end{figure}

Note that for vanishing nuclear binding energy $\epsilon_d \rightarrow
0$, the deuteron can be regarded as two nucleons sharing the
deuteron four-momentum (see Fig. \ref{fig:RED} (a)). In the zero-binding
limit one can show that the nuclear light-cone wave function properly
decomposes into a product of uncorrelated nucleon
wavefunctions  
\cite{jib86,lis92}.
The momentum $\ell$ is limited by the binding
and can thus be neglected, and to first approximation, the proton
and neutron share the deuteron's momentum equally. Since the
deuteron form factor contains the probability amplitudes for the
proton and neutron to scatter from $p/2$ to $p/2+q/2$, it is natural
to define the reduced deuteron form
factor {\cite{bmp93}, \cite{bjl83}, \cite{jib86}}:
\begin{equation}
f_d(Q^2) \equiv {F_d(Q^2)\over F_{1N} \left(Q^2\over 4\right)\,
F_{1N}\,\left(Q^2\over 4\right)}.
\end{equation}
The effect of nucleon
compositeness is removed from the reduced form factor. QCD then
predicts the scaling
\begin{equation} 
f_d(Q^2) \sim {1\over Q^2} 
\end{equation}
 {\it i.e.} \ the same scaling law as a meson form factor. Diagrammatically,
the extra power of $1/Q^2$ comes from the propagator of the struck
quark line, the one propagator not contained in the nucleon form
factors. Because of hadron helicity conservation, the prediction is
for the leading helicity-conserving deuteron form factor
$(\lambda=\lambda'= 0.)$ As shown in Fig. \ref{fig:FYP}, 
this scaling is consistent with experiment for $Q= p_T \sim$ 1 GeV.

The data are summarized in Ref. \cite{brh83}
The distinction between the QCD and other treatments of nuclear
amplitudes is particularly clear in the reaction $\gamma d
\rightarrow n p$;  {\it i.e.} \ photo-disintegration of the deuteron at fixed
center of mass angle. Using dimensional counting  \cite{brf75},
the leading power-law prediction from QCD is simply
${d\sigma\over dt}(\gamma d \rightarrow n p) \sim F(\theta_{\rm
cm})/s^{11}$. A comparison of the QCD prediction with the recent
experiment of Ref. \cite{bel94} is shown in Fig. \ref{fig:FYP}, 
confirming the validity of the QCD scaling prediction up to $E_\gamma
\simeq 3$ GeV. One can take into account much of the finite-mass,
higher-twist corrections by using the reduced amplitude formalism 
\cite{brh83}. The  photo-disintegration amplitude contains the
probability amplitude ( {\it i.e.} \ nucleon form factors) for the proton and
neutron to each remain intact after absorbing momentum transfers
$p_p-1/2 p_d$ and $p_n-1/2 p_d,$ respectively (see Fig. \ref{fig:RED} (b)).
After the form factors are removed, the remaining ``reduced" amplitude
should scale as $F(\theta_{\rm cm})/p_T$. The single inverse power of
transverse momentum $p_T$ is the slowest conceivable in any theory, but
it is the unique power predicted by PQCD.

\begin{figure}\centerline{
\epsfysize=80mm\epsfbox{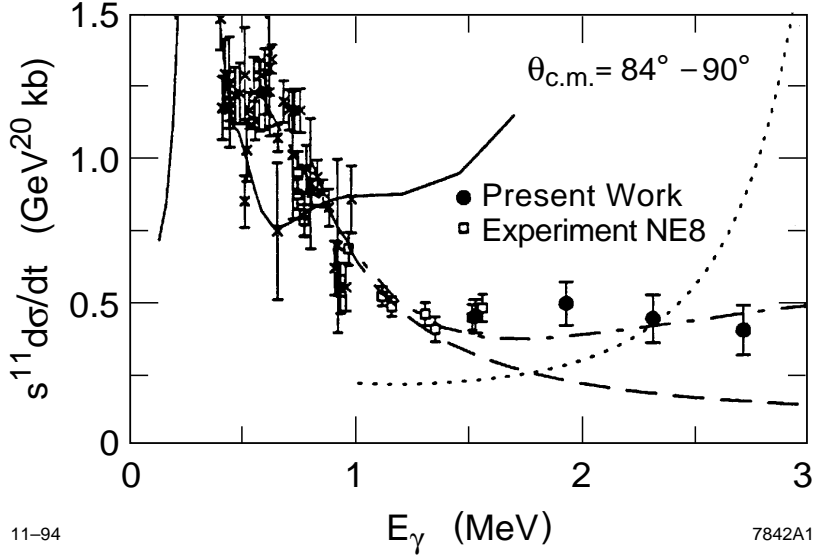}
}\caption{\label{fig:FYQ}
Comparison of deuteron photo-disintegration data with
the scaling prediction which requires 
$s^{11} d\sigma/dt(s,\theta_{cm})$
to be at most logarithmically dependent on energy 
at large momentum transfer. 
}\end{figure}

The data and predictions from conventional nuclear theory
in  are summarized in \cite{dry70}.
There are a number of related tests of QCD and reduced amplitudes
which require $\bar p$ beams  \cite{jib86}, such as $\bar
pd \rightarrow \gamma n$ and $\bar p d\rightarrow \pi  p$ in the fixed
$\theta_{\rm cm}$ region. These reactions are particularly
interesting tests of QCD in nuclei. Dimensional counting rules
predict the asymptotic behavior ${d\sigma\over dt}\ (\bar pd\rightarrow
\pi  p)\sim {1\over (p_T^2)^{12}} \ f(\theta_{\rm cm})$ since there
are 14 initial and final quanta involved. Again one notes that the
$\bar p d \rightarrow \pi  p$ amplitude  contains a factor representing
the probability amplitude ( {\it i.e.} \ form factor) for the proton to
remain intact after absorbing momentum transfer squared $\widehat t
= (p-1/2 p_d)^2$ and the $\bar NN$ time-like form factor at
$\widehat s = (\bar p + 1/2 p_d)^2$. Thus ${\cal M}_{\bar pd\rightarrow
\pi  p}\sim F_{1N}(\widehat t)\ F_{1N}(\widehat s)\, {\cal M}_r,$
where ${\cal M}_r$ has the same QCD scaling properties as quark
meson scattering. One thus predicts
\begin{equation}
 {{d\sigma\over d\Omega}\
(\bar pd\rightarrow \pi  p)\over F^2_{1N}(\widehat t)\,
F^2_{1N}(\widehat s)} \sim {f(\Omega)\over p^2_T} \ . 
\end{equation}
Other work has been done by Cardarelli {\it et al.} \cite{cgn95}.

\subsection{Conclusions}

As we have emphasized in this chapter, QCD and relativistic
light-cone Fock methods provide a new perspective on nuclear
dynamics and properties. In many some cases the covariant approach
fundamentally contradicts standard nuclear assumptions. More
generally,  the synthesis of QCD with the standard non-relativistic
approach can be used to constrain the analytic form and unknown
parameters in the conventional theory, as in Bohr's correspondence
principle. For example,  the reduced amplitude formalism and PQCD
scaling laws provide analytic constraints on the nuclear amplitudes
and potentials at short distances and large momentum transfers.

%% file: 07Exclu.tex
\section{Exclusive Processes and Light-Cone Wavefunctions}
\label{sec:exclusive}
\setcounter{equation}{0}

One of the major advantages of the light-cone formalism is that many
properties of large momentum transfer exclusive reactions can be
calculated without explicit knowledge of the form of the
non-perturbative light-cone wavefunctions. The main ingredients of
this analysis are asymptotic freedom, and the power-law scaling
relations and quark helicity conservation rules of perturbative QCD.
For example, consider the light-cone expression (\ref{CAI}) for a
meson form factor at high momentum transfer $Q^2.$  If the internal
momentum transfer is large then one can iterate the gluon-exchange
term in the effective potential for the light-cone wavefunctions.
The result is the hadron form factors can be written  in a
factorized form as a convolution of quark ``distribution amplitudes"
$\phi(x_i,Q)$, one for each hadron involved in the amplitude, with a
hard-scattering amplitude $T_H$ \cite{leb79a,leb79b,leb80}.  The
pion's electro-magnetic form factor, for example, can be written as
\begin{equation}
  F_\pi(Q^2) = \int_0^1 dx \int_0^1 dy\,
    \phi^*_\pi(y,Q)\,T_H(x,y,Q)\,\phi_\pi(x,Q)\,
    \left(1 + {\cal O}\left({1\over Q}\right)\right) .
  \label{fpi}
\end{equation}
Here $T_H$ is the scattering amplitude for the form factor but with
the pions replaced by collinear $q\bar q$ pairs---{\it i.e.}\ the
pions are replaced by their valence partons. We can also regard
$T_H$ as the free particle matrix  element of the order $1/q^2$ term
in the effective Lagrangian for $\gamma^* q \bar q \to q \bar q$.
 
The process-independent distribution amplitude \cite{leb79a,leb79b,leb80}
$\phi_\pi(x,Q)$ is the probability amplitude for finding the $q\bar
q$ pair in the pion with $x_q = x$ and $x_{\bar q} = 1-x$. It is
directly related to the light-cone valence wave function:
\begin{eqnarray}
\phi_\pi(x,Q) & = & \int {d^2\vec k_\perp \over 16\pi^3}\,
   \psi^{(Q)}_{q\bar q/\pi} (x,\vec k_\perp)  
   \label{phidefone} \\
           {} & = &  P^+_\pi \int {dz^-\over 4\pi}\, 
   e^{i x P+ \pi z^-/2} \,\left\langle 0 \right| 
\overline\psi(0)\,{\gamma^+\gamma_5 \over 2\sqrt{2n_c}}
\,\psi(z)\left|\pi\right\rangle^{(Q)} 
  _{\textstyle z^+=\vec z_\perp=0}  .
   \label{phideftwo}
\end{eqnarray}
The $\vec k_\perp$ integration in Eq. (\ref{phidefone}) is cut off
by the ultraviolet cutoff $\Lambda=Q$ implicit in the wave function;
thus only Fock states with invariant mass squared ${\cal M}^2< Q^2$
contribute. We will return later to the discussion of ultraviolet
regularization in the light-cone formalism.
 
It is important to note that the distribution amplitude is gauge
invariant. In gauges other than light-cone gauge,  a path-ordered
`string operator' $P\exp(\int_0^1 ds\,ig\,A(sz)\cdot z)$
must be included between the $\overline\psi$ and $\psi$. The line
integral vanishes in light-cone gauge because 
$A\cdot z = A^+z^-/2 =0$ and so the factor can be omitted 
in that gauge. 
This (non-perturbative) definition of $\phi$ uniquely fixes the
definition of $T_H$ which must itself then be gauge invariant.
 
The above result is in the form of a factorization theorem; all of
the non-perturbative dynamics is factorized into the
non-perturbative distribution amplitudes, which sums all internal
momentum transfers up to the scale $Q^2.$ On the other hand, all
momentum transfers higher than $Q^2$ appear in $T_H$, 
which, because of asymptotic freedom, 
can be computed perturbatively in powers of
the QCD running coupling constant $\alpha_s(Q^2)$. 
 
Given the factorized structure, one can read off a number of general
features of  the PQCD predictions; {\it e.g.} the dimensional
counting rules, hadron helicity conservation, color transparency,
etc. \cite{brl89}.  In addition, the scaling behavior of the
exclusive amplitude is modified by the logarithmic 
dependence of the distribution amplitudes in $\ell n~Q^2$  
which is in turn determined
by QCD evolution equations \cite{leb79a,leb79b,leb80}.
 
\begin{figure} [t]
\epsfysize=100mm\epsfbox{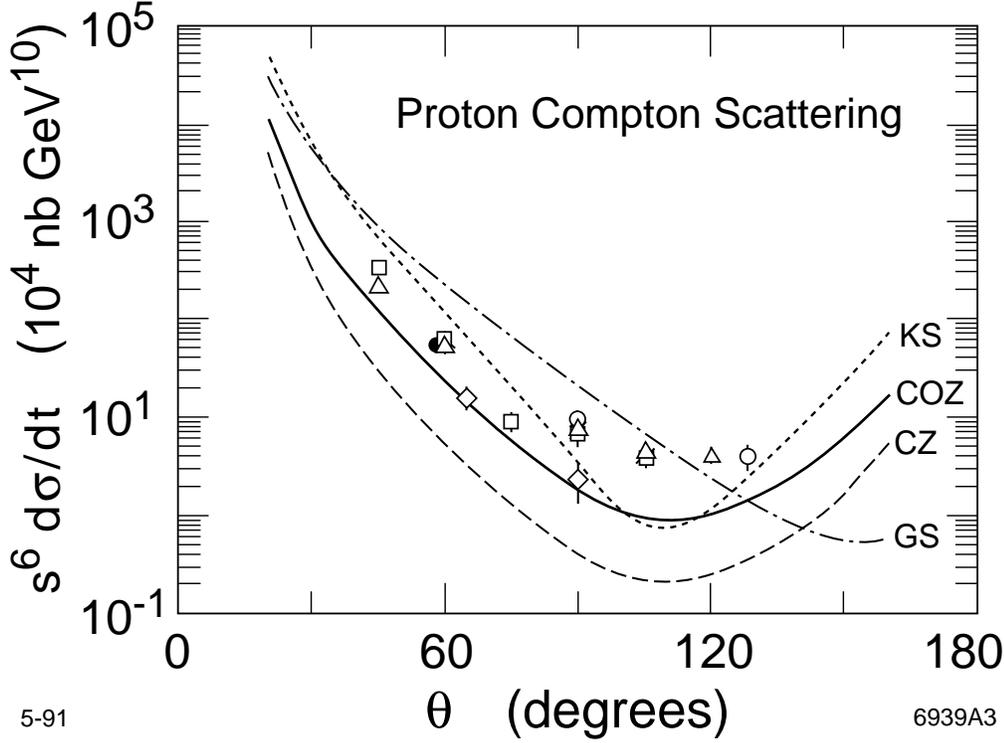} 
\caption{\label{fig:XXX}
Comparison 
of the order $\alpha_s^4 / s^6$ 
PQCD prediction for proton Compton scattering with the 
available data.
The calculation assumes PQCD factorization and 
distribution amplitudes computed from QCD sum rule 
moments. 
 } \end{figure}
 
An important application of the PQCD analysis is exclusive Compton
scattering and the related cross process $\gamma \gamma \to \bar p
p.$ Each helicity amplitude for $\gamma p \to \gamma p$ can be
computed at high momentum transfer from the convolution of the
proton distribution amplitude with the ${\cal O}(\alpha_s^2)$
amplitudes for $qqq \gamma \to qqq \gamma$. The result is a cross
section which scales as
\begin{equation}
{d\sigma \over dt}(\gamma p \to \gamma p) =
{F(\theta_{CM}, \ell n~s) \over s^6 }
\end{equation}
if the proton helicity is conserved. The helicity-flip amplitude and
contributions involving more quarks or gluons in the proton
wavefunction are power-law suppressed. The nominal  $s^{-6}$ fixed
angle scaling follows from dimensional counting rules
\cite{brf75}.  It is modified logarithmically due to the
evolution of the proton distribution amplitude and the running of
the QCD coupling constant \cite{leb79a,leb79b,leb80}.  The normalization,
angular dependence, and phase structure are highly sensitive to the
detailed shape of the non-perturbative form of $\phi_p(x_i, Q^2).$
Recently Kronfeld and Nizic \cite{krn91}  have calculated the
leading Compton amplitudes using  model forms for $\phi_p$ predicted
in the QCD sum rule analyses \cite{chz84};  the calculation is
complicated by the presence of integrable poles in the
hard-scattering subprocess $T_H.$  The results for the unpolarized
cross section are shown in Fig.~\ref{fig:XXX}.
 
There also has been important progress testing PQCD  experimentally
using measurements of the $p \to N^*$ form factors. In an 
analysis of existing SLAC data, Stoler \cite{sto91}  has obtained
measurements of several transition form factors of the proton to
resonances at $W=1232, 1535,$ and $1680~MeV.$ As is the case of the
elastic proton form factor, the observed behavior of the transition
form factors to the $N^*(1535)$ and $N^*(1680)$ are each consistent
with  the $Q^{-4}$ fall-off and dipole scaling predicted by PQCD and
hadron helicity conservation over the measured range $1 < Q^2 <
21~GeV^2$. In contrast, the $p \to \Delta(1232)$ form factor
decreases faster than $1/Q^4$ suggesting that non-leading processes
are dominant in this case. Remarkably, this pattern of scaling
behavior is what is expected from PQCD and the QCD sum rule analyses
\cite{chz84},  since, unlike the case of the proton and its other
resonances, the distribution amplitude $\phi_{N^*}(x_1,x_2,x_3,Q)$
of the $\Delta$ resonance is predicted to be nearly symmetric in the
$x_i$, and a symmetric distribution leads to a strong cancelation
\cite{cap88}  of the leading helicity-conserving terms in the matrix
elements of the hard scattering amplitude for $qqq \to \gamma^* qqq$.
 
These  comparisons of the proton form factor and Compton scattering
predictions with experiment are very encouraging, showing agreement
in both the fixed-angle scaling behavior predicted by PQCD and the
normalization predicted by QCD sum rule forms for the proton
distribution amplitude. Assuming one can trust the validity of the
leading order analysis, a systematic series of polarized target and
beam Compton scattering measurements on proton and neutron targets
and the corresponding two-photon reactions $\gamma \gamma \to p \bar
p$ will strongly constrain a fundamental quantity in QCD, the
nucleon distribution amplitude $\phi(x_i, Q^2)$. It is thus
imperative for theorists to develop methods to calculate the shape
and normalization of the non-perturbative distribution amplitudes
from first principles in QCD.
 
\subsection{Is PQCD Factorization Applicable to Exclusive Processes?}
 
One of the concerns in the derivation of the PQCD results for
exclusive amplitudes is whether the momentum transfer carried by the
exchanged gluons in the hard scattering amplitude $T_H$ is
sufficiently large to allow a safe application of perturbation
theory \cite{isl89}.  The problem appears to be especially serious if
one assumes a form for the hadron distribution amplitudes
$\phi_H(x_i, Q^2)$ which has strong support at the endpoints, as in
the QCD sum rule model forms suggested by Chernyak and Zhitnitskii
and others \cite{chz84,zhi85}.
 
This problem  has now been clarified by two groups: Gari {\em et
al.} \cite{gas86}  in the case of baryon form factors, and Mankiewicz
and Szczepaniak \cite{szm91},  for the case of meson form factors.
Each of these authors has pointed out that the assumed
non-perturbative input for the distribution amplitudes must vanish
strongly in the endpoint region; otherwise, there is a
double-counting problem for momentum transfers occurring in the hard
scattering amplitude and the distribution amplitudes. Once one
enforces this constraint, ({\it e.g.}\ by  using exponentially
suppressed wavefunctions \cite{leb80}) on the basis functions used
to represent the QCD moments, or uses a sufficiently large number of
polynomial basis functions, the resulting distribution amplitudes do
not allow significant contribution to the high $Q^2$ form factors to
come from  soft gluon exchange region. The comparison of the PQCD
predictions with experiment thus becomes phenomenologically and
analytically consistent. An analysis of exclusive reactions on the
effective Lagrangian method   is also consistent with
this approach. In addition, as discussed by Botts \cite{bot91},
potentially soft contributions to large angle hadron-hadron
scattering reactions from Landshoff pinch contributions
\cite{lan74}  are strongly suppressed by Sudakov form factor
effects.
 
The empirical successes of the PQCD approach, together with the
evidence for color transparency in quasi-elastic $pp$ scattering
\cite{brl89}  gives strong support for the validity of PQCD
factorization for exclusive processes at moderate momentum transfer.
It seems difficult to  understand this pattern of  form factor
behavior if it is due to simple convolutions of soft wavefunctions.
Thus it should be possible to use these processes to empirically
constrain the form of the hadron distribution amplitudes, and thus
confront non-perturbative QCD in detail.
For recent work, see \cite{abs94,dkn97,jps95,mil97}. 

\begin{figure} [t]
\epsfysize=80mm\epsfbox{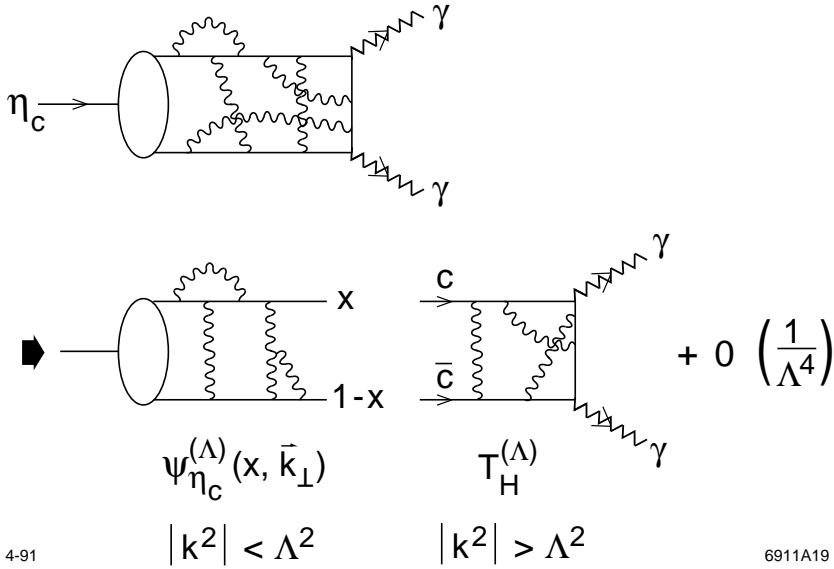}
\caption{\label{fig:LLH}
Factorization of perturbative and non-perturbative contributions to
the decay $\eta_c \to \gamma \gamma$. 
}\end{figure}

\begin{figure} [t]
\epsfbox{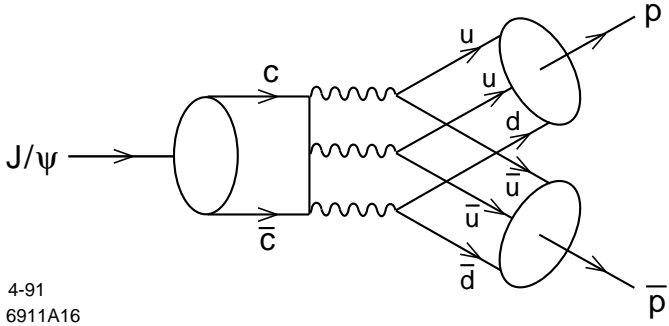}
\caption{\label{fig:LLE}
Calculation of $J/\psi \to p \bar p$ in PQCD.
}\end{figure}

\subsection{Light-Cone Quantization and Heavy Particle Decays}

One of the most interesting applications of the light-cone PQCD
formalism is to large momentum transfer exclusive processes to heavy
quark decays. For example, consider the decay $\eta_c \to \gamma
\gamma$. If we can choose the Lagrangian cutoff $\Lambda^2 \sim
m_c^2,$ then to leading order in $1/m_c,$ all of the bound state
physics and virtual loop corrections are contained in the $c \bar c$
Fock wavefunction $\psi_{\eta_c}(x_i,k_{\perp i}).$ The hard
scattering matrix element of the effective Lagrangian coupling $c
\bar c \to \gamma \gamma$ contains all of the higher corrections in
$\alpha_s(\Lambda^2)$ from virtual momenta $|k^2| > \Lambda^2.$ Thus
\begin{eqnarray} 
{\cal M}(\eta_c\rightarrow\gamma\gamma) 
  &=&\int d^2k_\perp\int^1_0dx\,\psi^{(\Lambda)}_{\eta_c}(x,k_\perp)\
    T^{(\Lambda)}_H (c\bar c\rightarrow\gamma\gamma) 
\nonumber \\
&\Rightarrow& \int^1_0dx\,\phi(x,\Lambda)\,
    T^{(\Lambda)}_H(c\bar c\rightarrow\gamma\gamma)
\end{eqnarray}
where $\phi(x,\Lambda^2)$ is the $\eta_c$ distribution amplitude.
This factorization and separation of scales is shown in 
Fig.~\ref{fig:LLH}.  
Since the $\eta_c$ is quite non-relativistic, its
distribution amplitude is    peaked at $x = 1/2$, and its integral
over $x$ is essentially equivalent to the wavefunction at the
origin, $\psi(\vec r = \overrightarrow 0).$
 
Another interesting calculational example  of quarkonium decay in
PQCD is the annihilation of the $J/\psi$ into baryon pairs. The
calculation requires the convolution of the hard annihilation
amplitude $T_H(c \bar c \to ggg \to uud~uud)$ with the $J/\psi$,
baryon, and anti-baryon distribution amplitudes
\cite{leb79a,leb79b,leb80}. (See Fig. \ref{fig:LLE}.~) The magnitude
of the computed decay amplitude for $\psi \to {\bar p} p$ is consistent
with experiment assuming the proton distribution amplitude computed
from QCD sum rules \cite{chz84}, see also Keister \cite{kei91}.  
The angular distribution of the
proton in $e^+ e^- \to J/\psi \to p \bar p$ is also consistent with
the hadron helicity conservation rule predicted by PQCD; {\it i.e.}\
opposite proton and anti-proton helicity.
The spin structure of hadrons has been investigated by
Ma \cite{mab93,mab97}, using light-cone methods.

The effective Lagrangian method was used by Lepage, Caswell, and
Thacker \cite{let87}  to systematically compute  the order
$\alpha_s(\widehat Q)$ corrections to the hadronic and photon decays
of quarkonium. The scale $\widehat Q$ can then be set by
incorporating vacuum polarization corrections into the running
coupling constant \cite{blz81}.   A summary of the results  can be
found in Ref. \cite{kmr88}.
   
\subsection{Exclusive Weak Decays of Heavy Hadrons}
 
\begin{figure} [t]
\epsfysize=80mm\epsfbox{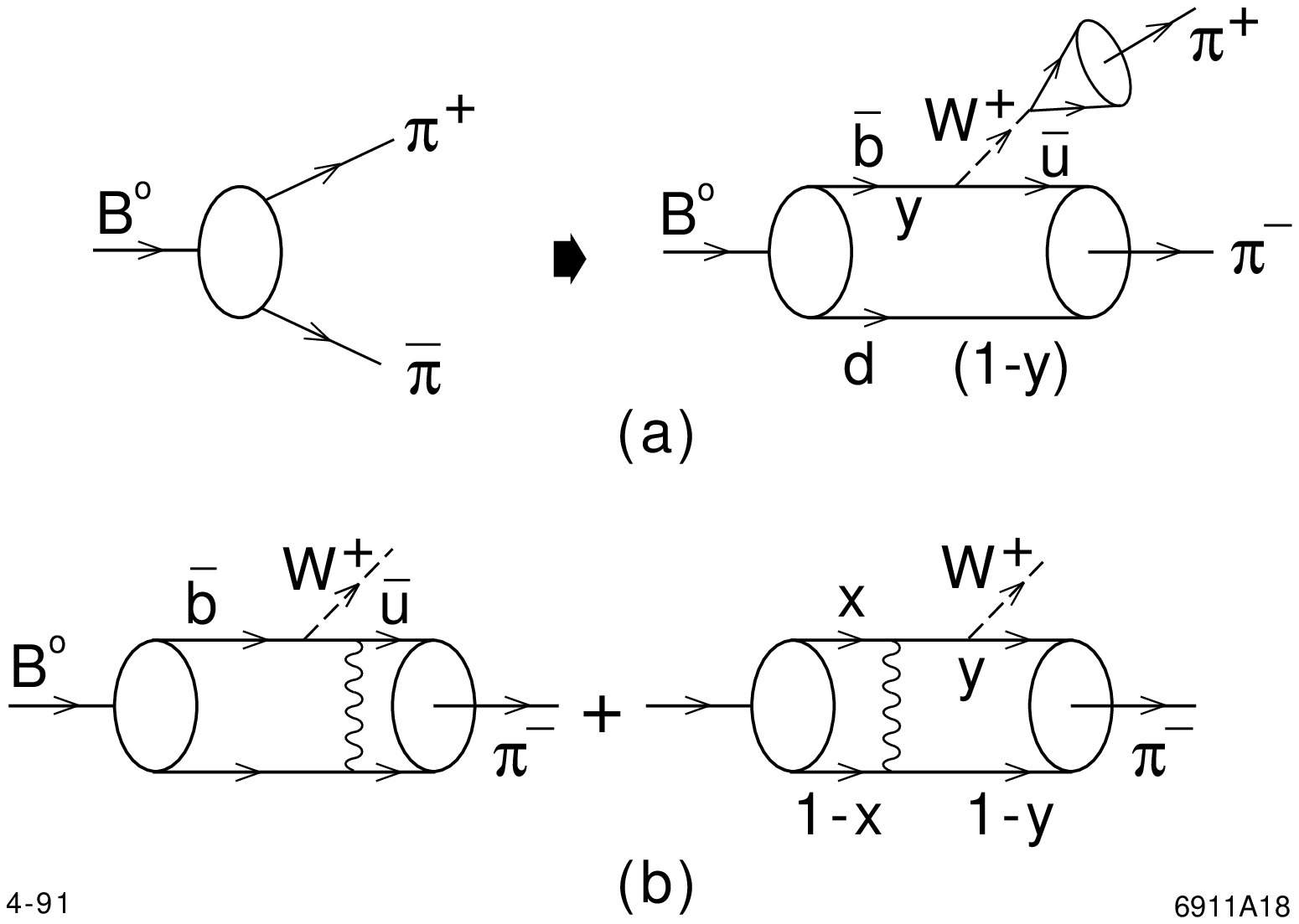}
\caption{\label{fig:LLG}
Calculation of the weak decay ${\cal B} \to \pi \pi$ in the 
PQCD formalism of Ref. \protect\cite{shb90}.
The gluon exchange kernel of the hadron wavefunction 
is exposed where hard momentum transfer is required.
}\end{figure}

An important application of the PQCD effective Lagrangian formalism
is to the exclusive decays of heavy hadrons to light hadrons, such
as $B^0 \to \pi^+ \pi^-,\ K^+, K^-$ \cite{shb90}.  To a good
approximation, the decay amplitude $\cal M$=
$\left\langle{B|H_{Wk}|\pi^+ \pi^-}\right\rangle$ is caused by the
transition $\bar b \to W^+ \bar u$; thus ${\cal M}$ $=f_\pi
p^\mu_\pi {G_F\over\sqrt 2} \left\langle
{\pi^-|J_\mu|B^0}\right\rangle$ where $J_\mu$ is the $\bar b \to
\bar u$ weak current. The problem is then to recouple the spectator
$d$ quark and the other gluon and possible quark pairs in each $B^0$
Fock state to the corresponding Fock state of the final state
$\pi^-$  (see Fig.~\ref{fig:LLG}). The kinematic constraint that
$(p_B-p_\pi)^2= m^2_\pi$ then demands that at least one quark line
is far off shell: $p_{\bar u}^2=(y p_B-p_\pi)^2 \sim -\mu m_B \sim
-1.5~GeV^2$, where we have noted that the light quark takes only a
fraction $(1-y) \sim \sqrt{(k_\perp^2+m_d^2)}/m_B$ of the heavy
meson's momentum  since all of the valence quarks must have nearly
equal velocity in a bound state. In view of the successful
applications \cite{sto91}  of PQCD factorization to form factors at
momentum transfers in the few $GeV^2$ range, it is reasonable to
assume that $\left\langle {|p^2_{\bar u}|}\right\rangle$ is
sufficiently large that we can begin to apply perturbative QCD
methods.
 
The analysis of the exclusive weak decay amplitude can be carried
out in parallel to the PQCD analysis of electro-weak form factors
\cite{blz81}  at large $Q^2$. The first step is to iterate the
wavefunction equations of motion so that the large momentum transfer
through the gluon exchange potential is exposed. The heavy quark
decay amplitude can then be written as a convolution of the hard
scattering amplitude for $Q\bar q \to W^+ q \bar q$ convoluted with
the $B$  and $\pi$ distribution amplitudes. The minimum number
valence Fock state of each hadron gives the leading power law
contribution. Equivalently, we can choose the ultraviolet cut-off
scale in the Lagrangian at $(\Lambda^2 < \mu m_B)$ so that the hard
scattering amplitude $T_H(Q\bar q \to W^+ q \bar q)$ must be
computed from the matrix elements of the order $1/\Lambda^2$ terms
in $\delta {\cal L}$. Thus $T_H$ contains all perturbative virtual
loop corrections of order $\alpha_s(\Lambda^2).$  The result is the
factorized form: 
\begin{equation} 
{\cal M}(B\to \pi \pi)= 
   \int^1_0 dx \int^1_0 dy \phi_B(y,\Lambda)T_H
   \phi_{\pi}(x,\Lambda)
\end{equation} 
be correct up to terms of  order $1/\Lambda^4.$ All of the
non-perturbative corrections with momenta $|k^2| < \Lambda^2$ are
summed in the distribution amplitudes.
 
In order to make an estimate of the size of the $B \to \pi \pi$
amplitude, in Ref. \cite{shb90}  we have taken the simplest possible 
forms for the required wavefunctions
\begin{equation}
\phi_\pi (y) \propto \gamma_5 \not p_\pi y(1-y)
\end{equation}
for the pion and
\begin{equation}
\phi_{\cal B} (x)\propto{\gamma_5[\not p_B+ m_B g(x)]\over  
\left[1-{1\over x}-{\epsilon^2\over (1-x)}\right]^{2}}
\end{equation}
for the B, each normalized to its meson decay constant. The above
form for the heavy quark distribution amplitude is chosen so that
the wavefunction peaks at equal velocity; this is consistent with
the phenomenological forms used to describe heavy quark
fragmentation into heavy hadrons. We estimate $\epsilon \sim$ 0.05
to 0.10. The functional dependence of the mass term $g(x)$ is
unknown; however, it should be reasonable to take $g(x) \sim 1$
which is correct in the weak binding approximation.

One now can compute the leading order PQCD decay amplitude
\begin{equation} 
{\cal M}(B^0\rightarrow\pi^-\pi^+) 
  = {G_F\over\sqrt2}\, V^*_{ud}\, V_{ub}\,P^\mu_{\pi^+} 
    \left\langle{\pi^-\,|\,V^\mu\,|\,B^0}\right\rangle
\end{equation}
where
\begin{eqnarray} 
\left\langle{\pi^-\,|\,V^\mu\,|\,B^0}\right\rangle &=&
{8\pi\alpha_s(Q^2)\over3}\int^1_0dx\int^{1-\epsilon}_0dy\,
\phi_B(x)\, \phi_\pi(y)\nonumber \\
&&\quad\times
{{\rm Tr}[\not P_{\pi^-}
\gamma_5\gamma^\nu\not k_1\gamma^\mu(\not P_B +
M_B g(x))\gamma_5\gamma_\nu]\over k^2_1q^2} \nonumber \\
&&\quad + {{\rm Tr}[\not P_{\pi^-} \gamma_5\gamma^\nu(\not
k_2+M_B)\gamma^\nu
(\not P_B+M_B g(x))\gamma_5\gamma_\nu]\over (k^2_2-M^2_B)Q^2}
\end{eqnarray} 
Numerically, this gives the branching ratio
\begin{equation}
BR(B^0 \to \pi^+ \pi^-) \sim 10^{-8} \xi^2 N
\end{equation}
where $\xi = 10 |V_{ub}/V_{cb}|$ is probably
less than unity,
and $N$ has strong dependence on the value of $g$:
$N = 180$ for $g=1$ and $N = 5.8$ for $g=1/2.$
The present experimental limit \cite{bar89}  is
\begin{equation}
BR(B^0 \to \pi^+ \pi^-) < 3 \times 10^{-4}.
\end{equation}
A similar PQCD analysis can be applied to other two-body decays of
the $B$; the ratios of the widths will not be so sensitive to the
form of the distribution amplitude, allowing tests of the flavor
symmetries of the weak interaction. Semi-leptonic decay rates
can be calculated \cite{cch97,dxt95,gns96,jau90,sim96}, and
the construction of the heavy quark wave functions 
\cite{czl95,zlc96} can be helpful for that. 

\subsection{Can light-cone wavefunctions be measured?}

Essential information on the shape and form of the valence light-cone
wavefunctions can be obtained empirically through measurements of
exclusive processes at large momentum transfer.  In the case of the
pion, data for the scaling and magnitude of the photon transition form
factor $F_{\gamma\pi^0}(q^2)$ suggest that the distribution amplitude
of the pion $\phi_\pi(x,Q)$ is close in form to the asymptotic form
$\phi^\infty_\pi(x) = \sqrt 3\, f_\pi(1-x)$, the solution to the
evolution equation for the pion at infinite resolution 
$Q\rightarrow\infty$, \cite{leb80}.  
Note that the pion distribution amplitude is constrained by
$\pi \rightarrow \mu\nu$ decay,
\begin{equation}
    \int^1_0 dx\, \phi_\pi(x,Q) = {f_\pi \over 2\sqrt 3} 
\,.\end{equation}
The proton distribution amplitude as determined by the proton 
form factor at large momentum transfer,
and Compton scattering is apparently highly asymmetric as 
suggested by QCD sum rules and SU(6) flavor-spin symmetry.

The most direct way to measure the hadron distribution wavefunction is
through the diffractive dissociation of a high energy hadron to jets or
nuclei; {\it e.g.} $\pi A \rightarrow {\rm Jet} + {\rm Jet} + A^\prime$,
where the final-state nucleus remains intact 
\cite{bbg81,flm96}.
The incoming hadron is a sum over all of its $H^0_{LC}$ fluctuations.
When the pion fluctuates into a $q\bar q$ state with small impact
separation $b_\perp^0\, (1/Q)$, its color interactions are minimal 
the ``color transparency'' property of QCD \cite{brm88}.  Thus this fluctuation will
interact coherently throughout the nucleus without initial or final state
absorption corrections.  The result is that the pion is coherently
materialized into two jets of mass $\cal M^2$ with minimal momentum 
transfer to the nucleus
\begin{equation}
   \Delta Q_L = \frac{\cal M^2-m^2_\pi}{2E_L} 
\,.\end{equation}
Thus the jets carry nearly all of the momentum of the pion.
The forward amplitude at $Q_\perp, Q_L \ll R^{-1}_\pi$ is linear
in the number of nucleons.  The total rate integrated over the forward
diffraction peak is thus proportional to
$\frac{A^2}{R^2_\pi} \propto A^{1/3} \ $.

The most remarkable feature of the diffractive $\pi A \rightarrow
{\rm Jet} + {\rm Jet} + X$ reactions is its potential to measure the
shape of the pion wavefunction.  The partition of jet longitudinal
momentum gives the $x$-distribution; the relative transverse momentum
distribution provides the $\vec k_\perp$-distribution of
$\psi_{q\bar q/\pi}(x,\vec k_\perp)$.  Such measurements are now
being carried out by the E791 collaboration at Fermilab.  In principle
such experiments can be carried out with a photon beam, which should
confirm the $x^2+(1-x)$ $\gamma \rightarrow q\bar q$ distribution of
the basic photon wavefunction.  Measurements of $pA \rightarrow
{\rm Jet} + {\rm Jet} A$ could, in principle, provide a direct
measurement of the proton distribution amplitude $\phi_p(x_i;Q)$.

%% file: 08Vacuu.tex
\section{The Light-Cone Vacuum}
\label{sec:Vacuum}
\setcounter{equation}{0}

The unique features of `front form' or light-cone quantized field
theory \cite{dir49} provide a powerful tool for the study of QCD. 
Of primary importance in this approach is the existence of a vacuum 
state that is the ground state of the full theory. 
The existence of this state gives a firm   basis for the
investigation of many of the complexities that must exist in QCD. 
In this picture the rich structure of vacuum is transferred to the 
zero modes of the theory. 
Within this context the long range physical phenomena of 
spontaneous symmetry breaking \cite{hkw91a,hkw91b,hkw91c}
\cite{bpv93,piv94,hpv95,rob93,pin93a,pin93b} 
as well as the topological structure
of the theory \cite{kap94,pin94,pin96,pik96,pir96,kpp94} 
can be associated with the
zero mode(s) of the fields in a quantum field theory defined 
in a finite spatial
volume and quantized at equal light-cone time \cite{leb80}.

\subsection {Constrained Zero Modes}
 
As mentioned previously, the light-front vacuum state is simple; 
it contains no particles in a massive theory.  
In other words, the Fock space vacuum is the physical vacuum.  
However, one commonly associates important long range properties
of a field theory with the vacuum:  spontaneous symmetry breaking, 
the Goldstone pion, and color confinement.  
How do these complicated phenomena manifest
themselves in light-front field theory?  

If one cannot associate long range phenomena with the vacuum state 
itself, then the only alternative is the zero momentum components 
or ``zero modes'' of the
field (long range $\leftrightarrow$ zero momentum).  
In some cases, the zero mode operator is not an independent degree 
of freedom but obeys a constraint equation. 
Consequently, it is a complicated operator-valued function of all 
the other modes of the field~\cite{may76}.  

This problem has recently been attacked from several directions. 
The question of whether boundary conditions can be consistently 
defined in  light-front quantization has been discussed by 
McCartor and 
Robertson~\cite{mcc88,mcc91,mcr92,mcr94,mcc94,mcr95,mcr97},
and by Lenz~\cite{len90,ltl91}. 
They have shown that for massive theories the energy and
momentum derived from 
light-front quantization are conserved and are equivalent 
to the energy and momentum one would normally write down in 
an equal-time theory. 
In the analyses of Lenz {\em et al.}~\cite{len90,ltl91} and  
Hornbostel~\cite{hor92} one traces the fate of the equal-time 
vacuum in the limit  $P^3 \to \infty$ and
equivalently  in the limit $\theta \to \pi/2$ 
when rotating the evolution parameter 
$\tau= x^0 \cos \theta + x^3 \sin \theta$ from the instant 
parametrization to the front parametrization.
Heinzl and Werner 
{\em etal.}~\cite{hkw91a,hkw91b,hkw91c,hkw92a,hkw92b,hei96b}  
considered 
$\phi^4$ theory in (1+1)--dimensions and attempted to solve 
the zero mode constraint equation by truncating the equation 
to one  particle.  
Other authors~\cite{hav87,hav88,rob93} find that, 
for  theories  allowing spontaneous symmetry breaking, 
there is a degeneracy  of light-front vacua and the true vacuum
state can differ from the perturbative vacuum through the addition 
of  zero mode quanta.   
In addition to these approaches there are many  others, 
like \cite{bur93,pau92,kal95}, \cite{bjs93,ctj95,hul93,juj94}, 
or \cite{cop94,hei96b,kss96}. 
Grange {\it et al.} \cite{bgw93,bgw95} have dealt with a broken 
phase in such scalar models, see also \cite{cha93,gho92,krp93}.

An analysis of the zero mode constraint  equation for (1+1)--dimensional
$\phi^4$ field theory  [$\left(\phi^4\right)_{1+1}$] with symmetric  
boundary conditions shows how spontaneous symmetry breaking occurs 
within the context of this model.  
This theory has a $Z_2$ symmetry $\phi \rightarrow - \phi$ 
which is spontaneously broken for some values of the mass and coupling.   
The approach of Pinsky, van de Sande and 
Bender \cite{bpv93,piv94,hpv95} is to apply a
Tamm-Dancoff truncation to the Fock space.  
Thus operators are finite matrices and
the operator valued constraint equation can be solved numerically. 
The truncation assumes that states  with a large number of particles 
or large  momentum do not have
an important  contribution to the zero mode.

\begin{figure} [t]
\begin{minipage}[t]{75mm}
\epsfxsize=50mm\epsfbox{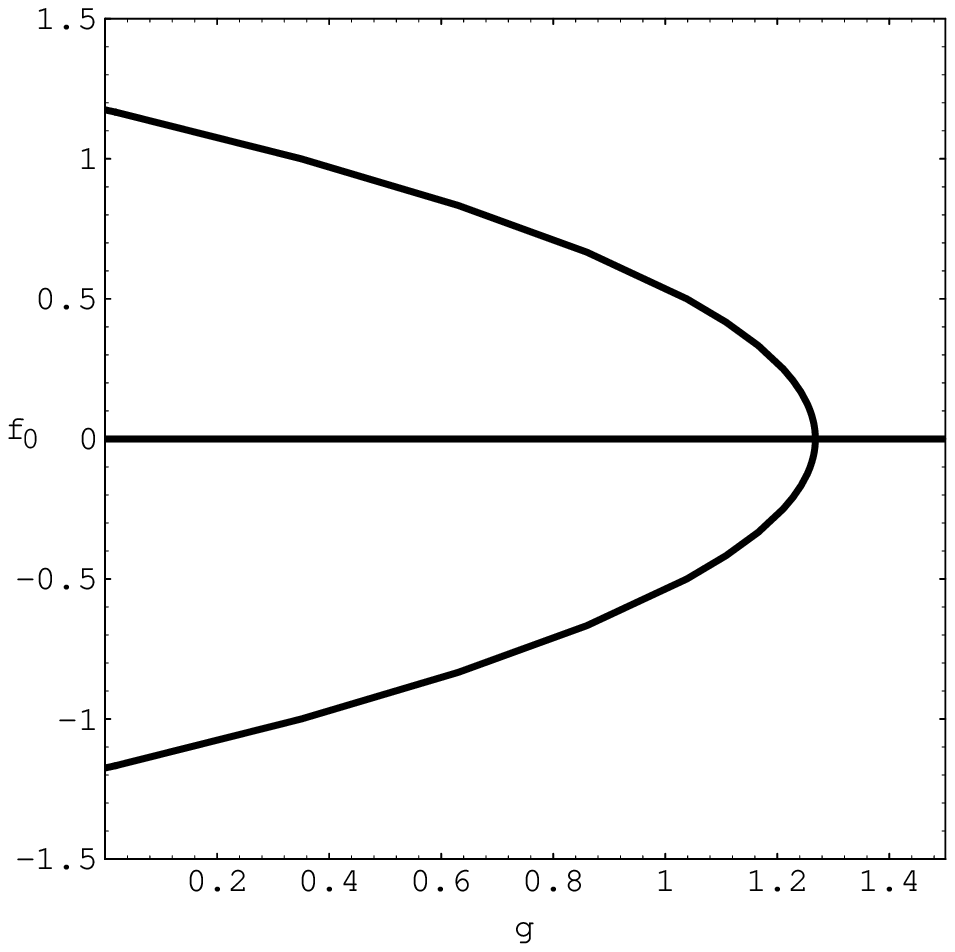} 
\caption{\label{ff1}
$f_0=\protect\sqrt{4\pi} \langle 0|\phi| 0\rangle$ vs. 
$g=24\pi\mu^2/\lambda$ in the one mode case with $N=10$.}
\end{minipage}
\hfill
\begin{minipage}[t]{75mm}
\epsfxsize=75mm\epsfbox{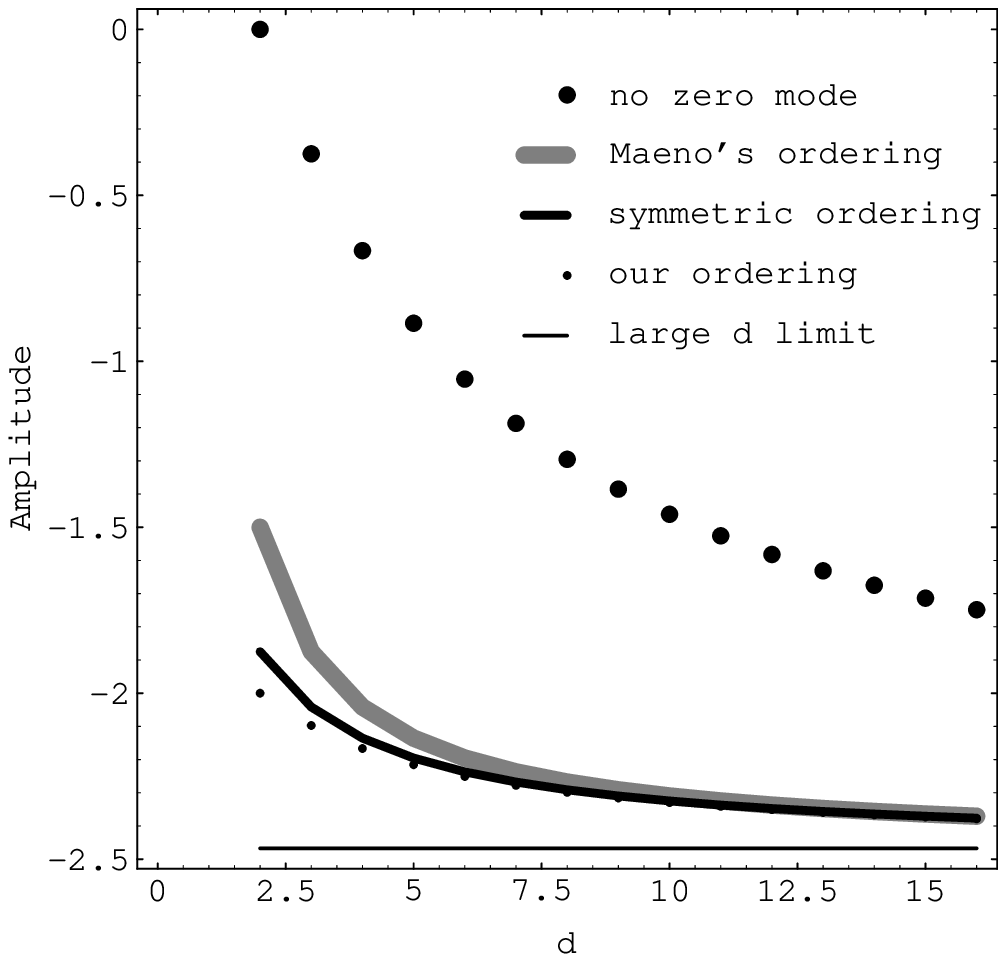}
\caption{\label{ff14} 
Convergence to the large $d$ limit of $1\protect\to 1$ 
setting $E=g/p$ and dropping any constant terms.}
\end{minipage}
\end{figure}

Since this represents a completely new paradigm for spontaneous 
symmetry breaking we
will present this calculation in some detail. 
One finds the following general behavior:  
for small coupling 
(large $g$, where  $g \propto 1/{\rm coupling}$) 
the constraint equation has a single solution and the field has no
vacuum expectation value (VEV). As one increase the coupling 
(decrease $g$) to  the ``critical coupling'' $g_{\rm critical}$, 
two  additional solutions which give the
field a nonzero VEV appear.   
These solutions differ only infinitesimally from the
first solution near the  critical coupling, 
indicating  the presence of a second order phase transition. 
Above the critical coupling ($g < g_{\rm critical}$), there
are three solutions:  one with zero VEV, the ``unbroken phase,'' 
and two with nonzero VEV, the ``broken phase''.   
The ``critical curves'' shown in Figure~\ref{ff1}, is a
plot the VEV as a function of $g$.  
%
%

Since the vacuum in this theory is trivial, all of the long range  properties 
must occur in the operator structure of the Hamiltonian. Above the critical
coupling ($g < g_{\rm critical}$) quantum oscillations spontaneously break the 
$Z_2$ symmetry of the theory. In a loose analogy with a symmetric  double well
potential,  one has two new Hamiltonians for the  broken phase, each producing
states localized in one of the wells.   The structure of the two Hamiltonians is
determined from the broken phase  solutions of the zero mode constraint
equation.   One finds that the two Hamiltonians have  equivalent spectra.   In a
discrete theory without zero modes it is well known that, if one  increases  the
coupling sufficiently, quantum correction will generate tachyons  causing the 
theory to break down near the critical coupling.   Here the zero mode generates
new interactions that prevent  tachyons from developing.   
In effect what happens is that, while  quantum  corrections attempt to 
drive the mass negative, they also change the  vacuum energy  
through the zero mode and the  mass eigenvalue
can never catch the  vacuum eigenvalue.   
Thus, tachyons never appear in the spectra.

In the weak coupling limit ($g$ large) the  solution to the constraint equation
can be obtained in perturbation theory.  This solution does not break the $Z_2$
symmetry and is believed to simply insert 
the missing zero momentum contributions into internal propagators.  
This must  happen if light-front perturbation theory
is to agree with equal-time perturbation  
theory~\cite{cry73a,cry73b,chy73}.  

Another way to investigate the zero mode is to study the spectrum of the field
operator  $\phi$. Here one finds a picture that agrees with the symmetric double
well potential analogy.  
In the broken phase, the field is localized in one of the
minima of the  potential and there is tunneling to the other minimum.

\subsubsection{Canonical Quantization}

For a classical field the $(\phi^4)_{1+1}$  Lagrange density is
\begin{equation} {\cal L} = \partial_+\phi\partial_-\phi 
- {{\mu^2}\over 2} \phi^2
-  {\lambda
\over 4!} \phi^4\;. 
\end{equation}
One puts the system in a box of length $d$ and impose periodic  boundary
conditions.   Then
\begin{equation}
\phi\!\left(x\right) = {1\over{\sqrt d}} 
\sum_n q_n(x^+) e^{i k_n^+ x^-}\; , 
\end{equation}
where $k_n^+ =2\pi n/d$ and summations run over all integers 
unless otherwise noted.

It is convenient to define the integral 
$\int dx^- \, \phi(x)^n-(zero modes) =
\Sigma_n$. 
In term of the modes of the field it has the form, 
\begin{equation}
\Sigma_n = {1\over{n!}} \sum_{i_1, i_2, \ldots, i_n \neq 0} q_{i_1}  q_{i_2}
\ldots q_{i_n}\, \delta_{i_1 + i_2 + \ldots + i_n, 0}. 
\end{equation}
Then the canonical Hamiltonian is
\begin{equation} P^- = {{\mu^2 q_0^2}\over 2} + \mu^2 \Sigma_2 + {{\lambda 
q_0^4}\over{4! d}}
 + {{\lambda q_0^2 \Sigma_2}\over{2! d}} + {{\lambda q_0 \Sigma_3}\over  d} +
{{\lambda \Sigma_4}\over d}\;. 
\end{equation}
Following the Dirac-Bergman prescription, 
described in Appendix~\ref{sec:Bergmann},
one identify  first-class constraints which define the conjugate momenta
\begin{equation} 0 = p_n - i k_n^+ q_{-n}\;, 
\end{equation}
where
\begin{equation}
\left[q_m,p_n\right] = {{\delta_{n, m}}\over 2}\;,\quad m ,n \neq 0
\;. 
\end{equation}

The secondary constraint is 
\cite{wit89},
\begin{equation} 0 = \mu^2 q_0 + {{\lambda q_0^3}\over{3! d}} 
+ {{\lambda q_0 
\Sigma_2}\over d} + {{\lambda \Sigma_3}\over d}\; , 
\end{equation}
which determines the zero mode $q_0$. 
This result can also  be obtained by
integrating the equations of motion. 

To quantize the system one replaces the classical fields with the  
corresponding field operators, and the Dirac bracket by  
$i$ times a commutator. One must choose
a regularization and an operator-ordering prescription 
in order to make the system well-defined.
 
One begin by defining creation and annihilation operators 
$a_k^\dagger$ and $a_k$,
\begin{equation} q_k = \sqrt {d\over{4 \pi \left| k \right|}} \: a_k\;,
\quad a_k = a_{-k}^\dagger\;,\quad  k\neq 0\; , 
\end{equation}
which satisfy the usual commutation relations
\begin{equation}
\left[a_k,a_l\right] =0\;,
\quad\left[a_k^\dagger,a_l^\dagger\right] =0\;,\quad
\left[a_k,a_l^\dagger\right] =\delta_{k, l}\; ,\quad k,l > 0\; .
\end{equation}
Likewise, one defines the zero mode operator
\begin{equation} q_0 = \sqrt{d\over{4 \pi}} \: a_0\;.
\end{equation}
In the quantum case, one  normal orders the operator $\Sigma_n$. 

General arguments suggest that the Hamiltonian  should be symmetric
ordered~\cite{bmp86}. However, it is not clear how one should treat the 
zero mode since it is not a dynamical field. As an {\em ansatz} one treats
$a_0$ as an ordinary field operator when symmetric ordering the Hamiltonian.  
The tadpoles are removed from the symmetric ordered Hamiltonian by  
normal ordering
the terms having no zero mode factors and by subtracting,
\begin{equation} {3\over{2}}\; a_0^2\sum_{n\neq 0} {1\over{|n|}} \; .
\end{equation}
In addition, one subtracts a constant so that the VEV of $H$ is zero. 
Note that this renormalization prescription is equivalent to a conventional mass
renormalization and  does not introduce any new operators 
into the Hamiltonian. The constraint equation for the zero mode can
be obtained by taking a derivative of $P^-$ with respect to $a_0$. One finds,
\begin{equation} 0 = g a_0 + a_0^3 +  \sum_{n\neq 0} {1\over{|n|}}
\left( a_0 a_n a_{-n} + a_n a_{-n} a_0 + a_n a_0 a_{-n} -
\frac{3 a_0}{2} \right)+ 6 \Sigma_3 \; . \label{constraint}
\end{equation}
where $g= 24 \pi \mu^2/\lambda$. 
It is clear from the general structure of 
(\ref{constraint}) that $a_0$ as a
function of the other modes is not necessarily odd under the 
transform $a_k \to
-a_k$,  ($k \neq 0$ )associated with the $Z_2$ symmetry of the system. 
Consequently, the zero mode can induce $Z_2$ 
symmetry breaking in the Hamiltonian.

In order to render the problem tractable, 
we impose a Tamm-Dancoff truncation on the Fock space. 
One defines $M$ to be the number of nonzero modes and $N$ to be the 
maximum number of allowed particles.  
Thus, each state in the truncated Fock space
can be represented by a vector of length 
$S=\left(M+N\right)!/\left(M! N! \right)$
and  operators  can be represented by $S \times S$ matrices.   
One can define the
usual Fock space basis, $\left|n_1,n_2,\ldots,n_M\right\rangle$.
where $n_1 + n_2 + \ldots +n_M \leq N$.  In matrix form, $a_0$ 
is real and symmetric. 
Moreover, it is  block diagonal in states of equal $P^+$ eigenvalue. 
\subsubsection{Perturbative Solution of the Constraints}
 
In the limit of large $g$, one can solve the constraint equation
perturbatively  . Then one substitutes the
solution back into the Hamiltonian and calculates various 
amplitudes to arbitrary
order in $1/g$ using Hamiltonian  perturbation theory.  

It can be shown that the solutions of the constraint equation
and the  resulting Hamiltonian are divergence free to all orders 
in perturbation
theory for both the broken and unbroken phases.
To do this one starts with the perturbative solution for the zero 
mode in the unbroken  phase,
\begin{equation}
 a_0 = - \frac{6}{g} \Sigma_3 +  \frac{6}{g^2}\left(2 \Sigma_2
\Sigma_3+ 2
\Sigma_3  \Sigma_2+\sum_{k=1}^M \frac{a_k \Sigma_3 a_k^\dagger 
+ a_k^\dagger
\Sigma_3 a_k -\Sigma_3}{k}\right)  + O(1/g^3) \; .
\label{pzm}
\end{equation}
and substitutes this into the Hamiltonian to obtain a complicated 
but well defined expression for the Hamiltonian in terms of the 
dynamical operators.

The finite volume box acts as an infra-red regulator and the only 
possible divergences are ultra-violet.  
Using diagrammatic language, any loop  of momentum
$k$ with $\ell$ internal lines has asymptotic form $k^{-\ell}$. 
Only the case of
tadpoles ${\ell}=1$ is divergent.  
If there are multiple loops, the effect is to
put factors of $\ln(k)$ in the numerator and the divergence 
structure is unchanged.  
Looking at Equation~(\ref{pzm}), the only possible tadpole is from the
contraction in the term
\begin{equation}
\frac{a_k \Sigma_3 a_{-k}}{k}
\end{equation}
which is canceled by the $\Sigma_3/k$ term.  This happens to all orders in
perturbation theory:  each tadpole has an associated term which cancels it.  
Likewise, in the Hamiltonian one has similar cancelations to all orders in perturbation
theory.  

For the unbroken phase, the effect of the zero mode should vanish in the  
infinite volume limit, giving a ``measure zero'' 
contribution to the continuum Hamiltonian.
However, for finite box volume the zero mode does contribute,  compensating
for the fact that the longest wavelength mode has been  removed from the system.
Thus, inclusion of the zero mode improves convergence to the infinite  volume
limit.  In addition, one can use
the perturbative expansion of the zero mode to study the operator ordering
problem.  One can directly compare our operator ordering ansatz with 
a truly Weyl ordered Hamiltonian and with Maeno's operator 
ordering ansatz \cite{mae94}.

As an example, let us examine $O(\lambda^2)$ contributions to the 
processes $1\rightarrow 1$. 
As shown in Figure~\ref{ff14}, including the zero
mode greatly  improves convergence to the large volume limit. 
The zero mode compensates 
for the fact that one  have removed the longest
wavelength mode from the system.

\subsubsection{Non-Perturbative Solution: One Mode, Many Particles}
 
Consider the case of one mode $M=1$ and many particles.   
In this case, the zero-mode is diagonal and can be written as
\begin{equation} a_0 = f_0 \left| 0 \right\rangle\left\langle 0 \right| +
{\sum_{k=1}^N f_{k} \left| k \right\rangle\left\langle k \right|}\; .
\label{onemode}
\end{equation}

Note that $a_0$ in (\ref{onemode}) is even under 
$a_k \to -a_k$, $k \neq 0$ and any non-zero solution  
breaks the $Z_2$ symmetry of
the original Hamiltonian. The VEV is given by
\begin{equation}
\langle 0 | \phi | 0 \rangle ={1\over{\sqrt{4 \pi}}}
\langle 0| a_0 | 0\rangle = {1\over{\sqrt{4 \pi}}} f_0 \; .
\end{equation}
Substituting (\ref{onemode}) into the constraint equation  (\ref{constraint})
and sandwiching the constraint equation between Fock states, 
one get a recursion
relation for $\left\{f_{n}\right\}$:
\begin{equation} 0 =  g f_{n} + {f_{n}}^{3} + (4n - 1) f_{n} +
\left(n+1\right) f_{n+1} + n f_{n-1} \label{recursion}
\end{equation}

where $n \leq N$, and one define $f_{N+1}$ to be unknown. Thus, 
$\left\{ f_{1}, f_{2}, \ldots, f_{N+1} \right\}$  is uniquely determined by a 
given choice of $g$ and $f_0$. In particular, if $f_{0}=0$ all the $f_{k}$'s are
zero independent of $g$. This is the unbroken phase. 

Consider the asymptotic 
behavior for large $n$.  If $f_{n}\gg 1$, the ${f_{n}}^3$ 
term will  dominate and
\begin{equation} f_{n+1} \sim {{f_{n}^3}\over n}\;,  
\end{equation}
thus,
\begin{equation} {\lim_{n\to\infty} f_{n}} \sim (-1)^n \exp\!\left(3^n  
{\rm constant}\right)\;.  
\end{equation}
One must reject this rapidly growing solution. 
One only seek solutions where $f_{n}$ is small for large $n$. 
For large $n$, the terms linear in $n$ 
dominate and  Equation~(\ref{recursion}) becomes
\begin{equation} f_{n+1} + 4 f_{n} + f_{n-1} = 0\; .
\end{equation}
There are two solutions to this equation:
\begin{equation} f_{n} \propto \left(\sqrt{3} \pm 2\right)^{n}\; .
\end{equation}
One must reject the plus solution because it grows with $n$.    
This gives the condition
\begin{equation} -{{\sqrt{3} - 3 + g}\over{2\sqrt 3}} = K\;,
\quad K= 0, 1, 2\ldots \label{criticals}
\end{equation}
Concentrating on the $K=0$ case, one find a critical coupling
\begin{equation} g_{\rm critical} = 3 - \sqrt{3}
\end{equation}
or
\begin{equation}
\lambda_{\rm critical}=4\pi\left(3+\sqrt{3}\right) \mu^2 \approx 60\mu^2,
\end{equation}
In comparison, values of 
$\lambda_{\rm critical}$ from $ 22 \mu^2 $ to $ 55 \mu^2 $ 
have  been reported for
equal-time quantized calculations \cite{cha76,aep76,fkk87,krg87}. 
The solution to the
linearized equation is an approximate solution to the full
Equation~(\ref{recursion})  for
$f_0$ sufficiently small. Next, one need to determine solutions of 
the full nonlinear equation which converge for large $n$.

One can study the critical curves by looking for numerical solutions to
Equation~(\ref{recursion}). 
The method used here is to find values of $f_{0}$ and
$g$ such that $f_{N+1}=0$. Since one seeks a solution where $f_{n}$ 
is decreasing with $n$,  this is a good approximation.  
One finds that for $g>3-\sqrt 3$ the only
real solution is $f_{n}=0$ for all $n$.  
For $g$ less than $3-\sqrt{3}$ there are two additional solutions.  
Near the critical point $\left| f_{0} \right|$ is small and
\begin{equation} f_{n} \approx f_{0} \left(2 - \sqrt 3\right)^n\; .
\end{equation}
The critical curves are shown in Figure~\ref{ff1}. These solutions converge
quite rapidly with $N$. The critical curve for the broken phase is 
approximately parabolic in shape:
\begin{equation} g \approx  3-\sqrt 3 - 0.9177 f_{0}^2 \; . 
\end{equation}

\begin{figure} [t]
\begin{minipage}[t]{60mm}
\makebox[0mm]{}
\epsfxsize=60mm\epsfbox{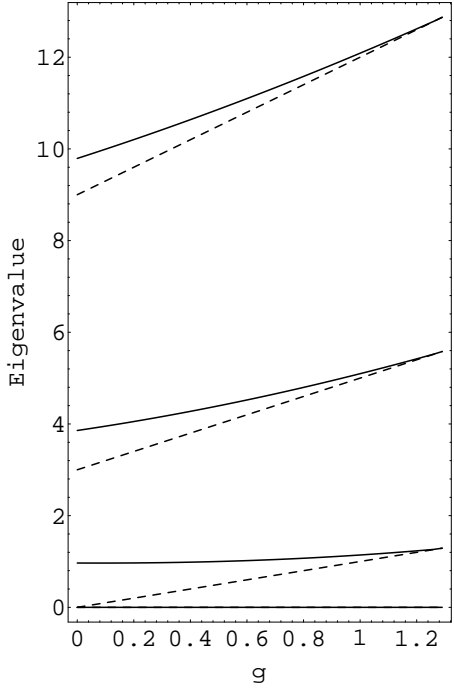} 
\end{minipage}
\hfill
\begin{minipage}[t]{70mm}
{\makebox[0mm]{}
\caption{\label{ff4} 
The lowest three energy eigenvalues for the one mode case as
a function of $g$ from the numerical solution 
of Equation~(\protect\ref{onehamiltonian}) 
with $N=10$.  
The dashed lines are for the unbroken phase $f_0 = 0$  
and the solid lines are for the broken phase $f_0\neq 0$.}
}\end{minipage}
\end{figure}

One can also study the eigenvalues of the Hamiltonian for the one mode case.  
The Hamiltonian is diagonal for this Fock space truncation and,
\begin{equation}
\left\langle n \right| H \left| n \right\rangle = {3\over 2} n (n-1) + n g
-{f_n^4\over 4} - {{2 n+1}\over 4} f_n^2 +{{n+1}\over 4}  f_{n+1}^2 + {n\over 4}
f_{n-1}^2 - C\; .
\label{onehamiltonian}
\end{equation}
The invariant mass eigenvalues are given by
\begin{equation}
   P^2 | n \rangle =
   2 P^+ P^- | n \rangle = 
\frac{n \lambda  \langle n | H | n \rangle}{24 \pi} | n \rangle
\end{equation}
In Figure~\ref{ff4} the dashed lines show
%
%
the first few eigenvalues as a function of $g$ without the zero-mode.   
When one includes the broken phase of the zero mode, 
the energy levels shift as shown by the solid curves.  
For $g < g_{\rm critical}$  the energy levels increase  above the
value they had without the zero mode.  The higher levels change very little 
because $f_n$ is small for large $n$.

In the more general case of many modes and
many particles many of the features that were seen in the one mode and one
particle cases remain.   
In order to calculate the zero mode for a given value of $g$ one converts the
constraint equation (\ref{constraint}) into an $S \times S$ matrix equation 
in the truncated Fock space. 
This becomes a set of $S^2$ coupled cubic equations and one
can solve for the matrix elements of $a_0$  numerically. Considerable
simplification occurs because $a_0$ is symmetric and is block diagonal in states
of equal momentum.   
For example, in the case $M=3$, $N=3$, the number of coupled
equations is 34 instead of $S^2=400$.   In order to find the critical coupling,
one take $\langle 0|a_0|0\rangle$ as given and $g$ as unknown and solve the
constraint  equation for $g$ and the other  
matrix elements of $a_0$ in the limit
of small but nonzero $\langle 0|a_0|0\rangle$.  
One sees that the solution   
quick convergence as $N$ increases and that there is 
a logarithmic divergence as $M$ increases. 
The logarithmic divergence of $g_{critical}$ is the major remaining
remaining missing part of this calculation and requires a careful
non-perturbative renormalization 
\cite{kww80}.
  
When one substitutes the solutions for the broken phase of 
$a_0$ into the Hamiltonian 
one gets two Hamiltonians $H^+$ and $H^-$
corresponding to the two signs of 
$\langle 0 | a_0 | 0 \rangle$ and the two branches of the curve in 
Figure~\ref{ff1}. This is the new paradigm for spontaneous symmetry breaking:  
multiple vacua  are replaced by multiple Hamiltonians.   Picking the Hamiltonian
defines the  theory in the same sense that picking the vacuum defines the theory
in  the equal-time paradigm.   The two solutions for $a_0$ are related to each
other  in a very specific way.  
Let $\Pi$ be the unitary operator associated with
the $Z_2$ symmetry of the system; 
$\Pi a_k \Pi^\dagger = -a_k$, $k \neq 0$.   
One breaks up $a_0$ into an even part
$\Pi a_0^E \Pi^\dagger = a_0^E$  and an odd part $\Pi a_0^O \Pi^\dagger =
-a_0^O$.   The even part $a_0^E$ breaks the 
$Z_2$ symmetry of the theory. For $g<g_{\rm critical}$,  the three solutions of
the constraint equation are:  $a_0^O$  corresponding to  the unbroken phase,
$a_0^O+a_0^E$ corresponding to the 
$\langle 0 | a_0 |0 \rangle >0$ solution, and
$a_0^O-a_0^E$ for the 
$\langle 0 | a_0 |0 \rangle <0$ solution. Thus, the two Hamiltonians are 
\begin{equation} H^+ = H\left(a_k, a_0 ^O + a_0^E\right) 
\end{equation}
and
\begin{equation} H^- = H\left(a_k, a_0^O - a_0^E\right) 
\end{equation}
where H has the property
\begin{equation} H\left(a_k, a_0\right) = H\left(-a_k, -a_0\right)
\end{equation}
and $a_k$ represents the nonzero modes. Since $\Pi$ is a unitary operator, if
$|\Psi\rangle$ is an eigenvector of $H$ with eigenvalue $E$ then $\Pi
|\Psi\rangle$ is an eigenvalue of $\Pi H \Pi^\dagger$  with eigenvalue $E$. 
Since,
\begin{eqnarray}
\Pi H^- \Pi^\dagger &=& 
\Pi H\left(a_k,a_0^O - a_0^E\right) \Pi^\dagger = 
          H\left(-a_k ,-a_0^O-a_0^E\right) \nonumber\\
    &=& H\left(a_k,a_0^O +a_0^E\right) = H^+  \; ,
\end{eqnarray}
$H^+$ and $H^-$ have the same eigenvalues.

Consider the $M=3$, $N=3$ case as an example and 
let  us examine the spectrum of $H$.
For large $g$ the eigenvalues are obviously: $0$, $g$, $g/2$, $2g$,
$g/3$, $3g/2$ and $3g$.   However as one decreases $g$ one of the last three 
eigenvalues will be driven negative.   This signals the breakdown of the theory
near  the critical coupling when the zero mode is not included.

%
%
%
\begin{figure} [t]
\begin{minipage}[t]{50mm}
\epsfxsize=50mm\epsfbox{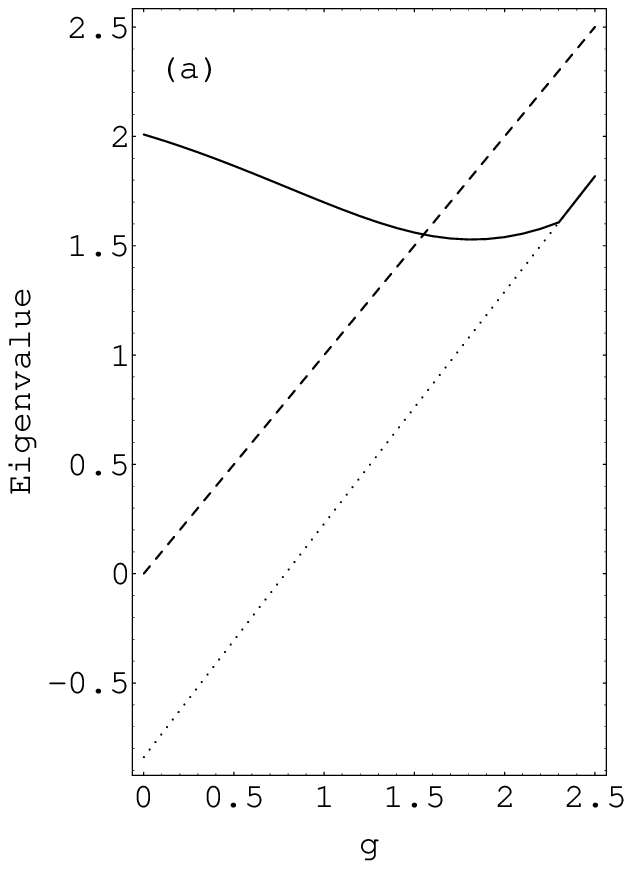}
\end{minipage}
\begin{minipage}[t]{50mm}
\epsfxsize=50mm\epsfbox{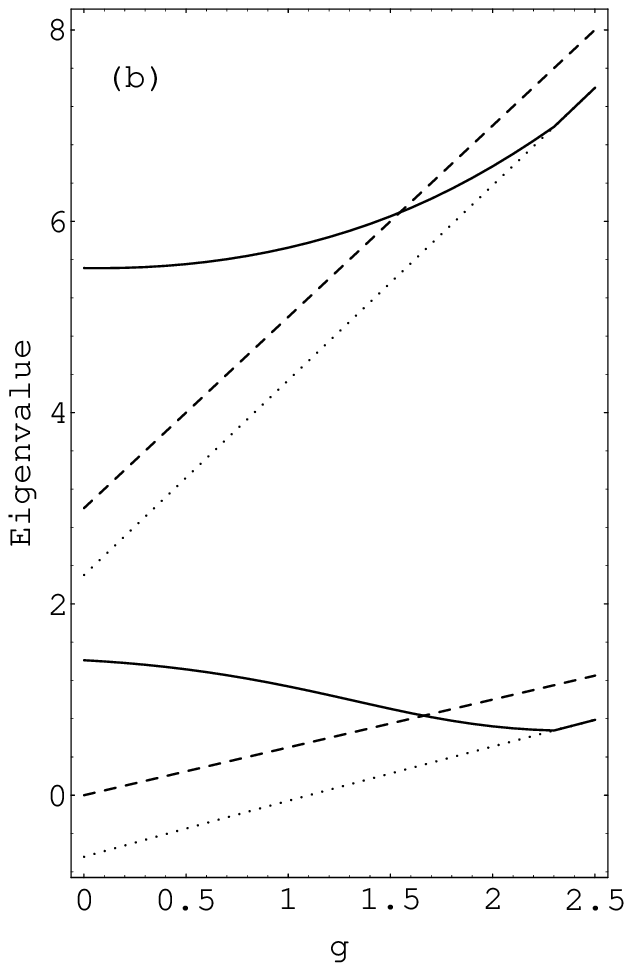}
\end{minipage}
\begin{minipage}[t]{50mm}
\epsfxsize=50mm\epsfbox{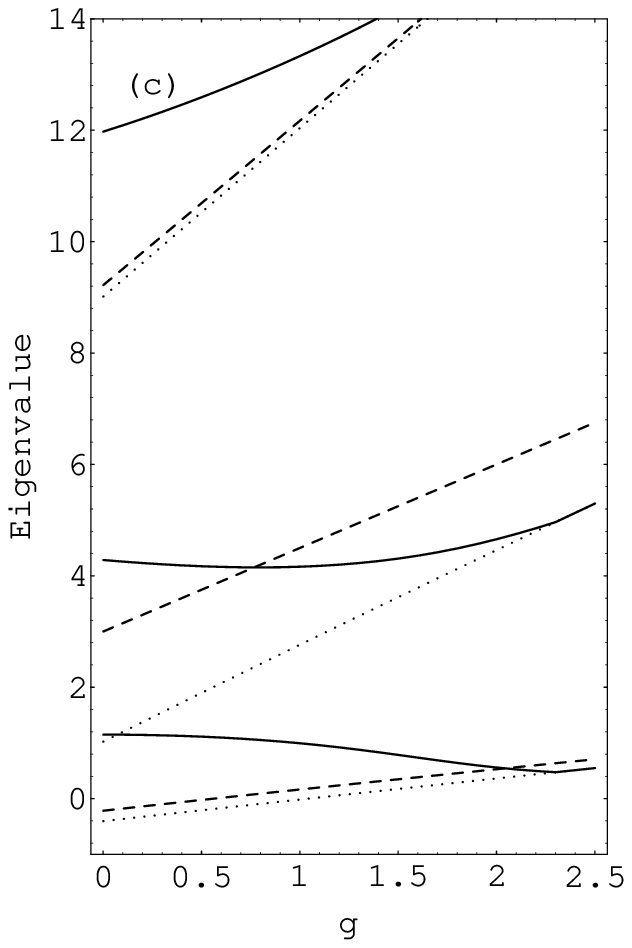}
\end{minipage}
\caption{\label{ff9a} 
The spectrum for 
(a) $P^{+}=2\pi/d$, (b) $P^{+}=4\pi/d$, and
(c) $P^{+}=6\pi/d$, all with $M=3$, $N=3$.
The dashed line shows the spectrum with no zero mode. 
The dotted line is the unbroken phase and the 
solid line is the broken phase. }
\label{ff9}
\end{figure}

Including the zero mode  fixes this problem. 
Figure~\ref{ff9} shows the spectrum
for the three lowest nonzero momentum sectors.   
This spectrum illustrates several
characteristics which  seem to hold generally (at least for truncations
 that have been examined,  $N+M \leq 6$).   
For the broken phase, the vacuum is the
lowest energy  state, there are no level crossings as a function of $g$,  
and the theory does not break down  in the vicinity of the critical point. 
None of these are true for the spectrum with the zero mode removed or 
for the unbroken phase  below the critical coupling.

One can also investigate the shape of the critical curve near the critical
coupling as a function of the cutoff $K$.  In scalar field theory, 
$\langle 0 |\phi|0\rangle$ acts as  the order parameter of the
theory.  Near the critical coupling,  one can fit the VEV to some
power of 
$g-g_{\rm critical}$; this will give us the associated critical exponent 
$\beta$,
\begin{equation}
\langle 0 | a_0 | 0 \rangle \propto \left(g_{\rm critical}-g\right)^\beta \; .
\end{equation}
Pinsky, van de Sande and Hiller \cite{hpv95} have calculated this 
as a function of cutoff and found a result consistent with $\beta=1/2$, 
independent of cutoff $K$. 
The theory
$\left(\phi^4\right)_{1+1}$ is in the same universality class as the 
Ising model in 2 dimensions and the correct critical exponent 
for this universality class is $\beta=1/8$. 
If one were to use the mean field approximation to calculate the
critical exponent,  the  result would be $\beta=1/2$. This is what was
obtained in this calculation.  
Usually, the presence of a mean field result
indicates that one is not probing all length scales properly. 
If one had a cutoff $K$ large enough to include many length scales, 
then the critical exponent should approach the correct value.   
However, one cannot be certain that this is the
correct explanation of our result since no evidence that 
$\beta$ decreases with increase $K$ is seen.

\subsubsection{Spectrum of the Field Operator}
How does the zero mode affect the field itself? Since $\phi$ is a
Hermitian operator it is an observable of  the system and one can measure
$\phi$ for a given  state $|\alpha\rangle$.   $\tilde{\phi}_i$ and 
$|\chi_i\rangle$ are the eigenvalue and eigenvector respectively of 
$\sqrt{4 \pi}\phi\,$:
\begin{equation}
 \sqrt{4 \pi} \phi \, |\chi_i\rangle = 
\tilde{\phi}_i|\chi_i\rangle \;,
\;\;\;\; \langle \chi_i | \chi_j \rangle = \delta_{i,j}\; .
\end{equation}
The expectation value of $\sqrt{4 \pi} \phi$ in the state 
$|\alpha\rangle$ is
$\left|\langle
\chi_i |
\alpha \rangle \right|^2$.  
%
%
\begin{figure} [t]
\centerline{
\epsfbox{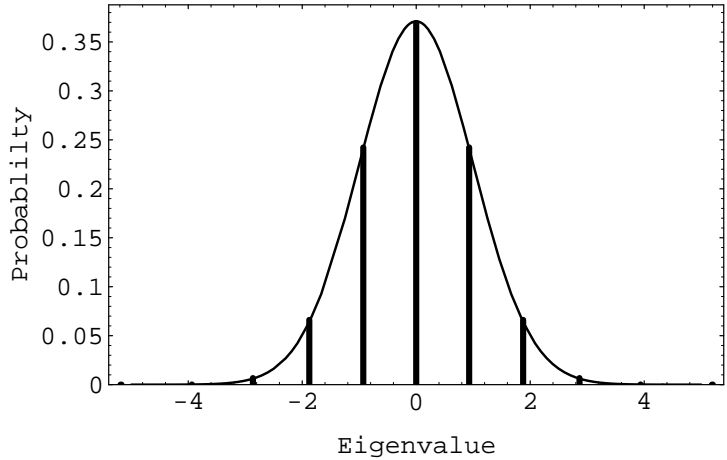}
}\caption{\label{ff10} 
Probability distribution of eigenvalues  of 
$\protect\sqrt{4 \pi}\phi$ 
for the vacuum with  $M=1$, $N=10$, and no zero mode.  
Also shown is the infinite $N$ limit from
Equation~(\protect\ref{gaussian}). }
\end{figure}

In the limit of large $N$, the probability distribution becomes continuous.  
If one ignores the zero mode, the probability of obtaining 
$\tilde{\phi}$ as the result of a measurement of 
$\sqrt{4 \pi} \phi$ for the vacuum state is
\begin{equation} P\left(\tilde{\phi}\right) = \frac{1}{\sqrt{2 \pi \tau}}\, 
\exp\left(-\frac{\tilde{\phi}^2}{2 \tau}\right)\, d\tilde{\phi} 
\label{gaussian}
\end{equation}
where $\tau = \sum_{k=1}^M 1/k$.  The probability distribution comes  from the 
ground state wave function of the Harmonic oscillator where one identify $\phi$
with the position operator. This is just the Gaussian fluctuation of a free
field.   Note that the width of the Gaussian diverges logarithmically in $M$.  
When $N$ is finite, the distribution becomes discrete as  shown in
Figure~\ref{ff10}.
%
%
\begin{figure} [t]
\begin{minipage}[t]{60mm}
\epsfxsize=60mm\epsfbox{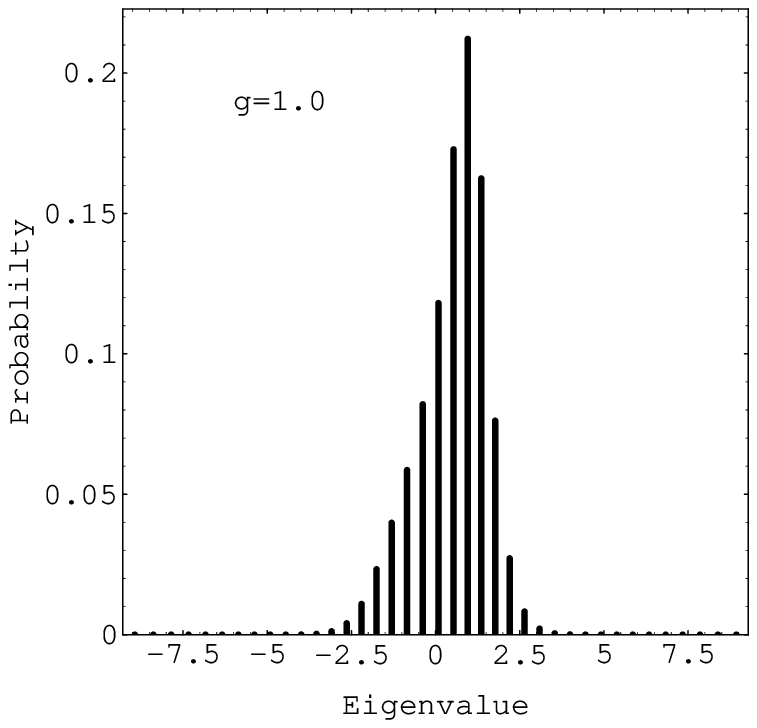} 
\end{minipage}
\hfill
\begin{minipage}[t]{60mm}
\epsfxsize=60mm\epsfbox{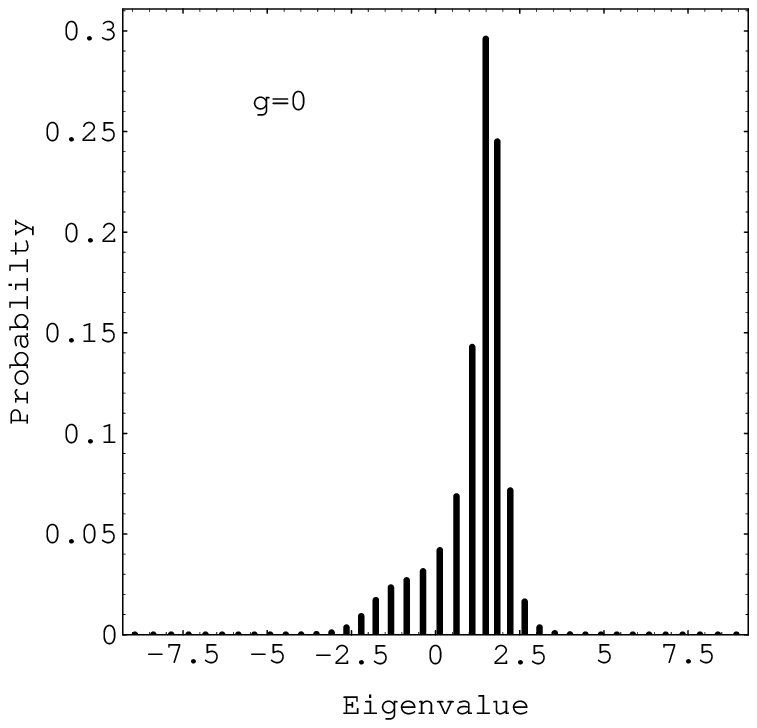}
\end{minipage}
\begin{minipage}[b]{60mm}
\epsfxsize=60mm\epsfbox{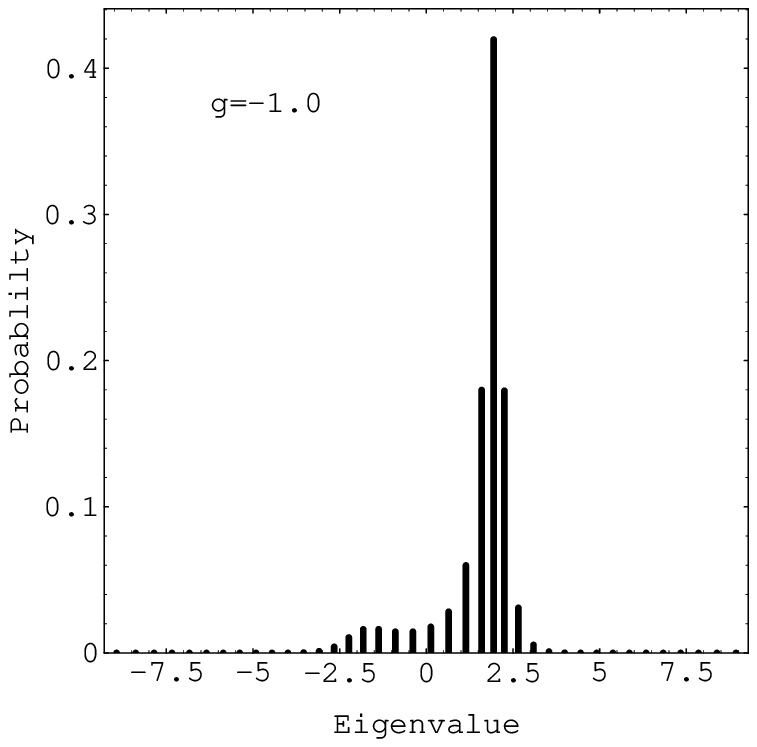}
\end{minipage}
\hfill
\begin{minipage}[b]{60mm}
\epsfxsize=60mm\epsfbox{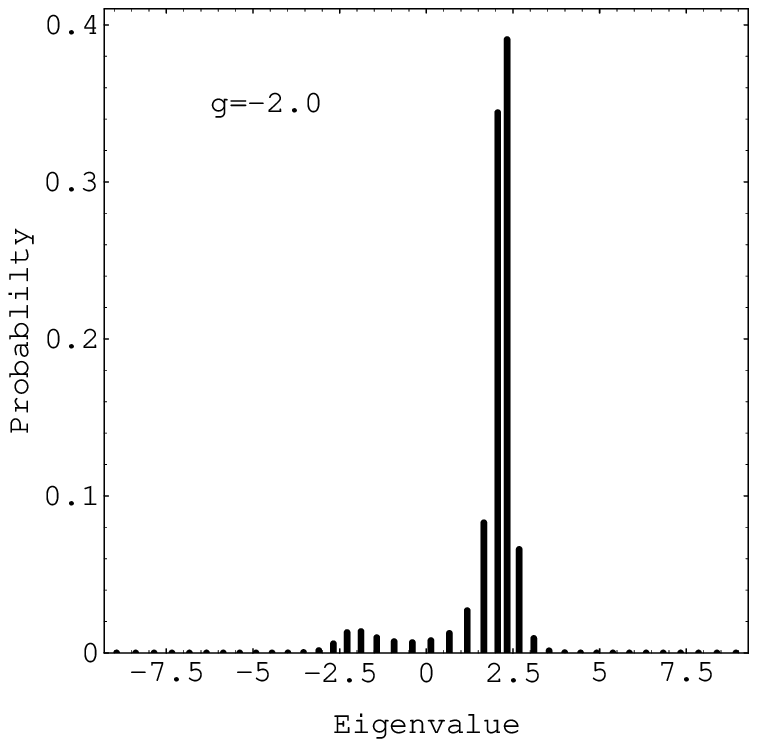}
\end{minipage}
\caption{ \label{ff11} 
Probability distribution of eigenvalues of 
$\protect\sqrt{4\pi}\phi$ 
for the vacuum with couplings 
$g=1$, $g=0$, $g=-1$, and $g=-2$, all for 
$M=1$ and $N=50$. The positive VEV solution to 
the constraint equation is used. } 
\end{figure}

In general, there are $N+1$ eigenvalues such that 
$\langle \chi_i | 0 \rangle\neq 0$, independent of $M$.   
Thus if one wants to examine the spectrum of the field
operator for the vacuum state, it is better to choose Fock space 
truncations where $N$ is large.   
With this in mind, one examines the $N=50$ and
$M=1$ case as a  function of $g$ in Figure~\ref{ff11}. 
Note that near the critical point, Figure~\ref{ff11}a, 
the  distribution is approximately equal to the free
field case shown in Figure~\ref{ff10}.  
As one moves away from the critical point,
Figures~\ref{ff11}b--d, the distribution  becomes increasingly 
narrow with a peak
located at the VEV of what would be the minimum of the symmetric double well 
potential in the equal-time paradigm. In addition, there is a small peak
corresponding to minus the VEV.  In the language of the equal-time paradigm, 
there is tunneling between the two minima of the potential. The
spectrum of $\phi$ has been examined for other values of $M$ and
$N$; the results are consistent with the example discussed here.
\subsection {Physical Picture and Classification of Zero Modes } 

When considering a gauge theory, there is a ``zero mode'' problem
associated with the choice of gauge in the compactified case.  This
subtlety, however, is not particular to the light cone; indeed, its
occurrence is quite familiar in equal-time quantization on a
torus \cite{man85,mip94,las92}.  
In the present context, the difficulty is that
the zero mode in $A^+$ is in fact gauge-invariant, so that the
light-cone gauge $A^+=0$ cannot be reached.  Thus we have a pair of
interconnected problems: first, a practical choice of gauge; and
second, the presence of constrained zero modes of the gauge field.  In
two recent papers\cite{kap93,kap94} these problems were separated and
consistent gauge fixing conditions were introduced to allow isolation
of the dynamical and constrained fields.  In ref.\cite{kap94} the  generalize
gauge fixing is described, and  the Poincare\'e generators are constructed
in perturbation theory.  

One observes that in the traditional treatment, choosing the light-cone
gauge $A^+=0$ enables Gauss's law to be solved for $A^-$. In any
case the spinor projection $\psi_-$ is constrained and determined from
the equations of motion.

Discretization is achieved by putting the theory in a light-cone
``box,'' with $-L_\perp\leq x^i\leq L_\perp$ and $-L\leq x^-\leq L,$
and imposing boundary conditions on the fields.  $A_\mu$ must be taken to
be periodic in both $x^-$ and
$x_\perp$.  It is most
convenient to choose the Fermion fields to be periodic in $x_\perp$ and
anti-periodic in
$x^-.$ This eliminates the zero longitudinal momentum mode while still
allowing an expansion of the field in a complete set of basis
functions.

The functions used to expand the fields may be taken to be plane
waves, and for periodic fields these will of course include
zero-momentum modes.  Let us define, for a periodic quantity $f$, its
longitudinal zero mode
\begin {equation}
\langle f \rangle_{o} \equiv {1\over2L}\int_{-L}^L dx^- f(x^-,x_\perp)
                                        \label{zeromode}
\end {equation}
and the corresponding normal mode part
\begin {equation}
\langle {f} \rangle_{n} \equiv f - \langle {f} \rangle_{o}\; .
                                        \label{normal}
\end {equation}
We shall further denote the ``global zero mode''---the mode
independent of all the spatial coordinates---by $\langle  {f} \rangle$:
\begin {equation}
\langle  {f} \rangle \equiv {1\over\Omega} \int_{-L}^{L} dx^-   
\int_{-L_\perp}^{L_\perp}d^2x_\perp f(x^-,x_\perp)\; .
                                        \label{ozm}
\end {equation}
Finally, the quantity which will be of most interest to us is the
``proper zero mode,'' defined by
\begin {equation}
{f_0} \equiv \langle {f} \rangle_{o} - \langle {f}  \rangle\; .
                                        \label{pzmdef}
\end {equation}

By integrating over the appropriate direction(s) of space, we can
project the equations of motion onto the various sectors.  The global
zero mode sector requires some special treatment, 
and will not be discussed here.

We concentrate our attention on the proper zero mode sector, in
which the equations of motion become
\begin {equation}
-\partial_\perp^2{A_0}^+ = g{J_0}^+ 
                                        \label{zm2}
\end {equation}
\begin {equation}
-2(\partial_+)^2{A_0}^+-\partial_\perp^2{A_0}^-
-2\partial_i\partial_+{A_0}^i = g{J_0}^-
\label{zm3}
\end {equation}
\begin {equation}
-\partial_\perp^2{A_0}^i+\partial_i\partial_+{A_0}^+
+\partial_i\partial_j{A_0}^j = g{J_0}^i
\; . 
\label{zm1}
\end {equation}
We first observe that Eq.(\ref{zm2}), the projection of Gauss' law, is a
constraint which determines the proper zero mode of $A^+$ in terms of
the current $J^+$:
\begin {equation}
{A_0}^+ = -g{1\over\partial_\perp^2}{J_0}^+\; .
                                        \label{zm2solved}
\end {equation}
Eqs.(\ref{zm3}) and (\ref{zm1}) then determine the
zero modes ${A_0}^-$ and ${A_0}^i$.

Equation (\ref{zm2solved}) is clearly incompatible with the strict light-cone
gauge $A^+=0,$ which is most natural in light-cone analyses of gauge
theories.  Here we encounter a common problem in treating axial gauges
on compact spaces, which has nothing to do with
light-cone quantization {\it per se.} The point is that any
$x^-$-independent part of $A^+$ is in fact gauge invariant, since
under a gauge transformation
\begin {equation}
A^+\rightarrow A^+ + 2\partial_-\Lambda\; ,
                                        \label{gt}
\end {equation}
where $\Lambda$ is a function periodic in all coordinates.
Thus it is not possible to bring an arbitrary gauge field
configuration to one satisfying $A^+=0$ via a gauge transformation,
and the light-cone gauge is incompatible with the chosen boundary
conditions.  The closest we can come is to set the normal mode part of
$A^+$ to zero, which is equivalent to
\begin {equation}
\partial_-A^+=0\; .
                                        \label{gauge1}
\end {equation}
This condition does not, however, completely fix the gauge---we are
free to make arbitrary $x^-$-independent gauge transformations without
undoing Eq.(\ref{gauge1}).  We may therefore impose further conditions
on $A_\mu$ in the zero mode sector of the theory.

To see what might be useful in this regard, let us consider solving
Eq.(\ref{zm1}).  We begin by acting on Eq.(\ref{zm1}) with $\partial_i$.  The
transverse field ${A_0}^i$ then drops out and we obtain an expression
for the time derivative of ${A_0}^+$:
\begin {equation}
\partial_+{A_0}^+ = g{1\over\partial_\perp^2}\partial_i {J_0}^i\; .
                                        \label{dpap}
\end {equation}
[This can also be obtained by taking a time derivative of 
Eq.(\ref{zm2solved}), and using current conservation to re-express the right
hand side in terms of $J^i.$]  Inserting this back into Eq.(\ref{zm1}) we
then find, after some rearrangement,
\begin {equation}
-\partial_\perp^2
\Bigl(\delta^{i}_j - 
{\partial_i\partial_j \over \partial_\perp^2}\Bigr) {A_0}^j =
g \Bigl(\delta^{i}_j -
{\partial_i\partial_j \over \partial_\perp^2}\Bigr){J_0}^j\;.
\label{Atrans}
\end {equation}
Now the operator $(\delta^i_j- \partial_i\partial_j/\partial_\perp^2)$ 
is nothing
more than the projector of the two-dimensional transverse part of the
vector fields ${A_0}^i$ and ${J_0}^i$.  No trace remains of the
longitudinal projection of the field $(\partial_i\partial_j /
\partial_\perp^2){A_0}^j$ in Eq.(\ref{Atrans}).  This reflects precisely the
residual gauge freedom with respect to $x^-$-independent
transformations. To determine the longitudinal part, an additional
condition is required.

More concretely, the general solution to Eq.(\ref{Atrans}) is
\begin {equation}
{A_0}^i = -g{1\over\partial_\perp^2}{J_0}^i 
+ \partial_i\varphi(x^+,x_\perp)\; ,
\label{secsoln}
\end {equation}
where $\varphi$ must be independent of $x^-$ but is otherwise
arbitrary.  Imposing a condition on, say, $\partial_i{A_0}^i$ will
uniquely determine $\varphi$.

In ref.\cite{kap94}, for example, the condition
\begin {equation}
\partial_i{A_0}^i=0
                                        \label{compg}
\end {equation}
was proposed as being particularly natural. This choice, taken with
the other gauge conditions we have imposed, has been called the
``compactification gauge.''  In this case
\begin {equation}
\varphi = g{1\over(\partial_\perp^2)^2}\partial_i {J_0}^i\; .
                                        \label{compgp}
\end {equation}
Of course, other choices are also possible.  For example, we might
generalize Eq.(\ref{compgp}) to
\begin {equation}
\varphi = \alpha g{1\over(\partial_\perp^2)^2}\partial_i{J_0}^i\; ,
                                        \label{gencompgp} 
\end {equation}
with $\alpha$ a real parameter.  The gauge condition corresponding to this
solution is
\begin {equation}
\partial_i {A_0}^i=-g(1-\alpha){1\over\partial_\perp^2}\partial_i {J_0}^i\; .
                                        \label{zmgaugecond}
\end {equation}
We shall refer to this as the ``generalized compactification gauge.''
An arbitrary gauge field configuration $B^\mu $ can be brought to one
satisfying Eq.(\ref{zmgaugecond}) via the gauge function
\begin {equation}
\Lambda(x_\perp) =- {1\over\partial_\perp^2} 
\Bigl[ g (1 - \alpha) {1\over\partial_\perp^2} \partial_i {J_0}^i
+ \partial_i {B_0}^i \Bigr]\; .
                                        \label{gaugetrans}
\end {equation}
This is somewhat unusual in that $\Lambda(x_\perp)$ involves the
sources as well as the initial field configuration, but this is
perfectly acceptable.  More generally, $\varphi$ can be any
(dimensionless) function of gauge invariants constructed from the
fields in the theory, including the currents $J^\pm$. For our purposes
Eq.(\ref{zmgaugecond}) suffices.

We now have relations defining the proper zero modes of $A^i$,
\begin {equation}
{A_0}^i = -g{1\over\partial_\perp^2}\Bigl(
\delta^i_j-\alpha{\partial_i\partial_j\over\partial_\perp^2}\Bigr) 
{J_0}^j\; ,
\label{aizeromode}
\end {equation}
as well as ${A_0}^+$ [Eq.(\ref{zm2solved})].  All that remains 
is to use the final constraint Eq.(\ref{zm3}) to determine ${A_0}^-$.  
Using Eqs.(\ref{dpap}) and (\ref{zmgaugecond}), 
we find that Eq.(\ref{zm3}) can be written as
\begin {equation}
\partial_\perp^2{A_0}^- = -g{J_0}^--2\alpha g{1\over\partial_\perp^2}
\partial_+\partial_i {J_0}^i\; .
                                        \label{solvingam0}
\end {equation}
After using the equations of motion to express $\partial_+{J_0}^i$ in
terms of the dynamical fields at $x^+=0$, this may be
straightforwardly solved for ${A_0}^-$ by inverting the
$\partial_\perp^2.$ In what follows, however, we shall have no need of
${A_0}^-.$ It does not enter the Hamiltonian, for example; as usual,
it plays the role of a multiplier to Gauss' law Eq.(\ref{zm1}), 
which we are able to implement as an operator identity.

We have shown how to perform a general gauge fixing of Abelian gauge
theory in DLCQ and cleanly separate the dynamical from the
constrained zero-longitudinal momentum fields.  The various zero mode
fields {\it must} be retained in the theory if the equations of motion
are to be realized as the Heisenberg equations.  We have further seen
that taking the constrained fields properly into account renders the
ultraviolet behavior of the theory more benign, in that it results in
the automatic generation of a counter term for a non-covariant
divergence in the fermion self-energy in lowest-order perturbation
theory.  

The solutions to the constraint relations for the ${A_0}^i$ are all
physically equivalent, being related by different choices of gauge in
the zero mode sector of the theory.  There is a gauge which is
particularly simple, however, in that the fields may be taken to
satisfy the usual canonical anti-commutation relations.  This is most
easily exposed by examining the kinematical Poincare\'e generators and
finding the solution for which these retain their free-field forms.
The unique solution that achieves this is $\varphi=0$ in 
Eq.(\ref{secsoln}).
For solutions other than this one, complicated commutation relations
between the fields will be necessary to correctly translate them in
the initial-value surface.

It would be interesting to study the structure of the operators
induced by the zero modes from the point of view of the light-cone
power-counting analysis of Wilson\cite{wwh94}.  As noted
in the Introduction, to the extent that DLCQ coincides with reality,
effects which we would normally associate with the vacuum must be
incorporated into the formalism through the new, non-canonical
interactions arising from the zero modes.  Particularly interesting is
the appearance of operators that are nonlocal in the transverse
directions .  These are interesting because the
strong infrared effects they presumably mediate could give rise to
transverse confinement in the effective Hamiltonian for QCD.  There is
longitudinal confinement already at the level of the canonical
Hamiltonian; that is, the effective potential between charges
separated only in $x^-$ grows linearly with the separation.  This
comes about essentially from the non-locality in $x^-$ (i.e. the
small-$k^+$ divergences) of the light-cone formalism.

It is clearly of interest to develop non-perturbative methods for
solving the constraints, since we are ultimately interested in
non-perturbative diagonalization of $P^-.$ Several approaches to this
problem have recently appeared in the 
literature\cite{hkw92a,hkw92b,bpv93,piv94},
in the context of scalar field theories in 1+1 dimensions.  For QED
with a realistic value of the electric charge, however, it might be
that a perturbative treatment of the constraints could suffice; that
is, that we could use a perturbative solution of the constraint to
construct the Hamiltonian, which would then be diagonalized
non-perturbatively.  An approach similar in spirit has been proposed in
ref.\cite{wwh94}, where the idea is to use a perturbative realization
of the renormalization group to construct an effective Hamiltonian for
QCD, which is then solved non-perturbatively.  There is some evidence
that this kind of approach might be useful.  Wivoda and Hiller have
recently used DLCQ to study a theory of neutral and interacting
charged scalar fields in 3+1 dimensions\cite{wih93}.  They discovered that
including four-fermion operators precisely analogous to the
perturbative ones appearing in $P^-_Z$ significantly improved the
numerical behavior of the simulation.

The extension of the present work to the case of QCD is complicated by
the fact that the constraint relations for the gluonic zero modes are
nonlinear, as in the $\phi^4$ theory.  A perturbative solution of the
constraints is of course still possible, but in this case, since the
effective coupling at the relevant (hadronic) scale is large, it is
clearly desirable to go beyond perturbation theory.  In addition,
because of the central role played by gauge fixing in the present
work, we may expect complications due to the Gribov
ambiguity\cite{gri78}, which prevents the selection of unique
representatives on gauge orbits in non-perturbative treatments of
Yang-Mills theory.  As a step in this direction, work is
in progress on the pure glue theory in 2+1 dimensions \cite{kap94}. There
it is expected that some of the non-perturbative techniques used
recently in 1+1 dimensions 
\cite{bpv93,piv94,pim96,pin97a,pin97b,mcr97} can be applied.

\subsection{Dynamical Zero Modes}
 
Our concern in this section is with zero modes that are
true dynamical independent fields. They can arise  
due to the boundary conditions in gauge theory one cannot fully implement the 
traditional light-cone gauge $A^{+}=0$. The development of the
understanding of this problem in DLCQ can be traced in Refs.
\cite{hkw91a,hkw91b,hkw91c,hkw92a,mcc88,mcc91}.   
The field $A^+$ turns out to have a zero mode
which cannot be gauged away 
\cite{kap93,kap94,kpp94,pim96,pin97a,pin97b,mcr97}. 
This mode is indeed dynamical, 
and is the object we study in
this paper.   It has its analogue in instant form approaches to gauge theory. 
For example, there exists a large body of work on Abelian and non-Abelian
gauge theories in 1+1 dimensions quantized on a cylinder geometry
\cite{man85,het93}.  
There indeed this dynamical zero mode plays an important role.  
We too shall concern ourselves in the present section 
with non-Abelian gauge theory
in 1+1 dimensions, revisiting the model introduced 
by 't Hooft \cite{tho75}. 

The specific task we undertake here is to understand the zero mode 
{\it subsector} of
the pure glue theory, namely where only zero mode external sources
excite only zero mode gluons. We shall see that this is not an approximation
but rather a consistent solution, a sub-regime within the complete
theory. A similar framing of the problem lies behind the work of L\"uscher 
\cite{lus83}
and van Baal \cite{van92} using the instant form Hamiltonian approach 
to pure glue gauge theory in 3+1 dimensions. 
The beauty of this reduction in the 1+1 dimensional
theory is two-fold. First, it yields a theory which is exactly soluble. 
This is useful given the dearth of soluble models in field theory.  
Secondly, the zero mode theory represents a paring down to the point 
where the front and instant forms are manifestly
{\it identical}, which is nice to know indeed. 
We solve the theory in this specific
dynamical regime and find a discrete spectrum of states 
whose wavefunctions can be
completely determined. These states have the quantum numbers of the vacuum.

We consider an SU(2) non-Abelian gauge theory 
in 1+1 dimensions with classical sources 
coupled to the gluons. The Lagrangian density is
\begin{equation}
{\cal L} = {1\over 2}\, {\rm Tr}\, (F_{\mu \nu} F^{\mu \nu}) 
+ 2\, {\rm Tr}\, (J_\mu A^\mu) 
\end{equation}
where 
$F_{\mu \nu} = \partial _{\nu} A_{\nu} - \partial_{\nu} A_{\mu} -g[A_{\mu}, 
A_{\nu}]$. 
With a finite interval in $x^-$ from $-L$ to $L$, we impose periodic
boundary conditions  on all gauge potentials $A_\mu$.

we cannot eliminate the zero mode of the gauge potential. The
reason is evident: it is {\it invariant} under periodic gauge
transformations. 
But of course we can always perform a rotation in color space.  
In line with other authors \cite{apf93,prf89,fnp81a,fnp81b,fnp82}, 
we choose this so that
$\stackrel{o}{A_3^+}$  is the only non-zero element, 
since in our representation only $\sigma^3$ is diagonal.  
In addition, we can impose the subsidiary gauge condition
$ \stackrel{o}{A^-_3} = 0$. 
The reason is that there still remains freedom to perform
gauge transformations that depend only on light-cone time 
$x^+$ and the color matrix $\sigma^3$.

The above procedure would appear to have enabled complete 
fixing of the gauge. This is still not so. Gauge transformations
\begin{equation}
V = \exp\{i x^- ({{n\pi} \over {2L}}) {\sigma}^3\}
\label{GribU}
\end{equation}
generate shifts, according to Eq.(\ref{gaugetrans}), in the zero mode component
\begin{equation}
\stackrel{o}{A^+_3} \rightarrow \stackrel{o}{A^+_3} + {{n\pi}\over{gL}}
\;.
\end{equation}
All of these possibilities, labelled by the integer $n$, 
of course still satisfy $\partial_- A^+=0$, 
but as one sees $n=0$ should not really be included. One can verify
that the transformations $V$ also preserve the subsidiary condition.   
One notes that the transformation is 
$x^-$-dependent and {\it $Z_2$ periodic}. It is
thus a simple example of a Gribov copy \cite{gri78} in 1+1 dimensions.   
We follow the conventional procedure by demanding 
\begin{equation}
\stackrel{o}{A^+_3} \neq {n \pi \over gL}\;, \quad  n= \pm1, \pm2, \ldots 
\;.
\end{equation}
This eliminates singularity points at the Gribov `horizons' which
in turn correspond to a vanishing Faddeev-Popov determinant
\cite{van92}.

For convenience we henceforth use the notation
\begin{equation} 
\stackrel{o}{A^+_3} = v \;, \quad  
 x^+ = t \;, \quad  
w^2  =  { {\stackrel{o}{J^+_+} \stackrel{o}{J^+_-}} \over {g^2} } 
\quad {\rm{and}} \quad  
\stackrel{o}{J^-_3} = {B\over2}  
\;. 
\end{equation}
We pursue a Hamiltonian formulation. The only conjugate momentum is 
\begin{equation} 
p\, \equiv\, \stackrel{o}{\Pi^-_3} \, = \, \partial^- 
\!\! \stackrel{o}{A^+_3}\,  =\, \partial^- v  
\;.
\end{equation}
The Hamiltonian density  
$ T^{+ -}\, = \, \partial^- \!\! \stackrel{o}{A^+_3} \Pi^-_3 -
\cal L $  leads to the Hamiltonian 
\begin{equation}
 H =  {1 \over 2} [ {p}^2  +  {w^2 \over v^2} + B v] (2L)  
\;.
\label{qmHamil} 
\end{equation}
Quantization is achieved by imposing a commutation relation 
at equal light-cone time 
on the dynamical degree of freedom. Introducing the variable 
$ q = 2L v $, 
the appropriate commutation relation is    $[q(x^+), p(x^+)] = i .$  
The field theoretic problem reduces to quantum mechanics of a 
single particle as in Manton's treatment of the Schwinger model in
Refs.\cite{man85}.  One thus has to solve the Schr\"odinger equation 
\begin{equation}
{1 \over 2} (- {d^2 \over dq^2} + {(2Lw)^2 \over q^2} 
+ {{B q} \over {2L}})\psi = 
{\cal E}  \psi, 
\label{schrodeq} 
\end{equation}
with the eigenvalue ${\cal E} = E/(2L)$ actually being an energy density. 

All eigenstates $\psi$ have the quantum numbers of the naive vacuum 
adopted in standard front form field theory: 
all of them are eigenstates of the light-cone 
momentum operator $P^+$ with zero eigenvalue. 
The true vacuum is now that state with lowest $P^-$ eigenvalue. 
In order to get an exactly soluble system we eliminate the source $2B =
\stackrel{o}{J^-_3}$.  

The boundary condition that is to be imposed comes 
from the treatment of the Gribov problem. 
Since the wave function vanishes at $q=0$ we must
demand that the wavefunctions vanish at the first 
Gribov horizon $q = \pm 2\pi / g$. 
The overall constant $R$ is then fixed by normalization. This leads to   
the energy density only assuming the discrete values
\begin{equation}
 {\cal E}_m^{(\nu)} = {g^2 \over {8 {\pi}^2}} (X_m^{(\nu)})^2 
, \quad   m = 1,2, \dots, 
\label{Evals}
\end{equation}
where $X_m^{(\nu)}$ denotes 
the m-th zero of the $\nu$-th Bessel function $J_\nu$.
In general, these zeroes can only be obtained numerically. 
Thus
\begin{equation}   
\psi_m (q) = R\sqrt{q} J_{\nu} (\sqrt{2 {\cal E}_m^{(\nu)}} q)  
\end{equation}
is the complete solution. 
The true vacuum is the state of lowest energy namely with $m=1$.

The exact solution we obtained is genuinely non-perturbative in 
character. It describes
vacuum-like states since for all of these states $P^+=0$. 
Consequently, they all have
zero invariant mass $M^2 = P^+ P^-$. The states are labelled
by the eigenvalues of the operator $P^-$. 
The linear dependence on $L$ in the result
for the discrete {\it energy} levels is also consistent with what one would
expect from a loop of color flux running around the cylinder. 

In the source-free equal time case Hetrick \cite{het93,het94} uses a wave 
function that is symmetric about $q = 0$. 
For our problem this corresponds to
\begin{equation}
\psi_m(q) = N \cos (\sqrt{2 \epsilon_m} q )
\;.
\end{equation}
where N is fixed by normalization. 
At the boundary of the fundamental modular region
$q =2\pi /g$ and $\psi_m = (-1)^m N $, 
thus $\sqrt{2 \epsilon_m} 2 \pi /g =m \pi$ and
\begin{equation}
\epsilon = {g^2 (m^2-1)\over{8}}
\;.
\end{equation}
Note that $ m=1 $ is the lowest energy state and has as 
expected 
one node in the allowed region $ 0\le g \le 2 \pi /g $. 
Hetrick \cite{het93} discusses the connection
to the results of Rajeev \cite{raj92} but it amounts to a shift in
$\epsilon$ and a redefining of $m \rightarrow m/2$. 
It has been argued by van Baal that the correct boundary condition 
at $q=0$ is $\psi(0) = 0 $. This would
give a sine which matches smoothly with the Bessel function solution. 
This calculation offers the lesson that even in a front form approach, 
the vacuum might not be just the simple Fock vacuum. 
Dynamical zero modes do imbue the vacuum with a rich structure.

%% file: 09Renor.tex
\section {Regularization and\\ Non-Perturbative Renormalization}
\label{sec:renormalization}
\setcounter{equation}{0}

The subject of renormalization is a large one and high energy theorists have
developed a standard set of renormalization techniques based on perturbation
theory (see, for example, Collins~\cite{col84}).  However, many of these
techniques are poorly suited for light-front field theory.  Researchers in
light-front field theory must either borrow techniques from condensed matter
physics~\cite{pin92,vap92,wil75} or nuclear physics or come up with
entirely new approaches.  Some progress in this direction has already been
made,  see for example \cite{wwh94,weg72a,weg72b,weg76}.
A considerable amount of work is focusing on these questions 
\cite{amm94,atw93,bur97,dks95,guw97,hye94,zha94}, 
particularly see the work of 
Bassetto {\it et al.} \cite{acb94,bas93,bar93,bkk93,bas96}, 
Bakker {\it et al.} \cite{lib95a,lib95b,scb97}, and
Brisudova {\it et al.} \cite{brp95,brp96,bpw97}.

The biggest challenge to renormalization of light-front field theory is the
infra-red divergences that arise.  Recall that the Hamiltonian for a free
particle is
\begin{equation}
P^- = \frac{{\bf P}_\perp^2+m^2}{2 P^+} \; \; .
\end{equation}
Small longitudinal momentum $P^+$ is associated with large energies. Thus,
light-front field theory is subject to infra-red longitudinal divergences. 
These divergences are quite different in nature from the infra-red divergences
found in equal-time quantized field theory. In order to remove small $P^+$
states, one must introduce non-local counter terms into the Hamiltonian.  
Power counting arguments allow arbitrary {\em functions} of transverse
momenta to be associated with these counter terms.   This is in contrast to
more conventional  approaches where demanding locality strongly constrains
the number of allowed operators.

One hopes to use light-front field theory to perform bound state
calculations.  In this case one represents a bound state by a finite number
of particles (a Tamm-Dancoff truncation) whose momenta are restricted to some
finite interval.  This has a number of implications.  In particular, momentum
cutoffs and Tamm-Dancoff truncations both tend to break various symmetries of
a theory .  Proper renormalization must  restore these symmetries.  In
contrast, conventional calculations  choose regulators (like dimensional
regularization) that do not break many symmetries. 

In conventional approaches, one is often concerned simply whether the system
is renormalizable, that is, whether the large cutoff limit is well defined.  
In bound state calculations, one is also interested in how quickly the results
converge as one increases the cutoffs since  numerical calculations must be
performed with a finite cutoff.  Thus, one is potentially interested in the
effects of irrelevant operators along with the usual marginal and relevant
operators.

Conventional renormalization is inherently perturbative in nature. However, 
we are interested in many phenomena that are essentially non-perturbative:
bound states, confinement, and spontaneous symmetry breaking.   The bulk of
renormalization studies in light-front field theory to date have used
perturbative techniques~\cite{wwh94,ghp93}. Non-perturbative techniques
must be developed.

Generally, one expects that renormalization will produce a  large number of
operators in the light-front Hamiltonian.   A successful approach to
renormalization must be able to produce these operators automatically (say,
as part of a numerical algorithm).  In addition, there should be only a
few free parameters which must be fixed phenomenologically.  Otherwise, the 
predictive power of a theory will be lost.

\subsection{Tamm-Dancoff Integral Equations} 

Let us start by looking at a simple toy model
that has been studied by a number of authors
\cite{vap92,ghp93,pin92,tho79a,tho79b,jac91}. In fact, it is the famous Kondo
problem truncated to one particle states~\cite{kon69}. Consider the
homogeneous integral equation
\begin{equation}
\left(p - E\right) \phi\left(p\right) + g \int_0^\Lambda
dp^\prime\;\phi\left(p^\prime\right) = 0 
\end{equation}
with eigenvalue $E$  and eigenvector $\phi\left(p\right)$.  This is a 
model for Tamm-Dancoff equation of a single particle of momentum $p$ with
Hamiltonian $H\left(p,p^\prime\right) = p\;\delta\left(p - p^\prime\right) +
g$.  We will focus on the $E < 0$ bound state solution:
\begin{equation}
 \phi\left(p\right) = {{\rm constant}\over {p - E}}\;,\;\;\;\;\;\; E =
{\Lambda\over{1 - e^{-1/g}}}\;.  
\end{equation}
Note that the eigenvalue diverges in the limit $\Lambda\to\infty$.  
Proper renormalization involves modifying the system to make $E$ and
$\phi\left(p\right)$ independent of $\Lambda$ in the limit 
$\Lambda\to\infty$. Towards this end, we add a counter term $C_\Lambda$ to the
Hamiltonian.  Invoking the high-low analysis \cite{wil76}, we divide the
interval  $0<p<\Lambda$ into two subintervals:  $0<p<L$, a ``low momentum
region,''  and  $L<p<\Lambda$, a``high momentum region,'' where the momentum
scales characterized by $E$, $L$, and $\Lambda$ are assumed to be widely
separated.  The idea is that the eigenvalue and eigenvector should be
independent of the behavior of the  system in the high momentum region.  The
eigenvalue equation can be written as  two coupled equations
\begin{equation}
p\ \epsilon\left[0,L\right]\;\;\;\; \left(p - E\right) 
\phi\left(p\right) 
+ \left(g + C_\Lambda\right) \int_0^L 
dp^\prime\;\phi\left(p^\prime\right) +
\left(g + C_\Lambda\right) \int_L^\Lambda 
dp^\prime\;\phi\left(p^\prime\right) 
= 0
\label {re13a}
\end{equation}
\begin{equation}
p\ \epsilon\left[L,\Lambda\right]\;\;\;\; \left(p - E\right)
\phi\left(p\right) + \left(g + C_\Lambda\right) \int_L^\Lambda 
dp^\prime\;
\phi\left(p^\prime\right) + \left(g + C_\Lambda\right) \int_0^L
dp^\prime\;\phi\left(p^\prime\right) = 0\;. \label {re13b}
\end{equation}
Integrating Equation~(\ref{re13b}) in the limit $L,\Lambda\gg E$,
\begin{equation}
\int_L^\Lambda dp\;\phi\left(p\right) = -{{\left(g + C_\Lambda\right) 
\log
{\Lambda\over L}}\over{1+ \left(g + C_\Lambda\right) \log {\Lambda\over 
L}}}
\int_0^L dp^\prime\;\phi\left(p^\prime\right)\;, 
\end{equation}
and substituting this expression into Equation~(\ref{re13a}), we obtain an
eigenvalue  equation with the high momentum region integrated out
\begin{equation}
p\ \epsilon\left[0,L\right]\;\;\;\; \left(p - E\right) 
\phi\left(p\right) 
+ {{\left(g + C_\Lambda\right)}\over{1+ \left(g + C_\Lambda\right) \log
{\Lambda\over L}}}\int_0^L dp^\prime\;\phi\left(p^\prime\right) = 
0\;. \label {re15}
\end{equation}
If we demand that this expression be independent of $\Lambda$,
\begin{equation}
 {d\over{d\Lambda}}\left({{\left(g + C_\Lambda\right)}\over{1+ \left(g 
+
C_\Lambda\right) \log {\Lambda\over L}}}\right) = 0\;, 
\end{equation}
we obtain a differential equation for $C_\Lambda$
\begin{equation}
{{d C_\Lambda}\over{d\Lambda}} = {{\left(g +
C_\Lambda\right)^2}\over\Lambda}\;. 
\end{equation}
Solving this equation, we are free to insert an arbitrary constant 
$-1/A_\mu - \log\mu$
\begin{equation}
g + C_\Lambda = {A_\mu\over{1 - A_\mu \log{\Lambda\over\mu}}}\;. 
\label {re18}
\end{equation}
Substituting this result back into Equation~(\ref{re15}),
\begin{equation}
p\ \epsilon\left[0,L\right]\;\;\;\; \left(p - E\right) 
\phi\left(p\right) +
{{A_\mu}\over{1- A_\mu \log {L\over \mu}}}\int_0^L
dp^\prime\;\phi\left(p^\prime\right) = 0 \label {re19}
\end{equation}
we see that $\Lambda$ has been removed from the equation entirely.  
Using Equation~(\ref{re18}) in the original eigenvalue equation
\begin{equation}
\left(p - E\right) \phi\left(p\right) + {A_\mu\over{1 - A_\mu
\log{\Lambda\over\mu}}} \int_0^\Lambda 
dp^\prime\;\phi\left(p^\prime\right) 
= 0  
\end{equation}
gives the same equation as (\ref{re19}) with $L$ replaced by $\Lambda$.  The 
eigenvalue is now,
\begin{equation}
E = {\Lambda\over{1 -{\Lambda\over \mu} e^{-1/A_\mu}}}\;\;\;\;\;
\lim_{\Lambda\to\infty}\; E = - \mu e^{1/A_\mu}\;. \label {re111}
\end{equation}
Although the eigenvalue is still a function of the cutoff for finite 
$\Lambda$, 
the eigenvalue does become independent of the cutoff in the limit
$\Lambda\to\infty$, and the system is properly renormalized.

One can think of $A_\mu$ as the renormalized coupling constant and $\mu$  as
the  renormalization scale.  In that case, the eigenvalue should depend on 
the choice of $A_\mu$ for a given $\mu$ but be independent of $\mu$ itself.  
Suppose, for Equation~(\ref{re111}), we want to change $\mu$ to a new value,
say  $\mu^\prime$.  In order that the eigenvalue remain the same, we must also
change the  coupling constant from $A_\mu$ to $A_{\mu^\prime}$
\begin{equation}
\mu e^{1/A_\mu} = \mu^\prime e^{1/A_{\mu^\prime}}\;. 
\end{equation}
In the same manner, one can write down a $\beta$-function for $A_\mu$
\cite{vap95}
\begin{equation}
\mu\;{d\over{d\mu}} A_\mu  = A_\mu^2\;. \label {re113}
\end{equation}
Using these ideas one can examine the 
general  case.  Throughout, we will be working with operators projected onto
some Tamm-Dancoff subspace (finite particle number) of the full Fock space.  
In addition, we will regulate the system by demanding that each component  of
momentum of each particle lies within some finite interval.  One defines  the
``cutoff'' $\Lambda$ to be an operator which projects onto this subspace  of
finite particle number and finite momenta.  Thus, for any operator $O$,  $O
\equiv \Lambda O \Lambda$.  Consider the Hamiltonian
\begin{equation}
H = H_0 + V + C_\Lambda 
\end{equation}
where, in the standard momentum space basis, $H_0$ is the diagonal part  of 
the Hamiltonian, $V$ is the interaction term, and $C_\Lambda$ is the  
counter term which is to be determined and is a function of the cutoff.   Each
term of the Hamiltonian is Hermitian and compact.  Schr\"{o}dinger's  equation
can be written                                              
\begin{equation}
\left(H_0 - E \right) |\phi\rangle + \left( V + C_\Lambda\right)|\phi\rangle =
0 
\label {re22}
\end{equation}
with energy eigenvalue $E$ and eigenvector $|\phi\rangle$.  The goal is to 
choose $C_\Lambda$ such that $E$ and $|\phi\rangle$ are independent of
$\Lambda$ in the   limit of large cutoff.

One now makes an important assumption:  the physics of 
interest is characterized by energy scale $E$ and is independent of physics
near  the boundary of the space spanned by $\Lambda$.  Following the approach
of  the previous section, one define two  projection
operators, ${\cal Q}$ and ${\cal P}$, where $\Lambda = {\cal Q} + {\cal P}$,
${\cal Q P} = {\cal P Q} = 0$, and ${\cal Q}$ and ${\cal P}$ commute  with
$H_0$.    ${\cal Q}$ projects onto a ``high momentum region'' which  contains
energy scales one does not care about, and ${\cal P}$ projects onto a ``low 
momentum region'' which contains energy scales characterized by $E$. 
Schr\"{o}dinger's equation (\ref{re22}) can be rewritten as two coupled
equations:
\begin{equation}
\left(H_0 - E \right){\cal P}|\phi\rangle + {\cal P}\left( V + C_\Lambda 
\right) {\cal P}|\phi\rangle  + {\cal P}\left( V + 
C_\Lambda\right){\cal Q}|\phi\rangle = 0 
 \label {re23a}
\end{equation}
and
\begin{equation}
\left(H_0 - E \right){\cal Q}|\phi\rangle + {\cal Q}\left( V + C_\Lambda 
\right){\cal Q}|\phi\rangle  + {\cal Q}\left( V + C_\Lambda\right){\cal 
P}|\phi\rangle = 0\;.  
\label {re23b}
\end{equation}
Using Equation~(\ref{re23b}), one can formally solve for ${\cal
Q}|\phi\rangle$ in  terms  of ${\cal P}|\phi\rangle$
\begin{equation}
{\cal Q}|\phi\rangle = {1\over {{\cal Q}\left(E - H\right){\cal Q}}} \left( V 
+ C_\Lambda \right){\cal P}|\phi\rangle\;. 
\end{equation}
The term with the denominator is understood to be defined in terms of  its
series expansion in $V$.  One can substitute this result back into 
Equation~(\ref{re23a})
\begin{equation}
\left(H_0 - E \right){\cal P}|\phi\rangle + {\cal P}\left( V + C_\Lambda 
\right){\cal P}|\phi\rangle  
+ {\cal P}\left( V + C_\Lambda\right) {1\over {{\cal Q}\left(E -
H\right){\cal Q}}} \left( V + C_\Lambda \right){\cal P}|\phi\rangle = 0\;.
\label {re25}
\end{equation}
In order to properly renormalize the system, we could choose $C_\Lambda$  such
that Equation~(\ref{re25}) is independent of one's choice of $\Lambda$ for a
fixed  ${\cal P}$ in the limit of large cutoffs.  However, we will make a
stronger demand:   that Equation~(\ref{re25}) should be equal to
Equation~(\ref{re22}) with the cutoff $\Lambda$ replaced  by  ${\cal P}$. 

One can express  $C_\Lambda$ as the solution of an 
operator equation, the ``counter term equation,''
\begin{equation}
V_\Lambda = V -  V F V_\Lambda\;. 
\end{equation}
where $V_\Lambda  = V + C_\Lambda$, and
provided that we can make the approximation
\begin{equation}
V {{\cal Q}\over{E - H_0}}V \approx  V{\cal Q} F V\;. \label {re29}
\end{equation}
This is what we will call the ``renormalizability condition''.  A system  is 
properly renormalized if, as we increase the cutoffs $\Lambda$ and  ${\cal P}$,
Equation~(\ref{re29}) becomes an increasingly good approximation. 
In the standard momentum space basis, this 
becomes a set of coupled inhomogeneous integral equations.  Such  equations
generally have a unique solution, allowing us to renormalize systems  without
having to resort to perturbation theory.  This includes cases where the
perturbative  expansion diverges or converges slowly.

There are many possible choices for $F$ that satisfy the  renormalizability 
condition.  For instance, one might argue that we want $F$ to resemble
$1/\left(E - H_0\right)$ as much as possible and choose
\begin{equation}
F = {1\over{\mu  - H_0}} \label {re212}
\end{equation}
where the arbitrary constant $\mu$ is chosen to be reasonably close to  $E$. 
In this case, one might be able to use a smaller cutoff in numerical 
calculations.

One might argue that physics above some energy scale $\mu$ is simpler  and
that it is numerically too difficult to include the complications of the 
physics at  energy scale $E$ in the solution of the counter term equation. 
Thus one  could choose
\begin{equation}
F = -{{\theta\left(H_0 - \mu\right)}\over{H_0}} \label {re213}
\end{equation}
where the arbitrary constant $\mu$ is chosen to  be somewhat larger than  $E$
but smaller than the energy scale associated with  the cutoff.  The 
$\theta$-function is assumed to act on each diagonal element in the  standard
momentum space basis.  The difficulty with this renormalization scheme  is
that it involves three different energy scales, $E$, $\mu$ and the cutoff 
which might make the numerical problem more difficult.
One can relate our approach to conventional renormalization group  concepts.  
In renormalization group language, $V_\Lambda$ is the bare interaction  term
and $V$ is the renormalized interaction term.  In both of the  renormalization
schemes introduced above, we introduced an arbitrary energy scale $\mu$;  this 
is the renormalization scale.  Now, physics (the energy eigenvalues and
eigenvectors) should not depend on this parameter  or  on the renormalization
scheme itself, for that matter.  How does one  move from one renormalization
scheme to another?  Consider a particular choice of renormalized interaction
term $V$ associated with a renormalization  scheme which uses $F$ in the
counter term equation.  We can use the counter term equation to find the bare
coupling $V_\Lambda$ in terms of $V$.  Now, to  find the renormalized
interaction term $V^\prime$ associated with a different renormalization scheme
using a different operator $F^\prime$ in the  counter term equation, we simply
use the counter term equation with  $V_\Lambda$  as given and solve for
$V^\prime$
\begin{equation}
V^\prime = V_\Lambda  +  V_\Lambda F^\prime V^\prime\;. 
\end{equation}
Expanding this procedure order by order in $V$ and summing the result,  we can
obtain an operator equation relating the two renormalized interaction  terms
directly
\begin{equation}
V^\prime  = V + V\left(F^\prime - F\right) V^\prime\;. 
\end{equation}
The renormalizability condition ensures that this expression will be
independent of the cutoff in the limit of large cutoff.

For the the two particular renormalization schemes mentioned above, 
(\ref{re212})  and (\ref{re213}), we can regard the renormalized interaction
term $V$ as an  implicit function of $\mu$.  We can see how the renormalized
interaction term  changes with $\mu$ in the case (\ref{re212}):
\begin{equation}
\mu\;{d\over{d\mu}}V  = - V{{\mu}\over{\left(H_0 - 
\mu\right)^2}}V 
\label {re33}
\end{equation}
and in the case (\ref{re213}):
\begin{equation}
\mu\;{d\over{d\mu}}V  = V\delta\left(H_0 - \mu\right)V\;. 
\end{equation}
This is a generalization of the $\beta$-function.
The basic idea of asymptotic  and box counter term renormalization in the 3+1
Yukawa model calculation in an earlier section can be illustrated with a simple
example. Consider an eigenvalue equation of the 
form \cite{tho79a,tho79b},
\begin{equation}
 k \;  \phi(k) - g \int_{0}^{\Lambda} \;  dq \;
V(k,q) \;  \phi(q) \;  = \;  E \; \phi(k) \; .
\end{equation}  
Making a high-low analysis of this equation as above and assuming that
\begin{equation}   
V_{LH}(k,q) \; = \; V_{HL}(k,q) \; = \; V_{HH}(k,q) \; = \; 
f \; \; 
\end{equation}  
Then one finds the following renormalized equation;
\begin{equation}   
k \; \phi(k) - g \int_{0}^{\Lambda}  dq \;
[ V(k,q) - f] \; \phi(q) - {A_{\mu} \over 1 + A_{\mu} ln{\Lambda
\over \mu}} \int_{0}^{\Lambda} dq \phi(q) = E \phi(k) \; .
\end{equation}  
One has renormalized the original equation in the sense that the low-energy
eigenvalue $E$ is independent of the high energy cutoff and we have an
arbitrary parameter $C$ which can be adjusted to fit the ground state energy
level. 

One can motivate both the asymptotic counter term and one-box counter term
in the Yukawa calculation as different choices in our analysis.  For a fixed
$\mu$ we are free to chose $A_\mu$ at will.  The simple asymptotic counter term
corresponds to $A_\mu = 0$.  However subtracting the asymptotic behavior of
the kernel with the term
$gf$ causes the wavefunction to fall off more rapidly than it would otherwise
at large $q$.  As a result the ${A_\mu \over {1 + A_\mu \ln \Lambda/\mu}}\int
\phi dq$ is finite, and this term can be retained as an arbitrary
adjustable finite counter term.

The perturbative counter terms correspond to $A_\mu = gf$ then expanding in $g
\ln \Lambda/\mu$. Then one finds  
\begin{equation}  
k\phi \left(k\right) - g \int^\Lambda_0 dq \left [ V\left( k,q\right) - f
\right ] \phi \left(q\right) - gf \sum^\infty_{n=0} 
\left( -g f \ln \Lambda/\mu
\right)^n \int_0^\Lambda dq\; \phi \left( q\right) = E \; .
\end{equation}  
\noindent
Keeping the first two terms in the expansion one gets the so called ``Box
counter term''
\begin{equation}  
k\phi \left( k\right) - g \int_0^\Lambda dq\; V \left( k,q \right) 
\phi \left(q \right) + g^2 f^2 \ln \Lambda/\mu \int^\Lambda_0 dq\; \phi
\left(q\right) = E
\end{equation}  
\noindent Note that the box counter term contains $f^2$ indicating that it
involves the kernel at high momentum twice.  Ideally, one would like to carry
out the non-perturbative renormalization program rigorously in the sense that
the cutoff independence is achieved for any value of the coupling constant
and any value of the cutoff.   In practical cases, either one may not have the
luxury to go to very large cutoff or the analysis itself may get too
complicated. For example, the assumption of a uniform high energy limit was
essential for summing up the series. In reality $V_{HH}$ may differ from
$V_{LH}$. 

The following is a simplified two-variable problems that are more closely related
to the equations and approximations used in the Yukawa calculation.  The form
of the asymptotic counter term that was used can be understood by considering
the following equation,
\begin{equation}
{k \over {x\left(1-x\right)}} \phi \left( k,x\right) - g
\int^\Lambda_0 dq
\int^1_0 dy\; K \left( k,q\right) \phi \left(q,y\right) = E \phi \left(
k,x\right)\; .
\end{equation}
\noindent This problem contains only $x$ dependence associated with the free
energy, and no $x$ dependence in the kernel.  It is easily solved
using the high-low analysis used above and one finds
\begin{eqnarray}
k\phi \left(kx\right) - g \int^\Lambda_0 dq \int^1_0 dy \left(
K\left( k,q\right) - f \right) \phi \left( q,y\right) 
\noindent\\
- {A_\mu \over {1 + {1
\over 6} A_\mu \ln \Lambda/\mu}} \int^\Lambda_0 dq \int^1_0 dy \phi
\left(q,y\right) = E\phi \left(kx\right) 
\end{eqnarray}
\noindent The factor of $1/6$ comes from the integral
$\int^1_0 dx\, x(1-x)$.  This result motivates our choice for $G_\Lambda$ in
the Yukawa calculation.

\subsection{Wilson Renormalization and Confinement}  
 
QCD was a step backwards in the sense that it forced upon us a complex
and mysterious vacuum.  In QCD, because the effective coupling grows
at long distances, there is always copious production of low-momentum
gluons, which immediately invalidates any picture based on a few
constituents.  Of course, this step was necessary to understand the
nature of confinement and of chiral symmetry breaking, both of which
imply a nontrivial vacuum structure. But for 20 years we have avoided
the question: Why did the CQM work so well that no one saw any need
for a complicated vacuum before QCD came along?

A bridge between equal-time quantized QCD and the equal-time CQM would
clearly be extremely complicated, because in the equal-time formalism
there is no easy {\it non-perturbative} way to make the vacuum simple.
Thus a sensible description of constituent quarks and gluons would be
in terms of quasiparticle states, i.e., complicated collective
excitations above a complicated ground state.  Understanding the
relation between the bare states and the collective states would
involve understanding the full solution to the theory.  Wilson and
collaborator argue that on the light front, however, simply implementing a
cutoff on small longitudinal momenta suffices to make the vacuum completely
trivial.  Thus one immediately obtain a constituent-type picture, in which all
partons in a hadronic state are connected directly to the hadron.
The price one pays to achieve this constituent framework is that the
renormalization problem becomes considerably more complicated on the
light front \cite{glp92a,glp92b,glw93a,glw93b,glw94}.

Wilson and collaborators also included a mass term for the gluons as well as
the quarks (they include only transverse polarization states for the
gluons) in $H_{free}$.  They have in mind here that all masses that occur in
$H_{free}$ should roughly correspond to constituent rather than
current masses.  There are two points that should be emphasized in
this regard.

First, cutoff-dependent masses for both the quarks and gluons will be
needed anyway as counter terms.  This occurs because all the cutoffs one
has for a non-perturbative Hamiltonian calculations
violate both equal-time chiral symmetry and gauge invariance.  These
symmetries, if present, would have protected the quarks and gluons
from acquiring this kind of mass correction.  Instead, in the
calculations  discussed here both the fermion and gluon self-masses
are quadratically divergent in a transverse momentum cutoff $\Lambda$.

The second point is more physical. When setting up perturbation theory
(more on this below) one should always keep the zeroth order problem
as close to the observed physics as possible.  Furthermore, the
division of a Hamiltonian into free and interacting parts is always
completely arbitrary, though the convergence of the perturbative
expansion may hinge crucially on how this division is made.  Nonzero
constituent masses for both quarks and gluons clearly comes closer to
the phenomenological reality (for hadrons) than do massless gluons and
nearly massless light quarks.

Now, the presence of a nonzero gluon mass has important consequences.
First, it automatically stops the running of the coupling below a
scale comparable to the mass itself.  This allows one to (arbitrarily)
start from a small coupling at the gluon mass scale so that
perturbation theory is everywhere valid, and only extrapolate back to
the physical value of the coupling at the end.  The quark and gluon
masses also provide a kinematic barrier to parton production; the
minimum free energy that a massive parton can carry is ${m^2\over
p^+}$, so that as more partons are added to a state and the typical
$p^+$ of each parton becomes small, the added partons are forced to
have high energies.  Finally, the gluon mass eliminates any infrared
problems of the conventional equal-time type.

In there initial work they use a simple cutoff on
constituent energies, that is, requiring
\begin{equation}
{p_\perp^2+m^2\over p^+} < {\Lambda^2\over P^+}
                                        \label {eq:cutoff}
\end{equation}
for each constituent in a given Fock state. 

Imposing (\ref{eq:cutoff}) does not completely regulate the theory,
however; there are additional small-$p^+$ divergences coming from the
instantaneous terms in the Hamiltonian.  They regulate these by
treating them as if the instantaneous exchanged gluons and quarks were
actually constituents, and were required to satisfy condition

Having stopped the
running of the coupling below the constituent mass scale, one
arbitrarily take it to be small at this scale, so that perturbation
theory is valid at all energy scales.  Now one can use power counting
to identify all relevant and marginal operators (relevant or marginal
in the renormalization group sense).  Because of the cutoffs one must
use, these operators are not restricted by Lorentz or gauge
invariance.  Because we have forced the vacuum to be trivial, the
effects of spontaneous chiral symmetry breaking must be manifested in
explicit chiral symmetry breaking effective interactions. This means the
operators are not restricted by chiral invariance either.  There are thus a
large number of allowed operators.  Furthermore, since transverse divergences
occur for any longitudinal momentum, the operators that remove transverse
cutoff dependence contain functions of dimensionless ratios of all available
longitudinal momenta.  That is, many counter terms are not
parameterized by single coupling constants, but rather by entire
functions of longitudinal momenta.  A precisely analogous result
obtains for the counter terms for light-front infrared divergences;
these will involve entire functions of transverse momenta.
The counter term functions can in principle be determined by requiring
that Lorentz and gauge invariance be restored in the full theory.

The cutoff Hamiltonian, with renormalization counter terms, will thus
be given as a power series in $g_\Lambda$:
\begin{equation}
H(\Lambda) = H^{(0)} + g_\Lambda H^{(1)}+g_\Lambda^2 H^{(2)}
+ \dots\; ,
\end{equation}
where all dependence on the cutoff $\Lambda$ occurs through the
running coupling $g_\Lambda,$ and cutoff-dependent masses.  

The next stage in building a bridge from the CQM to QCD is to
establish a connection between the {\it ad hoc} $q\overline{q}$
potentials of the CQM and the complex many-body Hamiltonian of QCD.

In lowest order the canonical QCD Hamiltonian contains gluon emission
and absorption terms, including emission and absorption of high-energy
gluons.  Since a gluons energy is ${k_\perp^2+\mu^2 \over k^+}$ for
momentum $k$, a high-energy gluon can result either if $k_\perp$ is
large or $k^+$ is small.  But in the CQM, gluon emission is ignored
and only low-energy states matter.  How can one overcome this double
disparity?  The answer is that we can change the initial cutoff
Hamiltonian $H(\Lambda)$ by applying a unitary transformation to it.
We imagine constructing a transformation $U$ that generates a new
effective Hamiltonian $H_{\rm eff}$:
\begin{equation}
H_{\rm eff} = U^\dagger H(\Lambda) U\; .
                                        \label {eq:simtrans}
\end{equation}
We then {\it choose} $U$ to cause $H_{\rm eff}$ to look as much like a
CQM as we can \cite{blo58,weg72a,weg72b,weg76}.

The essential idea is to start out as though we were going to
diagonalize the Hamiltonian $H(\Lambda),$ except that we stop short of
computing actual bound states.  A complete diagonalization would
generate an effective Hamiltonian $H_{\rm eff}$ in diagonal form; all
its off-diagonal matrix elements would be zero.  Furthermore, in the
presence of bound states the fully diagonalized Hamiltonian would act
in a Hilbert space with discrete bound states as well as continuum
quark-gluon states.  In a confined theory there would only be bound
states.  What we seek is a compromise: an effective Hamiltonian in
which some of the off-diagonal elements can be nonzero, but in return
the Hilbert space for $H_{\rm eff}$ remains the quark-gluon continuum
that is the basis for $H(\Lambda)$.  No bound states should arise. All
bound states are to occur through the diagonalization of $H_{\rm
eff}$, rather than being part of the basis in which $H_{\rm eff}$
acts.

To obtain a CQM-like effective Hamiltonian, we would ideally eliminate
all off-diagonal elements that involve emission and absorption of
gluons or of $q\overline{q}$ pairs.  It is the emission and absorption
processes that are absent from the CQM, so we should remove them by
the unitary transformation.  However, we would allow off-diagonal
terms to remain within any given Fock sector, such as $q\overline{q}
\rightarrow q\overline{q}$ off-diagonal terms or
$qqq \rightarrow qqq$ terms.  This means we allow off-diagonal
potentials to remain, and trust that bound states appear only when the
potentials are diagonalized.

Actually, as discussed in Ref. \cite{wwh94}, we cannot remove all the
off-diagonal emission and absorption terms.  This is because the
transformation $U$ is sufficiently complex that we only know how to
compute it in perturbation theory.  Thus we can reliably remove in
this way only matrix elements that connect states with a large energy
difference; perturbation theory breaks down if we try to remove, for
example, the coupling of low-energy quark to a low-energy quark-gluon
pair.  They therefore introduce a second cutoff parameter
${\lambda^2\over P^+}$, and design the similarity transformation to
remove off-diagonal matrix elements between sectors where the energy
{\it difference} between the initial and final states is greater than
this cutoff.  For example, in second order the effective Hamiltonian
has a one-gluon exchange contribution in which the intermediate gluon
state has an energy above the running cutoff.  Since the gluon energy
is ${k_\perp^2+\mu^2\over k^+}$, where $k$ is the exchanged gluon
momentum, the cutoff requirement is
\begin{equation}
{k_\perp^2+\mu^2\over k^+}> {\lambda^2\over P^+}\; .
\end{equation}
This procedure is known as the ``similarity renormalization group''
method.  
For a more detailed discussion and for connections to
renormalization group concepts see Ref. \cite{wwh94}.

The result of the similarity transformation is to generate an
effective light-front Hamiltonian $H_{\rm eff},$ which must be solved
non-perturbatively.  Guided by the assumption that a constituent
picture emerges, in which the physics is dominated by potentials in
the various Fock space sectors, we can proceed as follows.

We first split $H_{\rm eff}$ anew into an unperturbed part $H_0$ 
and a perturbation $V$.  The principle guiding this new division is that
$H_0$ should contain the most physically relevant operators, e.g.,
constituent-scale masses and the potentials that are most important
for determining the bound state structure.  All operators that change
particle number should be put into $V$, as we anticipate that
transitions between sectors should be a small effect.  This is
consistent with our expectation that a constituent picture results,
but this must be verified by explicit calculations.  Next we solve
$H_0$ non-perturbatively in the various Fock space sectors, using
techniques from many-body physics.  Finally, we use bound-state
perturbation theory to compute corrections due to $V$.

We thus introduce a second perturbation theory as part of building the
bridge.  The first perturbation theory is that used in the computation
of the unitary transformation $U$ for the incomplete diagonalization.
The second perturbation theory is used in the diagonalization of
$H_{\rm eff}$ to yield bound-state properties.  Perry in particular
has emphasized the importance of distinguishing these two different
perturbative treatments \cite{per94a}.  
The first is a normal field-theoretic
perturbation theory based on an unperturbed free field theory.  In the
second perturbation theory a different unperturbed Hamiltonian is
chosen, one that includes the dominant potentials that establish the
bound state structure of the theory.  Our working assumption is that
the dominant potentials come from the lowest-order potential terms
generated in the perturbation expansion for $H_{\rm eff}$ itself.
Higher-order terms in $H_{\rm eff}$ would be treated as perturbations
relative to these dominant potentials.

It is only in the second perturbative analysis that constituent masses
are employed for the free quark and gluon masses.  In the first
perturbation theory, where we remove transitions to high-mass
intermediate states, it is assumed that the expected field theoretic
masses can be used, i.e., near-zero up and down quark masses and a
gluon mass of zero.  Because of renormalization effects, however,
there are divergent mass counter terms in second order in $H(\Lambda).$
$H_{\rm eff}$ also has second-order mass terms, but they must be
finite---all divergent renormalizations are accomplished through the
transformation $U$.  When we split $H_{\rm eff}$ into $H_0$ and $V$,
we include in $H_0$ both constituent quark and gluon masses and the
dominant potential terms necessary to give a reasonable qualitative
description of hadronic bound states.  Whatever is left in $H_{\rm
eff}$ after subtracting $H_0$ is defined to be $V$.

In both perturbation computations the same expansion parameter is
used, namely the coupling constant $g$.  In the second perturbation
theory the running value of $g$ measured at the hadronic mass scale is
used.  In relativistic field theory $g$ at the hadronic scale has a
fixed value $g_s$ of order one; but in the computations an expansion
for arbitrarily small $g$ is used.  It is important to realize that
covariance and gauge invariance are violated when $g$ differs from
$g_s$; the QCD coupling at any given scale is not a free parameter.
These symmetries can only be fully restored when the coupling at the
hadronic scale takes its physical value $g_s$.

The conventional wisdom is that any weak-coupling Hamiltonian derived
from QCD will have only Coulomb-like potentials, and certainly will
not contain confining potentials.  Only a strong-coupling theory can
exhibit confinement.
This wisdom is wrong \cite{wwh94}.  
When $H_{\rm eff}$ is constructed by the unitary transformation of 
Eq.(\ref{eq:simtrans}), with $U$ determined by the ``similarity
renormalization group'' method, $H_{\rm eff}$ has an explicit
confining potential already in second order!  We shall explain this
result below.  However, first we should give the bad news.  If quantum
electrodynamics (QED) is solved by the same process as we propose for
QCD, then the effective Hamiltonian for QED has a confining potential
too.  In the electro dynamic case, the confining potential is purely an
artifact of the construction of $H_{\rm eff}$, an artifact which
disappears when the bound states of $H_{\rm eff}$ are computed.  Thus
the key issues, discussed below, are to understand how the confining
potential is cancelled in the case of electrodynamics, and then to
establish what circumstances would prevent a similar cancelation in
QCD.

%% file: 10Chira.tex
\section {Chiral Symmetry Breaking} 
\label{sec:chiral}
\setcounter{equation}{0}

In the mid-70's QCD emerged from Current Algebra and the 
Parton Model.  
In Current Algebra one makes use of the Partially Conserved 
Axial-Current hypothesis (PCAC), 
which states that light hadrons would be subjected to a fermionic 
symmetry called 'chiral symmetry' if only the pion mass was zero. 
If this were the case, the symmetry would be spontaneously broken, 
and the pions and kaons would be the
corresponding Goldstone bosons. The real world slightly misses this state of
affairs by effects quantifiable in terms of the pion mass and decay constant. This
violation can be expressed in terms of explicit symmetry breaking due to the
nonzero masses of the fundamental fermion fields, quarks of three light flavors,
and typically one assigns values of 4 MeV for the up-quark, 7 MeV for the
down-quark and 130 MeV for the strange-quark \cite{gal75,shv79}.
Light-Front field theory is particularly well suited to study these
symmetries  \cite{fri93}. This section follows closely the review of Daniel Mustaki
\cite{mus94}.

\subsection{Current Algebra}

To any given transformation of the fermion field we associate a
current
\begin{equation}
{\delta \cal L \over\delta(\partial_{\mu}\psi)}{\delta\psi\over\theta}
=i\bar\psi \gamma^\mu {\delta\psi \over \theta}~~, 
\end{equation}
where $\delta\psi$ is the infinitesimal variation parameterized
by $\theta$. Consider first the free Dirac theory in space-time and light-front
frames. For example the vector transformation is defined in space-time by
\begin{equation}
\psi\mapsto e^{-i\theta} \psi~~,~~\delta\psi =-i\theta\psi~~, 
\end{equation}
whence the current
\begin{equation}
j^{\mu} =\bar\psi \gamma^{\mu} \psi~. 
\end{equation}
In a light-front frame the vector
transformation will be defined as
\begin{equation}
\psi_{+}\mapsto e^{-i\theta} \psi_{+}~~,~~\delta\psi_{+} 
=-i\theta\psi_{+}
{}~~,~~\delta\psi=\delta\psi_{+} +\delta\psi_{-}~~, 
\end{equation}
where $\delta\psi_{-}$ is calculated in section II. The distinction
in the case of the vector is of course academic:
\begin{equation}
\delta\psi_{-}=-i\theta\psi_{-}~~\Longrightarrow~~\delta\psi=
-i\theta\psi~~. 
\end{equation}
Therefore for the free Dirac theory the light-front current 
$\tilde{j} ^{\mu}$ is,
\begin{equation}
\tilde{j} ^{\mu}=j^{\mu}~~. 
\end{equation}
One checks easily
that the vector current is conserved:
\begin{equation}
\partial_{\mu}j^{\mu}=0~~. 
\end{equation}
therefore the space-time and light-front vector charges, 
which measure fermion number
\begin{equation}
Q\equiv\int d^3 {\bf x}~j^0 (x)~~,~~\tilde{Q}\equiv\int d^3 \tilde{x}~
j^+ (x) 
\ ,\end{equation}
are equal \cite{mcc88}.  

The space-time chiral transformation is defined by
\begin{equation}
\psi\mapsto e^{-i\theta\gamma_5} \psi~~,~~\delta\psi=-i\theta
\gamma_5 \psi~~, 
\end{equation}
where $\gamma_5 \equiv i\gamma^{0}\gamma^{1}\gamma^{2}\gamma^{3}$. From
the Hamiltonian, one sees that the space-time theory with nonzero
fermion masses is not chirally symmetric.
The space-time axial-vector current associated to the transformation is
\begin{equation}
j^{\mu}_5 =\bar\psi \gamma^{\mu}\gamma_5 \psi ~~. 
\end{equation}
and
\begin{equation}
\partial_{\mu}j^{\mu}_5 =2im\bar\psi \gamma_5 \psi ~~. 
\end{equation}
As expected, this current is not conserved for nonzero fermion mass. The
associated charge is
\begin{equation}
Q_5 \equiv\int d^3 {\bf x}~ j^{0}_{5}=\int d^3 {\bf x}~\bar\psi \gamma^0
\gamma_5 \psi~~~. 
\end{equation}

The light-front chiral transformation is
\begin{equation}
\psi_{+}\mapsto e^{-i\theta\gamma_5} \psi_{+}~~,~~\delta\psi_{+}=
-i\theta\gamma_5 \psi_{+}~~. 
\end{equation}
This {\it is} a symmetry of the light-front theory
without requiring zero bare masses. Using $\{\gamma^{\mu},\gamma_5\}=0$,
one finds
\begin{equation}
\delta\psi_{-} (x)=-\theta\gamma_5 \int dy^- \,{\epsilon (x^- -y^- )
\over 4}(i\vec{\gamma}_{\perp}\cdot\vec{\partial}_{\perp}-m)
\gamma^+ \psi_+ (y)~~. 
\end{equation}
This expression differs from
\begin{equation}
-i\theta\gamma_5 \psi_{-} =-\theta\gamma_5 \int dy^- \,{\epsilon (x^- -y^- )
\over 4}(i\vec{\gamma}_{\perp}\cdot\vec{\partial}_{\perp}+m)
\gamma^+ \psi_+ (y) 
\end{equation}
therefore ${\tilde j}^{\mu}_5 \not= j^{\mu}_5$
(except for the plus component, due to $(\gamma^+)^2 =0$). To be precise,
\begin{equation}
{\tilde j}^{\mu}_5 = j^{\mu}_5 +im\bar\psi \gamma^{\mu} \gamma_5 \int
dy^- \,{\epsilon(x^- -y^- )\over 2}\gamma^+ \psi_+ (y)~~. 
\end{equation}
A straightforward calculation shows that
\begin{equation}
\partial_{\mu} {\tilde j} ^{\mu}_5 =0 ,
\end{equation}
as expected. Finally the light-front chiral charge is
\begin{equation}
{\tilde Q}_5 \equiv\int d^3 {\tilde x}~ {\tilde j}^{+}_{5}=\int d^3 {\tilde x}
{}~\bar\psi \gamma^{+} \gamma_5 \psi 
\label{eqn:35} \end{equation}
From the canonical anti-commutator
\begin{equation}
\{\psi(x),\psi^{\dagger}(y)\}_{x^0 =y^0}=\delta^3 ({\bf x}-{\bf y})~~,
\end{equation}
\noindent one derives
\begin{equation}
[\psi,Q_5 ]=\gamma_5 \psi~~\Longrightarrow~~[Q,Q_5]=0~~, 
\end{equation}
so that fermion number, viz., the number of
quarks minus the number of anti-quarks is conserved by the chiral charge.
However, the latter are not conserved {\it separately}. This
can be seen by using the momentum expansion of the field
one finds,

\begin{eqnarray}
Q_5 =\int {d^3 {\bf p}\over 2p^0} \sum_{s=\pm 1}s [
{|{\bf p}|\over p^0}\biggl( b^{\dagger}({\bf p},s)b({\bf p},s)+d^{\dagger}
({\bf p},s)d({\bf p},s \biggr) \nonumber \\
+{m\over p^0}\biggl( d^{\dagger}(-{\bf p},s)b^{\dagger}({\bf p},s)
e^{2ip^0 t}+b({\bf p},s)d(-{\bf p},s)e^{-2ip^0 t}\biggr) ]~~.
\end{eqnarray}

This implies that when $Q_5$ acts on a hadronic state, it will add
or absorb a continuum of
quark-antiquark pairs (the well-known pion pole) with a probability
amplitude proportional to the fermion mass and inversely proportional
to the energy of the pair. Thus $Q_5$ is most unsuited
for classification purposes.

In contrast, the light-front chiral charge conserves not only fermion
number, but also the number of
quarks and anti-quarks separately. In effect, the canonical anti-commutator
is
\begin{equation}
\{\psi_+ (x),\psi^{\dagger}_+ (y)\}_{x^+ =y^+}={\Lambda_+ \over\sqrt{2}}
\,\delta^3 ({\tilde x}-{\tilde y})~~,
\end{equation}
hence the momentum expansion of the field reads
\begin{equation}
\psi_{+}(x)=\int {d^3 \tilde{p} \over (2\pi)^{3/2} 2^{3/4} \sqrt{p^{+}}}
\sum_{h=\pm {1\over 2}}\Biggl[w(h)e^{-ipx}b(\tilde{p},h)
+w(-h)e^{+ipx}d^{\dagger}(\tilde{p},h)\Biggr]~~,
\label{eqn:43}\end{equation}
and
\begin{equation}
\{b(\tilde{p},h),b^{\dagger}(\tilde{q},h')\}=2p^{+}\delta^3 (\tilde{p}-
\tilde{q})\delta_{hh'}=\{d(\tilde{p},h),d^{\dagger}(\tilde{q},h')\}~~,
\end{equation}
\begin{equation}
\sum_{h=\pm{1\over 2}}w(h)w^{\dagger}(h)=\Lambda_+ ~~.
\end{equation}
In the   rest frame of a system, its total angular momentum along the $z$-axis is
called 'light-front helicity'; the helicity of an elementary particle is just the
usual spin projection; we label the eigenvalues of helicity with the letter 'h'. It
is easiest to work in the so-called 'chiral representation' of Dirac matrices, where
\begin{equation}
\gamma_5 =\left[\matrix{
1&0&0&0\cr
0&1&0&0\cr
0&0&-1&0\cr
0&0&0&-1\cr}\right]~,~w(+{1\over 2})=\left[\matrix{1\cr 0 \cr 0\cr 0\cr}
\right]~,~w(-{1\over 2})=\left[\matrix{0\cr 0\cr 0 \cr 1\cr}\right]
\end{equation}
\begin{equation}
\Longrightarrow ~w^{\dagger}(h)\gamma_5 w(h')=2h\delta_{hh'}~~.
\end{equation}
Inserting Eq.(\ref{eqn:43}) into Eq.(\ref{eqn:35}), one finds
\begin{equation}
\tilde{Q}_5 =\int {d^3 \tilde{p}\over 2p^{+}}\sum_h 2h\, \Biggl[ b^{\dagger}
(\tilde{p},h)b(\tilde{p},h)+d^{\dagger}(\tilde{p},h)d(\tilde{p},h)\Biggr]~~.
\end{equation}
This is just a superposition of fermion and anti-fermion
number operators, and thus our claim is proved.
This expression also shows that $\tilde{Q}_5$ annihilates the vacuum,
and that it simply measures (twice) the sum
of the helicities of all the quarks and anti-quarks  of
a given state. Indeed, in a light-front frame, the handedness of an
individual fermion is automatically determined by its helicity. To show
this, note that
\begin{equation}
\gamma_5 w(\pm {1\over 2})=\pm w(\pm {1\over 2})~~\Longrightarrow ~~
{1\pm\gamma_5 \over 2} w(\pm{1\over 2})=w(\pm{1\over 2})~,~{1\pm\gamma_5
\over 2}w(\mp{1\over 2})=0~~. 
\end{equation}
Defining as usual
\begin{equation}
\psi_{+R}\equiv {1+\gamma_5 \over 2}\psi_{+}~,~\psi_{+L}\equiv {1-\gamma_5
\over 2}\psi_{+}~~,
\end{equation}
it follows from Eq.(\ref{eqn:43}) that $\psi_{+R}$ contains only
fermions of helicity $+{1\over 2}$ and anti-fermions of helicity
$-{1\over 2}$, while $\psi_{+L}$ contains only fermions of helicity
$-{1\over 2}$ and anti-fermions of helicity $+{1\over 2}$.
Also, we see that when acted upon by the right- and left-hand charges
\begin{equation}
\tilde{Q}_{R} \equiv {\tilde{Q} +\tilde{Q}_5 \over 2}~,~\tilde{Q}_{L} \equiv
{\tilde{Q} -\tilde{Q}_5 \over 2}~~,
\end{equation}
a chiral fermion (or . anti-fermion) state may have eigenvalues
$+1$ (resp. $-1$) or zero.

In a space-time frame, this identification between helicity and chirality
applies only to massless fermions.

\subsection{Flavor symmetries}

We proceed now to the theory of three flavors of free fermions $\psi_f$,
where $f=u,d,s$, and
\begin{equation}
\psi\equiv\left[\matrix{\psi_u \cr \psi_d \cr \psi_s \cr}\right]
{}~,~{\rm and}~~M\equiv\left[\matrix{
m_u& 0 & 0 \cr
0 & m_d & 0 \cr
0 & 0 & m_s \cr
}\right]~~.
\end{equation}
The vector, and axial-vector, flavor non-singlet transformations are defined
respectively as
\begin{equation}
\psi\mapsto e^{-i{\lambda^{\alpha}\over 2}\theta^{\alpha}} \psi~~,~~\psi
\mapsto e^{-i{\lambda^{\alpha}\over 2}\theta^{\alpha}\gamma_5}\psi~~,
\end{equation}
where the summation index $\alpha$ runs from 1 to 8. The space time
Hamiltonian $P^0$ is invariant under vector transformations if the quarks have
equal masses ('$SU(3)$ limit'), and invariant under chiral transformations if
all masses are zero ('chiral limit').

The light-front Hamiltonian is
\begin{eqnarray}
P^{-}=\sum_f {i\sqrt{2}\over 4}\int d^3 \tilde{x}\int dy^- \,\epsilon(x^-
-y^-)\,\psi^{\dagger}_{f+} (y)\,(m_f ^2 -\Delta_{\perp})\psi_{f+}(x) \nonumber \\
={i\sqrt{2}\over 4}\int d^3 \tilde{x}\int dy^- \,\epsilon(x^-
-y^-)\,\psi^{\dagger}_{+} (y)\,(M^2 -\Delta_{\perp})\psi_{+} (x)~~.
\end{eqnarray}
Naturally,  $P^-$ is not invariant under the vector transformations
\begin{equation}
\psi_+ \mapsto e^{-i{\lambda^{\alpha}\over 2}\theta^{\alpha}} \psi_{+}
\end{equation}
unless the quarks have equal masses. But if they do, then
$P^-$ is also invariant under the chiral transformations
\begin{equation}
\psi_+ \mapsto e^{-i{\lambda^{\alpha}\over 2}\theta^{\alpha}\gamma_5}
\psi_{+}~~, 
\end{equation}
whether this common mass is zero or not.

One finds that the space-time currents
\begin{equation}
j^{\mu\alpha} =\bar\psi \gamma^{\mu} {\lambda^{\alpha}\over 2}\psi~~,~~
j^{\mu\alpha}_5 =\bar\psi \gamma^{\mu}\gamma_5 {\lambda^{\alpha}\over 2}
\psi ~~, 
\end{equation}
have the following divergences:
\begin{equation}
\partial_{\mu}j^{\mu\alpha} =i\bar\psi \biggl[M,{\lambda^{\alpha}\over 2}
\biggr]\psi~~,~~\partial_{\mu}j^{\mu\alpha}_5 =i\bar\psi \gamma_5
\biggl\{ M,{\lambda^{\alpha}\over 2}\biggr\}\psi~~. 
\end{equation}
These currents have obviously the expected conservation properties.

Turning to the light-front frame we find
\begin{equation}
\tilde{j}^{\mu\alpha}=j^{\mu\alpha}-i\bar\psi \biggl[ M,{\lambda^{\alpha}
\over 2}\biggr]\gamma^{\mu}\int dy^- \,{\epsilon(x^- -y^-)\over 4}\,
\gamma^+ \psi_+ (y)~~. 
\end{equation}
So $\tilde{j}^{\mu\alpha}$ and $j^{\mu\alpha}$ may be equal for all $\mu$
only if the quarks have equal masses. The {\it vector}, flavor non-singlet
charges in each frame are two different octets of operators,
except in the $SU(3)$ limit.

For the light-front current associated with axial transformations, we get
\begin{equation}
\tilde{j}^{\mu\alpha}_5 =j^{\mu\alpha}_5 -i\bar\psi \biggl\{ M,
{\lambda^{\alpha}\over 2}\biggr\}\gamma_5 \gamma^{\mu}\int dy^- \,
{\epsilon(x^- -y^-)\over 4}\,\gamma^+ \psi_+ (y)~~. 
\end{equation}
Hence $\tilde{j}^{\mu\alpha}_5$ and $j^{\mu\alpha}_5$ are not equal
(except for $\mu =+$), even in the $SU(3)$ limit, unless all quark masses
are zero. Finally, one obtains the following divergences:
\begin{eqnarray}
\partial_{\mu}\tilde{j}^{\mu\alpha}=\bar\psi \biggl[ M^2 ,{\lambda^{\alpha}
\over 2}\biggr]\int dy^- \,{\epsilon(x^- -y^-)\over 4}\,\gamma^+
\psi_+ (y)~~, \nonumber \\
\partial_{\mu}\tilde{j}^{\mu\alpha}_5 =-\bar\psi \biggl[ M^2 ,
{\lambda^{\alpha}\over 2}\biggr]\gamma_5 \int dy^- \,{\epsilon(x^- -y^-)
\over 4}\,\gamma^+ \psi_+ (y)~~.
\end{eqnarray}
As expected, both light-front currents are conserved in the $SU(3)$ limit,
without requiring zero masses. Also note how light-front relations often seem
to involve
the masses {\it squared}, while the corresponding space-time relations are
{\it linear} in the masses. The integral operator
\begin{equation}
\int dy^- \,{\epsilon(x^- -y^-)\over 2}~\equiv~{1\over \partial^{x}_-}
\end{equation}
compensates for the extra power of mass.

The associated light-front charges are
\begin{equation}
\tilde {Q}^{\alpha} \equiv \int d^3 {\tilde x} \,\bar{\psi} \gamma^{+}
{\lambda^{\alpha}\over 2}\psi~,~
\tilde {Q}^{\alpha}_5  \equiv \int d^3 {\tilde x} \,\bar{\psi} \gamma^{+}
\gamma_5 {\lambda^{\alpha}\over 2}\psi~~. 
\end{equation}
Using the momentum expansion of the fermion triplet 
Eq.(\ref{eqn:43}), 
where now
\begin{equation}
b(\tilde{p},h)\equiv\left[\matrix{b_u (\tilde{p},h), b_d (\tilde{p},h),
 b_s (\tilde{p},h)}\right]~,~{\rm and}~~d(\tilde{p},h)\equiv\left[
\matrix{d_u (\tilde{p},h),
d_d (\tilde{p},h), d_s (\tilde{p},h)}\right]~~,
\end{equation}
one can express the charges as
\begin{equation}
\tilde{Q}^{\alpha} =\int {d^3 \tilde{p}\over 2p^{+}}\sum_h \Biggl[ b^{\dagger}
(\tilde{p},h){\lambda^{\alpha}\over 2}b(\tilde{p},h)-d^{\dagger}(\tilde{p},h)
{\lambda^{\alpha T}\over 2}d(\tilde{p},h)\Biggr]~~,
\end{equation}
\begin{equation}
\tilde{Q}^{\alpha}_5 =\int {d^3 \tilde{p}\over 2p^{+}}\sum_h 2h\,
\Biggl[ b^{\dagger}
(\tilde{p},h){\lambda^{\alpha}\over 2}b(\tilde{p},h)+d^{\dagger}(\tilde{p},h)
{\lambda^{\alpha T}\over 2}d(\tilde{p},h)\Biggr]~~,
\end{equation}
where the superscript $T$ denotes matrix transposition. Clearly all
sixteen charges annihilate the vacuum
\cite{ida75a,jes68,jes69,leu68,leu69,sas75}. 
As $\tilde {Q}^{\alpha}$ and $\tilde {Q}^{\alpha}_5$ conserve the number
of quarks and anti-quarks separately, these charges are well-suited for
classifying hadrons in terms of their valence constituents, {\it whether the
quark masses are equal or not} \cite{des74}. Since the
charges commute with
$P^+$ and ${\bf P}_{\perp}$, all hadrons belonging to the same multiplet
have the same momentum. But this common value of momentum is arbitrary, because
in a light-front frame one can boost between any two values of momentum, using
only kinematic operators.

One finds that
these charges generate an $SU(3)\otimes SU(3)$ algebra:
\begin{equation}
[\tilde{Q}^{\alpha},\tilde{Q}^{\beta}]=i\,f_{\alpha\beta\gamma}
\,\tilde{Q}^{\gamma}~,~
[\tilde{Q}^{\alpha},\tilde{Q}^{\beta}_5 ]=i\,f_{\alpha\beta\gamma}
\,\tilde{Q}^{\gamma}_5 ~,~
[\tilde{Q}^{\alpha}_5 ,\tilde{Q}^{\beta}_5 ]=i\,f_{\alpha\beta\gamma}
\,\tilde{Q}^{\gamma}~~,
\end{equation}
and the corresponding right- and left-hand charges generate two commuting
algebras denoted
$SU(3)_R$ and $SU(3)_L$
\cite{jes68,jes69,leu68,leu69,leu74b,leu74c}
\cite{bel74,buc70,dag66,dea73,des74,efw73,fei73}
\cite{ida74,ida75a,ida75b,ida75c,mel74,osb74}
\cite{cah75,cat76,sas75}.
Most of these papers in fact study a larger algebra of light-like
charges, namely $SU(6)$, but the sub-algebra $SU(3)_R\otimes SU(3)_L$ suffices for
our purposes.

Since
\begin{equation}
[\psi_+ ,\tilde{Q}^{\alpha}_5 ]=\gamma_5 {\lambda^{\alpha} \over 2}
\psi_+ ~~,
\end{equation}
the quarks form an irreducible representation of this algebra. To be
precise, the
quarks (resp. anti-quarks) with helicity $+{1\over 2}$ (resp. $-{1\over
2}$) transform as a triplet of $SU(3)_R$ and a singlet of $SU(3)_L$, the
quarks (resp. anti-quarks) with helicity $-{1\over 2}$ (resp. $+{1\over
2}$) transform as a triplet of $SU(3)_L$ and a singlet of $SU(3)_R$. Then
for example the ordinary vector $SU(3)$ decuplet of $J={3\over 2}$ baryons
with $h=+{3\over 2}$ is a pure right-handed (10,1) under $SU(3)_R\otimes
SU(3)_L$.
The octet ($J={1\over 2}$) and decuplet ($J={3\over 2}$) with $h=+{1\over 2}$
transform together as a (6,3). For bosonic states we expect both chiralities
to contribute with equal probability. For example, the
octet of pseudo-scalar mesons arises from a superposition
of irreducible representations of $SU(3)_R\otimes SU(3)_L$:
\begin{equation}
|J^{PC}=0^{-+}>={1\over \sqrt{2}}\,|(8,1)-(1,8)>~~,
\end{equation}
while the octet of vector mesons with zero helicity corresponds to
\begin{equation}
|J^{PC}=1^{--}>={1\over \sqrt{2}}\,|(8,1)+(1,8)>~~,
\end{equation}
and so on. These low-lying states have $L_z =0$, where
\begin{equation}
L_z=-i\int d^3 \tilde{x}~\bar\psi \gamma^+ (x^1 \partial_2 -x^2
\partial_1)\psi 
\end{equation}
is the orbital angular momentum along $z$.

In the realistic case of unequal masses, the chiral charges are not conserved.
Hence they generates multiplets which
are not mass-degenerate --- a welcome feature.
The fact that the invariance of the vacuum
does not enforce the 'invariance of the world' (viz., of energy), in sharp
contrast with the order of things in space-time (Coleman's theorem), is yet
another remarkable property of the light-front frame.

In contrast with the space-time picture, free light-front {\it current}
quarks are also {\it constituent} quarks because:

\noindent$\bullet$ They can be massive without preventing chiral
symmetry, which we know is (approximately) obeyed by hadrons.

\noindent$\bullet$ They form a basis for a classification of hadrons
under the light-like chiral algebra.

\subsection{Quantum Chromodynamics}

In the quark-quark-gluon vertex $gj^{\mu}A_{\mu}$, the transverse
component of the vector current is
\begin{equation}
{\bf j_{\perp}}(x)=\cdots +{im\over 4}\int dy^{-}\, \epsilon(x^{-}
-y^{-})\, \Biggl[\bar{\psi}_{+}(y)\gamma^{+}\vec{\gamma}_{\perp}
\psi_{+}(x)+\bar{\psi}_{+}(x)\gamma^{+}\vec{\gamma}_{\perp}\psi_{+}(y)
\Biggr]~~,
\end{equation}
where the dots represent chirally symmetric terms, and where color, as well as
flavor, factors and indices have been omitted for clarity. The term explicitly
written out breaks chiral symmetry for nonzero quark mass. Not surprisingly, it
generates vertices in which the two quark lines have opposite helicity.

The canonical anti commutator for the bare fermion fields
still holds in the interactive theory (for each flavor). The momentum
expansion of $\psi_+(x)$ remains the same except that now the $x^+$ dependence in
$b$ and $d$.
and
\begin{equation}
\{b(\tilde{p},h,x^+),b^{\dagger}(\tilde{q},h',y^+)\}_{x^+ =y^+}
=2p^{+}\delta^3 (\tilde{p}-
\tilde{q})\delta_{hh'}=\{d(\tilde{p},h,x^+),d^{\dagger}(\tilde{q},h',y^+)\}
_{x^+ =y^+}~~
\end{equation}
The momentum expansions of the light-like charges remain the same 
(keeping in mind that the creation and annihilation
operators are now unknown functions of 'time'). Hence the charges still
annihilate the Fock vacuum, and are suitable for classification purposes.

We do not require annihilation of the physical vacuum (QCD
ground state). The successes of CQM's suggest that to understand the
properties of the hadronic spectrum, it may not be necessary to take
the physical vacuum into account. This is also the point of view taken
by the authors of a recent paper on the renormalization of QCD \cite{ghp93}. Their
approach consists in imposing an 'infrared' cutoff in longitudinal
momentum, and in compensating for this suppression by means of Hamiltonian
counter terms. Now, only terms that annihilate the {\it Fock vacuum} are
allowed in their Hamiltonian $P^-$. Since all states in the truncated
Hilbert space have strictly positive longitudinal momentum except for the
Fock vacuum (which has $p^+ =0$), the authors hope to be able to adjust the
renormalizations in order to fit the observed spectrum, without having to
solve first for the physical vacuum.

Making the standard choice of gauge: $A_- =0$~, one finds that the properties of
vector and axial-vector currents  are also unaffected by the inclusion of QCD
interactions, except for the replacement of the derivative by the covariant
derivative. The divergence of the renormalized, space-time, non-singlet axial
current is anomaly-free
\cite{col84}. As $j^{\mu\alpha}_{5}$ and $\tilde{j}^{\mu\alpha}_{5}$ become equal
in the chiral limit, the divergence of the light-front current is also anomaly-free
(and goes to zero in the chiral limit). The corresponding charges, however, {\it do
not} become equal in the chiral limit. This can only be due to contributions at
$x^-$-infinity coming from the Goldstone boson fields, which presumably cancel the
pion pole of the space-time axial charges. Equivalently, if one chooses periodic
boundary conditions, one can say that this effect comes from the longitudinal zero
modes of the fundamental fields.

From soft pion physics we know that the chiral limit of $SU(2)\otimes SU(2)$ is
well-described by PCAC. Now, using PCAC one can show that in the chiral limit
$Q^{\alpha}_5~(\alpha=1,2,3)$ is conserved, but $\tilde{Q}^{\alpha}_5$ is not
\cite{cah75,ida74}. In other words, the renormalized light-front charges are
sensitive to spontaneous symmetry breaking, although they do annihilate the vacuum.
It is likely that this behavior generalizes to $SU(3)\otimes SU(3)$, viz., to the
other five light-like axial charges. Its origin, again, must lie in zero modes.

In view of this 'time'-dependence, one might wonder whether the light-front axial
charges are observables. From PCAC, we know that it is indeed the case: their
matrix elements between hadron states are directly related to off-shell pion
emission \cite{fei73,cah75}. For a hadron $A$ decaying into a hadron $B$ and a pion,
one finds
\begin{equation}
<B|\tilde{Q}^{\alpha}_5 (0)|A>=-{2i(2\pi)^3 p^{+}_A \over
m^{2}_A -m^{2}_B}<B,\pi^{\alpha}|A>\delta^3 (\tilde{p}_A -
\tilde{p}_B)~~. 
\end{equation}
Note that in this reaction,
the mass of hadron $A$ must be larger than the mass of $B$  due to the
pion momentum.

\null

\subsection{Physical multiplets}

Naturally, we shall assume  that real hadrons fall into representations of an
$SU(3)\otimes SU(3)$ algebra. We have identified the generators of this algebra
with the light-like chiral charges. But this was done in the artificial case of the
free quark model. It remains to check whether this identification works in the real
world.

Of course, we already know that the predictions based on isospin ($\alpha =1,2,3$)
and hyper-charge ($\alpha=8$) are true. Also, the nucleon-octet ratio
$D/F$ is correctly predicted to be 3/2, and several relations between magnetic
moments match well with experimental data.

Unfortunately, several other predictions are in disagreement with observations
\cite{clo79}. For example, $G_A/G_V$ for the nucleon is expected to be equal to 5/3,
while the experimental value is about 1.25. Dominant decay channels such as $N^*
\rightarrow N\pi$, or $b_1 \rightarrow\omega\pi$, are forbidden by the light-like
current algebra. The anomalous magnetic moments of nucleons, and all form factors
of the rho-meson would have to vanish. De Alwis and Stern \cite{des74} point out
that the matrix element of $\tilde{j}^{\mu\alpha}$ between two given hadrons would
be equal to the matrix element of $\tilde{j}^{\mu\alpha}_5$ between the same two
hadrons, up to a ratio of Clebsch-Gordan coefficients. This is excluded though
because vector and axial-vector form factors have very different analytic
properties as functions of momentum-transfer.

In addition there is, in general, disagreement between the values of $L_z$ assigned
to any given hadron. This comes about because in the classification scheme, the
value of $L_z$ is essentially an afterthought, when group-theoretical
considerations based on flavor and helicity have been taken care of. On the other
hand, at the level of the current quarks, this value is determined by covariance
and external symmetries. Consider for example the $L_z$ assignments in the case of
the pion, and of the rho-meson with zero helicity. As we mentioned earlier, the
classification assigns to these states a pure value of $L_z$, namely zero. However,
at the fundamental level, one expects these mesons to contain a wave-function
$\phi_1$ attached to $L_z =0$ (anti-parallel $q\bar q$ helicities), and also a
wave-function $\phi_2$ attached to $L_z =\pm 1$ (parallel helicities). Actually,
the distinction between the pion and the zero-helicity rho is only based on the
different momentum-dependence of $\phi_1$ and $\phi_2$
\cite{leu74b,leu74c}. If the interactions were turned off, $\phi_2$ would vanish
and the masses of the two mesons would be degenerate (and equal to $(m_u +m_d)$).

We conclude from this comparison with experimental data, that if indeed real
hadrons are representations of some $SU(3)\otimes SU(3)$ algebra, then the
generators $G^{\alpha}$ and $G^{\alpha}_5$ of this classifying algebra must be
different from the current light-like charges $\tilde{Q}^{\alpha}$ and
$\tilde{Q}^{\alpha}_{5}$ (except however for $\alpha=1,2,3,8$). Furthermore, in
order to avoid the phenomenological discrepancies discussed above, one must forego
kinematical invariance for these generators; that is, $G^{\alpha}(\tilde{k})$ and
$G^{\alpha}_5 (\tilde{k})$ must depend on the momentum $\tilde k$ of the hadrons in
a particular irreducible multiplet.

Does that mean that our efforts to relate the physical properties of hadrons to the
underlying field theory turn out to be fruitless? Fortunately no, as argued by De
Alwis and Stern \cite{des74}. The fact that these two sets of generators (the
$\tilde{Q}$'s and the $G$'s) act in the same Hilbert space, in addition to
satisfying  the same commutation relations, implies that they must actually be
unitary equivalent (this equivalence was originally suggested by Dashen, and by
Gell-Mann, \cite{gel72}). There exists a set of momentum-dependent unitary
operators $U(\tilde {k})$ such that
\begin{equation}
G^{\alpha}(\tilde {k})=U(\tilde {k})\tilde {Q}^{\alpha}U^{\dagger}(\tilde{k})
{}~,~G^{\alpha}_{5}(\tilde {k})=U(\tilde {k})\tilde {Q}^{\alpha}_{5}
U^{\dagger}(\tilde{k})~~.
\end{equation}
Current quarks, and the real-world hadrons built out of them, fall into
representations of this algebra. Equivalently (e.g., when calculating electro-weak
matrix elements), one may consider the original current algebra, and define its
representations as 'constituent' quarks and 'constituent' hadrons. These quarks
(and antiquark s) within a hadron of momentum $\tilde {k}\,$ are represented by a
'constituent fermion field'
\begin{equation}
\chi ^{\tilde {k}}_{+}(x){\Bigl\vert}_{x^+ =0}\,\equiv U(\tilde {k})\psi_{+}(x)
{\Bigl\vert}_{x^+ =0}~U^{\dagger}(\tilde {k})~~,
\end{equation}
on the basis of which the physical generators
can be written in canonical form:
\begin{equation}
\tilde {G}^{\alpha} \equiv \int d^3 {\tilde x} \,\bar{\chi} \gamma^{+}
{\lambda^{\alpha}\over 2}\chi~,~
\tilde {G}^{\alpha}_5  \equiv \int d^3 {\tilde x} \,\bar{\chi} \gamma^{+}
\gamma_5 {\lambda^{\alpha}\over 2}\chi~~. 
\end{equation}
it follows that the constituent annihilation/creation
operators are derived from
the current operators via
\begin{equation}
a^{\tilde {k}}\,(\tilde {p},h)\equiv U(\tilde {k})b (\tilde {p},h)
U^{\dagger}(\tilde {k})~~,~~{c^{\tilde {k}}\,}^{\dagger}(\tilde {p},h)
\equiv
U(\tilde {k})d^{\dagger}(\tilde {p},h)U^{\dagger}(\tilde {k})~~.
\end{equation}

Due to isospin invariance, this unitary transformation cannot mix flavors,
it only mixes helicities. It can therefore be represented by three unitary
$2\times 2$ matrices $T^{f}(\tilde {k},\tilde {p})$ such that
\begin{equation}
a^{\tilde {k}}_{f}(\tilde {p},h)=\sum_{h'=\pm{1\over 2}}
T^{f}_{hh'}(\tilde {k},\tilde {p})\,b_f
(\tilde {p},h')~,~c^{\tilde {k}}_{f}(\tilde {p},h)=\sum_{h'=\pm{1\over 2}}
T^{f\,*}_{hh'}(\tilde {k},\tilde {p})\,d_f (\tilde {p},h')~~, 
\end{equation}
one for each flavor $f=u,d,s$.
Since we need the transformation to be unaffected when
$\tilde {k}$ and $\tilde {p}$ are boosted along $z$
or rotated around $z$ together,
the matrix $T$ must actually be a
function of only kinematical invariants. These are
\begin{equation}
\xi \equiv {p^+ \over {k^+}}~~{\rm and}~~{\bf\kappa}_{\perp}\equiv
{\bf p}_{\perp}-\xi{\bf k}_{\perp}~~,~~{\rm where}
\sum_{\rm constituents}\xi =1~~,~~\sum_{\rm constituents}{\bf\kappa}_{\perp}
={\bf 0}~~.
\end{equation}
Invariance under time reversal ($x^+ \mapsto -x^+$) and parity ($x^1
\mapsto -x^1$) further constrain its functional form, so that finally
\cite{leu74b,leu74c}
\begin{equation}
T^{f}(\tilde {k},\tilde {p})=\exp\,[-i\,{{\bf\kappa}_{\perp}\over |{\bf\kappa}
_{\perp}|}\cdot{\bf\sigma}_{\perp}~\beta _{f}(\xi,{\bf\kappa}_{\perp}^{2})]
\ .\end{equation}

Thus the relationship between current and constituent quarks is embodied
in the three functions $\beta _f (\xi,{\bf\kappa}_{\perp}^{2})$, which we
must try to extract from comparison with experiment. (In first approximation
it is legitimate to take $\beta _u$ and $\beta _d$ equal since $SU(2)$
is such a  good symmetry.)

Based on some assumptions abstracted from the free-quark model 
\cite{frg72,leu74a,leu74b,leu74c} has derived a set
of sum rules obeyed by mesonic wave-functions. Implementing then the transformation
described
above, Leutwyler finds various relations involving form factors and scaling
functions of mesons, and computes the current quark masses. For example, he
obtains
\begin{equation}
F_{\pi}<F_{\rho}~~,~~F_{\rho}=3\,F_{\omega}~~,~~F_{\rho}<{3\over {\sqrt 2}}\,
|F_{\phi}|~~, 
\end{equation}
and the $\omega /\phi$ mixing angle is estimated to be about 0.07 rad.
\cite{leu74a} also shows that the average transverse momentum of a quark inside a
meson is substantial ($|{\bf p}_{\perp}|_{\rm rms} > 400$ MeV), thus justifying
{\it a posteriori} the basic assumptions of the relativistic CQM (e.g.,  Fock space
truncation and relativistic energies). This large value also provides an
explanation for the above-mentioned failures of the
$SU(3)\otimes SU(3)$ classification scheme \cite{clo79}. On the negative side, it
appears that the functional dependence of the $\beta _f$'s cannot be easily
determined with satisfactory precision.

%% file: 11Prosp.tex
\section{The prospects and challenges}
\setcounter{equation}{0}

Future work on light-cone physics can be discussed in terms of 
developments along two distinct lines.  
One direction  focus on solving phenomenological 
problems while the other will  focus on  the use of light-cone 
methods to understand various properties of quantum field theory. 
Ultimately both point towards understanding the physical world. 

An essential features of  relativistic quantum field theories such as
QCD is that particle number is not conserved; {\it i.e.} if we examine the
wavefunction of a hadron at fixed-time $t$ or light-cone time
$x^+$, any number of particles can be in flight.  The expansion of a 
hadronic eigenstate of the full Hamiltonian has to be represented as 
a sum of amplitudes representing the fluctuations over particle number, 
momentum, coordinate configurations, color partitions, and helicities.
The advantage of the light-cone Hamiltonian formalism is that
one can conceivably predict the individual amplitudes for each of these
configurations.  As we have discussed in this review, the basic
procedure is to diagonalize the full light-cone Hamiltonian 
in the free light-cone Hamiltonian basis.  The eigenvalues
are the invariant mass squared of the discrete and continuum eigenstates 
of the spectrum.  The  projection of the eigenstate on the free Fock basis 
are the light-cone wavefunctions and provide a rigorous
relativistic many-body representation in terms of its degrees of freedom. 
Given the light-cone wavefunction one can compute the structure
functions and distribution amplitudes.  More generally, the light-cone
wavefunctions provide the interpolation between hadron scattering
amplitudes and the underlying parton subprocesses. 

The unique property of light-cone quantization that makes the  
calculations of light cone wavefunctions particularly useful is that they
are independent of the reference frame. Thus when one does 
a non-perturbative bound-state calculation  of a light-cone
wave function, that same wavefunction can be use in many different 
problems. 

Light-cone methods have been quite successful in understanding
recent experimental results, as we discussed in 
Section~\ref{sec:impact} and Section~\ref{sec:exclusive}.
We have seen that light-cone methods are very useful for understanding
a number of properties of nucleons as well as many exclusive
processes.  We also saw that these methods can be applied 
in conjunction with perturbative QCD calculations. 
Future  phenomenological application will continue to address
specific experimental results that  have a distinct non-perturbative 
character and which are therefore difficult to address by other methods. 

The simple structure of the light-cone Hamiltonian can be used as a basis to
infer information on the non-perturbative and perturbative structure of
QCD.  For example, factorization theories separating hard and soft
physics in large momentum transfer exclusive and inclusive reactions \cite{leb80}. 
Mueller {\it et al.} \cite{bhm92,chm95} have pioneered the 
investigation of structure
functions at $x \rightarrow 0$ in the light-cone Hamiltonian formalism.
Mueller's approach is to consider the light-cone wavefunctions of heavy
quarkonium in the large $N_c$ limit.   The resulting structure functions
display energy dependence related to the Pomeron. 

One can also consider the hard structure of the light-cone wavefunction.
The wavefunctions of a hadron contain fluctuations which are
arbitrarily far off the energy shell.  In the case of light-wave
quantization, the hadron wavefunction contains partonic states
of arbitrarily high invariant mass. If the light-cone wavefunction is known 
in the domain of low invariant mass, then one can use the projection 
operators formalism to construct the wavefunction for large 
invariant mass by integration of the hard interactions.  
Two types of hard fluctuations emerge:
``extrinsic'' components associates with gluon splitting
$g\rightarrow q\bar q$ and the $q \rightarrow qg$ bremsstrahlung
process and ``intrinsic'' components associated with multi-parton 
interactions within the hadrons, $gg \rightarrow Q\bar Q$; etc..  

One can use the probability of the intrinsic  contribution 
to compute the $x \rightarrow 1$ power-law
behavior of structure functions, the high relative transverse momentum
fall-off of the light-cone wavefunctions, and the probability for high
mass or high mass $Q\bar Q$ pairs in the sea quark distribution of the
hadrons \cite{bhm92}.  The full analysis of the hard components of hadron 
wavefunction can be carried out systematically using an effective 
Hamiltonian operator approach.

If we contrast the light-cone approach with lattice calculations we see 
the potential power of the light-cone method. In the lattice approach one 
calculates a set of numbers, for example a set of operator product 
coefficients \cite{mas89}, and then one uses  them to calculate a physical observable 
where the expansion is valid. This should be contrasted with the 
calculation of a light-cone wave function which gives predictions 
for all physical observables independent of the reference frame.  
There is a further advantage in that the shape of the light-cone 
wavefunction can provide a deeper 
understanding of the  physics that underlies a particular experiment. 

The focus is then on how to find reasonable approximations to light-cone 
wavefunctions that  make non-perturbative calculations
tractable.  For many problems it is not necessary to know 
everything about  the wavefunction  to make physically interesting 
predictions.  Thus one attempts to isolate and calculate the important aspects of 
the light-cone wavefunction. We saw in the discussion of the properties of nuclei 
in this review, that spectacular results can be obtained this way 
with a minimal input. Simply incorporating the angular momentum 
properties lead to very 
successful result almost independent of the rest of the structure 
of the light-cone wavefunction.

Thus far there has been remarkable success in applying the light-cone
method to theories in one-space and one-time dimension.  
Virtually any 1+1 quantum
field theory can be solved using light-cone methods. 
For calculation in 3+1 dimensions the essential problem is that  
the number of degrees of freedom
needed to specify each Fock state even in a discrete basis quickly
grows since each particles' color, helicity, transverse momenta
and light-cone longitudinal momenta  have
to be specified.  Conceivably advanced computational algorithms for
matrix diagonalization, such as the Lanczos method could allow the
diagonalization of sufficiently large matrix representations to give 
physically meaningful results.  A test of this
procedure in QED is now being carried out by J. Hiller {\it et al.} 
\cite{hbo95} for the
diagonalization of the physical electron in QED. 
The goal is to compute the electron's anomalous moment at large
$\alpha_{QED}$ non-perturbatively.

Much of the current work in this area attempts to find approximate 
solution to problems in 3+1 dimensions by starting from a 
1+1 dimensionally-reduce versions of that theory. 
In some calculations this reduction is very explicit while 
in others it is hidden. 

An interesting approach has been proposed by
Klebanov and coworkers \cite{bdk93,dkb94,dak93}.  
One decomposes the Hamiltonian into two classes
of terms. Those which have the matrix elements  that are at 
least linear in the transverse momentum (non-collinear) and those
that are independent of the transverse momentum (collinear).
In the collinear  models one discards the nonlinear 
interactions and calculates 
distribution functions which do not explicitly depend on 
transverse dimensions.  These can then be directly compared with data. 
In this approximation QCD (3+1) reduces to a 1+1 theory in which all
the partons move along $\vec k_{\perp i}=0$.  However, the
transverse polarization of the dynamical gluons is retained.  In effect
the physical gluons are replaced by two scalar fields representing
left- and right-handed polarized quanta.

Collinear QCD has been solved in detail by Antonuccio {\it et al.}
\cite{and95,and96a,and96b,anp97}.  The result is hadronic eigenstates
such as mesons with a full complement of $q\bar q$ and $g$ light-cone Fock
states.  Antonuccio and Dalley also obtain a glueball spectrum which
closely resembles the gluonium states predicted by lattice gauge theory
in 3+1 QCD.  They  have also computed the wavefunction and structure
functions of the mesons, including the quark and gluon helicity
structure functions.  One interesting result,  shows
that the gluon helicity is strongly correlated with the helicity of the parent
hadron, a result also expected in 3+1 QCD \cite{bbs94}.
While collinear QCD is a drastic approximation to physical QCD, it
provides a solvable basis as a first step to actually theory.

More recently Antonuccio {\it et al.} \cite{and95,and96a,and96b,anp97} 
have noted that Fock states differing by 1 or 2 gluons are coupled 
in the form of ladder relations
which constrain the light-cone wavefunctions at the edge of phase space.  
These relations in turn allow one to construct the leading behavior of the
polarized and unpolarized gluon structure function at 
$x  \rightarrow 0$.

The transverse lattice method includes the 
transverse behavior approximately through a lattice that only
operates in the transverse directions. In this method which was 
proposed by Bardeen, Pearson and Rabinovici
\cite{bap76,bpr80}, the transverse degrees of freedom of the gauge theory
are represented by lattice variables and the longitudinal degrees of
freedom are treated with light-cone variables.  Considerable progress
has been made in recent years on the integrated method by 
Burkardt \cite{bur93,bur94}, Griffin \cite{gri92} and van de 
Sande {\it et al.} \cite{dav96,van96,vab96,vad96}, 
see also Gaete {\it et al.} \cite{ggs93}.  This method is
particularly promising for analyzing confinement in QCD.

The importance of renormalization is seen in the Tamm-Dancoff solution 
of the Yukawa model. We present some simple examples of 
non-perturbative renormalization 
in the context of integral-equations which seems to have 
all the ingredients one would want. However, the method has not been 
successfully transported to a 3+1 dimensional field theory.
We also discussed the Wilson approach which focuses on
this issue as a guide to developing their light-cone method. They use a unique
unitary transformation to band-diagonalize the theory on the way to 
renormalization.  The method,  however, is perturbative at its core which
calls into question its applicability as a true non-perturbative renormalization. 
They essentially starts from the confining potential one gets from the 
longitudinal confinement that is fundamental to lower dimensional 
theories and then builds the three-dimensional structure on that. 
The methods has been successfully applied to solving for the low-lying
levels of  positronium and their light-cone wavefunctions.
Jones and Perry \cite{jop96a,jop96b} 
have also shown how the Lamb shift and its associated
non-perturbative Bethe-logarithm arises in the light cone Hamiltonian
formulation of QED.

There are now many examples, some of which were reviewed here,
that show that DLCQ as a numerical method provides excellent 
solutions to almost all two dimensional theories with a minimal effort.  
For models in 3+1 dimensions,  the method is also applicable, 
while much more complicated.  
To date only QED has been solved with a high degree of precision 
and some of those results are presented in this review 
\cite{kpw92,kap92,trp96,trp97a,trp97b}.  
Of course there one has high order perturbative results to check  against. 
This has proven to be an important laboratory for developing 
light-cone methods. Among the most interesting results of these calculations is the fact, 
that rotational symmetry of the result appears in spite of the fact that
the approximation must necessary break that symmetry.

One can use light-cone quantization to study the structure 
of quantum field theory. The theories considered are often 
not physical, but are selected to help in the understanding 
a particular non-perturbative phenomenon.
The relatively simple vacuum properties  of light-front field theories
underlying many of these `analytical' approaches.
The relative simplicity of the light-cone vacuum provides a firm 
starting point to attack many non perturbative issues. 
As we saw in this review in two dimensions not only are the problems
tractable from the outset, but in many cases, like the Schwinger model, 
the solution gives a unique insight and understanding. In the Schwinger model
we saw that the Schwinger particle indeed has the simple parton
structure that one hopes to see QCD.

It has been know for some time that light-cone field theory is uniquely suited for
address problems in string theory.
In addition recently new developments in formal field theory associated with
string theory, matrix  models and  M-theory have appeared which also
seem particularly well suited to the light-cone approach
\cite{sus97}. 
Some issues in formal field theory which have proven to intractable 
analytically, such as the density of the states at high energy,  
have been  successfully addressed with numerical light-cone methods.

In the future one hopes to address a number of outstanding issues, 
and one of the most interesting is spontaneous symmetry breaking.  
We have already seem in this review that the light cone provides a new 
paradigm for spontaneous symmetry breaking in $\phi^4$ in 2 dimensions.
Since the vacuum is simple in the light-cone approach the physics of 
spontaneous symmetry breaking must reside in the zero modes operators.
It has been known for some time that these operators satisfy a constraint
equation. We reviewed here the now well-known fact that the solution of this 
constraint equation can spontaneously break a symmetry. In fact, in 
the simple $\phi^4$-model the numerical results for the 
critical coupling constant and the critical exponent are quite good.

The light cone has a number of unique properties with respect 
to chiral symmetry. It has been known for a long time, for example, 
that the free theory of a fermion with a mass still has a chiral symmetry 
in a light-cone theory.   
In Section~\ref{sec:chiral} we reviewed chiral symmetry on the light cone. 
There has recently been a few applications of light-cone methods to solve 
supersymmetry but as yet no one has addressed the issue of dynamical 
supersymmetry breaking.

Finally, let us highlight the intrinsic advantages  of light-cone field theory:
\begin{itemize}
\item
The light-cone wavefunctions are independent of the
momentum of the bound state --
only relative momentum coordinates appear.
\item
The vacuum state is simple and in many cases trivial.
\item
Fermions and fermion derivatives
are treated exactly; there is no fermion doubling problem.
\item
The minimum number of physical degrees of freedom are used because of
the light-cone gauge. No Gupta-Bleuler or Faddeev-Popov
ghosts occur and unitarity is explicit.
\item
The output is the full color-singlet spectrum of the theory,
both bound states and continuum, together with their respective
wavefunctions.
\end{itemize}

%% file: 12Appen.tex
\section{General Conventions}
\label{app:General}
\setcounter{equation}{0}
For completeness notational conventions  are collected 
in line with the textbooks \cite{bjd65,itz85}.  

 \noindent\textbf{Lorentz vectors.} 
We write contravariant four-vectors of position $x^\mu$  
in the instant form as
\begin {equation}
       x^\mu 
       = (x^0, x^1, x^2, x^3)
       = (t, x, y, z)
       = (x^0, \vec x_{\!\perp}, x^3)
       = (x^0, \vec x)
\ . \end {equation}
The covariant four-vector $x_\mu$ is given by
\begin{equation}
       x_\mu 
       = (x_0, x_1, x_2, x_3)
       = (t, -x, -y, -z)
       = g_{\mu\nu} x^\nu
\ ,\end{equation}
and obtained from the contravariant vector by 
the metric tensor 
\begin {equation}
       g_{\mu\nu}=\pmatrix{+1&0&0&0\cr 0&-1&0&0\cr 
       0&0&-1&0\cr 0&0&0&-1\cr}
\ .\end {equation}
Implicit summation over repeated 
Lorentz ($\mu,\nu,\kappa$) or  space ($i,j,k$) indices 
is understood.
Scalar products are  
\begin {equation}
       x\cdot p = x^\mu p_\mu 
       = x^0 p_0 + x^1 p_1+ x^2 p_2 +x^3 p_3
       = tE - \vec x\cdot \vec p
\ ,\end{equation}
with four-momentum 
$ p^\mu = (p^0, p^1, p^2, p^3) = (E, \vec p)$. 
The metric tensor $g^{\mu\nu}$ raises the indices.

\noindent\textbf{Dirac matrices.}
Up to unitary transformations, the $4\times4$ Dirac 
matrices $\gamma^\mu$ are defined by  
\begin{equation}
       \gamma^\mu\gamma^\nu +
       \gamma^\nu\gamma^\mu
       =2g^{\mu\nu}
\ .\end{equation}
$\gamma^0$ is hermitean and $\gamma^k$ anti-hermitean.
Useful combinations are $\beta=\gamma^0$ 
and $\alpha^k = \gamma^0 \gamma^k$, as well as
\begin{equation} 
      \sigma^{\mu\nu} = 
      {i\over2}\left(
       \gamma^\mu\gamma^\nu -
       \gamma^\nu\gamma^\mu \right)
\ , \quad
      \gamma_5 =  \gamma^5 = 
      i \gamma^0\gamma^1\gamma^2\gamma^3
\ . \end{equation}
They usually are expressed in terms of the 
$2\times2$ Pauli matrices 
\begin{equation} 
      I= \left[ \begin{array} {lr}
      1 & 0 \\ 0 & 1 \end{array} \right]
\ , \quad
      \sigma^1= \left[ \begin{array} {lr}
      0 & 1 \\ 1 & 0 \end{array} \right]
\ , \quad
      \sigma^2= \left[ \begin{array} {lr}
      0 & -i \\ i & 0 \end{array} \right]
\ , \quad
      \sigma^3= \left[ \begin{array} {lr}
      1 & 0 \\ 0 & -1 \end{array} \right]
\ .\end{equation}
In \underbar{Dirac representation} \cite{bjd65,itz85} 
the matrices are
\begin{eqnarray} 
      \gamma^0 &=& \left( \begin{array} {l@{\,}r}
      I & 0 \\ 0 & -I \end{array} \right)
\ , \quad
      \gamma^k = \left(\begin{array} {l@{\,}r}
      0 & \sigma^k \\ -\sigma^k & 0 \end{array} \right)
\ , \\
      \gamma_5  &=& \left( \begin{array} {l@{\,}r}
      0 & +I \\ I & 0 \end{array} \right) 
\ , \quad
     \alpha^k  = \left( \begin{array} {l@{\,}r}
      0 & \sigma^k \\ +\sigma^k& 0 \end{array} \right) 
\ , \quad
      \sigma^{ij}  = \left(\begin{array} {l@{\,}r}
      \sigma^k & 0 \\ 0 & \sigma^k \end{array} \right) 
\ . \end{eqnarray}
In \underbar{chiral representation} \cite{itz85} 
$\gamma_0$ and $\gamma_5$ are interchanged:
\begin{eqnarray}
      \gamma^0 &=& \left( \begin{array} {l@{\,}r}
      0 & +I \\ I & 0 \end{array} \right)
\ , \quad
      \gamma^k = \left( \begin{array} {l@{\,}r}
      0 & \sigma^k \\ -\sigma^k & 0 \end{array} \right)
\ , \\
      \gamma_5  &=& \left( \begin{array} {l@{\,}r}
      I & 0 \\ 0 & -I \end{array} \right) 
\ , \quad
      \alpha^k  = \left( \begin{array} {l@{\,}r}
      \sigma^k & 0 \\ 0 & -\sigma^k \end{array} \right) 
\ , \quad
      \sigma^{ij}  = \left(\begin{array} {l@{\,}r}
      \sigma^k & 0 \\ 0 & \sigma^k \end{array} \right) 
\ .\end{eqnarray}
$(i,j,k)=1,2,3$ are  used cyclically. 

\noindent\textbf{Projection operators.}
Combinations of Dirac matrices like the hermitean
matrices 
\begin{equation}
      \Lambda_+ = {1\over2} (1+\alpha^3)
      = {\gamma^0\over2} (\gamma^0+\gamma^3)
      \quad{\rm and}\quad
      \Lambda_-= {1\over2} (1-\alpha^3)
      = {\gamma^0\over2} (\gamma^0-\gamma^3)
\end{equation}
often have projector properties, particularly
\begin{equation}
      \  \quad \Lambda_++\Lambda_-={\bf 1}
      \ ,\quad \Lambda_+\Lambda_-=0
      \ ,\quad \Lambda^2_+=\Lambda_+
      \ ,\quad \Lambda^2_-=\Lambda_-
\ .\end{equation}
They are diagonal in the chiral  and
maximally off-diagonal in the Dirac representation:
\begin{equation}
      (\Lambda_+)_{\rm chiral} =
      \pmatrix{1&0&0&0\cr 0&0&0&0\cr
                        0&0&0&0\cr 0&0&0&1\cr}
      \ ,\quad 
      (\Lambda_+)_{\rm Dirac} = {1\over2}
      \pmatrix{1&0&1&0\cr 0&  1&0&-1\cr
                        1&0&1&0\cr 0&-1&0&  1\cr}
\ .\label{eq:A14}\end{equation}

\noindent\textbf{Dirac spinors.} 
The spinors 
$u_\alpha(p,\lambda)$ and $v_\alpha(p,\lambda)$
are solutions of the Dirac equation
\begin{equation}
       (/\!\!\!p-m)\, u(p,\lambda)=0\ ,
     \ (/\!\!\!p+m)\,v(p,\lambda)=0
\ . \end{equation} 
They are orthonormal and complete:
\begin{eqnarray}
       \bar u(p,\lambda)u(p,\lambda')
       = - \bar v(p,\lambda')v(p,\lambda)
       = 2m\,\delta_{\lambda\lambda'} \ , \\
       \sum_\lambda u(p,\lambda)\bar u(p,\lambda)
       = /\!\!\!p+m 
       \ ,\ \sum_\lambda v(p,\lambda)\bar v(p,\lambda)
       = /\!\!\!p-m
\ .\end{eqnarray}
Note the different normalization as compared to 
the textbooks \cite{bjd65,itz85}. 
The `Feynman slash' is 
$/\!\!\!p= p_\mu \gamma^\mu$.
The Gordon decomposition of the currents is useful:
\begin{equation}
          \bar u(p,\lambda)\gamma^{\mu} u(q,\lambda')
       =\bar v(q,\lambda')\gamma^{\mu} v(p,\lambda) 
       = {1\over 2m}\bar u(p,\lambda) \Bigl( 
       (p+q)^\mu + i \sigma^{\mu\nu} (p-q)_\nu 
       \Bigr) u(q,\lambda')
\ .\end{equation}
With $\lambda=\pm 1$, the spin projection 
is $ s=\lambda/2$. 
The relations
\begin{eqnarray}
      \gamma^\mu /\!\!\!a \gamma_\mu = -2a
\ ,\\
      \gamma^\mu /\!\!\!a /\!\!\!b \gamma_\mu = 4ab 
\ ,\\
      \gamma^\mu /\!\!\!a /\!\!\!b /\!\!\!c \gamma_\mu
      = /\!\!\!c /\!\!\!b /\!\!\!a
\end{eqnarray}
are useful.

\noindent\textbf{Polarization vectors.}
The two polarization four-vectors 
$\epsilon_\mu(p,\lambda)$
are labeled by the spin projections $\lambda=\pm1$.
As solutions of the free Maxwell equations they are
orthonormal and complete: 
\begin{equation}
       \epsilon^\mu(p,\lambda)       \,
       \epsilon^\star_\mu(p,\lambda^\prime)
       = -\delta_{\lambda\lambda^\prime} 
       \ ,\qquad
       p^\mu\,\epsilon_\mu (p,\lambda)
       =0
\ .\end{equation}
The star ($^\star$) refers to complex conjugation. 
The polarization sum is 
\begin{equation}
       d_{\mu\nu}(p)=\sum\limits_{\lambda}
       \epsilon_\mu (p,\lambda)
       \epsilon^\star_\nu (p,\lambda)
       = - g_{\mu\nu}
       +{\eta_\mu p_\nu+\eta_\nu p_\mu
       \over p^\kappa \eta_\kappa}
\ ,\end{equation}
with the null vector $\eta^\mu\eta_\mu = 0$ given below.

\section{The Lepage-Brodsky convention (LB)}
\label{app:Lepage-Brodsky}
\setcounter{equation}{0}

This section summarizes the conventions which have
been used by Lepage, Brodsky and others 
\cite{brp91,lbh83,leb80}.

\noindent\textbf{Lorentz vectors.} 
The contravariant four-vectors of position $x^\mu$
are written as 
\begin {equation}
       x^\mu 
       = (x^+, x^-, x^1, x^2)
       = (x^+, x^-, \vec x_{\!\perp})
\ . \end {equation}
Its time-like and space-like components are related to
the instant form by \cite{brp91,lbh83,leb80}
\begin {equation}
       x^+ = x^0 + x^3
\quad{\rm and}\quad
       x^- = x^0 - x^3
\ , \end {equation}
respectively, and referred to as the `light-cone time'
and `light-cone position'. The covariant vectors are
obtained by $x_\mu=g_{\mu\nu}x^\nu$, 
with the metric tensor(s) 
\begin {equation}
       g^{\mu\nu}=\pmatrix{0&2&0&0 \cr 
       2&0&0&0 \cr 0&0&-1&0 \cr 0&0&0&-1 \cr}
       \qquad{\rm and}\qquad
       g_{\mu\nu}=\pmatrix{0&{1\over2}&0&0\cr 
       {1\over2}&0&0&0\cr  0&0&-1&0\cr 0&0&0&-1\cr}
\ .\end {equation}
Scalar products are 
\begin {equation}
       x\cdot p = x^\mu p_\mu 
       = x^+ p_+ + x^- p_- + x^1 p_1+ x^2 p_2 
       =  {1\over2}(x^+ p^- + x^- p^+) 
       - \vec x_{\!\perp} \vec p_{\!\perp}
\ .\end{equation}
All other four-vectors including $\gamma^\mu$ 
are treated correspondingly.

\noindent\textbf{Dirac matrices.}
The Dirac representation of the $\gamma$-matrices is used,
particularly
\begin {equation}
       \gamma^+\gamma^+ =
       \gamma^- \gamma^-  = 0
\ .\end {equation}
Alternating products are for example
\begin{equation}
       \gamma^+\gamma^-\gamma^+ = 4\gamma^+
       \qquad{\rm and}\qquad
       \gamma^-\gamma^+\gamma^-  = 4\gamma^- 
\ .\end{equation}

\noindent\textbf{Projection operators.}
The projection matrices become
\begin{equation}
       \Lambda_+={1\over2}\gamma^0\gamma^+ 
       ={1\over4}\gamma^-\gamma^+ 
       \qquad{\rm and}\qquad
       \Lambda_- ={1\over2}\gamma^0\gamma^- 
       ={1\over4}\gamma^+\gamma^- 
\ .\end{equation}

\noindent\textbf{Dirac spinors.} 
Lepage and Brodsky \cite{brp91,lbh83,leb80}
use a particularly simple spinor representation
\begin{eqnarray}
       u(p, \lambda) = {1\over \sqrt{p^+} } 
       \left(p^+ + \beta m + \vec \alpha_{\!\perp} 
       \vec p_{\!\perp}\right) \times
       \cases{ \chi(\uparrow), &for $\lambda=+1$, \cr
       \chi(\downarrow), &for $\lambda=-1$, \cr} 
\\
       v(p, \lambda) = {1\over \sqrt{p^+} } 
       \left(p^+ - \beta m + \vec \alpha_{\!\perp} 
       \vec p_{\!\perp}\right) \times
       \cases{\chi(\downarrow), &for $\lambda=+1$, \cr
       \chi(\uparrow), &for $\lambda=-1$. \cr}
\end{eqnarray}
The two $\chi$-spinors are
\begin{equation}
       \chi(\uparrow) = {1\over\sqrt{2} } \, 
      \left(\begin{array} {r}
      1 \\  0 \\ 1 \\ 0 \end{array}\right)
       \qquad{\rm and}\qquad
       \chi(\downarrow) = {1\over\sqrt{2}} \, 
      \left(\begin{array} {r}
      0 \\  1 \\ 0 \\ -1 \end{array}\right)
\ .\label{eq:B10}\end{equation}

\noindent\textbf{Polarization vectors}
The null vector is 
\begin{equation}
       \eta^\mu =
       \left( 0, 2, \vec 0  \right)
\ . \end{equation}
In Bj\o rken-Drell convention \cite{bjd65}, one works 
with circular polarization, with spin projections 
$\lambda=\pm1= \uparrow\downarrow$. 
The transversal polarization vectors are
$\vec\epsilon_{\!\perp} (\uparrow) = -1/\sqrt{2} \,(1,i)$ and 
$\vec\epsilon_{\!\perp}(\downarrow) = 1/\sqrt{2} \,(1,-i)$,
or collectively
\begin{equation}
       \vec \epsilon_{\!\perp} (\lambda)   =
      {-1\over\sqrt{2}}(\lambda\vec e_x + i \vec e_y)
\ ,\end{equation}
with $\vec e_x$ and $\vec e_y$ as unit vectors in 
$p_x$- and $p_y$-direction, respectively.
With $\epsilon^+ (p, \lambda) = 0$, induced by the 
light-cone gauge, the polarization vector is 
\begin{equation}
       \epsilon^\mu (p, \lambda) =
       \left( 0, {2\vec \epsilon_{\!\perp} (\lambda)
       \vec p_{\!\perp}\over p^+}, 
       \vec \epsilon_{\!\perp}(\lambda)  \right)
\ , \end{equation}
which satisfies $ p_\mu\epsilon^\mu (p, \lambda) $.

\section{The Kogut-Soper convention (KS)}
\label{app:Kogut-Soper}
\setcounter{equation}{0}

\noindent\textbf{Lorentz vectors.} 
Kogut and Soper \cite{kos70,sop71,bks71,kos73} have used 
\begin {equation}
       x^+ = {1\over\sqrt{2} } \left( x^0 + x^3 \right)
\quad{\rm and}\quad
       x^- = {1\over\sqrt{2} }\left( x^0 - x^3 \right)
\ , \end {equation}
respectively, referred to as the `light-cone time'
and `light-cone position'. The covariant vectors are
obtained by $x_\mu=g_{\mu\nu}x^\nu$, 
with the metric tensor
\begin {equation}
       g^{\mu\nu}=g_{\mu\nu}=\pmatrix{0&1&0&0 \cr 
       1&0&0&0 \cr 0&0&-1&0 \cr 0&0&0&-1 \cr}
\ .\end {equation}
Scalar products are 
\begin {equation}
       x\cdot p = x^\mu p_\mu 
       = x^+ p_+ + x^- p_- + x^1 p_1+ x^2 p_2 +
       =  x^+ p^- + x^- p^+
       - \vec x_{\!\perp} \vec p_{\!\perp}
\ .\end{equation}
All other four-vectors including $\gamma^\mu$ 
are treated correspondingly.

\noindent\textbf{Dirac matrices.}
The chiral representation of the $\gamma$-matrices 
is used, particularly
\begin {equation}
       \gamma^+\gamma^+ =
       \gamma^- \gamma^-  = 0
\ .\end {equation}
Alternating products are for example
\begin{equation}
       \gamma^+\gamma^-\gamma^+ = 2\gamma^+
       \qquad{\rm and}\qquad
       \gamma^-\gamma^+\gamma^-  = 2\gamma^- 
\ .\end{equation}

\noindent\textbf{Projection operators.}
The projection matrices become
\begin{equation}
       \Lambda_+={1\over \sqrt{2}}\gamma^0\gamma^+ 
       ={1\over2}\gamma^-\gamma^+ 
       \qquad{\rm and}\qquad
       \Lambda_- ={1\over\sqrt{2}}\gamma^0\gamma^- 
       ={1\over2}\gamma^+\gamma^- 
\ .\end{equation}
In the chiral representation the projection matrices
have a particularly simple structure, see 
Eq.(\ref{eq:A14}).

\noindent\textbf{Dirac spinors.} 
Kogut and Soper \cite{kos70} use as Dirac spinors
\begin{eqnarray}
   u(k, \uparrow) &=& {1\over 2^{1/4}\sqrt{k^+}} 
   \left(\begin{array}{c}\sqrt{2}k^+\\ \phantom{-}k_x+ik_y\\m\\0 
            \end{array}\right) ,\quad   
   u(k, \downarrow) = {1\over 2^{1/4}\sqrt{k^+}}
   \left(\begin{array}{c}0\\m\\-k_x+ik_y\\ \sqrt{2}k^+  
            \end{array}\right) 
,\nonumber\\
   v(k, \uparrow) &=& {1\over 2^{1/4}\sqrt{k^+}}
   \left(\begin{array}{c}0\\-m\\-k_x+ik_y\\ \sqrt{2}k^+  
            \end{array}\right) , \quad 
   v(k, \downarrow) = {1\over 2^{1/4}\sqrt{k^+}} 
   \left(\begin{array}{c}\sqrt{2}k^+\\ \phantom{-}k_x+ik_y\\-m\\0 
            \end{array}\right) 
.\end{eqnarray}

\noindent\textbf{Polarization vectors.}
The null vector is 
\begin{equation}
       \eta^\mu =
       \left( 0, 1, \vec 0  \right)
\ . \end{equation}
The polarization vectors of Kogut and Soper \cite{kos70} 
correspond to linear polarization $\lambda=1$ and 
$\lambda=2$:
\begin{eqnarray}
   \epsilon^\mu(p,\lambda=1)
   &=& (0,{p_x\over p^+}, 1, 0)
,\nonumber\\
   \epsilon^\mu (p,\lambda=2) 
   &=&  (0,{p_y\over p^+}, 0, 1)
.\end{eqnarray}
The following are useful relations
\begin{eqnarray}
   \gamma^\alpha\gamma^\beta 
    d_{\alpha\beta}(p) &=& -2
\ ,\nonumber\\ 
   \gamma^\alpha\gamma^\nu\gamma^\beta 
    d_{\alpha\beta}(p) &=& {2\over p^+}
   (\gamma^+ p^\nu+g^{+\nu} /\!\!\!p)
\ ,\nonumber\\ 
   \gamma^\alpha\gamma^\mu\gamma^\nu
   \gamma^\beta d_{\alpha\beta}(p) &=& -4g^{\mu\nu} +
\nonumber    \\  
   +2{p_\alpha\over p^+} \Big\{
      g^{\mu\alpha}\gamma^\nu\gamma^+ 
    - g^{\alpha\nu}\gamma^\mu\gamma^+
    &+& g^{\alpha+}\gamma^\mu\gamma^\nu
    - g^{+\nu}\gamma^\mu\gamma^\alpha
   +g^{+\mu}\gamma^\nu\gamma^\alpha\Big\}
\ . \end{eqnarray}
The remainder is the same as in 
Appendix~\ref{app:General}

\section{Comparing BD- with LB-Spinors}
\label{app:F}
\setcounter{equation}{0}

The Dirac spinors $u_\alpha(p,\lambda)$ and 
$v_\alpha(p,\lambda)$ 
(with $\lambda=\pm 1$) 
are the four linearly independent solutions 
of the free Dirac equations 
$\left( /\!\!\!p -m\right)\,u(p,\lambda) = 0$
and 
$\left( /\!\!\!p +m\right)\,v(p,\lambda) = 0$.
Instead of $u(p,\lambda)$ and $v(p,\lambda)$, 
it is sometimes convenient \cite{bjd65}
to use spinors $w^r(p)$ defined by
\begin{equation} 
       w^1_\alpha(p) = u_\alpha(p,\uparrow)  
       ,\quad w^2_\alpha(p) = u_\alpha(p,\downarrow) 
       ,\quad w^3_\alpha(p) = v_\alpha(p,\uparrow) 
       ,\quad w^4_\alpha(p) = v_\alpha(p,\downarrow) 
\ .\end{equation} 
With $p^0=E=\sqrt{m^2 + \vec p ^{\,2}}$ 
holds quite in general 
\begin{eqnarray} 
       u(p,\lambda) &=& {1\over\sqrt{N}} 
       \left(E + \vec\alpha\cdot\vec p +\beta m\right) \chi^r
       \ , \quad{\rm for}\quad r=1,2 
\ ,   \\
       v(p,\lambda) &=& {1\over\sqrt{N}} 
       \left(E + \vec\alpha\cdot\vec p -\beta m\right) \chi^r
       \ ,\quad{\rm for}\quad r=3,4 
\ .\end{eqnarray} 
Bj\o rken-Drell (BD) \cite{bjd65} choose 
$ \chi^r_\alpha =\delta_{\alpha r}$. 
With $N=2m(E+m)$, the four spinors are then explicitly:
\begin{equation} 
       w^r_\alpha (p) = {1\over\sqrt{N}} 
       \left(\begin{array} {cccc}
       E+m & 0 & p_z & p_x-ip_y \\
       0  &  E+m & p_x+ip_y & -p_z \\
       p_z & p_x-ip_y & E+m & 0 \\
       p_x+ip_y & -p_z & 0 & E+m  \end{array}\right)
\ .\end{equation} 
Alternatively (A), one can choose 
\begin{equation} 
      \chi^1_\alpha  = \chi(\uparrow) \ , \quad
      \chi^2_\alpha  = \chi(\downarrow) \ , \quad
      \chi^3_\alpha  = \chi(\uparrow)  \ , \quad
      \chi^4_\alpha  = \chi(\downarrow)  
\ ,\label{eq:F5} \end{equation} 
with $ $ given in Eq.(\ref{eq:B10}).
With $N = 2m(E+p_z)$, the spinors become explicitly: 
\begin{equation} 
      w^r_\alpha (p) 
      = {1\over \sqrt{2N}} \left(\begin{array} {cccc}
      E+p_z+m & -p_x+ip_y &  E+p_z-m & -p_x+ip_y  \\
      p_x+ip_y  &  E+p_z+m & p_x+ip_y & E+p_z-m \\
      E+p_z-m & p_x-ip_y & E+p_z+m & p_x-ip_y  \\
      p_x+ip_y & -E-p_z+m & p_x+ip_y  & -E-p_z-m  
      \end{array}\right)
. \end{equation} 
One verifies that both spinor conventions (BD) and (A) 
satisfy orthogonality and completeness
\begin{equation} 
\sum_{\alpha=1}^4 \overline w ^r_\alpha w ^{r^\prime}_\alpha = 
\gamma^0 _{r r^\prime} \ , \quad
\sum_{r=1}^4 \gamma^0_{rr} w ^r_\alpha \overline w ^{r}_\beta = 
\delta _{\alpha\beta} 
\ , \end{equation} 
respectively, with $\overline w = w^\dagger \gamma^0$.
But the two {\em do not have} the same form for a particle
at rest, $\vec p = 0$, namely
\begin{equation} 
      w^r_\alpha (m) _{\rm BD}
      = \left(\begin{array} {cccc}
      1 & 0 & 0 & 0 \\
      0 & 1 & 0 & 0 \\
      0 & 0 & 1 & 0 \\
      0 & 0 & 0 & 1 \\
      \end{array}\right)
\ , \quad{\rm and}\quad
      w^r_\alpha (m) _{\rm A}
      = \left(\begin{array} {rrrr}
      1 & 0 & 0 & 0 \\
      0 & 1 & 0 & 0 \\
      0 & 0 & 1 & 0 \\
      0 & 0 & 0 & -1 \\
      \end{array}\right)
, \end{equation} 
respectively, but they have the same spin projection:
\begin{equation} 
      \sigma^{12}\,u(m,\lambda) = \lambda\,u(m,\lambda)
\ ,\ {\rm and} \qquad 
      \sigma^{12}\,v(m,\lambda) = \lambda\,v(m,\lambda)
\ . \label{eq:F9} \end{equation} 
Actually, Lepage and Brodsky \cite{leb80} have {\em not}
used Eq.(\ref{eq:F5}), but rather 
\begin{equation} 
      \chi^1_\alpha  = \chi(\uparrow) \ , \quad
      \chi^2_\alpha  = \chi(\downarrow) \ , \quad
      \chi^3_\alpha  = \chi(\downarrow) \ , \quad
      \chi^4_\alpha  = \chi(\uparrow)  
\ , \end{equation} 
by which reason Eq.(\ref{eq:F9}) becomes
\begin{equation} 
      \sigma^{12}\,u(0,\lambda) = \lambda\,u(0,\lambda)
      \ ,\ {\rm and} \qquad 
      \sigma^{12}\,v(0,\lambda) = -\lambda\,v(0,\lambda)
\ .  \end{equation} 
In the LC formulation the $\sigma/2$ operator is a helicity 
operator
which has a different spin for fermions and anti-fermions.

\input 03Quant.tex

%% file: 03Quant.tex
\section{The Dirac-Bergmann Method}
\label{sec:Bergmann}
\setcounter{equation}{0}

The dynamics of a classical, non-relativistic system with N degrees of 
freedom can be derived from the Lagrangian. Obtained from an action 
principle,
this Lagrangian is a function of the velocity phase space variables:
\begin{equation}
    L=L(q_n,{\dot q}_n),\qquad n=1,\ldots,N
\,,\end{equation}
where the
$q$'s and ${\dot q}$'s are the generalized coordinates and velocities
respectively. For simplicity we consider only Lagrangians without 
explicit
time  dependence. The momenta conjugate to the generalized 
coordinates are defined by
\begin{equation}
p_n={\partial L\over \partial{\dot q}_n}.
\label{ra.2}
\end{equation}
Now it may turn out that not all the momenta may be expressed 
as independent functions of the velocities. If this is the case, the 
Legendre transformation that takes us from the Lagrangian 
to the Hamiltonian is not defined uniquely over the whole phase 
space ($q,p$). 
There then exist a
number of {\it constraints} connecting the $q$'s and $p$'s:
\begin{equation}
\phi_m(q,p)=0,\qquad m=1,\ldots,M.
\label{ra.3}
\end{equation}
These constraints restrict the motion to a {\it subspace} of 
the full $2N$-dimensional phase space defined by the ($p,q$).

Eventually, we would like to formulate the dynamics in terms 
of Poisson brackets defined for any two dynamical quantities 
$A(q,p)$ and $B(q,p)$:
\begin{equation}
\{ A,B\}={\partial A\over\partial q_n}{\partial B\over\partial p_n}
-{\partial A\over\partial p_n}{\partial B\over\partial q_n}.
\label{ra.4}
\end{equation}
The Poisson bracket (PB) formulation is
the stage from which we launch into quantum mechanics. Since
the PB is defined over the whole phase space only for  {\it independent}
variables ($q,p$), we are faced with  the problem of extending the PB
definition (among other things) onto a constrained phase space.

The constraints are a consequence of
the form of the Lagrangian alone. 
Following Anderson and Bergmann \cite{anb51}, 
we  will call the $\phi_m$ {\it primary constraints}. 
Now to develop the theory, consider the quantity 
$p_n{\dot q}_n-L$. If we make
variations in the quantities $q$, ${\dot q}$ and $p$ we obtain
\begin{equation}
\delta(p_n{\dot q}_n-L)=\delta p_n{\dot q}_n-{\dot p}_n\delta q_n
\label{rb.1}
\end{equation}
using Eq. (\ref{ra.2}) and the Lagrange equation 
${\dot p}_n={\partial L\over \partial q_n}$. 
Since the right hand side of 
Eq.(\ref{rb.1}) is independent of $\delta {\dot q}_n$ we will call 
$p_n{\dot q}_n-L$ the Hamiltonian $H$. 
Notice that this Hamiltonian is not unique. 
We can add to $H$ any linear combination of the primary 
constraints and the resulting new Hamiltonian
is just as good as the original one.

How do the primary constraints affect the equations of motion? 
Since not all the $q$'s and $p$'s are independent, 
the variations in Eq.(\ref{rb.1}) cannot be made 
independently. Rather, for Eq.(\ref{rb.1}) to hold, the variations 
must preserve the conditions (\ref{ra.3}). The result is
\cite{sun82}
\begin{equation}
{\dot q}_n={\partial H\over \partial p_n}+u_m{\partial\phi_m\over 
\partial p_n}
\label{rb.2a}
\end{equation}
and 
\begin{equation}
{\dot p}_n=-{\partial H\over\partial q_n}-
u_m{\partial\phi_m\over\partial q_n}
\label{rb.2b}
\end{equation}%
where the $u_m$ are unknown coefficients. The $N$ ${\dot q}$'s 
are fixed by the $N$ $q$'s, the $N-M$ independent $p$'s 
{\it and} the $M$  $u$'s.
Dirac takes the variables $q$, $p$ and $u$ as the 
Hamiltonian variables.

Recalling the definition of the Poisson bracket 
Eq.(\ref{ra.4}) we can write,
for any function $g$ of the $q$'s and $p$'s
\begin{equation}
{\dot g}={\partial g\over\partial q_n}{\dot q}_n+
{\partial g\over\partial p_n}{\dot p}_n =\{ g,H\}+u_m\{ g,\phi_m\}
\label{rb.3}
\end{equation}
using Eqs.(\ref{rb.2a}) and (\ref{rb.2b}). 
As mentioned already,
the Poisson bracket has meaning only for two dynamical 
functions defined uniquely over the whole phase space. 
Since the $\phi_m$ restrict the independence of some of 
the $p$'s, we must not use the condition $\phi_m=0$
within the PB. 
The PB should be evaluated based on the functional form of
the  primary constraints. After all PB's have been calculated, 
then we may impose $\phi_m=0$. 
From now on, such restricted relations will be denoted
with a squiggly equal sign:
\begin{equation}
\phi_m\approx 0.
\label{rb.4}
\end{equation}
This is called a {\it weak equality }.
The equation of motion for $g$ is now
\begin{equation}
{\dot g}\approx\{ g,H_T\}
\label{rb.5}
\end{equation}
where
\begin{equation}
H_T=H+u_m\phi_m
\label{rb.6}
\end{equation}
is the {\it total} Hamiltonian \cite{dir64}. If we take $g$ in
Eq.(\ref{rb.5}) to be one of the ${\phi}$'s we will get some 
consistency conditions since the primary constraints 
should remain zero throughout all time:
\begin{equation}
\{ \phi_m,H\}+u_{m^\prime}\{ \phi_m,\phi_{m^\prime}\}\approx 0.
\label{rb.7}
\end{equation}
What are the possible outcomes of Eq.(\ref{rb.7}) ? 
Unless they all reduce to $0=0$ i.e., are identically satisfied, 
we will get more conditions
between the  Hamiltonian variables $q$, $p$ and $u$. 
We will exclude the case where an inappropriate Lagrangian 
leads to an inconsistency like $1=0$.
There are then two cases of interest. The first possibility is that
Eq.(\ref{rb.7}) provides no new information but imposes conditions 
on the $u$'s. The second  possibility is that we get an equation 
independent of $u_m$ but relating the $p$'s and $q$'s. 
This can happen if the $M\times M$ matrix
$\{\phi_m,\phi_{m^\prime}\}$ has any rows (or columns) 
which are linearly dependent. 
These new conditions between the $q$'s and $p$'s are called
{\it secondary constraints}
\begin{equation}
\chi_{k^\prime}\approx 0, \qquad k^\prime = 1,\ldots ,K^\prime
\label{rb.8}
\end{equation}
by Anderson and Bergmann \cite{anb51}. Notice that primary
constraints follow from the form of the Lagrangian alone whereas 
secondary constraints involve the equations of motion as well.
These secondary constraints, like the primary constraints,
must remain zero throughout all time so
we can perform the same consistency operation on the $\chi$'s:
\begin{equation}
{\dot\chi}_k=\{ \chi_k,H\}+u_m\{ \chi_k,\phi_m\}\approx 0.
\label{rb.9}
\end{equation}
This equation is treated in the same manner as 
Eq.(\ref{rb.7}). If it leads to more conditions on the $p$'s and $q$'s 
the process is  repeated again. 
We continue like this until either all the
consistency  conditions are exhausted or we get an identity. 

Let us write all the constraints obtained in the above manner 
under one index as
\begin{equation}
\phi_j\approx 0,\qquad j=1,\ldots ,M+K\equiv J
\label{rb.10}
\end{equation}
then we obtain the following matrix equation for the $u_m$ 
\begin{equation}
\{\phi_j,H\}+u_m\{\phi_j,\phi_m\}\approx 0.
\label{rb.11}
\end{equation}
The most general solution to Eq.(\ref{rb.11}) is 
\begin{equation}
u_m=U_m+v_aV_{am},\qquad a=1,\ldots ,A
\label{rb.12}
\end{equation}
where $V_m$ is {\it a} solution of the homogeneous part of
Eq.(\ref{rb.11}) and $v_aV_{am}$ is a linear combination 
of {\it all} such independent solutions. 
The coefficients $v_a$ are arbitrary.

Substitute Eq.(\ref{rb.12}) into Eq.(\ref{rb.6}). This gives
\begin{eqnarray}
H_T&=H+U_m\phi_m+v_{a}V_{am}\phi_m  \nonumber \\ 
&=H^\prime+v_a\phi_a
\label{rb.13} 
\end{eqnarray}
where
\begin{equation}
H^\prime=H+U_m\phi_m
\label{rb.14}
\end{equation}
and
\begin{equation}
\phi_a=V_{am}\phi_m .
\label{rb.15}
\end{equation}
Note that the $u$'s must satisfy consistency requirements 
whereas
the $v$'s are totally arbitrary functions of time. 
Later, we will have more to say about the appearance 
of these arbitrary features in our theory.

To further classify the quantities in our theory, consider 
the following  definitions given by Dirac \cite{dir50}. 
Any dynamical variable, $F(q,p)$, is called {\it first class} if
\begin{equation}
\{ F,\phi_j\}\approx 0, \qquad j=1,\ldots ,J
\label{rb.16}
\end{equation}
i.e., $F$ has zero PB with {\it all} the $\phi$'s. 
If $\{F,\phi_j\}$  is not weakly zero $F$ is called 
{\it second class}. Since the $\phi$'s are the only 
independent quantities which are weakly zero, we can
write the following {\it strong} equations when $F$ is first class:
\begin{equation}
\{ F,\phi_j\}=c_{jj^\prime}\phi_{j^\prime}.
\label{rb.17}
\end{equation}
Any quantity which is weakly zero is strongly equal to some 
linear 
combination of the $\phi$'s. Given Eq.(\ref{rb.16}) and Eq.(\ref{rb.17}) 
it
is easy to show that $H^\prime$ and $\phi_a$ 
(see Eq.(\ref{rb.14}) and (\ref{rb.15}) are first class quantities. 
Since $\phi_a$ is a linear combination of primary 
constraints Eq.(\ref{rb.15}), it too is a primary constraint. 
Thus, the total Hamiltonian Eq.(\ref{rb.13}),
which is   expressed as the sum of a first class Hamiltonian 
plus a linear combination of  primary, first class constraints, 
is a first class quantity. 

Notice that the number of arbitrary functions of the time 
appearing in our theory is equivalent to the number of 
independent primary first class constraints. 
This can be seen by looking at Eq.(\ref{rb.12}) where 
all the  independent first class primary constraints are 
included in the sum.  This same number will also appear 
in the general equation of motion because of Eq.(\ref{rb.13}).
Let us make a small digression on the role of these 
arbitrary functions of time.

The physical state of any system is determined by the $q$'s and
$p$'s only and not by the $v$'s. However, if we start out at $t=t_0$
with fixed initial values ($q_0$,$p_0$) we arrive at different values of
($q$,$p$) at later times depending on our choice of $v$. 
The physical state does not uniquely determine a set of 
$q$'s and $p$'s but a given set of $q$'s and $p$'s must 
determine the physical state. We thus have the situation 
where there may be several sets of the dynamical variables 
which correspond to the same physical state.

To understand this better consider two functions $A_{v_a}$ 
and $A_{{v^\prime}_a}$ of the dynamical variables which 
evolve from some $A_0$ with different multipliers. 
Compare the two functions after a short time interval
$\Delta t$ by considering a Taylor expansion to first order 
in $\Delta t$:
\begin{eqnarray}
A_{v_a}(t) = & A_0 + {\dot A}_{v_a}\Delta t 
            = A_0 + \{ A_0,H_T\}   \Delta t \nonumber \\
            &= A_0 + \Delta t [\{ A_0,H^\prime\} + v_a\{ A_0,\phi_a\}].
\label{rb.18}
\end{eqnarray}
Thus,
\begin{equation}
A_{v_a}-A_{{v^\prime}_a}
=\Delta t (v_a-{v^\prime}_a)\{ A_0,\phi_a\}
\end{equation}
or
\begin{equation}
\Delta A = \epsilon_a\{ A_0,\phi_a\}
\label{rb.19}
\end{equation}
where
\begin{equation}%
\epsilon_a=\Delta t (v_a-{v^\prime}_a)
\label{rb.20}
\end{equation}
is a small, arbitrary quantity. This relationship between 
$A_{v_a}$ and $A_{{v^\prime}_a}$ tells us that the two 
functions are related by an
infinitesimal canonical transformation (ICT) \cite{gol50} 
whose generator is a first  class primary constraint $\phi_a$. 
This ICT leads to changes in the $q$'s and 
$p$'s which do no affect the physical state.

Furthermore, it can also be shown \cite{dir64} that by 
considering successive ICT's
that the generators need not be primary but can be 
secondary as well. To be completely general then, 
we should allow for such variations which do not change
the physical state in our equations of motion. 
This can be accomplished by redefining $H_T$ to  include 
the first class secondary constraints with arbitrary
coefficients. Since the distinction between first class
primary and first class secondary is not significant \cite{sun82} 
in what follows we will not make any explicit  changes.

For future considerations let us call those transformations which do not 
change the
physical state {\it gauge transformations}. The ability to perform gauge
transformations is a sign that the mathematical framework of our theory 
has some
arbitrary features. Suppose we can add conditions to our theory that 
eliminate our
ability to make gauge transformations. These conditions would enter as 
secondary
constraints since they do not follow from the form of the  Lagrangian. 
Therefore
upon imposing these conditions, all constraints  become second class. 
If there {\it were} any more first class constraints  
we would have generators for gauge transformations which, 
by assumption, can no longer be made. 
This is the end of the digression although we will see 
examples of gauge transformations later.

In general, of the $J$ constraints, some are first class and some are 
second class.
A linear combination of constraints is again a constraint so we can 
replace the
$\phi_j$ with independent linear combinations of them. In doing so, we 
will try to
make as many of the constraints first class as possible. Those 
constraints which
cannot be brought into the first class through appropriate linear 
combinations are labeled by $\xi_s,\quad s=1,\ldots ,S.$
Now form the PB's of all the $\xi$'s with each other and arrange them 
into a matrix:
\begin{equation}
\Delta\equiv\pmatrix{
0&\{\xi_1,\xi_2\}&\ldots&\{\xi_1,\xi_s\}\cr
\{\xi_2,\xi_1\}&0&\ldots&\{\xi_2,\xi_s\}\cr
\vdots&\vdots&\ddots&\vdots\cr
\{\xi_s,\xi_1\}&\{\xi_s,\xi_2\}&\ldots&0\cr }
. \end{equation}
Dirac has proven that the determinant of $\Delta$ is non-zero 
(not even weakly zero).
Therefore, the inverse of $\Delta$ exists:
\begin{equation}
({\Delta}^{-1})_{ss^\prime}\{\xi_{s^\prime},\xi_{s^{\prime\prime}}\}
=\delta_{ss^{\prime\prime}}.
\label{rb.22}
\end{equation}
Define the Dirac bracket (DB) 
(Dirac called them `new Poisson brackets') between 
any two dynamical quantities $A$ and $B$ to be
\begin{equation}
{\{ A,B\}}^*=\{ A,B\}-\{ A,\xi_s\}({\Delta}^{-1})_{ss^\prime}\{\xi
_{s^\prime},B\}.
\label{rb.23}
\end{equation}
The DB satisfies all the same algebraic properties 
(anti-symmetry, 
linearity, product law, Jacobi identity) as the ordinary PB. Also, the 
equations of motion can be written in terms of the DB since for any
$g(p,q)$,
\begin{equation}
{\{ g,H_T\}}^*=\{ g,H_T\}-\{ g,\xi_s\}({\Delta}^{-1})_{ss
^\prime}\{\xi_{s^\prime},H_T\} \approx\{ g,H_T\}.
\label{rb.24}
\end{equation}
The last step follows because $H_T$ is first class.

Perhaps the most important feature of the DB is the way 
it handles second class constraints. 
Consider the DB of a dynamical quantity with one of
the (remaining) $\xi$'s:
\begin{eqnarray}%
{\{ g,\xi_{s^{\prime\prime}}\}}^*&=\{ g,\xi_{s^{\prime\prime}}\}-
\{ g,\xi_s\}({\Delta}^{-1})_{ss^\prime}\{\xi_{s^\prime},\xi_{s^{\prime
\prime}}\}\cr &=\{ g,\xi_{s^{\prime\prime}}\}-
\{ g,\xi_s\}\delta_{ss^{\prime\prime}} =0.
\label{rb.25}
\end{eqnarray}
The definition Eq.(\ref{rb.22}) was used in the second step 
above. Thus the 
$\xi$'s may be set {\it strongly} equal to zero {\it before}  
working out the Dirac bracket. Of course we must still be 
careful that we do not set $\xi$ 
strongly to zero within a Poisson bracket.  
If we now replace all PB's by DB's
(which is legitimate since the dynamics  can be written 
in terms of DB's via  Eq.(\ref{rb.24})) any second class 
constraints in $H_T$ will appear in the DB in Eq.(\ref{rb.24}). 
Eq.(\ref{rb.25}) then tells us that those constraints can be 
set to zero. Thus all we are left with in our  Hamiltonian are 
first class constraints:
\begin{equation}
{\tilde H}_T=H+v_i\Phi_i\, ,\qquad i=1,\dots ,I
\ , \end{equation}
where the sum is over the remaining constraints which 
are first class.  It must be emphasized that this is possible 
only because we have reformulated the theory in
terms of the  Dirac brackets. 
Of course, this reformulation in terms of the DB's
does not uniquely determine the dynamics for us since 
we still have arbitrary functions of the time accompanying 
the first class constraints. 
If the  Lagrangian is such it exhibits no first class constraints 
then the dynamics are completely defined.

Before doing an example from classical field theory, 
we should note some features of a field theory that 
differentiate it from point mechanics.
In the classical theory with a finite number of degrees of 
freedom we had constraints which were functions of the 
phase space variables. Going over to field theory these 
constraints become {\it functionals} which in 
general may depend upon the spatial derivatives of the 
fields and conjugate 
momenta as well as the fields and momenta themselves:
\begin{equation}
\phi_m=\phi_m[\varphi(x),\pi(x),\partial_i\varphi,\partial_i\pi]
\ .\end{equation}
The square brackets indicate a functional relationship 
and $\partial_i\equiv \partial /\partial x^i$.
A consequence of this is that the constraints are 
differential equations in general. 
Furthermore, the constraint itself is no longer the only 
independent weakly vanishing quantity. 
Spatial derivatives of $\phi_m$ and integrals of constraints 
over spatial variables are weakly zero also.

Since there are actually an infinite number of constraints 
for each $m$ (one at each space-time point $x$) we write,
\begin{equation}
H_T=H+\int\! d{\vec x}\,u_m(x)\phi_m(x).
\label{rc.1}
\end{equation}
Consistency requires that the primary constraints be 
conserved in time:
\begin{equation}%
0\approx\{\phi_m(x),H_T\}=\{\phi_m,H\}+\int\! d{\vec y}\,u_n(y)
\{\phi_m(x),\phi_n(y)\}.
\label{rc.2}
\end{equation}
The field theoretical Poisson bracket for any two phase space
functionals  is given by
\begin{equation}%
\{ A,B\}_{x^0=y^0}({\vec x},{\vec y})=\int\! d{\vec z}\,\left(
{\delta A\over\delta\varphi_i(z)}{\delta B\over\delta\pi_i(z)}-
{\delta A\over\delta\pi_i(z)}{\delta B\over\delta\varphi_i(z)}\right)
\label{rc.3}
\end{equation}
with the subscript $x^0=y^0$ reminding us that the bracket is
defined for  equal times only. Generally, there may be a 
number of fields present hence  the discrete label $i$. 
The derivatives appearing in the PB above are 
{\it functional} derivatives. 
If $F[f(x)]$ is a functional its
derivative with respect to a function $f(y)$ is defined to be:
\begin{equation}
{\delta F[f(x)]\over \delta f(y)} = \lim_{\epsilon\to0}{1\over\epsilon}
\left[ F[f(x)+\epsilon\delta (x-y)]-F[f(x)]\right].
\end{equation}
Assuming Eq.(\ref{rc.3}) has a non-zero determinant 
we can define an inverse:
\begin{equation}
\int\! d{\vec y}\, P_{lm}({\vec x},{\vec y})P_{mn}^{-1}({\vec
y},{\vec z} )=
\int\! d{\vec y}\, P_{lm}^{-1}({\vec x},{\vec y})P_{mn}({\vec y},{\vec 
z})=
\delta_{ln}\delta({\vec x}-{\vec z})
\label{rc.4}
\end{equation}
where
\begin{equation}
P_{lm}({\vec x},{\vec y})\equiv \{\phi_l({\vec x}),\phi_m({\vec y})\}
_{x^0=y^0}.
\label{rc.5}
\end{equation}
Unlike the discrete case, the inverse of the PB matrix 
above is not unique in  general. 
This introduces an arbitrariness which was not present
in  theories with a finite number of degrees of freedom. 
The arbitrariness 
makes itself manifest in the form of {\it differential} 
(rather than algebraic) equations for the multipliers. 
We must then supply boundary 
conditions to fix the multipliers \cite{sun82}.

The Maxwell theory for the free electro-magnetic field is 
defined by the action 
\begin{equation}
S=\int\! {\rm d}^4x\,{\cal L}(x)
\ ,\end{equation}
where $\cal L$ is the Lagrangian {\it density} 
Eq.(\ref{eq:2.7}).
The action is invariant under local gauge transformations.
The ability to perform such gauge transformations \
indicates the presence of first class constraints. 
To find them, we first obtain the momenta conjugate 
to the fields $A_\mu$:
$\pi^\mu=-F^{0\mu}$ as defined in Eq.(\ref{eq:2.11}).
This gives us a primary constraint, namely $\pi^0(x)=0$.
Using Eq.(\ref{eq:2.23}), 
we can write the canonical Hamiltonian density as
\begin{equation}
P_{lm}({\vec x},{\vec y})\equiv \{\phi_l({\vec x}),
\phi_m({\vec y})\}_{x^0=y^0}
.\end{equation}
where the velocity fields ${\dot A}_i$ have been expressed 
in terms of the momenta $\pi_i$. 
After a partial integration on the second term, the 
Hamiltonian becomes
\begin{eqnarray}
H &= \int\! {\rm d}^3x\, \left( {1\over 2}
\pi_i\pi_i-A_0\partial_i\pi_i+{1\over 
4}F_{ik}F_{ik}\right)\cr\Rightarrow\! 
H_T &= H+\int\! {\rm d}^3x\, v_1(x)\pi^0(x).
\end{eqnarray}
Again, for consistency, the primary constraints must be 
constant in time so that
\begin{equation}
0\approx\{\pi^0,H_T\}=-\left\{\pi^0,\int\! {\rm d}^3x\, 
A_0\partial_i\pi_i\right\}=\partial_i\pi_i.
\label{2.7}
\end{equation}
Thus, $\partial_i\pi_i\approx 0$ is a secondary constraint. 
We must then  check to see if Eq.(\ref{2.7}) leads to further 
constraints by also requiring that $\partial_i\pi_i$ is 
conserved in time:
\begin{equation}
0\approx\{\partial_i\pi_i,H_T\}.
\end{equation}
The PB above vanishes identically however so there are 
no more constraints which follow from consistency 
requirements. So we have our two {\it first class} constraints:
\begin{equation}
\phi_1=\pi^0\approx 0\qquad 
\label{primary}
\end{equation}
and
\begin{equation}
\chi\equiv\phi_2=\partial_i\pi_i\approx 0\qquad
\ . \label {secondary}
\end{equation}
In light of the above statements 
the first class {\it secondary} constraints should be included 
in $H_T$ as well
(Some authors call the Hamiltonian with first class 
secondary constraints included the {\it extended} 
Hamiltonian):
\begin{equation}
H_T=H+\int\! {\rm d}^3x\, (v_1\phi_1+v_2\phi_2)
\ . \label{2.8} \end{equation}
Notice that the fundamental PB's among the $A_\mu$ 
and $\pi^\mu$,
\begin{equation}
\{ A_\mu(x),\pi^\nu(y)\}_{x^0=y^0}=\delta_\mu^\nu
\delta({\vec x}-{\vec y})
\end{equation}
are incompatible with the constraint $\pi^0\approx 0$ so we 
will modify them using the Dirac-Bergmann procedure.
The first step towards this end is to impose certain 
conditions to break the local gauge invariance.
Since there are two first class constraints, we need two 
gauge conditions imposed as second class constraints. 
The traditional way to implement this
is by imposing the {\it radiation gauge} conditions:
\begin{equation}
\Omega_1 \equiv A_0\approx 0 \quad and \quad 
\Omega_2  \equiv \partial_iA_i\approx 0
.\end{equation}
It can be shown \cite{sun82} that the radiation gauge 
conditions completely break the gauge invariance 
thereby bringing all constraints into the second class.

The next step is to form the matrix of second class 
constraints with matrix elements $\Delta_{ij}=
\{\Omega_i,\phi_j\}_{x^0=y^0}$ and $i$, $j=1$,$2$:
\begin{equation}
\Delta=\pmatrix{
0 & 0 & 1 & 0 \cr
0 & 0 & 0 & -\nabla^2 \cr
-1 & 0 & 0 & 0 \cr
0 & \nabla^2 & 0 & 0 \cr}\delta({\vec x}-{\vec y})
\end{equation}
To get the Dirac bracket we need the inverse of $\Delta$. Recalling the 
definition Eq.(\ref{rc.4}) we have 
\begin{equation}
\int\! d{\vec y}\, \Delta_{ij}({\vec x},{\vec y})(\Delta^{-1})_{jk}
({\vec y},{\vec z})=\delta_{ik}\delta({\vec x}-{\vec z}).
\end{equation}
With the help of $\nabla^2({1\over |{\vec x}-{\vec y}|})
=-4\pi\delta({\vec x}-{\vec y})$ we can easily perform 
(2.11) element by element to obtain
\begin{equation}
\Delta^{-1}=\pmatrix{
0 & 0 & -\delta({\vec x}-{\vec y}) & 0 \cr
0 & 0 & 0 & -{1\over 4\pi|{\vec x}-{\vec y}|} \cr
\delta({\vec x}-{\vec y}) & 0 & 0 & 0 \cr
0 & {1\over 4\pi|{\vec x}-{\vec y}|} & 0 & 0 \cr}
\end{equation}
Thus, the Dirac bracket in the radiation gauge is 
(all brackets are at equal times),
\begin{equation}
\{ A(x),B(y)\}^\ast=\{ A(x),B(y)\}-\int\!\!\!\int\! d{\vec u}\,d{\vec v}
\, \{ A(x),\psi_i(u)\}(\Delta^{-1})_{ij}({\vec u},{\vec v})\{\psi_j(v),
B(y)\}
\end{equation}
where $\psi_1=\Omega_1$, $\psi_2=\Omega_2$, 
$\psi_3=\phi_1$ and $\psi_4=\phi_2$. 
The fundamental Dirac brackets are
\begin{eqnarray}
\{ A_\mu(x),\pi^\nu(y)\}^\ast 
&= (\delta^\nu_\mu+\delta^0_\mu
g^{\nu 0})\delta ({\vec x}-{\vec y})
-\partial_\mu\partial^\nu
{1\over 4\pi|{\vec x}-{\vec y}|} \cr 
\{ A_\mu(x),A_\nu(y)\}^\ast =0= 
&\{\pi^\mu(x),\pi^\nu(y)\}^\ast. 
\end{eqnarray}
From the first of the above equations we obtain,
\begin{equation}
\{ A_i(x),\pi_j(y)\}^\ast=
\delta_{ij}\delta({\vec x}-{\vec y})-\partial_i
\partial_j{1\over 4\pi|{\vec x}-{\vec y}|}.
\end{equation}
The right hand side of the above expression is often 
called the `transverse delta function' in the context 
of canonical quantization of the electro-magnetic field 
in the radiation gauge. 
In nearly all treatments of that subject, however, 
the transverse delta function is introduced 
`by hand' so to speak. 
This is done after realizing that the standard
commutation relation 
$[A_i(x),\pi_j(y)]=i\delta_{ij}\delta({\vec x}-{\vec y})$ 
is in contradiction with Gauss' Law. 
In the Dirac-Bergmann approach the familiar equal-time
commutator relation 
is obtained without any hand-waving arguments.

The choice of the radiation gauge in the above example 
most naturally reflects the splitting of ${\vec A}$ and 
${\vec \pi}$ into transverse and longitudinal parts. 
In fact, the gauge condition $\partial_i A_i=0$ implies 
that the longitudinal part of ${\vec A}$ is zero. 
This directly reflects the observation that no longitudinally 
polarized photons exist in nature.

Given this observation, we should somehow be able to 
associate the true degrees of freedom with the 
transverse parts of ${\vec A}$ and ${\vec\pi}$.
Sundermeyer \cite{sun82} shows that this is indeed 
the case and that, for the true degrees of freedom, 
the DB and PB coincide.

We have up till now concerned ourselves with constrained 
dynamics at the classical level. 
Although all the previous developments have occurred 
quite naturally in the classical context, 
it was the problem of {\em quantization} which originally 
motivated Dirac and others to develop 
the previously described techniques. 
Also, more advanced techniques 
incorporating constraints into the path integral 
formulation of quantum theory have been developed.

The general problem of quantizing theories with 
constraints is very formidable especially when considering 
general gauge theories. 
We will not attempt to address such problems. 
Rather, we will work in the non-relativistic framework of 
the Schr\"odinger equation where quantum states
are described by a wave function.

As a first case, let us consider a classical theory where 
all the constraints are first class. 
The Hamiltonian is written then as the sum of 
the canonical Hamiltonian $H=p_i{\dot q}_i-L$ plus a 
linear combination of the first class constraints:
\begin{equation}%
H^\prime=H+v_j\phi_j.
\label{rd.1}
\end{equation}
Take the $p$'s and $q$'s to satisfy,
\begin{equation}
\{ q_i,p_j\}\Longrightarrow 
{i\over\hbar}[{\hat q}_i,{\hat p}_j]
\label{rd.2}
\end{equation}
where the hatted variables denote quantum operators and
$[{\hat q}_i,{\hat p}_j]={\hat q}_i{\hat p}_j-{\hat p}_j{\hat q}_i$ 
is the commutator. The Schr\"odinger equation reads
\begin{equation}
i\hbar{d\psi\over dt}=H^\prime\psi
\label{rd.3}
\end{equation}
where $\psi$ is the wave function on which the dynamical 
variables operate.
For each constraint $\phi_j$ impose supplementary 
conditions on the wave function:
\begin{equation}
{\hat \phi}_j\psi=0.
\label{rd.4}
\end{equation}
Consistency of the  Eq.(\ref{rd.4}) with one another 
demands that
\begin{equation}
[{\hat \phi}_j,{\hat \phi}_{j^\prime}]\psi=0.
\label{rd.5}
\end{equation}
Recall the situation in the classical theory where anything 
that
was  weakly zero could be written strongly as a linear 
combination of the $\phi$'s:
\begin{equation}
\{\phi_j,\phi_{j^\prime}\}=c_{jj^\prime}^{j^{\prime\prime}}
\phi^{j^{\prime\prime}}.
\label{rd.6}
\end{equation}
Now if we want Eq.(\ref{rd.5}) to be a {\it consequence} 
of Eq.(\ref{rd.4}), an analogous relation to Eq.(\ref{rd.6}) 
must hold in the quantum theory namely,
\begin{equation}%
[{\hat\phi}_j,{\hat\phi}_{j^\prime}]
={\hat c}_{jj^\prime}^{j^{\prime
\prime}}{\hat\phi}^{j^{\prime\prime}}.
\label{rd.7}
\end{equation}
The problem is that the coefficients $\hat c$ in the quantum 
theory are in general functions of the {\it operators} 
${\hat p}$ and ${\hat q}$  and
do not necessarily commute with the ${\hat\phi}$'s. 
In order for consistency then, we must have the 
coefficients in the quantum theory all
appearing to the {\it left} of the ${\hat\phi}$'s. 

The same conclusion follows if we consider the 
consistency of Eq.(\ref{rd.4}) with the 
Schr\"o\-din\-ger equation. 
If we cannot arrange to have the coefficients to the left 
of the constraints in the quantum theory then as Dirac 
 says `we are out of luck' \cite{dir64}.

Consider now the case where there are second class 
constraints, $\xi_s$.
The problems encountered when there are second class 
constraints are similar in nature to the first class case but 
appear even worse. This statement follows simply from 
the definition of second class. If we try to 
impose a condition on $\psi$ similarly to Eq.(\ref{rd.4}) 
but with a second
class  constraint we must get a contradiction since already
$\{\xi_s,\phi_j\}\ne 0$ for all $j$ at the {\it classical} level.

Of course if we imposed ${\hat\xi}_s=0$ as an operator 
identity then there is no contradiction. 
In the classical theory, the analogous constraint condition 
is the strong equality $\xi=0$. We have seen that strong 
equalities for second class constraints emerge in the 
classical theory via the Dirac-Bergmann method. 
Thus it seems quite suggestive to postulate
\begin{equation}
\{ A,B\}^\ast\Longrightarrow {i\over\hbar}[\hat A,\hat B]
\label{rd.8}
\end{equation}
as the rule for quantizing the theory while imposing
${\hat\xi}_s=0$ as an operator identity. 
Any remaining weak equations are all first class and 
must then be treated as in the first case using
supplementary  conditions on the wave function. 
Hence the operator ordering ambiguity still exists in general.

We have seen that there is no definite way to guarantee 
a well defined quantum theory given the corresponding 
classical theory. It is possible, since the Dirac bracket 
depends on the gauge constraints imposed by hand, 
that we can choose such constraints in such a way as to 
avoid any problems.
For a general system however, such attempts would at 
best be difficult to implement. We have seen that there 
is a consistent formalism for determining
(at least as much one can) the dynamics of a 
generalized Hamiltonian system.
The  machinery is as follows:

$\bullet$ Obtain the canonical momenta from the Lagrangian.

$\bullet$ Identify the primary constraints and construct the total
Hamiltonian.

$\bullet$ Require the primary constraints to be conserved in time.

$\bullet$ Require any additional constraints obtained by step 3 to 
also be conserved in time.

$\bullet$ Separate all constraints into first class or second class.

$\bullet$ Invert the matrix of second class constraints.

$\bullet$ Form the Dirac bracket and write the equations of motion in 
terms of them.

$\bullet$ Quantize by taking the DB over to the quantum commutator.

Of course there are limitations throughout this program; especially
in steps six and eight. If there any remaining first class constraints
it is a sign that we still have some gauge freedom left in our theory.
Given the importance of gauge field theory in today's physics it is
certainly worth one's while to understand the full implications of
constrained dynamics. The material presented here is 
meant to serve as a primer for further study.

%% file: 0000rev.bbl
\begin{thebibliography}{99}

\bibitem {aep76} J.  Abad,  J. G. Esteve,  and A. F. Pacheco,  
Phys.Rev. {\bf D32},  2729 (1985).

\bibitem{atw93} O. Abe, K. Tanaka, K.G. Wilson, 
Phys.~Rev. {\bf D48}, 4856-4867 (1993). 

\bibitem{aat97} O. Abe, G.J. Aubrecht, K. Tanaka,
``Mesons in two-dimensional QCD on the light cone.'' 
OHSTPY-HEP-T-96-032, Mar 1997. 16pp. hep-th/9703074. 

\bibitem{abh90} N.A. Aboud, J.R. Hiller
Phys.~Rev. {\bf D41}, 937-945 (1990).  

\bibitem{acb94} C. Acerbi, A. Bassetto, 
Phys.~Rev. {\bf D49}, 1067-1076 (1994).  

\bibitem {adl94} S. Adler,  
Nucl. Phys. {\bf  B415}, 195 (1994).

\bibitem{abs94} A. Ali, V.M. Braun, H. Simma, 
Z.~Physik {\bf C63}, 437-454 (1994). 

\bibitem{amm94} E.A. Ammons
Phys.~Rev. {\bf D50}, 980-990 (1994).  

\bibitem {and95} F. Antonuccio and S. Dally, 
Phys. Lett. {\bf B348}, 55 (1995).

\bibitem {and96a} F. Antonuccio and S. Dally, 
Nucl. Phys. {\bf B461}, 275 (1996).

\bibitem {and96b} F. Antonuccio and S. Dally, 
Phys. Lett. {\bf B376}, 154 (1996).

\bibitem {anp97} F. Antonuccio and S. Pinsky 
{\it Matrix Theories fron Reduced SU(N)
Yang-Mills with Adjoint Fermions},
 hep-th/9612021, to appear in Phys. Lett.

\bibitem {apf93} A.M.~Annenkova, E.V.~Prokhatilov, 
and V.A.~Franke,
Phys.~Atom.~Nucl. {\bf 56}, 813-825 (1993).

\bibitem {anb51}  J. L. Anderson and P. G. Bergman,  
Phys. Rev. {\bf 83},  1018 (1951).

\bibitem {bah68}  K.~Bardakci and M.B.~Halpern,
Phys.Rev. {\bf 176}, 1786 (1968).
   
\bibitem {bap76}  W.A. Bardeen and R. B. Pearson,  
Phys. Rev.{\bf  D13}, 547 (1976).

\bibitem {bpr80} W.A.~Bardeen, R.B.~Pearson, 
and E.~Rabinovici,
Phys.Rev.{\bf D21},  1037 (1980). 

\bibitem {bar48} V.~Bargmann,   
Proc. Natl. Acad. Sci. (USA) {\bf 34}, 211 (1948).

\bibitem {bag78} I.~Bars, and M.B.~Green,    
Phys. Rev. {\bf D17}, 537 (1978).

\bibitem{bar89} D. Bartoletta {\em et al.}, 
Phys. Rev. Lett. {\bf 62}, 2436 (1989).

\bibitem {bas91}  Bassetto,  G.Nardelli and R. Soldati,  
{\it Yang-Mills Theories in Algebraic Non-covariant Gauges}  
(World Scientific,Singapore,  1991).

\bibitem{bas93} A. Bassetto,
Phys.~Rev. {\bf D47}, 727-729 (1993).  

\bibitem{bar93} A. Bassetto, M. Ryskin, 
Phys.~Lett. {\bf B316}, 542-545 (1993).  

\bibitem{bkk93} A. Bassetto, I.A. Korchemskaya, 
G.P. Korchemsky, G. Nardelli,
Nucl.~Phys. {\bf B408}, 62-90 (1993).  

\bibitem{bas96} A. Bassetto, 
Nucl.~Phys.~Proc.~Suppl. {\bf 51C}, 281-288 (1996).  

\bibitem{bgn96} A. Bassetto, L. Griguolo, and G. Nardelli,
Phys.Rev. {\bf D54}, 2845-2852 (1996). 

\bibitem{ban97}  A. Bassetto and  G. Nardelli, 
Int.~J.~Mod.~Phys. {bf  A12}, 1075-1090 (1997).  

\bibitem {bap96}R.~Bayer and H.C.~Pauli, 
Phys.Rev. {\bf D53}, 939 (1996). 

\bibitem {bel74}   J.S. Bell,    
Acta Phys. Austriaca,  Suppl. {\bf 13}, 395 (1974).

\bibitem {bel94} E. Belz, (1994) ANL preprint.

\bibitem {bmp86} C. M. Bender,  L.R. Mead,  and S. S. Pinsky, 
Phys. Rev. Lett. {\bf  56},  2445 (1986).

\bibitem {bpv93} C.M.~Bender, S.S.~Pinsky, B.~Van~de~Sande,
Phys. Rev. {\bf D48}, 816 (1993).

\bibitem {ber77}  H. Bergknoff,  
Nucl. Phys. {\bf  B122},  215 (1977).

\bibitem {blo58}  C. Bloch,  
Nucl. Phys. {\bf  6},  329 (1958).

\bibitem{bbg81} 
G. Bertsch, S. J. Brodsky, A. S. Goldhaber, and J. F. Gunion,
Phys. Rev. Lett. {\bf 47},  297 (1981).

\bibitem {beh55} H. A. Bethe and F. de Hoffman,  
{\it  Mesons and  Fields},
(Row, Peterson and Company,  Evanston,  Illinois,  1955),  Vol.II.

\bibitem{bdk93}G.~Bhanot, K.~Demeterfi, and I.R.~Klebanov,
Phys. Rev. {\bf D48},  4980 (1993). 

\bibitem {bjd65}  J. D. Bj\o rken and S. D. Drell,
{\it Relativistic Quantum Mechanics}  
(McGraw-Hill,  New York,  1964);
J.D.Bj\o rken and S. D. Drell,  {\it Relativistic Quantum Fields}
 (McGraw-Hill,NewYork, 1965).

\bibitem {bjp69}  J. D. Bj\o rken and E. A. Paschos,  
Phys. Rev. {\bf  185},1975 (1969).

\bibitem {bks71} J. D. Bj\o rken,  J. B. Kogut, D. E. Soper,  
Phys.Rev. {\bf D3},   1382 (1971).

\bibitem {bbf93}
B. Blaettel, G. Baym, 
L. L. Frankfurt, H. Heiselberg, and M.Strikman, 
Phys. Rev. {\bf D47}, 2761 (1993).

\bibitem {bbg73} 
R. Blankenbecler, S. J. Brodsky, J. F. Gunion, and R. Savit, 
Phys. Rev. {\bf  D8},  4117 (1973).

\bibitem {byg85}   G. T. Bodwin,  D. R. Yennie,  and M. A. Gregorio,  
Rev.Mod. Phys. {\bf 56},  723 (1985).

\bibitem{bgw93} A. Borderies, P. Grange, E. Werner, 
Phys.~Lett. {\bf B319}, 490-496 (1993).  

\bibitem{bgw95} A. Borderies, P. Grange, E. Werner, 
Phys.~Lett. {\bf B345}, 458-468 (1995).  

\bibitem {bot91} J. Botts, Nucl. Phys. {\bf B353},  20 (1991).

\bibitem{brp95} M.~Brisudova, R.J.~Perry, 
Phys.~Rev. {\bf D54}, 1831-1843 (1996).  

\bibitem{brp96} M.~Brisudova, R.J.~Perry, 
Phys.~Rev. {\bf D54}, 6453-6458 (1996).  

\bibitem{bpw97} M.~Brisudova, R.J.~Perry, K.G.~Wilson,
Phys.~Rev.~Lett. {\bf  78},  1227-1230 (1997).  

\bibitem {brp60} S.J. Brodsky and J.R.~Primack,   
Annals Phys. {\bf 52},315 (1960). 

\bibitem {brp68} S.J. Brodsky and J.R.~Primack,   
Phys. Rev. {\bf 174}   2071 (1968).

\bibitem {brs73}  S.J.~Brodsky, R.~Roskies and R.~Suaya,
Phys. Rev. {\bf D8},  4574 (1973).

\bibitem {brf75} S.J. Brodsky  and G. R. Farrar,  
Phys. Rev. {\bf D11},  1309(1975).

\bibitem {brc76} S.J. Brodsky and B. T. Chertok,
Phys. Rev. {\bf  D14},   3003  (1976).

\bibitem {brd80} S.J. Brodsky and S.D.~Drell,  
Phys. Rev. {\bf D22}, 2236(1980).

\bibitem{blz81}
S. J. Brodsky, G. P. Lepage and S.A.A. Zaidi, 
Phys. Rev. {\bf D23}, 1152 (1981).

\bibitem {brh83} S. J. Brodsky and J.R.~Hiller,
Phys.~Rev. {\bf C28}, 475 (1983).

\bibitem {bjl83} S. J. Brodsky, C.-R. Ji and G. P. Lepage,
Phys. Rev. Lett. {\bf 51}, 83 (1983). 

\bibitem {brm88} S. J. Brodsky and A. H. Mueller,
Phys. Lett. {\bf 206B}, 685 (1988).

\bibitem {brt88} S. J. Brodsky and G. F. de Teramond, Phys. 
Rev. Lett. {\bf60},  1924  (1988).

\bibitem {brl89}  S. J. Brodsky and G. P. Lepage,  in 
{\it Perturbative QuantumChromodynamics} ,  A. H. Mueller,  
Ed.(World Scientific, Singapore, 1989)

\bibitem {brs90} S.J.~Brodsky and I.A.~Schmidt,    
Phys. Lett. {\bf B234} (1990) 144.

\bibitem {brt90} S. J. Brodsky, G. F. de Teramond 
and I. A. Schmidt, 
Phys.Rev. Lett. {\bf 64},  1011 (1990).

\bibitem {brs91} S.J.~Brodsky and I.~A. Schmidt,     
Phys. Rev.{\bf D43} (1991) 179.

\bibitem {brp91}  S.J. Brodsky and H.C. Pauli,  
published in {\it Recent Aspect of Quantum Fields},  
H. Mitter and H. Gausterer,  Eds.; 
Lecture Notes in Physics,Vol.  396  
(Springer-Verlag,  Berlin, Heidelberg, 1991).

\bibitem {bhm92} S. J. Brodsky, P. Hoyer, 
A. H. Mueller and W-K. Tang,
Nucl. Phys. {\bf B369},  519 (1992). 

\bibitem {bmp93}  S.J. Brodsky,  G. McCartor,  
H.C. Pauli  and S.S. Pinsky,
Particle World  {\bf  3},  109 (1993).

\bibitem {brt93} S. J. Brodsky,  W-K. Tang, and C. B. Thorn, 
Phys. Lett. {\bf B318}, 203 (1993).

\bibitem {bbs94} S. J. Brodsky, M. Burkardt, and I. Schmidt,
Nucl. Phys. {\bf B441}, 197 (1994).

\bibitem {brs94} S.J. Brodsky and F. Schlumpf, Phys. 
Lett. {\bf B329},  111 (1994).

\bibitem{brs95} S.J. Brodsky, F. Schlumpf,
Prog.~Part.~Nucl.~Phys. {\bf 34}, 69-86 (1995).  

\bibitem{bjs93} R.W. Brown, J.W. Jun, S.M. Shvartsman, C.C. Taylor,
Phys.~Rev. {\bf D48}, 5873-5882 (1993).  

\bibitem {buc70}  F. Buccella,  E. Celeghin,  H. Kleinert,  
C.A. Savoy and E. Sorace,  
Nuovo Cimento  {\bf A69},  133 (1970).

\bibitem {bur89}  M. Burkardt,  
Nucl. Phys. {\bf A504}, 762  (1989).

\bibitem {bul91a}   M. Burkardt and A. Langnau, 
 Phys. Rev. {\bf D44}, 1187(1991).

\bibitem {bul91b}   M. Burkardt and A. Langnau,  
Phys. Rev. {\bf D44}, 3857 (1991).

\bibitem {bur93} M. Burkardt,  
Phys. Rev. D {\bf  47},  4628 (1993).

\bibitem {bur94} M. Burkardt,  
Phys. Rev. {\bf D49},  5446 (1994).

\bibitem {bur95a}  M. Burkardt,
Adv.Nucl.Phys. {\bf 23}, 1-74 (1996).

\bibitem {bur95b}  M. Burkardt, 
``Light front hamiltonians and confinement,''
hep-ph/9512318 

\bibitem {bue96}  M. Burkardt and  H. El-Khozondar, 
``A (3+1)-dimensional light front model with spontaneous
breaking of chiral symmetry,''
hep-ph/9609250 

\bibitem {bur96}  M. Burkardt,
Phys.Rev. {\bf  D54}, 2913-2920 (1996).  

\bibitem {buk97}  M. Burkardt and  B. Klindworth, 
Phys.Rev. {\bf  D55}, 1001-1012 (1997).  

\bibitem{bur97} M. Burkardt, 
``Finiteness conditions for light front hamiltonians.'' 
hep-th/9704162.

\bibitem{cgn95} F. Cardarelli, I.L. Grach, I.M. Narodetskii, G. Salme, and 
S. Simula,
front constituent quark model. 
Phys.~Lett. {\bf B349}, 393-399 (1995).  

\bibitem {cah75}  Carlitz,  R.,  D. Heckathorn,  
J. Kaur and W.-K. Tung,  
Phys. Rev. {\bf  D11},  1234 (1975).

\bibitem {cat76}  Carlitz,  R.,  and W.-K. Tung,      
Phys. Rev. {\bf  D13}, 3446 (1976).

\bibitem{cap88} C. E. Carlson and J. L. Poor,
Phys. Rev. {\bf D38}, 2758 (1988).

\bibitem {car93} A. Carroll,  
 Lecture  at the {\it Workshop on Exclusive Processes 
at High Momentum Transfer}, Elba, Italy (1993).

\bibitem {cas76}  A. Casher,   
Phys.Rev. {\bf  D14},  452 (1976).

\bibitem {chm69} S.~Chang and S.~Ma,
Phys.Rev. {\bf 180}, 1506 (1969).
   
\bibitem {cha76} S. J. Chang,  Phys. Rev. D {\bf  13},  2778 (1976).

\bibitem {chy73}  S.J. Chang and T.M. Yan,   
Phys.Rev. {\bf D7},1147 (1973).

\bibitem {cry73a}  S.J. Chang,  R.G. Root, and T.M. Yan,    
Phys.Rev. {\bf D7}, 1133 (1973).

\bibitem {cry73b}  S.J. Chang,  R.G. Root,  and T.M. Yan,  
Phys.Rev. {\bf  D7}, 1133 (1973).

\bibitem{cha93} L. Chao, 
Mod.~Phys.~Lett. {\bf A8}, 3165-3172 (1993).  

\bibitem {chm95} 
Z.~Chen, and  A.H.~Mueller 
Nucl.Phys. {\bf B451}579 (1995). 

\bibitem{cch97} H.Y.~Cheng, C.Y.~Cheung, C.W.~Hwang,
Phys.~Rev. {\bf D55}, 1559-1577 (1997).  

\bibitem {chz84} V. L. Chernyak,  A.R. Zhitnitskii, 
Phys. Rept. {\bf 112}, 173(1984).

\bibitem{czl95} C.Y.~Cheung, W.M.~Zhang, G.-L.~Lin, 
Phys.~Rev. {\bf D52}, 2915-2925 (1995).  

\bibitem {cpc88} P.L. Chung, W.N. Polyzou, 
F. Coester and B.D. Keister,
Phys. Rev. {\bf C37}, 2000 (1988).

\bibitem {chc91} P.L. Chung and F.~Coester,   
Phys. Rev.  {\bf D44}, 229 (1991).

\bibitem {clo79} F. E. Close,  
{\em An Introduction to Quarks and Partons} ,
(Academic Press,  New York.  1979);

\bibitem {cop82} F. Coester and W.N. Polyzou, 
Phys. Rev.  {\bf D26}, 1349 (1982).  

\bibitem {coe92}  F. Coester,   
Prog. Nuc. Part. Phys. {\bf 29}, 1 (1992).

\bibitem{cop94} F. Coester, W. Polyzou. 
Found.~Phys. {\bf  24},  387-400 (1994).  

\bibitem {col73} S. Coleman,    
Comm. Math. Phys. {\bf 31}, 259 (1973).

\bibitem {cjs75}  S. Coleman, R. Jackiw and L. Susskind,   
Ann. Phys.(N.Y.) {\bf  93},  267 (1975).

\bibitem {col76}  S. Coleman,  
Ann. Phys. (N.Y.) {\bf  101},  239 (1976).

\bibitem {col84}  J.~Collins,  {\em Renormalization},  
(CambridgeUniversity Press, New York. 1984).

\bibitem{ctj95} M.E. Convery, C.C. Taylor, J.W. Jun, 
Phys.~Rev. {\bf D51}, 4445-4450 (1995).  

\bibitem {crh80} D. P. Crewther and C.J. Hamer, 
Nucl. Phys. {\bf B170}, 353 (1980).

\bibitem {dad55}  R.H. Dalitz and F.J. Dyson,   
Phys.     Rev.     {\bf  99},301(1955).

\bibitem {dak93} S.~Dalley and I.R.~Klebanov,
Phys. Rev {\bf D47}, 2517 (1993). 

\bibitem {dav96}  S.~Dalley and B.~van~de~Sande,
Proceedings of Lattice 96: $14^{th}$ International Symposium on 
Lattice Field Theory, St. Louis MO. 1996.  hep-ph/9602291.
 
\bibitem {dan50}  S. M. Dancoff, 
Phys. Rev. {\bf  78}, 382 (1950).

\bibitem {dag66}  R. Dashen and M. Gell-Mann,   
Phys.     Rev.     Lett.{\bf 17}, 340 (1966).

\bibitem {dea73}  S.P. De Alwis,   
Nucl. Phys. {\bf  B55},  427 (1973).

\bibitem {des74}  S.P. De Alwis,   and J. Stern,  
Nucl. Phys. {\bf B77},  509  (1974).

\bibitem {dkb94} K.~Demeterfi, I.R.~Klebanov and G.~Bhanot, 
Nucl. Phys. {\bf B418},  15 (1994).

\bibitem{dkn97} N.B. Demchuk, P.Yu. Kulikov, I.M. Narodetskii, 
 P.J. O'Donnell, 
``The light front model for exclusive semileptonic b and d decays.''
UTPT-97-1, Jan 1997. 27pp. hep-ph/9701388.

\bibitem {dir49}  P.A.M. Dirac,     
Rev. Mod. Phys. {\bf  21},  392 (1949).

\bibitem {dir50}  P.A.M. Dirac, 
Can. Jour. Math. {\bf 2}, 129 (1950).

\bibitem {dir58}  P.A.M. Dirac,       
{\it The Principles of Quantum Mechanics}, 4th ed.,  
 Oxford Univ. Press, Oxford (1958).

\bibitem {dir64}  P.A.M. Dirac, 
{\it Lectures on Quantum Mechanics}, 
Belfer Graduate School of Science, 
Yeshiva University (1964).

\bibitem {dir81}  P.A.M. Dirac,  in 
{\it Perturbative QuantumChromodynamics},
D.W. Duke and J.F. Owens,  Eds. 
(Am. Inst. Phys.,  NewYork,  1981).

\bibitem{dxt95} P.J. O'Donnell, Q.P. Xu, H.K.K. Tung,
Phys.~Rev. {\bf D52}, 3966-3977 (1995).   
 
\bibitem {drh66} S. D. Drell and A.~C.~Hearn,  
Phys. Rev. Lett. {\bf 16},  908 (1966).

\bibitem {dly69}  S. D. Drell,  D. Levy and  T. M. Yan,  
Phys. Rev. {\bf 187}, 2159 (1969). 

\bibitem {dly70}  S. D. Drell,  D. Levy and  T. M. Yan,
Phys. Rev. {\bf D1}, 1035 (1970).

\bibitem {dly70b}  S. D. Drell,  D. Levy and  T. M. Yan,
Phys. Rev. {\bf D1}, 1617 (1970).

\bibitem {dry70} S.D. Drell and T.M. Yan,  
Phys. Rev. Lett. {\bf 24}, 181(1970).

\bibitem{dks95} A.Yu. Dubin, A.B. Kaidalov, Yu.A. Simonov, 
Phys.~Lett. {\bf B343}, 310-314 (1995).  

\bibitem {efw73}  Eichten,  E.,  F. Feinberg and J.F. Willemsen, 
Phys. Rev. {\bf  D8},  1204 (1973).

\bibitem {ein76}   M.B. Einhorn, 
Phys. Rev. {\bf D14}, 3451(1976).

\bibitem {epb87}  T. Eller,  H.C. Pauli and S.J. Brodsky,   
Phys. Rev. {\bf D35}, 1493 (1987).

\bibitem {elp89}  T. Eller  and H.C. Pauli,   
Z. Physik {\bf C42}, 59 (1989).

\bibitem {els94} S. Elser, 
   {\it The Spectrum of QED$_{1+1}$ in the framework
   of the DLCQ~method},
   Proceedings of {\em Hadron structure `94}, 
   Kosice, Slowakia (1994);  
and Diplomarbeit, U. Heidelberg (1994). 

\bibitem {elk96} S. Elser and A.C. Kalloniatis,   
Phys. Lett. {\bf B375}, 285-291 (1996). 

\bibitem {fan93} G. Fang, et. al., 
presented  at the INT -Fermilab Workshop on 
{\it Perspectives of High Energy Strong Interaction Physics at Hadron
Facilities} (1993).

\bibitem {fei73}  F.L. Feinberg,    
Phys. Rev. {\bf  D7}, 540 (1973).

\bibitem {fgv95} T.J. Fields, K.S. Gupta, and J.P. Vary, 
``Renormalization of effective Hamiltonians,''
ISU-NP-94-15, Jun 1995. 10pp. hep-ph/9506320 

\bibitem {fey69}  R. P. Feynman,   
Phys.     Rev.     Lett. {\bf  23},1415(1969)

\bibitem {fey72}  R. P. Feynman,  
{\it Photon-hadron interactions}, 
(Benjamin,Reading,  Mass.,  1972)

\bibitem {fnp81a} V.A.~Franke, Yu.A.~Novozhilov
and  E.V.~Prokhvatilov,
Lett.Math.Phys. {\bf 5},  239 (1981). 

\bibitem {fnp81b} V.A.~Franke, Yu.A.~Novozhilov
and  E.V.~Prokhvatilov,
Lett.Math.Phys. {\bf 5},  437 (1981). 

\bibitem {fnp82} V.A.~Franke, Yu.A.~Novozhilov 
and E.V.~Prokhvatilov,
in {\it Dynamical Systems and Microphysics}, 
Academic Press, 1982, p.389-400. 

\bibitem {flm96}  L.~Frankfurt, T.S.H.~Lee, 
   G.A.~Miller, M.~Strikman,
   {\it Chiral transparency},
   nucl-th/9608059. 

\bibitem {frg72}  H. Fritzsch and M. Gell-Mann,   in 
{\it Proc.16th. Int. Conf. on HEP,  Batavia,  IL} (1972).

\bibitem {fri90} H. Fritzsch,  
Mod. Phys. Lett.  {\bf A5},  625 (1990).

\bibitem {fri93}  H. Fritzsch,   `
`Constituent Quarks, Chiral Symmetry and the Nucleon Spin'',  
talk  given at the Sep.1993 Leipzig Workshop on
Quantum Field Theory Aspects of High Energy Physics 
CERN preprintTH. 7079/93, (1993).

\bibitem {fuf65}  S. Fubini and G. Furlan,  
Physics {\bf  1},  229 (1965). 

\bibitem {fhj73} S. Fubini, A.J. Hanson and R. Jackiw,    
Phys. Rev. {\bf D7}, 1732 (1973).

\bibitem {fud87} M.G. Fuda, 
Phys.Rev. {\bf C36}, 702-709 (1987). 

\bibitem {fud90} M.G. Fuda, 
Phys.Rev. {\bf D42}, 2898-2910 (1990). 

\bibitem {fud91} M.G. Fuda,
Phys.Rev. {\bf D44}, 1880-1890 (1991). 

\bibitem {fud92} M.G. Fuda, 
Nucl.Phys. {\bf A543}, 111c-126c (1992).

\bibitem {fud94} M.G. Fuda, 
Annals Phys. {\bf 231}, 1-40 (1994).  

\bibitem {fud95} M.G. Fuda, 
Phys.Rev. {\bf C52}, 1260-1269  (1995).  

\bibitem {fud96} M.G. Fuda, 
Phys.Rev. {\bf D54}, 5135-5147 (1996).  

\bibitem {fuo93} T. Fujita and A. Ogura, 
Prog.Theor.Phys. {\bf 89}, 23-36,1993 

\bibitem {fio94} T. Fujita, C. Itoi, A. Ogura, and M. Taki,
J.Phys. {\bf G20}, 1143-1157 (1994). 

\bibitem {fus95a} T. Fujita and Y. Sekiguchi,
Prog.Theor.Phys. {\bf 93}, 151-160 (1995). 

\bibitem {fus95b} M.~Fujita and Sh.M. Shvartsman, 
``Role of zero modes in quantization of QCD in light-cone coordinates,''
 CWRU-TH-95-11, Jun 1995. 21pp. hep-th/9506046 

\bibitem {fht96a} T. Fujita, M. Hiramoto, and H. Takahashi, 
``Bound states of (1+1)-dimensional field theories,''
hep-th/9609224 

\bibitem {fht96b} T. Fujita, Y. Sekiguchi, and K. Yamamoto ,  
``A new interpretation of Bethe ansatz solutions for massive
Thirring model,'' 
Preprint NUP-A-96-7, Jul 1996. 53pp. hep-th/9607115

\bibitem {fkk87}  M. Funke,  V. Kaulfass,  and H. Kummel, 
Phys. Rev. D {\bf 35}, 621 (1987);

\bibitem{ggs93} P. Gaete, J. Gamboa,I. Schmidt,
Phys.~Rev. {\bf D49}, 5621-5624 (1994).   

\bibitem {gas86}  M. Gari and N. G. Stephanis, P
Phys. Lett. {\bf B175},  462(1986). 

\bibitem {gal75}  A. Gasser  and H. Leutwyler,    
Nucl. Phys. B {\bf 94},  269 (1975). 

\bibitem {gel64}  M. Gell-Mann,  
Phys. Lett. {\bf  8},  214 (1964). 

\bibitem {gel72}   M. Gell-Mann,  
Lectures given at the1972 Schladming Winter School, 
Acta Phys. Austriaca,  Suppl. {\bf 9},  733 (1972).

\bibitem {ger65} S. B. Gerasimov, 
Yad. Fiz. {\bf 2},  598 (1965)
[ Sov.J. Nucl. Phys. {\bf 2},  430 (1966)].

\bibitem{gho92} S.N. Ghosh, 
Phys.~Rev. {\bf D46}, 5497-5503 (1992).  

\bibitem {gla84}  S.~G{\l}azek,  
Acta Phys. Pol. {\bf  B15},  889 (1984).

\bibitem {gla88} S.~G{\l}azek,  
Phys. Rev. {\bf D38},  3277 (1988).

\bibitem {glp92a} S.~G{\l}azek and R.J.~Perry,
Phys. Rev. {\bf D45}, 3734 (1992).

\bibitem {glp92b} S.~G{\l}azek, R.J.~Perry,
Phys. Rev. {\bf  D45}, 3740 (1992).

\bibitem {glw93a}  S.~G{\l}azek,  and K. G. Wilson,  
Phys. Rev.  {\bf  D47}, 4657 (1993).

\bibitem {glw93b}  S.~G{\l}azek and K.G. Wilson,   
Phys.Rev. {\bf  D48}, 5863 (1993).

\bibitem {ghp93} S.~G{\l}azek, A.~Harindranath, S.~Pinsky, 
J.~Shigemitsu and K.~Wilson,
Phys. Rev. {\bf  D47},  1599 (1993).

\bibitem {glw94}  S.~G{\l}azek and K.G. Wilson,  
Phys.Rev. {\bf D49},  4214 (1994).

\bibitem {gla95} 
{\it Theory of Hadrons and Light-front QCD}, 
S.~Glazek, Ed., 
(World Scientific Publishing Co., Singapore, 1995).

\bibitem {gol50} H. Goldstein, 
{\it Classical Mechanics},  
Addison-Wesley, Reading, Mass. (1950). 

\bibitem{gns96} I.L. Grach, I.M. Narodetskii, S. Simula,
Phys.~Lett. {\bf B385}, 317-323 (1996).   
 
\bibitem {gri78} V.N. Gribov, 
Nucl. Phys. {\bf B139} 1 (1978).

\bibitem {gri92}  P. A. Griffin,  
Nucl. Phys. {\bf  B372},  270 (1992).

\bibitem{gri92b} P.A. Griffin,
Phys.~Rev. {\bf D46}, 3538-3543 (1992).  

\bibitem {grs74} D. Gromes, H.J. Rothe and B. Stech 
Nucl. Phys. {\bf B75}, 313 (1974).

\bibitem {grk96} D. J. Gross, I. R. Klebanov, 
A, V. Matytsin and A. V. Smilga
{\it Screening vs. Confinement in 1+1 Dimensions}, 
hep-th/9511104

\bibitem{guw97} E.~Gubankova and F.~Wegner,
``Exact renormalization group analysis in Hamiltonian theory:
   I. QED Hamiltonian on the light front,''
   hep-th/9702162.

\bibitem {ham77} C.J. Hamer, 
Nucl. Phys. {\bf B121}, 159 (1977).

\bibitem {ham78} C.J. Hamer, 
Nucl. Phys. {\bf B132},  542 (1978).

\bibitem {ham82} C.J. Hamer, 
 Nucl. Phys. {\bf B195}, 503 (1982).

\bibitem {hot94} K.~Harada, A.~Okazaki, M.~Taniguchi, and M.~Yahiro, 
Phys.Rev. {\bf  D49}, 4226-4245 (1994). 

\bibitem {hot95} K.~Harada, A.~Okazaki, and M.~Taniguchi, 
Phys.Rev. {\bf  D52}, 2429-2438 (1995). 

\bibitem {hao96} K.~Harada and A.~Okazaki,  
``Perturbative Tamm-Dancoff renormalization,''\\
KYUSHU-HET-35, Oct 1996. 20pp. hep-th/9610020 

\bibitem {hot96} K.~Harada, A.~Okazaki, and M.~Taniguchi, 
Phys.~Rev. {\bf  D54}, 7656-7663 (1996). 

\bibitem {hot97} K.~Harada, A.~Okazaki, and M.~Taniguchi, 
Phys.~Rev. {\bf  D55}, 4910-4919 (1997). 

\bibitem {hav87}  A. Harindranath and J. P. Vary,      
Phys. Rev. {\bf D36}, 1141 (1987).

\bibitem {hav88}  A. Harindranath and J. P. Vary, 
Phys. Rev. {\bf  D37}, 1064-1069 (1988). 

\bibitem {hap91}  A. Harindranath,  
R.J. Perry and J. Shigemitsu,
Ohio State preprint, (1991). 

\bibitem {hkw91a} T.~Heinzl, S.~Krusche, E.~Werner and B.~Zellermann,
Phys. Lett. B {\bf 272},  54 (1991).

\bibitem {hkw91b} T. Heinzl,  S. Krusche and E. Werner,  
Phys.Lett. {\bf B256},  55 (1991).

\bibitem {hkw91c}  T. Heinzl,  S. Krusche  and E. Werner, 
Nucl. Phys.  {\bf A532}, 4290 (1991).

\bibitem {hkw92a}  T. Heinzl,  S. Krusche,  S. Simburger  and E. Werner, 
Z.Phys. {\bf  C56},  415 (1992).

\bibitem {hkw92b}  Heinzl,  S. Krusche  and E. Werner, 
Phys. Lett. {\bf B275}, 410 (1992).

\bibitem {hew94} T. Heinzl, and E. Werner,
Z.Phys. {\bf  C62}, 521-532 (1994).  

\bibitem {hei95} T. Heinzl,
hep-th/9409084, 
Nucl.Phys.Proc.Suppl. {\bf  B39C}, 217-219 (1995).   

\bibitem {hei96a} T. Heinzl,
``Hamiltonian formulations of Yang-Mills quantum theory and
the Gribov problem,'' 
hep-th/9604018 

\bibitem{hei96b} T. Heinzl, 
Phys.~Lett. {\bf B388}, 129-136 (1996).   

\bibitem {hei97} T. Heinzl,
Nucl.Phys.Proc.Suppl. {\bf  54A}, 194-197 (1997).   

\bibitem {hep90} S. Heppelmann, 
Nucl. Phys. B, Proc. Suppl. {\bf  12}, 159 (1990);
and references therein.

\bibitem {het93} J.E.~Hetrick,
{\it Nucl.Phys.} {\bf B30} (1993) 228.

\bibitem {het94} J.E.~Hetrick,
Int. J. Mod. Phys. {\bf A9} (1994) 3153.  

\bibitem {hek95} M. Heyssler, and A.C. Kalloniatis, 
Phys. Lett. {\bf B354}, 453 (1995).

\bibitem {hil91a} J.R. Hiller,  
Phys. Rev. {\bf D43}, 2418 (1991).

\bibitem {hil91b} J.R. Hiller,  
Phys.Rev. {\bf D44}, 2504 (1991).

\bibitem {hbo95} J.R. Hiller, S.J. Brodsky and Y. Okamoto, 
(1997) (in progress).

\bibitem {hpv95}   J.~ Hiller, S.S.~Pinsky and B.~van~de~Sande, 
Phys.Rev. {\bf D51} 726 (1995). 

\bibitem {hhw91} L.C.L. Hollenberg, K. Higashijima, 
R.C.~War\-ner   and B.H.J.~McKellar,    
Prog. Theor. Phys. {\bf 87},  3411 (1991).

\bibitem {how94}  L.C.L. Hollenberg and N.S. Witte,  
Phys. Rev. {\bf D50}, 3382 (1994). 

\bibitem {hbp88} K.~Hornbostel, S.J.~Brodsky, and H.C.~Pauli,  
Phys. Rev. {\bf D38}, 2363 (1988).

\bibitem {hbp90}  K. Hornbostel,  S. J. Brodsky,  and H. C. Pauli,
Phys. Rev. {\bf  D41},  3814 (1990).

\bibitem {hor90} K.~Hornbostel,    
{\it Constructing Hadrons on the Light Cone},    
Workshop on 
{\it From Fundamental fields to nuclear phenomena},   
Boulder, CO, Sept 20-22, 1990;   
Cornell preprint CLNS 90/1038 (1990).   

\bibitem {hor91}    K. Hornbostel,    
Cornell Preprint CLNS 91/1078, Aug1991.

\bibitem {hor92}  K. Hornbostel,  
Phys. Rev. {\bf  D45},  3781 (1992).

\bibitem {hot93} S.~Hosono and  S.~Tsujimaru,
Int.J.Mod.Phys. {\bf  A8}, 4627-4648 (1993).

\bibitem{hul93} S.Z. Huang, W. Lin, 
Annals Phys. {\bf  226},  248-270 (1993).  

\bibitem{hye94} T. Hyer, 
Phys.~Rev. {\bf D49}, 2074-2080 (1994).  

\bibitem {ida74}  M. Ida,   
Progr. Theor. Phys. {\bf  51},  1521 (1974).

\bibitem {ida75a}  M. Ida,   
Progr. Theor. Phys. {\bf  54},  1199 (1975).

\bibitem {ida75b} M. Ida,   
Progr. Theor. Phys. {\bf  54},  1519 (1975).

\bibitem {ida75c} M. Ida,   
Progr. Theor. Phys. {\bf  54},  1775 (1975).

\bibitem {isl89} N. Isgur and C.H. Llewellyn Smith,
Phys. Rev. {\bf B217}, 2758 (1989).

\bibitem {ita96a} K. Itakura,
Phys.Rev. {\bf D54}, 2853-2862 (1996).  

\bibitem {ita96b} K.~Itakura,
``Dynamical symmetry breaking in light front Gross-Neveu model,''
UT-KOMABA-96-15, Aug 1996. 12pp. hep-th/9608062 


\bibitem {itm97} K.~Itakura and S.~Maedan,
``Spontaneous symmetry breaking in discretized light cone
quantization,''  
UT-KOMABA-96-29, Mar 1997. 24pp. hep-th/9703107 

\bibitem {itz85} C.~Itzykson and J.B.~Zuber, 
{\it Quantum Field Theory'},   
McGraw-Hill, New York, 1985.

\bibitem {jac62} J.D.~Jackson, 
{\it Classical Electrodynamics},   
Wiley, New York (1962).

\bibitem {jac91}  R. Jackiw 
{\it Delta-Function Potentials 
in Two and Three-Dimensional Quantum Mechanics}, 
in M.A.B.B\'eg Memorial Volume. 
A.Ali and P. Hoodbhoy,  eds.\ World Scientific,Singapore,1991.

\bibitem {jam80} R.~Jackiw and N.S.~Manton,   
Ann.Phys.(N.Y.) {\bf 127}, 257 (1980). 

\bibitem{jau90} W. Jaus, 
Phys.~Rev. {\bf D41}, 3394 (1990).   

\bibitem {jes68}  J. Jers\'ak and J. Stern,  
Nucl. Phys. {\bf B7}, 413 (1968).

\bibitem {jes69}  J. Jers\'ak and J. Stern,  
Nuovo Cimento {\bf 59}, 315 (1969).

\bibitem {jib86} C. R. Ji and S.~J.~Brodsky,
Phys. Rev. {\bf D34}, 1460 (1986); {\bf D33},  1951, 1406, 2653 (1986);

\bibitem{jix93} X.D.~Ji,
Comments Nucl.Part.Phys. {\bf 21} 123-136 (1993).  

\bibitem{jic94} C.R.~Ji, 
Phys.~Lett. {\bf B322}, 389-396 (1994).  

\bibitem{jhm95} C.R.~Ji, G.H. Kim, D.P. Min, 
Phys.~Rev. {\bf D51}, 879-889 (1995).  

\bibitem{jir95} C.R.~Ji, S.J.~Rey,
Phys.~Rev. {\bf D53}, 5815-5820 (1996).  

\bibitem{jps95} C.R. Ji, A. Pang, A. Szczepaniak, 
Phys.~Rev. {\bf D52}, 4038-4041 (1995).  

\bibitem {jop96a} B.~Jones and R.~Perry,
{\it The Lamb-shift in a light-front Hamiltonian approach,}
hep-th/9612163.

\bibitem {jop96b} B.~Jones and R.~Perry,
{\it Analytic treatment of positronium spin-splitting in light-front
QED,} hep-th/9605231.

\bibitem{juj94} J.W. Jun, C.K. Jue, 
Phys.~Rev. {\bf D50}, 2939-2941 (1994).  

\bibitem {kap93} A.C. Kalloniatis and H.C.Pauli, 
   Z.Phys.  {\bf C60}, 255 (1993).

\bibitem {kap94}
   A.C. Kalloniatis and H.C.Pauli, 
   Z.Phys. {\bf C63}, 161 (1994).

\bibitem {kar94} A.C. Kalloniatis and D.G. Robertson, 
Phys. Rev. {\bf D50}, 5262 (1994).

\bibitem {kpp94} A.C. Kalloniatis, H.C. Pauli and S.S. Pinsky,
Phys. Rev.  {\bf D50}, 6633 (1994).

\bibitem {kal95} A.C. Kalloniatis, 
Phys. Rev. {\bf D54}, 2876 (1996). 

\bibitem {kar96} A.C. Kalloniatis and D.G. Robertson, 
Phys.Lett. {\bf B381}, 209-215 (1996).

\bibitem {kap92} M.~Kalu\v za and H.C.~Pauli, 
   Phys. Rev. {\bf D45}, 2968 (1992). 

\bibitem {kap93b}  M. Kalu\v za and H.-J. Pirner,   
Phys. Rev. {\bf D47}, 1620 (1993).

\bibitem {kar92} G.  Karl, 
Phys. Rev.  {\bf D45},  247  (1992).

\bibitem {kar80} V. A. Karmanov, 
Nucl. Phys. {\bf B166},  378 (1980).

\bibitem {kar81} V. A. Karmanov, 
Nucl.  Phys. {\bf A362},  331 (1981).

\bibitem{kei91} B.D. Keister, 
Phys.~Rev. {\bf C43}, 2783-2790 (1991).   

\bibitem{kei94} B.D. Keister,
Phys.~Rev. {\bf D49}, 1500-1505 (1994).   

\bibitem {kty95} Y.~Kim, S.~Tsujimaru, and K.~Yamawaki,
Phys.Rev.Lett. {\bf 74}, 4771-4774 (1995);
Erratum-ibid.{\bf 75}, 2632 (1995).

\bibitem {klp90}  D. Klabu\v car and H.C. Pauli, 
Z. Phys.\ {\bf C47}, 141(1990).

\bibitem {kll69}  J. R. Klauder,  H. Leutwyler,  and L. Streit, 
Nuovo Cimento {\bf 59},  315 (1969).

\bibitem {kos70} J. B. Kogut and D. E. Soper,  
Phys. Rev. {\bf D1}, 2901(1970).

\bibitem {kos73}  J. B. Kogut and L. Susskind,  
Phys. Rep.{\bf  C8}, 75(1973).

\bibitem{kss96} S. Kojima, N. Sakai, T. Sakai, 
Prog.~Theor.~Phys. {\bf  95},  621-636 (1996).  

\bibitem {kon69} J. Kondo,  
Solid State Physics {\bf  23},  183 (1969).

\bibitem{kou95} V.G. Koures, 
Phys.~Lett. {\bf B348}, 170-177 (1995).   

\bibitem {kpw92} M.~Krautg\"artner, H.C.~Pauli and F.~W\"olz,
Phys. Rev. {\bf D45}, 3755 (1992). 

\bibitem {kri92} A. D. Krisch,   
Nucl. Phys. B (Proc. Suppl.) {\bf  25B},   285 (1992).

\bibitem {kww80}  H. R. Krishnamurthy,  J. W. Wilkins,  
and K. G. Wilson,Phys. Rev. B {\bf  21,}  1003 (1980).

\bibitem {krg87}  H. Kr\"oger and  R. Girard,  
and G. Dufour, Phys. Rev. D {\bf 35},  3944 (1987).

\bibitem{krp93} H. Kr\"oger, H.C. Pauli, 
Phys.~Lett. {\bf B319}, 163-170 (1993).  

\bibitem {krn91}  A.S.~Kronfeld  and B.~Nizic, 
Phys. Rev. {\bf D44},  3445 (1991).

\bibitem{kru93} S. Krusche, 
Phys.~Lett. {\bf B298}, 127-131 (1993).  

\bibitem {kmr88} W. Kwong, P. B. Mackenzie, 
R. Rosenfeld and J. L. Rosner 
Phys. Rev. {\bf D37}, 3210 (1988). 

\bibitem {lal51} L.D.~Landau and E.M.~Lifshitz,  
{\it Classical Theory of Fields},   
Addison-Wesley, Reading, Mass. (1951). 

\bibitem {lal60} L.D.~Landau and E.M.~Lifshitz,  
{\it Electrodynamics of Continuous Media},   
Addison-Wesley, Reading, Mass. (1960). 

\bibitem {lan74} P.V. Landshoff, 
 Phys. Rev. {\bf D10},  10241 (1974).

\bibitem {las92} E.~Langmann and G.W.~Semenoff, 
Phys. Lett.{\bf B296} 117 (1992).  

\bibitem{lab93a} A. Langnau, S.J. Brodsky, 
J.~Comput.~Phys. {\bf  109}, 84-92 (1993).  

\bibitem{lab93b} A. Langnau, M. Burkardt, 
Phys.~Rev. {\bf D47}, 3452-3464 (1993).  

\bibitem {lee54}  T.D. Lee,  
Phys. Rev. {\bf  95},  1329 (1954).

\bibitem {len90} F. Lenz,  in 
{it Nonperturbative Quantum Field Theory},  
D. Vautherin,  F. Lenz,  and J.W.Negele, Eds., 
Plenum Press,  N. Y.  (1990).

\bibitem {ltl91}  F. Lenz,  M. Thies,  S. Levit  and K. Yazaki, 
Ann. Phys.{\bf 208},  1 (1991).

\bibitem {leb79a}  G.P. Lepage and S.J. Brodsky, 
 Phys. Lett. {\bf 87B}, 359 (1979)

\bibitem {leb79b}  G. P. Lepage and S. J. Brodsky,  
Phys. Rev. Lett. {\bf 43}, 545,  1625(E) (1979).

\bibitem {leb80}  G.P. Lepage and S.J. Brodsky,   
Phys. Rev. {\bf D22}, 2157 (1980). 

\bibitem {lbh83}  G. P. Lepage,  S. J. Brodsky,  
T. Huang and P.B.Mackenzie, 
in {\it  Particles and Fields 2} ,  A. Z. Capri and A.N.Kamal,
 Eds. (PlenumPress,  New York,  1983)

\bibitem {let87} G. P. Lepage and B. A. Thacker,
CLNS-87, (1987).

\bibitem {leu68}  H. Leutwyler, 
Acta Phys. Austriaca,  Suppl. {\bf 5} , 320 (1968).

\bibitem {leu69} H. Leutwyler, 
in Springer Tracts in Modern Physics, No. 50,  
(Springer,  New York),  p. 29 (1969).

\bibitem {leu74a} H. Leutwyler, 
Phys. Lett.  {\bf  B48},  45 (1974).

\bibitem {leu74b}  H. Leutwyler,  
Phys. Lett. {\bf  B48},  431 (1974).

\bibitem {leu74c} H. Leutwyler,  
Nucl. Phys. {\bf  B76},  413 (1974).

\bibitem {les78} H. Leutwyler and J. Stern,    
Annals of Physics {\bf 112},  94 (1978).

\bibitem {lis92}  H.~Li and G.~Sterman, 
Nucl. Phys. {\bf  B381}, 129 (1992). 

\bibitem{lib95a} N.E. Ligterink, B.L.G. Bakker,
`
Phys.~Rev. {\bf D52}, 5917-5925,1995 

\bibitem{lib95b} N.E. Ligterink, B.L.G. Bakker,
Phys.~Rev. {\bf D52}, 5954-5979 (1995). 

\bibitem {los71}  J. Lowenstein and A. Swieca,  
Ann. Phys.(N.Y.) {\bf  68}, 172 (1971).

\bibitem {lsg91} 
      W.~Lucha,  F.F.~Sch\"oberl,  and D.~Gromes, 
      Phys.~Rept. {\bf  200}, 127 (1991). 

\bibitem {lms92} M. Luke,  A. V. Manohar, M. J. Savage, 
Phys. Lett. {\bf B288}, 355 (1992).

\bibitem {lus83} M. L\"uscher, 
Nucl.Phys. {\bf B219} 233(1983).

\bibitem {mah89} Y. Ma and J.R. Hiller,  
J. Comp. Phys. {\bf  82}, 229 (1989).

\bibitem {ma91} B.~Q. Ma,  J.~Phys.  {\bf G17}  L53 (1991).

\bibitem {maz93} B.~Q. Ma and Qi-Ren Zhang,  
Z.~Phys. {\bf C58} 479 (1993).

\bibitem{mab93} B.Q. Ma,  
Z.~Physik {\bf A345}, 321-325 (1993).  

\bibitem{mab97} B.Q. Ma,  
``The proton spin structure in a light cone quark spectator
diquark model.'' hep-ph/9703425.  

\bibitem {mae94} N. Makins, et. al. , 
NE-18 Collaboration. MIT preprint (1994).

\bibitem {man85}  N.S.~Manton, 
Ann.Phys.(N.Y.) {\bf 159} 220 (1985) .

\bibitem {mas89} G. Martinelli and C. T. Sachrajda, 
Phys. Lett. {\bf B217}, 319 (1989).

\bibitem {max54} J.C.~Maxwell,  
{\it Treatise on Electricity and Magnetism}, 3rd ed., 2 Vols.,  
reprint by Dover, New York (1954).

\bibitem {may76}  T. Maskawa  and K. Yamawaki,  
Prog. Theor. Phys. {\bf 56},  270 (1976).

\bibitem {mcc88} G. McCartor,  
Z. Phys. {\bf C41},  271 (1988).

\bibitem {mcc91}  G. McCartor,  
Z. Phys. {\bf  C52},  611 (1991).

\bibitem {mcr92} G.~McCartor and D.G.~Robertson,
Z. Phys. {\bf C53} (1992) 679.

\bibitem{mcr94} G. McCartor, D.G. Robertson, 
Z.~Physik {\bf C62}, 349-356 (1994).  

\bibitem{mcc94} G. McCartor, 
Z.~Physik {\bf C64}, 349-354 (1994).  

\bibitem{mcr95} G. McCartor, D.G. Robertson, 
Z.~Physik {\bf C68}, 345-351 (1995).  

\bibitem {mcr97} G. McCartor, D. Robertson, and S. Pinsky, 
"Vacuum Structure of Two-Dimensional
Gauge Theories on the Light Front", hep-th/96112083.

\bibitem {mel74}  H.J. Melosh,        
Phys. Rev. {\bf  D9},1095 (1974).

\bibitem {mes62} A. Messiah,    
{\it Quantum Mechanics}, 
2 Vols., North Holland, Amsterdam, 1962.

\bibitem{mil97} G.A.~Miller,
``A light front treatment of the nucleus implications for deep
inelastic scattering.'' 
SLAC-PUB-7412, Feb 1997. 11pp. nucl-th/9702036.

\bibitem {mip94} J.A.~Minahan and A.P.~Polychronakos,  
Phys.Lett. {\bf B326}288 (1994).

\bibitem{mis96} A. Misra, 
Phys.~Rev. {\bf D53}, 5874-5885 (1996).   

\bibitem {mof53} P.M.~Morse and H.~Feshbach,   
{\it Methods of Theoretical Physics}, 2Vols.,  
McGraw-Hill, New York (1953).

\bibitem {mop93}  Y. Mo and R.J. Perry,  
J. Comp. Phys. {\bf 108}, 159 (1993).

\bibitem {mps91} D. Mustaki,  S. Pinsky,  
J. Shigemitsu and K. Wilson, 
Phys.Rev.{\bf D43},  3411 (1991).

\bibitem {mup92}  D. Mustaki and S. Pinsky,  
Phys. Rev. {\bf D45}, (1992).

\bibitem {mus94}  D. Mustaki,  
{\it  Chiral symmetry and the constituent quark model: 
A null-plane point of view},  
Bowling Green State Univ. preprint (1994).

\bibitem {mut87} T.~Muta, 
{\it Foundations of Quantum Chromodynamic}.  
Lecture notes in Physics -- Vol.5, 
World Scientific, Singapore, 1987.

\bibitem {nac86} O.~Nachtmann, 
{\it Elementarteilchenphysik}.  Vieweg,Braunschweig, 1986. 

\bibitem {ntt95} T. Nakatsu, K. Takasaki, and S. Tsujimaru,
Nucl. Phys. {\bf  B443},  155-200 (1995).

\bibitem {nam84}  J. M. Namyslowski,  
Prog. Part. Nuc.Phys. {\bf  74}, 1(1984)

\bibitem {noe18} E.~Noether, 
Kgl.~Ges.~d.~Wiss. Nachrichten,   
Math.-phys. Klasse, G\"ottingen (1918).

\bibitem {nak83}  Y. Nakawaki,  
Prog. Theor.  Phys. {\bf 70},  1105 (1983).

\bibitem {npf97} H.W.L. Naus, H.J. Pirner, T.J.Fields, and J.P. Vary,
``QCD near the light cone,''
hep-th/9704135 

\bibitem {otf95} A. Ogura, T. Tomachi, and T. Fujita,
Annals Phys. {\bf 237}, 12-45 (1995). 

\bibitem {osb74}  H. Osborn,    
 Nucl. Phys. {\bf  B80},  90 (1974).

\bibitem {pdg92} Particle~Data~Group,  
Phys. Rev. {\bf D45},  Part 2,  1 (1992).

\bibitem {pau81}
        H.C.~Pauli,
        Nucl.Phys. {\bf A396}, 413 (1981). 

\bibitem {pau84} H.C.~Pauli,  
Z. Phys. {\bf A319},  303 (1984). 

\bibitem {pab85a}  H. C. Pauli and S. J. Brodsky,      
Phys. Rev. {\bf D32}, 1993 (1985).

\bibitem {pab85b}  H. C. Pauli and S. J. Brodsky,     
Phys. Rev. {\bf D32}, 2001 (1985).

\bibitem {pau92} H.C. Pauli,  
Nucl. Phys. {\bf A560}, 501 (1993).

\bibitem {pau93} H.C. Pauli,    
{\it The Challenge of Discretized Light-Cone Quantization},   
in: ``Quantum Field Theoretical Aspects of High Energy Physics'',  
B.~Geyer and E.~M.~Ilgenfritz, Eds.,
Natur\-wissen\-schaftlich\-Theoretisches Zentrum der 
Universit\"at Leipzig, 1993.

\bibitem {pkp95} H.C.~Pauli, A.C.~Kalloniatis and S.S.~Pinsky,
Phys.~Rev. {\bf D52}, 1176 (1995).

\bibitem {pam95} H.C.~Pauli and J.~Merkel, 
     Phys. Rev. {\bf D55}, 2486-2496 (1997).

\bibitem {pab96} H.C.~Pauli and R.~Bayer, 
    Phys.Rev. {\bf D53}, 939 (1996).
   
\bibitem {pau96} H.C. Pauli,          
    ``Solving Gauge Field Theory by 
    Discretized Light-Cone Quantization'',
    Heidelberg Preprint MPIH-V25-1996, 
    hep-th/9608035.

\bibitem {pau96a} 
      H.C. Pauli,  
      {\it Discretized Light-Cone Quantization,}
      in: {\it  Neutrino mass, Monopole Condensation,
      Dark matter, Gravitational waves, and Light-Cone
      Quantization,}
      B.N. Kursunoglu, S. Mintz, and A. Perlmutter, Eds.,
       Plenum Press, New York, 1996; p.183-204.

\bibitem {phw90}  R. J. Perry,  A. Harindranath  and K. G. Wilson, 
Phys. Rev. Lett. {\bf  65},  2959 (1990).

\bibitem {peh91} R.J.~Perry and A.~Harindranath,
Phys. Rev. {\bf D43} (1991) 4051.

\bibitem {per93a}  R. J. Perry,  
Phys. Lett. {\bf  300B},  8 (1993).

\bibitem {pew93}  R. J. Perry and K. G. Wilson,  
Nucl. Phys.{\bf  B403}, 587(1993).

\bibitem{per94a} R. J. Perry
"Hamiltonian Light-Front Field Theroy and Quantum Chromodynamics"
{\it Hadron 94}, Gramado, Brasil, April 1994

\bibitem {per94b}  R.J. Perry,  
Ann. Phys. {\bf  232},  116-222 (1994).

\bibitem {per90}  V. N. Pervushin, 
Nucl. Phys. {\bf B15},  197 (1990).

\bibitem {pes95a} I. Pesando, 
Mod.Phys.Lett. {\bf A10}, 525-538 (1995). 

\bibitem {pes95b} I. Pesando, 
Mod.Phys.Lett. {\bf A10}, 2339-2352 (1995). 

\bibitem{pin91a} S. S. Pinsky
``Null-Plane QED" Proceedings of the 4th Conference on the
Intersections between Particle and Nuclear Physics. 
Tucson, AZ, May 1991, World Scientific
(1991). 

\bibitem{pin91b} S.S. Pinsky,
``The Tamm-Dancoff Method in Gauge Field Theory"
Proceeding of Division of Particles and Fields of the APS
Vancouver B.C. Canada, Aug. 18-22 1991, World Scientific 

\bibitem{pin92} S.S. Pinsky,
``Renormalization of Light-Cone Tamm-Dancoff Integral Equations"
Proceeding of the 27th Recontre de Moriond, Les Arcs, Savoie, France
March 22-28 1992, Editions Frontieres Ed. J. Tran Thanh Van.

\bibitem{pin93a} S.S. Pinsky,
``Spontaneous Symmetry Breaking on the Light-Cone"
Proceeding of Orbis Scientiae 1993 Jan. 25-27 1993 Coral Gables FL.
Nova Science.

\bibitem{pin93b} S.S. Pinsky,
``The Light-Cone Field Theory Paradigm for Spontaneous Symmetry Breaking."
Proceedings of ``Hadron Structure '93"
Banska Stiavnica Slovakia, Sept. 5-10, 1993
Institute of Physics, Slovak Academy of Science 
Ed. by S. Dubnicka and
A. Dubnickova.  

\bibitem{pin94} S.S. Pinsky, 
``Topology and Confinement In Light-Front QCD"
Proceedings of ``Theory of Hadrons and Light-Front QCD"
Polona Zgorselisko Poland, Aug. 1994. Published by World Scientific, Ed.
St. Glazek.

\bibitem{pin96} S. S. Pinsky
``Dirac's Legacy: Light-Cone Quantization"
Proceeding of Orbis Scientiae 1996 Jan. 25-28 1996 Coral Gables FL.
Nova Science.

\bibitem{pim96} S. Pinsky and R. Mohr,
"The condensate of SU(2) Yang -Mills Theory in 1+1
Dimensions Coupled to Massless Adjoint Fermions", to appear in the 
proceedings of The Conference on Low Dimensional Field Theory, 
Telluride Summer Research Institute, August 1996, International 
Journal of Modern Physics A.

\bibitem{pin97a} S. Pinsky,
 "Wilson loop on a Light-cone cylinder", 
hep-th/9702091.

\bibitem{pin97b} S. Pinsky,
"(1+1)-Dimensional Yang-Mills Theory Coupled to Ajoint Fermions 
on the Light Front," hep-th/9612073.

\bibitem {piv94}  S.S. Pinsky,  and B. van de Sande, 
Phys. Rev. {\bf  D49},  2001 (1994).

\bibitem{pik96}  S.S. Pinsky  and  A.C. Kalloniatis,
Phys. Lett. {\bf  B365}, 225-232 (1996).

\bibitem{pir96} S.S. Pinsky  and D. G. Robertson
Phys. Lett. {\bf  B379}, 169-178 (1996).

\bibitem {prf89} E. V. Prokhvatilov and V. A. Franke,  
Sov. J. Nucl. Phys. {\bf 49},  688 (1989)

\bibitem {pnk96} J. Przeszowski, H.W.L. Naus,  A.C. Kalloniatis, 
Phys.Rev. {\bf D54}, 5135-5147 (1996).

\bibitem {raj92} S.G. Rajeev, 
Phys. Lett.  {\bf  B 212} 203 (1988).

\bibitem {rom92} D.G.~Robertson and G.~McCartor, 
Z. Phys. {\bf C53}, 661(1992).

\bibitem {rob93} D.G. Robertson,    
Phys. Rev. {\bf  D47}, 2549-2553 (1993).

\bibitem {roh71}  F. Rohrlich,  
Acta Phys. Austriaca,  Suppl. VIII, 2777(1971).

\bibitem{rud94} R.E. Rudd, 
Nucl.~Phys. {\bf B427}, 81-110 (1994).   

\bibitem {saw85}  M. Sawicki,   
Phys. Rev. {\bf  D32},  2666 (1985). 

\bibitem {saw86}  M. Sawicki,   
Phys. Rev. {\bf  D33},  1103 (1986). 

\bibitem {sas75}  H. Sazdjian  and J. Stern,  ,  
Nucl. Phys. B {\bf 94}, 163 (1975). 

\bibitem {sch93a} F. Schlumpf,    
Phys. Rev. {\bf D47}, 4114 (1993).    

\bibitem {sch93b} F. Schlumpf,    
Phys. Rev.  {\bf D48} 4478 (1993).

\bibitem {sch93c} F. Schlumpf,    
Mod.  Phys. Lett.  {\bf A8}  2135 (1993).    

\bibitem {sch94} F. Schlumpf,    
J. Phys. {\bf G20}, 237 (1994).

\bibitem{scb97} 
N.C.J. Schoonderwoerd, B.L.G. Bakker,
``Equivalence of renormalized covariant and light front
   perturbation theory.'' 
Feb 1997.11pp. hep-ph/9702311. 

\bibitem {sch61}   S. Schweber,  
{\it Relativistic Quantum Field Theory},   
Harper and Row, New York (1961).  

\bibitem {sch62a}  J. Schwinger,   
Phys. Rev. {\bf  125},  397 (1962). 

\bibitem {sch62b}  J. Schwinger,   
Phys. Rev. {\bf  128},  2425 (1962).

\bibitem {shv79}  M.A. Shifman,  
A.I. Vainshtein and V.I. Zakharov,
Nucl. Phys. {\bf B147},  38 (1979).

\bibitem{sim96} S. Simula 
Phys.~Lett. {\bf B373}, 193-199 (1996).   
 
\bibitem {som73} C.M.~Sommerfield,   
{\it Quantization on space-time Hyperboloids},   
Yale preprint (July, 1973).   

\bibitem {sop71} D.E.~Soper, Ph.D Thesis (1971),
SLAC Report No. 137, 1971.

\bibitem {sos94} M. G. Sotiropoulos and G. Sterman, 
Nucl. Phys. {\bf B425}, 489 (1994). 

\bibitem{sri94} P.P. Srivastava,
Nuovo~Cim. {\bf  107A}, 549-558 (1994).  

\bibitem {sto91} P. Stoler, 
Phys. Rev. Lett. {\bf 66}, 1003 (1991).

\bibitem{smy94} T. Sugihara, M. Matsuzaki, M.Yahiro,
Phys.~Rev. {\bf D50}, 5274-5288 (1994).  

\bibitem {suy96} T.~Sugihara, M.~Yahiro,
Phys.Rev. {\bf  D53}, 7239-7249 (1996). 

\bibitem {suy97} T.~Sugihara, M.~Yahiro,
Phys.Rev. {\bf  D55}, 2218-2226 (1997). 

\bibitem {sun82}  K. Sundermeyer,  
{\it Constrained Dynamics}, 
(Springer-Verlag, New York,  1982).

\bibitem {suf67}  L. Susskind and  G. Frye,  
Phys. Rev. {\bf 164},  2003 (1967).

\bibitem {sus68}  L. Susskind,  
Phys. Rev. {\bf 165},  1535 (1968). 

\bibitem {sus97}  L. Susskind, 
``Another conjecture about M(atrix) Theory'', 
Stanford preprint SU-ITP-97-11, Apr. 1997,
 hep-th/9704080; hep-th/9611164.

\bibitem {sul80}  K. Suzuki and S. Y. Lee,  
Prog. Theor. Phys. {\bf  64}, 2091 (1980).

\bibitem{shb90}
A. Szczepaniak, E. M. Henley and S. J. Brodsky, 
Phys.Lett. {\bf B243}, 287 (1990).

\bibitem{szm91}
A. Szczepaniak and L. Mankiewicz, 
Univ. of Florida Preprint (1991).

\bibitem {thy95}  A. Tam, C.J. Hamer, and C.M. Yung,
J.Phys. {\bf G21}, 1463-1482 (1995). 

\bibitem {tam45} I. Tamm,  
J. Phys. (USSR) {\bf  9},  449 (1945). 

\bibitem {tbp91}  A.C. Tang,  S.J. Brodsky,  and H.C. Pauli,     
Phys. Rev. {\bf D44},  1842 (1991). 

\bibitem {tac95} M.~Tachibana,
Phys.Rev. {\bf  D52}, 6008-6015 (1995). 

\bibitem {tho75} G.~'t~Hooft,
Published in Erice Subnucl. Phys. 1975:261. 

\bibitem{tho93} M. Thies, K. Ohta, 
Phys.~Rev. {\bf D48}, 5883-5894 (1993).  

\bibitem {tho79a}  C. B. Thorn,   
Phys. Rev. {\bf  D19},  639 (1979).

\bibitem {tho79b}  C. B. Thorn,   
Phys. Rev. {\bf D20}, 1934 (1979). 

\bibitem {tof91} T. Tomachi and  T. Fujita,
``Massive Schwinger model: 1 / K expansion method vs. Bogolyubov
transformation,'' 
NUP-A-91-10, Sep 1991. 45pp. 

\bibitem {trp96} U.~Trittmann and H.C.~Pauli,
``Quantum electrodynamics at strong coupling'',
Heidelberg preprint MPI H-V4-1997, Jan. 1997.

\bibitem {trp97a} U.~Trittmann and H.C.~Pauli,
``On rotational invariance in front form dynamics'',
Heidelberg preprint MPI H-V7-1997, Apr. 1997.

\bibitem {trp97b} U.~Trittmann and H.C.~Pauli,
``On the role of the annihilation channel in front front positronium'',
Heidelberg preprint MPI H-V17-1997, Apr. 1997.

\bibitem {tun68} W.K. Tung,  
Phys. Rev. {\bf 176}, 2127 (1968).

\bibitem {tun85} W.K.~Tung,   
{\it Group Theory in Physics},   
(World Scientific, Singapore, 1985).

\bibitem {van92} P. van Baal, 
Nucl.Phys. {\bf B369} 259 (1992).  

\bibitem {vap92}  B. van de Sande and S. Pinsky,  
Phys. Rev. {\bf  D46}, 5479 (1992).

\bibitem {vap95}     B.~van~de~Sande and S.S.~Pinsky, 
Phys. Rev.{\bf D49}  2001 (1994).

\bibitem {vab96}  B.~van~de~Sande,  and M.~Burkardt,
   Phys.Rev. {\bf D53}, 4628 (1996). 

\bibitem {van96}  B.~van~de~Sande, 
Phys.Rev.{\bf D54} 6347 (1996). 
 
\bibitem {vad96}  B.~van~de~Sande,  and S.~Dalley,
{ \it The transverse lattice in 2+1 dimensions} 
The proceedings of Orbis Scientiae: Neutrino
Mass, Dark Matter, Gravitational Waves, Condensation of Atoms and
Monopoles, Light-Cone Quantization, Miami Beach, FL, 25-28 Jan 1996. .
 
\bibitem {swh93} J.B. Swenson, and J.R.~Hiller,
Phys. Rev. {\bf D48}, 1774 (1993).

\bibitem{tsy97} S.~Tsujimaru, and K.~Yamawaki,
   ``Zero mode and symmetry breaking on the light front,''
   MPI H-V19-1997, Apr 1997. 47pp. hep-th/9704171.

\bibitem {vfp94} J.P. Vary, T.J. Fields, and H.J. Pirner, 
``Massive Schwinger model for small fermion mass,''
ISU-NP-94-14, Aug 1994. 5pp. hep-ph/9411263 

\bibitem {vfp96} J.P. Vary, T.J. Fields, and H.-J. Pirner,
Phys.Rev. {\bf  D53}, 7231-7238 (1996). 

\bibitem {weg72a}  F.J. Wegner,   
Phys. Rev. {\bf  B5}, 4529 (1972).

\bibitem {weg72b}  F.J. Wegner,   
Phys. Rev. {\bf  B6},  1891 (1972).

\bibitem {weg76}  F.J. Wegner,  in 
{\it  Phase Transitions and CriticalPhenomena},
C. Domb and M.S. Green,  Eds.,  Vol. 6 
(AcademicPress,  London, 1976).

\bibitem {wei66}  S. Weinberg,   
Phys. Rev. {\bf  150},  1313 (1966).

\bibitem {wey29} H.~Weyl, 
Z. Phys. {\bf 56} (1929) 330.

\bibitem {wig39} E. Wigner,  
Ann. Math. {\bf 40},  149 (1939).

\bibitem {wil65}  K.G. Wilson,   
Phys. Rev. {\bf  140},  B445 (1965).

\bibitem {wil70}  K. G. Wilson,   
Phys. Rev. {\bf  D2},  1438 (1970).

\bibitem {wil75}  K. G. Wilson,     
Rev.    Mod. Phys. {\bf  47},  773 (1975).

\bibitem {wil76}  K.G. Wilson,    in 
{\it  New Pathways in High Energy Physics},  
A.Perlmutter,  ed. (Plenum Press,  New York,  1976), Vol. II, pp.243-264.

\bibitem {wil89}  K. Wilson,  in {\em Lattice '89},  
Proceedings ofthe International Symposium,  Capri,  Italy, 
1989,  edited by R. Petronzio{\em et al} .
[Nucl. Phys. B (Proc. Suppl.) {\bf  17},  (1989)].

\bibitem {wwh94}  K. G. Wilson,  T. S. Walhout,  A. Harindranath, 
W.-M.Zhang,R. J. Perry,  St. D. G{\l}azek, 
Phys. Rev.{\bf  D49}, 6720-6766 (1994).

\bibitem {wit89}  R. S. Wittman, in 
{\em Nuclear and Particle Physics on the Light Cone} , 
M. B. Johnson and L. S. Kisslinger, Eds.
(World Scientific, Singapore,  1989).

\bibitem {wih93} J.J.~Wivoda and J.R.~Hiller,
Phys. Rev. {\bf D47},  4647 (1993).

\bibitem {wor92}  P.M. Wort,  
Carleton University preprint, February 1992.

\bibitem {yuh91}  C.M. Yung and C.J. Hamer,  
Phys. Rev. {\bf D44}, 2598 (1991).

\bibitem {zah93a}  W. M. Zhang and A. Harindranath,  
Phys. Rev. {\bf  D48}, 4868-4880 (1993).

\bibitem {zah93b}  W. M. Zhang and A. Harindranath,  
Phys. Rev. {\bf  D48}, 4881 (1993).

\bibitem {zah93c}  W. M. Zhang and A. Harindranath,  
Phys. Rev. {\bf  D48}, 4903  (1993).

\bibitem{zha94} W.M.~Zhang, 
Phys.~Lett. {\bf B333}, 158-165 (1994). 

\bibitem{zlc96} W.M.~Zhang, G.L.~Lin, C.Y.~Cheung, 
Int.~J.~Mod.~Phys. {bf  A11},  3297-3306 (1996).  

\bibitem {zvb93} 
        D.C.~Zheng, J.P.~Vary, and B.R.Barret,
       Nucl.Phys. {\bf A560}, 211 (1993). 

\bibitem {zhi85} A.R. Zhitnitskii,    
Phys. Lett. {\bf 165B}, 405 (1985).

\bibitem {zwe64}  G. Zweig,  
CERN Reports Th. 401 and 412 (1964), 
and in Proc. Int. School of Phys. 
``Ettore Majorana'',  Erice, Italy (1964), 
A.Zichichi,Ed.,  p.192 (Academic,  New York).

\end{thebibliography}
